\RequirePackage[l2tabu]{nag}		% Warns for incorrect (obsolete) LaTeX usage
\documentclass[a4paper,11pt,reqno,openbib,oldfontcommands]{memoir}

\usepackage{datetime}
\usepackage{ifpdf}
\ifpdf
\fi
\ifdraftdoc 
	\usepackage{draftwatermark}				%Sets watermarks up.
	\SetWatermarkScale{0.3}
	\SetWatermarkText{\bf Draft: \today}
\fi
\newsubfloat{figure}
\newsubfloat{table}
\settrimmedsize{297mm}{210mm}{*}
\settypeblocksize{634pt}{448.13pt}{*} 
\setulmargins{4cm}{*}{*} 
\setlrmargins{*}{*}{1} %1 is equal odd and even pages. Good for PDF reading. 1.5 is flushed right even pages and left odd pager, good for bookprinting.
\setmarginnotes{17pt}{51pt}{\onelineskip} 
\setheadfoot{\onelineskip}{2\onelineskip} 
\setheaderspaces{*}{2\onelineskip}{*} 
\checkandfixthelayout

\usepackage{fouriernc}
\usepackage[T1]{fontenc}
\OnehalfSpacing 
\usepackage[font=footnotesize]{caption}

\usepackage{pdfpages} %Needed to include PDF front cover

\setsecnumdepth{subsection} 
\usepackage{calc,soul,fourier}
\makeatletter 
\newlength\dlf@normtxtw 
\newsavebox{\feline@chapter} 
\newcommand\feline@chapter@marker[1][4cm]{%
	\sbox\feline@chapter{% 
		\resizebox{!}{#1}{\fboxsep=1pt%
			\colorbox{gray}{\color{white}\thechapter}% 
		}}%
		\rotatebox{90}{% 
			\resizebox{%
				\heightof{\usebox{\feline@chapter}}+\depthof{\usebox{\feline@chapter}}}% 
			{!}{\scshape\so\@chapapp}}\quad%
		\raisebox{\depthof{\usebox{\feline@chapter}}}{\usebox{\feline@chapter}}%
} 
\newcommand\feline@chm[1][4cm]{%
	\sbox\feline@chapter{\feline@chapter@marker[#1]}% 
	\makebox[0pt][c]{% aka \rlap
		\makebox[1cm][r]{\usebox\feline@chapter}%
	}}
\makechapterstyle{daleifmodif}{

	\renewcommand\printchapternum{\null\hfill\feline@chm[2.5cm]\par}

} 
\makeatother 
\chapterstyle{daleifmodif}

\usepackage{emptypage}
\makepagestyle{myvf} 
\makeoddfoot{myvf}{}{\thepage}{} 
\makeevenfoot{myvf}{}{\thepage}{} 
\makeheadrule{myvf}{\textwidth}{\normalrulethickness} 
\makeevenhead{myvf}{\small\textsc{\leftmark}}{}{} 
\makeoddhead{myvf}{}{}{\small\textsc{\rightmark}}
\newcommand{\clearemptydoublepage}{\newpage{\thispagestyle{empty}\cleardoublepage}}
\makeindex
\usepackage{import}

\usepackage{lipsum}					%Needed to create dummy text
\usepackage{amsfonts} 					%Calls Amer. Math. Soc. (AMS) fonts
\usepackage[centertags]{amsmath}			%Writes maths centred down
\usepackage{stmaryrd}					%New AMS symbols
\usepackage{amssymb}					%Calls AMS symbols
\usepackage{amsthm}					%Calls AMS theorem environment
\usepackage{newlfont}					%Helpful package for fonts and symbols
\usepackage{layouts}					%Layout diagrams
\usepackage{graphicx}					%Calls figure environment
\usepackage{longtable,rotating}			%Long tab environments including rotation. 
\usepackage[utf8]{inputenc}			%Needed to encode non-english characters 
\usepackage{colortbl}					%Makes coloured tables
\usepackage{wasysym}					%More math symbols
\usepackage{mathrsfs}					%Even more math symbols
\usepackage{float}						%Helps to place figures, tables, etc. 
\usepackage{verbatim}					%Permits pre-formated text insertion
\usepackage{upgreek }					%Calls other kind of greek alphabet
\usepackage{latexsym}					%Extra symbols
\usepackage{url}						%Supports url commands
\usepackage[spanish,english]{babel}		%For languages characters and hyphenation
\usepackage{color}                    				%Creates coloured text and background

\usepackage{memhfixc}					%Must be used on memoir document 
\usepackage{enumerate}					%For enumeration counter
\usepackage{footnote}					%For footnotes
\usepackage{microtype}					%Makes pdf look better.
\usepackage{rotfloat}					%For rotating and float environments as tables, 
\usepackage{alltt}						%LaTeX commands are not disabled in 
\usepackage[version=0.96]{pgf}			%PGF/TikZ is a tandem of languages for producing vector graphics from a 
\usepackage{tikz}						%geometric/algebraic description.
\usetikzlibrary{arrows,shapes,snakes,
		       automata,backgrounds,
		       petri,topaths}				%To use diverse features from tikz		
\newcommand{\pgftextcircled}[1]{                                                                    %Defines encircled text
    \setbox0=\hbox{#1}%
    \dimen0\wd0%
    \divide\dimen0 by 2%
    \begin{tikzpicture}[baseline=(a.base)]%
        \useasboundingbox (-\the\dimen0,0pt) rectangle (\the\dimen0,1pt);
        \node[circle,draw,outer sep=0pt,inner sep=0.1ex] (a) {#1};
    \end{tikzpicture}
}

\newcommand{\blackged}{\hfill$\blacksquare$}
\newcommand{\whiteged}{\hfill$\square$}
\newcounter{proofcount}

\let\oldsqrt\sqrt
\def\sqrt{\mathpalette\DHLhksqrt}
\def\DHLhksqrt#1#2{%
\setbox0=\hbox{$#1\oldsqrt{#2\,}$}\dimen0=\ht0
\advance\dimen0-0.2\ht0
\setbox2=\hbox{\vrule height\ht0 depth -\dimen0}%
{\box0\lower0.4pt\box2}}
\newcommand{\mycaption}[2][\@empty]{
	\captionnamefont{\scshape} 
	\changecaptionwidth
	\captionwidth{0.9\linewidth}
	\captiondelim{.\:} 
	\indentcaption{0.75cm}
	\captionstyle[\centering]{}
	\setlength{\belowcaptionskip}{10pt}
	\ifx \@empty#1 \caption{#2}\else \caption[#1]{#2}
}
\newcommand{\mysubcaption}[2][\@empty]{
	\subcaptionsize{\small}
	\hangsubcaption
	\subcaptionlabelfont{\rmfamily}
	\sidecapstyle{\raggedright}
	\setlength{\belowcaptionskip}{10pt}
	\ifx \@empty#1 \subcaption{#2}\else \subcaption[#1]{#2}
}

\usepackage{wrapfig}
\usepackage{parskip} %Used so that double line breaks with enter yield a separated new paragraph without need of \\ or similar
\usepackage{indentfirst}%Used to indent paragraph after section header
\usepackage{cite}%Group citations

\usepackage[linktocpage=true]{hyperref}%hyperlinks and formatting
\hypersetup{
    colorlinks,
    linkcolor={red!50!black},
    citecolor={blue!50!black},
    urlcolor={blue!80!black}
}

\usepackage{mathtools}
\usepackage{amssymb}%Needed for some mathematical symbols: mathbb
\usepackage{mathrsfs}%Needed for some mathematical symbols: mathscr
\usepackage{amsmath}%Needed for some mathematical environments like split
\usepackage{xcolor}%Needed for using varied colors in the text font
\usepackage{physics}%Braket notation

\newcommand{\eg}{{\it e.g. }}
\newcommand{\ie}{{\it i.e., }}

\newcommand{\cc}[1]{\mathcal{#1}}
\newcommand{\cm}{\mathcal{M}}
\newcommand{\cn}{\mathcal{N}}
\newcommand{\ct}{\mathcal{T}}
\newcommand{\cf}{\mathcal{F}}
\newcommand{\cd}{\mathcal{D}}
\newcommand{\cl}{\mathcal{L}}
\newcommand{\cR}{\ensuremath\mathcal{R}}
\newcommand{\cP}{\ensuremath\mathcal{P}}

\newcommand{\cS}{\ensuremath\mathcal{S}}
\newcommand{\cU}{\ensuremath\mathcal{U}}
\newcommand{\cB}{\ensuremath\mathcal{B}}
\newcommand{\cL}{\ensuremath\mathcal{L}}
\newcommand{\cD}{\ensuremath\mathcal{D}}
\newcommand{\cA}{\ensuremath\mathcal{A}}
\newcommand{\cZ}{\ensuremath\mathcal{Z}}
\newcommand{\cG}{\ensuremath\mathcal{G}}
\newcommand{\cF}{\ensuremath\mathcal{F}}
\newcommand{\cC}{\ensuremath\mathcal{C}}

\newcommand{\bb}[1]{\mathbb{#1}}
\newcommand{\rn}{\mathbb{R}^n}
\newcommand{\bbr}{\mathbb{R}}
\newcommand{\bbc}{\mathbb{C}}
\newcommand{\bbI}{\mathbb{I}}

\newcommand{\tpm}{\mathcal{T}_p\mathcal{M}}
\newcommand{\ctpm}{\mathcal{T}_p^\ast\mathcal{M}}
\newcommand{\tb}{\mathcal{T}\mathcal{M}}
\newcommand{\ctb}{\mathcal{T}^\ast\mathcal{M}}

\newcommand{\fr}{\textbf{e}}
\newcommand{\df}{\reflectbox{\textbf{e}}}

\newcommand{\dif}{\mathrm{d}}
\renewcommand{\div}{\mathrm{Div}}
\renewcommand{\dv}{{\dif V}}

\newcommand{\lr}[1]{\left(#1\right)}

\newcommand{\lrsq}[1]{\left[#1\right]}

\newcommand{\omg}{{{}^g\omega}}
\newcommand{\Gamg}{{{}^g\Gamma}}
\newcommand{\Gamh}{{{}^h\Gamma}}
\newcommand{\bpsi}{\ensuremath{\bar{\psi}}}

\newcommand{\al}{\ensuremath\alpha}
\newcommand{\be}{\ensuremath\beta}
\newcommand{\ga}{\ensuremath\gamma}
\newcommand{\Ga}{\ensuremath\Gamma}
\newcommand{\del}{\ensuremath\delta}
\newcommand{\la}{\ensuremath\lambda}
\newcommand{\om}{\ensuremath\omega}

\newcommand{\beq}{\begin{equation}}
\newcommand{\eeq}{\end{equation}}

\newcommand{\na}{\nabla}
\newcommand{\pa}[1]{\partial_{#1}}

\newcommand{\bL}{\ensuremath\tilde{L}}
\newcommand{\bGa}{\ensuremath\bar{\Gamma}}

\newcommand{\cO}{\ensuremath\mathcal{O}}

\newcommand{\Upsilonh}{\hat{\Upsilon}}
\newcommand{\Deltah}{\hat{\Delta}}
\newcommand{\hT}{h^\mathrm{T}}
\newcommand{\mpl}{\mathrm{M_P}}
\newcommand{\mg}{\mathrm{M_G}}
\newcommand{\mq}{\mathrm{M_Q}}
\newcommand{\mbi}{\mathrm{M_{BI}}}
\newcommand{\Omegah}{\hat{\Omega}}
\newcommand{\Id}{\mathbb 1}
\newcommand{\sq}{\bar{q}}

\definecolor{darkred}{rgb}{0.55, 0.0, 0.0}
\definecolor{darkmagenta}{rgb}{0.55, 0.0, 0.55}
\definecolor{lincolngreen}{rgb}{0.11, 0.35, 0.02}
\definecolor{mygreen}{rgb}{0.05, 0.35, 0.1}

\newcommand\dfR{\boldsymbol{R}}

\newcommand\cofr{\boldsymbol{e}}
\newcommand\dimM{D}
\newcommand\CosmC{\Lambda_0}
\newcommand\KMaxSym{\Lambda}
\newcommand{\bgar}{\bar\Gamma_\mathrm{Ric}}
\newcommand{\tgar}{\tilde\Gamma_\mathrm{Ric}}

\usepackage{lettrine}
\newcommand{\initial}[1]{%
	\lettrine[lines=3,lhang=0.33,nindent=0em]{
		\color{gray}
     		{\textsc{#1}}}{}}
\theoremstyle{plain}

\theoremstyle{plain}

\theoremstyle{plain}
\theoremstyle{definition}

\theoremstyle{plain}

\theoremstyle{plain}

\theoremstyle{plain}

\begin{document}
% UoB guidlines:
%
% Preliminary pages
% 
% The five preliminary pages must be the Title Page, Abstract, Dedication
% and Acknowledgements, Author's Declaration and Table of Contents.
% These should be single-sided.
% 
% Table of contents, list of tables and illustrative material
% 
% The table of contents must list, with page numbers, all chapters,
 % sections and subsections, the list of references, bibliography, list of
% abbreviations and appendices. The list of tables and illustrations
% should follow the table of contents, listing with page numbers the
% tables, photographs, diagrams, etc., in the order in which they appear
% in the text.
% 

\frontmatter
\pagenumbering{roman}
\includepdf[pages=-]{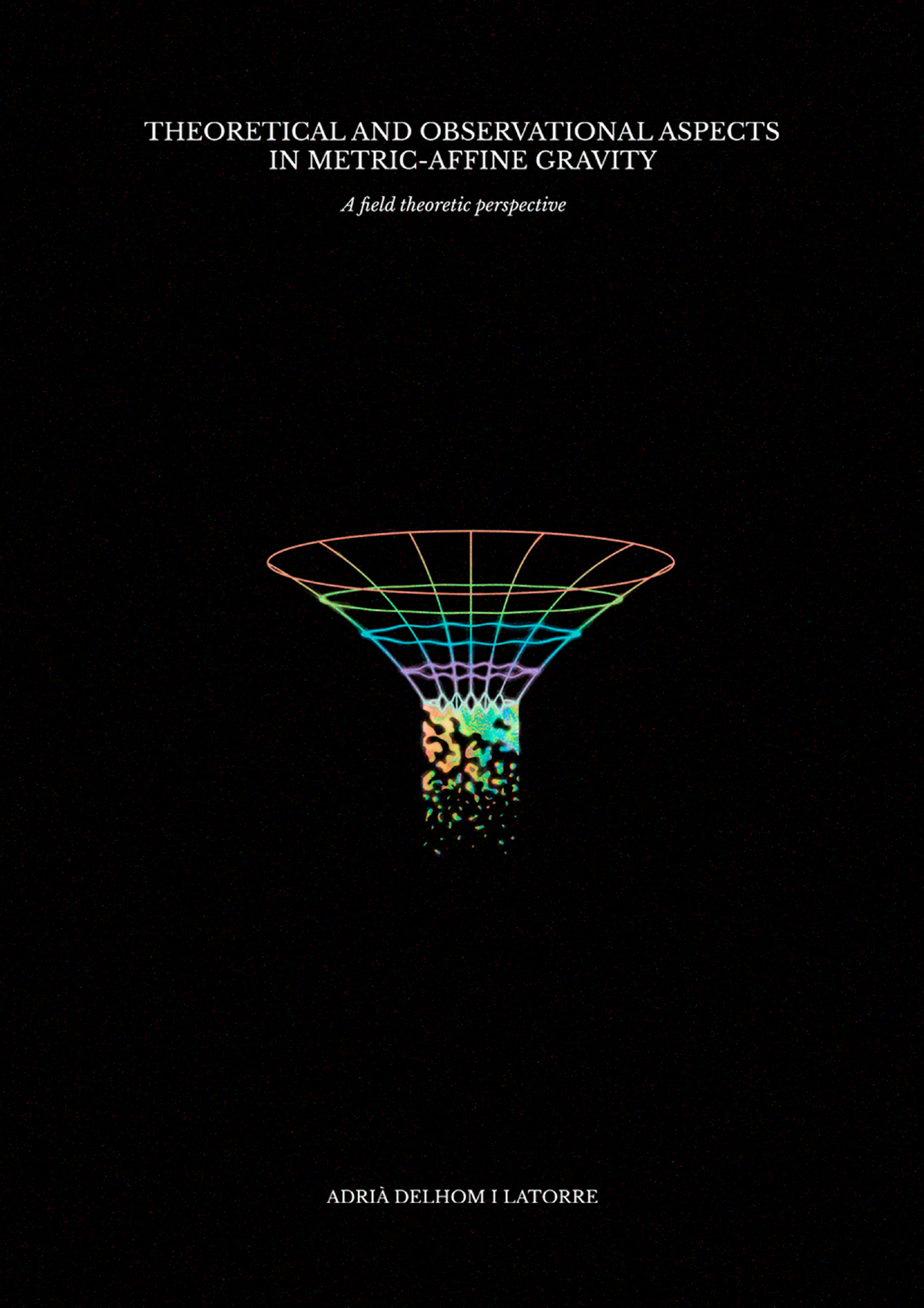}
\clearemptydoublepage
%
%
% File: Title.tex
% Author: V?ctor Bre?a-Medina
% Description: Contains the title page
%
% UoB guidelines:
% 
% At the top of the title page, within the margins, the dissertation should give the title and, if 
% necessary, sub-title and volume number. If the dissertation is in a language other than English, the 
% title must be given in that language and in English. The full name of the author should be in the 
% centre of the page. At the bottom centre should be the words ?A dissertation submitted to the 
% University of Bristol in accordance with the requirements for award of the degree of ? in the 
% Faculty of ...?, with the name of the school and month and year of submission. The word count of 
% the dissertation (text only) should be entered at the bottom right-hand side of the page.
%
%

\begin{titlingpage}
\clearemptydoublepage

\begin{SingleSpace}
\calccentering{\unitlength} 
\begin{adjustwidth*}{\unitlength}{-\unitlength}
\vspace*{-9mm}
\begin{center}
\rule[0.5ex]{\linewidth}{2pt}\vspace*{-\baselineskip}\vspace*{3.2pt}
\rule[0.5ex]{\linewidth}{1pt}\\[\baselineskip]
{\HUGE Theoretical and Observational Aspects in Metric-Affine Gravity}\\[4mm]
{\Large \textit{A field theoretic perspective}}\\
\rule[0.5ex]{\linewidth}{1pt}\vspace*{-\baselineskip}\vspace{3.2pt}
\rule[0.5ex]{\linewidth}{2pt}\\
\vspace{10mm}
{\large By}\\
\vspace{10mm}
{\Large\textsc{Adri\`a Delhom I Latorre}}\\
\vspace{15mm}
\includegraphics[scale=0.6]{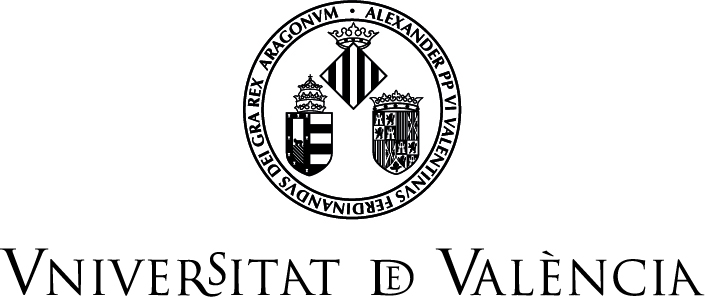}\\
\vspace{15mm}
{\large PhD program in Physics\\
\textsc{Department of Theoretical Physics}}\\
\vspace{15mm}
\begin{minipage}{10cm}
\centering
\large PhD Thesis\\
\vspace{5mm}
supervised by\\
\vspace{1mm}
\textsc{Prof. Gonzalo J. Olmo}
\end{minipage}\\
\vspace{14mm}
July 2021
\end{center}

\newpage
\clearpage
\thispagestyle{empty} 

\newpage
\clearpage
\vspace*{\fill}

\end{adjustwidth*}
\end{SingleSpace}
\end{titlingpage}
\clearemptydoublepage
%
%
%
% file: dedication.tex
% author: V?ctor Bre?a-Medina
% description: Contains the text for thesis dedication
%
\clearpage
\thispagestyle{empty} 

\newpage

\clearpage
\thispagestyle{empty} % optional -- suppress showing of page number
\begin{quotation}
 % optional -- to switch to emphasis (italics) mode
{\em Oscilando en este moog de infinitas octavas que es el universo.}

\medskip

Erik Urano
\end{quotation}
\vspace*{\fill}

\chapter*{Dedication and acknowledgements}
\begin{SingleSpace}

A totes les persones que han contribu\"it a aquesta tesi. Des de la primera profe d'El Trenet, fins al meu director de tesi, passant pels meus professors a l'Eliseu Vidal i l'Enric Valor, en particular a Joan Soler i Pep Ballester. Gr\`acies per haver-me ensenyat part del que sabeu. En particular vull agrair a Gonzal Olmo per ser un excepcional guia i mestre en aquest viatge que acaba d'escomen\c{c}ar. No puc agrair suficient el teu suport aix\'i com les teves ensenyances, tant cient\'ifiques com pragm\`atiques i personals, les quals crec (i espere) que em premetran continuar dedicant-me a la investigaci\'o per molts anys. A Jose Beltr\'an Jim\'enez per la teva paci\`encia i ajuda en l'aprenentatge de nous conceptes i pel teu exemple com a investigador. Ha estat un plaer treballar junts i espere que segueixca sent-ho. A la resta de col$\cdot$laboradors/es amb els que he treballat, Diego, Lu\'is, Caio, Param, Albert, Paulo... per les vostres particulars contribucions al meu aprenentatge; Julio, Alejandro, Ana i V\'ictor per fer possible la fusi\'o d'amistat i treball i, en especial, a Joan, per les meravelloses nits rient i imaginant junts. A la resta de companys/es amb qui he discutit de f\'isica o de qualsevol altra cosa; en especial a Iv\`an i Sergi.  

Tampoc vull oblidar la part menys visible, per\`o la m\'es important. A totes les persones amb les qui he compartit la vida. Gr\`acies als meus pares pel suport incondicional i per l'educaci\'o i l'amor que m'han oferit. M' haveu donat suficients ferramentes i carinyo per ser feli\c{c} i desitjar la felicitat als dem\'es, i crec que no es pot demanar m\'es. Al meu germ\`a pels moments divertits, les baralles, i l'enteniment mutu. Espere que aconsegueixques el que et proposses. I per supossat, gr\`acies pel disseny de la portada.  Als Iaios i els Abuelos, els tios, Juanjo, Jordi i Ram\'on. Als amics, N\'estor, Miquel, Mart\'in,... per estar ah\' i des del principi. A Quique, Alex, pels grans moments que hem passat i passarem junts, a Tori, Carlos, Juan, Pablo i, en especial, a Vicente, per acollir-me i ensenyar-me tant sobre Rock \& Roll i sobre la vida. A Jon, Livia i Tom\'as; ens veiem poc pero val la pena!. A Javi Olmedo pel suport i els \`anims en moments dif\'icils. A Alberto, Stef, i Marc, per les noves amistats i el Liberal-Determinisme, ja sabeu que la vostra pres\`encia aci era una decisi\'o inevitable. A Ana, pels bons moments que hem compartit i el que hem apr\'es, junts o separats, per exist\`encia de l'altre. A Andrea, per haver accedit a compartir la vida amb mi. Per ensenyar-me tantes coses, i per les ingents quantitats d'amor, comprensi\'o i suport durant aquestos anys, ja siga convivint o separats per la dist\`ancia. Espere que mai deixem de compartir les nostres vides, d'una forma o altra. T'estime.

No puc nomenar a tothom, per\`o us tinc presents als pensaments. Gr\`acies per la vostra companyia i ensenyaments. Aix\'i dona gust viure. Pels anys passats i pels que estan per vindre.

\end{SingleSpace}
\clearpage
\clearemptydoublepage
%
%
% File: declaration.tex
% Author: V?ctor Bre?a-Medina
% Description: Contains the declaration page
%
% UoB guidelines:
%
% Author's declaration
%
% I declare that the work in this dissertation was carried out in accordance
% with the requirements of the University's Regulations and Code of Practice
% for Research Degree Programmes and that it has not been submitted for any
% other academic award. Except where indicated by specific reference in the
% text, the work is the candidate's own work. Work done in collaboration with,
% or with the assistance of, others, is indicated as such. Any views expressed
% in the dissertation are those of the author.
%
% SIGNED: .............................................................
% DATE:..........................
%
\chapter*{Author's declaration}
\begin{SingleSpace}

I declare that the work in this dissertation was carried out in accordance with the requirements of  the University's Regulations and Code of Practice for Research Degree Programmes and that it  has not been submitted for any other academic award. Any views expressed in the dissertation are those of the author.

\vspace{0.3cm}
This thesis is based on the following works by the candidate (and collaborators):

\begin{enumerate}
\item A. Delhom, Minimal coupling in presence of non-metricity and torsion, Eur. Phys. J. C 80 (2020), no. 8 728

\item A. Delhom, G. J. Olmo, and E. Orazi, Ricci-Based Gravity theories and their impact on
Maxwell and nonlinear electromagnetic models, JHEP 11 (2019) 149

\item J. Beltr\'an Jim\'enez and A. Delhom, Instabilities in metric-affine theories of gravity with higher
order curvature terms, Eur. Phys. J. C 80 (2020), no. 6 585

\item J. Beltr\'an Jim\'enez, D. De Andr\'es, and A. Delhom, Anisotropic deformations in a class of
projectively-invariant metric-affine theories of gravity, Class. Quant. Grav. 37 (2020),
no. 22 225013

\item A. Delhom, C. F. B. Macedo, G. J. Olmo, and L. C. B. Crispino, Absorption by black hole
remnants in metric-affine gravity, Phys. Rev. D 100 (2019), no. 2 024016

\item J. Beltr\'an Jim\'enez and A. Delhom, Ghosts in metric-affine higher order curvature gravity,
Eur. Phys. J. C79 (2019), no. 8 656,

\item J. Beltr\'an Jim\'enez, A. Delhom, G. J. Olmo, and E. Orazi, Born-Infeld Gravity: Constraints from Light-by-Light Scattering and an Effective Field Theory Perspective,

\item A. Delhom, G. J. Olmo, and M. Ronco, Observable traces of non-metricity: new constraints
on metric-affine gravity, Phys. Lett. B780 (2018) 294–299

\item A. Delhom, V. Miralles, and A. Pe\~nuelas, Effective interactions in Ricci-Based Gravity
below the non-metricity scale, Eur. Phys. J. C 80 (2020), no. 4 340

\item A. Delhom and J. Ruiz Vidal, To appear

\item A. Delhom, G. J. Olmo, and P. Singh, To appear

\item A. Delhom, J. R. Nascimento, G. J. Olmo, A. Y. Petrov, and P. J. Porf\'irio, Metric-affine
bumblebee gravity: classical aspects, Eur. Phys. J. C 81 (2021), no. 4 287

\item A. Delhom, J. R. Nascimento, G. J. Olmo, A. Y. Petrov, and P. J. Porf\'irio, Metric-affine
bumblebee gravity: quantum aspects, arXiv:2010.06391

\item A. Delhom, I. P. Lobo, G. J. Olmo, and C. Romero, A generalized Weyl structure with
arbitrary non-metricity, Eur. Phys. J. C 79 (2019), no. 10 878

\item A. Delhom, I. P. Lobo, G. J. Olmo, and C. Romero, Conformally invariant proper time with
general non-metricity, Eur. Phys. J. C 80 (2020), no. 5 415

\item J. Arrechea, A. Delhom, and A. Jim\'enez-Cano, Comment on “Einstein-Gauss-Bonnet
Gravity in Four-Dimensional Spacetime”, Phys. Rev. Lett. 125 (2020), no. 14 149002

\item J. Arrechea, A. Delhom, and A. Jim\'enez-Cano, Inconsistencies in four-dimensional
Einstein-Gauss-Bonnet gravity, Chin. Phys. C 45 (2021), no. 1 013107

\item C. Bejarano, A. Delhom, A. Jim\'enez-Cano, G. J. Olmo, and D. Rubiera-Garcia, Geometric
inequivalence of metric and Palatini formulations of General Relativity, Phys. Lett. B
802 (2020) 135275
\end{enumerate}

The works on which each chapter is based are made explicit in the introduction, though the order of the works in the above list grossly corresponds to their order of appearance in the thesis. The order of appearance of the authors is always by alphabetical order except in Phys. Rev. D 100 (2019), no. 2 024016.

\vspace{0.1cm}

The text presented here should be understood only as a dissertation submitted to the University of Valencia as required to obtain the degree of Doctor of Philosophy in Physics. Except where indicated by specific reference in the text, this is the candidate’s own work, done in collaboration with, and/or with the assistance of, the candidate’s supervisors and collaborators. Any views expressed in the thesis are those of the author.
\vspace{1.5cm}
\flushright Adri\`a Delhom I Latorre\\
\flushright Val\`encia, July 2021

\end{SingleSpace}
\clearpage
%\clearemptydoublepage
%
\renewcommand{\contentsname}{Table of Contents}
\maxtocdepth{subsection}
\tableofcontents*
\addtocontents{toc}{\par\nobreak \mbox{}\hfill{\bf Page}\par\nobreak}
\clearemptydoublepage
%
%\listoftables
%\addtocontents{lot}{\par\nobreak\textbf{{\scshape Table} \hfill Page}%\par\nobreak}
%\clearemptydoublepage
%
%\listoffigures
%\addtocontents{lof}{\par\nobreak\textbf{{\scshape Figure} \hfill Page}%\par\nobreak}
%\clearemptydoublepage
%
%
% The bulk of the document is delegated to these chapter files in
% subdirectories.
\mainmatter
%
%
% File: chap01.tex
% Author: Victor F. Brena-Medina
% Description: Introduction chapter where the biology goes.
%
\let\textcircled=\pgftextcircled

\chapter*{Introduction and motivation}
\markboth{Introduction and motivation}{}
\addcontentsline{toc}{chapter}{Introduction and motivation}

\initial{G}ravitation combines some of the most intuitive phenomena for humans, and probably other species \cite{CambridgeDeclaration,BIRCH2020789}, with the fact of being the only known interaction for which we do not have a satisfactory ultraviolet (UV) complete theory yet. General Relativity (GR) is our first successful relativistic theory of gravitation, and has passed all observational tests up to date, predicting as well several phenomena such as \eg the recently detected gravitational waves \cite{Abbott:2016blz,TheLIGOScientific:2017qsa,LIGOScientific:2021qlt} or the correct light bending during the 1919 solar eclipse \cite{Crispino:2019yew} (see also \cite{Crispino:2020txj}). GR has a natural interpretation in geometrical terms, where the gravitational interaction is actually understood as the dynamics of the spacetime geometry upon which matter fields evolve. From this perspective, the gravitational field is traditionally encoded in the metric of a (pseudo-)Riemannian manifold and related to its corresponding curvature tensor, although there are (apparently) equivalent interpretations in terms of other geometrical objects such as the torsion or nonmetricity tensor of particular types of affine connections, as done in the teleparallell frameworks \cite{Aldrovandi:2013wha,Nester:1998mp,BeltranJimenez:2017tkd,Jimenez:2019ghw,BeltranJimenez:2019tjy}. 

From this perspective, gravity is a theory of the dynamics of spacetime itself, a view which led to fruitful developments such as the birth of cosmology as a scientific discipline with the pioneering works by Slipher, Lemaitre, and Hubble \cite{Slipher,Lemaitre,Hubble:1929ig}. Furthermore, it naturally accommodates the Friedman-Lema\^itre-Robertson-Walker (FLRW) metric, which provides the best description of cosmological observations up to date through the $\Lambda$CDM model, though the presence of unobserved components in the stress-energy tensor of our universe is required \cite{Joyce:2014kja} to describe the standard cosmological model, and some tensions with observations have arisen recently \cite{Battye:2014qga,Birrer:2018vtm, Wong:2019kwg,Riess:2019cxk}. As well, it predicted the existence of compact objects from which nothing could ever escape after crossing certain spacetime region, namely black holes and their event horizon. Both, the study of cosmology and of compact objects are nowadays established and active disciplines within gravitational physics, and both signal one of the main caveats of GR as a fundamental theory for the gravitational interactions, namely the presence of singularities both at early times and at the center of black hole spacetimes. 

From the classical perspective, these singularities signal a breakdown of spacetime which physical observers can reach in a finite proper time. Though this is not inconsistent at the classical level, it is extremely unpleasant to accept the idea that observers can disappear from the universe if they fall into a singularity. From the quantum point of view, this is even worse due to the fact that it would imply information loss after the black holes have evaporated via Hawking radiation, which is incompatible with unitary evolution as required by quantum physics. Taking seriously the quantum nature of the gravitational field, however, offers a way out to this problem. Indeed, though non-renormalisable, GR is a well behaved quantum effective field theory of the gravitational field up to the Planck scale \cite{Burgess:2003jk,Donoghue:2012zc,Ruhdorfer:2019qmk}, where it looses unitarity. Hence, classical solutions containing singularities with unbounded curvature scalars are physically meaningless beyond the Planck scale, where quantum effects of gravity are expected to dominate, thus changing the nonperturbative structure of the theory. In that way, the singular backgrounds present in GR would differ strongly from the exact solutions of the UV complete theory at scales beyond the Planck mass, rendering the classical singularities as unphysical by pushing them out of the regime of validity of the classical theory. Indeed, it is generally believed that the correct UV completion of GR will heal those singularities due to quantum effects. This happens in some candidates to be the UV completion of GR such as Loop Quantum Gravity, which apparently\footnote{These findings generally involve Loop quantisation of simmetry reduced spacetimes. Though these effects are expected to occur also in full Loop Quantum Gravity, I use the word apparently to emphasise the fact that they hve not been proved rigorously in the full theory yet. } regularises the Big Bang and Schwartzschild singularities by corresponding bounces \cite{Agullo:2016tjh,Li:2021mop,Gambini:2013hna,Ashtekar:2018lag,Ashtekar:2018cay,Gambini:2020qhx,Ashtekar:2020ckv,Kelly:2020lec,Kelly:2020uwj}. 

Though there are reasons to search for departures of GR in its infrared (IR) regime to see if any of the effects commonly attributed to Dark Matter and Dark Energy can be accounted for in this way \cite{Heisenberg:2018vsk}, the strongest motivation to look for departures of GR is the finding of a UV complete theory for quantum gravity, since we know that GR needs modifications in the UV to be physically meaningful at high energies. One of the possible ways of doing so is to explore the landscape of effective theories that can encode Quantum Gravity (QG) effects below the QG scale and reduce to GR in the low energy limit. To that end, there are several paths to follow. On the one hand, semiclassical corrections to the Einstein-Hilbert (EH) action arise to guarantee renormalisability of matter fields in curved spacetimes \cite{Parker:2009uva}. Furthermore, quadratic curvature corrections yield a renormalisable theory of gravity at the expense of loosing unitarity \cite{Stelle:1976gc,Antoniadis:1986tu,Johnston:1987ue}. Indeed, higher-order curvature corrections generally lead to the propagation of ghostly degrees of freedom around arbitrary backgrounds due to the presence of non-degenerate terms with second order time derivatives of the metric in the action, which unleashes the Ostrogradski instability (see chapter \ref{sec:UnstableDOF}). A possible way to avoid this would be to resort to the metric-affine formalism, where an independent affine structure is introduced as part of the spacetime geometry. 

The metric-affine framework consists on extending GR by allowing more general spacetime geometries to arise. This is done by introducing an independent affine connection so that the spacetime is a post-Riemannian\footnote{Here we will use Riemannian (referred to the spacetime manifold) as a synonym of manifold with a metric and its canonical affine structure, see chapter \ref{sec:DifferentialGeometry}. However, bear in mind that the metric of this spacetime will always be Lorentzian and not Reimanian. This is commonly denoted by writting (pseudo-)Riemannian, but I think that (pseudo-)post-Riemannian is too much, and we will generally omit the post- prefix through this chapter to lighten the text. We will assume that the metric is always of Lorentzain signature.} manifold, namely a smooth manifold with affine and metric structure which are independent from each other. This independence is encoded in two geometrical objects dubbed as nonmetricity and torsion tensors, which measure departures from Riemannianity. In this framework, the dynamics of both metric and affine connection are derived from an action as usual. The original path to this framework came  from geometric considerations shortly after the formulation of GR. Weyl formulated the first metric-affine theory where he tried to unify gravity and electromagnetism by relating both to a metric-affine spacetime which had a nonmetricity of the Weyl kind \cite{Weyl:1918ib}. The seminal works by Cartan \cite{Cartan1,Cartan2,Cartan3,Cartan4} established a formulation of a theory of connections without relating them to a metric structure, thus showing their independent nature. Later, works by Utiyama, Kibble and Sciama \cite{Utiyama:1956sy,Kibble:1961ba,Sciama:1964wt} showed how a gauge theory of the Lorentz group leads to a theory which was equivalent to GR except for a coupling between fermions and spacetime torsion which generated a four fermion effective interaction. The theory is known as Einstein-Cartan-Sciama-Kibble (ECKS) theory. This line of work was continued by Hehl and collaborators, who developed a gauge theory of the Poincar\'e and General Linear groups \cite{Hehl:1976kj,Hehl:1994ue}. Parallely, other metric-affine theories that were not based in gauging any local symmetry were also considered. Indeed, the first-order (or Palatini) formulation of GR is described precisely by the metric-affine version of the Einstein-Hilbert action. As is well known, this formulation is equivalent to GR in the absence of fermionic fields, and to ECKS in presence of them, due to the fact that the connection is an auxiliary field whose equations force it to be the Levi-Civita connection of the metric (up to a choice of projective gauge) plus a nondynamical torsion term in presence of fermions. Many relevant results within gravitation, such as the ADM formalism or Deser's argument to show that GR is a consistent nonlinear theory of a massless spin-2 field (see chapter \ref{sec:GravityAsGeometry}) have been derived from this starting point. 

Shortly after the results by Stelle that there is a renormalisable gravity theory which is quadratic in curvature invariants, it was shown that it contains ghostly degrees of freedom in its spectrum due to the presence of higher-order derivatives of the metric in the Lagrangian of the theory \cite{Stelle:1976gc,Antoniadis:1986tu,Johnston:1987ue}. In the metric-affine framework, the Riemann tensor does not feature derivatives of the metric, and has only first derivatives of the affine connection. Hence, metric-affine higher order curvature theories do not have higher derivatives in the Lagrangians, and there was hope that this would be enough to avoid the Ostrogradskian instability \cite{Ostrogradski:1850fid,Woodard:2015zca}. One of the central topics of this thesis is to address this issue, as explained below in more detail. More recently, higher-order curvature metric affine theories have been studied in both cosmological and astrophysical contexts with interesting results (see below). 

Another reason to explore metric-affine theories of gravity is the possibility of them being able to encode QG effects below the Planck scale. Indeed, it has been argued that, by an analogy with crystals, which can be described by a smooth metric-affine manifold in the continuum limit, a quantum spacetime could lead to nontrivial nonmetricity and/or torsion torsion tensors in the effective geometry below the Planck scale. Indeed, in the case of crystals, a perfect lattice without defects leads to a continuum limit where the crystal's macroscopic properties can be described by a Riemannian manifold but, in a crystal which features some defects in its crystalline structure, the corresponding continuum limit is described by a post-Riemannian manifold that may develop nontrivial nonmetricity and torsion tensors \cite{ArashDefects,KupfermanDefects,DONGDefects,KronerDefects,KleinertDefects}. In a parallel way, were QG described in terms of some discretisation of spacetime which is subject to quantum fluctuations, these could be seen as dynamical defects that would end up being described by nonmetricity and/or torsion in the appropriate continuum limit \cite{Lobo:2014nwa,Hossenfelder:2017rub}. Indeed, crystalline defects always arise at finite temperature due to entropic reasons, since they increase the number of available microscopic configurations, and perfect crystalline structures do not exist in nature. On the other hand, the continuum limit of a fluctuating quantum geometry could be described by similar principles where, as the energy density increases \cite{Latorre:2017uve}, limiting configurations which encode spacetime defects would be entropically favoured.

Before going into the dynamical aspects of metric-affine theories, let us elaborate on the  subtleties that arise by allowing for an independent affine connection. Riemannian spacetimes can be seen as post-Riemannian spacetimes where the metric-compatibility (or metricity) and torsion-free conditions are imposed to the connection {\it a priori}. The relaxation of the metricity and torsion-free conditions in a general metric-affine setup introduces some ambiguities in the way matter fields couple to the geometry, specially spinor fields. This ambiguities are often treated naively and, in my opinion, there is a lack of understanding about the degree of arbitrariness of some of the prescriptions employed to bypass these ambiguities. This will be the topic concerning the first part of the thesis. 

We will start in chapter \ref{sec:DifferentialGeometry} where we will introduce basic notions of differential geometry and the theory of connections. The aim of the chapter is to bring the question of what structures and relations are canonical with respect to one another, in the sense that having one mathematical structure implies having the other, and which ones are arbitrary. The final goal is to show that there is a canonical way of defining the affine covariant derivative of Dirac spinor fields in a general post-Riemannian spacetime. Though this problem admits other solutions besides the canonical one, we expect to understand what is the degree of arbitrariness behind them. This question will be tackled from the formulation of connections in the theory of fiber bundles, which will also allow us to formalise the notion of matter fields and gauge fields as sections and connections in a given fiber bundle. This will also help us in providing a solution to  another ambiguity typically present in the metric-affine framework, namely, the way in which matter fields couple minimally to geometry, which will be the content of chapter \ref{sec:MinimalCoupling}, based on \cite{Delhom:2020hkb}. There, we will show that the usual minimal coupling recipe of replacing Minkowski metric by spacetime metrics and partial by covariant derivatives leads to nonminimal couplings to the affine connection in presence of nontrivial nonmetricity and/or torsion. We will also provide a precise definition of what I understand as minimal coupling, as well as an algorithm to implement minimal coupling in generic metric-affine theories which is compatible with the given definition. Then we will explicitly work out the cases of scalar, Dirac and 1-form fields, showing the differences between the usual recipe and the one that we propose. After having discussed these issues, we will also go through the question of what paths do freely falling particles follow within metric-affine theories. Regarding this question, it is sometimes assumed in the literature that they will follow affine geodesics, which we argue that cannot be the case provided that matter fields evolve according to an action principle. This provides a partial answer to an issue that still seems to be confusing through the literature.

After having the tools to deal with a precise formulation of metric-affine theories and deal with the ambiguities that appear in the framework, we will dwell into the dynamical aspects of the theories, emphasising the understanding of their mathematical structure and theoretical or phenomenological aspects that allow us to constrain the landscape of viable metric-affine theories. To that end, we will analyse in depth Ricci-Based theories of gravity, a subclass of metric-affine theories whose action is built in terms of the metric and Ricci tensor of the independent affine connection. As we will see, understanding the features of these theories leads to valuable insights on the properties of more general metric-affine theories. We will start this analysis in chapter \ref{sec:RBGTheory}, based on\footnote{The way in which the structure of RBG theories synthesises several results present in the literature, but in chapter \ref{sec:RBGTheory} have used a completely general approach which cannot be found explicitly elsewhere.} \cite{Delhom:2019zrb,Jimenez:2020dpn}, where the general structure of Ricci-Based theories and their field equations will be analysed. As we will see, there always exist an Einstein-like frame for this theories. The case wihout projective symmetry propagates ghosts degrees of freedom, as will be seen in chapter \ref{sec:UnstableDOF}. However, if projective symmetry is enforced thus forbidding the antisymmetric piece of the Ricci tensor in the action, the corresponding Lagrangian in the Einstein frame takes the metric-affine Einstein-Hilbert form. This allows to define a mapping procedure in which the corresponding RBG\footnote{We will use RBG as an acronym for projectively invariant Ricci-Based theories.} theory coupled to a given matter sector can be written as GR coupled to a nonlinearly modified version of the {\it same}\footnote{In the sense of having the same fields with different interactions.} matter sector. This mapping procedure is then explicitly worked out for RBG theories coupled to an abelian gauge field, and as an explicit example we will show how Eddington-inspired Born-Infeld (EiBI) gravity coupled to Maxwell electrodynamics is equivalent to GR coupled to Born-Infeld electrodynamics. This opens the door to the study of some interesting exact solutions found in EiBI as solutions of GR with the corresponding matter sector.

Once the structure of the theories and their field equations has been understood, we will follow by studying some nontrivial aspects of their solution space in chapter \ref{sec:SolutionsDeformation}, based on \cite{BeltranJimenez:2020guo}. As it turns out, the mapping procedure is possible due to the fact that the connection field equations are an algebraic constraint which can be solved in terms of a new metric, obtained from an on-shell field redefinition of the original metric. It is this new metric the one which obeys Einstein's equations coupled to a modified matter sector. The algebraic equations that relate both metrics are nonlinear and, though there always exists one solution which reduces to vacuum GR, there might be other solutions that are typically overlooked in the literature. We will give conditions for the existence of anisotropic solutions in the Einstein frame when the original metric and matter sector are isotropic and homogeneous. We will also provide a no-go theorem for the presence of these solutions in EiBI gravity and study the behaviour of the existing anisotropic solutions in quadratic curvature theories. We find that they are pathological in general, thus providing solid grounds to ignore them in the literature. We also elaborate on the consequences in spherically symmetric spacetimes, and square the results with the no-hair theorem in cosmological backgrounds that must be satisfied in the Einstein frame of the theory.

In some subclasses of metric-affine theories of gravity, singularities in cosmological as well as spherically symmetric scenarios are solved without the need of adding extra degrees of freedom. Therefore, understanding the structure of these theories could offer some insight into the plethora of possible theories and solutions to the gravity-matter field equations which are free of singularities \cite{Banados:2010ix,Barragan:2009sq,Poplawski:2010kb,Olmo:2013gqa,Olmo:2015dba,Lobo:2013adx,Lobo:2013vga,Lobo:2014zla,Olmo:2015bya,Olmo:2016fuc,Olmo:2017fbc}. These results are generally at the background level and, though tensor perturbations have been seen to develop instabilities \cite{EscamillaRivera:2012vz,Yang:2013hsa,BeltranJimenez:2017uwv}, apparently, there are ways in which this problem can be ameliorated \cite{Avelino:2012ue,Albarran:2019ssh}, and further research in this direction is needed. Having studied the nontrivial structure of the solution space, and practically ruled out the nontrivial solutions to the relation between the original and the Einstein frame metric, we are ready to build in this direction. In chapter \ref{sec:Absorption}, based on \cite{Delhom:2019btt}, we study the absorption spectrum of scalar waves by black hole remnants which behave as wormholes. These solutions arise as spherically symmetric electrovacuum spacetimes occurring in some RBG theories. Due to the presence of the throat, we observe resonant absorption lines similar to those occurring in other exotic compact objects (ECOs) which could be used to distinguish them from regular black hole solutions. 

We then turn back to the general properties of RBG theories and beyond. In chapter \ref{sec:UnstableDOF}, based on \cite{BeltranJimenez:2019acz,Jimenez:2020dpn}, we tackle the longstanding issue of whether higher-order curvature and more general metric-affine theories of gravity are ghost-free due to their apparent lack of higher derivatives in the Lagrangian. Our results show how ghost degrees of freedom are a generic feature of the metric-affine framework. To do that, we explore the particular case of Ricci-Based theories. We will see that projective symmetry plays a key role in avoiding pathological degrees of freedom within such class and, when dropped, five extra ghostly degrees of freedom appear through a 2-form and a vector field that represents the dynamical projective mode. Besides imposing projective symmetry, we will also analyse geometrical constraints that can be placed in the theories to render them ghost-free. We will finish the chapter by arguing how the appearance of these degrees of freedom is not a feature of the particular subclass of Ricci-Based theories, but a rather general characteristic of the metric-affine framework, though some subclasses may be ghost-free. This poses a serious drawback to consider generic metric-affine theories as physically viable and shows that one must build metric-affine actions with great care if one wants to avoid the presence of instabilities.

Though the geometric view is the predominant one within the metric-affine literature, we should not forget that these theories can also be studied from the field theoretic perspective, where the nonmetricity and torsion fields constitute two extra matter fields that interact in particular ways with the massless spin-2 depending on the particular theory under consideration. This viewpoint allows for a systematic study of the metric-affine landscape by resorting to the Effective Field Theory framework (EFT). In chapter \ref{sec:MetricAffineEFT}, based on\footnote{Part of this chapter has been developed by the author while writing the thesis, and it remains unpublished up to date.} \cite{BeltranJimenez:2021iqs}, we will analyse whether RBG theories fit into the EFT framework finding a negative answer, though they are perfectly fine effective theories below a given UV scale that controls the induced nonlinearities in the matter sector. We then elaborate on several aspects of generic metric-affine theories, arguing that, in the most general case, symmetrised Ricci terms in the action are redundant in the sense that they only introduce further interactions among the propagating degrees of freedom of the theory, without exciting any new degrees of freedom. To that end, we build a generalised Einstein-like frame for general theories and study the perturbative form of the corresponding Einstein frame metric. By similar reasonings to those in section \ref{sec:OnMoreGeneral}, this allows us to argue why ghostly degrees of freedom will plague generic metric-affine theories of gravity. 

The results regarding the perturbative form of the corresponding Einstein frame metric in terms of the original metric found in the previous chapter show how there are some terms due to nonlinear symmetrised Ricci operators which are also related to the nonmetricity tensor in general metric-affine theories. These terms source new interactions in the matter Lagrangian and are suppressed by a UV scale which controls deviations from standard GR and as well it is the scale at which nonmetricity becomes nonperturbative. In chapter \ref{sec:ObservableTraces}, based on \cite{Latorre:2017uve,Delhom:2019wir,JoanAdri}, we exploit these new interactions in the matter sector to constrain the coupling parameters of nonlinear symmetrised Ricci operators in general metric-affine theories. Explicit constraints for RBG models are also derived, finding an improvement of six orders of magnitude compared to the next most stringent constraints known up to date. Concretely, we find that the UV scale controlling deviations of GR in theories with nonlinear symmetrised Ricci terms in the action should be above $\sim$100 GeV. To our knowledge, this constitutes the first generic effect that, from the geometric viewpoint, can be unambiguously related to a piece of the nonmetricity tensor in generic metric-affine theories.

In the third part of the thesis, dubbed as Funhouse,\footnote{Note the nod to the wonderful homonimous record by  \href{https://en.wikipedia.org/wiki/Fun_House_(The_Stooges_album)}{The Stooges}.} we present a miscellanea of works generally related to metric-affine theories but without a strong link to the study of their generic properties. In chapter \ref{sec:LQC}, based on work yet to be published \cite{LSU}, we tackle the problem of finding an effective explicitly covariant action that describes the background evolution of Loop Quantised cosmological backgrounds. We manage to find a family of metric-affine $f(R)$ theories which can fit standard Loop Quantum Cosmology (LQC) and other two models of Loop quantised cosmologies, namely mLQC-I and mLQC-II \cite{Li:2018opr}, that arise due to ambiguities in the quantisation procedure. 

In chapter \ref{sec:SLSB}, based on \cite{Delhom:2019wcm,Delhom:2020gfv}, we present an explicit model which includes spontaneous breaking of Lorentz symmetry by a vacuum expectation value (VEV) of a vector field in the metric affine formalism. This model is a metric-affine version of the well known bumblebee model \cite{Kostelecky:2003fs}, and can be encoded within the class of RBG theories with nonminimally couplings between matter and geometry. We study the stability of the nontrivial vacua that break Lorentz symmetry at a perturbative level in the nonminimal coupling to the geometry, finding a classically stable spacelike VEV. We then find the effective Lorentz breaking coefficients for such background. 

The problem of finding a scale invariant notion of proper time is suggested by the idea that the laws of nature may be scale invariant in the deep UV, and the proper time defined as the spacetime length of timelike worldlines is not a conformally invariant notion of time. This problem was partially solved by Perlick, who defined a Weyl invariant notion of proper time, namely, a scale invariant notion of proper time in presence of Weyl-like nonmetricity \cite{PerlickTime}. In chapter \ref{sec:GeneralisedTime}, based on \cite{Delhom:2019yeo,Delhom:2020vpe}, we will deal with the possibility of generalising this notion in presence of arbitrary nonmetricity, which is done in a straightforward way. After the definition and its basic properties are presented, we discuss the conditions for this notion of proper time to be equivalent to that given by Ehlers, Pirani, and Schild (EPS) by considering compatibility between the conformal structure defined by light rays and the affine structure defined by the trajectories of massive particles. We then study the presence of a second clock effect within this definition of time, and discuss about its (unlikely) measurability.

As a last attraction of the Funhouse, in chapter \ref{sec:4DEGB}, based on the works \cite{Arrechea:2020gjw,Arrechea:2020evj}, we will argue why the recently presented four-dimensional Einstein-Gauss-Bonet theory (4DEGB) is not well defined. To show that we will first outline why the $D\rightarrow 4$ {\it limit} in which this work is based is not a well defined limit in the mathematical sense unless one considers maximally symmetric backgrounds from start. This leads to undefined field equations in backgrounds which are not maximally symmetric. We then explicitly compute second order perturbations around maximally symmetric backgrounds to show that there appears a $0/0$ indetermination in the field equations after the $D\rightarrow 4$ prescription is enforced, contrary to what was claimed by the authors of \cite{Glavan:2019inb}. We then suggest a way to regularise these field equations and argue why no diffeomorphism invariant action can lead to the regularised field equations. We finish by showing how the spherically symmetric geometries presented in \cite{Glavan:2019inb} as a solution of the ill-defined field equations, which were also claimed to be geodesically complete, are neither a solution of these field equations, nor of the regularised field equations, nor geodesically complete. Finally, we will conclude the thesis with a brief outlook on the achievements presented through it and the possible research windows that they suggest.

%=========================================================
\clearemptydoublepage
%
%
% File: chap01.tex
% Author: Victor F. Brena-Medina
% Description: Introduction chapter where the biology goes.
%
\let\textcircled=\pgftextcircled

\part{Gravitation and the Metric-Affine Framework}
\markboth{Part II}{}

{\noindent\Large\textbf{Part I - Outline}}
\vspace{0.5cm}
\newline
This part is a general introduction to the metric affine framework,  as well as to some mathematical aspects that are relevant to have a detailed understanding of some subtle issues arising within it. We will begin with a somewhat odd chapter where we will review the renowned debate of geometry vs. force field that is usually at the heart of many misunderstandings between the two sides that compose the community of gravitational and theoretical physicists. To do that I will follow a path in which the relevant aspects of the two views and their relation to each other will be emphasised, with the aim of reconciling these two views, as well as showing their strengths and limitations. In passing, my thoughts (and doubts) on these matters will lay wide open to the reader, which will be of use for them to understand my perspective on the rest of the work carried on through this thesis. We will then continue with an exposition of the necessary mathematical framework and some subtle aspects regarding the coupling between matter fields and metric-affine geometries, which will be of use to start the main part of the thesis on the same page with respect to these issues.

\chapter{Gravity: force field or geometry?}
\label{sec:GravityAsGeometry}

\initial{T}he ideas of Aristoteles regarding motion and free falling bodies were rejected already by Filoponos around the VI century, who greatly influenced Galileo in his thinking, leading to the modern concepts of inertia and to the realisation that all bodies fall with the same acceleration provided that there are no frictional forces. This property of the gravitational interaction is usually referred to as universality of freefall. An equivalent statement, which is one of the formulations of the Weak Equivalence Principle (WEP), is that the trajectory of a freely falling body\footnote{By freely falling we mean that it only interacts gravitationally.} in a gravitational field is determined completely by its initial position and velocity and the gravitational field, being thus oblivious to the characteristics of the body. Within the framework of Newtonian mechanics, universality of freefall (\ie the WEP) has a straightforward implementation in Newtonian gravity, where, the force felt by material bodies due to a gravitational field\footnote{The concept of field may have been introduced much later than the time when these findings occurred, but the seed of this idea was already latent in those findings.} is proportional to the field, and the proportionality constant is the gravitational charge of the body (usually called gravitational mass). In order for all bodies to feel an equal acceleration if seen from an inertial frame, according to Newton's second law, gravitational charge must be proportional to inertial mass with the same proportionality constant for all bodies (and equal to 1 in appropriate units). Though it was later discovered that proportionality to the field also occurs in the way that bodies respond to other known forces, such as the electric force on a test body, given by its electric charge times the background electric field, in these interactions the proportionality constants (charges) had nothing to do with inertial mass. Hence, although for other forcefields one needs to measure both the acceleration felt by a test body {\it and} its mass ratio in order to know the value of the field at a given point, this is not the case for the gravitational force, for which it suffices to know the value of the acceleration of any test body at a given point in order to know the field at that point just by measuring, without knowing anything about the body's composition and structure. 

Note, as well, that the proportionality constants that calibrate the response of a body to a force field, namely the charges, are also typically the sources for that force field, whose strength is also proportional to the charge of the source. From the experimental viewpoint, this raises the following question. Though force fields like the electric one can be observationally distinguished by their charges even if they describe the same $r^{-2}$ behavior, can there exist two distinct $r^{-2}$ force fields like the gravitational one, namely fields which propel all test bodies with an equal amount of acceleration independently of its characteristics, so that their charge is proportional to inertial mass? Interestingly, we can elaborate the following argument: If two {\it a priori} different such fields existed, their values would be proportional to the inertial mass of the source and therefore to each other. Hence, there would be no physical scenario in which one of these fields vanishes but the other is present, and they could only be potentially distinguished by the proportionality of the corresponding charges to inertial mass. If two such fields have proportionality constants $\alpha_1$ and $\alpha_2$, then because of the property that these fields affect equally to all bodies, I cannot come up with any empirical way of discriminating a scenario where these two fields exist from another scenario with only one such field with proportionality constant $\alpha_1+\alpha_2$. 
Note that this argument\footnote{Actually this applies for $n$ fields provided that they have the same functional behavior.} relies only on universality of freefall. Thus we see that, in any observational regime where the WEP is backed up by observations, gravity stands out as a special interaction because of its universality, which has the direct consequence that one only needs to measure the acceleration of a test body at a point to know the gravitational field at that point, as opposed to acceleration, mass, and the corresponding charges for other nonuniversal interactions. At the same time, this guarantees that the trajectories of test bodies affected only by gravitation will be determined only by their initial position and velocity, independently of any characteristics of the body. On the other hand, the gravitational charge being inertial mass implies that any existing body will feel and source gravitational interaction, so that, strictly, the closest that a body can be to a free particle is if it interacts only with gravity. This fact, together with universality, leads to the following question: If the trajectory of a {\it closest to free} test body is not straight due to gravitational interactions, but at the same time we know that any test particle with the same initial conditions would follow that very same trajectory, is it not reasonable to interpret the resulting trajectories as properties of the space on which the test bodies propagate, instead of their reaction to a force field?

Another consequence of universality in the above sense is the following: the effects of some special types of gravitational field cannot be told apart from those of describing motion from an accelerated frame. This is the conclusion of the well known elevator thought experiment by Einstein, where it is argued that a freely falling observer  in a uniform gravitational field would see no gravitational field at all, as any other freely falling body would fall with the same acceleration as the observer. Hence, the observer will measure the effects of other interactions among the bodies as if the gravitational field did not exist. Of course, this would not be true if the gravitational field is not uniform, as the observer would then measure differences in the accelerations described by freely falling bodies due to the difference in the field strength at different points. These effects, which cannot be mimicked by an accelerated frame, are known as tidal forces and, for a freely falling observer in a general gravitational field, their size increases with the nonuniformity of the field and with the distance to the observer. Indeed, even in highly nonuniform gravitational fields, these effects can be made arbitrarily small in a sufficiently small neighbourhood of a freely falling observer. Hence, locally, freely falling observers will see bodies around them behave as if there was no gravitational field. This idea can be carried even further as, if the observer is not freely falling, this will be equivalent to a uniform gravitational field in a sufficiently local neighbourhood, which will not have any effect on the outcomes of local experiments disregarding of whether they test gravitational interactions between the test bodies or any other phenomena. This is commonly known as the Strong Equivalence Principle (SEP), and it provides a further link between gravitation and geometry, namely, the local validity of Special Relativity (SR) provides a chronometric interpretation for the metric tensor in GR by relating it locally to the special relativistic chronometric interpretation of the Minkowskian metric\footnote{In coordinates adapted to the freely falling observer, in a small enough neighbourhood around the observer, the metric looks approximately Minkowskian. Thus should the Minkowski metric have a chronometric interpretation, this is easily lifted to the GR metric through the SEP. For an explicit operational construction of the Minkowski metric as encoding the information in clocks and rods built only with timelike and null trajectories as well as Lorentz covariance, see \cite{MarzkeWheeler,Pauri:2000cr}.}. In turn, the fact that we can make a chronometric interpretation of the Minkowskian metric in special relativity is due to the fact that the matter fields known to exist behave universally in a Lorentz covariant way. Note that, should this universality of Lorentz covariance be violated within the matter sector, spacetime intervals could be relative to the fundamental constituents which a given observer is made of. As a remark, note that the relativistic version of the WEP requires the gravitational charge to be energy-momentum as opposed to inertial mass, and the SEP then implies that the gravitational field must also couple to itself through its own energy-momentum.

Note that, in the above discussion, we can distinguish two different aspects in which the gravitational interaction can be geometrised, with universality playing an enabling role in both cases. On the one hand, the universality of freefall provided by the WEP allows to think of freely falling trajectories as {\it straightest} paths so that, within this geometric interpretation, their bending indicates a property of the spacetime where trajectories take place, rather than reaction to a force. On the other hand universality of Lorentz covariance allows for a {\it clocks and rods} interpretation of the Minkowski metric which together with the SEP allows to lift this chronometric interpretation of the metric to the metric in GR. Thus the fact that the metric $g_{\mu\nu}$ encodes information about lengths and time intervals is tied to universality of approximate Lorentz covariance in a small enough neighbourhood of each spacetime event. Given that GR fulfils both the WEP and the SEP, it is hard to avoid the temptation of a geometric interpretation of gravitation within this theory, as well as other theories satisfying these principles. Adopting this viewpoint, then we now ought to clarify the meaning of {\it gravitational field} within GR. The answer is actually not so obvious, and it was a matter of philosophical debate for quite some years, though currently there appears to be a consensus in the way in which gravitational physicists think of the gravitational field. 

On the one hand, we have Einstein's view, for whom one of the main achievements of GR (if not the greatest) was the unification of gravity and inertia into a single theory, which he expressed through the Einstein Equivalence Principle (EEP), formulated in \cite{Einstein1918} with the statement that gravitation and inertia are {\it wesensgleich}, translated by Lehmkuhl \cite{LehmkuhlEquivalence} as `the same in their very essence'. Thus, in his view, the gravitational field and the old inertial fictitious forces are the same thing, a sort of unified {\it gravito-inertial} field in analogy to the (recent by then) unification of electric and magnetic forces in Maxwell's theory. Hence, two observers in relative nonuniform motion that insist on measuring the gravitational field, will differ in their measurements in such a way that compensates the corresponding fictitious forces. In this view, it does not make sense of talking about absence of gravity in any context, including Minkowski space, because inertia can be understood as gravity for some Minkowskian observers, and Minkowsi spacetime makes as much of a solution with a nontrivial gravito-inertial field as any spacetime with nonvanishing curvature. Furthermore, it does not make sense for an accelerated observer to talk about fictitious gravitational fields. The gravito-inertial field would then be associated to the Christoffel symbols of the Levi-Civita connection of the metric, which do not transform covariantly under changes of frame so that they can always be made vanishing at a point in the appropriate (locally freely falling) coordinates. On the other hand, the modern perspective adopted by most gravitational physicists is that the gravitational field is precisely related to the presence of these tidal gravitational forces that cannot be mimicked by any particular state of motion for a given observer. These effects are typically measured through geodesic deviation, which is sensitive to the local value of the Riemann tensor. Thus, in this view, a gravitational field is related to a nonvanishing Riemann tensor,\footnote{In this case, we mean the Riemann tensor of the Levi-Civita connection of the metric, and not of an arbitrary connection.} which being a tensor under changes of frame cannot be made vanishing anywhere only for some observers: either it vanishes or it does not for all of them. In this language, the SEP suggests that spacetime should be a locally Lorentzian smooth manifold, so that the corresponding gravitational theory is diffeomorphism invariant and local experiments enjoy a local Lorentz symmetry. In this view, there are thus fictitious gravitational fields which depend on the motion of the observer in much the same way as there are fictitious forces for accelerated observers. However, the presence of {\it true} gravitational fields do not depend on the observer's state of motion. We will stick to this later view of the gravitational field for the rest of the thesis.

Whatever of the geometric interpretations one might prefer, both cast the gravitational phenomena as the dynamics of a (pseudo-)Riemannian manifold, and therefore of its Lorentzian metric, on top of which matter fields evolve. Wheeler coined this view of the gravitational phenomena as {\it geometrodynamics}. Adopting this perspective, one migh wonder about how many geometrodynamical theories are there that are physically viable to describe the gravitational phenomena as a geometric effect. This very same question was famously answered by Lovelock in \cite{Lovelock:1971yv,Lovelock:1972vz}, but to better understand the answer, let us clarify some aspects beforehand. By physically viable, it is meant that the theory does not have higher order field equations, so that it is free from Ostrogradskian instabilities (see chapter \ref{sec:UnstableDOF}). As well, if an action for such theory is assumed to exist, the Bianchi identities due to diffeomorphism symmetry imply that the variation of the action with respect to the metric needs to be divergence-free. Lovelock was able to prove that, in four spacetime dimensions, GR is the unique theory satisfying the assumptions of divergence-free second-order field equations\footnote{As is well known, in higher spacetime dimensions there are other theories which also satisfy the requirements, known as Lovelock theories.}. The divergence-free condition is consistent with generic non-vacuum cases: if the action of the full theory (should it exist) is separated into gravitational and matter sectors, and both sectors are required to be diffeomorphism invariant on their own, the variation of the matter action with respect to the metric yields a divergence-free stress-energy tensor to which the gravitational part of the action couples. Though this might seem in contradiction with the above formulation of the WEP that gravitational charge equals gravitational mass, note that for universality of freefall to be consistent with SR in the appropriate limit as required by the SEP, the gravitational charge cannot be inertial mass anymore, but rather its Lorentz covariant generalisation, \ie energy-momentum. The WEP thus generalises in a straightforward manner to the relativistic case through a coupling through the stress-energy tensor.

We have presented a line of thought in which universality of both freefall and (local) Lorentz invariance is a necessary and sufficient condition to geometrise gravity. Indeed, these requirements allow to describe gravitational phenomena in terms of diffeomorphism invariant dynamics of a (pseudo-)Riemannian metric and its coupling to the stress-energy tensor of the matter sector. This is the geometric view of the gravitational interaction and is the picture accepted by part of the community of gravitational physicists, being most popular among those who study nonperturbative aspects of the theory or have a stronger background in classical GR. On the other hand, there is a completely different picture that describes gravity as an interaction mediated by a massless spin-2 particle. Let us now comment on this view and in what sense this relates to the geometric one. To start with, we assume Lorentz invariance and face the empirical fact that gravity is an $r^{-2}$ long range force, so that it must be mediated by a massless particle. Because of Lorentz invariance, we can make use of Wigner's classification to pin down the type of particle that the mediator of the gravitational interaction can be. Among fermion or boson, we ought to choose the later if we want to allow classical (tree-level) emission of the mediator, or exchange with any other particle, while maintaining conservation of angular momentum. Then, because of the masslessness due to long-range and Lorentz invariance, we are only left with spins 0, 1 and 2; as there are no Lorentz invariant theories of massless  fields of spin 3 or higher that couple nontrivially in the soft (\ie macroscopic) limit so that they produce a long-range force \cite{Weinberg:1964ew,Weinberg:1965rz,Weinberg:1995mt}. The attractive-only nature of the gravitational interaction leaves out of the game spin 1, which lead to attractive and repulsive forces. Finally, we know that a relativistic theory of gravitation satisfying the WEP must couple to stress-energy. The leading order coupling of the stress-energy tensor to a spin-0 field must be through its trace. Since the electromagnetic stress-energy tensor is traceless, yet we have observed light bending due to gravitational effects, this option is also ruled out on experimental grounds, leaving only the option of a massless spin-2 field, which can be represented by a symmetric two-index Lorentz tensor that couples to the full stress-energy tensor and not only to its trace. 

We are thus led to the construction of a Lorentz invariant theory of a massless spin-2 field which couples consistently with matter. We should start by finding the appropriate kinetic term for a symmetric Lorentz (0,2)-tensor $h_{\mu\nu}$, which will yield a second order equation of motion of the generic form $\cD^{\mu\nu}(h)$. Given that this object has 10 independent components, our kinetic term must also be such that it yields only the two degrees of freedom associated to a massless spin-2 field. To find such kinetic term, we note that there is a unique Lorentz invariant kinetic term for a spin-2 field (massless or not) which does not lead to the propagation of pathological ghost degrees of freedom. This term is the  Fierz-Pauli Lagrangian $\cl_{\mathrm{FP}}$ (see chapter \ref{sec:UnstableDOF}), which leads to the well known kinetic operator
\beq
\cD_{\mu\nu}(h)=\partial^2 h+\partial_\mu\partial_\nu h-2\partial^\lambda \partial_{(\mu}h_{\nu)\lambda}-\eta_{\mu\nu}\lr{\partial^2 h -\partial_\lambda\partial_\sigma h^{\lambda\sigma}}
\eeq
where indices are risen and lowered with the Minkowski metric and $h=h^\mu{}_\mu$. If we focus on the coupling to the matter stress-energy tensor at the linear level, the lowest order coupling to the stress-energy tensor is of the form 
\beq\label{eq:LinearisdGravity}
\cD^{\mu\nu}(h)=\kappa T_\mathrm{m}^{\mu\nu},
\eeq
where $\kappa$ is a coupling constant with appropriate dimensions. The ghost-free condition completely specifies the kinetic term which, as a consequence, satisfies the off-shell constraint $\partial_{\mu}\cD^{\mu\nu}(h)=0$, tied to the Bianchi identities due to a symmetry of the kinetic operator\footnote{And the Fierz-Pauli action up to a total derivative.} under transformations of the form $\delta h_{\mu\nu}=-2\partial_{(\mu}\xi_{\nu)}$. This implies a consistency condition on the choice of stress-energy tensor to which the spin-2 can couple,\footnote{Note that there are several definitions of stress-energy tensor that we could have chosen. See chapter 2 of \cite{Ortin:2004ms} for a nice discussion.} pointing towards the Belinfante-Rosenfeld stress-energy tensor due to its symmetry and on-shell vanishing divergence. To study the consistency of these couplings, let us argue in the following line. Given a Lorentz invariant matter Lagrangian $\cl_\mathrm{m}[\Psi_i]$, we need to add to the Fierz-Pauli action for the spin-2 field a coupling of the form $h_{\mu\nu}T_\mathrm{m}^{\mu\nu}$, so that the total action is\footnote{Note that we make explicit the dependence of both $T^{\mu\nu}_\mathrm{mat}$ and $\cl_\mathrm{m}$ only the matter fields and their derivatives (contracted with the Minkowski metric) appear there, and not $h_{\mu\nu}$.}
\beq
\cl_{\mathrm{FP}}+\cl_\mathrm{m}+\kappa h_{\mu\nu}T_\mathrm{m}^{\mu\nu}\equiv\cl_{\mathrm{FP}}+\tilde\cl_\mathrm{m}[h,\Psi_i].
\eeq
Due to the fact that in presence of the spin-2 field only the total stress-energy tensor of matter plus the spin-2 (and not $T^{\mu\nu}_\mathrm{m}$) will satisfy the on-shell divergence-free constraint. This can be seen by noticing that, if we have that $\partial_\mu T^{\mu\nu}_\mathrm{m}=0$ when the old matter field equations $\delta\cl_\mathrm{m}/\delta\Psi_i=0$ are satisfied, it will not be true in general on-shell for the updated matter field equations\footnote{Note that here the variational derivative acounts for the derivative terms too. See section 3.2.4 of \cite{Ortin:2004ms} for an explicit example of the nonvanishing divergence of $T_\mathrm{m}^{\mu\nu}$ on-shell for the updated equations after adding the coupling $h_{\mu\nu}T_\mathrm{m}^{\mu\nu}$.}
\beq
\frac{\delta\cl_\mathrm{m}}{\delta\Psi_i}+\kappa h_{\mu\nu}\frac{\delta T_\mathrm{m}^{\mu\nu}}{\delta\Psi_i}\equiv\frac{\delta\tilde\cl_\mathrm{m}}{\delta\Psi_i}=0.
\eeq
Allowing for self-coupling of the gravitational field, which is required for \eg explaining Mercury's perihelion precession, will not alleviate the problem unless a gravitational stress-energy Lorentz tensor $t^{\mu\nu}$ such that
\beq\label{eq:SelfCoupling}
\cD^{\mu\nu}(h)=\kappa\lr{\tilde T_\mathrm{m}^{\mu\nu}[h,\Psi_i]+t^{\mu\nu}},
\eeq
Proceeding as above, we could naively build yet another Lagrangian as $\cl_{\mathrm{FP}}+\cl_\mathrm{m}+\kappa h_{\mu\nu}(T_\mathrm{m}^{\mu\nu}+t^{\mu\nu})$, but this does not lead to the desired equation given that $t^{\mu\nu}$ must depend on $h_{\mu\nu}$ and its derivatives at least quadratically. This would introduce (at least) second order terms in the field equations, so that sticking with the linear level, we are fine. Following this line, higher order terms could be aded so that \eqref{eq:SelfCoupling} is satisfied order by order, but nothing guarantees that we will find the definite answer in a finite number of steps. A more systematic way to do this would be to exploit the symmetries of the problem, and resort to the Noether method, which provides a systematic way to couple theories with a gauge symmetry to external sources in a consistent manner.\footnote{At least to a given order, see \eg \cite{Ortin:2004ms} for details.}  This can be seen to yield a similar result, in the sense that despite being able to find the necessary order-by-order corrections to consistently couple the spin-2 field to matter and itself through the (canonical) stress-energy tensor, the method does not end in a finite number of iterations, unlike the case for coupling a spin-1 gauge field to an external source. Happily, Deser came up with a solution to the problem of finding a consistent theory of a self-coupled spin-2 field coupled to matter by resorting to a first-order form of the Fierz-Pauli action, written in terms of the fields $\tilde h_{\mu\nu}$ and $\gamma^\alpha{}_{\mu\nu}$ as
\beq\label{eq:FirstOrderFP}
\cl^{(1)}_{\mathrm{FP}1st}=\frac{1}{\kappa^{2}}\lr{\eta^{\mu\nu} 2 \gamma_{\lambda[\mu}{ }^{\rho} \gamma_{\rho] \nu}{ }^{\lambda}-\kappa \tilde h^{\mu \nu} 2 \partial_{[\mu} \gamma_{\rho]\nu}{ }^{\rho}}
\eeq
which is invariant under local transformations of the form $\delta \tilde h_{\mu\nu}=-2\partial_{(\mu}\xi_{\nu)}+\eta_{\mu\nu}\partial_\alpha\xi^\alpha$ and $\delta\gamma_{\alpha\mu\nu}=-\kappa\partial_\alpha\partial_\mu\xi_\nu$, and can be seen to be on-shell equivalent to the 4-dimensional FP Lagrangian for the redefined field variable
\beq
 h_{\mu\nu}= \tilde h_{\mu\nu}-\frac{1}{2}\tilde h\eta_{\mu\nu}.
\eeq
To see this, note that the equation for the $\gamma$ field is a constraint equation which can be written as
\beq
\gamma_{(\alpha|\mu|\nu)}=\frac{\kappa}{2}\partial_\alpha h_{\mu\nu}.
\eeq
This equation is linear, and can be uniquely solved by adding and subtracting the same equation with suitable cyclic permutations of its indices, yielding
\beq
\gamma_{\alpha\mu\nu}=\frac{\kappa}{2}\lr{\partial_{\alpha}h_{\mu\nu}+\partial_{\mu}h_{\nu\alpha}-\partial_{\nu}h_{\alpha\mu}}.
\eeq
Plugging the solution to the constraint for the $\gamma$ field into the field equations of $\tilde h$ written in terms of the new field variable $h$ leads, after some manipulations, to
\beq
\cD_{\mu\nu}(h)=0.
\eeq
Being convinced that \eqref{eq:FirstOrderFP} is dynamically equivalent to the usual FP action, before adding matter, we now need to find an extra term $\cl^{(2)}_{\mathrm{FP}1st}$ such that it leads to the desired equation $\cD_{\mu\nu}(h)=\kappa t_{\mu\nu}$ where $t_{\mu\nu}$ is the $\tilde h_{\mu\nu}$ stress-energy tensor associated to $\cl^{(1)}_{\mathrm{FP}1st}$. The expected correction is
\beq
\cl^{(2)}_{\mathrm{FP}1st}=-\frac{2}{\kappa} \tilde h^{\mu \nu} \gamma_{\rho[\mu}{ }^{\lambda} \gamma_{\lambda]\nu}{ }^{\rho}
\eeq
which can be seen to provide a full solution to the problem once the new constraint equation for $\gamma$ is taken into account. Indeed, though in terms of the variables $h$ or $\tilde h$ the solution to constraint equation is not known in compact form, by redefining again our field variable $\tilde h_{\mu\nu}$ by
\beq
\sqrt{|g|}{g}^{\mu\nu}
=\eta^{\mu\nu}-\kappa\tilde h^{\mu\nu}
\eeq
we are led to a solution of the constraint equation for $\gamma$ in the compact form
\beq
\gamma^\nu{}_{\alpha\mu}=\frac{1}{2}g^{\nu\rho}\lr{\partial_{\alpha}g_{\mu\rho}+\partial_{\mu}g_{\rho\alpha}-\partial_{\rho}g_{\alpha\mu}}
\eeq
where, in the process, indices are risen and lowered with the new field\footnote{Note that any nondegenerate symmetric 2-tensor defines an isomorphism between the vector space it acts upon and its dual.} $g^{\mu\nu}$ . Using this solution for the constrained $\gamma$ into the new field equations of $\tilde h_{\mu\nu}$ we can write them in terms of the redefined field variable as $R_{\mu\nu}(g)=0$ where we say that $R_{\mu\nu}(g)$ is the Ricci tensor of the object\footnote{Technicaly, $R_{\mu\nu}(g)$ has te exact functional dependence on the symmetric object $g^{\mu\nu}$ and its first and second derivatives as the Ricci tensor of a metric $g^{\mu\nu}$ would have. Hence, in short, we say that $R_{\mu\nu}(g)$ is the Ricci tensor of $g^{\mu\nu}$.} $g^{\mu\nu}$. To see that that this is the full solution to a consistent self-coupled spin-2 theory, we need that the field equation for $\tilde h^{\mu\nu}$ given by $\cl^{(1)}_{\mathrm{FP}1st}+\cl^{(2)}_{\mathrm{FP}1st}$ is indeed consistent with $\cD^{\mu\nu}(h)=\kappa t^{\mu\nu}$. This can be verified by undoing the field redefinition of $\tilde h$ in terms of $g^{\mu\nu}$ and expanding its field equation, namely  $R_{\mu\nu}(g)=0$, in terms of the former (see \eg \cite{Ortin:2004ms} for a detailed derivation of the whole process). Hence, this is indeed a consistent extension of the FP theory including self-couplings of the spin-2 field through its own divergenceless (Belinfante-Rosenfeld) stress-energy tensor, which features an infinite number of coupling terms of growing dimension, and reduces to FP when the coupling $\kappa$ is set to zero. This extension can be done in the presence of matter leading as well to a consistent result, and it ends up having the same field equations for the redefined field variable $g^{\mu\nu}$ as the metric field equations in GR which we can interpret as the metric, so that the theories are equivalent with an appropriate field-redefinition which allows to encode the infinite coupling terms in a compact form using the field variable $g^{\mu\nu}$, which has a natural geometric interpretation as the spacetime metric as we argued above. Furthermore, it appears that the original gauge symmetry of the FP theory that is obtained by demanding absence of ghosts in the kinetic term of $h_{\mu\nu}$, given by $\delta h_{\mu\nu}=\partial_{(\mu}\xi_{\nu)}$ is now extended to general covariance. 

Although this extension of the FP theory to a nonlinear theory introduced by Deser\footnote{See also the work of Ogievetsky and Polubarinov in \cite{Ogievetsky:1965zcd}.} in \cite{Deser:1969wk} leads to GR, it needs not be unique, and there is a result by Wald constraining the possible extensions to be either generally covariant or having `normal spin-2 gauge invariance', namely $\delta h_{\mu\nu}=\partial_{(\mu}\xi_{\nu)}$, although this last possibility could be in danger if the spin-2 field couples to matter through the stress-energy tensor \cite{Wald:1986bj}. We can go even further by following Weinberg and considering a quantum spin-2 particle described by a theory with a Lorentz invariant unitary and analytic S-matrix, so that the amplitude for the emission of a soft graviton in a process with N initial plus final particles (without counting the graviton) in a given scattering process with four-momentum $q$ will be proportional to a term like
\beq
\sum_{n=1}^N\frac{\eta_n\kappa_n p^\mu_n p^\nu_n}{q_\mu p_n^\mu-i\eta_n \epsilon}
\eeq
where $p_n$ is the four-momentum of some of the in or out particles, $\eta_n=\pm1$ with plus for out particles and minus for in particles, and $\kappa_n$ is the coupling to each of the particles to the massless spin-2. Lorentz invariance of the S-matrix requires that the contraction of this term (hence of the amplitude) with the graviton four-momentum vanishes, which in the soft $q\rightarrow0$ limit yields the condition
\beq
\sum_{n=1}^N \eta_n\kappa_n p^\nu_n=0.
\eeq
Lorentz invariance also requires that the total four momentum of the process is conserved so that $\sum_n \eta_n\kappa_n p^\nu_n=0$. The only way to satisfy both conditions at the same time is to have all $\kappa_n$ equal in value. This implies that, in the soft limit, massless spin-2 must couple to all particles, namely all forms of energy-momentum, with the same strength, even to itself\footnote{This result, in my opinion, implies that the graviton is unique by a similar argument that an universal interaction with an $r^{-1}$ potential is unique, namely, any theory with several massless spin-2 particles admits an equivalent formulation with only one massless spin-2 field and a redefined coupling.} \cite{Weinberg:1964ew,Weinberg:1964kqu}. Namely, any Lorentz-invariant quantum theory for a massless spin-2 field must satisfy the Strong Equivalence Principle in the low energy limit. Weinberg also proved that in such theory, the spin-2 must couple to a stress-energy tensor, which was later found by Boulware and Deser to be the Belinfante-Rosenfeld stress-energy tensor in the soft limit \cite{Boulware:1974sr}, so that the low energy theory for a quantum massless spin-2 must be GR. How does this square with the common lore that `GR cannot be quantised'? Well, it squares by noting that this statement is not accurate enough. To my knowledge, the strictly correct statement is that we have not found any UV complete quantisation of GR.\footnote{Of course there are candidates, but they still have their problems and there is no agreement that such UV complete theory exist} However, in much the same way as we can deal with a quantum theory the electromagnetic field below the electron mass described by the Euler-Heisenberg Lagrangian, we can perfectly make sense of GR as an effective quantum field theory below the Planck mass, where unitarity breaks down \cite{Donoghue:1994dn,Burgess:2003jk,Donoghue:2012zc}.

We have thus drawn a circle in which we have been able to find that some of the basic postulates that led Einstein to GR {\it must} be satisfied if there is a Lorentz invariant theory of quantum gravity. We have seen that from the point of view of a classical field theory for a spin-2 field in a Minkowskian spacetime that couples to itself and to matter we can arrive to GR, and we have also seen that GR is the unique low energy theory for a quantum massless spin-2 field, which remarkably must couple universally to stress-energy and satisfy the SEP in the low energy limit. We also argued above how geometrisation of GR and the chronometric interpretation of the metric tensor is enabled by the SEP. Hence, we can conclude that the existence of a quantum massless spin-2 particle implies that there is a universal interaction that can also be described in geometrical terms as the dynamics of a spacetime geometry influenced by the (other) fields and on top of which the (other) fields evolve. Which is the preferred picture? That is a matter for the reader to decide\footnote{I hope you were not expecting that I decide for you!}. However, we can raise some points that could be relevant for making this decision (take it easy though). On the one hand, the geometric picture allows for a simple generally covariant description of gravity, where all the nonperturbative effects of the theory are encoded in the spacetime metric, and one can think in terms of smooth manifolds and use the full machinery of differential geometry and topology to extract information about the features of the full theory in an easier way, such as the causal structure and the presence of singularities. Furthermore, taking seriously the geometric interpretation leads to different quantisation schemes that could offer insight on the UV completion of GR. A drawback of this interpretation is that there is no unambiguous way of defining a diffeomorphism covariant stress-energy tensor associated to the gravitational field. Moreover, even though GR can be interpreted in terms of curvature of a (pseudo-)Riemannian manifold, it can also be interpreted as the effects of nonmetricity/torsion in a flat manifold \cite{Aldrovandi:2013wha,Nester:1998mp,BeltranJimenez:2017tkd,Jimenez:2019ghw}. However, a common feature of all these geometrical interpretations is the fact that, from the field theoretic perspective, the degrees of freedom that they describe always correspond to those of a massless spin-2 field \cite{BeltranJimenez:2019tjy}. A common drawback against the field theory description is the failure of this viewpoint in describing nontrivial spacetime topologies. However, perturbations on top of a nontrivial background of the gravitational field can mimic the effects of nontrivial topologies. Besides, we know of many examples in nature, such as \eg the surface of a fluid, in which it could be argued that the topology can change dynamically.

Whether there is anything fundamental in the geometrisation of gravity, or it is just an artefact, is for nature to tell. In order to understand what possible behaviours can gravity have at higher energies, both the field theoretic and geometric viewpoints have been followed. The work in this thesis is inspired, in origin, by the geometric one, and we will mostly study metric-affine modifications of GR. However, during the course, the motivations that guided my latest research leaned closer to the field theoretic mindset. Indeed, thanks to the decomposition of any affine connection as in \eqref{eq:ConnectionDecomposition}, metric-affine theories can always be written as metric theories plus a bunch of other terms involving two tensorial fields, namely the nonmetricity and torsion tensors, and their metric-covariant derivatives as well as interactions with the Riemann tensor. 

The key results obtained in this thesis by thinking from this angle are twofold. On the one hand, we have found that  terms in the Lagrangian that are built with the symmetrised Ricci tensor induce effective interactions in the matter sector which can be used to constrain the theory. On the other hand, we have shown that terms with the symmetrised Ricci tensor in the action will lead to propagation of ghosts degrees of freedom which we argued that will be a generic feature of metric-affine gravity theories, in line with other research \cite{Aoki:2019snr,Percacci:2019hxn} and the common knowledge that it is not easy to modify a theory of a massless spin-2 field without running into the appearance of instabilities or strong coupling issues \cite{Jimenez:2020gbw}. This poses a drawback to consider metric-affine theories as fundamental theories, unless one is willing to tune the coefficients of the theory to evade these problems. Even in this case, quantum corrections could spoil the tunings and bring them\footnote{Allow me a homage to the wonderful Louisiana and the southern accents so well portrayed in {\it A Confederacy of Dunces}.} ghosts back. There are, however, better reasons to study metric-affine theories than hoping that they provide a solution to the UV completion of GR. For instance, in some theories, there are interesting kinds of exact solutions, most of the known ones being compact objects, which have nonperturbative features worth to be studied both at the theoretical and phenomenological level in order to better understand the landscape of possible phenomenology that can arise in gravity theories. Furthermore, by an analogy to how defects in crystals can be described in the continuum limit by effective nonmetricity and torsion tensors in a smooth post-Riemannian manifold \cite{ArashDefects,KupfermanDefects,DONGDefects,KronerDefects,KleinertDefects}, a possible spacetime microstructure at the quantum gravity scale could result in effective spacetimes with nontrivial post-Riemannian features at some intermediate UV scale. Though this last possibility is highly speculative, and a clear connection with spacetime granularity and post-Riemannian features is yet to be found, in my opinion, we should keep this interesting possibility open in the back of our minds. In fact, some promising insights in this direction come from the relation that seems to exist between the effective description of Loop-quantised geometries and metric-affine $f(R)$ theories, a topic to be discussed in this thesis. 

To close up, in my opinion, the geometrical view of a physical theory should always be guided by its interpretation in terms of the propagating degrees of freedom and their time evolution given some initial conditions. Indeed, this way of understanding the phenomena occurring in the universe still constitutes the basic paradigm of physics ever since Newton materialised it in his Principia in the XVII century \cite{Newton:1687eqk}. A field theoretic approach is generally closer to this view than a geometric one and, from this perspective, it does not make much sense to me to understand a theory on geometrical grounds unless there is a benefit from it, either because this view allows to do computations or extract physical conclusions in an easier way, thus shedding light into some aspects of the theory\footnote{For instance, if the is some kind of universality, a geometrical interpretation may offer simplicity in the understanding of some aspects of the theory, be them phenomenological or theoretical, as has happened with the understanding of GR historically.}, or because it allows one to think in existing problems in a different way, thus potentially leading to the exploration of new theories or paradigms that would have not been explored otherwise. In any case, the geometrical-or-not debate is an ontological one that should be irrelevant as soon as both interpretations agree on the observable phenomena predicted by the theory.

%%%%%%%%%%%%%%%%%%%%%%%%%
%%%%%%%%%%%%%%%%%%%%%%%%%
%%%%%%%%%%%%%%%%%%%%%%%%%

			%NEW CHAPTER

%%%%%%%%%%%%%%%%%%%%%%%%%
%%%%%%%%%%%%%%%%%%%%%%%%%
%%%%%%%%%%%%%%%%%%%%%%%%%

\chapter{Canonical and arbitrary geometric structures}\label{sec:DifferentialGeometry}

\initial{T}he primary topics of this thesis range within the realm of metric-affine theories of gravity. These theories are based on a formulation of gravitational theories in geometrical terms through a variational principle where the field content of the action is a metric, an affine connection and other fields usually regarded as matter. The key difference from metric theories is that while in the later the connection is taken to be the canonical connection associated to the metric, in the metric-affine formulation it is regarded as a fundamental field whose dynamics is dictated by extremising the action. In order to be as self-complete as possible, we will introduce the basic geometric notions of differential geometry that allow to build both Riemannian and post-Riemannian space-times. This presentation will have the intention of clarifying which structures are canonical and which are not, where by canonical we mean that can be constructed only with pre-existing mathematical structure or data and that are unique. In other words: A given pre-existing mathematical structure allows to build a {\it new} canonical structure if this {\it new} structure can be built without any arbitrariness either in the introduction of mathematical data not available in the pre-existing structure or in the builder's choices through building procedure. The reason why I write {\it new} in italics is because my viewpoint is that if a structure is canonical with respect to the pre-existing one, then it must be understood as {\it part of} the preexisting structure rather than a new structure on top of the old one. For instance, given a vector space with an inner product, one cannot say that the set of all orthonormal basis with respect to such inner product is a new structure associated to such vector space because there is one and only one (maximal) set of orthonormal basis associated to that inner product.

 Another instance, perhaps of more interest to us, is that of the Levi-Civita connection: Given a smooth manifold with a metric structure as preexisting structure (see \ref{sec:MetricStructure}), the Levi-Civita connection is canonical and cannot be seen as extra structure arbitrarily chosen by the designer of the manifold. This does not preclude, of course, the introduction of other noncanonical affine structures in a manifold with a metric. The reason why I mentioned this example here is because there is a folklore within the metric-affine community stating that the metric-affine formulation of a given action functional contains {\it less arbitrariness} in the sense that it does not assume any particular affine structure, but rather lets the action functional determine it. This would be opposed to the metric formulation, where the connection is chosen to be the Levi-Civita connection of the metric beforehand. According to our view there must be something wrong with that statement: a canonical structure can never be an arbitrary choice as it is already present in the original structure. Still, that folklore carries a hidden truth that our community has intuitively identified. Indeed, in my opinion, what is hazily believed to be arbitrariness in the choice of a particular affine connection, is actually arbitrariness in the very definition of what a spacetime is. 
 
 The part of the definition which we all agree upon is that a spacetime is the support\footnote{Here support refers to the set of points where a function is defined.}
 of the solutions of some set of field equations for some physical fields. Now, the arbitrary part of the definition is what are the variables of these field equations (\ie the physical fields). Although there can potentially be other choices, the dilemma that concerns us is the choice between these two options: We can either choose in regarding just a metric tensor as the variable of the field equations, known as metric formalism, or regarding a metric tensor and an affine connection as the variables of these field equations, known as metric-affine formalism.\footnote{Note that it is not quite right to talk about metric and connection for the variables if these are to {\it define} what the spacetime manifold is. However, this is a shorthand notation for a set of variables that will play the role of a metric and a connection in the space that is the support of the solutions of the corresponding field equations.} One could argue that, if a metric has always to be in the recipe, to add a connection looks like adding arbitrariness to the game. However, one should bear in mind that some authors have also considered theories with only an affine connection, where the metric is interpreted as derived from the connection \cite{EddingtonAffine1924,SchrodingerAffine1950,Catto:1980fs,Poplawski:2006zr,Kijowski:2016qbc,Castillo-Felisola:2015cqa,Castillo-Felisola:2019wcs,Chakraborty:2020yag} and, therefore, there are indeed other possible choices. As a remark, let us point out that although these theories typically find difficulties in defining their coupling to matter, there appears to be recent progress on that issue \cite{Castillo-Felisola:2015cqa,Chakraborty:2020yag}. 
 
 Thus, we see that rather than choosing or not an affine structure a priori, we have the choice to {\it define} the spacetime either as a smooth manifold with only a metric structure (which has a canonical affine structure), or as a smooth manifold with a metric and an arbitrary affine structure, or as a smooth manifold with only an affine structure and an emergent\footnote{I did not use the word canonical here because it is not clear to me if the emergent metric in purely affine theories of gravity is canonical or not.} metric for that matter. Once we have chosen this, there is no freedom left but to choose the preferred set of field equations\footnote{If we assume the dynamics to be dictated by an action principle, then this freedom is translated to the choice of a preferred action functional.} that determine the dynamics of the physical fields. Given that we have this freedom in both metric-affine and metric formalisms, we cannot say that there is more freedom (at this level) neither in one nor the other. Of course, as always, what is the correct definition of a physical spacetime is not for us physicists to decide, but for the universe to tell. Our role is, therefore, to find out whether there is any observable difference between these two (or any other) choices or if, instead, it is just a matter of pragmatism and/or aesthetics to chose one or the other.

It is true, however, that there are claims that the way in which some types of matter fields couple {\it minimally} to the spacetime geometry is rather arbitrary in the metric-affine formalism, specially regarding spinor fields and the spinor connection. Of course, this arbitrariness stems from the arbitrary definition of a minimal coupling prescription in this formalism. I have tried to give an as canonical as possible minimal coupling prescription in metric-affine geometries in \cite{Delhom:2020hkb} guided by the idea that a minimal coupling prescription when passing from one geometry to a more general one should couple the fields {\it as little as possible} to the elements of the new geometry. The aim of this section is to give the reader the necessary tools to judge by themselves whether or not this arbitrariness remains when this definition is employed. Of course, as happened with the definition of spacetime, the way in which physical fields couple to each other, or to geometry for that matter, is to be answered by empirical data.

To provide such tools, I will give a self contained approach starting with the definition of smooth manifold and all the associated canonical structures that it contains, passing through that of connection in vector and principal (frame) bundles and reaching to the definition of spinor bundle and spinor connection in a manifold with an arbitrary affine connection and a Lorentzian metric structure. I believe that the canonicality of the different objects and structures that arise while doing physics is often a useful guide to physicists, but while working on this thesis, I have often found myself with unanswered questions about what is canonical in the metric-affine formalism and what is not. Hence, I will try to put emphasis on making clear which constructions are canonical and with respect to what pre-existing structure and which are not. Should the reader have experience on these topics, they can perhaps skip over this chapter. Most of the information that is includded can be found on \eg \cite{Frankel,Marsh,Nakahara,Szekeres}.

\section{Differentiable manifolds}

The basic ingredient of any (space-time) geometry is that of a manifold. To introduce it, we will need the notion of topology and topological space. Given a set M, a topology for M  is a set $\tau$ whose elements are subsets of M such that
\begin{itemize}
\item[1.] Both the empty set and M are elements of $\tau$.
\item[2.] Any arbitrary union of elements of $\tau$ also belongs to $\tau$.
\item[3.] Any finite intersection of elements of $\tau$ also belongs to $\tau$.
\end{itemize}
A pair (M,$\tau$) is a topological space $\cm$. The elements of $\tau$ are called open sets of $\cm$, and as well a subset $X$ of M is closed in $\cm$ if its complementary (M-X) belongs to $\tau$. Note that a given set admits several topologies. In an intuitive sense, a topology provides a broad notion of proximity or spatial relation between different `parts' of a set. Indeed, a metric\footnote{In mathematics, a metric function $d$ on a set M is a function $d:\mathrm{M}^2\rightarrow\mathbb{R}$ that is symmetric, nondegenerated and obeys the triangle inequality. In physics we use the term metric more generally, as the Minkowski metric does not satisfy nondegeneracy on the lightcone.} function defined on a set canonically provides a topology on such set, called the metric topology, such as the usual n-ball topology in $\rn$. 

The notion of topology provides enough structure as to define continuity of functions, which indeed is the basic notion in topology. We say that a map between two topological spaces  is said to be {\bf continuous} if the inverse image of an open set is an open set. Intuitively, a continuous function maps `nearby' points to `nearby' points. A {\bf homeomorphism} is a continuous bijection with continuous inverse. Two homeomorphic spaces are identical from the topological point of view.

 Another relevant notion is that of Hausdorff space. A topological space is Hausdorff if for any two different points there are two disjoint open sets each containing one of the points. This provides a notion of the {\bf separability} of the space. Any metric space is Hausdorff. 
 
 With these notions, we are ready to define the notion of a manifold. An {\bf $n-$dimensional manifold} $\cm$ is a topological space which is Hausdorff and locally homeomorphic to $\rn$ by a set of maps $\phi_U:\cm\supset O_i\rightarrow B_i\subset\rn$ named {\bf atlas} $\{(\phi_U,O_U)\}$ such that 
\begin{itemize}
\item[1.] The union of all $O_U$ is M.
\item[2.] For any two of these maps $\phi_U(O_{UV})=\phi_V(O_{UV})$, where $O_{UV}\equiv O_U\cap O_V$ are called overlaps.
\item[3.] The transition maps $\phi_{UV}\equiv\phi_U\circ\phi^{-1}_V$ are continuous.
\end{itemize}
The pairs $(\phi_U,O_U)$ are called {\bf charts}, and they provide a way in which the subsets $O_U\subset \mathrm{M}$ can be seen as open sets of $\rn$. It is in this sense that $\cm$ locally looks like $\rn$. If the transition maps are $C^\infty$, this atlas is a $C^{\infty}$-structure in $\cm$, and it defines a {\bf smooth} $n$-dimensional manifold. Two $C^\infty$-structures over $\cm$ are said to be equivalent if the union of all their charts also forms a $C^\infty$-structure on $\cm$. As a curiosity, note that a given topological space can be endowed with several nonequivalent $C^{\infty}$-structures, thus giving rise to different smooth manifolds (see \eg the \href{https://en.wikipedia.org/wiki/Exotic_sphere}{exotic spheres}). From here onwards, the symbol $\cm$ will be used for an n-dimensional smooth manifold as defined above, which is the basic object of differential geometry.

Three basic notions in a smooth manifold are those of coordinates of a point, derivative of a map and curve on $\cm$. Given a chart $(\phi_U,O_U)$, the maps $x^\mu_U\equiv\mathrm{pr}^\mu\circ \phi_U:O_U\rightarrow\mathbb{R}$, where $\mathrm{pr}^\mu$ are the standard projection operators into a basis of $\rn$ and $\mu$ runs from $1$ to $n$, are called coordinate maps of the chart $(\phi_U,O_U)$. For a point $p\in O_U$, the real numbers $x_U^\mu(p)$ are the {\bf coordinates} of $p$ in this chart. Charts can also be denoted as $(\phi_U\sim x_U^\mu,O)$ or simply $(x_U^\mu,O)$, which is called a {\bf coordinate system} or {\bf chart} at $p$. We will use indices indistinctly both, to denote a particular element from the collection of elements indexed the corresponding index, and to denote the whole collection.  Thus, the notation $\phi\sim x^\mu$ means that $x^\mu$ is the collection of coordinate functions associated to the chart $\phi$ by the projectors $\mathrm{pr}^\mu$ of a given basis of $\rn$. Note, however, that in an abuse of the notation we will call $x_U^\mu$ a chart.

We say that $f:\cm\rightarrow\bbr$ is {\bf differentiable} or {\bf smooth} at\footnote{To facilitate readability, we will allow notations like $p\in\cm$ instead of $p\in$ M to denote an element $p$ of a set M endowed with a differentiable structure.} $p\in O_U\subset \cm$ if there is a coordinate system $\phi_U$ such that $\hat{f}_U\equiv f\circ\phi_U^{-1}:U\subset\rn\rightarrow\bbr$ is differentiable at $\phi_U(p)$. We will denote $\cf_p(\cm)$ to the spaces of differentiable functions at $p\in\cm$. If a function is differentiable over all the points of an open subset $O\subset\cm$ we say that it is differentiable at $O$ and belongs to $\cf(O)$. Differentiablility of functions does not depend on the chart. For two smooth manifolds $\cm$ and $\cn$ of dimensions $m$ and $n$ respectively, a map $F:\cm\rightarrow\cn$ is differentiable at $p\in\cm$ if there are two charts $(\phi_U,O_U\subset\cm)$ and $(\varphi_V,O_V\subset\cn)$ containing $p$ and $F(p)$ respectively, the map $\varphi_V\circ F\circ{\phi_U}^{-1}:\bbr^m\rightarrow\bbr^{n}$ is differentiable at $p$ in the usual sense. If furthermore, $F$ is one-to-one and with differentiable inverse, then $F$ is a {\bf diffeomorphism} between $\cm$ and $\cc{N}$. Diffeomorphic manifolds are understood as equivalent manifolds from the point of view of differential geometry.

A (parametrized smooth) {\bf curve} in $\cm$ is a differentiable map $\gamma$ mapping an open subset of the real line $(a,b)$ to $\cm$. The curve is said to pass through $p$ at $t_0\in (a,b)$ if $\gamma(t_0)=p$. A reparametrization for $\gamma$ is a monotonous function $f:\bbr\rightarrow\bbr$ such that $\tilde{\gamma}=\gamma\circ f:(f(a),f(b))\rightarrow\cm$ is a different (parametrized) curve with the same image as $\gamma$. The coordinate representation of $\gamma$ by a coordinate chart $x_U^\mu$ is $\gamma_U^\mu\equiv x_U^\mu\circ\gamma:\bbr\rightarrow\rn$. 

%%%%%%%%%%%%%%%%%%%%%%%
%%%%%%%%%%%%%%%%%%%%%%%
\section{Canonical structures on smooth manifolds}
%%%%%%%%%%%%%%%%%%%%%%%
%%%%%%%%%%%%%%%%%%%%%%%

As seen above, smooth charts allow to define a notion of differentiability of functions in $\cm$ by exploiting the notion of differentiability that we have in $\rn$. However, we have not yet defined any notion of derivative for such functions. The above definitions allow to do so relying exclusively on the smooth structure of the manifold. In this section we will present a bunch of structures that arise canonically on every smooth manifold from their smooth structure. More explicitly, we use the word canonical to imply that the construction is unique and completely determined by the structures already defined, so that no arbitrary choices have to be made and no extra piece of information has to be given. Needles to say, the canonical structures that arise in any smooth manifold are crucial in the formulations of modern physical theories. Let us briefly discuss their properties.

%%%%%%%%%%%%%
\subsection{Tangent space: vectors are derivative operators}\label{sec:TangentSpace}
%%%%%%%%%%%%%

The set of differentiable functions on $\cm$ forms a vector space over $\bbr$ which is a commutative algebra with respect to the product of functions. Hence, the directional derivatives of smooth functions at $x\in\bbr^m$ along $X\in\bbr^m$ can be seen as `algebraic derivations' over the algebra $\cf_x(\bbr^m)$. By `algebraic derivation' we mean a linear operator over an algebra which obeys the Leibniz rule with respect to its product. We will denote the set of directional derivatives at $x$ by $\cd\cf_x(\bbr^m)$. Due to their linearity, linear combinations of any two elements of $\cd\cf_x(\bbr^m)$ will also be in $\cd\cf_x(\bbr^m)$, which makes it a vector space over $\bbr$. From an algebraic perspective, we can consider the set of linear operators over the algebra $\cf_p(\cm)$ which obey the Leibniz rule with respect to the product of functions. Because their algebraic properties are identical to those of $\cd\cf_x(\bbr^m)$, they will also form a vector space over $\bbr$, which we will call $\cd\cf_p(\cm)$.

Having the above algebraic definition of directional derivatives in mind, let us now try to provide an `infinitesimal meaning' from the analysis point of view. To that end, let us define the derivative of a function $\cf(\cm)$ along a curve. Let be a map $f:\cm\rightarrow\bbr$ smooth at $p$ and $\gamma(t)$ a curve through $p$ at $t=t_0$. The derivative of $f$ along $\gamma$ at $p$ is
\begin{equation}
\left.\frac{d}{dt}f\big(\gamma(t)\big)\right|_{t=t_0}
\end{equation} 
which satisfies linearity, Leibniz and the chain rule. It can be seen that curves through $p$ can be classified in equivalence classes, where two curves are equivalent at $p$ if they lead to the same directional derivative of a function at $p$. The set of equivalence classes of curves at $p$ can be seen to form a vector space. Let us argue why this should be viewed as a vector space tangent to $\cm$ at $p$. Given a chart $x_U^\mu$ over $p$, define the operators $(\partial_{U\mu})_p:\cf_p(\cm)\rightarrow\bbr$ as 
\begin{equation}\label{DefCoordDerOp}
	(\partial_{U\mu})_p[f]=\left.\frac{\partial \hat f_U(x_U^1,...,x_U^n)}{\partial x_U^\mu}\right|_{x_U^\mu=\phi_U(p)}.
\end{equation}
With this definition, $(\partial_{U\mu})_p$ satisfy both linearity and Leibniz. Hence they are elements of $\cd\cf_p(\cm)$. This means that any linear combination of the form $X_p=(X_p)^\mu(\partial_{U\mu})_p$, where $X_p^\mu\in\rn$, is also in $\cd\cf_p(\cm)$. By linearity, $X_p$ acts on $\cf_p(\cm)$ as $X_p[f]=(X_p)^\mu (\partial_{U\mu})_p[f]$. By usnig the algebraic properties of $\cd\cf_p(\cm)$ as well as calculus on $\bbr^n$, it can be shown that the $n$ operators $(\partial_{U\mu})_p$ induced by a chart $x_U^\mu$ form a basis of $\cd\cf_p(\cm)$, which therefore has dimension $n$. This basis is called {\bf coordinate basis} of $x_U^\mu$, and the components of a vector $X_p\in\cd\cf_p(\cm)$ can be read off its action on the coordinate functions as $(X_p)^\mu=X_p[x_U^\mu]$. Now, let us deal with the notion of tangent vector to a curve at a point. Let $\gamma(t)$ be a curve such that $\gamma(t_0)=p$. Define the operator $\dot\gamma_p\in\cd\cf_p(\cm)$ as
\begin{equation}
\dot\gamma_p[f]=\left.\frac{d}{dt}f(\gamma(t))\right|_{t=t_0}
\end{equation} 
which given a chart $(\phi_U\sim x_U^\mu,O_U)$ over $p$ can also be written as
\begin{equation}
\dot\gamma_p[f]=\left.\frac{d}{dt}\Big[\big(f\circ\phi_U^{-1}\big)\circ\big(\phi_U\circ\gamma(t)\big)\Big]\right|_{t=t_0}=\left.\left.\frac{\partial\hat f(x_U^1,...,x_U^n)}{\partial x_U^\mu}\right|_{x_U^\mu=\phi_U(p)}\frac{d \gamma_U^\mu(t)}{d t}\right|_{t=t_0}.
\end{equation} 
The components of $\dot\gamma_p$ in the basis $(\partial_{x^\mu})$ are then
\begin{equation}
\dot\gamma_p^U=\dot\gamma_p[x_U^\mu]=\left.\frac{d \gamma_U^\mu(t)}{d t}\right|_{t=t_0}.
\end{equation}
For $\cm=\rn$, the components of the tangent vector to a curve can be read off the directional derivative of a function $f$ along the curve as the coefficients of the partial derivatives of $f$ with respect to the chosen coordinates. This motivates the definition of $\dot\gamma_p$ as the tangent vector of $\gamma(t)$ at $\gamma(t_0)=p$. Now note that we can label the equivalence classes of curves having $\dot\gamma_p$ as tangent vector at $p$ by its tangent, which leads to the conclusion that vector space spanned by these equivalence classes is the {\bf tangent space} of $\cm$ at $p$, usually denoted as $\tpm$. Moreover, since $\dot\gamma\in\cd\cf_p(\cm)$, we have that $\cd\cf_p(\cm)\cong\tpm$. This completes the picture of why tangent vectors to a manifold $\cm$ are nothing but directional derivatives on $\cf(\cm)$.

For any differentiable map between two $m-$ and $n-$dimensional smooth manifolds,  and given a smooth map $\alpha:\cm\rightarrow\cn$, the action of vectors on real valued functions canonically induces a map $\alpha_\ast:\tpm\rightarrow\ct_{\alpha(p)}\cn$ by
\begin{equation}\label{PushForward}
(\alpha_\ast(X_p)\big)[f]\equiv Y_{\alpha(p)}[f]=X_p[f\circ\alpha]
\end{equation}
where $f\in\cf_{\alpha(p)}(\cn)$. This map is called {\bf tangent map}, {\bf differential} or {\bf pushforward} of $\alpha$. Given two charts $(x^\mu_U,O_U\subset\cm)$ and $(y_V^\nu,O_V\subset\cn)$ such that $p\in O_U$ and $\alpha(p)\in O_V$, the components of $\alpha_\ast(X_p)$ read 
\begin{equation}
 \big(\alpha_\ast(X_p)\big)^\nu=\left.(X_p)^\mu\frac{\partial y_V^\nu}{\partial x_U^\mu}\right|_{x_U^\mu(p)}.
\end{equation}
Note that if $\cm$ and $\cn$ were both Euclidean, the action of $\alpha_\ast$ on $X_p$ is basically the right-multiplication by the Jacobian of $\alpha$. Hence this map generalises the notion of total derivative of $\alpha$ to general manifolds.

As a final remark, let us point out that the above constructions can be done at any point on the manifold. For instance, the operators $\partial_{U\mu}$ are smooth vector fields in $O_U$ and so on. Therefore, there is a tangent space to every point in $\cm$, which will allow to define further canonical structures stemming solely from the smooth structure of $\cm$. A vector field such that its coefficients in a basis are smooth functions over $O\subset\cm$ is called a {\bf smooth vector field} over $O$. Unless stated otherwise, we will refer to smooth vector fields over $\cm$ just as vector fields from here onwards. We will denote the space of smooth vector fields over $O\subset\cm$ as $\Gamma(\ct O)$. The meaning of $\ct O$ will later be clarified, but we can just advance that it is the piece of the tangent bundle $\tb$ above $O\subset\cm$.

%%%%%%%%%%%%%
\subsection{Lie bracket, flows and Lie derivatives}
%%%%%%%%%%%%%
Vector fields allow us to introduce a new notion of derivative on a manifold: the Lie derivative. The aim of this section is to canonically build several objects based on the properties of vector fields to reach an understanding of the Lie derivative. The first of this objects is the Lie bracket. Given two smooth vector fields $X$ and $Y$ in $\Gamma(O)$, their {\bf Lie bracket} is a third vector field $[X,Y]$ defined by its action on $f\in\cf(O)$ as
\begin{equation}\label{DefLieBracket}
[X,Y][f]=X\big[Y[f]\big]-Y\big[X[f]\big]
\end{equation}
As any other vector field, the Lie bracket acts linearly on $\cf(O)$. Furthermore, it satisfies the Jacobi identity, antisymmetry, linearity both in $X$ and $Y$, and also has the property $[X,f Y]=f[X,Y]+X[f]Y$. An important use of the Lie bracket is the characterisation of coordinate basis. Given a set of $k$ linearly independent vector fields $\{X_1,...,X_k\}$, these are elements of a coordinate basis if and only if they all have vanishing Lie brackets between them, namely iff $[X_a,X_b]=0$ for any pair $(a,b)$.

Given a fixed $X\in\Gamma(O)$, the Lie bracket defines a map $\cl_X:\Gamma(O)\rightarrow\Gamma(O)$ by $\cl_X (Y)\equiv\cl_X Y=[X,Y]$. Note that by defining also an action of $\cl_X$ on $f\in\cf(O)$ by $\cl_X(f)\equiv \cl_X f=X[f]$ this map has a `Leibnitz-like' property 
\begin{equation}
\cl_X (fY)=f \cl_X Y+(\cl_X f)Y.
\end{equation}
We will later see that this `Leibniz-like' property is a proper Leibniz rule with respect to a given algebra (the tensor algebra) to be introduced later, which makes this map an algebraic derivation. As well, we will see that it also has an infinitesimal meaning related to the rate of change of a vector field in a given direction, thus making sense as a derivative also from the analysis perspective. To that end we need to introduce the concept of flow associated to a vector field. Given a vector field $X\in\Gamma(O)$, the {\bf integral curve} of $X$ in $O\subset\cm$ is a parametrised curve $\gamma:(a,b)\rightarrow O$ whose tangent vector at each point $p\in\cm$ is $X_p$. The existence and uniqueness theorem of ordinary differential equations ensures that through each point there is only one (maximal) integral curve of $X$. A {\bf one-parameter group of transformations} or {\bf flow} is a map $\sigma:\bbr\times\cm\rightarrow\cm$ such that for each $t\in\bbr$ we have a diffeomorphism $\sigma_t:\cm\rightarrow\cm$ and $\sigma$ preserves the real sum, \ie $\sigma_{t+s}=\sigma_t\circ\sigma_s$. Due to the bijectivity of $\sigma_t$ and the preservation of the real sum, $\sigma_t^{-1}=\sigma_{-t}$ and $\sigma_0=\mathrm{id}_{\cm}$. The action of the flow on a given point $p\in\cm$ generates a parametrized curve $\gamma_p(t)=\sigma_t(p)$ such that $\gamma_p(0)=p$ called {\bf orbit} of $p$ under $\sigma$. Any vector field whose integral curves are the orbits of a flow $\sigma$ on $\cm$ is called a {\bf complete} vector field generating $\sigma$. Although not all vector fields are complete, they always generate a {\bf local} flow. This allows to prove that, for any vector field $X$, there is always a local coordinate system such that $X$ is an element of the coordinate basis.

Let $\sigma$ be the local flow on $\cm$ generated by the smooth vector field $X$. Then we can define the {\bf Lie derivative} of $Y$ with respect to $X$ as
\begin{equation}\label{DefLieDer}
\lim_{t\to0}\frac{Y-(\sigma_t)_\ast Y}{t}.
\end{equation}
Hence, the Lie derivative is a way to measure how a vector $Y$ changes along the integral curves of $X$ when push-forwarded with the flow generated by $X$. It can be shown that this limit equals to $[X,Y]=\cl_X Y$. Therefore, this shows that the algebraic operator defined above is also a derivative in the sense of analysis, as it describes a rate of change. 

%%%%%%%%%%%%%
\subsection{Co-tangent space, 1-forms, tensor fields and tensor algebra}
%%%%%%%%%%%%%
Using the smooth structure of $\cm$, we have been able to define a vector space at each of its points, namely $\tpm$. Any vector space canonically defines a dual space as well as a tensor algebra. In this section we will define the canonical dual and tensor spaces that stem solely from the smooth structure of $\cm$.

The {\bf cotangent space} at a point $p\in\cm$, dubbed as $\ctpm$, is the space of $\bbr$-valued linear operators on $\tb$, namely its dual space, which as always has the same dimension as $\tpm$. The elements of $\ctpm$ are called {\bf 1-forms} or {\bf covectors} at $p$. The double dual space to any finite dimensional vector space is canonically isomorphic to $V$, so that vectors at $p$ act canonically on 1-forms at $p$ by $X_p[\xi_p]=\xi_p[X_p]$ where $X_p\in\tpm$ and $\xi_p\in\ctpm$. 

As for $\tpm$, the cotangent space exists in every point of a smooth manifold. A map that assigns a 1-form in $\ct_p^\ast O$ at each $p\in O\subset\cm$ such that its action on any (smooth) vector field is a smooth function $O\rightarrow \bbr$ is called {\bf smooth 1-form field} over $O$. The space of smooth 1-form fields over $O$ is dubbed as $\Gamma(\ct^\ast O)$. 1-forms that are smooth over all $\cm$ will simply be called 1-forms. 

Given a chart $(x_U^\mu,O_U)$, the maps $\dif{x_U^\mu}: \cm\mapsto \Gamma(\ct^\ast O_U)$ are local 1-form fields that provide a local basis in $\ctpm$ for every $p\in O_U$. This basis is {\bf dual} to the coordinate basis of $\tpm$ as $\dif x_U^\mu[\partial_{U\nu}]=\delta^\mu{}_\nu$. By linearity, we have that if $X=X_U^\mu\partial_{U\mu}$ then we can read its components in the given chart by $\dif x_U^\mu[X]=X^\mu$.

Using the notion of pushforward map from the previous section, any smooth function $\alpha:\cm\rightarrow\cn$ defines a map between $\alpha^\ast:\ct_{\alpha(p)}\cn^\ast\rightarrow \ct_p\cm$ as 
\begin{equation}
\big(\alpha^\ast(\xi_{\alpha(p)})\big)[X_{p}]=\xi_{\alpha(p)}[\alpha_\ast (X_p)].
\end{equation}
This map is called {\bf pullback} map associated to $\alpha$. Unlike the pushforward, which might fail to be well defined for vector fields, the pullback map is always well defined for 1-form fields.

Some 1-forms can be understood as total derivatives of real-valued functions on $\cm$\footnote{In contexts where the Poincar\'e Lemma applies, all one forms can be understood as such.} in the following sense. There is a map $\dif: \cf(O)\rightarrow\ct^\ast O$ called {\bf differential} that associates a smooth 1-form over $O$ to every smooth function $f:\cm\rightarrow\bbr$. This 1-form dubbed $\dif f$ is defined by its action on vector fields as 
\begin{equation}
\dif f [X]=X[f].
\end{equation}
and is called {\bf differential} of f. When acting on a vector field, it gives the directional derivative of $f$ along $X$, as a generalisation of the gradient operator in $\rn$-calculus. Given a chart $x_U^\mu$, the maps $\dif x^\mu_U$ defined above are just the differentials of the coordinate functions $x^\mu_U$. Given a chart $x_U^\mu$, it is possible to see from the above definition that the components of $\dif f=\dif f_{U\mu} \dif x_U^\mu$ are
\begin{equation}
\dif f_{U\mu} = \frac{\partial \hat f_U}{\partial x_U^\mu}
\end{equation}

Having defined $\ctpm$, we can consider the space of real-valued multilinear operators acting on $r$ vectors and $s$ 1-forms. We call this the space of $r$-contravariant $s$-covariant tensors, tensors of rank $(r,s)$ or simply $(r,s)$ {\bf tensors} at $p$, denoted by $\ct_p^{(r,s)}\cm$. This space inherits the vector space structure from the tangent space and its dual and it has dimension $n^{(r+s)}$. Together with the tensor product 
\begin{equation}
\otimes: \ct_p^{(r,s)}\cm\times\ct_p^{(m,n)}\cm\rightarrow \ct_p^{(r+m,s+n)}\cm,
\end{equation}
which is linear and associative, the direct sum of all tensor spaces on $p$ forms an associative (graded) algebra named {\bf tensor algebra} at $p$. A map that assigns an $(r,s)$ tensor at each $p\in O$ such that it acts smoothly on all (smooth) vector and 1-form fields is a {\bf smooth tensor field} over $O$. The space of $(r,s)$ tensor fields over $O$ will be denoted by $\ct^{(r,s)}O$. A tensor field which is smooth over all $\cm$ will be called just tensor field. Note that $\ct^{(1,0)}O$ and $\ct^{(0,1)}O$ are $\Gamma(\ct O)$ and $\Gamma(\ct^\ast O)$ respectively.

 Given a chart $x_U^\mu$, the set of all elements of the form $\partial_{U\mu_1}\otimes\overset{r }{...}\otimes\partial_{U\mu_r}\otimes\dif x_U^{\nu_1}\otimes\overset{s}{...}\otimes \dif x_U^{\nu_s}$ forms a basis of $ \ct_p^{(r,s)}\cm$ at each $p\in O_U$. Hence, an $(r,s)$ tensor field $T$ can then be written as
\begin{equation}
T=T_U{}^{\mu_1...\mu_r}{}_{\nu_1...\nu_s} \partial_{U\mu_1}\otimes\overset{r }{...}\otimes \partial_{U\mu_r}\otimes\dif x_U^{\nu_1}\otimes\overset{s }{...}\otimes \dif x_U^{\nu_s},
\end{equation}

and $T_U{}^{\mu_1...\mu_r}{}_{\nu_1...\nu_s}\equiv T\big(\partial_{U\mu_1}...\partial_{U\mu_r}\dif x_U^{\nu_1}...\dif x_U^{\nu_s}\big)$ are its components on that basis. The components of the tensor product of $T\in \ct^{(r,s)}O$ and $S\in \ct^{(p,q)}O$ has components $$(T\otimes S)_U{}^{\mu_1...\mu_{r+p}}{}_{\nu_1...\nu_{s+q}}=T_U{}^{\mu_1...\mu_r}{}_{\nu_1...\nu_s}S_U{}^{\mu_{r+1}...\mu_{r+m}}{}_{\nu_{s+1}...\nu_{s+q}}.$$ 

An important operation on the tensor algebra is the {\bf contraction} $C^i_j:\ct_p^{(r,s)}\cm\rightarrow\ct_p^{(r-1,s-1)}\cm$ where $1\leq i\leq r$ and $1\leq j \leq s$ and such that 
\begin{equation}
(C^i_j T_p)^{\tilde  \mu_1...\tilde \mu_{r-1}}{}_{\tilde \nu_1...\tilde \nu-1}=T_p{}^{\mu_1...\mu_{i-1}\alpha\mu_{i+1}...\mu_r}{}_{\nu_1...\nu_{j-1}\alpha\nu_{j+1}...\nu_{s}}
\end{equation}
where $(\tilde  \mu_1...\tilde \mu_{r-1})=(\mu_1...\mu_{i-1}\mu_{i+1}...\mu_r)$ and $(\tilde  \nu_1...\tilde \nu_{s-1})=(\nu_1...j_{j-1}\nu_{j+1}...\nu_s)$.

The Lie derivative along a vector $X$ can be extended to tensor fields by suitably generalising the pushforward map of the flow generated by $X$ (see \eg \cite{Szekeres}). This leads to the notion of Lie derivative of any tensor field on $\cm$, which satisfies the Leibniz rule with respect to the tensor product as anticipated above.  In a chart $x^\mu$ the components of the Lie derivative of $T\in\ct^{(r,s)}O$ with respect to $X\in\Gamma(\ct O)$ can be written as
\begin{equation}
\begin{split}
&(\cl_X T_U)^{\mu_1...\mu_r}{}_{\nu_1...\nu_s}=\frac{\partial (T_U)^{\mu_1...\mu_r}{}_{\nu_1...\nu_s}}{\partial x_U^\alpha}(X_U)^\alpha-(T_U)^{\alpha...\mu_r}{}_{\nu_1...\nu_s}\frac{(X_U)^{\mu_1}}{\partial x_U^\alpha}-...\\&-(T_U)^{\mu_1...\alpha}{}_{\nu_1...\nu_s}\frac{(X_U)^{\mu_r}}{\partial x_U^\alpha}+(T_U)^{\mu_1...\mu_r}{}_{\alpha...\nu_s}\frac{(X_U)^{\alpha}}{\partial x_U^{\nu_1}}+...+(T_U)^{\mu_1...\mu_r}{}_{\nu_1...\alpha}\frac{(X_U)^{\alpha}}{\partial x_U^{\nu_s}}
\end{split}
\end{equation}
As for vectors, $\cl_X T$ is a tensor of the same degree as $T$. The Lie derivative is a derivation in the algebraic sense acting on the tensor algebra.

Now that we have defined tensor fields, of which 1-forms and vectors are an instance, we know that they have been defined in a coordinate invariant way as linear operators on the tangent and cotangent spaces. However, although they are invariant objects, their components in a given coordinate frame, by definition, depend on the coordinates employed. To find out how the components of a tensor in different coordinate basis on an overlap $O_{UV}\subset\cm$ are related, we shall make use of their covariance. Under a change of coordinates $x_V^\mu(x_U^\alpha)$ on $O_{UV}$, by definition of differential, the coordinate dual basis vectors change as
\begin{equation}
\dif x_
V^\mu=\frac{\partial x_V^\mu}{\partial x_U^\nu}\dif x_U^\nu
\end{equation}
where  $x_U^\mu(y_V^\alpha)$ only makes sense on the overlap $O_{UV}$. Since we know that $\dif x_V^\mu(\partial_{V\nu})=\delta^\mu{}_\nu$ the coordinate basis vectors must change with the inverse transformation
\begin{equation}
\partial_{V\mu}=\frac{\partial x_U^\nu}{\partial x_V^\mu}\partial_{U\nu}
\end{equation}
and if tensor fields are invariant under coordinate changes, their components must transform covariantly as
\begin{equation}\label{eq:TransformationTensorComponents}
T_V{}^{\mu_1...\mu_r}{}_{\nu_1...\nu_s}=\frac{\partial x_V^{\mu_1}}{\partial x_U^{\alpha_1}}\;\overset{r}{...}\;\frac{\partial x_V^{\mu_r}}{\partial x_U^{\alpha_r}}\frac{\partial x_U^{\nu_1}}{\partial x_V^{\beta_1}} \;\overset{s}{...}\;\frac{\partial x_U^{\nu_s}}{\partial x_V^{\beta_s}} T_U{}^{\alpha_1...\alpha_r}{}_{\beta_1...\beta_s}.
\end{equation}
As a remark, let us point out that it is common to introduce tensors as objects with indices that transform this way, but then their invariant nature under coordinate changes results rather obscure.

%%%%%%%%%%%%%
\subsection{Exterior algebra and differential forms}
%%%%%%%%%%%%%
Tensors which exhibit symmetry or antisymmetry under permutations of their arguments can form linear subspaces of the tensor algebra. Consider now the set $\Lambda^s_p\cm$ of totally antisymmetric $(0,s)$ tensors. This is a linear subspace of $\ct^{(0,s)}_p\cm$ with dimension $\binom{n}{s}$. We can also define the space of all totally antisymmetric linear operators on $\tpm$ as the direct sum of all\footnote{Note that $(\Lambda_{s>n})_p(\cm)=\{ 0\}$.} $\Lambda^s_p\cm$. This set is a linear subspace of the tensor algebra with dimension $2^n$, but it is not a subalgebra with respect to $\otimes$ because it is not closed under the tensor product. We can define the {\bf exterior product} 
\begin{equation}
\wedge:\Lambda^s_p\cm\times\Lambda^t_p\cm\rightarrow\Lambda^{s+t}_p\cm\quad \text{as} \quad T\wedge S= \mathrm{A}(T\otimes S),
\end{equation} 
where $\mathrm{A}(T)$ is the projection of any $T\in \ct_p^{(0,s)}\cm$ into $\Lambda^s_p\cm$, \ie the antisymmetrization operator. This operation is associative, linear and also has the property $\alpha_p\wedge\beta_p=(-1)^{rs}\beta_p\wedge\alpha_p$ where $\alpha_p$ and $\beta_p$ are $r$- and $s$-forms respectively. This product gives $\Lambda_p\cm$ the structure of a graded algebra called {\bf exterior algebra} of $\tpm$. Proceeding analogously to tensor fields, a map that assigns a (rank $k$) element of the exterior algebra for each $p\in O\subset\cm$ in a smooth way is a differential $k$-form field over $O$ and belongs to $\Lambda^k O$. The space of all differential forms over $O$ is denoted as $\Lambda O$. If the $k$-form field is smooth over all $\cm$ we will call it just $k$-form.

Given a chart $x_U^\mu$, it can be verified that  $\dif x_U^{\mu_1}\wedge...\wedge\dif x_U^{\mu_s}$ is a basis of $\Lambda^sO_U$. A $k$-form $\alpha$ can thus be written as
\begin{equation}
\alpha=\alpha_U{}_{\mu_1,...,\mu_k}\dif x_U^{\mu_1}\wedge...\wedge\dif x_U^{\mu_k}
\end{equation}
where $\alpha_U{}_{\mu_1,...,\mu_k}\equiv\alpha_U{}_{[k_1,...,k_s]}$ are totally antisymmetric coefficients called components of $\alpha$ in that basis. The differential operator defined above to act on $\cf(O)=\Lambda^0 O$ can be naturally extended to act over $\Lambda O$. We define the {\bf differential} or {\bf exterior derivative} as 
\begin{equation}
\dif:\Lambda O\rightarrow\Lambda O\quad \text{such that}\quad \dif\big(\Lambda^k O)\subseteq\Lambda^{k+1}O
\end{equation}
and it has the following properties:
\begin{itemize}
\item[1.] Linearity in each $\Lambda^k O$, this is $\dif(\alpha+\beta)=\dif \alpha+\dif \beta$ for any two $k$-forms $\alpha$ and $\beta$.
\item[2.] $\dif f[X]=X[f]$ for any vector field $X\in\Gamma(\tb)$ and scalar function $f\in\cf O$.
\item[3.] $\dif^2 f=0$ for any scalar function $f$.
\item[4.] $\dif(\alpha\wedge\beta)=\dif\alpha\wedge\beta+(-1)^k \alpha\wedge\dif\beta$ for any $\alpha\in\Lambda^kO$ and $\beta\in\Lambda O$.
\end{itemize}
 This last `sort of Leibniz rule' turns $\dif$ into an {\bf anti-derivation} on $\Lambda O$ in the algebraic sense. An important property that follows from these properties is that $\dif^2\beta=0$ for any differential form. Given a chart $x_U^\mu$ and a $k$-form $\alpha$ over $O_U$, the components of $\dif\alpha$ in such chart are
\begin{equation}
\dif\alpha_U^{\mu_1...\mu_{k+1}}=(-1)^k\frac{\partial\alpha_U{}_{\mu_1...\mu_{k}}}{\partial x^{\mu_{k+1}}}\dif x_U^{\mu_1}\wedge...\wedge\dif x_U^{\mu_{k+1}}
\end{equation}
From the respective definitions, it can be seen that both the pullback of a map $\phi:\cm\rightarrow\cn$ and the Lie derivative commute with the exterior differential
\begin{equation}
\dif (\phi^\ast\alpha)=\phi^\ast(\dif\alpha) \quad \text{and} \quad \cl_X(\dif\alpha)=\dif (\cl_X\alpha).
\end{equation}
 Another operation that we can define on the exterior algebra is the interior product. Given a vector field $X\in\Gamma(\tb)$, the {\bf interior product} with $X$ is the map 
 \begin{equation}\label{eq:InteriorProduct}
\mathrm{i}_{X}:\Lambda O\rightarrow\Lambda O \quad \text{such that}\quad \mathrm{i}_X\big(\Lambda^k O)\subseteq\Lambda^{k-1}O
\end{equation}
defined by $\mathrm{i}_{X} \alpha=C^1_1 (X\otimes\alpha)$. This is also an antiderivation in the sense that it satisfies 
\begin{equation}
\mathrm{i}_{X}(\alpha\wedge\beta)_p=\mathrm{i}_X\alpha\wedge\beta+(-1)^k \alpha\wedge\mathrm{i}_X\beta
\end{equation}
where $\alpha$ is an $r$-form and $\beta$ any differential form. An important property of the interior product is 
\begin{equation}
\cl_X=\mathrm{i}_X\circ \dif+\dif\circ\mathrm{i}_X\end{equation} 
for any vector $X\in\Gamma(\tb)$. 

%%%%%%%%%%%%%%%%%%%%%%%%%%%%%%%%
\subsection{Tangent, cotangent and tensor bundles}
%%%%%%%%%%%%%%%%%%%%%%%%%%%%%%%%

There is a very interesting class of manifolds whose main characteristic is that they look locally like the cartesian product of two manifolds called (fiber) bundles. These objects are the natural arena where connections live. We shall see below that the smooth structure of any manifold canonically provides several bundles related to the manifold which stem solely from its smooth structure. Among them, we will be interested in the tangent and co-tangent bundles and the frame bundle. The tangent and co-tangent bundles are instances of vector bundles which are somewhat simpler than principal bundles, the class of bundles to which the frame bundle belongs. Let us then start by describing the tangent bundle, whose features will let us gain some intuition for the later definitions of general vector and principal bundles.

The tangent bundle of $\cm$, dubbed $\tb$, as the set of all tangent vectors to $\cm$, \ie the disjoint union of all the tangent spaces to $\cm$. A point in $\tb$ is of the form $\tb\ni P=(p,X)=X_p$. Therefore, there is a canonical (smooth) map $\pi:\tb\rightarrow\cm$ such that $\pi(p,X)=p$ called  projection that maps all the vectors tangent to a given point in $\cm$ to that point. Therefore we have that $\pi^{-1}(p\in\cm)=\tpm$, which is called the fiber over $p$. This guarantees the existence of local diffeomorphisms $\Phi_U:O_U\times\rn\rightarrow\pi^{-1}(O_U)$, called trivialising functions over each trivialising patch $O_U\subset\cm$, which allow to see $\tb$ locally as a product space. This provides $\tb$ the structure of a vector bundle over $\cm$ as will be clarified below.

With the help of these local diffeomorphisms, coordinates $x_U^\mu$ in $O_U\subset\cm$ provide coordinates in $\pi^{-1}(O_U)\subset\tb$ by the smooth map $(x_U^\mu,X_U^\mu)\circ\Phi_U:\pi^{-1}(O_U)\rightarrow\bbr^{2n}$, where $(x_U^\mu,X_U^\mu)$ maps the point $(p,X)\in O_U\times\rn$ to the ordered $2n$-tuple $\lr{x_U^\mu(p),X^\mu_U(p)}$ formed by the $n$ coordinates of the point $p$ and the $n$ components of the vector $X_p$ in the coordinate basis $\partial_{U\mu}$ at $p$, which are called the fiber coordinates of $(p,X)$ in the trivialisation $\Phi_U$. In an overlap $O_{UV}$, where $x_V^\mu$ are smooth functions of $x_U^\mu$, we have that the fiber coordinates associated to the trivialisations $\Phi_U$ and $\Phi_V$ relate as
\begin{equation}
X_V^\mu=\frac{\partial x_V^\mu}{\partial x_U^\nu} X_U^\nu
\end{equation}
 which is a smooth and invertible transformation law as it is given by the Jacobian of the coordinate transformation. Note that this is just the transformation law for components of a vector field under a change of coordinate frame \eqref{eq:TransformationTensorComponents}. The Jacobian here plays the role of transition function between trivialising patches. Since the transition functions (Jacobians) belong to the matrix group\footnote{We will write $GL_n$ or $GL_V$ to denote the group of changes of frames on an n-dimensional vector space V. We will generally refer to real vector spaces, but most results apply to complex vector spaces as well.} $GL_n$, we say that $GL_n$ is the structure group of $\tb$. 
 
 With this definitions, we see that the smooth structure of $\cm$ canonically defines $\tb$ and provides a smooth structure to it, which makes it into a $2n$-dimensional manifold. Note that a vector field can be seen as a map from $\cm$ to $\tb$ which maps every point $p\in\cm$ to an element of the fiber above that point $\pi^{-1}(p)$ in a smooth fashion. Such map is called a (cross) section of $\tb$, which is indeed a vector field on $\cm$. The space of sections of the tangent bundle is denoted as $\Gamma(\tb)$.

This construction can be adapted straightforwardly to the cotangent spaces at each point in $\cm$, leading to the co-tangent bundle $\ctb$, as well as to the spaces of $(p,q)$ tensors at each point in $\cm$, which leads to the $(p,q)$ tensor bundle $\ct^{(p,q)}\cm$. However, both the cotangent and $(p,q)$ tensor bundles can also be canonically constructed in a more elegant fashion by applying the associated bundle construction to $\tb$ (see below). A section of the co-tangent bundle is a 1-form field on $\cm$, and a section of the $(p,q)$ tensor bundle is a $(p,q)$ tensor field on $\cm$. Although it would be more correct to call $\Gamma(\ct^{(p,q)}\cm)$ to the space of tensor fields on $\cm$, we will abuse of the notation and use the symbol $\ct^{(p,q)}\cm$ for both the $(p,q)$ tensor bundle and for the space of $(p,q)$ tensor fields over $\cm$ to simplify the notation.

\subsection{Frame bundle over $\cm$}
Consider now the collection of all frames of tangent vectors at each point $p\in\cm$, dubbed $F\cm$, and the (smooth) projection map $\pi:F\cm\rightarrow\cm$ assigning the point $p$ to every frame of $\tpm$. The fiber $\pi^{-1}(p)$ consists of all frames in $\tpm$. In contrast to the elements of $\tb$ (vectors), which can be acted upon by $GL_n$ only when a particular basis has been chosen, the group of $n\times n$ invertible matrices acts naturally on frames. Furthermore, this action is fiber-preserving, since it transforms a frame at $p$ into another frame at $p$, thus mapping each fiber onto itself. This mapping is one-to-one and, therefore, given a particular element $\textbf{e}\in\pi^{-1}(p)$ (a frame at $p$) we can write any other element $\textbf{f}$ of the fiber above $p$ as $\textbf{f}=\textbf{e} g$ for some $g\in GL_n$. Therefore, we can identify each fiber in $F\cm$ with its structure group $GL_n$, which acts on the right on each fiber by its action on itself. This provides $F\cm$ with a structure of principal $GL_n$ bundle, as will be clarified below. 

The local trivialisation of the frame bundle is made explicit by the diffeomorphisms $\Phi_U:O_U\times GL_n\rightarrow\pi^{-1}(O_U)$, which define the identity section $\sigma_U$ associated to the trivialisation $\Phi_U$ as $\sigma_U(p)=\Phi_U(p,e)$ where $e$ is the identity element of $GL_n$. Thus, we have that $\Phi_U(p,g)=\sigma_U(p)g$. There is an associated diffeomorphism $\Phi_{Up}:GL_n\rightarrow\pi^{-1}(p)$ mapping the abstract fiber $GL_n$ to each fiber $\pi^{-1}(p)$ by $\Phi_{Up}(g)=\Phi_U(p,g)$. In an overlap $O_{UV}$ we have that $\Phi_{Up}(g)=\sigma_U(p)g$ and $\Phi_{Vp}(g)=\sigma_V(p)g$ which are generally different elements of $F\cm$ but lay on the same fiber $\pi^{-1}(p)$. Therefore, we know that $t_{UV}(p)=\Phi_{Up}\circ\Phi^{-1}_{Vp}:GL_n\rightarrow GL_n$ must be an element of $GL_n$ called transition function such that $\Phi_V(p,g)=\Phi_U(p,t_{UV}(p)g)$. We say that $g$ are the fiber coordinates of $\Phi_U(p,g)$ in the local trivialisation $O_U\times GL_n$. As well, the transition functions satisfy $t_{UV}=t_{VU}^{-1}$ and $t_{UV}t_{VW}t_{WX}=t_{UX}$.

%%%%%%%%%%%%%%%%%%%%%%%
\section{Bundles and connections}
%%%%%%%%%%%%%%%%%%%%%%%
The tangent, co-tangent and frame bundles are examples of the various fiber bundles that are canonically associated to any smooth manifold. In those instances, the original manifold acts as the base space of the bundle, and the tangent spaces at each point as the fibers. These notions can be generalised to construct general fiber bundles over a manifold. Since principal bundles are the natural spaces where connections live, understanding their properties will help us in understanding what are general affine connections and why they act the way they do on the different physical fields.
\subsection{Vector and frame bundles}\label{sec:VectorFrameBundles}

A (smooth) {\bf fiber $G$-bundle}, usually denoted as $(B,\pi,\cm,F,G)$, consists of three smooth manifolds $\cm$, $B$ and $F$ respectively called {\bf total (or bundle) space, base space and (abstract) fiber}; a Lie group $G$ that acts on the left on the abstract fiber $F$ and is called {\bf structure group}; and a (smoth) projection map $\pi:B\rightarrow\cm$ such that $\pi^{-1}(p)$ is diffeomorphic to $F$ for any point $p\in B$. We say that $\pi^{-1}(p)$ is the {\bf fiber above} $p\in B$. For an open cover of $\cm$ by $\{O_U\}$ there exist {\bf local trivialisation} diffeomorphisms $\Phi_U:O_U\times F\rightarrow \pi^{-1}(O_U)$, and we say that the point $\Phi_U(p,f)$ has the {\bf fiber coordinates} $f$ on such trivialization. Naively, we shall see trivializations in a similar way as we see coordinates on a regular manifold: they allow to express a point on the bundle in a more familiar way which we know how to operate with. On an overlap $O_{UV}$, the diffeomorphisms $\Phi_{Up}:F\rightarrow\pi^{-1}(p)$ defined by $\Phi_{Up}(f)=\Phi_U(p,f)$ define the transition functions by $t_{VU}(p)f=\phi_{Up}^{-1}\circ\phi_{Vp}(f)$, where $t_{UV}$ is an element of $G$ which corresponds to the transformation of $F$ given by $\phi_{Up}^{-1}\circ\phi_{Vp}$. The fiber coordinates satisfy $\Phi_{Vp}(f)=\Phi_{Up}\lr{t_{VU}(p)f}$, and the transition functions must satisfy 
\begin{equation}\label{eq:ConditionsTransitionFunctions}
t_{UV}=t_{VU}^{-1} \qquad\text{and}\qquad t_{UV}t_{VW}t_{WX}=t_{UX}.
\end{equation}
Note that given a base space and a fiber, their cartesian product forms a trivial bundle. Any fiber bundle which admits a global trivialisation of the form $\cm\times F$ is said to be trivial.

A {\bf (cross) section} of $B$ is a map $\sigma:\cm\rightarrow B$ such that $\sigma\circ\pi=id(\cm)$. Hence, it associates to a point $p\in B$ an element of the fiber $\pi^{-1}(p)$ in a smooth way. The reason why it is called a cross section is because it `goes across each fiber' exactly one time as the argument travels through the base space. All these constructions apply to any fiber bundle, but we are mostly interested in bundles with fibers of particular types, namely vector spaces and Lie groups.

A  {\bf rank $k$ vector bundle} is a fiber bundle $E$ with fiber $\bbr^k$ (or any k-dimensional vector space) equipped with its standard basis and with the structure group being $GL_k$. The local trivialisations are diffeomorphisms $\Phi_U:O_U\times\bbr^k\rightarrow\pi^{-1}(O_U)$. Given a frame, the way in which the structure group acts on the fibers is by matrix multiplication on its elements. Hence, given two local frames $\textbf{e}_U=\lr{\textbf{e}_{U1},...,\textbf{e}_{Uk}}$ and $\textbf{e}_V=\lr{\textbf{e}_{V1},...,\textbf{e}_{Vk}}$ above the trivialising patches $O_U$ and $O_V$, the components of an element of $X\in\pi^{-1}(O_{UV})$ above the overlap will satisfy $X_V{}^a=t_{VU}{}^a{}_b X_U{}^b$ if the two frames are related by $\textbf{e}_{Va}= \textbf{e}_{Ub} t_{VU}{}^b{}_a$ where $t_{UV}\subseteq GL_k$. 

In general, rank $k$ vector bundles naturally have $GL_k$ as their structure group and if it is not explicitly stated otherwise, a rank $k$ vector bundle has structure group $GL_k$. However, if extra structure (a $G$-structure) is specified such that the structure group can be reduced to a linear subgroup  $G\subset GL_k$ then we say that $E$ is a rank $k$ vector  $G$-bundle and denote it by $E_G$. In a vector $G$-bundle, there is a preferred class of frames called {\bf $G$-frames} such that in an overlap $O_{UV}$ two such frames $\textbf{e}_U$ and $\textbf{e}_V$ are related by an element of $G$ as $\textbf{e}_{Va}=\textbf{e}_{Ub} g^b{}_a$ with $g^b{}_a\in G$. 

We will generally use latin indices for fiber coordinates and components of the elements of a general bundle and greek indices for coordinates on $\cm$ and components of elements of $\tb$ and $\ctb$ and the associated tensor spaces. A useful concept that allows to relate rank $n$ vector bundles over $\cm$ to the tangent bundle $\tb$ is that of a (vector bundle) {\bf soldering form}, which is basically a linear isomorphism `gluing' the abstract fiber of $E$ to the tangent bundle. The soldering can be seen either as a vector valued 1-form over $\tb$ or as a vector valued 1-form on $E$. Therefore its components will have a latin index and a greek one, allowing to relate the components of a vector in any $n$-vector bundle in a given frame to the components of the tangent bundle in a coordinate basis.\footnote{The structure group of the tangent bundle is a representation of the diffeomorphism group on $\cm$ (which can be seen as the group of changes of coordinates on $\cm$). Strictly, this is why we cannot consider arbitrary $G$-frames on $\tb$. To do that, one has to 'glue' an abstract rank $n$ vector $G$-bundle over $\cm$ to $\tb$ through the soldering form, which gives a canonical map of arbitrary $G$-frames from such vector $G$-bundle to $\tb$.} In particular, if the solder form is given by $\fr_U{}_a{}^\alpha$ it defines a frame $\fr_U{}_a$ from the coordinate frame $\partial_U{}_\mu$ by $\fr_U{}_a\equiv\fr_U{}_a{}^\alpha\partial_U{}_\alpha$. The inverse isomorphism will be denoted by $\df_U{}^a{}_\alpha$ and will define the dual frames $\df_U{}^a=\df_U{}^a{}_\alpha \dif x_U^a$.

The rank $k$ vector bundle $E^\ast$ over the same base space and with the same structure group as $E$ but with transition functions $t^\ast_{UV}=(t_{UV}^{-1})^\top$ is called the {\bf dual bundle} of $E$. As well, two vector $G$-bundles $E_G$ and $E'_G$ with ranks $k$ and  $k'$ which share base space allow to define the {\bf tensor product bundle} as the rank $k k'$ vector $G$-bundle with transition functions $t_{UV}\otimes t'_{UV}$. We will see that these bundles can also be built from $E$ via the associated bundle construction. Note that the co-tangent bundle is the dual bundle of $\tb$, and the bundle of $(p,q)$ tensors over $\cm$ is the tensor product bundle $\tb\otimes\overset{p}{...}\otimes\tb\otimes\ctb\otimes\overset{q}{...}\otimes\ctb$.

We will see that physical fields are sections over various vector bundles that can be built above the spacetime manifold. Thus, the concept of section in a vector bundle allows to generalise the concept of vector (and more general) fields on $\cm$ in a powerful manner. For instance, it allows to define vectors on $\cm$ that need not be tangent to $\cm$, or even belong to a vector space with the same dimension as $\cm$. We see then that the concept of vector bundle generalizes that of tangent bundle. The tangent bundle (and any tangent tensor bundle) is a particular case of vector bundle over $\cm$ with abstract fiber $\rn$ and the fiber over each point $\tpm\simeq\rn$. Its transition functions belong to $GL_n$ and a (smooth) section of the tangent bundle is what we defined as (smooth) vector field in $\cm$ back in section \ref{sec:TangentSpace}. It can be proven that a rank $k$ vector bundle is trivial $E\simeq\cm\times \bbr^k$ if and only if there exist $k$ linearly independent global sections, \ie if and only if it admits a global frame. When $\tb$ is trivial we say that $\cm$ is parallelizable (such a manifold admits a trivial connection, see section \ref{sec:ConnectionsOnBundles}).

A {\bf principal G-bundle} is a fiber G-bundle $P_G$ with its own structure group as abstract fiber $F=G$. The structure group acts on the left on each fiber via transition functions as in any $G$-fiber bundle, but there is also a right action $R_G:G\rightarrow \mathrm{Diff}(F=G)$ of the structure group on the abstract fiber given by the right action of $G$ into itself which is not possible in a general fiber $G$-bundle and provides principal bundles of richer structure. Given a local trivialization $\Phi_U:O_U\times G\rightarrow\pi^{-1}(O_U)$ such that $\Phi^{-1}_{Up}(u)=g^u_U$ for $u\in\pi^{-1}(O_U)$, this right action is defined by $R_h u\equiv u h=\Phi_{Up}(g^u_U h)$ for any $h\in G$.

The existence of this right action allows to define a local trivialisation from a local section in a principal bundle, leading to the concept of identity section. Given a local section $\sigma_U:\cm\rightarrow P_G$ over $ O_U\subset\cm$, there is a canonical local trivialisation defined as follows: to each $u_p\in\pi^{-1}(p\in O_U)$, because the action of a group on itself is regular, there is a unique element $g_U^u\in G$ such that $u=\sigma_U(p)g_U^u$. The map defined by $\Phi_U^{-1}(u_p)=(p,g^u_U)$ is a diffeomorphism mapping $\pi^{O_U}\rightarrow O_U\times G$, and therefore its inverse is a local trivialisation. Conversely given a local trivialisation, $\{\Phi_U,O_U\}$ of $P_G$, the {\bf identity section} associated to such trivialisation $\sigma_U$ is defined by $\sigma_U(p)=\Phi_U(p,e)$, where $e$ is the identity element of $G$. Note that the right action of $G$ on $\sigma_i(p)\in\pi^{-1}(p)$ is $\sigma_U(p) h=\Phi_U(p,e h)=\Phi_U(p,h)$, which is why $\sigma_U$ is called the identity section associated to $\Phi_U$.

A relevant instance of principal bundle is the bundle of frames of sections of a rank $k$ vector bundle $E$, namely the {\bf frame bundle} of $E$, denoted by $F E$. An element of $F E$ is a frame of $E$ over a point $p\in \cm$. Note that, whereas $GL_k$ does not have a natural action on vectors in the fibers of $E$ until a basis in $\bbr^k$ is chosen (then $GL_k$ acts on the fibers by matrix multiplication), this is not the case for $F E$ where $GL_k$ acts canonically on frames via matrix multiplication, without having to make any arbitrary choice.\footnote{Recall that general linear groups can be seen as the group of changes of frame.} Thus, the fibers of $F E$ can be identified with the matrix group $GL_k$ as follows: Consider a frame $\textbf{e}_U=(e_{U1},...,e_{Uk})$ on a trivialising neighbourhood $O_U$. Then, because the action of $GL_k$ on the space of frames is regular, to each element of the fiber over $p\in O_U$, $\textbf{f}_U\in\pi^{-1}(p)$ where $\textbf{f}_U=(\textbf{f}_{U1},...,\textbf{f}_{Uk})$ corresponds a unique $g^\textbf{f}\in GL_k$ such that $\textbf{f}_U$ can be written as $\textbf{f}_U= \textbf{e}_U g^\textbf{f}$ or if matrix multiplication is written explicitly as $\textbf{f}_{Ub}= e_{Ua} g^\textbf{f}{}^a{}_b$. This proves that the fibers of $F E$ can be identified with the abstract fiber $GL_k$, and $F E$ is a principal $GL_k$ bundle. In the next section we will see that given a vector $G$-bundle $E$, there is a unique (up to isomorphism) associated principal $G$-bundle which is $F_G E$. This allows to canonically build the frame bundle of any smooth manifold $F\cm$ as the unique principal $GL_n$-bundle associated to $\tb$. Note that, as usual in the literature, we use the notation $F\cm$ instead of $F \tb$ for the frame bundle of a manifold.

Note that, if provided with a $G$-structure, we can also consider the bundle of $G$-frames of $E$, dubbed as $F_G E$. It is possible to see in an analogous manner that $F_G E$ is a principal $G$-bundle. For instance, structures such as orientation, volume forms or (pseudo)-Riemannian metrics on $\cm$ allow to reduce the structure group of $F\cm$ from $GL_n$ to $GL_n^+$, $SL_n$ and $O_n$ respectively. We will write $F_G\cm$ for the $G$-frame bundle of $\cm$.

 \subsubsection{The associated bundle construction: relating frame and vector bundles}

Two G-bundles over the same base space and sharing trivializing neighbourhoods and transition function are said to be {\bf associated bundles}. Associated bundles are unique (up to bundle isomorphism): given a principal $G$-bundle and a left action of $G$ a manifold $F$, there is a unique associated  $G$-bundle with fiber $F$. As well, there is a unique principal $G$-bundle associated to a given $G$-bundle.

A common way of finding associated bundles is through representations $\rho(G)\subseteq GL_k$ of structure groups $G$, which associates a vector $G$-bundle to a $G$-frame bundle as follows. Let $P_G$ be a principal $G$-bundle over $\cm$ with transition functions $t_{UV}$ on overlaps $O_{UV}$, and $\rho:G\rightarrow GL_m$ a representation of $G$ into some matrix subgroup of a general linear group $\rho(G)\subseteq GL_m$. There is a canonical rank $m$ vector $G$-bundle over $\cm$ with transition functions $\rho(t_{UV})\equiv\rho_{UV}$ on the same overlaps $O_{UV}$. We denote that vector bundle as $E^\rho_G$ and we say that $P_G$ and $E_G^\rho$ are associated though $\rho(G)$.  A rank $k$ vector $G$-bundle $E_G^\rho$ is associated to its $G$-frame bundle $F_G E$. We also say that the rank $m$ and $k$ vector bundles $E_G^\rho$ and $E_G$ are associated through $\rho$.

A nice property of $G$-frame bundles of $\cm$ is that they have a canonical (principal) soldering form\footnote{The canonical soldering form of the frame bundle induces a canonical soldering of its associated vector $G$-bundle to $\tb$. A soldering form on the frame bundle is an $\bbr^n$-valued form on $F\cm$} of their tangent bundle $\ct F_G \cm$ to their associated vector $G$-bundle  given by the differential of the projection: $\pi_\ast:\ct F_G\cm\rightarrow \tb_G$. At a point $\textbf{e}_p$ on $F_G\cm$, which corresponds to a frame $\textbf{e}_U$ over the point $p\in O_U$, $\pi_\ast$ projects $V\in\ct_{e_p} F_G\cm$ down to $\tb_G$ and takes the components of the resulting vector\footnote{Note that the fibers of $\tb_G$ are still the tangent spaces to $\cm$ at its points, although the diference between $\tb$ and $\tb_G$ is that the second is a vector $G$-bundle while the former is a vector bundle with structure group a representation of $\mathrm{Diff}(\cm)$.} $\pi_\ast(\fr_p,V)\in\tpm$ in the $\bbr^n$-frame $\textbf{e}_U$. The definition using the projection makes the soldering form trivially horizontal. By pulling $\theta_G$ back to $\tb_G$ via the identity section over $O_U$ we find a $\tb_G$-valued 1-form in $\cm$, denoted by $\fr_U=\sigma_U^\ast\pi_\ast$. We say that $\fr_U:\tb_G\rightarrow\tb$ `solders' the elements in the vector $G$-bundle associated to $F\cm$ to the tangent bundle of $\cm$ with the canonical structure group of $\tb$, \ie a representation of $\mathrm{Diff}(\cm)$. As we will see, the existence of a soldering form will allow to define the torsion of a linear connection on $\tb$, which does not make sense for linear connections in general vector bundles unless they are also soldered to $\tb$.

Let us mention several instances of vector bundles associated to another vector bundle which are relevant in physics. Given a rank $k$ vector $G$-bundle $E_G$, consider the dual representation by $\rho^\ast:G\rightarrow G$ such that $\rho^\ast(t_{UV})=(t_{UV}^{-1})^\top$ where $G$ is a subgroup of $GL_k$. We have that $E_G^{\rho^\ast}\equiv E_G^\ast$. Particularly, note that the $\ctb$ is the associated bundle to $\tb$ through the dual representation. As well, the $(p,q)$ tensor representation $\rho\otimes \overset{p}{...}\otimes \rho\otimes \rho_\ast\otimes \overset{q}{...}\otimes \rho_\ast$ leads to the $(p,q)$ tensor bundle as $E_G\otimes \overset{p}{...}\otimes E_G\otimes E^\ast_G\otimes \overset{q}{...}\otimes E^\ast_G$. As a last example, consider the adjoint representation of a Lie group $G$ given by $Ad:G\rightarrow GL(\mathfrak{g})$ by $Ad(g)\equiv Ad_g=L_{g\ast}\circ R_{g^{-1}\ast}:\mathfrak{g}\rightarrow\mathfrak{g}$ where $\mathfrak{g}$ is the Lie algebra of $G$, namely $\mathfrak{g}\equiv\ct_e G$. The adjoint bundle $E_{Ad}$ is the vector $Ad(G)$-bundle with fiber $\mathfrak{g}$ and transition functions $Ad(t_{UV})\equiv Ad_{UV}=L_{t_{UV}\ast}\circ R_{t_{UV}^{-1}\ast}\in GL_\mathfrak{g}$.

\subsection{Connections on bundles}\label{sec:ConnectionsOnBundles}

In a general fiber bundle $B$, the projection $\pi:B\rightarrow \cm$ allows to define a canonical notion of verticality of its tangent vectors as vectors which are tangent to the fibers. Intuitively, since vectors are directional derivatives, a vector tangent to a fiber will be a directional derivative along the fiber, which given that moving along a fiber implies laying above the same base space point, should vanish when projected down to the base space. Hence, a vector is vertical if  its projection to the base manifold vanishes. More precisely, a tangent vector to the fiber bundle $X\in\Gamma(\ct B)$ is {vertical} if $\pi_\ast(X)=0$ where $\pi_\ast$ is the pushforward of $\pi$ (see section \ref{sec:TangentSpace}). The subset of vertical vectors at $q\in B$ forms a vector subspace called vertical subspace $V_q B$, and the disjoint union of all the vertical subspaces form the vertical bundle $V B$, which is a sub-bundle of $\ct B$ with the vertical subspaces as fibers, and can be seen as a distribution in $\ct B$. The notion of verticality of vectors tangent to $B$ allows to define a notion for horizontality for $p$-forms on $B$: A $p$-form on $B$ is a {\bf horizontal form} if it vanishes when any of its arguments is a vertical vector.

Note that the existence of a canonical notion of vertical tangent vectors relies in the projection to the base space that comes with the deffiniton of fiber bundle. We could then ask whether there is any canonical notion of horizontal tangent vectors. There would be one (built in full analogy to vertical vectors) if there was a canonical projection from the bundle to the abstract fiber $F$. Although for trivial bundles $B=\cm\times F$ we can define such canonical projection, that is not the case for general fiber bundles, and therefore, there is no canonical notion of horizontality in general fuber bundles. The extra structure needed to define a sensible notion of horizontality on a general fiber bundle is that of an Ehresmann connection. 

An {\bf Ehresmann connection} is a (smooth) assignation of a horizontal subspace $H_p B$ at each point $p\in B$ such that $\ct_p B=V_p B\oplus H_p B$. The smoothness of the separation, carried out by defining a vertical projection, allows to define the horizontal bundle or horizontal distribution $H B$ and have that $\ct B=V B\oplus HB$ as the direct sum of bundles. Given that $p$-forms define distributions in a natural manner, an Ehresmann connection is specified by a vector valued {\bf connection 1-form} on $B$, which is a (local) $V B$-valued 1-form on $B$ denoted by $\omega$, that defines the horizontal distribution as its kernel $HV=\mathrm{ker}(\omega)$, \ie that projects to he vertical subspace. Explicitly, from $\ct B=VB\oplus HB$ we can write any tangent vector to $B$ as $X=X^H+X^V$ and $\omega(X)=X^V$ where $X^H\in HB$ and $X^V\in VB$. Having this notion of horizontality, we can define vertical forms as those which vanish whenever any of the arguments are vertical. A type of Ehresmann connections that are interesting to us are linear Ehresmann connections, which are Ehresmann connections on vector bundles whose connection 1-form is linear in the fiber coordinates. We will see that this is equivalent to a Koszul connection which is a generalization of affine connection

In a principal $G$-bundle, a $G$-compatible splitting $\ct_p P_{G}=V_p P_G\oplus H_p P_G$ defines a {\bf principal $G$-connection}, where $G$-compatibility means that $H_{pg}=R_{g\ast}H_p$. This splitting is defined by the {principal connection 1-form} $\omega^G$, which is a $\mathfrak{g}$-valued vertical 1-form on $P_G$ that projects any element of $\ct_p P_G$ into $V_p P_G\simeq \mathfrak g$. The compatibility with the right $G$-action is expressed through the property $R_g^\ast\omega^G=Ad_{g^{-1}}\omega^G$. On a principal bundle $P_G$, an Ehresmann connection such that $H_{pg}=R_{g\ast}H_p$ defines a principal G-connection. As well, a principal connection 1-form $\omega^G$ defines a $G$-compatible Ehresmann connection 1-form $\omega$ through the pushforward of the right $G$-action $R_{G\ast}:\mathfrak{g}\rightarrow \ct P_G$ since $R_{G\ast}(\omega^G(X))=X^V=\omega(X)$ where $X\in\ct P_G$.

A connection on a rank $k$ vector bundle $E$, called {\bf linear} or {\bf Koszul connection}, is a linear map $\na:\Gamma(E)\rightarrow \Gamma(E\otimes \ctb)$ that maps sections on $E$ (vectors) to sections on $E\otimes \ctb$ (E-valued 1-forms on $\cm$) and which satisfies the following Leibniz rule
\begin{equation}
\na(f X)=f (\na X)+ X\otimes \dif f
\end{equation}
for any $f\in \cf(\cm)$ and $X\in\Gamma(E)$. Given a frame $\textbf{e}_U$ on $E$ over $O_U \subset\cm$ there exists a $\mathfrak{gl}(k,\bbr)$ matrix of 1-forms over $O_U$ called {\bf connection 1-forms} $\omega_U{}^a{}_b$ such that 
\begin{equation}\label{eq:ConnectionCoefficients}
\na \textbf{e}_{Ub}=\textbf{e}_{Ua}\otimes\omega_U{}^a{}_b.
\end{equation}
Using a chart $x^\mu_U$ we can write $\omega_U{}^a{}_b=\omega_{U\mu}{}^a{}_b \dif x_U^\mu$, and we call $\omega_{U\mu}{}^\alpha{}_\beta$ the {\bf connection coefficients} of $\na$ in the chart $x_U^\mu$ and the frame $\textbf{e}_U$. Thus, for a general section $X\in\Gamma(E)$, we have that 
\begin{equation}\label{eq:CovariantDifferentialComponents}
\na X=\textbf{e}_{Ua}\otimes\lr{\dif X_U^a+\omega_U^a{}_b X_U^b}\equiv \textbf{e}_{Ua}\otimes \na X_U^a,
\end{equation}
where $X=X^a \textbf{e}_{Ua}$. Equations \eqref{eq:ConnectionCoefficients} and \eqref{eq:CovariantDifferentialComponents} are also known as Cartan structure equations associated to $\na$.

We call {\bf covariant derivative} of $X\in\Gamma(E)$ in the $Y\in\Gamma(\tb)$ direction to $\na_Y X\equiv (\na X)[Y]$. Given a chart $x_U^\mu$ where $Y=Y_U^\mu \partial_{U\mu}$ we write
\begin{equation}\label{eq:CovariantDerivativeVectorBundleComponents}
\na_Y X=\textbf{e}_{Ua}\otimes Y_U^\mu \lr{\partial_{U\mu}X_U^a+\omega_{U\mu}{}^a{}_b X_U^b}\equiv e_{Ua}\otimes Y_U^\mu\na_\mu X_U^a
\end{equation} 
where $\na_\mu\equiv\na_{\partial_{U\mu}}$. Note that $\omega^a{}_b=\omega_{U\mu}{}^{a}{}_b\dif x_U^\mu $ depends linearly on the fiber coordinates. 
It is possible to show that a Koszul (and therefore affine) connection on a vector bundle is univocally related to a linear Ehresmann connection on that vector bundle and vice versa. If the vector bundle is the tangent bundle of the base space $\cm$, the covariant derivative seen as an operator $\na:\Gamma(\tb)\times\Gamma(\tb)\rightarrow\Gamma(\tb)$ is usually regarded as the definition of  affine connection on a manifold.

The matrix of {\bf curvature 2-forms} $\theta_U{}^a{}_b$ of the connection $\na$ is defined as
\begin{equation}\label{eq:Curvature2Form}
\theta_U{}^a{}_b= \dif \omega_U{}^a{}_b+\omega_U{}^a{}_c\wedge\omega_U{}^c{}_b\equiv \frac{1}{2}R_U{}^a{}_{b \mu\nu} \,\dif x_U^\mu\wedge\dif x_U^\nu
\end{equation}
where the coefficients $R_U{}^a{}_{b\mu\nu}$ are the components of the {\bf Riemann curvature tensor} of $\na$ in the given frame and chart. A linear connection univocally defines (and can be defined as) a notion to relate elements of the different fibers of $E$ that generalises that of parallelism known as {\bf parallel transport}. We say that a section $X\in\Gamma(E)$ is parallel along a curve $\gamma(t):\bbr\rightarrow\cm$ if $\na_{\dot\gamma(t)} X=0$. %We will see that the usual notion of affine connection on $\cm$ is equivalent to that of a linear connection in $\tb$, since the former is specified once a covariant derivative (which defines a linear connection) in $\tb$ is chosen.

We can generalise the action of a connection to act on $p$-form sections of $E$ (note that as defined above it acts on $0$-form sections of $E$). This generalization leads to the {\bf exterior covariant differential} $\na:\Gamma(E\otimes \ctb{}^p)\rightarrow \Gamma(E\otimes \ctb{}^{p+1})$  mapping $E$-valued $p$-forms to $E$-valued $(p+1)$-forms on $\cm$ by
\begin{equation}
\na (X\otimes\alpha)=(\na X) \wedge \alpha + X\otimes \dif\alpha.
\end{equation}

The fact that $\na X$ is a $1$-form section of $E$ implies that on an overlap $O_{UV}$ we have $(\na X)_U=(\na X)_V$, which requires that $\omega_V{}^a{}_b=t_{UV}^{-1}{}^a{}_c\omega_U{}^c{}_d t_{UV}{}^c{}_b+t_{UV}^{-1}{}^a{}_c\dif t_{UV}{}^c{}_b$ or if writing implicitly matrix multiplication
\begin{equation}\label{eq:TransformationConnection}
\omega_V=t_{UV}^{-1}\omega_U t_{UV}+t_{UV}^{-1}\dif t_{UV}
\end{equation}
where $\omega_U$ and $\omega_V$ are the local matrices of connection 1-forms over the patches $O_U$ and $O_V$. From this relation, it is straightforward to see that the difference of two Koszul connection 1-forms (associated to different connections) transforms as a global 1-form on $\cm$, namely $\omega_V-\tilde{\omega}_V=t_{UV}^{-1}\lr{\omega_U-\tilde{\omega}_U} t_{UV}$.

Supose now that we reduce the structure group of $E$ to some subgroup $G\subset GL_k$ so that $\na$ is a Koszul connection on a rank $k$ vector $G$-bundle. We say that $\na$ is a $G$-connection if the parallel transport of a $G$-frame along any curve $\gamma(t):\bbr\rightarrow O_U\subset\cm$ is also a $G$-frame. The matrix of 1-forms of a $G$-connection is a $\mathfrak{g}$-valued 1-form. To prove it, let $\textbf{e}_U$ be a $G$-frame over $O_U$ at and $\na$ a $G$-connection. Then, any parallelly transported $G$-frame $\textbf{f}_U(t)$ along $\gamma(t)$ must also be a $G$-frame, and therefore can be written as $\textbf{f}_U(t)=\textbf{e}_U g(t)$ where $g(t)\in G\; \forall t$. Because $\textbf{f}_U(t)$ is parallelly transported along $\gamma(t)$, we have that $(\na \textbf{f}_{Ua})[\dot\gamma]=0$ for each vector in the frame. Then
\begin{equation}
\textbf{f}_{Ub}\otimes\Big[g^{-1}{}^b{}_c\omega_U{}^c{}_d[\dot\gamma]g^d{}_a+g^{-1}{}^b{}_c\dif g^c{}_a[\dot\gamma]\Big]=0 \qquad \forall \;t
\end{equation}
Using that $g(t)$ is a 1-parameter subgroup of a Lie group we have that $dg[\dot\gamma]=g(t)g'(0)$ where $g'(0)\in \ct_e G\equiv\mathfrak{g}$, and therefore $g^{-1}{}^b{}_c\dif g^c{}_a[\dot\gamma]=g'(0)^b{}_a\in\mathfrak{g}$. For the above equation to vanish along any curve and for all parallelly transported frames, the first term has to cancel $g^{-1}{}^b{}_c\dif g^c{}_a[\dot\gamma]$ and therefore it is also in $\mathfrak{g}$. Finally, we have that $L_{\ast g}\circ R_{\ast g^{-1}}$ maps $\mathfrak{g}$ to itself, and therefore $L_{\ast g}\circ R_{\ast g^{-1}}\lr{g^{-1}\omega_U[\dot\gamma]g}=\omega_U[\dot\gamma]\in\mathfrak{g}$ \underline{c.v.d.} An Ehresmann connection will be an Ehresmann $G$-connection if it is associated to a Koszul $G$-connection.

\subsubsection{Associated connections}

Given a Koszul $G$-connection $\na$ on a rank $k$ vector  $G$-bundle $E_G$ and a (faithful) linear representation $\rho:G\rightarrow GL_m$, the associated bundle construction canonically induces a linear $G$-connection on the rank $m$ vector $G$-bundle $E_G^\rho$ in the following way. Let $\omega_U$ be the $\mathfrak{g}$-valued connection 1-form associated to a frame $\textbf{e}_U$ over a trivialising patch $O_U$ such that $\na \textbf{e}_{Ua}=\textbf{e}_{Ub}\otimes\omega_U{}^b{}_a$. Then the pushforward of the representation $\rho_\ast:\mathfrak{g}\rightarrow \mathfrak{gl}_m$ canonically defines the $\rho_\ast(\mathfrak{gl}_m)$-valued connection 1-form $\omega_U^\rho$ on $O_U$ by its action on $X\in\Gamma(\tb)$ as 
\begin{equation}\label{eq:AssociatedKoszulConnections}
\omega_U^\rho[X]=\rho_\ast(\omega_U[X]).
\end{equation}
It can be verified that in an overlap $O_{UV}$ the induced connection 1-form transforms as a connection one form
\begin{equation}\label{eq:TransformationAssociatedConnection}
\omega^\rho_V=\rho_{UV}^{-1}\,\omega^\rho_U\,\rho_{UV}+\rho_{UV}^{-1}\,\dif\rho_{UV}
\end{equation}
where $\rho_{UV}\equiv\rho(t_{UV})$ are the transition matrices of $E_G^\rho$. We call these (Koszul) connections {\bf associated connections} through $\rho$. 

There is as well a canonical principal connection on $F_G E$ defined by the $\mathfrak{g}$-valued connection 1-form on $F_G E$
\begin{equation}\label{eq:PrincipalConnectionFromVectorConnection}
\omega^\ast=g_U^{-1}\pi^\ast(\omega_U)g_U+g_U^{-1}\dif g_U
\end{equation}
where $g_U$ are the fiber coordinates of the point at which $\omega^\ast$ is considered. It is possible to verify that this satisfies the requirements for a principal connection 1-form, and that it does not depend on the trivialising patch, \ie that it is indeed a global object in $F_G E$, see \eg \cite{Nakahara,Frankel}. As well, given a $G$-connection on $F_G E$ the pullback of the principal connection 1-form by the identity section $\sigma_U:O_U\rightarrow F_G E$ over each trivialising neighbourhood $O_U$ induces a $G$-connection 1-form on $E_G$ over $O_U$. For these connections to be canonically associated (meaning that having one we have the other without any extra structure) we need that if the connection on $F_G E$ is induced from a $G$-connection on $E_G$ by \eqref{eq:PrincipalConnectionFromVectorConnection}, the pullback of $\omega^\ast$ by the identity sections lead to the original $G$-connection 1-form on $E_G$ over each trivialising patch. That this is the case can be seen as follows. $\sigma_i^\ast:\ct^\ast_{\sigma_i(p)} F_G E\rightarrow\ct^\ast_p \cm$ maps $\mathfrak{g}$-valued 1-forms on $F_GE$ to $\mathfrak{g}$-valued 1-forms on $\cm$. By definition of pullback we have
\begin{equation}
(\sigma^\ast_U\omega^\ast)[X]=\omega^\ast[\sigma_{i\ast}X]=g_U^{-1}(\pi^\ast\omega_U)[\sigma_{U\ast}X]g_U+g_U^{-1}\dif g_U[\sigma_{U\ast}X].
\end{equation}
By definition of identity section we have that the fiber coordinates of the point $\sigma_U(p)\in \pi^{-1}(p)$ are $g_U=e\;\forall\;p\in O_U$. Therefore $\dif g$ vanishes along $\sigma_{U\ast}X$, arriving at $(\sigma^\ast_U\omega^\ast)[X]=\pi^\ast(\omega_U)[\sigma_{U\ast}(X)]$ which by definition of pullback is $\omega_U[(\pi_\ast\circ\sigma_{U\ast})X]$. By definition of section $\pi\circ\sigma_U=\mathrm{id}(\cm)$ and using the properties of the pushforward map $\pi_\ast\circ\sigma_{U\ast}=(\pi\circ\sigma_U)_\ast=\mathrm{id}(\cm)_\ast=\mathrm{id}(T_p\cm) \;\forall\; p\in\cm$. Therefore $\omega_U[\pi_\ast\circ\sigma_{U\ast}(X)]=\omega_U[X]$ and we have arrived at
 \begin{equation}
 \sigma_U^\ast(\omega^\ast)=\omega_U
 \end{equation}
\underline{c.v.d.}\footnote{This is the Valencian way for q.e.d. A litlle hommage to my high school math teacher.} We say that the principal and Koszul connections described by $\omega^\ast$ and $\omega_U$ are also {\bf associated connections}. In general, we will talk about associated connections as being `the same connection acting on associated bundles', and we will denote all of them by $\na$. 
Some relevant examples of associated connections that are of common use are the connection on the dual and tensor bundles. Given a Koszul $G$-connection on $E_G$ described by the 1-form $\omega_U$ in the frame $\fr_U$ the trivialising patch $O_U$, the dual vector bundle $E_G^\ast$ was defined in section \ref{sec:VectorFrameBundles}. The above algorithm leads to an associated connection on $E_G^\ast$ which in the dual frame to $\fr_U$ is described by the connection 1-form
\begin{equation}
\omega^{\rho_\ast}_U[X]=-\omega_U[X]^\top
\end{equation}
where $X\in\Gamma(\tb)$. Therefore, when acting on a section of the dual bundle $X\in\Gamma(E^\ast_G)$ such that $X=X_U{}_a \df_U^a$ (where $\df_U$ is the dual frame to $\fr_U$) we can write
\begin{equation}
\na X_U{}_a=\dif X_U{}_a+\omega_U^\top{}^a{}_b X_U{}_b=\dif X_U{}_a-\omega_U{}^b{}_a X_U{}_b.
\end{equation}
As expected, this leads to the usual law for covariant differentiation of $(0,1)$ tensors. Now we can generalize this to the $(p,q)$ tensor representation, which leads to the usual law for covariant differentiation of $(p,q)$ tensors. We will write it for the particular case of $(1,1)$ tensors but the result is well known in general
\begin{equation}\label{eq:CovariatDerivativeComponentsGeneralTensor}
\na T_U{}^a{}_b=\dif T_U{}^a{}_b+\omega_U{}^a{}_c T_U{}^c{}_b-\omega_U{}^c{}_b T_U{}^a{}_c.
\end{equation}
%%%%%%%%%%%%%%%%%%%%%%%
%%%%%%%%%%%%%%%%%%%%%%%
\section{Noncanonical structures on a smooth manifold}
%%%%%%%%%%%%%%%%%%%%%%%
%%%%%%%%%%%%%%%%%%%%%%%

We have presented all the canonical structures that exist in any smooth manifold and that are based solely on its differential structure. However, in order to do physics, there are other structures, that are not canonical, that we wish to introduce on a manifold, in order for it to be what we know as a space-time.

%%%%%%%%%%%%%
\subsection{Orientation and volume element}
%%%%%%%%%%%%%

First of all, we would like to present the notion of orientability and that of volume element. This notion is fundamental in order to be able to define an arrow of time. We will also see that any orientation in an n-dimensional manifold provides a volume element.

Given an n-dimensional (real) vector space $V$, two basis in $V$ are said to have positive orientation if the $GL_n$ transformation relating them has positive determinant. An n-dimensional manifold $\cm$ is called {\bf orientable} if there exists an atlas for $\cm$ having positive Jacobians in each overlap of its charts. This is true if and only if there exists a smooth $n$-form that does not vanish anywhere in $\cm$. In an orientable manifold, it is possible to pick up an orientation in each $\tpm$ in a continuous manner. In such a manifold, a pseudo-tensorial object is a tensor that changes sign under a change in orientation.

Given that the space of $n$-forms in an $n$-dimensional manifold is a 1-dimensional vector space, an $n-$form that does not vanish anywhere on $\cm$ has definite sign. Note that if there is an $n$-form $\dv$ that does not vanish anywhere, then any $n-$form $a(p)\dv$ where $a\in\cf\cm$ is also nowhere vanishing also has definite sign. A choice for a particular nowhere vanishing $n$-form $\dv$ in $\cm$ assigns an {\bf orientation} to $\cm$, and the pair $(\cm,\dv)$ is called an oriented manifold, where $\dv$ is its {\bf volume element}. Any orientation of the form $a\dv$ with $a\leq0$ is said to assign opposite orientation to $\cm$ as compared to the one assigned by $\dv$. The volume element provides a notion of volume spanned by an $n$-tuple of linearly independent vectors at a point $p\in\cm$. This notion allows to define integration of $n$-forms in any $n$-dimensional smooth manifold (see \eg \cite{Lee2013}).

A volume form $\dv$ also defines a canonical isomorphism $\star_\dv:\cf(O)\rightarrow\Lambda^nO$ over $O\in\cm$ by 
\beq
\star_\dv f=f \dv, 
\eeq
with the property $\dv={\star_\dv} 1$. This isomorphism allows to define the {\bf divergence operator} $\div_\dv:\Gamma(\ct O)\rightarrow \cf(O)$ associated to that volume form by its action on a vector field $X\in\Gamma(\ct O)$ as
\beq\label{eq:DefinitionDivergence}
\dif\big(\mathrm{i}_X \dv\big)=\div_\dv(X)\dv.
\eeq
This operator plays a crucial role in defining conserved quantities in a covariant way through a particular instance of the celebrated Stokes' theorem that can be stated when the volume form is associated to a metric, namely the Gauss-Ostrogradski theorem (see chapter \ref{sec:MinimalCoupling} for more details).

%%%%%%%%%%%%%
\subsection{Metric structure}\label{sec:MetricStructure}
%%%%%%%%%%%%%

The metric structure is central in physics, as it allows to define a notion of distance between any two space-time points. Furthermore, it provides for a canonical choice of volume element as well as a connection on $\cm$ and its associated bundles, as we will see below.

In physics, a {\bf metric} structure on $\cm$ is a nonsingular $(0,2)$ tensor field $g\in\ct^{(0,2)}\cm$. Because it is nonsingular, a metric tensor has an inverse metric tensor $g^{-1}\in\ct^{(2,0)}\cm$ such that in any basis $(g^{-1}_U)^{\mu\alpha}g_U{}_{\alpha\nu}=\delta^\mu{}_\nu$. For simplicity we write $g_U{}^{\mu\nu}=(g_U^{-1})^{\mu\nu}$. The pair $(\cm, g)$ is called (pseudo-)Riemannian manifold.

The metric of physical space-time will be pseudo-Riemannian, meaning that its eigenvalues are $(1,-1,-1,-1)$. Thus, there is a frame $\fr_U$ on a neighbourhood of any point $p\in O_U$ such that $g(\textbf{e}_{Ua},\textbf{e}_{Ub})\equiv g_U{}_{ab}=\eta_{ab}$ where $\eta$ is the (mostly minus) Minkowski metric when written in cartesian coordinates. Given such a basis, any other basis $\tilde{\fr}_U$ related to it by a local Lorentz transformation, \ie an element of $O_{(1,3)}$, also satisfies $g(\tilde{\fr}_{Ua},\tilde{\fr}_{Ub})=\eta_{ab}$. Indeed any such transformation at $p$ leaves invariant $g(X,Y)$ for any two vectors defined at $p$. Therefore, a pseudo-Riemannian manifold is canonically endowed with a local $O_{(1,3)}$ symmetry\footnote{Any metric structure on a manifold provides a local symmetry group.} and a local lightcone structure that allows to label vectors as timelike, spacelike or null according to the sign of their norms. 

The existence of an $O(1,3)$ metric structure provides $\tb$ and associated bundles with an $SO^+_{(1,3)}$-structure, thus allowing to reduce the structure group of the frame bundles to $SO^+_{(1,3)}$, yielding the bundle of positively oriented orthonormal frames over $\cm$ and the associated tangent bundles with structure group $SO^+_{(1,3)}$. As we will clarify below, (Dirac) spinor fields will be sections of an associated vector bundle via the spinor representation of $SO^+_{(1,3)}$ called the spin bundle.

A metric structure on $\cm$ provides a local linear isomorphism $I_g:\tpm\rightarrow\ctpm$ by associating to a vector $X$ the 1-form $I_g X (Y)=g(X,Y)$. In components, it works as follows: given a frame $\fr_U$ on $\ct O_U\subset\tb$ and its dual frame $\df_U$, the components of $I_g X$ when written in the dual frame $\df_{U}$ are 
\begin{equation}\label{eq:MetricIsomorphism}
(I_g X)_\mu\equiv X_U{}_\mu=g_U{}_{\mu\alpha}X_U{}^\alpha.
\end{equation}
Whenever there is a metric, we will write the components of a vector and those of its associated 1-form with the same symbol without making explicit the use of this isomorphism. This isomorphism extends in a straightforward manner to $(p,q)$ tensor fields and $p$-forms, and is also used to rise and lower the indices of these objects.

A metric also provides us with a volume form as follows. The metric volume (pseudo) form $\dv_g$ is the unique $n$-form on $\cm$ that associates to an orientation of $\tpm$ and any positively oriented orthonormal basis the value $+1$. In an orthonormal coordinate system $x_U^\mu$ at $p\in O_U$ the volume form $\dv_g$ must be of the form $\pm \dif x_U^1\wedge...\wedge\dif x_U^n$ where the sign is defined by the choice of orientation. It can be seen that in general coordinates the expression for $\dv_g$ reads
\begin{equation}\label{DefMetricVolumeForm}
\dv_g=\sqrt{-g_U}\dif x_U^1\wedge...\wedge\dif x_U^n
\end{equation}
where $-g_U$ stands for the absolute value of the determinant of $g$ in the given chart.

A metric $g$ also induces a canonical isomorphism $\star_g:\Lambda^kO_U\rightarrow\Lambda^{n-k}O_U$ coined as {\bf Hodge dual} that generalises the $\star_\dv$ isomorphism defined by any volume form. We will give its definition only in local coordinates $x_u^\mu$, and it is as follows. Let $F=F_{U\mu_1...\mu_k}\dif x_U^{\mu_1}\wedge...\wedge\dif x_U^{\mu_k}$ be a $k$-form on $\cm$. Its Hodge dual is the (pseudo) $(n-k)$-form $\star_g F= (\star_g F)_{U\mu_{1}...\mu_{n-k}}\dif x_U^{i_1}\wedge...\wedge\dif x_U^{i_{n-k}}$ on $\cm$ defined by 
\begin{equation}\label{DefHodgeDual}
(\star_g F)_{U\mu_{1}...\mu_{n-k}}=\sqrt{-g_U}F_U^{\mu_1...\mu_k}\varepsilon_{\mu_1...\mu_k \mu_{k+1}...\mu_n}
\end{equation}
where $\varepsilon_{\mu_1...\mu_n}$ is the Levi-Civita symbol of $n$ indices. It can be seen that ${\star_g}{}\star_g F=(-1)^{k(n-k)+s}F$ where $s$ is the number of negative eigenvalues of the metric (0 for Riemannian metrics and 1 for Lorentzian metrics), generalising the analog property of $\star_\dv$.  The components of the Hodge-dual of a contravariant $k$-form can be seen to be
\begin{equation}
(\star_g F)_U{}^{\mu_{1}...\mu_{n-k}}=\frac{(-1)^s}{\sqrt{-g_U}}F_{U\mu_1...\mu_k}\varepsilon^{\mu_1...\mu_k \mu_{k+1}...\mu_n}.
\end{equation}
The Hodge dual allows to write the divergence operator associated to the volume element of the metric, namely $\div_g$ as
\beq
\div_g X=\star_g\dif\star_g I_g(X),
\eeq
which can be straightforwardly generalised to act on the covariant version of arbitrary $p$-forms. Apart from the divergence operator, the Hodge dual can also be used to define another differential operator that is canonical in presence of a metric, namely the \textbf{codifferential} operator $\delta_g:\Lambda^kO\rightarrow\Lambda^{k-1}O$ by its action on any $k$-form $F$ as
\beq\label{eq:Codiferential}
\delta_g F=(-1)^{n(k+1)+s+1}\star_g \dif\star_g F.
\eeq 
Because $\star_g{}^2\propto 1$ and $\dif^2=0$ it is straightforward to check that $\delta_g^2=0$ as well. This allows to define a generalisation of the Laplacian and d'Alembertian operators to arbitrary manifolds with metric called Laplace-de Rham operator and dubbed by $\Box_g:\Lambda^kO\rightarrow\Lambda^kO$ as
\beq
\Box_g=\dif\delta_g+\delta_g\dif.
\eeq 
These operators play a crucial role in physical theories where the physical fields are sections of the bundle of $k$-forms over spacetime, particularly in gauge theories where matter and gauge fields are 0- and 1-form sections of some vector bundle over spacetime.Through this work, we will generally use volume elements associated to a metric, so that, in general, we will drop the subindices of the Hodge operator to ease the readability of the notation.

Another relevant notion that is canonically defined by a metric structure is that of Killing vector fields. We say that $X\in\Gamma(\tb)$ is a Killing vector field if
\begin{equation}
\cl_X g=0.
\end{equation}
 The infinitesimal interpretation is that $X$ generates infinitesimal isometries of the metric. Namely, an infinitesimal diffeomorphism $x_U^\mu\mapsto x_U^\mu+\epsilon X_U{}^\mu$ generates a change in the metric given by $\delta_\epsilon g=\epsilon\cl_X g$. Therefore, Killing vector fields are associated to infinitesimal diffeomorphisms that leave $g$ invariant. The one parameter groups of transformations generated by Killing vector fields leave the metric (and therefore the local geometry of a (pseudo-)Riemannian manifold) invariant, and therefore their generators will be associated to symmetries of the metric.

\subsubsection{Spin structure and spin bundle $\cS\cm$}

A (pseudo)-Riemannan metric on $\cm$ is equivalent to a $G$-structure on $F\cm$, where $G$ is the orthogonal group associated to symmetries of the metric. For a Minkowskian metric, this group is $O_{(3,1)}$, and by choosing the canonical orientation and volume form associated to the metric, we can reduce it to an $SO^+_{(3,1)}$ structure on $F\cm$, which allows to define the frame of positively oriented (and time-oriented) orthonormal bundles over $\cm$ $F_{SO^+_{(3,1)}}\cm$. 

Spinor representations are typically understood as representations of the $Spin$ groups, which are the double cover groups of special orthogonal groups. Particularly, there is a 2 to 1 homomorphism $\Lambda^s:SL_{(2,\mathbb{C})}\rightarrow SO^+_{(3,1)}$ such that $\Lambda^s(A)=\Lambda^s(-A)$ yielding a two-valued spinor representation of $SO^+_{(3,1)}$ as the usual representation of $SL_{(2,\mathbb{C})}$ in terms of $2\times2$ complex matrices. In this representation, every Lorentz transformation is associated with two matrices $\pm A\in SL_{(2,\mathbb{C})}$. Choosing one of both matrices, this representation, usually dubbed as $\mathscr{D}(1/2,0)$, is the {\bf spinor representation} of $SO^+_{(3,1)}$, and acts on {\bf left handed} spinors $\Psi_L$. The (complex) dual representation (see \ref{sec:VectorFrameBundles}) using $A^{\dagger-1}$ and denoted $\mathscr{D}(0,1/2)$ is called {\bf cospinor representation} of $SO^+_{(3,1)}$ and acts on $\textbf{right handed}$ spinors.

 There is a representation $\rho^s:SL_{(2,\mathbb{C})}\rightarrow GL_{(4,\mathbb{C})}$ which we call Dirac representation defined by
\begin{equation}
\rho^s(A)=\begin{pmatrix}
A & 0 \\
0& A^{\dagger-1}
\end{pmatrix}
\end{equation}
which acts naturally in $\Psi=(\Psi_L,\Psi_R)^\top\in\mathbb{C}^4$ dubbed as $\mathscr{D}(1/2,1/2)$. The 4-component spinors acted upon by $\mathscr{D}(1/2,1/2)$ are called {\bf Dirac spinors}. In order for the Dirac operator to be well defined, Dirac spinors should transform under a Lorentz transformation $\Lambda(A)$ as $\Psi\mapsto\rho^s(A)\Psi$.

Consider now a manifold with a pseudo-Riemannian metric $g$. This induces an $SO^+_{(3,1)}$-structure on the frame bundle, which can be reduced to $F_{SO^+_{3,1}}\cm$ and over a trivialising patch $O_U\in\cm$, an orthonormal frame $\textbf{e}_U$ satisfies $g(\textbf{e}_{U\mu},\textbf{e}_{U\nu})\equiv g_{U\mu\nu}=\eta_{\mu\nu}$ where in an overlap $O_{UV}\ni p$ we have $\textbf{e}_V(p)=\textbf{e}_U(p) t_{UV}(p)$ whith $t_{UV}(p)\in SO^+_{3,1}$. While in Minkowski space the choice among the two possibilities $\pm A\in SL_{(2,\bbc)}$ for a given Lorentz transformation $\Lambda(\pm A)$ is global, the choice among $\pm A(p)=\tilde{t}_{UV}(p)\in SL_{(2,\bbc)}$ corresponding to $\Lambda(p)=t_{UV}(p)\in SO^+_{(3,1)}$ must be carried in a smooth manner. If this can be satisfied and the chosen matrices $\tilde{t}_{UV}(p)$ also satisfy the transition function conditions \eqref{eq:ConditionsTransitionFunctions} we say that there is a lift of the structure group of $\tb$ from $SO^+_{(3,1)}$ to $SL_{(2,\bbc)}$ and we say that $\cm$ has a {\bf spin structure}.

 If $\cm$ has a spin structure, the complex rank-4 vector bundle $\mathcal{S}\cm$ associated to the $SL_{(2,\bbc)}$ tangent bundle through $\rho^s$, which according to the associated bundle construction has transition functions $\rho^s_{UV}\equiv\rho^s(\tilde{t}_{UV})$, is the {\bf Dirac spinor bundle} over $\cm$.

%%%%%%%%%%%%%
\subsection{Connections on $\cm$}\label{sec:Connections}
%%%%%%%%%%%%%

Although the tangent spaces of a manifold are smoothly patched together in the tangent bundle, there is no canonical way in which to relate tangent spaces at different points {\it i.e} different fibers of $\ct\cm$. An affine connection on $\cm$ provides such a notion as a particular case of vector bundle Koszul connection. As we saw in section \ref{sec:ConnectionsOnBundles}, we can build connections on the associated bundles to $E_G$ from a Koszul $G$-connection on $E_G$. Thus, by giving an affine connection to $\cm$, we will be able to canonically induce connections in the tensor and spinor bundles of $\cm$.

\subsubsection{Affine connections as linear connections on $\tb$}

An {\bf affine connection} on $\cm$ is a Koszul connection in $\tb$. As seen in section \eqref{sec:ConnectionsOnBundles}, this is equivalent to a bilinear operator $\na:\tb\times\Gamma(\ct\cm)\rightarrow\Gamma(\ct\cm)$ called {\bf covariant derivative} that assigns (in a smooth way) a vector $\na_{Y_p} X\in\tpm$ to a vector $Y_p\in\tpm$ at p and a vector field $X$ near $p\in\cm$  in a way which satisfies the following Leibniz-like rule for smooth functions at $p$
\begin{equation}
\na_{X_p}(fY)=X_p[f]Y+f\na_{X_p} Y.
\end{equation}
By {\it in a smooth way} we mean that if $X$ is a smooth vector field, then $\na_X Y$ must also be smooth. In a coordinate frame $\partial_{U\mu}$ on $\ct O_U$, there exists a set of coefficients $\Gamma_U{}^{\alpha}{}_{\mu\nu}$ such that
\begin{equation}
\na_\mu \partial_{U\nu}=\Gamma_U{}^{\alpha}{}_{\mu\nu}\partial_{U\alpha}.
\end{equation}
 These symbols are called {\bf coefficients of $\na$} in the coordinate frame $\partial_{U\mu}$, and the analog of \eqref{eq:CovariantDerivativeVectorBundleComponents}
\begin{equation}\label{ComponentsCovDerVec}\na_\mu X_U ^\nu=\partial_{U\mu}X_U^\nu+\Gamma_U{}^{\nu}{}_{\mu\alpha}X_U^\alpha 
\end{equation}
 
By using the canonical soldering form of the frame bundle $F\cm$ into the tangent bundle $\tb$, we are able to use noncoordinate (or nonholonomic) frames in $\tb$ by $\fr_U{}_a=\fr_U{}_a{}^\alpha\partial_U{}_\alpha$ and $\df_U{}^a=\df_U{}^a{}_\alpha\dif x_U^\alpha$. We can then define the connection coefficients in a general frame as done in \eqref{eq:ConnectionCoefficients} by $\na \fr_U{}_a=\fr_U{}_b\omega_U{}_\mu{}^b{}_a \dif x_U^\mu$. Then, it is straightforward to see that 
\begin{equation}\label{eq:ConnectionCoefficientsArbitraryFrames}
\omega_{U\mu}{}^a{}_b=\df_U{}^a{}_\alpha\lr{\partial_{U\mu}\fr_U{}_b{}^\alpha+\fr_U{}_b{}^{\nu}\Gamma_U{}^{\alpha}{}_{\mu\nu}}
\end{equation}
where $\partial_{U\mu}\fr_U{}_b{}^\alpha$ is the component expresion of $\dif\fr_U{}_b{}^\alpha$. Given that $\omega^a{}_b=\omega_U{}_\mu{}^a{}_b\dif x_U^\mu$, we have that $\omega_U{}_k{}^a{}_b=\fr_U{}_k{}^\mu\omega_U{}_\mu{}^a{}_b$. From the general relation between the connection coefficients in trivialising charts of an overlap $O_{UV}$ \eqref{eq:TransformationConnection}, we can find the relation between the connection coefficients in different coordinate charts of an overlap $O_{UV}$ as
\begin{equation}
\Gamma_V{}^\alpha{}_{\mu\nu}=\frac{\partial x_V^\alpha}{\partial x_U^\beta}\lrsq{\frac{\partial x_U^\rho}{\partial x_V^\mu}\frac{\partial x_U^\sigma}{\partial x_V^\nu}\Gamma_U{}^\beta{}_{\rho\sigma}+\frac{\partial^2 x_U^\beta}{\partial x_V^\mu\partial x_V^\nu}}.
\end{equation}
Given any pair of vector fields, the  {\bf curvature transformation} of an affine connection is the linear transformation $R(X,Y)$ on each $\tpm$  defined by
\begin{equation}
R(X,Y)=[\na_X,\na_Y]-\na_{[X,Y]}.
\end{equation}
In a coordinate frame $\partial_{U\mu}$, this transformation can be written as a matrix $R(\partial_{U\mu},\partial_{U\nu})^\alpha{}_j\beta=R_U{}^\alpha{}_{\beta\mu\nu}$, and therefore $R_U{}^\alpha{}_{\beta\mu\nu}$ are the components of a $(1,3)$ tensor called {\bf Riemann curvature tensor} of the affine connection, which are given by
\begin{equation}
R_U{}^\alpha{}_{\beta\mu\nu}=2\partial_{[\mu}\Gamma_U{}^\alpha{}_{\nu]\beta}+2\Gamma_U{}^\alpha_{[\mu|\sigma|}\Gamma_U{}^\sigma{}_{\nu]\beta},
\end{equation}
Given the canonical soldering form, we can write the components of the Riemann tensor in an arbitrary frame $\fr_{U\mu}$, and it is possible to see that we can also define the (local) matrix of {\bf curvature 2-forms} $\theta_U{}^a{}_b$ defined in \eqref{eq:Curvature2Form} can be written as
\begin{equation}
\theta_U{}^a{}_b=\frac{1}{2}R_U{}^a{}_{bkl} \df_U^k\wedge\df_U^l.
\end{equation}

To each affine connection (or a connection in any vector bundle with a soldering to $\tb$), besides from the curvature tensor, we can always associate another geometric object named torsion tensor. This can be done using vector-valued $p$-form notation or usual tensorial notation. Given an affine connection, its vector valued {\bf torsion} 2-form $\tau$ is defined by its action on two vector fields $X,Y$ as
\begin{equation}
\tau[X,Y]=\na_X Y-\na_Y X -[X,Y].
\end{equation}
Given a frame $\fr_{Ua}$ we have that $\tau=\fr_{Ua} \otimes \tau_U{}^a$ where $\tau_U{}^a$ can be written as
\begin{equation}
\tau_U{}^a=\frac{1}{2}T_U{}^a{}_{ij} \df_U^i\wedge\df_U^j
\end{equation}
and $T_U{}^a{}_{ij}$ is the {\bf torsion tensor} which in an arbitrary frame can be written as\footnote{Here by $[indices]$ we denonte (normalised) antisymmetrisation with respect to such indices. As well, we use $(indices)$ for symmetrisation. An index between two vertical bars $|$ are omitted in the (anti)symmetrisation process. For instance, $2\omega^k{}_{[ij]}=\omega^k{}_{ij}-\omega^k{}_{ji}$,  and $2 \omega_{[k|i|j]}=\omega_{kij}-\omega_{jik}$.}
\begin{equation}\label{ComponentsTorsionTensor}
T_U{}^a{}_{ij}=2\omega_U{}^a{}_{[ij]}-2[\fr_U{}_i,\fr_U{}_j]^a.
\end{equation}
Note that this only makes sense if $X$ and $Y$ can be identified via soldering with objects belonging to $\tb$. Therefore, in a coordinate frame, we have that $T_U{}^\alpha{}_{\mu\nu}=2\Gamma_U{}^\alpha{}_{[\mu\nu]}$. This is the reason why a torsionless connection is also called symmetric. If we have a connection, there is a nice relation between its coefficients in a frame, the exterior derivative of the elements of the dual frame, and the torsion 2-form. Such relation completes Cartan's structure equations, and reads
\begin{equation}
\dif \df_U{}^a=-\omega_U{}^a{}_b\wedge\df_U{}^b+\tau_U{}^a.
\end{equation}

The curvature of a manifold provides information about a global notion called distant parallelism. We say that a manifold is {\bf parallelizable} if its tangent bundle is trivial, \ie if it can be covered by a single frame $\fr$. In such case, one can define a connection that vanishes in such a frame $\omega^i{}_j=0$ and therefore the frame fields are covariantly constant $\na \fr=0$, which by construction in \eqref{eq:Curvature2Form} means that the curvature 2-forms vanish in such a frame. Since the curvature 2-forms are covariant objects, we can conclude that a parallelizable manifold admits an affine connection with vanishing curvature. In such case, it is clear that the torsion 2-form measures the failure of $\df^\mu$ to be closed, as $\dif \df^a=\tau^a$.

Connections provide a way to relate different fibers in fiber bundles. Therefore, affine connections provide a recipe to relate vectors in tangent spaces at different points on $\cm$. This relation stems from the concept of parallel transport. Let $\na$ be an affine connection on $\cm$ and $\gamma(t)$ a curve through $p$ at $t=0$. Let $X_p$ be a vector at $p$. When expressed in a coordinate frame, the equation defining parallel transport $\na_{\dot\gamma}Y=0$ reads
\begin{equation}\label{eq:ParallelTransport}
\frac{d Y_U{}^\mu}{dt}+\Gamma_U{}^\mu{}_{\alpha\beta}\dot\gamma_U{}^\alpha Y_U{}^\beta=0,
\end{equation}
which is a 1st order differential ODE. Given the initial condition $Y(0)=X_p$, this equation has a unique solution for $Y(t)$. We say that $Y(t)$ is the {\bf parallel transport} of $X_p$ along the curve $\gamma$. In general, we say that a vector field $X$ is parallel along a curve $\gamma(t)$ if $\na_{\dot\gamma}X=0$ everywhere on the curve. This notion allows to single out a special class of curves called {\bf autoparallel} as those curves whose tangent vector is parallel along the curve. Note that this notion is not invariant under reparametrizations, for if $\na_{\dot\gamma}\dot\gamma=0$, then after a reparametrisation $\tilde\gamma(\tau)=\gamma(t(\tau))$ we have that $\dot{\tilde\gamma}=(dt/d\tau)\dot\gamma$. Therefore if $\gamma(t)$ is autoparallel then we have that $\na_{\dot{\tilde\gamma}}\dot{\tilde\gamma}=(d\tau/dt)(d^2t/d\tau^2)\dot{\tilde\gamma}$ which is generally different from zero, implying that $\tilde\gamma(\tau)$ is not autoparallel. However, it is well known that $\tilde\gamma(\tau)$ is autoparallel of a connection $\tilde\na$ that is related to $\na$ by the projective transformation $\tilde\omega_U{}^a{}_b=\omega_U{}^a{}_b+\delta^a{}_b\xi$ where $\xi$ is the 1-form such that $\xi[\dot{\tilde\gamma}]=-(d\tau/dt)(d^2t/d\tau^2)$. Projective symmetry has proven to be physically relevant in some gravitational theories, as we will see below.

\subsubsection{Spin connection on $\cS\cm$ from an $SO^+_{(3,1)}$-connection on $\tb$}

Equipped with the associated bundle construction, we are now in the position to understand the subtleties behind the relation between the spin connection in $\cS\cm$ and a general affine connection in $\cm$. To that end, note that a spin connection is a linear connection on the $\rho^s\lr{SL_{(2,\bbc)}}$-vector bundle $\cS\cm$, and therefore it will be described by an $\rho^s\lr{\mathfrak{sl}(2,\bbc)}$ connection 1-form on $\cm$. Now, note that while the spinor representation of $SO^+_{(3,1)}$ is given by the 2 to 1 homeomorphism $\Lambda^s:SO^+_{(3,1)}\rightarrow SL_{(2,\bbc)}$, the pushforward of the representation maps bijectively\footnote{Given that $\omega^{\Lambda^s}_U{}[X]\in \mathfrak{sl}(2,\bbc)$ is the tangent vector at the identity of $SL_{(2,\bbc)}$ of the 1-parameter subgroup $\mathrm{exp}(t\omega^{\Lambda^s}_U{})\subset SL_{(2,\bbc)}$, we have that $\Lambda^s_\ast (\omega^{\Lambda^s}_U{}[X])=\left.\frac{d}{dt}\Lambda^s\lr{\mathrm{exp}(t \omega^{\Lambda^s}_U{}[X])}\right|_{t=0}$. Although we know that $\Lambda^s\lr{\pm\mathrm{exp}(t\omega^{\Lambda^s}_U{}[X])}$ corresponds to the same Lorentz transfromation, the pushforward of $\Lambda^s$ is a map between Lie algebras only for tangent vectors at the identity of $SL_{(2,\bbc)}$. Given that only ${+\mathrm{exp}(t\omega^{\Lambda^s}_U{}[X])}$ is the identity at $t=0$, only the tangent vector at the identity of the 1-parameter subgroup ${+\mathrm{exp}(t\omega^{\Lambda^s}_U{}[X])}$ will be an element of $\mathfrak{sl}(2,\bbc)$. Therefore, $\Lambda^s_\ast$ maps the Lie algebras only when applied to the tangent vector at the identity of the 1-parameter subgroup $+\mathrm{exp}(t\omega^{\Lambda^s}_U{}[X])$. Hence the map is 1 to 1.}  their corresponding Lie algebras. Hence, for a given $\mathfrak{so}_{(3,1)}$ connection 1-form over $O_U\subset\cm$, there is a unique $\mathfrak{sl}(2,\bbc)$ connection 1-form $\omega^{\Lambda^s}_U$ such that $\Lambda^s_\ast\lr{\omega^{\Lambda^s}_U[X]}=\omega_U[X]$. This construction is a canonical lift of the connection in $\tb$ with structure group $SO^+_{(3,1)}$, \ie an $SO_{(3,1)}$-connection in $\tb$, to a connection in $\tb$ with structure group $SL_{(2,\bbc)}$, \ie an $SL_{(2,\bbc)}$-connection. Note that the lift has been carried out exactly in the same way as is done for inducing associated connections through the associated bundle construction. However, since the spin representation  is not strictly a representation ($\Lambda^s(-A)\neq -\Lambda^s(A)$) but a projective representation, it was not guaranteed that this construction would be successful in this case. Now that we have a linear connection in the $SL_{(2,\bbc)}$ tangent bundle, we can build a linear connection in the spinor bundle (recall it is a vector bundle as the fibers are $\bbc^4$) by means of the Dirac representation $\rho^s$ through the associated bundle construction which leads to a connection on $\cS\cm$ defined by
\begin{equation}
\omega_U^s[X]=\rho^s_\ast\lr{\omega_U^{\Lambda^s}[X]}.
\end{equation}
By construction, this connection is canonically associated to the original $SO^+_{(3,1)}$-connection through the associated vector bundle construction. In order to compute explicitly the spin connection 1-form in $\cS\cm$ in terms of the $SO^+_{(3,1)}$ connection 1-form in $\tb$ we would need to compute first the isomorphism $\Lambda_\ast$ between Lie algebras. Following \cite{Frankel}, this leads to $\omega^{\Lambda^s}_U=(1/2)\sum_{a<b}\sigma_a\sigma_b(\omega_i)^{ab}$ where $\sigma_0=1$ and $\sigma_i$ are the Pauli matrices. This is the $SL_{(2,\bbc)}$ connection 1-form on the $SL_{(2,\bbc)}$ tangent bundle built from the $SO^+_{(3,1)}$ tangent bundle. Using now $\rho_\ast^s$ we can now obtain the $\rho^s\lr{SL_{(2,\bbc)}}$-connection on $\cS\cm$ in terms of the $SO^+_{(3,1)}$ connection 1-form on $\tb$, which reads
\begin{equation}\label{eq:SpinConnection}
\omega^s_U=\frac{1}{8}\omega_U{}_{ab}[\gamma^a,\gamma^b],
\end{equation}
where $\omega_U{}_{ab}=-\omega_U{}_{ba}\in\mathfrak{so}_{3,1}$. Therefore, for a section over $O_U$ in the spinor bundle $\Psi\in\Gamma(\cS\cm)$ we have that its covariant exterior differential reads
\begin{equation}\label{eq:SpinorExteriorCovariantDifferential}
\na\Psi=\dif\Psi+\omega^s_U\Psi=\dif\Psi+\frac{1}{8}(\omega^s_U)_{ab}[\gamma^a,\gamma^b]\Psi,
\end{equation}
and therefore its covariant derivative along $X\in\tb$ is
\begin{equation}\label{eq:SpinorCovariantDerivative}
\na_X\Psi\equiv\na\Psi[X]=\dif\Psi[X]+\frac{1}{8}\omega_U{}_{ab}[X][\gamma^a,\gamma^b]\Psi,
\end{equation}
which in a coordinate frame $x^\mu$, by taking $X=\partial_\mu$  reads
\begin{equation}
\na_\mu\Psi=\partial_{U\mu}\Psi+\frac{1}{8}\omega_U{}_{\mu ab}[\gamma^a,\gamma^b]\Psi.
\end{equation}
Here $\omega_U{}_{\mu ab}=\omega_U{}_{ab}[\partial_\mu]$. Although it is important to keep in mind that components of objects in bundles over $\cm$ depend on the chosen trivialising patch, we will frequently drop the $U$ subindex to relax the notation, thus for instance writing just
\begin{equation}\label{eq:SpinorCovariantDerivativeCoordinateFrame}
\na_\mu\Psi=\partial_\mu\Psi+\frac{1}{8}\omega_{\mu ab}[\gamma^a,\gamma^b]\Psi.
\end{equation}
Through the associate bundle construction, we can also derive the canonical connection on the dual spinor bundle $\cS\cm^\ast$ from the affine connection. After some manipulations, this yields
\begin{equation}\label{eq:DualSpinorCovariantDerivativeCoordinateFrame}
\na_\mu\bar\Psi=\partial_\mu\bar\Psi-\bar\Psi\omega^{s}=\partial_\mu\bar\Psi-\frac{1}{8}\omega_{\mu ab}\bar\Psi[\gamma^a,\gamma^b].
\end{equation}
%%%%%%%%%%%%%
\section{Post-Riemannian space-times}
%%%%%%%%%%%%%
When gravity is viewed from the geometrical perspective, it can be said that the three independent noncanonical structures over a smooth manifold introduced above; namely the volume element, metric, and affine connection; play a crucial physical role. In classical GR, a space-time is a manifold with a metric structure which induces canonically a volume form and a connection (the Levi-Civita connection, see below), usually called (pseudo-)Riemannian manifold. Modifications of GR rooted in the geometrical view that do not want to give up the smooth structure can arise from introducing an independent affine connection or volume element (or both). Usually, in metric-affine gravity theories, the space-time is a manifold with a metric structure and its canonical volume form but an independent affine connection. The work carried out in this thesis concerns mainly metric-affine theories of gravity, and therefore we will here give a detailed account of several geometric objects that can be canonically defined in such space-times, which we will call post-Riemannian manifolds. However, to better understand what is new in post-Riemannian manifolds with respect to Riemannian ones, it will be useful to present a fundamental theorem in differential geometry which shows how a metric induces an affine connection in a canonical way.

\subsection{The Levi-Civita connection}

Given an $O_{1,3}$ metric structure $g$ on a manifold $\cm$, the pair $(\cm,g)$ is called a {\bf (pseudo-)Riemannian manifold}. The fundamental theorem of Riemannian geometry guarantees that there is a unique Koszul connection on $\tb$ dubbed as $\na^g$ with vanishing torsion and which is compatible with the metric in the sense that $\na^g g=0$ as a (0,2) tensor-valued 1-form.\footnote{(p,q) tensor-valued p-forms are defined in an analog fashion as vector p-forms. The covariant differential of a (p,q) tensro $T$ is a (p,q) tensor valued 1-form $\na T$ such that its action on a vector field $X$ is $(\na T)[X]=\na_X T$} This connection is called the Levi-Civitta (or Riemannian) connection of $g$ and is canonical in a Riemannian manifold (\ie it exists once the metric exists). As shown above, there is a unique principal $SO_{(1,3)}$-connection in $F\cm$ associated to the Koszul $SO_{(3,1)}$-connection called Levi-Civita connection. Therefore, by definition, an $O_{1,3}$ metric defines a canonical $SO_{(1,3)}$-compatible splitting of $\ct F\cm$ into horizontal and vertical spaces. 

Given a frame $\fr_{Ua}$, the metricity condition can be expressed as  \begin{equation}\label{MetricityCondition}
\dif g_{ab}-2\;\omg_{(ab)}=0
\end{equation}
where $\omg_{ab}\equiv\omg^k{}_b g_{ak}$. Therefore note that, when written in an orthonormal frame, the metricity condition is equivalent to skew-symmetry of the matrix of connection one-forms $\omg_{(ab)}|_\mathrm{orth.}=0$.

The components of the Levi-Civitta connection in a coordinate frame can be found as follows. By writing the `1-form components' and because the $\dif x_U^k$'s are linearly independent the above equation implies 
\begin{equation}
\partial_\mu g_{\alpha\beta}-2\Gamg_{(\alpha|\mu|\beta)}=0
\end{equation}
By \eqref{ComponentsTorsionTensor}, the torsionless condition in a coordinate frame implies $\Gamg^k{}_{[ij]}=0$.  By summing and subtracting suitable permutations of the above equation we arrive at
\begin{equation}
\partial_\alpha g_{\beta\mu}+\partial_\beta g_{\mu\alpha}-\partial_\mu g_{\alpha\beta}-2\Gamg_{\mu(\alpha\beta)}=0.
\end{equation}
Using again that $\Gamg_{\mu[\alpha\beta]}=0$ we arrive at the well known Christoffel symbols, which describe the coefficients of the Levi-Civita connection when written in a coordinate frame, namely
\begin{equation}\label{eq:Christoffel}
\Gamg^\alpha{}_{\mu\nu}=\frac{1}{2}g^{\alpha\beta}\lr{\partial_\mu g_{\nu\beta}+\partial_\nu g_{\beta\mu}-\partial_\beta g_{\mu\nu}}.
\end{equation}
The Levi-Civitta connection has the nice property that its autoparallel curves are the {\bf geodesics} of the metric $g$, \ie the curves of extremal length between fixed points in $\cm$. However, since the length of a curve is a reparametrization-invariant quantity, this property is shared with a family of connections related to the Levi-Civita connection by a projective transformation.\footnote{I have the intuition that this is behind the results by Ehlers Pirani and Schild where they find that the most general connection that is compatible with the conformal and pre-geodesic structure defined by null and timelike rays is a Weyl connection \cite{EPS}}

As a final remark, I would like to emphasise that a metric structure canonically induces an affine connection, namely the Levi-Civita connection. It is canonical in the typical sense that no extra structure has to be introduced, nor any arbitrary choice has to be made (it is unique). The canonical nature of this connection will become clearer later, where we will see that any affine connection can be written as the Levi-Civita connection plus other tensorial corrections involving the torsion and nonmetricity tensors. Hence, it will become apparent that introducing an extra affine connection is equivalent to choosing particular nonmetricity and torsion tensors. In this sense, one could think that adding an independent affine structure introduces more arbitrariness. Nevertheless, there are gravitational theories made only with an affine connection where the metric is derived from the connection. In the end, as we already pointed out in the introduction, the arbitrariness that exists in the design of our theories concerns only the choice of fundamental degrees of freedom (only metric, only connection, both, etc.) and the choice of a particular dynamics for them (either via a choice of action principle or field equations). Thus, in my opinion, any of these frameworks has the same degree of arbitrariness at a foundational level, and only experiments can allow us to determine if any of the chosen frameworks are able to provide valid or consistent physical descriptions. 

\subsection{A general affine connection}\label{sec:GeneralAffineConnection}

From the exposition in section \ref{sec:Connections}, it is clear that an affine connection can be introduced even if we have no metric structure on $\cm$. In this section we will first outline several properties of a general affine connection and build a plethora of geometric objects from it. I will try to emphasise which of these properties/objects do not need the notion of a metric and which of them do. However, bear in mind that in metric-affine theories of gravity, there is always a metric and a connection at play, so that we will be able to define and use all the objects and properties written below. 

Given a metric $g$ and a general affine connection $\na$, the tern $(\cm,g,\na)$ will generally be called a {\bf post-Riemannian} manifold.\footnote{Also known as metric-affine manifold, or non-Riemannian manifold. Although the later can introduce confusion with the terminology usually used by mathematicians.} In such spaces there is a canonical (0,2) tensor-valued 1-form,  called {\bf nonmetricity} 1-form $Q=\na g$, which measures the departures from metricity of the affine connection. The torsion tensor and nonmetricity tensors measure departures from Riemannian geometry in the following sense. In a coordinate frame, the connection coefficients can be decomposed as
\begin{equation}\label{eq:ConnectionDecomposition}
\Gamma^\alpha{}_{\mu\nu}=\Gamg^\alpha{}_{\mu\nu}+L^\alpha{}_{\mu\nu}+K^\alpha{}_{\mu\nu}
\end{equation}
where $L^\alpha{}_{\mu\nu}$ and $K^\alpha{}_{\mu\nu}$ are the {\bf distortion} and {\bf contortion} (1,2) tensors respectively. These objects are linear combinations of the nonmetricity and torsion tensors as follows
\begin{equation}\label{eq:ContortionDistortion}
\begin{split}
&L^\alpha{}_{\mu\nu}=\frac{1}{2}g^{\alpha\beta}\lr{Q_{\beta\mu\nu}-Q_{\mu\beta\nu}-Q_{\nu\beta\mu}}\\
&K^\alpha{}_{\mu\nu}=\frac{1}{2}g^{\alpha\beta}\lr{T_{\beta\mu\nu}+T_{\mu\beta\nu}+T_{\nu\beta\mu}}.
\end{split}
\end{equation}
Therefore, a Riemannian geometry can be seen as a particular case of metric-affine geometry with vanishing nonmetricity and torsion tensors. The tensorial nature of these objects stems from the fact that the difference of the connection coefficients of two different affine connections is always a (1,2) tensor (recall that the difference of two Koszul connection 1-forms is a global 1-form on $\cm$). Hence, although $\Gamma^\alpha{}_{\mu\nu}$ and $\Gamg^\alpha{}_{\mu\nu}$ do not transform as tensors and the expressions for their components will generally vary when expressed in two different frames, this will not be the case for $L^\alpha{}_{\mu\nu}$ and $K^\alpha{}_{\mu\nu}$, which will be written as above in any given frame. Particularly, although there generally exist frames in which $\Gamma^\alpha{}_{\mu\nu}$ or $\Gamg^\alpha{}_{\mu\nu}$ vanish at a given point, neither $L^\alpha{}_{\mu\nu}$ nor $K^\alpha{}_{\mu\nu}$ can be made to vanish at any point by a choice of frame. 

The above splitting of the connection coefficients can also be carried out in a general frame by means of $\eqref{eq:ConnectionDecomposition}$, leading to
\begin{equation}\label{eq:ConnectionDecompositionArbitraryFrames}
\omega_k{}^a{}_b=\omg_k{}^a{}_b+L^a{}_{kb}+K^a{}_{kb}
\end{equation}
where $\omg_k{}^a{}_b=\df^a{}_\alpha(\dif \fr_b{}^\alpha)_k+\Gamg^a{}_{kb}$ and we have defined $\Gamg^a{}_{kb}\equiv\df^a{}_\alpha\fr_k{}^\mu\fr_b{}^\beta\Gamg^\alpha{}_{\mu\beta}$.
 
We see that in a post-Riemannian space-time there are three basic covariant geometrical objects that one can construct from the connection, namely its  nonmetricity, torsion and Riemann curvature tensors (or their analog tensor-valued forms). Both the torsion and nonmetricity tensors (or their contortion and distortion combinations) measure the departure of a connection from the Riemannian connection and, therefore, they are post-Riemannian in nature. In light of the above decomposition \eqref{eq:ConnectionDecomposition}, it should be apparent why Riemannian geometries can be seen as a particular case of post-Riemannian ones. Another view to understand why the Levi-Civita conection is canonical is due to the fact that, given only a metric, the only canonical choice is the trivial choice for both torsion and nonmetricity, as any other choice would be arbitrary in the sense that it would require to endow $\cm$ with two additional tensor fields (\ie to add new structure). It is in this sense that, given a metric, the Levi-Civita connection is a canonical affine connection on $\cm$. 
 
 For completeness, let us also give the definition of other relevant objects that are built from the Riemann curvature tensor which, as a reminder, is given by
 \begin{equation}\label{eq:RiemannConnectionSymbols}
R{}^\alpha{}_{\beta\mu\nu}=2\partial_{[\mu}\Gamma{}^\alpha{}_{\nu]\beta}+2\Gamma{}^\alpha_{[\mu|\sigma|}\Gamma{}^\sigma{}_{\nu]\beta}.
\end{equation} 
These objects are the Ricci, homothetic, and co-Ricci curvature tensors respectively
 \begin{equation}\label{eq:RicciHomotheticCoRicci}
 \begin{split}
 R_{\mu\nu}=R^\alpha{}_{\mu\alpha\nu} \qquad , \qquad H_{\mu\nu}=R^\alpha{}_{\alpha\mu\nu} \qquad \text{and} \qquad P^\mu{}_\nu=g^{\alpha\beta}R^\mu{}_{\alpha\nu\beta}.
 \end{split}
 \end{equation}
 Note that while the Ricci and homothetic tensors do not require the existence of a metric, the co-Ricci does. As well, note that the homothetic tensor is the trace of the matrix of curvature 2-forms (hence it is a 2-form). Unlike the Riemann tensor of the Levi-Civita connection, the only index symmetry of the Riemann tensor of a general affine connection has only antisymmetry in its two later indices (which are the 2-form indices of the associated curvature 2-form). Hence, the Ricci tensor is not symmetric for a general affine connection, which will have important consequences regarding the stability of metric-affine theories (see chapter \ref{sec:UnstableDOF}).
 
  It will be useful to write down how the different objects that are defined from the connection coefficients relate when two connections are related by the most possible general transformation $\Gamma^\alpha{}_{\mu\nu}=\bar{\Gamma}^\alpha{}_{\mu\nu}+\delta\Gamma^\alpha{}_{\mu\nu}$. For the Riemann tensor we have that
\begin{align}\label{eq:TransformationRiemman}
R^\al{}_{\be\mu\nu}(\Ga)=R^\al{}_{\be\mu\nu}(\bar{\Ga})+2\bar{\na}_{[\mu}\delta\Ga^\al{}_{\nu]\be}+\bar T^\lambda{}_{\mu\nu}\delta\Ga^\al{}_{\lambda\beta}+2\delta\Ga^\al{}_{[\mu|\la|}\delta\Ga^\la{}_{\nu]\beta},
\end{align}
where the connection-related objects with an over-bar are defined in terms of the background connection $\bar\Ga^\al{}_{\mu\nu}$. By taking the corresponding traces we find
\begin{align}\label{eq:TransfRicciCoRicciHomothetic}
\begin{split}
&R_{\mu\nu}(\Ga)=R_{\mu\nu}(\bGa)+2\bar\na_{[\al}\delta\Ga^\al{}_{\nu]\mu}+\bar{T}^\lambda{}_{\al\nu}\delta\Ga^\al{}_{\la\mu}+2\delta\Ga^\al{}_{[\al|\la|}\delta\Ga^\la{}_{\nu]\mu}\;,\\\
&H_{\mu\nu}(\Ga)=H_{\mu\nu}(\bGa)+2\pa{[\mu}\delta\Ga^\al{}_{\nu]\al}\;,\\
&P^\mu{}_{\nu}(g,\Ga)=P^\mu{}_{\nu}(g,\bGa)+\bar\na_\nu\delta\Ga^{\mu\al}{}_{\al}-\bar\na_\al\delta\Ga^{\mu\;\;\al}_{\;\;\nu}+\bar{T}^\lambda{}_{\nu\al}\delta\Ga^{\mu}{}_{\la}{}^{\al}+2\delta\Ga^\mu{}_{[\nu|\la|}\delta\Ga^{\la}{}_{\al]}{}^{\al}\;.
\end{split}
\end{align}
The torsion and nonmetricity tensors of the two connections relate by
\begin{align}\label{eq:TransfTorsionNonmetricity}
\begin{split}
&T^\al{}_{\mu\nu}(\Ga)=\bar{T}^\al{}_{\mu\nu}+2\delta\Ga^\al{}_{[\mu\nu]}\;,\\
&Q_{\al\mu\nu}(g,\Ga)=Q_{\al\mu\nu}(g,\bGa)-2\delta\Ga_{(\mu|\al|\nu)}\;,
\end{split}
\end{align}
and therefore, the corresponding contortion and distortion tensors satisfy
\begin{align}\label{eq:TransfContortionDistortion}
\begin{split}
&K^\al{}_{\mu\nu}(\Ga)=K^\al{}_{\mu\nu}(\bGa)-\delta\Ga_{(\mu\nu)}{}^\al{}+\delta\Ga^\al{}_{[\mu\nu]}+\delta\Ga_{(\mu}{}^{\al}{}_{\nu)}\;,\\
&L^\al{}_{\mu\nu}(g,\Ga)=L^\al{}_{\mu\nu}(g,\bGa)-\delta\Ga_{(\mu}{}^{\al}{}_{\nu)}+\delta\Ga^\al{}_{(\mu\nu)}+\delta\Ga_{(\mu\nu)}{}^\al\;.
\end{split}
\end{align}
The properties of the different geometrical objects related to a general affine connection stated above are all that we will need in the rest of the thesis. Particularly, we will be mostly interested in the transformation properties of the Ricci tensor under a projective transformation, given by $\delta\Gamma^\alpha{}_{\mu\nu}=- \xi_\mu\delta^\alpha{}_\nu$, which are
\begin{align}\label{eq:TransfRicciProjective}
R_{\mu\nu}(\Gamma)=R_{\mu\nu}(\tilde\Gamma)-(\dif\xi)_{\mu\nu}.
\end{align}
Hence, while the symmetric part of the Ricci tensor is invariant under a projective transformation, its antisymmetric part changes proportionally to the fieldstrength of the projective mode. As a curiosity, let us point out that in metric-affine GR, which is invariant under projective transformations, the transformation properties of the full Riemann tensor under projective symmetry can be used to remove the curvature divergence of the metric-affine Krestchmann scalar $R^{\mu\nu\alpha\beta}R_{\mu\nu\alpha\beta}$ in a Schwartzschild geometry \cite{Bejarano:2019zco}. However, as could not be otherwise, the singularity in the metric structure is still there and appears through the projectively-invariant scalar $R^{\mu\nu\alpha\beta}R_{\alpha\beta\mu\nu}$.

We have all the metric-affine identities and definitions from this section that we will need through the thesis. Let us then carry on to elaborate on how a general affine connection defines a connection on the spin bundle. 
 
 \subsubsection{Spin connection on $\cS\cm$ from a general affine connection}
 
We have outlined above how to obtain explicitly the linear connection on the spin bundle canonically associated to the Levi-Civita connection, \ie an $SO^+_{(3,1)}$-connection on $\tb$. However, in metric-affine theories, we wish to work with more general affine connections, \ie linear connections on $\tb$ which need not be $SO^+_{(3,1)}$ connections. This might be seen as introducing some degree of arbitrariness in the process because of the following reason. While $\mathfrak{sl}_{(2,\bbc)}$ maps to $\mathfrak{so}_{(3,1)}$ in a one to one fashion, this is not possible in general for other general Lie subalgebras of $\mathfrak{gl}_{n}$. Hence, for affine connections more general than the Levi-Civita connection, it is not guaranteed that the whole connection can be lifted to the spin bundle in a canonical manner. This was explained in \cite{Hurley:1994cfa}, where they propose that one can always decompose the matrix of connection 1-forms $\omega_{ab}$ into its symmetric and antisymmetric pieces. The antisymmetric piece will be an element of $\mathfrak{so}_{(3,1)}$, which consists of antisymmetric $4\times4$ matrices. However, there is no canonical way to lift the symmetric part of the connection 1-form to the spinor bundle associated to the $SO^+_{(3,1)}$ tangent bundle, \ie to the Dirac spinor bundle $\cS\cm$. As explained in \cite{Hurley:1994cfa}, this does not mean that only the Levi-Civita piece of a general affine connection gets lifted canonically to $\cS\cm$. By computing the spin connection canonically associated to a general affine connection  by the same procedure used in \ref{sec:Connections}, we find that the canonical spin connection is still given by \eqref{eq:SpinConnection}, namely
\begin{equation}\label{eq:SpinConnectionGeneral}
\omega^s_U=\frac{1}{8}\omega_U{}_{ab}[\gamma^a,\gamma^b],
\end{equation}
which, by using the general decomposition of the affine connection 1-form \eqref{eq:ConnectionDecompositionArbitraryFrames}, can now be decomposed as
\begin{equation}\label{eq:SpinConnectionDecomposition}
\omega^s=\omg^s+\omega^s_\mathrm{PR},
\end{equation}
where $\omg^s$ is the piece corresponding to the $SO_{(1,3)}$ connection (the Levi-Civita connection) and $\omega^s_\mathrm{PR}$ encodes the post-Riemannian part. In terms of the torsion, nonmetricity and the Christoffel symbols these objects read
\begin{equation}\label{eq:CanonicalSpinConnection}
\omg^s_\mu=\frac{1}{8}\omg_\mu{}^a{}_b[\gamma^a,\gamma^b]\qquad\text{and}\qquad \omega^s_\mathrm{PR}{}_\mu=\frac{1}{8}\df^c{}_\mu\lr{K_{acb}+Q_{[ab]c}}[\gamma^a,\gamma^b].
\end{equation}
 Therefore, we see that there are other pieces of a general affine connection which are associated to nonmetricity and torsion which also get canonically lifted to the spin connection on $\cS\cm$. 

Nevertheless, note that while the canonical lift is able to lift all the torsion-related piece of the connection, it is blind to some pieces of the distortion tensor, and therefore, of the nonmetricity tensor, as it only lifts $Q_{[ab]c}$. This can be traced back to the fact that a (Dirac) spin connection 1-form must be $\mathfrak{so}_{(1,3)}$-valuated, and the other parts of the nonmetricity tensor are associated to pieces of the affine connection 1-form that are in $\mathfrak{gl}_n$ but not in $\mathfrak{so}_{(1,3)}$. Concretely, the piece $L_{(a|c|b)}$ is the piece of the connection 1-form which is not part of $\mathfrak{so}_{(1,3)}$, and therefore does not get lifted to the spin connection in $\cS\cm$. This part corresponds to the shear and expansion during parallel transport, which vanish if the nonmetricity vanishes. 

Even though we have seen that the canonical lift of a general affine connection to the Dirac spinor bundle lifts all the torsion-related and part of the nonmetricity related components, Dirac spinors evolve through the Dirac operator $\gamma^\mu\na_\mu$. When applied to a spinor field this operator reads $\gamma^\mu\na_\mu\Psi=\gamma^\mu(\dif\Psi)_\mu+\gamma^\mu\omega^s_\mu\Psi$. As we will see in more detail in section \ref{sec:MinimalCoupling} we have that $\gamma^\mu\omega^s_\mu=\gamma^\mu\omg^s_\mu-i/8 T^{abc}\epsilon_{abcd}\gamma^d\gamma_5$, which implies that the coupling of a Dirac spinor to the geometry of $\cm$ through the Dirac operator associated to the canonical spin connection \eqref{eq:CanonicalSpinConnection} is blind to all the post-Riemannian terms except for the totally antisymmetric part of the torsion.

In \cite{Hurley:1994cfa} a noncanonical lift for the expansion piece (the traceful part of $L_{(a|c|b)}$) was devised, but no lift was found for the shear (the traceless part). Inspired by the action of the expansion piece of the connection on vector fields, this noncanonical lift consists of adding a piece proportional to the identity to the canonical spin connection 1-form, leading to
\begin{equation}
\omega^s_{NC}{}_\mu=\omega^s_\mu+\frac{1}{8}L^\alpha{}_{\mu \alpha}.
\end{equation}
Note however that this lift is quite arbitrary, as we could have also chosen to lift other post-Riemannian pieces by adding terms such as for instance $Q^\alpha{}_{\alpha\mu}$, $T^\alpha{}_{\mu\alpha}$, $Q_{a\mu b}[\gamma^a,\gamma^b]$, or any other combination that we might think of.

Let us conclude with the following remark: although a canonical lift can be found that lifts a general affine connection to the spin bundle, some would feel that there is some degree of arbitrariness in the process due to the fact that we have to make a choice whether to lift the expansion due to nonmetricity or not. However, in my view, the canonical lift of a general affine connection from the tangent bundle to the spin bundle is unique and determined entirely by the associated bundle construction, which makes it laking of arbitrariness. On the other hand, the lift of the expansion piece proposed in \cite{Hurley:1994cfa} is arbitrary: we could well choose to multiply the lifted piece by any factor or add other post-Riemannian terms and the result would still be a connection in $\cS\cm$. Another way to couple spinor fields would be to consider spinor representations of $GL_{n}$ instead of the special orthogonal groups. The problem with this idea is that there are no finite-dimensional unitary spinor representations of $GL_n$. There are, however infinite-dimensional unitary spinor representations which lead to the concept of {\it world spinors}, discussed in \eg \cite{Hehl:1994ue}. Up to date, it is not clear to me how to end up with a Dirac spinor that correctly describes the observed Standard Model particles from these world spinors. Of course, in the end, whether our universe is post-Riemannian or not, what are the type of fields that describe the existing degrees of freedom, and whether noncanonical pieces should be a part of the physical spin connection when nonmetricity or torsion are nontrivial is a matter of our interpretation of experimental data, which must be the guide to the correct answer(s). These considerations, however, help in clarifying the notion of minimal coupling of spinor fields to the geometry of a spacetime with a general affine connection that I will introduce below.

%%%%%%%%%%%%%%%%%%%%%%%%%
%%%%%%%%%%%%%%%%%%%%%%%%%
%%%%%%%%%%%%%%%%%%%%%%%%%

			%NEW CHAPTER

%%%%%%%%%%%%%%%%%%%%%%%%%
%%%%%%%%%%%%%%%%%%%%%%%%%
%%%%%%%%%%%%%%%%%%%%%%%%%

\chapter{On coupling matter to an affine connection}\label{sec:MinimalCoupling}
%%%%%%%%%%%%%%%%%%%%%%%%%%%%%%%%%%%%%%%%%%%%%%%%%%%%%%%
%%%%%%%%%%%%%%%%%%%%%%%%%%%%%%%%%%%%%%%%%%%%%%%%%%%%%%%
%%%%%%%%%%%%%%%%%%%%%%%%%%%%%%%%%%%%%%%%%%%%%%%%%%%%%%%
%%%%%%%%%%%%%%%%%%%%%%%%%%%%%%%%%%%%%%%%%%%%%%%%%%%%%%%

\initial{A}s is known, the interaction mediated by a massless spin-2 field can be interpreted in geometrical terms due to the crucial fact that it has to satisfy the Equivalence Principle.\footnote{See chapter \ref{sec:GravityAsGeometry} for a quick review of the arguments that lead to these conclusions.} Indeed, the original view of GR by Einstein is as a geometrical theory, and this way of interpreting the gravitational interaction has led to powerful techniques both in the understanding of the structure of the theory as well as the phenomenology that it predicts. Following this view, a natural way to explore what kind of gravitational theories lay beyond GR is to enhance the geometrical framework in which it is formulated. Besides the usual interpretation of GR is in terms of curvature of a Riemannian manifold, where the affine structure is taken to be the canonical affine structure provided by the metric, there are other geometrical arenas in which this theory can be given a geometrical interpretation where an independent affine structure plays a central role. Paradigmatic examples are provided by the Teleparallel Equivalent \cite{Aldrovandi:2013wha}, Symmetric Teleparallel Equivalent to GR \cite{Nester:1998mp,BeltranJimenez:2017tkd}, and General Teleperallel equivalent to GR \cite{Jimenez:2019ghw}, where curvature vanishes and the geometrical interpretation of the gravitational interaction is realised through the torsion and/or nonmetricity of the affine structure. These different geometrical interpretations of GR suggest a particular direction in which to explore modifications to enhance the geometrical framework where one assumes that the gravitational interaction is not only described by a metric, but also by an independent affine connection. This is known as the metric-affine framework, and encodes a vast number of gravitational theories, ranging from gauge theories of gravity \cite{Blagojevic:2012bc} to Palatini theories\footnote{Here by Palatini theories we mean metric-affine gravity theories where the action is built only with the Riemann tensor and related invariants. I will not use this term in general, but rather write metric-affine curvature-based theories, which I find more instructive.} \cite{Olmo:2011uz} or the different Teleparallel classes mentioned above. 

A crucial task within the metric-affine framework, is to understand if it is possible to find experimental probes that can test the existence of post-Riemannian features in the geometrical interpretation of the gravitational interaction. One way to pursue this goal is to understand the different ways in which matter fields can couple to the independent connection, which would allow us to elaborate different tests to probe these couplings. In this direction, in Riemannian spacetimes, there is an algorithm that allows to build a matter sector that is minimally coupled to a Riemannian geometry starting from its Minkowskian counterpart which can be given a solid motivation from the use of general coordinates in Minkowski space. Through this chapter I will argue why, in my view, and despite the fact that it is usually employed in the literature, this algorithm is not well motivated from a coordinate independent point of view when general post-Riemannian geometries are considered, and that it leads to contradictions and undesired results. I will also propose a notion for coupling matter to a post-Riemannian spacetime minimally that does not suffer from these problems and has its roots in arguments relying on observer (coordinate) independence of physical equations. 

A side-effect of choosing a particular coupling of the matter fields to the spacetime geometry is the fact that this determines the type of trajectories that freely falling particles will follow. However, in the literature, it is sometimes assumed\footnote{This can be seen \eg in \cite{Avalos:2016unj,Lobo:2018zrz,Delhom:2020vpe}, where there are definitions of geometric clocks assuming that test bodies follow affine geodesics. Note that I have also made this assumption as a co-author of one of these works, although I later learnt that it is, at best, very optimistic.} that autoparallel paths (sometimes called affine geodesics) of the affine connection play an analog role as geodesics in Riemannian geometries do. We will also argue why, in general, nontrivial couplings between matter and connection deviate free particles from following metric geodesics, and they follow the autoparallel paths of an effective connection which depends on the particular coupling and is not, in general, the full affine connection. In fact, we will also argue that it is not clear that an action can be formulated so that the corresponding solutions to its field equations yield, in the eikonal limit, autoparallel paths of a general affine connection, which clashes with the common expectation that matter fields are described by an action principle. 
 
\section{A prescription for minimal coupling} \label{sec:MinimalCouplingPrescription}

To stand on the same ground, let me start by discussing the general meaning of minimal coupling between matter and geometry as I understand it, and the subtleties behind the usual algorithm that is regarded as a minimal coupling prescription. In general, one can define a notion for minimal coupling between matter and geometry whenever one promotes the geometry of a base space where a given physical theory is already formulated to a more general geometry. To my understanding, the idea is to couple the matter fields to the geometry {\it as little as possible}, namely, to leave things as they were {\it as much as possible}, without being inconsistent with any physical principle. Now in principle, one could just say that minimal coupling could consist on using the Minkowskian field equations and not coupling the matter fields to the geometry at all. However, that would be inconsistent with the principle of (general) relativity, namely, that the physical laws are formulated equally for any observer, and even with covariance.\footnote{Here coordinate invariant and observer-independent should be understood as equivalent. Covariant will imply a stronger condition, namely, besides being observer-independent, a covariant object (that takes values in some fiber-bundle) is an object which is not only coordinate-invariant, but also independent of the local trivialisations employed to describe the bundle. Given that the choice of trivialisation is just a purely mathematical choice due to the formalism employed, it must not have physical consequences. Hence, covariance must also be required.} We will see that taking seriously covariance already in Minkowski spacetime leads to a nontrivial minimal modification of the matter field equations (or Lagrangians) that introduces a coupling between the matter and the geometry when going to more general spacetimes. Having no other fundamental principles that force us to introduce additional couplings, this requirement together with the idea of modifying things as little as possible are the cornerstones behind the usual idea of minimal coupling to the spacetime geometry. 

When providing prescriptions for minimal minimal coupling for passing from a Minkowskian (or pseudo-Riemannian) spacetime to a general metric-affine spacetime, the crucial role played by requiring covariance in Minkowski space is usually overlooked, and this leads to the loss of the idea of coupling to the geometry {\it as little as possible}. In my view, this comes from the fact that, when passing from a Minkowskian spacetime to a Riemannian one, all that most texts will say about minimal coupling is just a statement giving the usual algorithm that defines minimal coupling to a (pseudo-)Riemanian geometry, which is usually similar to: \textit{Wherever you find a Minkowski metric $\eta$ or a partial derivative $\partial$ in flat space-times, substitute them by the (pseudo-)Riemannian metric $g$ and its covariant derivative $\na^g$ respectively.} This (century-old) algorithm stems from taking seriously the covariance of physical laws in Minkowski space, and then using the same coordinate-independent differential operators when formulating the theory in a Riemannian spacetime, \ie leaving things as they were {\it as much as possible}. Nevertheless, this fact is commonly overlooked in most treatments, with the consequence that this simple algorithm is usually elevated to the category of {\it definition} of minimal coupling to a geometry, and employed in some works on metric-affine theories as such. Though perfectly fine for (pseudo-)Riemannian spacetimes, this prescription leads to undesired consequences when passing from a Minkowskian or (pseudo-)Riemannian spacetimes to a metric-affine one, such as the loss of gauge invariance when applied to a gauge field, or to different results when applied directly to the field equations instead of the Lagrangian. In order to escape from this undesired results, let us try to understand how this prescription stems from taking seriously the covariance of physical laws in Minkowski spacetime. In the process, it will become clear how observer-independence is also enough to provide an algorithm for passing from a Minkowskian (or Riemannian) spacetime to a metric-affine one which respects the idea of coupling to the geometry {\it as little as possible}. The resulting minimal coupling prescription, besides respecting the idea of introducing minimal modifications, will be free from all the above mentioned issues, with the exception of giving different results for spin-1/2 fields if applied to the Lagrangian or to field equations. We will show how the application of the algorithm to the field equations leads to non-conservation of the fermionic vector current.

Going to the point, let me try to find a minimal coupling prescription by taking seriously covariance in Minkowski space. As explained in section \ref{sec:TangentSpace}, the operator $\partial$ with which field equations in Minkowski are usually written is associated to a given coordinate frame (or observer). Hence different coordinate systems will generally yield different $\partial$ operators and, therefore, $\partial$ will yield a non-coordinate-invariant object when applied over any covariant object like a tensor or spinor field. Hence, observer independence of physical theories implies that a particular choice of $\partial$ cannot be employed in the construction of physical theories even in Minkowski spacetime, where it will be a different operator for two non-inertially-related observers. Given that, usually, field theories are first formulated in a Minkowskian spacetime and for inertial observers, the symbol $\partial$ is commonly used. Thus, in order to apply a notion of minimal coupling that stems from covariance, one must formulate the Minkowskian field theories in terms of covariant operators. In chapter \ref{sec:DifferentialGeometry}, we introduced several coordinate-invariant differential operators, such as the exterior differential $\dif$, the co-differential $\delta_g$, or the covariant derivative $\na$. Suitable combinations of these operators, like the wave\footnote{Known as the Laplace-de-Rham operator, which generalizes the Laplacian to any  manifold with a metric structure.} operator $\Box_g\equiv\dif\delta_g+\delta_g\dif$, will also be coordinate-invariant operators, and therefore will yield coordinate invariant objects when applied to covariant matter fields. Recall that $\dif$ is defined in any differentiable manifold without adding extra structure, $\delta$ requires a metric structure that provides the notion of Hodge dual, and $\na$ requires an affine structure. In Minkowski spacetime, there is a metric structure $\eta$ and a canonical affine connection,\footnote{Minkowski space is parallelizable, and therefore there exist (inertial) frames in which the connection 1-form vanished $\omega=0$. It is possible to see that this affine structure is also the canonical one associated to the Minkowski metric, ans therefore $\na^\eta$ is its correpsonding covariant derivative.} and therefore the three mentioned operators are canonically defined in this spacetime. Thus, to define a minimal coupling prescription guided by covariance, we must unveil what are the covariant differential operators that are used in Minkowski spacetime and just use the same covariant operators generalised to a general spacetime. This can also be written as a simple algorithm: {\it Identify which are the covariant differential operators used in the Minkowski spacetime formulation of the corresponding matter Lagrangian, and then use the same Lagrangian with the same covariant operators as they are defined in the general metric-affine case}.\\

In order to see whether this algorithm is effective, let us see whether we can find out which are the covariant differential operators appearing in the Lagrangian and field equations of free scalar, spinor and 1-form (or vector) fields. To that end, it will be useful to take into account that only first-order differential operators appear in these Lagrangians. For scalar and spinor fields, given that they can be seen as 0-forms,\footnote{More precisely, 0-form sections of the trivial bundles $\cm\times\bbr$ or $\cm\times\bb{C}$ and the spin bundle $\cS\cm$ respectively} the $\delta_g$ operator annihilates them, so that it cannot play any role in the Lagrangian,  since it only depends on first derivatives of the fields. For a free scalar field $\phi$, the two remaining operators are equivalent $\dif\phi=\na\phi$ for any affine connection so that both can be used. However, using the exterior differential, in general metric-affine spacetimes, implies using only the metric structure and not the affine one, so we will do that to keep things minimal. Regarding the field equations, the kinetic term is typically written as $\partial^\mu\partial_\mu\phi$, which is equivalent to both $\eta^{\mu\nu}\na^\eta_\mu\na^\eta_\nu\phi$ or $\Box_\eta\phi$, where $\Box_\eta\equiv\dif\delta_\eta+\delta_\eta\dif$. Note that, since the canonical connection in Minkowski is the Levi-Civita connection of the Minkowski metric, both terms couple the scalar field only to the metric structure. However, we will find that the differential form notation is more transparent when generalised to post-Riemannian spacetimes, and we will use $\Box_\eta$ to write the free scalar field equations in Minkowski and we will see that the minimally coupled Lagrangian leads to the same field equations replacing it by $\Box_g$. As a remark, let me point out that when the scalar field is also fiber-valued for some nontrivial vector $G$-bundle over $\cm$, $\dif\phi$ is not covariant under local changes of trivialisation of the bundle, and therefore the covariant exterior differential must be used in its place. This operator features a {\it differential part} given by $\dif$ and a coupling to the $G$-connection 1-form of the corresponding $G$-bundle, which has nothing to do with the affine connection\footnote{This is the case of, {\it e.g.}, the Higgs field, which is an element of a vector $SU(2)\times U(1)$-bundle which interacts with the gauge fields corresponding to the $SU(2)\times U(1)$-connection 1-forms.} (see  section \ref{sec:ConnectionsOnBundles} for the definition of exterior covariant differential). 

For a 1-form field $A$, it can be seen that with this operators, the only kinetic term that is covariant and does not lead to ghost-like kinetic term for the longitudinal mode must be proportional to $(\dif A)^2$ up to boundary terms. This is usually encoded in the requirement of gauge invariance to the kinetic term of 1-form fields, which guarantees the propagation of two (three) healthy degrees of freedom for a massless (massive) 1-form field. In a Minkowski or (pseudo-)Riemannian spacetime, it is satisfied that $(\dif A)_{\mu\nu}=2\nabla^g{}_{[\mu} A_{\nu]}$, but this is not the case in more general spacetimes. Since the key point that leads to that kinetic term can be encoded into the requirement of gauge invariance, which is formulated naturally in the language of differential forms, we will stick to the use of $\dif$ for minimally coupling the vector field. This argument is clearer at the level of field equations, where again the covariant form of $\partial^\mu\partial_\mu A$ can be written as $\delta_\eta\dif A$ (or $\Box_\eta A$ in the Lorentz gauge). This kinetic structure leads to the divergence-free constraint for the fieldstrengh of the vector field, which is required if coupled to a conserved current. All the gauge structure of 1-form fields is formulated solely in terms of exterior differentials and Hodge duals, which require only the notion of a metric but not that of a connection.\footnote{However, recall that a canonical connection is given once we have a metric. Indeed, it can be seen that the coordinate-independent differential operators defined for $p$-forms can always be written in terms of the canonical covariant derivative associated to the metric if the Hodge dual operator is also the one defined by that metric. This interplay between the exterior-differential structure and the affine-structure only occurs for the Riemannian connection of the metric used both, to define the Hodge star operator and to identify the space of $p$-forms with its dual via a generalisation of the metric isomorphism defined in \eqref{eq:MetricIsomorphism}.} We will see below that, if one insists on using the affine covariant derivative to express observer-independence (instead of using the exterior differential), then, when going to a post-Riemannian spacetime, there will arise some extra couplings to the torsion that do not arise when using the exterior differential. These new terms do not respect the idea of coupling to the geometry {\it as little as possible}, which goes against the spirit of the idea that I have of minimal coupling. Furthermore, they break the gauge invariance, providing an effective mass term to the gauge fields and potentially unleashing the propagation of a longitudinal polarisation. This could give rise to strong coupling issues around vanishing torsion backgrounds, or instabilities around generic torsion backgrounds.

spin-1/2 fields $\Psi$ are a bit trickier because the kinetic term employed must be invariant under changes of local trivialisation in the spinor bundle, but $\partial\Psi$ is not in general. This forces us to use the covariant exterior differential associated to the Minkowskian affine connection seen as a linear connection in the spinor bundle or, in a more familiar language, the associated spinor covariant derivative (see section \ref{sec:Connections}). Then, our prescription for spin-1/2 fields will be to employ the covariant derivative $\nabla$ of the affine structure. As we will see, unlike the case of scalar and 1-form fields, this prescription will yield different results when applied to the Lagrangian instead of the field equations, leading to a violation of the conservation of the vector current. Furthermore, it also introduces a nontrivial minimal coupling of the spinor field to the post-Riemannian features of the geometry, particularly to the totally antisymmetric part of the torsion tensor. This shows how the algorithm facilitated above can indeed be implemented to yield the corresponding minimally coupled theories in a general metric-affine spacetime starting from a Minkowskian one. In the following, we will analyse in detail what are the differences between this prescription and the usual one of replacing $\eta$ by $g$ and $\partial$ by $\nabla$ everywhere. To that end, it will be useful for us to make a detour and derive the form of the Euler-Lagrange equation for any Lagrangian containing the derivatives of the matter fields inside affine covariant derivatives, and reproduce Noether's theorem that guarantees conservation of the fermionic vector current.

\section{Field equations and Noether currents}\label{sec:divopEL}

In order to derive the field equations and the corresponding Noether currents, it is useful to recall the Stokes' theorem and Gauss' law as a particular case. Given that we are admitting Lagrangians that use the differential operator $\na$, it will be useful to write the divergence operator $\div_g$ associated to the (volume form of the) metric defined in terms of the covariant derivative and the corresponding post-Riemannain corrections. Let $\mathcal{M}$ be a smooth orientable n-dimensional manifold with volume form $\dif V=\epsilon\, dx^{\nu_1}\wedge...\wedge dx^{\nu_n}$ in some chart. The divergence operator associated to a general volume form $\dif V$ is defined by its action on vector fields in \eqref{eq:DefinitionDivergence}, which in a particular coordinate chart $x^\mu$  can be written as
\begin{equation}\label{Divopcoord}
\div_{\dif V}(A)=\frac{1}{\epsilon}\pa{\mu}(\epsilon A^\mu),
\end{equation}
where $A=A^\mu\partial_{\mu}$. Note that this definition only requires the differential structure of $\mathcal{M}$ and a general volume form defined on it. Neither a metric tensor $\textbf{g}$ nor an affine structure ${\Ga}$ on $\mathcal{M}$ are necessary. The interest of this operator relies in that it satisfies a particular version of Stokes' theorem which relates the values of the vector field defined in a region $\mathcal{V}\subset\cm$ with its flux through the boundary of the region $\partial\mathcal{V}$.  In particular, the general Stokes' theorem states that 
\begin{equation}\label{eq:StokesTheorem}
\int_{\partial\mathcal{V}} \omega=\int_\mathcal{V}\dif\omega\,,
\end{equation}
For any $(k-1)$-form $\omega\in\Lambda^p\mathcal{V}$ and $k$-dimensional orientable submanifold $\mathcal{V}\subset\cm$. By taking $k=n$ and $\omega=\mathrm{i}_A\dif V$, and using the definition of the divergence operator $\eqref{eq:DefinitionDivergence}$ from Stokes' theorem one finds
\begin{equation}\label{Divtheogen}
\int_{\partial\mathcal{V}} \mathrm{i}_{A}\dif V
=\int_\mathcal{V}\dif V
 \div_{\dif V}({A})
\end{equation}
This is the generalised divergence theorem for $n$-dimensional manifolds. Now, if a metric structure $g$ is introduced, using the canonical volume form $\dif V_g$, it can be seen that $\left.\mathrm{i}_{A}\dif V_g\right|_{\partial V}=g(A,N)\dif V_{\tilde{g}}$, where $N$ is the unit normal to $\partial\mathcal{V}$ and $\dif V_{\tilde{g}}$ is the induced volume form on $\partial\mathcal{V}$ by $g$. The symbol $|_{\partial V}$ stands for restriction to $\partial V$. Therefore, when a metric is present  (and is chosen to define the volume element),  the right hand side of \eqref{Divtheogen} can be interpreted as the flux \textit{normal} to the boundary enclosing $\mathcal{V}$, and the generalised divergence theorem \eqref{Divtheogen} can be seen as a generalisation of the celebrated Gauss' law
\begin{equation}\label{Gaussgen}
\int_{\partial\mathcal{V}} g(A,N)\dif V_{\tilde{g}}=\int_\mathcal{V}\dif V_g\div_g(A)\;,
\end{equation}
 which relates the divergence of a vector field $A$ inside a closed volume $\mathcal{V}$ with the integration over $\partial\mathcal{V}$ of the component of $A$ normal to $\partial\mathcal{V}$. It can be seen that the divergence operator associated to a metric can also be written in terms of its canonical covariant derivative as $\div_g(A)=\nabla^g{}_\mu A^\mu$. In spaces where the covariant derivative is not the Riemannian one, it might be useful to find a similar relation in terms of the corresponding covariant derivative instead of the Riemannian one. The action of $\na$ on (the only component of) an n-form\footnote{Recll that, since the space of n-forms in an n-dimensional manifols is of dimension 1, any n-form is proportional to the trivial one $dx^1\wedge...\wedge dx^n$, and therefore is specified only by one component (the proportionality factor). This component can be seen as a tensor density of weight $+1$.} $\epsilon$, in a chart $x^\mu$ is given by $\nabla_\mu \epsilon =(\pa{\mu}\epsilon-\Ga^\al{}_{\mu\al} \epsilon)$, and therefore we can write, in general
\begin{equation}\label{Divtor}
Div_\dv(A)=\frac{1}{\epsilon}\nabla_{\mu}(\epsilon A^\mu)+T^\al{}_{\mu\al} A^\mu.
\end{equation}
Note that \eqref{Divtor} is, in general, independent of the metric structure. Indeed, we can generally provide $\cm$ with a volume form and an affine structure without having a metric structure, and in that case, \eqref{Divtor} is still valid. In a general metric-affine manifold, it is possible to show that 
\beq
\na_\mu \sqrt{-g}=\frac{1}{2}{Q_{\mu\al}}^\al \sqrt{-g}
\eeq
which using \eqref{Divtor} yields a general relation between the canonical divergence operator associated to the metric and the general affine covariant derivative as
\begin{equation}\label{Divgen}
\div_g({A})=\na_\mu A^\mu+\lr{\frac{1}{2}{Q_{\mu\al}}^\al+T^\al{}_{\mu\al}}A^\mu.
\end{equation}
which, if $\nabla$ is expanded using the decomposition \eqref{eq:ConnectionDecomposition}, reduces to the well-known expression $\div_g(A)=\na^g_\mu A^\mu$. The relation found above between the divergence operator and the affine structure can be useful to derive the matter field equations for any minimally coupled matter Lagrangian if the derivatives are written in terms of $\nabla$ (note that $\dif$ can always be written as $\nabla$ plus corrections depending on the particular affine connection). Thus, here we will be concerned with a matter action of the form
\begin{equation}\label{matteraction}
\cS_\textrm{m}\lrsq{\Psi_i,\na\Psi_i}=\displaystyle \int_\mathcal{V}\dif V_g\mathcal{L}[\Psi_i,\na\Psi_i] \ ,  \qquad i=1,...,N \ ;
\end{equation}
where $\mathcal{L}[\Psi_i,\na\Psi_i]$ is a covariant scalar and $i$ labels the different matter fields, which will be assumed to be sections of some vector bundle. The corresponding field equations are given by $\delta \cS_\textrm{m}=0$ for some arbitrary variations of the matter fields $\delta\Psi_i$ that vanish over $\partial\mathcal{V}$. A given variation of the field $\delta\Psi_i$ naturally introduces also a variation in its partial derivative $\delta(\partial\Psi_i)$, and the variational problems that one is used to solve are in terms of the field variables $\{\Psi_i, \partial\Psi_i\}$. Thus, we can treat $\nabla\Psi_i$ as a function of $(\Psi_i,\partial_\mu\Psi_i)$ and proceed with standard variational methods. Namely, we can rewritte \eqref{matteraction} as\begin{equation}\label{matteractionsubst}
\tilde \cS_\textrm{m}\lrsq{\Psi_i,\pa{\mu}\Psi_i}=\displaystyle \int_\mathcal{V}\dif V_g\mathcal{L}[\Psi_i,\partial\Psi_i+\omega^i\Psi_i],
\end{equation}
where $\omega^i$ is the connection 1-form in the vector bundle where $\Psi_i$ is defined. Now we can employ the standard methods of variational calculus. For an arbitrary variation $\delta\Psi_i$ we have
\begin{align}\label{deltatildeS3}
\delta \tilde \cS_\textrm{m}=\displaystyle \int_\mathcal{V}\dif V_g \lrsq{\lr{\frac{\pa{}\cL}{\pa{}\Psi_i}+\frac{\partial\cL}{\partial(\na_\mu\Psi_i)}\omega^i{}_\mu}\delta\Psi_i+\lr{\frac{\pa{}\cL}{\pa{}(\na_\mu\Psi_i)}}\delta(\partial_\mu\Psi_i)}=0,
\end{align}
where we have used that
\begin{equation}
\frac{\partial(\na_\mu\Psi_i)}{\partial\Psi_i}=\omega^i{}_\mu\qquad \text{and}\qquad\frac{\partial(\na_\nu\Psi_i)}{\partial(\partial_\mu\Psi_i)}=\delta^\mu{}_\nu.
\end{equation}
By writting $\partial\Psi_i$ as a function of $(\Psi_i, \na\Psi_i)$ by using $\partial_\mu\Psi_i=\na_\mu\Psi_i+\omega^i{}_\mu\Psi_i$, we can write an arbitrary variation of $\partial_\mu\Psi_i$ as $\delta(\pa{\mu}\tilde\Psi_i)=\delta(\na_\mu\Psi_i)-\omega^i{}_\mu\delta\Psi_i$, which using \eqref{deltatildeS3} leads to
\begin{align}\label{deltaSprimary}
\delta \cS_\textrm{m}=\displaystyle \int_\mathcal{V}\dif V_g \lrsq{\frac{\pa{}\cL}{\pa{}\Psi_i}\delta\Psi_i+\frac{\pa{}\cL}{\pa{}(\na_\mu\Psi_i)}\delta(\na_\mu\Psi_i)}=0,
\end{align}
which by means of \eqref{Divgen} can be recast into
\begin{align}\label{deltaScomplete}
\delta \cS_\textrm{m}&=\displaystyle\int_\mathcal{V} \dif V_g \lrsq{\frac{\pa{}\cL}{\pa{}\Psi_i}-\na_\mu\lr{\frac{\pa{}\cL}{\pa{}(\na_\mu\Psi_i)}}-\lr{\frac{1}{2}{Q_{\mu\al}}^\al+T^\al{}_{\mu\al}}\lr{\frac{\pa{}\cL}{\pa{}(\na_\mu\Psi_i)}}}\delta\Psi_i\\
&+\displaystyle\int_\mathcal{V} \dif V_g \,\div_g\lr{\frac{\pa{}\cL}{\pa{}(\na\Psi_i)}\delta\Psi_i}=0.
\end{align}
Since the last term is a boundary term by the generalised Gauss' law \eqref{Gaussgen}, it vanishes for variations $\delta \Psi_i$ vanishing on $\partial\mathcal{V}$. Given that physical solutions are those that extremise $\cS_\textrm{m}$ for fixed boundary values (namely for variations $\delta\Psi_i$ vanishing on $\partial\mathcal{V}$), the corresponding field equations are
 \begin{equation}\label{eom}
 \frac{\pa{}\cL}{\pa{}\Psi_i}-\na_\mu\lr{\frac{\pa{}\cL}{\pa{}(\na_\mu\Psi_i)}}-\lr{\frac{1}{2}{Q_{\mu\al}}^\al+T^\al{}_{\mu\al}}\lr{\frac{\pa{}\cL}{\pa{}(\na_\mu\Psi_i)}}=0.
\end{equation}
Note that while in the Riemannian limit we recover the usual covariant Euler-Lagrange equations, in the general case there are, apparently, explicit couplings between the nonmetricity and torsion tensors and the matter fields. However, these apparent couplings are indeed compensated by taking into account that the covariant derivative of the second term in \eqref{eom} is not the one associated to the canonical connection of $g$. To show this, we can use the decomposition of a general connection \eqref{eq:ConnectionDecomposition} and split the covariant derivative in front of the second term of \eqref{eom}. This allows us to re-write the above field equations \eqref{eom} as
\begin{equation}\label{eom2}
 \frac{\pa{}\cL}{\pa{}\Psi_i}-\na^g_\mu\lr{\frac{\pa{}\cL}{\pa{}(\na_\mu\Psi_i)}}+\lr{\frac{\pa{}\cL}{\pa{}(\na_\mu\Psi_i)}}(\omega^{i}_\mathrm{PR})_\mu=0,
\end{equation}
where $(\omega^{i}_\mathrm{PR})_\mu$ is the post-Riemannian part of the affine connection 1-form lifted to the vector bundle where $\Psi_i$ is defined, and where we have used that in order for $\cL$ to be covariant, $\partial\cL/\partial(\na_\mu\Psi_i)$ must be a $\tb$-valued section of the dual bundle\footnote{In general, it can be seen that for any section of a vector bundle $\phi\in\Gamma(V)$, if $O_{UV}\in V$ is an overlap with two local trivialisations, by computing $\frac{\partial \cl}{\partial\phi_U}=\frac{\partial \cl}{\partial\phi_V}\frac{\partial \phi_V}{\partial\phi_U}$, so that we see how $\frac{\partial \cl}{\partial\phi}$ transforms under change of trivialisation as an element of $V^\ast$. }, which is accounted by the sign in the $\omega^{i}_\mathrm{PR}$ term (see section \ref{sec:ConnectionsOnBundles}). 

As a by-product of the derivation of \eqref{eom}, we can explicitly check whether Noether currents  associated to matter fields described by Lagrangians like \eqref{matteraction} will be sensitive to nonmetricity and/or torsion corrections. For completeness, let me start by recalling the geometrical meaning of a conserved current. In an orientable manifold $\cm$ a vector field $J\in\Ga(\cO)$ is a conserved current over a region $\cO\subset\mathcal{M}$ with respect to the volume form $\dv$ if it satisfies 
\begin{equation}\label{defconscurrent}
\div_\dv(J)=0.
\end{equation}
This definition is only sensitive to the volume element, and not the metric or affine structures, which implies that the sentence {\it conserved with respect to $\nabla$} does not make sense in general\footnote{This sentence only makes sense when $\nabla$ is $\nabla^g$ and only if {\it conserved with respect to $\na^g$} is understood as conserved with respect to $\dv_g$.}. When the volume form is the canonical volume element given by the metric, the divergenceless condition has an intuitive geometrical meaning when interpreted through Gauss' law \eqref{Gaussgen}. From that perspective, having vanishing divergence over $\cO$ implies that there is no net flux of $J$ through $\partial\mathcal{V}$. If the metric is Lorentzian, by resorting to a Cauchy foliation, there is a covariant quantity $Q_{J}$ associated to any vector field that, for a conserved current, is invariant under change of spatial hypersurface, \ie under time translations. This is why a vector satisfying $\div_\dv(J)=0$ over $\cm$ is called a \textit{conserved current} and $Q_{J}$ its associated \textit{conserved charge}. 

\begin{figure}\label{fig:ConservedCharge}
\center\includegraphics[scale=0.2]{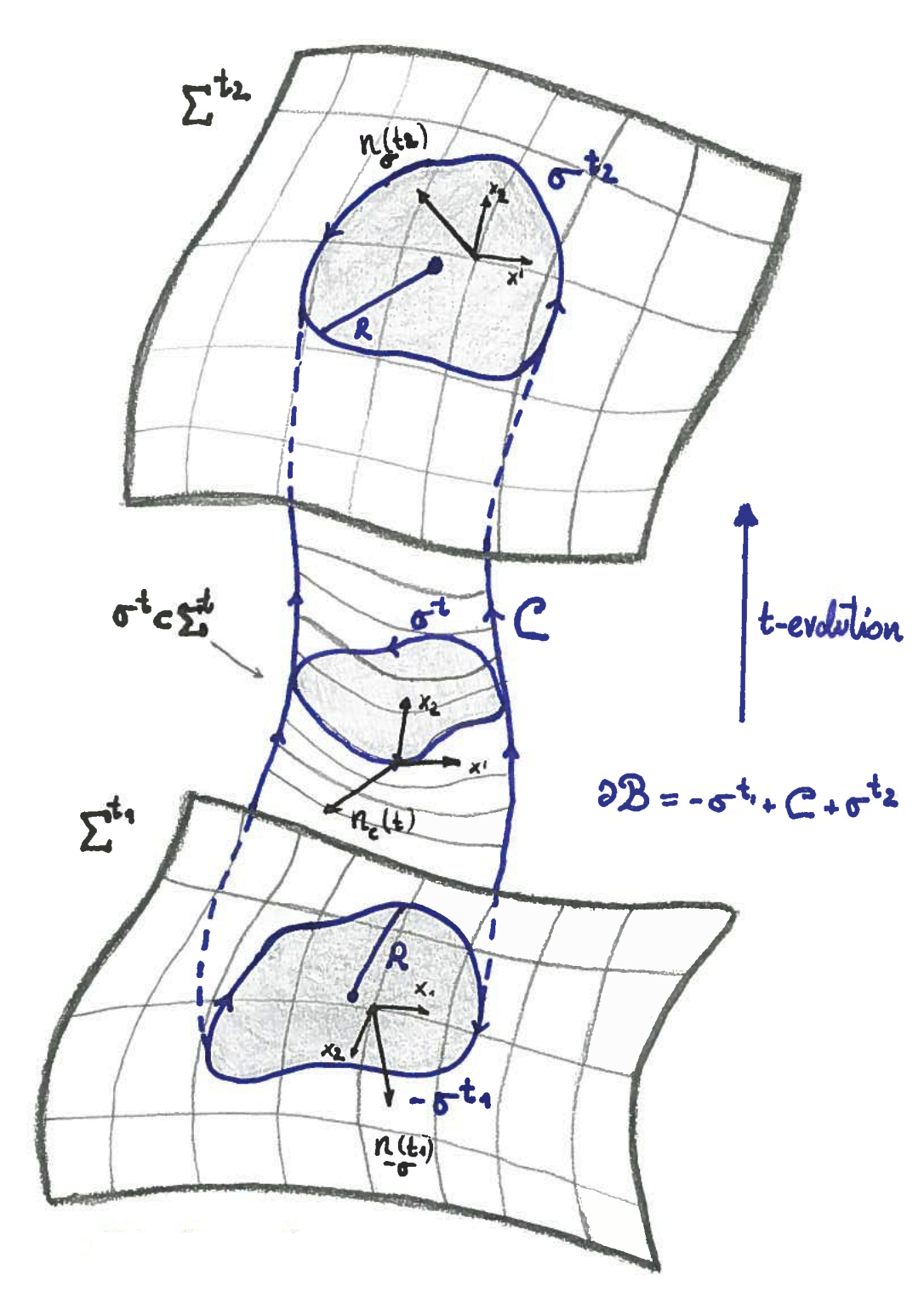}
\caption{Illustration of the different parts of $\cB$ and $\partial\cB$. The black oriented vector basis define the orientation of $\partial\cB$.}
\end{figure}

Charge conservation can also be understood in geometrical terms as follows. Consider coordinates $(x^0=t,x^i)$ adapted to a Cauchy foliation, namely, the Cauchy spatial hypersurfaces of the foliation are given by the one-parameter family of $(n-1)$-dimensional submanifolds ${\Sigma^t}\subset\cm$ normal to $\pa{t}$. Consider also an $(n-1)$-dimensional closed ball $\sigma^t$ defined on every $\Sigma^t$ by $(x^i x_i)^{1/2}\leq R$, with $R$ an arbitrary constant. Define the closed $n$-dimensional volume $\mathcal{B}(t_1,t_2)\subset\cm$ as the volume enclosed by $\sigma^{t_1}$, $\sigma^{t_2}$ and $\cC$; where $\cC$ is the union of the boundaries of each $\sigma^t$ for $t\in(t_1, t_2)$ (see figure  \ref{fig:ConservedCharge} for clarification). Any vector field $J$ defines a charge $ Q^t_J$ on each $\Sigma^t$ given by
\begin{equation}\label{ChargeDef}
Q^t_J=\underset{R\rightarrow\infty}{lim}\int_{\sigma^t}J^t \dif V_{\tilde{g}} \ ,
\end{equation}
where $J^t=g(J,\pa{t})$, $\pa{t}$ is the unit normal to $\sigma^t$ and $\dif V_{\tilde{g}}$  is the volume form induced on $\Sigma^t$ by $\dif V_g$. Using Gauss' law \eqref{Gaussgen}, decomposing $\partial\cB$ as\footnote{The sign infront of $\sigma^{t_1}$ is required for $\partial\mathcal{B}(t_1,t_2)$ to have the standard induced orientation from $\mathcal{V}(t_1,t_2)$.} $\partial\mathcal{B}(t_1,t_2)=(-\sigma^{t_1}+\mathcal{C}+\sigma^{t_2})$, and for configurations of $\Psi$ such that $J^t$ vanishes quickly enough at spatial infinity\footnote{The precise requirement is that $\Psi$ vanish quiclky enough with increasing $R$ so that the integal over $\cC$ vanishes when $R\rightarrow\infty$.} we find
 
\begin{equation}\label{ChargeDiv}
\int_{\mathcal{B}}\div_g(J)\dif V_g=Q^{t_2}_J-Q^{t_1}_J.
\end{equation}

This is valid for any value of $t_1$ and $t_2$. Therefore a charge defined by a conserved vector current remains constant under time-evolution, \ie it is conserved. The arguments within this section are independent of the choice of connection or topology of the smooth manifold. This puts forward the relevance of the condition $\div_g(J)=0$ instead of any condition involving any covariant derivative\footnote{Sometimes one reads that for $\nabla_\mu J^\mu=0$, $J$ is conserved with respect to $\nabla$}, which points out that the expression \textit{conserved with respect to a covariant derivative} is not accurate. Indeed, these arguments depend only on the metric (and its canonical volume form). Taking infinitesimally small $\delta t=t_2-t_1$, we see that
\beq
\mathcal{L}_t{Q}_{J}|_{t=t_1}\delta t=\lr{{Q}_{J}^{t_2}-{Q}_{J}^{t_1}}=\int_{\mathcal{B}}\div_g({J})\dif V_g.
\eeq
Hence, it is apparent that the condition \eqref{defconscurrent} implies that the change in the amount of charge $Q_J$ enclosed in the $(n-1)$-dimensional surface $\sigma^{t_1}$ is given exactly by the flux of the current ${J}$ through the boundary of the $n$-dimensional volume $\mathcal{B}$, \ie  the amount of charge that exits $\sigma^t_1$ in the time interval $t_2-t_1$. Therefore, the total charge associated to a conserved current cannot be created or destroyed.

Having understood the geometric meaning of conserved currents and charges, let me turn to Noether currents and whether they are sensitive to the affine structure in any sense. In general, Noether currents have proven to be very useful tools for analysing and extracting physical information out of field theories, and they lay at the heart of the definition of physical charges. Noether currents are defined from the properties of an action when acted upon by a continuous group of transformations of the matter fields, and are conserved for theories symmetric under the corresponding transformation. Given an action like \eqref{matteraction} and an infinitesimal transformation of the matter fields $\Psi\mapsto\Psi+\delta\Psi_i$ which  leaves the action invariant, we can work out the functional form for the Noether current associated to this symmetry by the following argument. The first term in \eqref{deltaScomplete} vanishes for solutions to the field equations \eqref{eom}. Thus, over physical solutions, a Lagrangian that is invariant under arbitrary infinitesimal transformations $\delta\Psi_i$ must satisfy
 \beq
 \div_g\lr{\frac{\pa{}\cL}{\pa{}(\na\Psi_i)}\delta \Psi_i}=0.
 \eeq
Therefore, as stated by the Noether theorem \cite{NoetherTheorem}, a continuous symmetry of the matter action implies the existence of an associated conserved current ${J}\in\Gamma({\tb})$ defined as
\begin{equation}\label{Noethercurrent}
J=\frac{\pa{}\cL}{\pa{}(\na\Psi_i)}\delta \Psi_i,
\end{equation}
with a corresponding conserved charge given by \eqref{ChargeDef}. A current defined from a Lagrangian as in \eqref{Noethercurrent} is called Noether current, and the corresponding charge is called  Noether charge. This definitions can be extended to infinitesimal transformations that leave the Lagrangian quasiinvariant, namely which change it only by a divergence $\delta L=\div_g (A)$, thus leaving the field equations invariant. In that case, the conserved Noether current is $J+A$. Physical charges are the Noether charges associated to some continuous global symmetry of the matter action. For instance, the electric and colour charges are associated to the global $U(1)_{EM}$ and $SU(3)_C$ symmetries of the Standard Model action (in spontaneously broken phase). The above equation \eqref{Noethercurrent} shows that Noether currents (and charges) might depend on the covariant derivative of the matter fields unless it enters linearly in the action (or does not enter). Thus, for matter Lagrangians of the form \eqref{matteraction}, nonmetricity and torsion might play a role in the definition of Noether currents. Nevertheless, we will see below that they are oblivious for minimally coupled fields according to the prescription given in section \ref{sec:MinimalCouplingPrescription}. 

\section{Analising the minimal coupling prescription}\label{sec:matterfields}

I will now proceed to show the different couplings to the geometry that arise for scalar, Dirac and 1-form fields, comparing the minimal coupling prescription defined in \ref{sec:MinimalCouplingPrescription} with the usual $(\eta,\partial)\mapsto(g,\nabla)$ prescription. We will also compare the results of applying both prescriptions on the Lagrangian or directly to the field equations for each of the fields. Through this section, we will use the name Usual Minimal Prescription (UMP) to denote the $(\eta,\partial)\mapsto(g,\nabla)$ prescription, and we will say Minimal Coupling Prescription (MCP) to denote the prescription defined in section \ref{sec:MinimalCouplingPrescription}. When an L or an F is added at the end of the acronyms, \eg MCPL or UCPF, it explicitly denotes that the particular prescription is being applied at the level of the Lagrangian (L) or at the level of the field equations (F). As a useful reminder for the rest of the section, let me state again the two guiding principles for the MCP, namely covariance and coupling to the geometry {\it as little as possible}. Effectively, this will be carried out by applying the following steps: 1) starting from a covariant (\ie observer and gauge independent) Minkowskian action, write an action using the same operators when generalised to an arbitrary post-Riemannian spacetime. 2) In case of being able to write the Minkowskian covariant action in equivalent ways using different covariant operators, choose the one that is {\it more minimal} in the sense that, when generalised to metric-affine spacetimes, couples {\it as little as possible} to the nonmetricity and torsion. If this choice is not well defined, then the prescription defined in section \ref{sec:MinimalCoupling} is not well defined for that case. We will see that, for scalar, Dirac and 1-form fields; this prescription gives a unique metric-affine minimally coupled theory.

\subsection{Minimally coupled scalar field}\label{sec:MinCoupScalar}

As argued in section \ref{sec:MinimalCouplingPrescription}, the action of a complex scalar field can be written in Minkowski space-time with any of the operators $\dif$ and $\nabla$, since they are all the same when acting on scalar fields (and equal to $\partial$ when written in a partiular coodinate frame). However, the covariant wave operator $\Box_\eta$ of the Minkowskian Klein-Gordon equation is defined, in general, only for $p$-forms, and it will prove useful to see $\phi$ as a 0-form section of the trivial bundle $\cm\times\bbc$. The Lagrangian for a complex scalar field in Minkowski spacetime written in an explicitly covariant form then reads
\begin{equation}
\cL^{(0)}_\mathrm{M}=\eta^{\mu\nu}(\dif\phi^\dagger)_\mu(\dif\phi)_\nu-m^2\phi^\dagger\phi.
\end{equation}
where $\dif\phi$ can also be covariantly written as $\nabla^\eta\phi$. This Lagrangian leads to the well known Klein-Gordon field equation, which in covariant form reads
\begin{align}\label{eomscalarcomplex}
\begin{split}
&\Box_{\eta}\phi+m^2\phi=0,\\
&\Box_{\eta}\phi^\dagger+m^2\phi^\dagger=0,\\
\end{split}
\end{align}
where recall that $\Box_\eta=\dif\delta_\eta+\delta_\eta\dif=\eta^{\mu\nu}\nabla^\eta{}_\mu\nabla^\eta{}_\nu$. Thus we see that the explicitly covariant form of the Minkowskian scalar Lagrangian and field equations can be equivalently written in terms of the exterior differential structure or the canonical affine structure, namely using $\dif$ and $\delta_\eta$ or using $\nabla^\eta$. Thus, we must check which of both forms is more minimal in its coupling to nonmetricity and torsion. Note that, if generalising the Minkowskian Lagrangian from the covariant derivative form, we end up with the same theory as if the UCP $(\eta,\partial)\mapsto(g,\nabla)$ is applied. Let us fist analyse the generalisation in terms of the exterior differential. If the MCPL is applied as explained in \ref{sec:MinimalCouplingPrescription}, in this case we have to make the substitution $(\eta,\dif)\mapsto(g,\dif)$, which leads to the Lagrangian
\begin{equation}\label{scalaraction}
\cL^{(0)}=g^{\mu\nu}(\dif\phi^\dagger)_\mu(\dif\phi)_\nu-m^2\phi^\dagger\phi.
\end{equation}
Computing now the field equations associated to this Lagrangian we find\footnote{A quick way of doing it is, for instance, to rewrite the above Lagrangian in terms of $\nabla\phi$, use \eqref{eom} or \eqref{eom2}, and the re-write again the results in terms of exterior differential operators.}
\begin{align}\label{eomscalar}
\begin{split}
&(\Box_g+m^2)\phi=0,\\
&(\Box_g+m^2)\phi^\dagger=0.
\end{split}
\end{align}
It is then straightforward to check that the MCPF and MCPL lead to the same metric-affine Lagrangian when applied to the Minkowskian Lagrangian of a scalar field as written in terms of exterior differential operators. Indeed, in this case, the MCPF is realised through the substitution $(\eta,\dif,\delta_\eta)\mapsto(g,\dif,\delta_g)$, which effectively consists on $(\eta,\Box_\eta)\mapsto(g,\Box_g)$. Thus the resulting theory that arises from this choice\footnote{Namely, the choice of using the MCP for the scalar Lagrangian and field equations written in terms of the exterior differential structure instead of the affine structure.} describes a scalar field in a metric-affine spacetime that does not couple to the connection at all. If we now do the same with the version of the Minkowskian scalar Lagrangian written in terms of $\nabla^\eta$, the MCP is realised through the substitution $(\eta,\nabla^\eta)\mapsto(\eta,\nabla)$. This choice would render the MCP and UCP as equivalent prescriptions for scalar fields. If applied to the Lagrangian, these prescriptions are also equivalent to the one where the exterior differential operator is used, so that the corresponding field equations are \eqref{eomscalar} and no couplings to the torsion or nonmetricity arise. In these cases, the new Lagrangian leads to a Noether current associated to $U(1)$ transformations (phase shifts), which written as a 1-form through the metric isomorphism (see section \ref{sec:MetricStructure}) reads
\beq
I_g J_{(0)}= i\lr{\phi^\dagger\dif\phi-(\dif\phi^\dagger)\phi}.
\eeq
subtractiong the corresponding field equations \eqref{eomscalar} we arrive at $\delta_g (I_g J_{(0)})=0$, which is equivalent to
\beq
\div_g(J_{(0)})=0.
\eeq
Thus these (equivalent) prescriptions maintain the global $U(1)$ invariance of the complex scalar field in Minkowski and the conservation of the associated current and charge. This could be easily generalised for scalar fields that are 0-form sections of a nontrivial vector $G$-bundle, which would lead to a global $G$-invariance and an associated Noether current(s) and charge(s).

If we now choose to view the usual partial derivatives appearing in the Minkowskian Klein-Gordon field equations as covariant derivatives $\na^\eta$ and apply the MCPF, or also if we applied the UCPF, the resulting scalar field theory is described by the field equations
\begin{align}\label{eomscalarUCPF}
\begin{split}
&(g^{\mu\nu}\na_\mu\na_\nu+m^2)\phi=0,\\
&(g^{\mu\nu}\na_\mu\na_\nu+m^2)\phi^\dagger=0.\\
\end{split}
\end{align}
By using the decomposition of the affine connection \eqref{eq:ConnectionDecomposition} we can show that
\begin{equation}\label{DecompBox}
 g^{\mu\nu}\na_\mu\na_\nu\phi=\Box_g\phi+\dif \phi[\Sigma],
 \end{equation}
where $\Sigma^\mu$ is the non-Riemannian vector current
\begin{equation}\label{NRcurrent}
\Sigma^\mu={{Q_\al}^\al}^\mu-\frac{1}{2}{Q^\mu{}_{\al}}^\al-{T^{\al\mu}{}_{\al}}.
\end{equation}
Hence the field equations corresponding to this choice can be written as
\beq
\begin{split}
&(\Box_g+\mathrm{i}_{\Sigma}\dif+m^2)\phi=0,\\
&(\Box_g+\mathrm{i}_{\Sigma}\dif+m^2)\phi^\dagger=0,
\end{split}
\eeq
where $\mathrm{i}_{\Sigma}\dif\phi$ is the interior product of $\dif\phi$ with the vector current $\Sigma$ defined in \eqref{eq:InteriorProduct}. These equations, therefore, contain a nontrivial coupling between the scalar field and the torsion and nonmetricity tensors. By subtracting both field equations, we find that the codiferential \eqref{eq:Codiferential} of the 1-form version of the scalar current is given by $\delta_g(I_g J_{(0)})=-I_g J[\Sigma]$, or equivalently in $D$ spacetime dimensions and for a Lorentzian metric
\beq
\div_g(J_{(0)})=(-1)^{D+1}I_g J_{(0)}[\Sigma].
\eeq
Thus, this prescription not only features extra couplings to the geometry, but it also has the undesired feature of spoiling the global invariance of scalar fields when passing from a Minkowskian to a general metric-affine spacetime. As explained above, these results generalise in a straightforward manner for any  scalar field that is a 0-form section of some vector $G$-bundle as, {\it e.g.}, the Higgs field, which in the Standard Model is a 0-form section of a vector $SU(2)\times U(1)$-bundle.

Given that the other choices (equivalent in this case), which also respect covariance, do not have any coupling to the nonmetricity and torsion, the rules defined in section \ref{sec:MinimalCouplingPrescription} for defining a minimal coupling prescription tell us that the theory without these couplings must be the minimally coupled one. Now, this theory is arrived at by UCPL and MCPL if the derivatives of the Minkowskian Lagrangian are written as $\na^\eta$ but not by UCPF or MCPF in this case. Namely, these two choices lead to different theories if applied at the level of the field equations or of the Lagrangian. On the other hand, if the derivatives in the Minkowskian Lagrangian and field equations are understood in terms of exterior calculus, the MCP leads to the same theory if applied to the field equations or to the Lagrangian. Though this is a purely aesthetic matter, it is more satisfying to see the scalar field as a 0-form acted upon by exterior differential operators and have the same results no matter what version of the MCP is applied. Furthermore, unlike in this case, we will see below that for a 1-form field the only choice leading to a minimal coupling prescription in the sense of \ref{sec:MinimalCouplingPrescription} is to view the differential operators of the corresponding field equations and Lagrangian as exterior differential operators. 

\subsection{Minimally coupled Dirac field}

A Dirac field is a 0-form section of the spinor bundle (see section \ref{sec:MetricStructure}). Hence, though $\dif$ is observer independent when applied to a Dirac field $\psi$ or its dual\footnote{Note that in order for $\bpsi\psi$ to be covariant, $\bpsi\in\cS\cm^\ast$ if $\psi\in\cS\cm$.} $\bpsi$, it is not covariant under choice of local trivialisations of the spinor bundle. Hence, contrary to the scalar field case, the only covariant derivative operator that acts on Minkowskian Dirac spinors is the covariant exterior differential (or the covariant derivative) $\na^\eta$. Thus, we conclude that the Minkowskian Lagrangian and field equations for a Dirac spinor field is written, in an explicitly covariant way, as 
\begin{align}
\cl^{(1/2)}_\mathrm{M}=\frac{i}{2}\lrsq{\bpsi\ga^\mu(\na^\eta_\mu\psi)-(\na^\eta_\mu\bpsi)\ga^\mu\psi}-m\bpsi \psi, 
\end{align}
which leads to the covariant version of the Dirac equations
\begin{align}\label{eomspinor}
\begin{split}
&\lrsq{i\ga^\mu\na^\eta_\mu-m}\psi=0 ,\\
&\bpsi\lrsq{i\overleftarrow{\na}^\eta{}_\mu\ga^\mu+m}=0.
\end{split}
\end{align}
Thus, apparently, in this case the MCP leads to a straightforward prescription $(\eta,\nabla^\eta)\mapsto(g,\na)$ which is equivalent to the UCP both when applied to the Lagrangian or to the field equations. However, we will see that, unlike for scalar fields, the MCPL and MCPF do not lead to the same results. Starting with the MCPL, the prescription $(\eta,\nabla^\eta)\mapsto(g,\na)$ leads to the following generalisation of the Dirac Lagrangian to metric-affine spacetimes
\begin{align}\label{SpinorActionH}
\cl^{(1/2)}=\frac{i}{2}\lrsq{\bpsi\ga^\mu(\na_\mu\psi)-(\na_\mu\bpsi)\ga^\mu\psi}-m\bpsi \psi, 
\end{align}
which, by means of \eqref{eom2} leads to the field equations
\begin{align}\label{preeomspinor}
\begin{split}
&\lrsq{i\ga^\mu\na_\mu+\frac{i}{2}\lr{(\na_\mu\ga^\mu)-\ga^\mu\sigma_\mu+\frac{1}{2}{Q_{\al\mu}}^\al\ga^\mu}-m}\psi=0 ,\\
&\bpsi\lrsq{i\overleftarrow{\na}_\mu\ga^\mu+\frac{i}{2}\lr{\lr{\na_\mu\ga^\mu}-\ga^\mu\sigma_\mu+\frac{1}{2}{Q_{\al\mu}}^\al\ga^\mu}+m}=0,
\end{split}
\end{align}
where we have defined the post-Riemannian 1-form current
\begin{equation}\label{NRcurrent2}
\sigma_\mu={{Q_{[\al\mu]}}^\al+{S_{\al\mu}}}^\al,
\end{equation}
which satisfies $\sigma_\mu=\Sigma_\mu+\frac{1}{2}Q_{\al\mu}{}^\al$. We have now to make sense of the covariant derivative of the Dirac matrices. This object is a $\tb$-valued section of the tensor product of the tangent bundle to $\cS\cm$ and its dual bundle, which can also be seen as a 0-form section of the tensor product bundle $\tb\otimes \ct^{(1,1)}(\cS\cm)$, which is also a vector bundle, and where $\ct^{(1,1)}(\cS\cm)$ is the bundle of $(1,1)$-tensors over $\cS\cm$. Thus, using \eqref{eq:CovariatDerivativeComponentsGeneralTensor} to obtain the connection on $\ct^{(1,1)}(\cS\cm)$ associated to the $\cS\cm$ connection 1-form $\omega^s$ through the associated bundle construction, we find that 
\beq
\nabla_\mu\gamma^\alpha=\partial_\mu\gamma^\alpha+\Ga^\alpha{}_{\mu\nu}\gamma^\nu+[\omega_\mu^{s},\gamma^\alpha]
\eeq
where the commutator comes from the $\ct^{(1,1)}(\cS\cm)$ connection. Resorting to the canonical lift, which allows to obtain a canonical connection in the spinor bundle from an affine connection given by \eqref{eq:SpinConnectionGeneral}, and using the decomposition \eqref{eq:SpinConnectionDecomposition}, we can show that the following relations hold
\begin{equation}\label{covdergamma}
\begin{split}
&\na_\mu\ga^\al=-\frac{1}{2}{Q_{\mu\nu}}^\al \ga^\nu,\\
&\ga^\mu\na_\mu\psi=\lrsq{\ga^\mu\na_\mu^g-\frac{i}{8}\epsilon^{\alpha\beta\mu\nu}T_{\alpha\beta\mu}\ga_\nu\ga_5+\frac{1}{2}\ga^\mu\sigma_\mu}\psi,\\
&(\na_\mu\bpsi)\ga^\mu=\bpsi\lrsq{\overleftarrow{\na}_\mu^g\ga^\mu+\frac{i}{8}\epsilon^{\alpha\beta\mu\nu}T_{\alpha\beta\mu}\ga_\nu\ga_5+\frac{1}{2}\ga^\mu\sigma_\mu},
\end{split}
\end{equation}
which, if applied to \eqref{preeomspinor}, lead to the following field equations for a Dirac field and its dual in a general post-Riemannian spacetime
\begin{align}\label{eomspinordecomposed}
\begin{split}
&\lrsq{i\ga^\mu\na^{g}_\mu+\frac{1}{8}\epsilon^{\alpha\beta\mu\nu}T_{\alpha\beta\mu}\ga_\nu\ga_5-m}\psi=0 ,\\
&\bpsi\lrsq{i\overleftarrow{\na}^{g}_\mu\ga^\mu-\frac{1}{8}\epsilon^{\alpha\beta\mu\nu}T_{\alpha\beta\mu}\ga_\nu\ga_5+m}=0.
\end{split}
\end{align}
Note that, unlike the case for spinor fields, the MCP applied to the Minkowskian Dirac Lagrangian leads to a minimal coupling between the Dirac fields and some post-Riemannian features of the geometry, namely the axial part of the torsion tensor. It is remarkable, however, that despite the spin connection obtained trough the canonical lift applied to a general affine connection is sensitive both to nonmetricity and torsion \eqref{eq:SpinConnectionGeneral}, only the axial part of the torsion couples to spinor fields described by the MCPed Lagrangian \eqref{SpinorActionH}. Indeed, this can be understood by decomposing the kinetic term in the Lagrangian into its Riemannian and post-Riemannian pieces, which by using \eqref{covdergamma} yields
\begin{align}\label{SpinorActionHdecomposed}
\cl^{(1/2)}=\frac{i}{2}\lrsq{\bpsi\ga^\mu(\na^g_\mu\psi)-(\na^g_\mu\bpsi)\ga^\mu\psi}+\frac{1}{8}\epsilon^{\alpha\beta\mu\nu}T_{\alpha\beta\mu}\bpsi\ga_\nu\ga_5\psi-m\bpsi \psi, 
\end{align}
thus getting rid of the rest of the post-Riemannian terms in \eqref{eq:SpinConnectionGeneral} and accounting for the interaction with the axial piece of the torsion. The above MCPed Lagrangian for Dirac fields \eqref{SpinorActionH} has a global $U(1)$ symmetry, with a corresponding Noether current given by
\beq
J_{(1/2)}=i\bpsi\gamma\psi.
\eeq
By adding both equations in \eqref{eomspinordecomposed} after multiplying each by the corresponding dual field, and using $\div_g(J)=\na^g_\mu J^\mu$ for any vector $J\in\Gamma(\tb)$, we end up with conservation of the Noether current 
\beq
\div_g(J_{(1/2)})\big|_\textrm{MCP(L)}=0,
\eeq
as it could not have been otherwise starting from a $U(1)$-symmetric Lagrangian. This can be extended in a straightforward manner if the spinors are also sections of a more general vector $G$-bundle for the corresponding Noether current due to global $G$-symmetry. This happens, {\it e.g.}, for the Standard Model quarks and leptons, which are $\cS\cm$-valued 0-form sections of a vector $SU(3)\times SU(2)\times U(1)$-bundle and $SU(2)\times U(1)$-bundle respectively.

We now want to compare the above results to the ones obtained if the MCP (equivalent to UCP in this case)  is applied directly to the Minkowskian Dirac equations \eqref{eomspinor}, which leads to
\begin{align}\label{MCPFspinoreom}
\begin{split}
&\lrsq{i\ga^\mu\na_\mu-m}\psi=0\; ,\\
&\bpsi\lrsq{i\overleftarrow{\na}_\mu\ga^\mu+m}=0.
\end{split}
\end{align}
If the covariant derivative terms are expanded using the connection decomposition of the spin connection  \eqref{eq:SpinConnectionDecomposition} derived from the canonical lift \eqref{eq:SpinConnectionGeneral}, we find
\begin{align}\label{eomspinordecomposedMCPF}
\begin{split}
&\lrsq{i\ga^\mu\na^{g}_\mu+\frac{1}{8}\epsilon^{\alpha\beta\mu\nu}T_{\alpha\beta\mu}\ga_\nu\ga_5+\frac{i}{2}\sigma_\mu\ga^\mu-m}\psi=0 ,\\
&\bpsi\lrsq{i\overleftarrow{\na}^{g}_\mu\ga^\mu-\frac{1}{8}\epsilon^{\alpha\beta\mu\nu}T_{\alpha\beta\mu}\ga_\nu\ga_5+\frac{i}{2}\sigma_\mu\ga^\mu+m}=0,
\end{split}
\end{align}
which shows how, if applied to the Minkowskian Dirac equation instead of the Lagrangian, besides the interaction with the axial part of the torsion tensor that appeared when the MCP(L) was applied, the MCP(F) introduces an extra coupling between the Dirac fields and the nonmetricity and torsion through the post-Riemannian current $\sigma^\mu$ defined in \eqref{NRcurrent2}. Guided by coupling to the geometry {\it as little as possible}, we conclude that the MCP(L) is the minimal coupling prescription for Dirac fields according to the criterion given in section \ref{sec:MinimalCouplingPrescription}. Another undesired feature of the MCP(F) is that, if the equations \eqref{eomspinordecomposedMCPF} are added (each multiplied by the corresponding dual field), we find 
\beq
\div_g( J_{(1/2)})\big|_\textrm{MCP(F)}=-\sigma[J_{(1/2)}].
\eeq
so that the Dirac equation looses the global $U(1)$ symmetry after the MCP(F) is implemented. This constituted an undesired result for this prescription beyond the criteria given for defining a minimal coupling prescription in section \ref{sec:MinimalCouplingPrescription}. Thus, it could be used to further discriminate between both prescriptions on physical grounds, as the MCP(F) would generally lead to violation of the Standard Model global symmetries through the interaction of the Standard Model quarks and leptons with nonmetricity and torsion, which offers a solution to discriminate between MCP(L) and MCP(F) on physical grounds, yielding a solution to the question raised in \cite{Formiga:2012ns}. 
 
Before finishing with Dirac spinors, let me elaborate on what would happen if we tried to apply the MCP to the nonhermitian version of the Dirac Lagrangian, and after applying the MCP to the Minkowskian version, in general metric-affine spacetimes reads
\begin{align}\label{SpinorActionNH}
\cl^{(1/2)}_\textrm{NH}=\bpsi\lrsq{\ga^\mu\na_\mu-m}\psi. 
\end{align}
This action is commonly used also in general Riemannian backgrounds (see \eg \cite{Parker:2009uva}) due to the fact that it differs from the Riemannian version of the hermitian action by a boundary term which is irrelevant to the dynamics, namely 
\beq
i\bpsi\ga^\mu\na^g_\mu\psi-\frac{i}{2}\lrsq{\bpsi\ga^\mu(\na^g_\mu\psi)-(\na^g_\mu\bpsi)\ga^\mu\psi}=\div_g\lr{\frac{J_{1/2}}{2}}.
\eeq
However, in post-Riemannian spacetimes, the kinetic terms of \eqref{SpinorActionH} and \eqref{SpinorActionNH} are not related by a divergence and therefore yield different dynamics. Particularly we have that in general
\beq
i\bpsi\ga^\mu\na_\mu\psi-\frac{i}{2}\lrsq{\bpsi\ga^\mu(\na_\mu\psi)-(\na_\mu\bpsi)\ga^\mu\psi}=\div_g\lr{\frac{J_{1/2}}{2}}+\frac{1}{2}\sigma[J_{(1/2)}]
\eeq
is satisfied. It is also possible to see how the dynamics generated by the nonhermitian Lagrangian \eqref{SpinorActionNH} is not consistent with the spin structure in the following sense. From \eqref{SpinorActionNH}, we find the following field equations for spinors and dual spinors
\beq\label{eomspinorNH}
\begin{split}
&\lrsq{i\ga^\mu\na_\mu-m}\psi=0 ,\\
&\bpsi\lrsq{i\overleftarrow{\na}_\mu\ga^\mu-i\sigma_\mu\ga^\mu+m}=0.
\end{split}
\eeq
Now, it is known that dual spinors must be related by $\bpsi=\psi^\dagger\gamma^0$ in order for them to have the transformation properties of the elements of $\cS\cm^\ast$. However, if we take the adjoint equation to the equation for $\psi$ in \eqref{eomspinorNH}, we arrive to
\begin{align}\label{eomadjointinconsistent}
\lr{\psi^\dagger\gamma^0}\lrsq{i\overleftarrow{\na}_\mu{\ga^\mu}+m}=0;
\end{align}
which is inconsistent with the identification $\bar\psi=\psi^\dagger\gamma^0$ if the second equation of \eqref{eomspinorNH} has to be satisfied unless the post-Riemannian terms drop.\footnote{Note that in that case the nonhermitian action is equivalent to the hermitean one, so this problem does not arise.} This proves that the nonhermitian Lagrangian is not only inequivalent to the hermitian one, but unable to describe a covariant theory for Dirac fields in presence of general torsion and/or nonmetricity. As a consistency check, it is possible to show that the adjoint equation to the first equation in \eqref{eomspinordecomposed} yields
\beq
(\psi^\dagger\gamma^0)\lrsq{i\overleftarrow{\na}^{g}_\mu\ga^\mu-\frac{1}{8}\epsilon^{\alpha\beta\mu\nu}T_{\alpha\beta\mu}\ga_\nu\ga_5+m}=0,
\eeq
which together with the second equation of \eqref{eomspinordecomposed} forces the identification $\bpsi=\psi^\dagger\gamma^0$. This also happens for the MCP(F) field equations \eqref{eomspinordecomposedMCPF} and, in this sense, both equations are adequate for describing a covariant theory for Dirac fields in presence of general torsion and/or nonmetricity. However, note that the MCP(F) dynamics features non-minimal couplings to the geometry in the sense of section \ref{sec:MinimalCouplingPrescription}.
 
As a final remark, let us comment on the following subtlety: The meaning of $\na\Psi$ in post-Riemannian spacetimes was discussed in section \ref{sec:GeneralAffineConnection}. There it was shown how the canonical lift of the affine connection (seen as a linear connection on the tangent bundle) to the spinor bundle leads to a particular form of the spinor connection which is sensitive to post-Riemannian corrections. Now, one could in principle, insist in using only the Riemannian piece of the connection, or to state it in more technical language, to lift the canonical connection associated to the metric instead of the affine connection, and it would do a perfect job in maintaining covariance. In this case, the Dirac Lagrangian and field equations would be 
\begin{align}
\cl^{(1/2)}=\frac{i}{2}\lrsq{\bpsi\ga^\mu(\na^g{}_\mu\psi)-(\na^g{}_\mu\bpsi)\ga^\mu\psi}-m\bpsi \psi, 
\end{align}
and
\begin{align}
\begin{split}
&\lrsq{i\ga^\mu\na^{g}_\mu-m}\psi=0 ,\\
&\bpsi\lrsq{i\overleftarrow{\na}^{g}_\mu\ga^\mu+m}=0,
\end{split}
\end{align}
and no coupling to the post-Riemannian features of the geometry would occur. Therefore, this would be preferred by our prescription for minimal coupling over lifting the affine connection, as it couples {\it less} to the post-Riemannian features of spacetime. From the bundle theory point of view, given that Dirac spinors are built using only the metric independently of the affine structure, lifting the horizontal distribution defined by the metric in $\tb$ (\ie the Riemannian connection) instead of the one defined by the affine connection could seem more natural. Nevertheless, from the perspective of building a gauge theory of gravity, to obtain GR one needs to consider gauge invariance with respect to the Poincar\'e group\footnote{$T_n$ is the n-dimensional translation group.}$SO^+_{(3,1)}\rtimes T_{4}$. In that case, the spin bundle inherits the $SO^+_{(3,1)}\rtimes T_{4}$-structure from the tangent bundle and the spin connection must be a $SO^+_{(3,1)}\rtimes T_{4}$-connection in $\cS\cm$ which contains a torsion piece due to the translational part. This torsion piece ends up yielding the same theory for Dirac fields as the one described by the Dirac Lagrangian after applying the MCP, namely it leads to a Lagrangian of the form \eqref{SpinorActionHdecomposed}. In this case, minimal coupling to torsion also acquires the meaning of being minimal in the sense that it arises from requiring gauge invariance.

\subsection{Minimally coupled 1-form field}\label{sec:vecfield}

The kinetic term for a 1-form field $A$,  be it massless or massive, is fixed by Lorentz invariance and absence of ghostly degrees of freedom, and in an explicitly covariant form it reads
\beq\label{1formAction}
\cl^{(1)}_\mathrm{M}=-\frac{1}{2\sqrt{-\eta}}\textrm{Tr}[\star_\eta\dif A\wedge\dif A].
\eeq
This kinetic term happens to be invariant under gauge transformations $A\mapsto A+\dif \xi$, which from the transformation properties of $G$-connections \eqref{eq:TransformationConnection} implies that $A$ can be understood as a $U(1)$-connection 1-form\footnote{This discussion can be generalised in a straightforward manner for the case when $A$ is any $G$-connection 1-form. In that case, the fieldstrength (or curvature 2-form \eqref{eq:Curvature2Form}) is $\dif A+A\wedge A$ and the gauge invariant kinetic term features self-interactions for nonabelian $G$, though this does not affect substantially the discussion of minimal coupling to the geometry.} unless further term in the Lagrangian  (\eg a mass) spoils its gauge invariance. For our discussion, it will not be relevant whether it is massless or massive, since we are only interested in the covariant form of the kinetic term. The corresponding field equations are
\beq\label{1formEqsMink}
\dif\star_\eta\dif A=0
\eeq
which are equivalent to $\div_\eta (\dif A)=0$ and, in the Lorentz gauge, given by $\div_g A=0$, they can also be written as $\Box_\eta A=0$. The MCP is straightforward to implement just by the prescription $\eta\mapsto g$ which implies $\star_\eta\mapsto \star_g$. Starting with the application of the MCP on the above Lagrangian \eqref{1formAction}, we find
\beq
\cl^{(1)}=-\frac{1}{2\sqrt{-g}}\textrm{Tr}[\star_g\dif A\wedge\dif A]
\eeq
with the corresponding field equations
\beq\label{1formEqs}
\dif\star_g\dif A=0,
\eeq
which are equivalent to $\div_g (\dif A)=0$ and, in the Lorentz gauge, given by $\div_g(A)=0$, they can also be written as $\Box_g A=0$. On the other hand, by applying the MCP directly to the field equations \eqref{1formEqsMink} we find exactly the same equations as \eqref{1formEqs}, showing how the MCP for 1-form fields yields exactly the same theory if applied to the Lagrangian or directly to the field equations, contrary to the claims in \cite{Chen:2019zhj}. Let me now compare the theory resulting from applying the MCP to that resulting from applying the UCP. The Minkowskian Lagrangian for a 1-form field \eqref{1formAction} is usually written as
\beq
\cl^{(1)}_\mathrm{M}=-\eta^{\mu\nu}\eta^{\alpha\beta}\partial_{[\mu}A_{\alpha]}\partial_{[\nu}A_{\beta]}.
\eeq
and the corresponding field equations \eqref{1formEqsMink} as
\beq\label{MaxwellEOMDer}
\partial_\mu \partial^\mu A^\alpha-\partial^\alpha\partial_\mu A^\mu=0.
\eeq
If we apply the UCP $(\eta,\partial)\mapsto(g,\nabla)$ to the above Lagrangian we get
\beq
\cl^{(1)}_{UCP}=-g^{\mu\nu}g^{\alpha\beta}\na_{[\mu}A_{\alpha]}\na_{[\nu}A_{\beta]}.
\eeq
This Lagrangian features two couplings to the torsion tensor of the schematic form $T A\partial A$ and $T^2A^2$ which break the gauge of the massless 1-form field Lagrangian. Using \eqref{eom} or \eqref{eom2}, we arrive at the following field equations
\begin{equation}\label{eomvector}
\nabla^g_\mu(\dif A)^{\mu\nu}+\frac{1}{2} T^{\nu}{}_{\al\be}(\dif A)^{\alpha\beta}+\lr{\na^g_\mu T^{\al\mu\nu}+\frac{1}{2}T^{\nu\mu\sigma}T^\al{}_{\mu\sigma}}A_\al=0,
\end{equation}
We see that if thought of as a background on top of which perturbations of $A$ propagate, spacetime torsion provides an effective mass term that breaks the gauge symmetry and unleashes the longitudinal polarisation, thus potentially giving rise to strong coupling issues around torsionless backgrounds. It also yields an interaction with the velocities of $A$ which can also, in general, be a source of instabilities.

The UCP(L) thus leads to extra interactions between the torsion and the 1-form field. We can also apply the UCP as $(\eta,\partial)\mapsto(g,\nabla)$ directly to the field equations \eqref{MaxwellEOMDer}, and we find
\begin{align}
\begin{split}
&\na^{g}_\mu (\dif A)^{\mu\al}+\Gamma_\mathrm{PR}{}^\mu{}_{\mu\be}(\dif A)^{\al\be}+2\Gamma_\mathrm{PR}{}^{\al\mu\be}\na^g_{\mu}A_\be+\\
&\lr{\na^g_\mu\Gamma_\mathrm{PR}{}^{\al\mu}{}_\be-\na^{g\al}\Gamma_\mathrm{PR}{}^{\mu}{}_{\mu\be}+\Gamma_\mathrm{PR}{}^\al{}_{\mu\sigma}\Gamma_\mathrm{PR}{}^{\sigma\mu}{}_\beta+\Gamma_\mathrm{PR}{}^\mu{}_{\mu\sigma}\Gamma_\mathrm{PR}{}^{\alpha\sigma}{}_\beta}A^\beta=0.
\end{split}
\end{align}
where $\Gamma_\mathrm{PR}{}^{\al}{}_{\mu\nu}=L^{\al}{}_{\mu\nu}+K^{\al}{}_{\mu\nu}$ is the post-Riemannian part of the affine connection in the decomposition \eqref{eq:ConnectionDecomposition}. We see that the UCP(F) yields a different result from the UCP(L), introducing extra interactions with both the nonmetricity and torsion tensors which also break the gauge symmetry. Thus, while for the vector field the MCP yields the same results if applied to the Lagrangian or field equations, this is not the case for the UCP. Furthermore, the MCP leads to a covariant theory with no interactions with the post-Riemannian terms, whereas the UCP, in both cases, features extra interactions that break the gauge symmetry of the kinetic term. Thus, sticking to the principles that define minimal coupling to the geometry given in section \ref{sec:MinimalCouplingPrescription}, we find that the MCP is the appropriate prescription.
 
\section{Freefall in metric-affine theories}\label{sec:geodesics}

By taking the appropriate limits, the matter field equations describe how classical particles propagate through spacetime. Thus, once the matter Lagrangian for a free\footnote{Note that in presence of gravity, free means interacting only with gravity. In the context of GR (or other theories with only a metric) this can be seen as interacting with a background Riemannian geometry, but for meric-affine theories, free fields can interact with other features of the geometry.} matter field in a metric-affine spacetime is specified, the paths followed by its associated particles are specified by the corresponding field equations. In a purely Riemannian background, free fields follow metric geodesics, which coincide with autoparallel paths for the Riemannian connection. However, in post-Riemannian spacetimes, this coincidence is lost: curves that extremise length are not autoparallel with respect to the affine connection. Dwelling into the literature on metric-affine theories, one can find works in which it is assumed that free particles could follow the autoparallel curves of the affine connection in order to uncover phenomenological aspects of the corresponding theories. Nevertheless, this assumption is usually not backed by finding a set of matter field equations which describes propagation through autoparallel curves in the appropriate limits. We will finish this chapter by elaborating on why this assumption does not seem to be compatible with a matter theory that is derived from a Lagrangian. To that end, let me start by reviewing the differences between metric and autoparallel curves (or affine geodesics).

The notion of parallely transporting a vector along a curve in an arbitrary manifold is only defined in terms of an affine connection, which defines a parallel transport equation \eqref{eq:ParallelTransport}. A curve $x^\mu(t)$ is said to be an autoparallel of a given affine connection if its tangent vector satisfies the parallel transport equation along itself, namely if
 \beq
 \ddot{x}^\alpha+\Gamma^\alpha{}_{\mu\nu}\dot{x}^\mu\dot{x}^\nu=0.
 \label{eq:autoparallel1}
 \eeq
This equation describes the {\it straightest} paths defined as those whose acceleration along the tangent direction vanishes, while the paths that extremise the spacetime interval are described by the metric geodesic equation
\beq
\ddot{x}^\alpha+{}^g\Gamma^\alpha{}_{\mu\nu}\dot{x}^\mu\dot{x}^\nu=0.
\label{eq:geodesic1}
\eeq
Unlike the autoparallel equation \eqref{eq:autoparallel1}, the metric geodesic equation is oblivious to the general affine structure, and is entirely determined by the metric\footnote{More precisely, it is determined by the canonical affine structure of the metric, or equivalently, by the Levi-Civita piece in the decomposition of the affine connection \eqref{eq:ConnectionDecomposition}.}, as it should because the length of curves only depends on the metric. The difference between both equations is accounted for by the post-Riemannian terms in the decomposition of a general affine connection \eqref{eq:ConnectionDecomposition}, so that the autoparallel equation reads
\beq
\ddot{x}^\alpha+{}^g\Gamma^\alpha{}_{\mu\nu}\dot{x}^\mu\dot{x}^\nu=-\Gamma_\mathrm{PR}{}^\alpha{}_{\mu\nu}\dot{x}^\mu\dot{x}^\nu.
\label{eq:autoparallel2}
\eeq
Only experiments can tell us whether particles follow metric geodesic paths, auto-parallel curves of an independent affine connection, or otherwise. In other words, we can only constrain the $\Gamma_\mathrm{PR}-$sector by resorting to experiments. However, we can argue which one seems more {\it natural}, with all the caveats that this word might induce, from a theoretical perspective. Let me state the conclusion that we will reach right away: metric geodesic trajectories seem better aligned with our current understanding of physics. In the following, I will present the arguments leading to that conclusion.

Firstly, the most natural action for a test particle on a gravitational field (that may include a general connection) is given by its line element. If the trajectory of the particle is $x^{\alpha}=x^{\alpha}(\lambda)$ for some affine parameter $\lambda$, we can expect its action to be
\beq
\cS_g=\int g_{\mu\nu}(x)\dot{x}^\mu\dot{x}^\nu\dd\lambda,
\label{eq:ppaction1}
\eeq
which leads to the metric geodesic equation and not to the affine autoparallel one. One might object that the naturalness and our expectation is crucially biased by our prejudice so some more motivation would seem desirable. That \eqref{eq:ppaction1} is the natural action for the gravitational interaction of the particle can be motivated by the fact that the particle's motion should be described by its velocity $\dot{x}^\alpha$ and, in compliance with the Equivalence Principle, it should reduce to $\eta_{\mu\nu}\dot{x}^\mu \dot{x}^\nu$ in a freely falling frame. Furthermore, once we accept that the particle dof's are described by $\dot{x}^\alpha$, \eqref{eq:ppaction1} can be regarded as the lowest order interaction with the metric tensor from an effective theory perspective. There could be other higher order interactions but they will be suppressed by some appropriate scale. In fact, we do expect higher order corrections of this type. The same reasoning can be applied to determine the coupling to the affine connection. If we stick to the Equivalence Principle for gravity in the geometrical sense\footnote{From the field theoretic perspective, the Equivalence Principle must be satisfied as a consistency condition for the couplings of a massless spin-2 field. Other fields do not need to satisfy it. From the geometrical perspective, gravity is associated to geometry and, if it has to satisfy the Equivalence Principle, the interaction of matter with spacetime geometry, namely metric and connection, must be universal. From a field theoretic view, what we regard as geometry and as matter is arbitrary. Indeed, by virtue of \eqref{eq:ConnectionDecomposition}, any metric-affine theory of gravity can be seen as a metric theory with nominimally coupled fields which do not need to satisfy universal coupling for consistency of the theory. Indeed, from this view, the statement of whether gravity satisfies the equivalence principle amounts to whether we associate any degrees of freedom to the gravitational sector beyond the massless spin-2.}, the connection cannot couple directly to the particle unless it couples universally (though it is not clear how to do it). On the other hand, from the field theoretic perspective this could be too restrictive because the Equivalence Principle is only a required consistency coupling prescription for the massless spin-2 sector of the theory \cite{Weinberg:1995mt,Weinberg:1965nx}, and the connection sector could contain additional propagating degrees of freedom that do not need to comply with the Equivalence Principle, so that there would be no reason to impose universal coupling to the connection. However, note that any of these degrees of freedom associated to the connection should be compatible with current bounds on fifth-force experiments. In this line, if we let the connection couple to the particle, the lowest order interaction is given by
\beq
\cS_\Gamma=\int \Upsilon_\mu\dot{x}^\mu\dd\lambda,
\label{eq:ppaction2}
\eeq
where $\Upsilon_\mu$ is some arbitrary combination of traces of the connection. The correction to the field equations coming from this coupling is of the form
\beq
\frac{\delta\cS_\textrm{$\Gamma$}}{\delta x^\alpha}\supset \big(\partial_\alpha \Upsilon_\mu-\partial_\mu\Upsilon_\alpha\big)\dot{x}^\mu,
\eeq
which contributes with a Lorentz-like force and, certainly, it does not lead to the affine autoparallel equation. Again, we can expect higher order corrections, but they will be suppressed by some suitable scale and it will contain higher powers of the particles velocity. Thus, obtaining the autoparallel equation for the full connection from an appropriate action is substantially more contrived than obtaining the metric geodesic equation, which in turn appears quite naturally from the lowest order interactions. In fact, the autoparallel equation \eqref{eq:autoparallel1} cannot be obtained from a standard variational principle in general. Within the context of teleparallel theories where the curvature vanishes identically, one can design an appropriate variational principle to obtain the corresponding autoparallel equation as suggested in \cite{Fiziev:1995te,Kleinert:1996yi}. One can always resort to suitable constraints and more or less involved couplings leading to the desired equations (whenever this is possible), but this procedure seems artificial to eventually produce the equations in a somewhat ad-hoc manner. An objection to the argument could be that there is no fundamental principle stating that physical equations should follow from an action. After all, not all field equations can be derived from an action principle. Thus, we could regard Eqs. \eqref{eq:autoparallel2} as Lagrange equations of the second kind with some generalised velocity-dependent force precisely given by $\Gamma_\mathrm{PR}{}^\alpha{}_{\mu\nu}\dot{x}^\mu\dot{x}^\nu$ that go beyond the usual friction forces linear in the velocities and derivable from a Rayleigh dissipation function. However, our current understanding of physics at the most fundamental level can be formulated in terms of the path integral whose primary ingredient is the action (besides an appropriate measure). Let us recall that the standard model of the fundamental interactions including gravity is indeed described by an action so it is natural, though not mandatory, to expect that physical equations should follow from an action principle and, in particular, the motion of particles in a gravitational background.

We will finalise our digression by going back to the idea that a particle is just an idealisation (or approximation in some cases) of some more fundamental classical or quantum field. As we have seen above, via the prescription for minimal coupling given in \ref{sec:MinimalCouplingPrescription}, standard bosonic fields like a scalar or spin-1 fields only couple to the metric structure, so it is difficult to justify the appearance of the connection (other than its Levi-Civita part) in their field equations and, consequently, on the propagation of the associated point-like particles unless nonminimal interactions are allowed. Even in this case, the propagation of these fields is usually obtained by applying the eikonal or geometric optics approximation to the corresponding hyperbolic equation describing their dynamics. As we saw in the previous section, the field equations of minimally coupled fields reduce to a wave equation, with the highest order differential operator given by the d'Alembertian associated to the metric (or a suitable combination operators $\dif$ and $\star_g$). In that approximation, the corresponding particle trajectories arises as the curve whose tangent vector is parallel transported with the canonical affine structure of the metric, namely its Levi-Civita connection. On the other hand, if we include nonminimal couplings to the connection bosonic field equations, these will modify the paths of the associated particles in the corresponding approximation, but ensuring that such modifications will lead to the affine autoparallel equation \eqref{eq:autoparallel1} will require a certain amount of artificiality, if possible at all. When considering fermions that do couple minimally to the connection, the conclusion is similar. In that case the eikonal approximation will exhibit additional torsional forces, but they will not mimic the effect of the affine autoparallel propagation \cite{HEHL1971225,Rumpf:1979vh,Audretsch:1981xn,Nomura:1991yx,Cembranos:2018ipn}.

This discussion is relevant, for instance, concerning the physical importance of {\it geodesically complete} spacetimes in metric-affine theories, meaning spacetimes where the solutions of \eqref{eq:autoparallel1} can be extended to the entire manifold. The incompleteness of these curves can be associated to the existence of singularities in the affine structure of spacetime. However, it is then crucial to discern the class of trajectories that carry physical information on the propagation of actual particles. In view of our discussion, it is most natural to consider the solutions of \eqref{eq:geodesic1} as the relevant ones in order to draw physical consequences regarding freely falling observers even if we are in a metric-affine framework. If our matter sector couples to the connection directly, then the geodesic equations \eqref{eq:geodesic1} cease to be valid to describe the dynamics of particles because we will need to include the corresponding {\it affine forces}, but these will not, in general, be encapsulated in an autoparallel equation and a case by case study would be required since, as commented above, universality is no longer a required property of the interactions with an affine connection. However, geodesic completeness should neither be understood as a sufficiency criterion for absence of singularities even in Riemannian spacetimes \cite{Geroch:1968ut}. Indeed, even if free particles do not follow autoparallel curves, we can think of observers which do by propelling them with the appropriate 4-acceleration that compensates the Lorentz-force like term in \eqref{eq:autoparallel1}. In a nonsingular spacetime, not only freely falling observers should be complete, but also accelerated ones, so that autoparallel completeness is also a necessary (though not sufficient) criterion for absence of singularities in metric-affine theories. As a final remark, note that the metric determining the trajectories of different particles could depend on the species around nontrivial backgrounds, as it is the case for projectively invariant RBG where gravitational waves follow the geodesics of an Einstein frame metric $q_{\mu\nu}$, while matter fields travel according to the RBG frame metric $g_{\mu\nu}$  (see chapter \ref{sec:RBGTheory} or \cite{BeltranJimenez:2017doy}).

%=========================================================
\clearemptydoublepage
%
%
% File: chap01.tex
% Author: Victor F. Brena-Medina
% Description: Introduction chapter where the biology goes.
%
\let\textcircled=\pgftextcircled

\part{Physical Aspects of Metric-Affine Theories: Ricci-based gravity and beyond}
\markboth{Part II}{}

{\noindent\Large\textbf{Part II - Outline}}
\vspace{0.5cm}

This part constitutes the core of the thesis, where we develop several arguments that allow to better understand several aspects of the metric-affine framework both at the theoretical and observational level. To that end, we will first resort to a subfamily of metric-affine theories that serve as a proxy that illustrates some features of generic metric-affine theories in a cleaner manner. This family, dubbed as Ricci-Based Gravity theories, is defined by metric-affine actions whose geometrical part depends only on the Ricci tensor. This might give the impression of an unnecessary restraint given the huge freedom permitted by the the general metric-affine formalism. Let us recall at this point that we have a plethora of different geometrical objects that could be used and which should indeed enter the action, unless some additional guiding principle is invoked. However, we will see that some of the features of these theories can be generalised to the full metric-affine family. Studying these features within this simplified class of theories allows for a clearer understanding of their implications both at the theoretical and phenomenological levels, which provides valuable insights on the structure of more general metric-affine theories. Particularly, we will achieve two main results that illuminate the physical properties of generic metric-affine theories. The first one consists on showing the pathological nature of generic metric-affine theories due to the presence of ghostly degrees of freedom (see chapters \ref{sec:UnstableDOF} and \ref{sec:ObservableTraces}). The other result is related to an observable effect that occurs in all metric-affine theories provided that they contain an operator in the action with $R_{(\mu\nu)}$ besides the Einstein-Hilbert term. These effects are better understood from the field theoretic perspective, where they take the form of effective interactions suppressed by a UV scale at which the perturbative expansion breaks down. Nonetheless, from the geometric perspective, they are related to a particular piece of the nonmetricity tensor which is precisely related to these $R_{(\mu\nu)}$ operators that appear outside the Einstein-Hilbert term in the action. To our knowledge, this is the first time that a universal effect that can be linked to nonmetricity in generic metric-affine theories,\footnote{Of course this link only makes sense from the geometric point of view.} without depending on the way it couples to matter, is found.

\chapter{Structure of Ricci-Based Gravity theories}\label{sec:RBGTheory}

Apart from helping in the understanding of the structure of generic metric-affine theories, Ricci-Based theories have received considerable attention in the literature \cite{Afonso:2017bxr,BeltranJimenez:2017doy,Afonso:2018bpv,Afonso:2018hyj,Afonso:2018mxn,Delhom:2019zrb,Latorre:2017uve,Delhom:2019wir,Shao:2020weq,Annala:2021zdt,Gialamas:2021rpr} due to useful properties that make them appealing and more tractable than other more general metric-affine theories, as well as the presence of interesting nonperturbative phenomenology both in astrophysical and cosmological contexts. In particular, as we will see in chapter \ref{sec:UnstableDOF}, these theories are known to be a ghost-free subclass of metric-affine theories with projective symmetry, and are able to heal singularities that arise in cosmological as well as astrophysical scenarios at the classical level. In the process of threshing its properties to understand which are relevant to generic metric-affine theories, we will also make some progress in the understanding of some theoretical and phenomenological aspects of these theories.

\section{Ricci-Based Gravity field equations}\label{sec:StructureRBG}
%%%%%%%%%%%%%%%%%%%%%%%%%%%%%%%%%%%%%%%%%%%%%%%%%%%%%%%

Ricci-Based gravity theories in $D$ spacetime dimensions are described by any diffeomorphism invariant action of the form
\beq\label{eq:GeneralRBGAction}
\cS[g_{\mu\nu},\Ga^\al{}_{\mu\nu},\Psi_i]=\frac12 {\mpl}^{2} \int^{}_{} \dd^Dx \sqrt{-g}\,\lrsq{\cL\big(g^{\mu\nu},R_{\mu\nu}\big)+\frac{2}{\mpl^2}\cL_\textrm{m}\big(g_{\mu\nu},\Gamma^\alpha{}_{\mu\nu},\Psi_i\big)},
\eeq
where $\cL$ is an arbitrary scalar function that depends on the (inverse) metric $g^{\mu\nu}$ and the Ricci tensor $R_{\mu\nu}$ of an arbitrary connection $\Gamma^\alpha{}_{\mu\nu}$ (see section \ref{sec:GeneralAffineConnection}) that is to be determined by the field equations. Here $\cL$ is of mass dimension 2, so that in $D$ spacetime dimensions it includes some (typically heavy) mass scale ${\mg}$ suppressing the couplings of $n$-th order curvature terms as ${\mg}^{D-2(n+1)}$ for $n\geq2$.  The second term is the matter action, also dubbed as $\cS_\textrm{m}[g_{\mu\nu},\Gamma^\alpha{}_{\mu\nu},\Psi_i]$, where $\Psi_i$ stands for the collection of matter fields (and their derivatives), which can in principle couple to the connection in an arbitrary way. In unveiling the general structure of Ricci-Based theories, we will see that those with projective symmetry have particularly nice properties. However, we can make some general statements on form of the field equations and the existence of an Einstein frame in general Ricci-Based theories before talking about projective symmetry.

 Let us start by analysing its field equations, and then showing how they always admit and Einstein frame representation in the sense that they can be described by a Lagrangian given by the `Ricci scalar' of a rank-2 tensor that needs not be symmetric in general. We will then particularize to the cases with and without projective symmetry. In the former, this rank-2 tensor will be symmetric and the gravitational sector in the Einstein frame representation will be seen to be equivalent to GR. In the later case, the rank-2 tensor develops an antisymmetric part due to the explicit breaking of projective symmetry, making the Einstein frame of the theory mirror the Nonsymmetric Gravity Theory introduced by Moffat in \cite{Moffat:1978tr}. In this chapter, we will mainly focus on the structure of Ricci-Based theories with projective symmetry, which we will denote by RBGs\footnote{Sometimes I will explicitly write RBGs with/without projective symmetry to avoid possible confusion.}, which will allow us to understand many results on particular models of this class which have been published mainly over the last 20 years. The field equations of general RBG theories admit several forms that can be reached through algebraic manipulation and which allow to gain some understanding of the general structure of the theories. We will first consider metric and connection field equations separately in full generality, including arbitrary couplings between matter and connection. This will allow us to see how both sets of equations take a simpler form when projective symmetry is a requirement for the gravitational action. For this case, we will mainly focus on the subcases of minimal coupling between matter and geometry and some particular kinds of nonminimal coupling that essentially shares the main features of the minimally-coupled case. We will leave the detailed analysis of the properties of RBGs without projective symmetry for chapter \ref{sec:UnstableDOF}. There we will see how they can teach us the valuable lesson that metric-affine theories generally propagate ghost degrees of freedom unless one explicitly tries to avoid them in the construction of a particular theory, or resorts to a particular subclass of the metric-affine family that are known to be ghost-free, such as \eg RBGs with projective symmetry. We will then comment on the cases with and without projective symmetry in the gravitational action particularising as well for certain types of couplings between matter and geometry. Let us start with the metric field equations first.
\subsection{Metric field equations}\label{sec:GeneralRBGMetricFieldEqs}

The metric field equations obtained by varying \eqref{eq:GeneralRBGAction} with respect to the metric, are
\begin{equation}\label{eq:GeneralRBGMetricFieldEquations}
\frac{\partial \cL}{\partial g^{\mu\nu}}-\frac{1}{2}\cL g_{\mu\nu}={\mpl^{-2}}T_{\mu\nu},
\end{equation}
where 
\beq\label{eq:StressEnergyTensor}
T_{\mu\nu}\equiv-\frac{2}{\sqrt{-g}}\frac{\delta\sqrt{-g}\cL_\textrm{m}}{\delta g^{\mu\nu}}
\eeq
 is the usual stress-energy tensor of the matter sector. Note that these field equations are algebraic for $g^{\mu\nu}$, which suggests that there could be some nontrivial constraints that would complicate the analysis of the propagating degrees of freedom propagated by these theories when written in the field variables $(g,\Gamma)$. We will later see that a field redefinition clarifies the analysis by leading to the Einstein frame of the theories.
 
 Diffeomorphism symmetry has a consequence on the allowed dependence of $\cL$ on its variables: given that $\cL$ is a scalar under diffeomorphisms, it can only depend on traces of powers of 
\beq\label{eq:PandZRBGLagrangianInvariants}
P^\mu{}_\nu=g^{\mu\alpha}R_{(\alpha\nu)}\qquad \text{and} \qquad Z^\mu{}_\nu=g^{\mu\alpha}R_{[\alpha\nu]}.
\eeq
Another possible decomposition for these two independent pieces is $\cP^{\mu}{}_\nu=g^{\mu\alpha}R_{\alpha\nu}$ and $\cZ^{\mu}{}_\nu=g^{\mu\alpha}R_{\nu\alpha}$, where $\cP^{\mu}{}_\nu=P^{\mu}{}_\nu+Z^{\mu}{}_\nu$ and $\cZ^{\mu}{}_\nu=P^{\mu}{}_\nu-Z^{\mu}{}_\nu$. Using the objects $\cZ$ and $\cP$ we can write the Lagrangian as 
\beq\label{eq:RBGLagrangianMatrixForm}
\cl\big(g^{\mu\nu},R_{\mu\nu}\big)=\cF\big[\cP^\mu{}_\nu(g^{\mu\nu},R_{\mu\nu}),\cZ^\mu{}_\nu(g^{\mu\nu},R_{\mu\nu})\big]
\eeq
which leads to the general relation
\beq\label{eq:RelationsPartialDerivativesRBGLagrangiangRPZ}
\begin{split}
&\frac{\partial\cl}{\partial g^{\mu\nu}}=\frac{\partial \cF}{\partial \cP^\mu{}_\alpha}R_{\nu\alpha}+\frac{\partial \cF}{\partial \cZ^\mu{}_\alpha}R_{\alpha\nu}\\
&\frac{\partial\cl}{\partial R_{\mu\nu}}=\frac{\partial \cF}{\partial \cP^\alpha{}_\mu}g^{\nu\alpha}+\frac{\partial \cF}{\partial \cZ^\alpha{}_\mu}g^{\alpha\nu}.
\end{split}
\eeq 
Note that for the case with projective symmetry, where the Ricci is symmetric, $Z^\mu{}_\nu$ vanishes and, therefore, $\cP^\mu{}_\nu=\cZ^{\mu}{}_\nu=P^\mu{}_\nu$. In that case, we must drop the $\cZ$ dependence from $\cF$, and the partial derivatives $\partial \cF/\partial \cZ$ cannot even be defined. As a rule of thumb, we can use the above relations and all the relations that we derive below with $\partial \cF/\partial \cZ=0$. After some manipulations we arrive to the general relation
\beq\label{eq:RelationDerivativesLagrangianRBG}
\frac{\partial\cl}{\partial R_{\mu\sigma}}R_{\nu\sigma}=g^{\mu\alpha}\frac{\partial\cl}{\partial g^{\alpha\nu}}+\frac{\partial \cF}{\partial \cZ^\alpha{}_\beta}\big(g^{\mu\alpha}\delta^\sigma{}_\beta R_{\sigma\nu}+g^{\alpha\sigma}\delta^\mu{}_\beta R_{\nu\sigma}\big),
\eeq 
which allows us to write the metric field equations \eqref{eq:GeneralRBGMetricFieldEquations} and its trace as
\beq\label{eq:IntermediateStepFieldEqs}
\begin{split}
\frac{\partial\cl}{\partial R_{\mu\sigma}}R_{\nu\sigma}&=\frac{\partial \cF}{\partial \cZ^\alpha{}_\beta}\big(g^{\mu\alpha}\delta^\sigma{}_\beta R_{\sigma\nu}+g^{\alpha\sigma}\delta^\mu{}_\beta R_{\nu\sigma}\big)+\frac{1}{2}\cL\delta^\mu{}_\nu+{\mpl^{-2}}T^\mu{}_\nu,\\
\frac{\partial\cl}{\partial R_{\mu\nu}}R_{\mu\mu}&=2\frac{\partial \cF}{\partial \cZ^\alpha{}_\beta}\cZ^\alpha{}_\beta+\frac{D}{2}\cL+{\mpl^{-2}}T.
\end{split}
\eeq 
We can now define a new rank-2 tensor field, which will later be seen to play the role of a metric, by
\beq\label{eq:DefinitionMetricq}
\begin{split}
\sqrt{-q}q^{\mu\nu}=\sqrt{-g}\frac{\partial\cl}{\partial R_{\mu\nu}}.
\end{split}
\eeq
Combining the two equations in \eqref{eq:IntermediateStepFieldEqs} in a clever way, this new tensor allows us to write the metric field equations as
\beq
\begin{split}\label{eq:GeneralRBGMetricFieldEquationsEinsteinTensor}
q^{\mu\sigma}R_{\nu\sigma}-\frac{1}{2}q^{\alpha\sigma}R_{\alpha\sigma}\delta^\mu{}_\nu=&{\mpl^{-2}}\sqrt{\frac{-g}{-q}}\Bigg[T^\mu{}_\nu-\lr{\frac{D-2}{4}{{\mpl^2}}\cl+\frac{T}{2}}\delta^\mu{}_\nu \\
&+{{\mpl^2}}\frac{\partial \cF}{\partial \cZ^\alpha{}_\beta}\big(g^{\mu\alpha}\delta^\sigma{}_\beta R_{\sigma\nu}+g^{\alpha\sigma}\delta^\mu{}_\beta R_{\nu\sigma}-\cZ^\alpha{}_\beta\delta^\mu{}_\nu\big)\Bigg].
\end{split}
\eeq
Now, we can define a metric-affine version of the Einstein tensor as $\cG^\mu{}_\nu(q,\Gamma)=q^{\mu\sigma}R_{(\sigma\nu)}-\frac{1}{2}q^{\alpha\sigma}R_{(\alpha\sigma)}\delta^\mu{}_\nu$ and write the metric field equations in the convenient form
\beq\label{eq:GeneralRBGFieldEqAntisymmetricRicci}
\begin{split}
\cG^\mu{}_\nu&(q,\Gamma)={\mpl^{-2}}\sqrt{\frac{-g}{-q}}\Bigg[T^\mu{}_\nu-\lr{\frac{D-2}{4}{{\mpl^2}}\cl+\frac{T}{2}}\delta^\mu{}_\nu \\
&+{{\mpl^2}}\frac{\partial \cF}{\partial \cZ^\alpha{}_\beta}\big(g^{\mu\alpha}\delta^\sigma{}_\beta R_{\sigma\nu}+g^{\alpha\sigma}\delta^\mu{}_\beta R_{\nu\sigma}-\cZ^\alpha{}_\beta\delta^\mu{}_\nu\big)\Bigg]+q^{\mu\sigma}R_{[\sigma\nu]}+\frac{1}{2}q^{\alpha\sigma}R_{[\alpha\sigma]}\delta^\mu{}_\nu.
\end{split}
\eeq
Note the following argument: from \eqref{eq:DefinitionMetricq} we have an algebraic relation that can be used to write $g^{\mu\nu}$ as a function of $q^{\mu\nu}$, $R_{(\mu\nu)}$ and $R_{[\mu\nu]}$ formally. By writting $\cP$ and $\cZ$ again in terms of $P$ and $Z$, the metric field equations \eqref{eq:IntermediateStepFieldEqs} give an algebraic relation between $R_{(\mu\nu)}$, $R_{[\mu\nu]}$, $g^{\mu\nu}$ and the matter fields through $T^\mu{}_\nu$. Both relations could be used to write $R_{(\mu\nu)}$, and by extension $\cL$, formally as a function of $q^{\mu\nu}$, $R_{[\mu\nu]}$ and the matter fields. The right hand side would then be a function of the new tensor $q^{\mu\nu}$, $R_{[\mu\nu]}$ and the matter fields. This would allow to define a new matter sector as $\tilde{\cL}_\textrm{m}\big(q^{\mu\nu},\Gamma^\alpha{}_{\mu\nu},R_{[\mu\nu]},\Psi_i\big)$, and $R_{[\mu\nu]}$ can acquire dynamics due to the terms $g^{\alpha\sigma}R_{\nu\sigma}$ in \eqref{eq:GeneralRBGFieldEqAntisymmetricRicci}. If such matter sector is found, then the metric field equations take formally the form
\beq
\cG^\mu{}_\nu(q,\Gamma)={\mpl^{-2}}\tilde{T}^\mu{}_\nu,
\eeq
where $\tilde{T}^\mu{}_\mu=q^{\mu\alpha}\tilde{T}_{\alpha\nu}$ and $\tilde{T}$ is defined as in \eqref{eq:StressEnergyTensor} making the substitution $(g,\cL_\textrm{m})\mapsto(q,\tilde{\cL}_\textrm{m})$. These are identically the metric field equations obtained for GR with a matter sector featuring additional nonlinear terms due to solving $g^{\mu\nu}$ and $R_{(\mu\nu)}$ in terms of $q^{\mu\nu}$, $R_{[\mu\nu]}$ and the original matter fields. As we said above, these formally defined new matter sector can also contain extra degrees of freedom due to $R_{[\mu\nu]}$ and the antisymmetric part of $q^{\mu\nu}$ (see chapter \ref{sec:UnstableDOF}). We will see that this is the case in general in chapter \ref{sec:UnstableDOF}, and we will now turn to the case when projective symmetry is required. In this case, the above arguments become more apparent and there are no ghostly degrees of freedom. This is (by far) the class containing most of the models studied in the literature with interesting nonperturbative properties, and we will devote this chapter to its study. Before turning into the subset of RBGs with projective symmetry, let us define the deformation matrix $\Omega^\mu{}_\nu$, which will be of great use later as it encodes the relation between the new rank-2 tensor $q^{\mu\nu}$ and the original metric $g^{\mu\nu}$ as  
\beq\label{eq:DeformationMatrixRelation}
q^{\mu\nu}=\left(\Omega^{-1}\right)^\mu{}_\rho g^{\rho\nu}. 
\eeq
Taking the determinant on both sides we find $\Omega=\det(gq^{-1})$, and using \eqref{eq:DefinitionMetricq} we can write the inverse deformation matrix as
\beq\label{eq:DeformationMatrixDefinition}
\sqrt{\Omega}\left(\Omega^{-1}\right)^\mu{}_\nu = \frac{\partial{\cL}}{\partial R_{\mu\rho}} g_{\rho\nu}\,,
\eeq
where $\Omega$ is the determinant of the deformation matrix. Taking the determinant of both sides of \eqref{eq:DeformationMatrixDefinition}, one finds
\beq\label{eq:DeterminantDeformation}
\Omega = \det\lrsq{\frac{\partial{\cL}}{\partial R_{\mu\rho}} g_{\rho\nu}}\,.
\eeq
and therefore the inverse deformation matrix is given by
\beq\label{eq:DefinitionInverseOmega}
\left(\Omega^{-1}\right)^\mu{}_\nu = \frac{1}{\sqrt{\det\lrsq{\frac{\partial{\cL}}{\partial R_{\alpha\lambda}}g_{\lambda\beta}}}}\frac{\partial{\cL}}{\partial R_{\mu\rho}} g_{\rho\nu}\,.
\eeq
By the same argumentation as above, it is possible to conclude that the deformation matrix can also formally be written as an on-shell function of $q^{\mu\nu}$, $R_{[\mu\nu]}$ and the matter fields. A crucial property that will be exploited later is the fact that if the symmetries of $g^{\mu\nu}$ and $\Omega^\mu{}_\nu$ are not the same, then the two metrics will describe spacetimes with different symmetries. We will later see (for the case with projective symmetry) that this is in general possible due to the nonlinearities if the deformation matrix. To that end it will be useful to rewrite the above definition of the deformation matrix in a way that can be interpreted as a matrix product, which can be done by using \eqref{eq:RelationsPartialDerivativesRBGLagrangiangRPZ}, and reads
\beq\label{eq:DefinitonGeneralInverseOmegaMatrix}
(\Omega^{-1})^\mu{}_\nu=\frac{1}{\sqrt{\det\lrsq{\frac{\partial \cF}{\partial \cP^\nu{}_\mu}
+\frac{\partial \cF}{\partial \cZ^\nu{}_\mu}}}}\lr{\frac{\partial \cF}{\partial \cP^\nu{}_\mu}+\frac{\partial \cF}{\partial \cZ^\nu{}_\mu}}.
\eeq
Let us now turn or attention to RBG theories with projective symmetry.

\subsubsection{The case with projective symmetry}

In RBGs projective symmetry has a very simple implementation: given that a projective transformation $\delta_\xi\Gamma^\alpha{}_{\mu\nu}=\xi_\mu\delta^\alpha{}_\nu$ leaves the symmetric part of the Ricci invariant \eqref{eq:TransfRicciProjective} and changes its antisymmetric part as $\delta_\xi R_{\mu\nu}\propto(\dif \xi)_{\mu\nu}$, we have that in RBGs with projective symmetry only $P^\mu{}_\nu=g^{\mu\alpha}R_{(\alpha\nu)}$ enters the Lagrangian while $Z^\mu{}_\nu=g^{\mu\alpha}R_{[\alpha\nu]}$ is banned from it, as it would explicitly break the projective symmetry. Thus in this case $\cZ^\mu{}_\nu=\cP^\mu{}_\nu=P^\mu{}_\nu$, and we can use the above relations setting $\partial \cF/\partial \cZ\mapsto0$. It can be seen that, in this case, we can rewrite \eqref{eq:RelationsPartialDerivativesRBGLagrangiangRPZ} and \eqref{eq:RelationDerivativesLagrangianRBG} as
\beq\label{eq:RelationDerivativesLagrangianPIRBG}
\frac{\partial \cL}{\partial R_{\mu\nu}}g_{\nu\alpha}=\frac{\partial\cF}{\partial P^\alpha{}_\mu}\qquad\text{and}\qquad g^{\mu\alpha}\frac{\partial\cL}{\partial g^{\alpha\nu}}=\frac{\partial \cF}{\partial P^\alpha{}_\mu}P^\alpha{}_\nu
\eeq 
where we have multiplied the first by $g_{\nu\alpha}$. The right hand side of these equations can be written in matrix notation respectively as
\beq
\frac{\partial\cF}{\partial\hat P} \qquad \text{and} \qquad \frac{\partial \cF}{\partial\hat P}\hat P.
\eeq
Therefore, using \eqref{eq:RelationsPartialDerivativesRBGLagrangiangRPZ}, the metric field equations \eqref{eq:GeneralRBGMetricFieldEquations} can be rewritten as well in matrix notation as
\begin{equation}\label{eq:PIRBGMetricFieldEquationsMatrixForm}
\frac{\partial \cF}{\partial \hat P}\hat P-\frac{1}{2}\cF\,\bbI={\mpl^{-2}}\hat T.
\end{equation}
By writting the relation between $g_{\mu\nu}$ and $q_{\mu\nu}$ given in \eqref{eq:DeformationMatrixRelation} in matrix form and inverting it, we find the relation  \beq\label{eq:RelationBothMetricsMatrxForm}
\hat q=\hat g\, \hat\Omega \qquad \text{or in tensorial form} \qquad q_{\mu\nu}=g_{\mu\rho}\Omega^\rho{}_\nu,
\eeq
where we have used that $\hat g$ is symmetric and $\hat\Omega$ is given by inverting the matrix form of the definition of $\hat\Omega^{-1}$ in \eqref{eq:DefinitionInverseOmega} a
\beq\label{eq:OmegaMatrixMatrixForm}
 \hat \Omega=\sqrt{\det\lr{\frac{\partial \cF}{\partial \hat P}}}
\lr{\frac{\partial \cF}{\partial \hat P}}^{-1},
\eeq
After taking $\partial\cF/\partial\cZ\rightarrow 0$ and $\cP^\mu{}_\nu=P^\mu{}_\nu$ as corresponds to the case with projective symmetry. Note that the (projective invariant) metric field equations in matrix form \eqref{eq:PIRBGMetricFieldEquationsMatrixForm} yield an algebraic relation between $\hat P$ and $\hat T$. In general, given the nonlinear nature of the left hand side of \eqref{eq:PIRBGMetricFieldEquationsMatrixForm}, there will generally exist several algebraic solutions $\hat P(\hat T)$ that solve the metric field equations and, substituting the solution on the definition on the expression for the deformation matrix \eqref{eq:OmegaMatrixMatrixForm}, they will lead to different on-shell expressions $\hat\Omega(T)$, one for each branch of solutions $\hat P(\hat T)$ of \eqref{eq:PIRBGMetricFieldEquationsMatrixForm}. The requirement that the gravitational Lagrangian is an analytic function of $R_{(\mu\nu)}$, namely that $\cF(\hat P)$ is an analytic function of $\hat P$, implies the existence of at least one solution, which satisfies
\beq
\left.\frac{\partial\cF}{\partial\hat P}\right|_{\hat P\rightarrow 0}\propto\bbI
\eeq
with a constant proportionality factor. For this branch, we have that $\hat\Omega^{-1}|_{\hat P\rightarrow0}\propto \bbI$ and Einstein frame metric will be related by a proportionality factor (up to $\cO(\hat P)$ corrections) that can be reabsorbed as a cosmological constant term on the right hand of the metric field equations. Note that this branch of solutions would then be identical to vacuum GR if the solution for the connection is given by the Levi-Civita connection of $q^{\mu\nu}$, as will be seen to be the case for RBG theories with projective symmetry. However, though the existence of this branch is guaranteed by the analyticity requirement, one should keep in mind that other nontrivial solutions, each characterised by its own expression for $\hat\Omega(\hat T)$, can exist due to the nonlinearities of the metric field equations. However, we will see in section \ref{sec:SolutionsDeformation} that these nontrivial solutions are typically unphysical.

Another relevant consequence of projective symmetry is that, due to the fact that only the symmetric part of the Ricci tensor enters the action, the definition of $q^{\mu\nu}$ given in \eqref{eq:DefinitionMetricq} now yields a symmetric rank-2 tensor that can be properly used as a metric. The metric field equations for RBGs with projective symmetry can now be written in terms of $q^{\mu\nu}$, $\Gamma^\alpha{}_{\mu\nu}$, and the matter fields as
\beq\label{eq:MetricFieldEquationsRBGEinsteinForm}
\cG^\mu{}_\nu(q,\Gamma)={\mpl^{-2}}\Omega^{-1/2}\lrsq{T^\mu{}_\nu-\lr{\frac{D-2}{4}{{\mpl^2}}\cl+\frac{T}{2}}\delta^\mu{}_\nu}
\eeq
where the possible dependence of the right hand side on the metric $g^{\mu\nu}$ can be eliminated by finding $\hat g(\hat q,\Psi_i)$ as a solution to $\eqref{eq:DeformationMatrixRelation}$ after substituting the deformation matrix by one of the solutions $\hat\Omega(\hat T)$. Note that the dependence on $\Psi_i$ of the solution $\hat g(\hat q,\Psi_i)$ enters through the stress energy tensor but it might be that a more general dependence on the matter fields occurs after solving for $g^{\mu\nu}$ in terms of $q^{\mu\nu}$ and the matter, as $g^{\mu\nu}$ is usually present inside of $\hat T$. Using this argument also in the solutions $\hat P(\hat T)$, we will be able to write $R_{(\mu\nu)}$, and therefore $\cL$, as an on-shell function of the matter fields and the metric $q^{\mu\nu}$. This leads to the possibility of writing the metric field equations again as the metric field equations of first order GR for some stress-energy tensor $\tilde T^\mu{}_\nu$ that is defined by writing the right hand side of \eqref{eq:MetricFieldEquationsRBGEinsteinForm} as a function of $q^{\mu\nu}$ and the matter fields, leading to
\beq\label{eq:RBGPIMetricEquationsEinsteinFrame}
\cG^\mu{}_\nu(q,\Gamma)={\mpl^{-2}}\tilde{T}^\mu{}_\nu
\eeq
Unlike the case without projective symmetry, the stress-energy tensor $\tilde{T}^\mu{}_\nu$ does not contain new matter degrees of freedom, though it generally incorporates new highly nonlinear interaction terms between the original matter degrees of freedom. In any case, for the above equation to be equivalent to metric-affine (or first-order) GR, we need that the field equations of the connection admit the Levi-Civita connection of $q^{\mu\nu}$ as its unique solution (up to a projective gauge mode). We will see that this is indeed the case for RBG theories with projective symmetry, but not the general one, where the solution to the connection involves new degrees of freedom that are not present in the case with projective symmetry. These degrees of freedom are described by $R_{[\mu\nu]}$ and a (now physical) projective mode. Though these new degrees of freedom spoil the equivalence to GR, we will see that there is a similar correspondence with Nonsymmetric Gravity Theory. 

\subsection{Connection field equations}

By varying \eqref{eq:GeneralRBGAction} with respect to the connection, using the definition of $q^{\mu\nu}$ introduced for the metric field equations in \eqref{eq:DefinitionMetricq}, the connection field equations can be written as
\begin{equation}\label{eq:GeneralRBGConnectuonEquations}
\begin{split}
{\nabla_{\lambda}\left[\sqrt{-q} q ^{\nu\mu}\right]-\del^\mu{}_\la \nabla_{\rho}\left[\sqrt{-q} q ^{ \nu\rho}\right]}{=\Delta_{\lambda}{}^{\mu \nu}+\sqrt{-q}\left[{T}^{\mu}{}_{\lambda \alpha} q ^{\nu\al}+{T}^{\alpha}{}_{\alpha \lambda} q ^{\nu\mu}-\delta_{\lambda}^{\mu} {T}^{\alpha}{}_{\alpha \beta} q ^{\nu\be}\right]}.
\end{split}
\end{equation}
where $\Delta_{\lambda}^{\mu \nu}$ is the hypermomentum defined as
\begin{equation}\label{eq:DefHypermomentum}
\Delta_{\lambda}{}^{\mu \nu} \equiv \frac{2}{{\mpl^2}} \frac{\delta \sqrt{-g}\cL_\textrm{m}}{\delta \Gamma^{\lambda}{}_{\mu \nu}}.
\end{equation}
The above equation can be recast in a more convenient form by introducing a new connection $\hat{\Gamma}^\alpha{}_{\mu\nu}$ obtained by subtracting a projective mode from the original one as
\beq
\hat{\Gamma}^\alpha{}_{\mu\nu}=\Gamma^\alpha{}_{\mu\nu}+\frac{2}{D-1}\Gamma^\lambda{}_{[\lambda\mu]}\delta^\alpha{}_\nu.
\label{Eq:Gammaheq}
\eeq
This connection identically satisfies $\hat{\Gamma}^\lambda{}_{[\lambda\mu]}=0$. The hypermomentum of the original connection $\Delta_\al{}^{\mu\nu}$ and the one related to the shifted connection  $\Deltah_\al{}^{\mu\nu}$ are related by
\beq
\Delta_\al{}^{\mu\nu}=\Deltah_\al{}^{\mu\nu}+\frac{2}{D-1}\delta_\al{}^{[\mu}\Deltah_\be{}^{\nu]\be},
\eeq
where the hypermomentum corresponding to the shifted connection is defined as in \eqref{eq:DefHypermomentum}. Note that this relation implies that the hypermomentum of projectively invariant matter fields satisfies $\Deltah_\be{}^{\mu\be}=0$. The explicit splitting of the connection into a projective mode done in this way will prove to be very convenient for analysing the physical content of RBGs both with and without projective symmetry. The main reason for this is that this splitting allows us to write the general connection field equations \eqref{eq:GeneralRBGConnectuonEquations} as
\beq\label{eq:GeneralRBGConnectuonEquationsSplitted}
\frac{1}{\sqrt{-q}}\partial_\la(\sqrt{-q}q^{\mu\nu})+\hat{\Gamma}^{\mu}{}_{\la\al}q^{\al\nu}+\hat{\Gamma}^{\nu}{}_{\al\la}q^{\mu\al}-\hat{\Gamma}^{\al}{}_{\la\al}q^{\mu\nu}=\frac{1}{\sqrt{-q}}\lr{\Deltah_\al{}^{\mu\nu}+\frac{2}{D-1}\delta_\al{}^{[\mu}\Deltah_\be{}^{\nu]\be}},
\eeq
and by taking its traces, we can derive from it the conditions
\beq
\begin{split}
&\partial_\mu\Big(\sqrt{-q} q^{[\mu\nu]}\Big)=\Deltah_\lambda{}^{[\lambda\nu]}+2\Deltah_\lambda{}^{[\nu\lambda]}\\
&\partial_{\alpha}\sqrt{-q}=\hat\Gamma^\lambda{}_{\alpha\lambda}+\frac{1}{\sqrt{-q}}q_{\mu\nu}\lr{\Deltah_\al{}^{\mu\nu}+\frac{2}{D-1}\delta_\al{}^{[\mu}\Deltah_\be{}^{\nu]\be}}
\end{split}
\eeq
where $q^{\mu\alpha}q_{\mu\beta}=q^{\alpha\mu}q_{\beta\mu}=\delta^\alpha{}_\beta$ defines $q_{\mu\nu}$ (see \cite{Damour:1992bt}).  After some manipulations, these conditions allow to write the above equation \eqref{eq:GeneralRBGConnectuonEquationsSplitted} as
\beq
\label{eq:Gammah}
\resizebox{0.9\hsize}{!}{$\partial_{\lambda} q^{\mu\nu}+\hat{\Gamma}^\mu{}_{\lambda\al}q^{\al\nu}+\hat{\Gamma}^\nu{}_{\al\lambda}q^{\mu\al}=\frac{1}{\sqrt{-q}}\Bigg[\Deltah_\al{}^{\mu\nu}-q^{\mu\nu}\Deltah_\al{}^{\alpha\beta}q_{\alpha\beta}+\frac{2}{D-1}\lr{\delta_\al{}^{[\mu}\Deltah_\be{}^{\nu]\be}-q_{[\lambda\beta]}\Deltah_\alpha{}^{\beta\alpha}}\Bigg].$}
\eeq
For a symmetric metric, or for vanishing hypermomentum, these equations are formally identical to the connection field equations obtained from the metric-affine version of GR coupled to a general matter sector. For vanishing hypermomentum and a symmetric $q^{\mu\nu}$, these equations are algebraic and linear in $\hat\Gamma^{\alpha}{}_{\mu\nu}$, and are known to admit a unique solution given by the Levi-Civita connection of $q^{\mu\nu}$ up to a choice of projective mode \cite{Bernal:2016lhq}. This result, however, needs not be extendible to a nonsymmetric $q^{\mu\nu}$, although it provides valuable insight in attempting to find a formal solution in this case. Consider splitting $q^{\mu\nu}$ into its symmetric and antisymmetric parts as
\beq\label{eq:MetricSplitting}
\sqrt{-q}q^{\mu\nu}=\sqrt{-h}\lr{h^{\mu\nu}+B^{\mu\nu}}
\eeq
where $\sqrt{-q}q^{(\mu\nu)}=\sqrt{-h}h^{\mu\nu}$ and $\sqrt{-q}q^{[\mu\nu]}=\sqrt{-h}B^{\mu\nu}$. We can now use $h_{\mu\nu}$ defined by $h^{\mu\alpha}h_{\alpha\nu}=\delta^\mu{}_\nu$ as a (symmetric) metric, and split the connection as
\beq\label{Eq:GammasplittingLeviCivitah}
\hat\Gamma^\alpha{}_{\mu\beta}={}^h\Gamma^{\alpha}{}_{\mu\beta}+\hat\Upsilon^\alpha{}_{\mu\beta}
\eeq
where $^h\Gamma^{\alpha}{}_{\mu\beta}$ are the Christoffel symbols of $h_{\mu\nu}$ defined as in \ref{eq:Christoffel}. The homogeneous equation corresponding \eqref{eq:Gammah} (namely the vanishing hypermomentum case) is trivially satisfied by the symmetric part of $q^{\mu\nu}$ and ${}^h\Gamma^\alpha{}_{\mu\nu}$, namely
\beq
\partial_{\lambda} h^{\mu\nu}+{}^h\Gamma^\mu{}_{\lambda\al}h^{\al\nu}+{}^h\Gamma^\nu{}_{\al\lambda}h^{\mu\al}=0
\eeq
identically. Then, by performing the usual trick of adding and subtracting \eqref{eq:Gammah} with suitably permuted indices, we can write a formal solution for the connection as
\begin{equation}
\resizebox{0.9\hsize}{!}{$\Upsilonh^{\alpha}{}_{\mu\nu}=\frac{1}{2} h^{\kappa \lambda}\Bigg[\left(\nabla_{\beta}^{h} B_{\ga \lambda}+\nabla_{\ga}^{h} B_{\lambda \beta}-\nabla_{\lambda}^{h} B_{\beta \ga}\right)+\frac{1}{\sqrt{-h}} \left(\Deltah_{\beta\ga \lambda}+\Deltah_{\ga\lambda \beta}+\Deltah_{\lambda\beta \ga}+\frac{2}{D-1}h_{\lambda[\gamma}\Deltah^\al{}_{\be]\al}\right)\Bigg]\left(A^{-1}\right)_{\kappa}{}^{\alpha}{}_{\mu\nu}{}^{\beta \gamma},$}
\end{equation}
where by definition ${A}^\kappa{}_{\al'}{}^{\mu'\nu'}{}_{\be\ga} ({A}^{-1})_\kappa{}^\al{}_{\mu\nu}{}^{\be\ga}= \delta^{\al'}{}_{\al} \delta^{\mu'}{}_{\mu} \delta^{\nu'}{}_{\nu}$. Here ${A}^\kappa{}_{\al'}{}^{\mu'\nu'}{}_{\be\ga}$ is linear in $B_{\mu\nu}$ and is given by
\begin{align}\label{Operatorsolcon}
 &{A}^\kappa{}_{\al}{}^{\mu\nu}{}_{\be\ga} = a^\kappa{}_{\al}{}^{\mu\nu}{}_{\be\ga}+b^\kappa{}_{\al}{}^{\mu\nu}{}_{\be\ga}{}^{\rho\sigma}B_{\rho\sigma}\nonumber\\
 &a^\kappa{}_{\al}{}^{\mu\nu}{}_{\be\ga} =
 \delta^\kappa{}_\al \delta^\mu{}_\be \delta^\nu{}_\ga +\frac{1}{2} \delta^\mu{}_{\alpha}\left(h^{\nu\kappa} h_{\beta \gamma}-2 \delta^{\nu}{}_{(\beta} \delta^\kappa{}_{\gamma)}\right)\\
 &b^\kappa{}_{\al}{}^{\mu\nu}{}_{\be\ga}{}^{\rho\sigma}=\frac{1}{2}\left[h_{\alpha\gamma} h^{\mu\sigma} \delta^\nu{}_{\beta} h^{\rho\kappa}+\delta^{\beta}{}_\rho h_{\alpha\gamma} h^{\mu\kappa} h^{\nu \sigma}+\delta^{\rho}{}_\gamma \delta^\mu{}_{\alpha} \delta^\nu{}_{\beta} h_{\kappa \sigma}-h^{\rho\kappa} \delta^{\mu}{}_\gamma h^{\nu\sigma} h_{\alpha \beta}\right.\nonumber\\
 &\hspace{1cm}\left.-\delta^{\rho}{}_ \beta h^{\sigma\kappa} \delta^\mu{}_{\alpha}\delta^{\nu}{}_\gamma-\delta^{\rho}{}_ \beta \delta^{\sigma}{}_\gamma\delta^{\mu}{}_\alpha h^{\nu\kappa}-\delta^{\rho}{}_\gamma h_{\alpha \beta} h^{\mu{\sigma}} h^{\nu{\alpha}}-\delta^{\rho}_{\gamma} \delta^{\kappa}{}_{\alpha} \delta^{\mu}{}_\beta h^{\nu\sigma}+\delta^{\rho}{}_\beta \delta^{\kappa}{}_ \alpha h^{\mu \sigma} \delta^{\nu}{}_\gamma\right].\nonumber
\end{align}
Note that in the most general case, the hypermomentum will depend on the connection as well, and the above formal solution will still be an implicit equation for $\hat\Upsilon^\alpha{}_{\mu\nu}$. However, for matter fields whose hypermomentum does not depend on the connection (\eg minimally coupled matter fields), this formal solution will do the job. As a remark, let us mention that if the matter fields have couplings to the connection through $R_{\mu\nu}$, instead of including them in the hypermomentum, we can define it as in \eqref{eq:DefHypermomentum} for variations which keep the Ricci constant, and include the variation of the matter Lagrangian with respect to the Ricci in the definition of the metric $q^{\mu\nu}$. This would lead to add the matter Lagrangian to $\cF$ in \eqref{eq:OmegaMatrixMatrixForm}, thus modifying the dependence of  the deformation matrix (and therefore the relation between $q^{\mu\nu}$, $g^{\mu\nu}$) on the matter fields, but keeping the structure of the connection field equations \eqref{eq:Gammah}.

In any case, we see that the antisymmetric part of the effective nonsymmetric metric $q^{\mu\nu}$ introduces deviations in the connection from being the Levi-Civita connection of $h^{\mu\nu}\propto q^{(\mu\nu)}$ even in vacuum. Remarkably, although these deviations are more general than a simple projective mode, they are due to the explicit breaking of projective symmetry in \eqref{eq:GeneralRBGAction}, and will be one of the sources of the pathologies present in RBGs without projective symmetry and in general metric-affine theories if gravity, as will be shown in chapter \ref{sec:UnstableDOF}. For the moment, now that we have analysed the algebraic structure of the connection field equations in general RBGs, let us discuss further the details of these equation and its solution in RBGs with projective symmetry.

\subsubsection{The case with projective symmetry}
As explained in section \ref{sec:GeneralRBGMetricFieldEqs}, the requirement of projective symmetry in RBG theories reduces to forbid the antisymmetric part of the Ricci tensor from entering the action. In this case we have that 
\beq
\frac{\partial\cL}{\partial R_{\mu\nu}}=\frac{\partial\cL}{\partial R_{(\mu\nu)}}
\eeq 
which implies that the object $q^{\mu\nu}$ defined in \eqref{eq:DefinitionMetricq} is symmetric and, as a consequence, the 2-form $B^{\mu\nu}\propto q^{[\mu\nu]}$ drops from the splitting \eqref{eq:MetricSplitting}. The results that follow  \eqref{eq:MetricSplitting} were derived in full generality and it can be seen that they are valid for RBGs with projective symmetry simply by performing the substitution $B^{\mu\nu}\mapsto0$. Concretely, we have that for RBGs with projective symmetry the only deviations of the connection from being the Levi-Civita connection of\footnote{Note that in this case $q^{\mu\nu}=h^{\mu\nu}$ exactly.} $q^{\mu\nu}$ are introduced by the coupling between the matter fields and the connection as
\begin{equation}
\Upsilonh^{\alpha}{}_{\mu\nu}=\frac{1}{2\sqrt{-h}} h^{\kappa \lambda}\lr{\Deltah_{\beta\ga \lambda}+\Deltah_{\ga\lambda \beta}+\Deltah_{\lambda\beta \ga}+\frac{2}{D-1}h_{\lambda[\gamma}\Deltah^\al{}_{\be]\al}}\left(A^{-1}\right)_{\kappa}{}^{\alpha}{}_{\mu\nu}{}^{\beta \gamma},
\end{equation}
where now ${A}^\kappa{}_{\al}{}^{\mu\nu}{}_{\be\ga} = a^\kappa{}_{\al}{}^{\mu\nu}{}_{\be\ga}$. Particularly, we see that in presence of matter fields that do not couple to the connection (such as minimally coupled bosonic fields) the connection is the Levi-Civita connection of $q^{\mu\nu}$ up to a projective (gauge) mode\footnote{This is also true if the matter fields couple to the connection only through the symmetric part of the Ricci tensor. In this cases, the analysis is identical if we include those terms in the definition of $q^{\mu\nu}$.}. This is a remarkable feature of RBGs with projective symmetry since, in this case, the connection is an auxiliary field that can be solved algebraically as the Levi-Civita connection of $q^{\mu\nu}$. In the case of including linear couplings to the connection, then the corresponding hypermomentum depends only on the matter fields and the connection can be solved algebraically as the canonical connection of $q^{\mu\nu}$ plus corrections involving the matter fields. In any of these cases, we can integrate the connection out and the resulting theory can be cast only in terms of a metric.  

Given that by definition the metric-affine version of the Einstein tensor introduced in \eqref{eq:GeneralRBGFieldEqAntisymmetricRicci} satisfies $\cG^\mu{}_\nu(q,{}^q\Gamma)\equiv G^\mu{}_\nu(q)$, where $G^\mu{}_\nu(q)$ is the usual Einstein tensor for the metric $q^{\mu\nu}$, the metric field equations \eqref{eq:RBGPIMetricEquationsEinsteinFrame} are exactly identical to the Einstein equations for the metric coupled to a modified stress-energy tensor
\beq
G^\mu{}_\nu(q)={\mpl^{-2}}\tilde T^\mu{}_\nu
\eeq
where $\tilde T^\mu{}_\nu$ is given by 
\beq\label{eq:RelationStressEnergyTensorsFramesRBG}
\tilde T^\mu{}_\nu=\sqrt{\frac{-g}{-q}}\lrsq{T^\mu{}_\nu-\lr{\frac{D-2}{4}{{\mpl^2}}\cl+\frac{T}{2}}\delta^\mu{}_\nu}
\eeq
written only in terms of the matter fields, their derivatives and $q^{\mu\nu}$ as explained in \ref{sec:GeneralRBGMetricFieldEqs}. We then see that the analogy with the Einstein field equations, though purely formal for the case without projective symmetry (there are even extra degrees of freedom), is fully realised in the case with projective symmetry, and the gravitational sector is indeed equivalent to that of GR. As noted in \cite{Jimenez:2015caa}, this implies that the only gravitational degrees of freedom present in RBG theories with projective symmetry are those corresponding to a spin-2 massless field described by the perturbations to the metric $q^{\mu\nu}$, which seems to be the appropriate object to describe the gravitational dynamics of the theory.\footnote{Thus, the characteristics of the propagation of gravitational waves will be related to geodesics of $q^{\mu\nu}$. However, note that minimally coupled matter fields will follow geodesics of $g^{\mu\nu}$.} In the next section we will see that this equivalence can be proved in a more transparent way at the level of the action by choosing the appropriate field variables that describe the degrees of freedom of RBG theories.

%%%%%%%%%%%%%%%%%%%%%%%%%%%%%%%%%%%%
\subsection{The Einstein frame of Ricci-Based Gravity theories}\label{sec:EinsteinFrame}

As discussed above, it is possible to obtain the main properties of general RBG theories by working with the field equations. However, it is more illuminating to re-write the action so that the gravitational sector looks more familiar and, consequently, the physical content of the theory is more apparent. In this section we will see that both projectively and nonprojectively invariant RBG theories admit an Einstein frame representation. In the case with projective symmetry, the action of the theory will be identical to the Einstein-Hilbert action. On the contrary, when projective symmetry is explicitly broken, the action for the theory in its Einstein frame will be equivalent to that of the Nonsymmetric Gravity Theory (NGT) proposed in \cite{Moffat:1978tr}. We will follow the procedure presented in \cite{BeltranJimenez:2017doy,Afonso:2017bxr} for the projectively invariant theories, extending it to the general nonprojectively invariant case. 

Let us start by performing a Legendre transformation in order to linearise the action for general RBG theories \eqref{eq:GeneralRBGAction} in the Ricci tensor as follows
\beq\label{eq:ModifiedAction}
\cS=\frac12{\mpl}^{2}\int \dd^Dx \sqrt{-g}\left[\cL(g^{\mu\nu},\Sigma_{\mu\nu})+\frac{\partial \cL}{\partial \Sigma_{\mu\nu}}\big(R_{\mu\nu}-\Sigma_{\mu\nu}\big)\right]
+\cS_\textrm{m}[g,\Gamma,\Psi_i],
\eeq
where $\Sigma_{\mu\nu}$ is an auxiliary field.\footnote{We will not write explicitly the dependence of $\cL$ but in this section it should be assumed that $\cL$ means $\cL(g^{\mu\nu},\Sigma_{\mu\nu})$} In order to put our action in a more familiar form, we can perform the following field redefinition
\beq\label{eq:DefMetricqSigma}
\sqrt{-q}q^{\mu\nu}=\sqrt{-g}\frac{\partial \cL}{\partial \Sigma_{\mu\nu}}.
\eeq
This definition will allow to express the auxiliary field $\Sigma_{\mu\nu}$ in terms of the metric and the object $q^{\mu\nu}$ by inverting the algebraic relation $\Sigma_{\mu\nu}=\Sigma_{\mu\nu}(q,g)$ defined by \eqref{eq:DefMetricqSigma}. The dynamics of this new auxiliary field is given by the constraint $\Sigma_{\mu\nu}=R_{\mu\nu}$, so that the above field redefinition looks exactly like the definition for $q^{\mu\nu}$ given in \eqref{eq:DefinitionMetricq} in section \ref{sec:GeneralRBGMetricFieldEqs} when $\Sigma$ is on-shell. After this field redefinition, we can then express the general RBG action in the form
\beq\label{eq:NonSymmetricActionTwoMetrics}
\cS=\frac12{{\mpl^2}}\int \dd^Dx\Big[\sqrt{-q}q^{\mu\nu}R_{\mu\nu}+\cU(q,g)\Big]+\cS_\textrm{m}[g,\Gamma,\Psi_i],
\eeq
where we have introduced the potential term
\beq
\cU(q,g)=\sqrt{-g}\lrsq{\cL-\frac{\partial \cL}{\partial\Sigma_{\mu\nu}}\Sigma_{\mu\nu}}_{\Sigma=\Sigma(q,g)}.
\eeq
The action \eqref{eq:NonSymmetricActionTwoMetrics} already features the standard Einstein-Hilbert term in the first order formalism, but for the object $q^{\mu\nu}$ instead of the spacetime metric $g_{\mu\nu}$. As a matter of fact, we can notice that $g_{\mu\nu}$ appears algebraically in the potential $\cU$ and the matter action so that it is simply an auxiliary field that we can integrate out by solving its field equations, which are given by
\beq\label{eq:MetricFieldEqEinsteinFrame}
\frac{\partial \cU}{\partial g^{\mu\nu}}=\sqrt{-g}\,{\mpl^{-2}}T_{\mu\nu}.
\eeq
From this equations we can obtain the spacetime metric $g_{\mu\nu}$ in terms of the object $q^{\mu\nu}$ and the stress-energy tensor of the matter sector, computed as the variation of the matter action with respect to $g_{\mu\nu}$ as defined in \eqref{eq:StressEnergyTensor}. We will see below that there is another stress-energy tensor that we can introduce to make the resemblance with the first-order formulation of GR even more apparent. Once we have obtained the corresponding solution to \eqref{eq:MetricFieldEqEinsteinFrame}, we can use it to finally express \eqref{eq:NonSymmetricActionTwoMetrics} as
\beq\label{eq:EHframe1}
\cS=\frac12{{\mpl^2}}\int \dd^Dx \Big[\sqrt{-q}q^{\mu\nu}R_{\mu\nu}+\cU(q,T)\Big]+\tilde\cS_\textrm{m}[q,\Gamma,\Psi_i].
\eeq
$\tilde\cS_\textrm{m}[q,\Psi_i]=\cS_\textrm{m}[g(q,T),\Gamma,\Psi_i]$. This is the desired appearance of the theory where the gravitational sector reduces to the well-known Einstein-Hilbert action in the first order formalism for a nonsymmetric metric $q^{\mu\nu}$. 

In addition to the purely gravitational sector, we also see how we have generated new couplings between the object $q^{\mu\nu}$ and the matter sector. Such couplings arise after integrating out the spacetime metric both from the potential $\cU$ generated when linearising in the Ricci tensor, and from the explicit couplings of the matter sector to $g^{\mu\nu}$. Notice that matter only enters the metric field equations \eqref{eq:MetricFieldEqEinsteinFrame} through the stress-energy tensor obtained as the reaction to variations of the metric $g^{\mu\nu}$. This further implies that all the newly generated matter couplings will only depend on $T_{\mu\nu}$, which guarantees the preservation of the symmetries in the original matter sector. Notice that since $g_{\mu\nu}$ appears in $T_{\mu\nu}$ not as $T_{\mu\nu}\propto g_{\mu\nu}$ but in a more involved form, it could be that if we truly want to eliminate $g_{\mu\nu}$ in favour of $q_{\mu\nu}$ and the matter fields, the dependence could also be more general than through $T_{\mu\nu}$ (we have to solve the corresponding equation for $g_{\mu\nu}$). However, the new couplings will still surely have the same symmetries as the matter action. It is important to emphasise that the resemblance is purely formal at this point and, in fact, solving for the connection will fail to recover GR owed to the lack of any symmetries of $q^{\mu\nu}$ as showed in the previous section
%%%%%%%%%%%%%%%%%%%%%%%%%%%%%%%%%%%%%%%%%%%%
\subsubsection{The case with projective symmetry: Equivalence to GR}
%%%%%%%%%%%%%%%%%%%%%%%%%%%%%%%%%%%%%%%%%%%%
As explained in the previous section, in the case of RBG theories, enforcing projective symmetry is equivalent to require that only the symmetric part of the Ricci tensor appears in the action. Therefore, in this case, we can take the auxiliary field $\Sigma_{(\mu\nu)}$ to be symmetric in its two indices without loss of generality, so that $q^{\mu\nu}$ will inherit the symmetric character of the Ricci tensor. Being a symmetric rank-2 tensor, $q^{\mu\nu}$ is then entitled to claim its status as a proper metric tensor so that the gravitational sector in (\ref{eq:EHframe1}) is actually the first order formulation of GR . The corresponding solution for the connection will then be given by the Christoffel symbols of the metric $q^{\mu\nu}$ (up to the projective mode entering as a gauge mode \cite{Bernal:2016lhq}) instead of those of the metric $g_{\mu\nu}$. In the Einstein frame we thus recover the usual form of the Einstein equations, but the right hand side is now given by the stress-energy tensor $\tilde T_{\mu\nu}$ describing the reaction of the matter action $\tilde\cS_\textrm{m}$ to the metric $q^{\mu\nu}$, defined as in \eqref{eq:StressEnergyTensor}. This stress-energy tensor is highly nonlinearly related to $T_{\mu\nu}$ \cite{Afonso:2018bpv}, and will feature new interactions between all the matter fields \cite{Latorre:2017uve,Delhom:2019wir}. We will latter devote a chapter to the analysis of these interactions, which are the origin of the different phenomenology and solutions that deviate from the {\it usual GR} behaviour. The apparent differences between RBGs and GR are simply due to the fact that a matter sector coupled to a projectively invariant RBG corresponds to another matter sector (obtained as a nonlinear deformation of the previous one) coupled to GR. The peculiar property of the RBG with projective symmetry is that the interactions in the matter sector present a somewhat universal form (that of course depends on the specific theory). The new interactions will be generated through the total stress-energy tensor \cite{Afonso:2018bpv,Delhom:2019wir}. Assuming that the most relevant interactions in the gravitational sector of RBG appear at some specific scale ${\mg}$, which means that the function $\cL$ only contains one additional parameter with nontrivial mass dimension, then all the new interactions in the matter sector will not only be universally constructed in terms of $T_{\mu\nu}$, but they all will in turn have the same coupling constant. This means that, if an effect is seen at a given scale in some sector of the standard model, effects at the same scale will arise in the remaining sectors. From this perspective, we can interpret RBG theories as a procedure to encapsulate a universally interacting matter sector in an auxiliary field that plays the role of a nondynamical affine connection. In particular, this property is precisely what permits to study the dynamics in terms of a metric $g_{\mu\nu}$ for all matter fields at the same time. Let us elaborate on this point a bit more.

The physical meaning of the two metrics is also apparent in the Einstein frame: the metric $g_{\mu\nu}$ will determine the trajectories of the particles, which will follow the corresponding geodesics\footnote{It is perhaps convenient to explicitly state the physical situation we have in mind and what we mean by particles and geodesics. We assume that there is some background configuration both for the gravitational sector and the matter fields. Then, there will be perturbations on top of this background configuration and these perturbations are what we will call {\it particles}, possibly with an unfortunate abuse of language. These perturbations are the ones that will follow geodesics of a given metric when we consider their {\it free} propagation. Of course, living on a nontrivial background, the propagation will occur in a medium with which these perturbations will interact.} (provided that they do not couple to the connection). One may then wonder why they do not follow the geodesics of $q_{\mu\nu}$ in the Einstein frame and how to square this with our statement that these theories are GR. The answer is quite simple. Around trivial matter backgrounds, both metrics are the same and therefore there is no possible confusion. In the presence of a matter background however both metrics are different and while matter fields follow the geodesics of $g_{\mu\nu}$, it is $q_{\mu\nu}$ that satisfies Einstein equations. There is no onus however because, also in GR when matter fields propagate on a nontrivial background (and are coupled to it) the propagation of the corresponding perturbations does not follow the geodesics of the metric, but those of an effective metric encoding the effects of the background instead. From the Einstein frame perspective, this effective metric encoding the effects of the nontrivial matter background is the RBG frame metric $g^{\mu\nu}$. Paradigmatic examples of this behaviour are for instance $K$-essence models of scalar fields or nonlinear electrodynamics (see e.g. \cite{Babichev:2007dw,ArmendarizPicon:2005nz,Plebanski:106680,Novello:1999pg,Gibbons:2000xe}). Indeed, as we will explicitly see below, starting from a standard canonical scalar field and usual Maxwellian electrodynamics in the RBG frame, the Einstein frame formulation will respectively be $K$-essence \cite{Afonso:2018hyj} and nonlinear electrodynamics \cite{Delhom:2019zrb}. This can be seen by explicitly constructing the matter action $\tilde\cS_\textrm{m}$ if a particular RBG theory and matter sector are specified. The explicit construction of this action allows to know what is the corresponding matter sector for which the dynamics of a given RBG theory is described by the Einstein equations
\beq\label{eq:EinsteinEquationsq}
G^\mu{}_\nu(q)={\mpl^{-2}}\tilde T^\mu{}_\nu.
\eeq
where, as we saw in the previous section, the Einstein frame stress energy tensor is related to the RBG frame stress energy tensor by \eqref{eq:RelationStressEnergyTensorsFramesRBG} which, if written in terms of the corresponding matter Lagrangians\footnote{This can be done from the definition of stress-energy tensor \eqref{eq:StressEnergyTensor}.} in both frames $\cL_\textrm{m}$ and $\tilde\cL_\textrm{m}$, leads to a relation between them of the form
\begin{equation}\label{eq:RelationMatterActionsRBG}
\left(g^{\rho \sigma} \frac{\delta \mathcal{L}_{m}}{\delta g^{\rho \sigma}}-\mathcal{L}_{G}-\mathcal{L}_{m}\right)\Omega^{-1/2}\delta^{\mu}{}_{\nu}-2 \Omega^{-1/2} g^{\mu \rho} \frac{\delta \mathcal{L}_{m}}{\delta g^{\rho \nu}}=\tilde\cL_\textrm{m}\delta^{\mu}{}_{\nu}-2 q^{\mu \rho} \frac{\delta \tilde{\mathcal{L}}_{m}}{\delta q^{\rho \nu}}
\end{equation}
where recall that $\Omega=\det(gq^{-1})$ is the determinant of the deformation matrix \eqref{eq:DeformationMatrixRelation}. Though this relation holds in general, the process of explictily building the Einstein frame Lagrangian (or action) is model dependent\footnote{In the above equation, this model dependence is encoded in $q^{\mu\nu}$ and $\Omega$.} and only a few cases have been worked out \cite{Afonso:2018mxn,Afonso:2018hyj,Delhom:2019zrb}. In the next section, we will show how to perform explicitly the mapping procedure of any RBG theory coupled to any nonlinear electrodynamics (NED), \ie a general theory for a $U(1)$ gauge field. This is, we will derive the most general form of the Einstein frame matter action $\tilde\cS_\textrm{m}$ corresponding to this physical system. Then we will explicitly construct $\tilde\cS_\textrm{m}$ for a particular RBG theory dubbed as Eddington-inspired Born-Infeld gravity coupled to Maxwell electrodynamics, which remarkably, maps to GR coupled to the well known Born-Infeld electrodynamics.  

There is a property of the deformation matrix that, besides being of great use through this procedure, gives powerful insights on the structure of the solution space and the phenomenology of the theory. Given that the deformation matrix is an on-shell function of the matter fields through the stress-energy tensor and one of the metrics (either $g^{\mu\nu}$ or $q^{\mu\nu}$), $\hat\Omega$ can always be expanded as a power series of the stress-energy tensor. By the Cayley-Hamilton theorem, in $D$ spacetime dimensions, this power series will be of the form
\beq\label{eq:OmegaExpansionTmunu}
\hat\Omega(\hat T)=\sum_{n=0}^{D-1}C_n\lr{\frac{T}{{\mq}^{D}}}\frac{\hat T^n}{{\mq}^{nD}}
\eeq
where  $C_n$ are arbitrary (analytic) functions of its argument and $T=\mathrm{Tr}(\hat T)$. Here $\mq$ should be seen as a UV scale that characterises deviations from the metricity condition as, given that the connection field equations guarantee that $\nabla q^{\mu\nu}=0$, the nonmetricity tensor will be of the schematic form $\nabla \hat g \sim (\nabla\hat\Omega) \hat q$ which will vanish for ${\mq}\rightarrow\infty$.  The requirement that a given RBG Lagrangian reduces to the Einstein-Hilbert action at low energies (up to a redefinition of the cosmological constant) implies that
\beq
\lim_{\hat T\rightarrow 0}C_{n\neq 0}=0 \qquad\text{and}\qquad \lim_{\hat T\rightarrow 0}C_{0}=\lambda
\eeq
where $\lambda$ is a finite constant. As explained in section \ref{sec:GeneralRBGMetricFieldEqs}, the requirement of analyticity of the Lagrangian as a function of the Ricci guarantees the existence of a branch of solutions with a deformation matrix satisfying these conditions. For this branch, the above expansion can also be written in tensor form as
\beq\label{Om-T-Exp}
\Omega^\mu{}_\nu=\lambda\delta^\mu{}_\nu+\cO^\mu{}_\nu\lr{\frac{T^\mu{}_\nu}{\mq^4}}
\eeq
where $\lambda$ is a constant and $\cO^\mu{}_\nu$ is an analytic function of the stress-energy tensor that vanishes in vacuum\footnote{Note that any Lagrangian that is an analytic function of the Ricci tensor will satisfy $\lim_{R_{\mu\nu}\rightarrow0}\partial \cL/\partial R_{\mu\nu}\propto\delta^\mu{}_{\nu}$. On shell, this translates into $\lim_{T^\mu{}_\nu\rightarrow0} \cO^\mu{}_\nu=0$ or also $\lim_{\mq\rightarrow0} \cO^\mu{}_\nu=0$ }. In this branch of solutions $\hat\Omega(\hat T)$ we have that in vacuum $g^{\mu\nu}=\lambda q^{\mu\nu}$, and both theories are exactly described by the Einstein-Hilbert action with a cosmological constant\footnote{The value of the cosmological constant described by the two metrics will be shifted proportionally to $(\lambda-1)$.}. Furthermore, we will see in chapter \ref{sec:SolutionsDeformation} that the other branches, if existing, are generally pathological from the physical perspective, thus providing a motivation for the choice of this branch in phenomenological analyses, which is rather established in the literature.

\section{Mapping RBGs to GR coupled to an abelian gauge field}\label{sec:Mapping}

As we explained in the previous section, there is a choice of field variables that allows to write RBG theories with projective symmetry\footnote{Through this section we will only consider RBG theories with projective symmetry even if it is not stated explicitly.  I might write only RBG theories to shorten the writing, but I will be referring to RBG theories with projective symmetry.} coupled to a given matter sector as GR coupled to a nonlinear deformation of that matter sector with the same degrees of freedom but a different set of interactions. In this section, we will explicitly build the correspondence for a matter sector consisting of a NED (\ie a $U(1)$ gauge field). To that end, it will be useful to introduce the basic invariants that can be built with an abelian gauge field. Given a 1-form field $A$ there are only two basic building blocks that can be used to build a Lagrangian that is both diffeomorphism and gauge invariant, namely its fieldstrength $F=\dif A$ and its Hodge dual $\star F$. Using the definition of Hodge dual given in section \ref{sec:MetricStructure} it is possible to derive the following relations
\begin{equation}\label{eq:FundamentalRelationsEMInvariants}
\begin{split}
&F^\mu{}_\lambda F^{\lambda}{}_\nu=K \delta^\mu{}_\nu  +(\star F)^\mu{}_\lambda {}(\star F)^\lambda{}_\nu\,,\\
&(\star F)^\mu{}_\lambda F^\lambda{}_\nu = -2G\delta^\mu{}_\nu.
\end{split}
\end{equation}
where  $K =-\frac12 F_{\alpha\beta}F^{\alpha\beta}$ and $G = \frac14 F_{\alpha\beta}(\star F)^{\alpha\beta}$.
Using these equations, the algebraic structure of products of even and odd number of field strength tensors can be reduced to a sum of four different tensorial structures, namely
\begin{equation}\label{eq:RelationsEMInvariants}
\begin{split}
&F^{\mu}{}_{\lambda_1}F^{\lambda_1}{}_{\lambda_2}\cdots F^{\lambda_{2k-2}}{}_{\lambda_{2k-1}} F^{\lambda_{2k-1}}{}_{\lambda_{2k}}= a_{2k}{(K,G)}\delta^\mu{}_{\lambda_{2k}} + b_{2k}{(K,G)}K^\mu{}_{\lambda_{2k}}\\
&F^{\mu}{}_{\lambda_1}F^{\lambda_1}{}_{\lambda_2}\cdots F^{\lambda_{2k-1}}{}_{\lambda_{2k}} F^{\lambda_{2k}}{}_{\lambda_{2k+1}}=a_{2k+1}{(K,G)}F^\mu{}_{\lambda_{2k+1}} + b_{2k+1}{(K,G)}(\star F)^\mu{}_{\lambda_{2k+1}}\,,
\end{split}
\end{equation}
where we have defined $K_{\mu\nu}=\partial K/\partial g^{\mu\nu}=F_{\mu\rho}{F^\rho}_\nu$, and
its trace is (unconveniently) given by $g^{\mu\nu}K_{\mu\nu}=2K$. In particular, the following identity will be useful
\begin{equation}\label{4FRed}
K^{\mu}{}_{\lambda}K^{\lambda}{}_{\nu}=G^2\delta^\mu{}_\nu +K K^\mu{}_{\nu} \,.
\end{equation}
The above relations \eqref{eq:RelationsEMInvariants} allow to write the most general diffeomorphism and gauge invariant Lagrangian that can be built out of a $U(1)$ field as a function only of the invariants $(K,G)$. Let us also note that, since $G$ is parity-odd, only even powers of $G$ are allowed in the Lagrangian in a parity preserving theory. In this section, we will assume a four-dimensional spacetime and a matter sector consisting of a general diffeomorphism invariant $U(1)$ gauge field. Therefore, the matter Lagrangian $\cL_\textrm{m}(K,G)$ will have an arbitrary dependence on $K$ and $G$. The stress-energy tensor \eqref{eq:StressEnergyTensor} for such a matter sector is given in general by
\beq\label{eq:StressEnergyNED}
{T^\mu}_\nu=\left(\cL_\textrm{m} - G\frac{\partial\cL_\textrm{m}}{\partial G}\right)\delta^\mu{}_\nu-2\frac{\partial\cL_\textrm{m}}{\partial K}K^{\mu}{}_\nu\,.
\eeq
The success of being able to carry out the mapping procedure explicitly depends on the ability to identify the most general tensorial dependence of $\Omega^\mu{}_\nu$ on the matter fields. Making use of \eqref{4FRed}, we can conclude that the most general deformation matrix (and inverse) for this matter sector will be of the form
\beq\label{eq:FormDeformationMatrixEM}
\begin{split}
&\Omega^\mu{}_\nu=A(K,G)\delta^\mu{}_\nu+B(K,G)K^\mu{}_\nu\,,\\
&(\Omega^{-1})^\mu{}_\nu=C(K,G)\delta^\mu{}_\nu+D(K,G)K^\mu{}_\nu\,,
\end{split}
\eeq 
where the relation between the coefficients is
\beq
A=\frac{C+DK}{C^2-D^2G^2+C D K}
 \qquad \text{and} \qquad
B=-\frac{D}{C^2-D^2G^2+C D K} \ ,\label{AB}
\eeq
or equivalently
\beq
C=\frac{A+BK}{A^2-B^2G^2+A B K}
 \qquad \text{and} \qquad
D=-\frac{B}{A^2-B^2G^2+A B K}\,,\label{CD}
\eeq
where we have omitted the functional dependence of $A$, $B$, $C$ and $D$ to lighten notation. Note that the particular form of $\Omega^\mu{}_\nu$, and therefore of the coefficients $A$, $B$, $C$ and $D$,  are completely specified once a particular gravitational Lagrangian and matter sector are chosen. Lastly, using again \eqref{4FRed} on \eqref{eq:FormDeformationMatrixEM} we can write the determinant of the deformation matrix as
\beq
\Omega = \left(A^2 - B^2 G^2 + A B K\right)^2 = \frac{1}{\left(C^2 - D^2 G^2 + C D K\right)^2}\,,\label{DetOmega}
\eeq
These results are in close analogy to the case where the matter sector is described by scalar fields \cite{Afonso:2018hyj}, where arbitrary powers of the scalar kinetic terms $K^{\mu}{}_\nu$ can be written as linear combinations of the $\delta^\mu{}_\nu$ and $K^{\mu}{}_\nu$, which allows to write the tensorial structure of a general deformation matrix in a closed form and compute its determinant. Therefore, we see that this key property of the deformation matrix that allowed to explicitly construct the mapping for a scalar matter sector transfers to a matter sector consisting of an   abelian gauge field as well.

Having developed the above results, we are now in position of writing the basic ingredients of the Einstein frame stress-energy tensor $\tilde T^\mu{}_\nu$. The dependence of the deformation matrix and the above relations imply that it will also be a NED, and therefore it will have the form
\beq
{\tilde T^\mu}{}_\nu=\left({\tilde\cL}_m - \tilde G\frac{\partial{\tilde\cL}_m}{\partial \tilde G}\right)\delta^\mu{}_\nu-2\frac{\partial{\tilde \cL}_m}{\partial \tilde K}\tilde K^{\mu}{}_\nu\,,
\eeq
where the tilded variables are defined in analogy to the ones without tilde substituting the RBG frame metric $g^{\mu\nu}$ by the Einstein frame metric $q^{\mu\nu}$. In general, through this section, tilded objects will imply that their indices are risen or lowered with the Einstein frame metric. For instance, we will write $\tilde F^\mu{}_\nu=q^{\mu\alpha}F_{\alpha\nu}=(\Omega^{-1})^\mu{}_\sigma g^{\sigma\alpha}F_{\alpha\nu}$. Using \eqref{eq:FundamentalRelationsEMInvariants} and \eqref{eq:FormDeformationMatrixEM} we can derive a relation between the fundamental objects in both frames, which in matrix notation reads
\begin{equation}
\left(\begin{array}{c} \tilde F^\mu{}_\nu \\ (\star \tilde F)^\mu{}_\nu \end{array}\right) = \left(\begin{array}{cc}  \left(C + D K\right) & - D\,G \\ D G& ( C + D K) \end{array}\right)\left(\begin{array}{c}  F^\mu{}_\nu \\ (\star F)^\mu{}_\nu \end{array}\right) \; ;
\end{equation}
which reminds of a duality rotation. From this relation it follows that
\begin{align}
\begin{split}
&\tilde K^\mu{}_\nu=D G^2 [2 C+D K] \delta^\mu{}_\nu+\left[(C+DK)^2+(DG)^2\right]K^\mu{}_\nu\\
&\tilde G^\mu{}_\nu  = - \Omega^{-1/2} G\delta^\mu{}_\nu = \left( D^2 G^2 - C^2 - C D K\right)G \delta^\mu{}_\nu \,
\end{split}
\end{align}
where $\tilde K^\mu{}_\nu=\tilde F^\mu{}_\rho\tilde F^\rho{}_\nu$ and we have introduced $\tilde G^\mu{}_\nu\equiv\tilde F^\mu{}_\rho(\star\tilde F)^\rho{}_\nu$. Tracing these equations, one finds the general relation between the scalar electromagnetic invariants in both frames given by
\begin{align}\label{eq:RelationBetweenScalarEMInvariantsInBothFrames}
\begin{split}
&\tilde K= \left[(C+DK)^2+3(DG)^2\right] K+ 4 C D G^2\\
&\tilde G = G \Omega^{-1/2} = G \left(C^2 - D^2 G^2 + C D K\right)
\end{split}
\end{align}
which explicitly proves that it is always possible to express the Einstein frame $U(1)$ invariant scalars $\tilde K$ and $\tilde{G}$ in terms of the ones in the RBG frame. For convenience let us also write the inverse relations, which are
\begin{equation}
\left(\begin{array}{c}  F^\mu{}_\nu \\ (\star F)^\mu{}_\nu \end{array}\right) = \left(\begin{array}{cc}  \left(\tilde A + \tilde B \tilde K\right) & - \tilde B\,\tilde G \\ \tilde B\,\tilde G &  \left(\tilde A + \tilde B \tilde K\right) \end{array}\right)\left(\begin{array}{c} \tilde F^\mu{}_\nu \\ (\star \tilde F)^\mu{}_\nu \end{array}\right) \ ,
\end{equation}
and also
\begin{align}
\begin{split}
&K^\mu{}_\nu=  \tilde B \tilde G^2 [2 \tilde A+\tilde B \tilde K] \delta^\mu{}_\nu+\left[(\tilde A+\tilde B \tilde K)^2+(\tilde B \tilde G)^2\right]\tilde K^\mu{}_\nu\,,\\
&G^\mu{}_\nu = -\tilde G \left[\tilde A\left(\tilde A +\tilde B\tilde K\right) - \tilde B^2 \tilde G^2\right] \delta^\mu{}_\nu\, .
\end{split}
\end{align}
which leads to
\begin{align}\label{eq:InverseRelationBetweenScalarEMInvariantsInBothFrames}
\begin{split}
&K= \left[(\tilde A+\tilde B \tilde K)^2+3(\tilde B \tilde G)^2\right] \tilde K+ 4 \tilde A \tilde B \tilde G^2\,,\\
&G= \tilde G \left[ \tilde A (\tilde A +  \tilde B \tilde K) - \tilde B^2 \tilde G^2 \right]=\tilde G \Omega^{1/2}\,.
\end{split}
\end{align}
This is an explicit manifestation that the mapping is an invertible construction, namely, one can also build an RBG frame for GR coupled to a given matter sector if the corresponding RBG matter sector is found. In other words, if the deformation matrix is known in terms of the Einstein frame, the corresponding field variables with which one would build the matter action in the RBG frame can be obtained in terms of those in the Einstein frame.

Once we have explicitly computed the relation between the appropriate field variables in each of the frames, we can now build the matter Lagrangian of the Einstein frame by means of \eqref{eq:RelationMatterActionsRBG}. For the particular case when both matter sectors are composed of an abelian gauge field,\footnote{Note that, as explained above, if the RBG frame matter Lagrangian $\cL_\textrm{m}$ describes an abelian gauge field, there is no assumption in asserting the Einstein frame matter Lagrangian $\tilde\cL_\textrm{m}$ also describes an abelian gauge field, as this is ensured by the tensorial structure of the deformation matrix \eqref{eq:FormDeformationMatrixEM} derived only from the requirement that $\cL_\textrm{m}$ describes an abelian gauge field.} after equating the coefficients of the two independent tensorial structures, we obtain that the following two equalities must be satisfied\footnote{There are some subtleties behind this argument. The interested reader is referred to Appendix B of \cite{Delhom:2018zrb} for details.}
\begin{align}
&\tilde \cL_\textrm{m} = \Omega^{-1/2}\left[2\left(K\frac{\partial \cL_\textrm{m}}{\partial K}+G\frac{\partial \cL_\textrm{m}}{\partial G}\right)-{\cL}_G-\cL_\textrm{m} \right] \ , \label{eq:Map1}\\
&\tilde K^\mu{}_\nu\frac{\partial \tilde \cL_\textrm{m}}{\partial \tilde K} + \frac12 \delta^\mu{}_\nu\tilde G\frac{\partial \tilde \cL_\textrm{m}}{\partial \tilde G}= \Omega^{-1/2}\,\left(K^\mu{}_\nu\frac{\partial \cL_\textrm{m}}{\partial K}+\frac12 \delta^\mu{}_\nu G\frac{\partial \cL_\textrm{m}}{\partial G}\right) \,.\label{eq:Map2}
\end{align}
The first of these equations provides a parametric representation of $\tilde\cL_\textrm{m}$ in terms of the $K$ and $G$ invariants of the RBG frame. The Lagrangian $\tilde \cL_\textrm{m}$ can generally be written as a function of the Einstein frame invariants $\tilde K$ and $\tilde G$ by means of \eqref{eq:InverseRelationBetweenScalarEMInvariantsInBothFrames}. However, note that the particular form of these relations depends on the particular gravitational model under consideration through the model dependent coefficients that define the deformation matrix $A$ and $B$ (or $C$ and $D$). The second equation leads to a relation between the partial derivatives of the matter Lagrangians which, by taking its trace and using the first equation leads to a parametrization of the matter Lagrangian in the RBG frame by the Einstein frame invariants given by
\beq
\cL_\textrm{m} = -{\cL}_G + \Omega^{1/2}\left[2\left(\tilde K\frac{\delta \tilde\cL_\textrm{m}}{\delta \tilde K} +\tilde G\frac{\delta \tilde\cL_\textrm{m}}{\delta \tilde G}\right) - \tilde \cL_\textrm{m} \right]\,,\label{eq:Map3}
\eeq
and which can be written in terms of the RBG frame invariants $K$ and $G$ by means of \eqref{eq:InverseRelationBetweenScalarEMInvariantsInBothFrames}. We would like to point out that the gravitational Lagrangian can be related to the Legendre transform of the matter Lagrangians by adding \eqref{eq:Map1} and \eqref{eq:Map3} as
\beq
L\left[\cL_\textrm{m}\right] + \Omega^{1/2}L\left[\tilde\cL_\textrm{m}\right] = {\cL}_G\,,
\eeq
which can also be written in terms of the traces of the stress-energy tensor in the two frames as
\beq
T + \Omega^{1/2}\tilde T = {\cL}_G\,.
\eeq
Let us remind the reader that these relations hold on the physical solutions of the theory. In order to make a quick check, let us recover a result that is already well known in the literature, namely, that the metric-affine version of quadratic (or Starobinski) $f(R)$ gravity is exactly equivalent to GR when both theories are coupled to a Maxwellian electrodynamics.

\subsection{f(R) theories coupled to a $U(1)$ gauge field}

Let us start by particularising the above arguments when the RBG model lies between the subset of $f(R)$ gravitational Lagrangians. Namely, we will take the gravitational sector of the action to be described by an action of the form
\begin{equation}\label{BIgrav}
\mathcal{S}_{f(R)} =\frac{1}{2}{\mpl}^{D-2} \int d^D x \sqrt{-g} f ( R ) \ .
\end{equation}
For this subclass of (projective invariant) RBG theories, the relation between the RBG and Einstein frame metric, and therefore the deformation matrix, take the particular form of a conformal transformation as
\beq
q^{\mu\nu} = f_R^{-\frac{2}{D-2}} g^{\mu\nu}\,\qquad\text{and}\qquad\Omega^{\mu}{}_{\nu} = f_R^{\frac{2}{D-2}} \delta^{\mu}{}_{\nu}\,.
\eeq
which in four dimensions reproduces the well known relation for metric-affine (or Palatini) $f(R)$ theories \cite{Olmo:2011uz}
\beq
q_{\mu\nu} = f_R g_{\mu\nu}\,.\label{Metric-Rel}
\eeq
Recall that as in all projectively invariant RBG theories, on the physical solutions of the theory, the gravitational Lagrangian and its derivatives can be written as a function of the trace of the stress-energy tensor, as is apparent from the metric field equations 
\beq
f_R  R - 2 f(R) = {\mpl}^{2-D} T\,\label{Alg-Eq}
\eeq
which give an algebraic relation between $R$ and $T$. In four spacetime dimensions, the matter action in the Einstein frame of $f(R)$ theories is thus provided by \eqref{eq:Map1}
\beq
\tilde \cL_\textrm{m} = f_R^{-2}\left[2\left(K\frac{\partial \cL_\textrm{m}}{\partial K}+G\frac{\partial \cL_\textrm{m}}{\partial G}\right)-\cL_\textrm{m}-{\mpl}^{2}f(R) \right]\label{f-par}\ ,
\eeq
where $K$ and $G$ are the electromagnetic invariants in the RBG frame. In this case, of the two coefficients that characterise a deformation matrix of a general (projective invariant) RBG coupled to a NED \eqref{eq:FormDeformationMatrixEM}, only the one in the $\delta^\mu{}_\nu$ term is nontrivial, which leads to the following relations between the electromagnetic invariants in the two frames
\begin{equation}\label{Inv-Rel}
K=f_{R} \tilde{K}\qquad\text{and}\qquad G=f_{R}^{2} \tilde{G}.
\end{equation}
Let us now illustrate the above discussion with a particular example example of a UV $f(R)$ correction, namely the metric-affine Starobinsky model, which is described by the action
\beq
S_s=\frac{1}{2}{{\mpl^2}}\int d^4 x\sqrt{-g}\left(R+\alpha R^2\right)\,,
\eeq
where $\alpha=(6{\mg^2})^{-1}$ has dimension of the inverse length squared in the International System of units. In this case $f(R)=R+\alpha R^2$ so that the metric field equations \eqref{Alg-Eq} imply that the curvature is proportional to the Legendre transform of the matter sector with respect to the electromagnetic invariants
\beq
R=-8{\mpl^{-2}}\left(\cL_\textrm{m} - G\frac{\partial\cL_\textrm{m}}{\partial G} -K\frac{\partial\cL_\textrm{m}}{\partial K}\right)\,,\label{Curv-f(R)}
\eeq
Using the above results, the relations between the electromagnetic invariants in different frames obtained by using \eqref{Inv-Rel} read
\beq
\tilde K = \frac{K}{1-16\alpha\kappa^2\left(\cL_\textrm{m} - G\frac{\partial\cL_\textrm{m}}{\partial G} -K\frac{\partial\cL_\textrm{m}}{\partial K}\right)}\,,\qquad \tilde G = \frac{G}{\left[1-16\alpha\kappa^2\left(\cL_\textrm{m} - G\frac{\partial\cL_\textrm{m}}{\partial G} -K\frac{\partial\cL_\textrm{m}}{\partial K}\right)\right]^2}\,.
\eeq
The above equations allow to write $K$ and $G$ as functions of $\tilde K$ and $\tilde G$, and therefore the Einstein frame matter Lagrangian \eqref{f-par} can be written in terms of the Einstein frame invariants. In the case in which the RBG frame matter sector is given by Maxwell electrodynamics, \ie $\cL_\textrm{m}=K/2$, the traceless property of the Maxwellian stress-energy tensor leads to the result $\tilde K=K$ and $\tilde G=G$. In that case, the Einstein frame matter Lagrangian that follows from \eqref{f-par} also describes a Maxwellian electrodynamics, namely
\beq
\tilde \cL_\textrm{m} = \frac{1}{2}\tilde K\,
\eeq
and the metrics in both frames are on-shell equivalent $q_{\mu\nu}=g_{\mu\nu}$ as well.
This proves that our framework recovers previous results derived for a particular RBG model, namely, that metric-affine Starobinsky theory gravity is equivalent to GR when both theories are minimally coupled to a free Maxwellian electromagnetic field. This motivates us to derive new results concerning other RBG theories that do not fit in the $f(R)$ subclass. To that end, in the next section, we will consider a popular RBG theory called Eddington-inspired Born-Infeld gravity, which in the weak field expansion recovers the Starobinsky theory plus additional quadratic and higher order corrections in the Ricci tensor.

%%%%%%%%%%%%%%%%%%%%%%%%%%%%%%%%%%%%%%%%%%%%%%%%%%%%%%%%%%%%%%%%%
\subsection{Eddington-inspired Born-Infeld gravity coupled to a $U(1)$ field}\label{sec:EiBIEMMapping}
%%%%%%%%%%%%%%%%%%%%%%%%%%%%%%%%%%%%%%%%%%%%%%%%%%%%%%%%%%%%%%%%%

In this section, we will explicitly derive the Einstein frame matter Lagrangian $\tilde{\mathcal{L}}_m$ resulting from an RBG frame matter Lagrangian $\mathcal{L}_m$ that describes a $U(1)$ gauge field coupled to Eddington-inspired Born-Infeld gravity theory (EiBI). This theory has inspiration in the purely affine model introduced by Eddington \cite{EddingtonAffine1924}, and a purely metric version was considered by Deser and Gibbons \cite{Deser:1998rj} which was seen to be plagued by ghosts. Then, Vollick \cite{Vollick:2003qp} introduced a more general metric-affine version of the theory without projective symmetry which was seen to be equivalent to an Einstein-Proca system. Later, a more restricted version with projective symmetry\footnote{The requirement of projective symmetry was not noticed by the authors at that time, but their requirement that only the symmetric part of the Ricci tensor appears in the action is equivalent to requiring projective symmetry to Vollick's version of the theory, as explained in section \ref{sec:StructureRBG}.} which got rid of the extra Proca field was considered by Ba\~nados and Ferreira in \cite{Banados:2010ix}.  

Since then, a lot of work on different aspects of the projective invariant version of the theory has been carried out. Some of the first works analysed collapsing matter, finding that EiBI admits a variety of compact objects not allowed in GR  \cite{Pani:2011mg,Pani:2012qb,Harko:2013wka}. In the subsequent years, several authors have shown that within EiBI there exist black holes solutions with central wormholes (also known as black bounces \cite{Barcelo:2014npa,Barcelo:2015rwa,Barcelo:2016hgb,Rovelli:2014cta,Haggard:2015iya,DeLorenzo:2015gtx,Christodoulou:2016vny,Malafarina:2017csn,Olmedo:2017lvt,Barrau:2018rts,Malafarina:2018pmv,Simpson:2018tsi,Simpson:2019cer,Lobo:2020ffi,Fran:2021pyi}). These objects turn out to be supported by matter which (in the RBG frame) does not violate the energy conditions, and describe spacetimes which are, in general, geodesically complete, thus offering new alternatives to address the issue of singularities in the classical theory \cite{Harko:2013aya,Olmo:2013gqa,Tamang:2015tmd,Shaikh:2015oha,Olmo:2015dba,Olmo:2015bya,Kim:2016pky,Olmo:2016fuc,Olmo:2017fbc,Menchon:2017qed,Shaikh:2018yku}. Recently, an analysis of the quasinormal modes of an AdS wormhole in EiBI has revealed interesting phenomenological properties \cite{Kim:2018ang}. As well, there are recent claims that GR coupled to Born-Infeld electrodynamics can sustain wormholes without violating the classical energy conditions \cite{Kim:2016pky,Kim:2018kpv}. The stability and properties of scalar perturbations of these objects were recently studied in \cite{KIM:2018kzv}. Cosmological and inflationary scenarios within EiBI have also been developed in \cite{EscamillaRivera:2012vz,Yang:2013hsa,Cho:2012vg,Jimenez:2014fla,BeltranJimenez:2017uwv,Li:2017ttl,Bouhmadi-Lopez:2014jfa,Bouhmadi-Lopez:2018tel,Albarran:2018mpg,Avelino:2012ue,Cho:2013pea,Cho:2014jta,Cho:2014xaa,Cho:2014ija,Kim:2013noa,Cho:2015yza,Jana:2016uvq,Jimenez:2015jqa}. Parallelly, several works aiming to constrain the energy scale at which EiBI deviates from GR were developed considering different physical scenarios. First astrophysical and cosmological constraints were worked out in  \cite{Avelino:2012ge},  then stronger constraints from nuclear physics phenomena were obtained in \cite{Avelino:2012qe,Avelino:2019esh}, and the most stringent constraints up to date come from particle collision experiments at LEP and LHC \cite{Latorre:2017uve,Delhom:2019wir,BeltranJimenez:2021iqs}. On the other hand, the work of Delsate and Steinhoff \cite{Delsate:2012ky,Delsate:2013bt} can be seen as a primitive version of the general mapping procedure for RBGs described here but restricted to isotropic perfect fluids in EiBI. An analogous approach had been used  before by Fatibene and Francaviglia \cite{Francaviglia12} in the context of $f(R)$ theories. For a recent up to date review on EiBI and generalizations see \cite{BeltranJimenez:2017doy}.

 In order to particularise the general discussion on explicitly performing the mapping to the Einstein frame of a general projective invariant RBG coupled to a $U(1)$ gauge field,  let us begin by defining the action of the EiBI theory as a UV (or high-curvature) modification of GR of the form
\begin{equation}\label{BIgrav}
\mathcal{S}_\textrm{EiBI}={{\mpl^2}}\mbi^2 \int d^4 x \left[\sqrt{-\left|g_{\mu\nu}+\mbi^{-2} R_{\mu\nu}\right|}-\lambda \sqrt{-g}\right],
\end{equation}
where $\mbi$ is a new scale that suppresses the higher curvature deviations from GR and we will work in four spacetime dimensions. The EiBI Lagrangian is the given by
\beq\label{eq:EiBILagrangian}
\cL_\textrm{EiBI}=2\mbi^2\left[\sqrt{\left|\delta^\mu{}_{\nu}+\mbi^{-2}g^{\mu\alpha} R_{\alpha\nu}\right|}-\lambda \right]
\eeq
In an expansion in inverse powers of $\mbi^2$, the metric-affine version of the Einstein-Hilbert action with a cosmological constant given by $\Lambda = (\lambda-1)/\mbi^2$ is recovered. By (conveniently) defining
\beq\label{eq:MetricqEiBI}
q_{\mu\nu} = g_{\mu\nu}+\mbi^{-2} R_{\mu\nu}\,,
\eeq
we can see that the EiBI metric field equations obtained from varying \eqref{BIgrav} with respect to the metric can be written as
\beq\label{eq:InvMetricqEiBI}
\sqrt{-q}q^{\mu\nu}=\sqrt{-g}\left(\lambda g^{\mu\nu}-{\mq^{-4}} T^{\mu\nu}\right)
\eeq
where remember that $q^{\mu\nu}$ is defined\footnote{Actually, we defined first $q^{\mu\nu}$ by \eqref{eq:DefinitionMetricq} and then $q_{\mu\nu}$ as its inverse. However, it can be seen that, by reformulating the EiBI action \eqref{BIgrav} with a suitably introduced auxiliary field $q_{\mu\nu}$, the field equations of the resulting equivalent formulation will tell us that $q_{\mu\nu}$ is given by \eqref{eq:MetricqEiBI} and its inverse by \eqref{eq:InvMetricqEiBI}. See section 2.6 of \cite{BeltranJimenez:2017doy} for a more detailed derivation.} by $q^{\mu\alpha}q_{\alpha\nu}=\delta^\mu{}_\nu$ and $\mq=({\mpl}{\mbi})^{1/2}$ is the geometric mean of the Planck mass and the mass scale $\mbi$ that will be seen to be associated to departures from metricity\footnote{Namely, the nonmetricity tensor will be proportional to inverse powers of $\mq$ (hence the suffix Q), and it would vanish if this scaled is pushed to infinity.} and to play a central role in phenomenological aspects of the theory (see chapter \ref{sec:ObservableTraces}). Using \eqref{eq:DeformationMatrixRelation}, implies that when the field equations are satisfied the (inverse) deformation matrix can be written on-shell as a function of the stress-energy tensor, which reads
\beq\label{eq:eq:DefinitionInverseOmegaegaEiBINED}
\sqrt{\Omega}\left(\Omega^{-1}\right)^\mu{}_\nu = \lambda\delta^\mu{}_\nu-{\mq^{-4}} T^\mu{}_\nu\,.
\eeq
From this relation, and using the stress-energy tensor for a general NED \eqref{eq:StressEnergyNED}, we can derive an explicit expression for $\Omega$ in EiBI coupled to a general electrodynamics,which reads
\beq\label{DetOmEiBI}
\begin{split}
\resizebox{0.9\hsize}{!}{$\Omega^{1/2}=\left[\lambda + {\mq^{-4}}\left(G\frac{\partial\cL_\textrm{m}}{\partial G}-
\cL_\textrm{m}\right)\right]^2 - \left[ 2{\mq^{-4}} \frac{\partial\cL_\textrm{m}}{\partial K}\right]^2 G^2+ 2 {\mq^{-4}} K\left[\lambda + {\mq^{-4}}\left(G\frac{\partial\cL_\textrm{m}}{\partial G}-
\cL_\textrm{m}\right)\right]\frac{\partial\cL_\textrm{m}}{\partial K}\,.$}
\end{split}
\eeq
Either from this result, or from \eqref{eq:eq:DefinitionInverseOmegaegaEiBINED}, we can derive the form of the coefficients of the tensorial decomposition of the deformation matrix \eqref{eq:FormDeformationMatrixEM}, which are given by
\beq\label{eq:InvCoefficientsDeformationEiBI}
C=\Omega^{-1/2}\left[\lambda + {\mq^{-4}}\left(G\frac{\partial\cL_\textrm{m}}{\partial G}-
\cL_\textrm{m}\right)\right]\,\qquad\text{and}\qquad
D=2{{\mq^{-4}}}\Omega^{-1/2} \frac{\partial\cL_\textrm{m}}{\partial K}\,.
\eeq
Having identified the structure of the deformation matrix of EiBI coupled to a general NED, we are now ready to build the Einstein frame matter Lagrangian $\tilde\cL_\textrm{m}$ that corresponds to an RBG frame matter Lagrangian $\mathcal{L}_\textrm{m}$ describing a general NED. Using the above results, the equations describing this correspondence \eqref{eq:Map1} and \eqref{eq:Map2} particularised to an EiBI coupled to a general NED read
\begin{align}
&\tilde\cL_\textrm{m}=\Omega^{-\frac12}\left[2\left(K\frac{\partial\cL_\textrm{m}}{\partial K}+G\frac{\partial\cL_\textrm{m}}{\partial G}\right)-\cL_\textrm{m}+\mq^{4}\lambda\right]-\mq^{4}\,,\label{CorrEiBIGR1}\\
&\tilde K^\mu{}_\nu\frac{\partial \tilde \cL_\textrm{m}}{\partial \tilde K} + \frac12 \delta^\mu{}_\nu\tilde G\frac{\partial \tilde \cL_\textrm{m}}{\partial \tilde G}= \Omega^{-1/2}\,\left(K^\mu{}_\nu\frac{\partial \cL_\textrm{m}}{\partial K}+\frac12 \delta^\mu{}_\nu G\frac{\partial \cL_\textrm{m}}{\partial G}\right)\, .\label{CorrEiBIGR2}
\end{align}
Using now \eqref{DetOmEiBI} we get
\beq\label{eq:TildeL}
\resizebox{.9\hsize}{!}{$
\tilde\cL_\textrm{m}=\frac{2\left(K\frac{\partial\cL_\textrm{m}}{\partial K}+G\frac{\partial\cL_\textrm{m}}{\partial G}\right)-\cL_\textrm{m}+\mq^{4}\lambda}{\left[\lambda + {\mq^{-4}}\left(G\frac{\partial\cL_\textrm{m}}{\partial G}-
\cL_\textrm{m}\right)\right]^2 - \left( 2 {\mq^{-4}} \frac{\partial\cL_\textrm{m}}{\partial K}\right)^2 G^2 + 2 {\mq^{-4}} K\left[\lambda + {\mq^{-4}}\left(G\frac{\partial\cL_\textrm{m}}{\partial G}-
\cL_\textrm{m}\right)\right]\frac{\partial\cL_\textrm{m}}{\partial K}}
-{\mq^4} \,.$}
\eeq
which provides a parametric representation of the Einstein frame matter Lagrangian $\tilde\cL_\textrm{m}$ in terms of the original RBG frame invariants $K$ and $G$. Given an explicit form for the RBG frame matter Lagrangian, we can write $K$ and $G$ in terms of $\tilde K$ and $\tilde G$ by means of \eqref{eq:InverseRelationBetweenScalarEMInvariantsInBothFrames}. Let us provide a particular example which will turn to be a new interesting result relating Born-Infeld electromagnetism and Born-Infeld gravity in a nontrivial way.
%%%%%%%%%%%%%%%%%%%%%%%%%%%%%%%
\subsubsection{A duality: Maxwell + EiBI \; as \; GR + BI electromagnetism}
%%%%%%%%%%%%%%%%%%%%%%%%%%%%%%
As a particular example of the mapping technique developed above, we will prove the conjecture that EiBI gravity coupled to Maxwell electromagnetism can be rewritten as GR coupled to Born-Infeld electrodynamics. This was partially proved only for static configurations in \cite{Afonso:2018mxn}, where the analogy of a general NED an anisotropic fluid (only valid for purely electric or magnetic cnfigurations) was used.  We will then start with the Maxwell Lagrangian $\cL_\textrm{m} = K/2$ in the RBG frame. Thus, using \eqref{eq:TildeL}, the Einstein frame Lagrangian in terms of the RBG frame invariants $K$ and $G$ will be given by
\beq\label{eq:Lraw}
\tilde\cL_\textrm{m} = \frac{2{\mq^4}\left(2\lambda + {\mq^{-4}} K\right)}{\left[4\lambda^2 - {\mq^{-8}} (K^2 + 4G^2)\right]} - {\mq^4}  \ .
\eeq
Now, we can use \eqref{eq:InverseRelationBetweenScalarEMInvariantsInBothFrames} to write the RBG frame invariants $K$ and $G$ in terms of the Einstein frame invariants $\tilde{K}$ and $\tilde{G}$, which will allow us to write the Einstein frame matter Lagrangian $\tilde\cL_\textrm{m}$ as a function of the Einstein frame metric $q^{\mu\nu}$ and the $U(1)$ gauge field. To that end, we need first to obtain the explicit form of the determinant of the deformation matrix, which from \eqref{DetOmEiBI} reads
\beq
\Omega = \left[\lambda^2 - {\mq^{-8}}\left(K^2 +4G^2\right)\right]^{2}
\eeq 
as well as the explicit form of the coefficients of the inverse deformation matrix (\ref{eq:FormDeformationMatrixEM}), which using \eqref{eq:InvCoefficientsDeformationEiBI} take the form
\beq
C=\frac{2\left(2\lambda - {{\mq^{-4}}} K\right)}{4\lambda^2 - {\mq^{-8}}\left(K^2 +4G^2\right)}\,,\qquad
D=\frac{4}{4\lambda^2{\mq^4} -  {\mq^{-4}}\left(K^2 +4G^2\right)} \,.
\eeq
Plugging these results back into (\ref{eq:RelationBetweenScalarEMInvariantsInBothFrames}), the Einstein frame invariants can be written in terms of the RBG ones as
\beq
\begin{split}
&\tilde K = 4\frac{4\lambda^2 K + \epsilon\kappa^2\left(4\lambda + \epsilon\kappa^2 K\right)\left(K^2 +4G^2\right)}{\left[4\lambda^2 - {{\mpl}^{-8}}\left(K^2 +4G^2\right)\right]^2}\\
&\tilde G = \frac{ 4G}{\left[4\lambda^2 - {{\mpl}^{-8}}\left(K^2 +4G^2\right)\right]}\,,
\end{split}
\eeq
which if inverted, give the RBG frame invariants $K$ and $G$ in terms of the Einstein frame invariants $\tilde K$ and $\tilde G$ as
\beq
\begin{split}
&K = \frac{2{\mq^4}\left({\mq^4}\tilde K - 4 \lambda\tilde G^2\right)\left(1+{\mq^{-4}}\lambda\tilde K \pm \sqrt{1+ 2{\mq^{-4}}\lambda\tilde K-4{\mq^{-8}} \lambda^2 \tilde G^2}\right)}{\left(\tilde K^2 + 4\tilde G^2 \right)}\\
&G = -\frac{2\tilde G\left[{\mq^{8}}-4{\mq^{4}}\lambda\tilde K + 2 \lambda^2 \tilde G^2   \pm \left({\mq^8}+{\mq^{4}}\lambda\tilde K\right) \sqrt{1+ 2{\mq^{-4}}\lambda\tilde K-4{\mq^{-8}}\lambda^2 \tilde G^2}\right]}{\left(\tilde K^2 + 4\tilde G^2 \right)} \ .
\end{split}
\eeq
Inserting these expressions into the above relation (\ref{eq:Lraw}), we obtain
\beq
\tilde\cL_\textrm{m} ={\mq^{4}} \frac{1-2\lambda\pm \sqrt{1+ 2{\mq^{-4}}\lambda\tilde K-4{\mq^{-8}}\lambda^2 \tilde G^2)}}{2\lambda}\, .
\eeq
In the asymptotically flat case ({\it i.e.} $\Lambda=0$, given by $\lambda\to 1$), taking the positive sign in front of the square root, defining a new mass scale 
\beq
\beta^2=-{\mq^4}/2,
\eeq
and writing $\tilde K$ and $\tilde G$ in terms of the gauge fieldstrength and its dual (see below \eqref{eq:FundamentalRelationsEMInvariants}); we find that the Einstein frame matter sector is described by the well known Lagrangian for Born-Infeld electromagnetism \cite{Born:1934gh}, namely
\beq
\tilde {\cL}_{BI} = \beta^2\left(1-\sqrt{1
+\frac{1}{2\beta^2}F_{\mu\nu}\tilde{F}^{\mu\nu} - \frac{1}{16\beta^4}(F_{\mu\nu}(\star \tilde{F})^{\mu\nu})^2}\right)\,.
\eeq
This constitutes the proof that EiBI gravity coupled to Maxwell electromagnetism can be written as GR coupled to Born-Infeld electromagnetism with the appropriate change of variables in field space. These type of correspondences can be used to establish a map to relate known properties of a Born-Infeld gravitational sector and unknown properties of GR, and vice-versa. Particularly, the existence of novel exact solutions in EiBI (or any RBG) could be used to unveil unknown solutions of GR coupled to the appropriate matter sector \cite{Afonso:2019fzv,Maso-Ferrando:2021ngp}.

%%%%%%%%%%%%%%%%%%%%%%%%%
%%%%%%%%%%%%%%%%%%%%%%%%%
%%%%%%%%%%%%%%%%%%%%%%%%%

			%NEWCHAPTER%

%%%%%%%%%%%%%%%%%%%%%%%%%
%%%%%%%%%%%%%%%%%%%%%%%%%
%%%%%%%%%%%%%%%%%%%%%%%%%

\chapter{Non-trivial aspects of the RBG solution space}\label{sec:SolutionsDeformation}

\initial{I}n section \ref{sec:StructureRBG} of the previous chapter we saw that an appealing feature of RBG theories with projective symmetry\footnote{Through this section we will only consider projectively invariant RBG theories. I might write only RBG theories to shorten the writing, but I will be referring to RBG theories with projective symmetry.} is that the independent affine connection turns out to be an auxiliary field that can be integrated out as the Levi-Civita connection of a metric tensor $q_{\mu\nu}$ that can differ from the spacetime metric $g_{\mu\nu}$ in a nontrivial way in presence of matter. This deviation is encoded in the deformation matrix, and will be different for each branch of (algebraic) solutions $\Omega(\hat T)$ of the metric field equations (see the discussion below \eqref{eq:OmegaMatrixMatrixForm}). As well, we have also seen that, through an involved field redefinition, RBG theories admit an Einstein frame representation with a nonlinearly modified matter Lagrangian when the gravitational field variables are written in terms of the metric $q_{\mu\nu}$ (which can always be done on-shell). In this Einstein frame it becomes apparent that the role of the connection is that of an auxiliary field which, when integrated out, provides new effective interactions among the degrees of freedom of the matter sector. At a perturbative level, these interactions can be used to constrain the theories (see chapter \ref{sec:ObservableTraces}) and, in some cases, the full matter Lagrangian after integrating out the connection can be solved, establishing an equivalence between a given RBG coupled to a matter sector with GR coupled to a nonlinearly modified version of the same matter sector, as we have shown explicitly in the previous section for EiBI coupled to Maxwell electrodynamics, which can be written as GR coupled to Born-Infeld electrodynamics by appropriate field redefinitions after the connection has been integrated out. In the Einstein frame representation, exact and numerical solutions can be found by standard methods \cite{Afonso:2019fzv,Afonso:2020giy,Orazi:2020mhb,Olmo:2020fnk,Shao:2020weq,Guerrero:2020azx,Maso-Ferrando:2021ngp}. The key aspects that allow to pass from the RBG frame to the Einstein frame can be traced back to the existence of the deformation matrix $\Omega^\mu{}_\nu$ which, as explained in section \ref{sec:StructureRBG} relates both metrics as\footnote{Though the following equations are already in the text, see \eqref{eq:RelationBothMetricsMatrxForm} and \eqref{eq:OmegaMatrixMatrixFormAniso}, I have rewritten them here because they will play a key aspect in this section, and I think that doing so will facilitate the reader if she wants to access them quickly.}
\beq\label{eq:RelationBothMetricsAniso}
\hat q=\hat g\, \hat\Omega \qquad \text{or in tensorial form} \qquad q_{\mu\nu}=g_{\mu\rho}\Omega^\rho{}_\nu,
\eeq
where the deformation matrix is given by 
\beq\label{eq:OmegaMatrixMatrixFormAniso}
 \hat \Omega=\sqrt{\det\lr{\frac{\partial \cF}{\partial \hat P}}}
\lr{\frac{\partial \cF}{\partial \hat P}}^{-1},
\eeq
and where the different on-shell expressions $\hat\Omega(\hat T)$ arise through the different solutions $\hat P(\hat T)$ of the metric field equations for RBG theories with projective symmetry \eqref{eq:PIRBGMetricFieldEquationsMatrixForm} seen as an algebraic equation for $\hat P$. Though a solution that is perturbatively close to GR at sufficiently low energies is guaranteed by analyticity (see section \ref{sec:GeneralRBGMetricFieldEqs}), the nonlinear terms\footnote{Note that, for the particular case of the metric-affine Einstein-Hilbert action there are no nonlinear terms, so that there is a unique solution given by a trivial deformation matrix $\hat\Omega=\bbI$.} of \eqref{eq:PIRBGMetricFieldEquationsMatrixForm} involving higher powers of $\hat P$   may allow for other solutions. The existence of these solutions that might deviate nonperturbatively from vacuum GR will depend on each particular RBG model. 

In this section, our aim is to show the existence of other solutions in generic RBG models. To do this, we will resort to a particular case where the matter sector and the Einstein frame metric $g^{\mu\nu}$ are isotropic. To that end, we will derive the necessary conditions that have to be satisfied by an RBG model for the existence of solutions $\hat\Omega(\hat T)$ that do not realise the symmetries of the stress-energy tensor and the Einstein frame metric. We can as well take these conditions as the sufficient conditions that have to be satisfied by a given RBG model so that, in presence of an isotropic matter sector, the nonlinearities of the equations that determine the deformation matrix as a function of $\hat T$ do not allow for other solutions apart from the one that is perturbatively close to the isotropic solution of vacuum GR at sufficiently low energies. In cosmological applications of RBG models, it is typically assumed that the deformation matrix has the same symmetries as the energy-momentum tensor and the spacetime metric, so that both metrics share the same symmetries. The existence of this solution is guaranteed by demanding that the nonlinear corrections amount to at most a cosmological constant in the low energy limit. As explained in section \ref{sec:StructureRBG}, in RBG theories with projective symmetry, the matter degrees of freedom evolve in the background given by the spacetime metric and gravitational waves can be associated to perturbations of the metric $q^{\mu\nu}$. Hence gravitational waves propagate in the background defined by the Einstein frame metric (see \cite{Jimenez:2015caa}) and, therefore, the possible existence of anisotropic deformation matrices for an isotropic cosmological fluid could introduce interesting effects in gravitational wave propagation that may be worth studying. This puts forward that, besides its relevance from a purely formal view of understanding the overall structure of the solution space of RBG theories with projective symmetry, these results could also be of physical interest in cosmological scenarios, and could be extended as well to astrophysical scenarios with spherical or axisymmetric symmetries.

To pursue our aim, it will be useful to reformulate the Lagrangian of a projective invariant RBG theory as follows. As explained in section \ref{sec:StructureRBG}, given that the Lagrangian of an RBG with projective symmetry must be a scalar built in terms of $g^{\mu\nu}$ and $ R_{(\mu\nu)}$, it can only depend on the matrix $P^\mu{}_\nu=g^{\mu\alpha} R_{(\alpha\nu)}$ or $\hat P$ in matrix notation. The Cayley-Hamilton theorem guarantees that we can express an $N\times N$ matrix in terms of its first $N-1$ powers and the identity matrix. This implies that we can express any scalar\footnote{Here, that the Lagrangian is a scalar function implies that it has to be a function of traces of products of $\hat P$.} function of $\hat P$ in terms of the traces $ X_n=\mathrm{Tr}(\hat P^n)$ of its first $N$ powers. In this section we will particularise to four spacetime dimensions so that we need only to consider $X_n$ with $n=1,...,4$. With this setup, we can write a generic Lagrangian for an RBG theory with projective symmetry\footnote{Through this section we will only consider RBG theories with projective symmetry even if it is not stated explicitly.  I might write only RBG theories to shorten the writing, but I will be referring to RBG theories with projective symmetry.} as a function of these four invariants. Through this section, it will be useful to redefine $\hat P$ as the matrix form of $\,{\mg^{-2}}g^{\mu\alpha} R_{(\alpha\nu)}$ so that now $\hat P$ and $X_n$ have vanishing mass dimension. We will also assume that all the higher-curvature corrections are controlled by the scale $\mg$, so that it will be useful to use the following dimensionless parametrisation for the RBG Lagrangian
\beq\label{eq:PIRBGLagrangianMatrixInvariants}
\, F\lrsq{X_1(\hat P),X_2(\hat P),X_3(\hat P),X_4(\hat P)}=\cF(\hat P)\,, 
\eeq
which are related to the original form of the Lagrangian \eqref{eq:GeneralRBGAction} as in \eqref{eq:RBGLagrangianMatrixForm}, namely
\beq
{\mg^2}\cF\big[{\mg^{-2}}g^{\mu\alpha} R_{(\alpha\nu)}]=\cL\big(g^{\mu\nu},R_{(\mu\nu)}\big),
\eeq 
where recall that $\cL$ has mass dimension 2 and an implicit dependence on ${\mg}$. Note that an explicit dependence on a UV energy scale has been extracted from the Lagrangian so that $F$ has zero mass dimension and we have defined the dimensionless invariants $X_n=\tilde X_n/\mq^{2n}$. Through the section we will consider that the matter action has no dependence on the connection, which is true for minimally coupled bosonic fields. Fermionic fields would only introduce a 4-fermion interaction that (presumably) would not alter our conclusions qualitatively, though the results might loose clarity if these interactions are included. We will also make extensive use of the metric field equations in matrix form given in \eqref{eq:PIRBGMetricFieldEquationsMatrixForm}, which in this parametrisation read\beq\label{metriceqsP}
\hat{P}\frac{\partial \cF}{\partial\hat{P}}-\frac12 \cF\bbI=({\mpl}{\mg})^{-2}\hat{T}.
\eeq
Note that the Einstein-Hilbert term in this parametrisation is ${\mg^{-2}}R$ so that any RBG Lagrangian which introduces corrections to GR at quadratic or higher orders in the curvature tensor will couple the graviton to the stress energy tensor with the inverse of the squared Planck mass and not with ${\mpl^{-2}}{\mg^{-2}}$. Nevertheless, the departures from metricity and the new interactions that arise in the matter sector are controlled by the effective scale ${\mq}=({\mpl}{\mq})^{1/2}$ and become nonperturative when the energies reach $\mq$, see chapter \ref{sec:ObservableTraces}.

\section{Anisotropic deformation in isotropic backgrounds}\label{sec:AnisoDefSolutions}

We will use the particular case when the matter sector is described by a perfect fluid with isotropic pressure to illustrate the possibility that the nonlinearities of the field equations admit other solutions for $\hat\Omega(\hat T)$ besides the one that is perturbatively close to vacuum GR at low energies. In most of the cases treated in the literature, the isotropy of the stress-energy tensor is assumed to be inherited by the deformation matrix which, since the existence of that solution is guaranteed by analitycity, is a consistent assumption. Our interest in this work is, however, to go beyond this assumption and explore whether solutions with a deformation matrix  that does not inherit the isotropy of the matter sector are possible. This would imply that RBG and Einstein frame metrics do not share the same symmetries either. The existence of such solutions is plausible due to the nonlinear nature of the equations (were they linear, the symmetries of the stress-energy tensor must always be inherited by the gravitational sector), in close analogy to the existence of Bianchi I solutions in a universe filled with an isotropic fluid. We will expand on this analogy in Section \ref{sec:AnisoEinstein}. 

Our ansatz for the stress-energy tensor and the matrix $\hat{P}$ will then be
\beq
T^\mu{}_\nu=
\left(
\begin{array}{cccc}
  -\rho&  0 \;& 0\; & 0\; \\
  0 \;&  p & 0 \;  & 0 \;\\
 0\; &  0\; & p   & 0\;\\
  0\;& 0\; & 0\; &p
\end{array}
\right)\quad\text{and}\quad P^\mu{}_\nu=
\left(
\begin{array}{cccc}
  P_0& 0 & 0 & 0\\
  0 &   P_1&  0 & 0\\
  0 & 0 &  P_2 & 0\\
  0 & 0 & 0 &P_3
\end{array}
\right),
\eeq
which leads to a diagonal deformation matrix
\beq
\Omega^\mu{}_\nu=
\left(
\begin{array}{cccc}
  \Omega_0& 0\; &0 \; &0\; \\
 0 \;&  \Omega_1 &0 \;  &0\;\\
 0\; &  0\; &\Omega_2   &0\;\\
 0\;& 0\; & 0\; &\Omega_3
\end{array}
\right)\,;
\eeq
and also to a simple relation between the $X_n$ and the eigenvalues of $\hat{P}$, given by
\beq
X_n=\sum_{i=0}^3 P_i^n.
\eeq 
In this scenario, the metric field equations as given in \eqref{metriceqsP} read 
\begin{align}
&P_0\frac{\partial \cF}{\partial P_0}=\frac12 \cF-\bar{\rho},\label{eq:P0}\\
&P_i\frac{\partial \cF}{\partial P_i}=\frac12 \cF+\bar{p}\quad {\textrm{for}}\quad i=1,2, 3;\label{eq:Pi}
\end{align}
where no summation over $i$ is taking place, and where we have normalised the density and pressure as $\bar{\rho}=\rho/({{\mpl^2}} {\mq^2})$ and $\bar{p}=p/({{\mpl^2}} {\mq^{2}})$. We can split the spatial equations (\ref{eq:Pi}) into the isotropic part given by the trace
\beq
\frac13\sum_{i=1}^3P_i\frac{\partial \cF}{\partial P_i}=\frac12 \cF+\bar{p}
\label{eq:trace}
\eeq
and the anisotropic part given by
\beq\label{eq:anisotropiccondition1}
P_i\frac{\partial \cF}{\partial P_i}-P_j\frac{\partial \cF}{\partial P_j}=0\quad {\textrm{for}}\quad i\neq j
\eeq
We can alternatively use the parametrization of the RBG Lagrangian in terms of the invariants $X_n$ given in \eqref{eq:PIRBGLagrangianMatrixInvariants} to rewrite the anisotropic part of the field equations \eqref{eq:anisotropiccondition1} as
\beq
\sum_{n=1}^4a_n(P_i^n-P_j^n)=0,\quad{\textrm{for}}\quad i\neq j
\label{eq:anisotropiccondition2}
\eeq
where $a_n=n\partial F/\partial X_n$. Out of these three conditions, only two of them are independent because the sum of the three equations identically vanishes. Moreover, since the equations are invariant under permutations of $\Omega_1$, $\Omega_2$ and $\Omega_3$, we can take the two independent conditions to be
\begin{align}\label{eq:anisotropiccondition3}
\begin{split}
&a_1(P_1-P_2)+a_2(P^2_1-P^2_2)+a_3(P^3_1-P^3_2)+a_4(P^4_1-P^4_2)=0\,,\\
&a_1(P_1-P_3)+a_2(P^2_1-P^2_3)+a_3(P^3_1-P^3_3)+a_4(P^4_1-P^4_3)=0
\end{split}
\end{align}
From these equations we can easily obtain a set of necessary conditions for the existence of  solutions with a nonisotropic deformation matrix. A remarkable result is that, since these equations do not depend on the matter content, it is only the precise form of RBG Lagrangian theory what will determine whether anisotropic solutions are possible or not. The way to proceed then is to solve (\ref{eq:anisotropiccondition3}) for two of the components of $\Omegah$ for the anisotropic branch of solutions (if any) and, then, use (\ref{eq:trace}) and (\ref{eq:P0}) to obtain the full solution with the components of the matrix $\hat{P}$ in terms of the $\bar{\rho}$ and $\bar{p}$. Obviously, the isotropic solution with $\Omega_1=\Omega_2=\Omega_3$ satisfies (\ref{eq:anisotropiccondition3}). However, given the nonlinearity of the conditions, it is possible to have multiple isotropic branches. It is guaranteed by construction that for one of these branches the nonlinearities will become irrelevant at low energies. The next nontrivial example is the case with axisymmetry, \ie two components are equal and different from the third. Without loss of generality we can assume $\Omega_1=\Omega_2\neq\Omega_3$, which implies that $P_1=P_2\neq P_3$.  In that case, the first of the two conditions in \eqref{eq:anisotropiccondition3} is trivially satisfied, but the second one still represents a constraint. In the general case eqs. \eqref{eq:P0}, \eqref{eq:trace} and \eqref{eq:anisotropiccondition3} will also be contraints that should be interpreted as necessary but not sufficient conditions that a particular theory of isotropic matter plus gravity has to fulfil in order to admit at least one anisotropic solution. Besides finding nontrivial anisotropic solutions from those equations, one needs to further corroborate that they can be physical. for instance, the resulting $\hat{\Omega}$ must be positive definite so that the metric of both frames have the same causal character. In the following we will illustrate these considerations with some explicit examples.

\subsection{Anisotropic deformations in vacuum}

Let us see whether there is any theory within the projectively invariant RBG class which admits an anisotropic deformation matrix in vacuum. The interest is twofold: 1) because if there is no such theory, all the anisotropic solutions that can be constructed in the presence of matter will not have a well behaved infrared behavior. 2) Because any theory within the RBG class that admits an anisotropic vacuum deformation, since it also admits an isotropic one by construction, will have a nontrivial vacuum structure that could potentially introduce vacuum instabilities. The metric field equations in vacuum are given by \eqref{eq:P0} and \eqref{eq:Pi} with $\bar{\rho}=\bar{p}=0$, which can be written as
\begin{align}
P_\mu\frac{\partial \cF}{\partial P_\mu}=\frac12 \cF,\label{eqvacuum}
\end{align}
where $\mu=0,1,2,3$ and no sumation over $\mu$ is understood here. In general, the above equation implies an on-shell relation of the form $P_0(P_1,P_2,P_3)$. For the particular cases of isotropic ($P_1=P_2=P_3$) and axisymmetric ($P_1\neq P_2=P_3$) deformations, this dependence is reduced to $P_0(P_1)$ and $P_0(P_1,P_2)$ respectively. By using the definition of the deformation matrix in matrix form \eqref{eq:OmegaMatrixMatrixFormAniso}, from \eqref{eqvacuum} we also arrive to another on-shell condition that must be satisfied by any vacuum anisotropic solution, that is
\begin{equation}
\frac{\Omega_\mu}{\Omega_\nu}=\frac{P_\mu}{P_\nu} \quad\forall\; \mu,\nu.
\end{equation}
Since we are demanding that all the eigenvalues of $\Omegah$ are positive, the above equation implies that the $P_\mu$'s must all have the same sign when the field equations of the corresponding theory are satisfied. Yet another condition imposed by the positivity of the $\Omega_\mu$'s and the dynamics of RBG is that the following relation
\beq
\frac{\cF}{P_\mu}>0 \quad\forall\;\mu
\eeq
must hold on-shell. This implies that the Lagrangian must also have the same sign as the $P_\mu$ when the field equations are satisfied. Thus, in principle, an RBG satisfying this conditions could have anisotropic vacuum solutions. Let us now turn to the analysis of some sub-classes of theories that are of particular interest.
%%%%%%%%%%%%%%%%%%%%%%%%%%%%
\subsection{No anisotropic deformations in EiBI and inspired theories}
%%%%%%%%%%%%%%%%%%%%%%%%%%%%
Let us now analyse the particular class of EiBI, which is one of the most extensively analysed metric-affine theories (see \cite{BeltranJimenez:2017doy} and also section \ref{sec:EiBIEMMapping}). The EiBI Lagrangian \eqref{eq:EiBILagrangian}, written in matrix form and in the parametrisation employed in this section, reads
\beq
\cF_\textrm{EiBI}=\det(\Id+\hat{P})^{1/2}-\lambda\,,
\eeq 
When particularised to this type of Lagrangian, the anisotropic part of the necessary conditions for the existence of solutions with anisotropic deformation matrix in an isotropic background \eqref{eq:anisotropiccondition1} yields\beq
\det\lr{\bbI+\hat P}^{1/2}\left(\frac{P_i}{1+P_i}-\frac{P_j}{1+P_j}\right)=0,\quad i\neq j.
\eeq
Since $\det(\Id+\hat{P})$ must be nonvanishing and $P_i>0$ in order to have a regular deformation matrix, the only solution to the above equation is $P_i=P_j$ and, therefore, the solution must be isotropic. This implies that no anisotropic solutions exist in presence of isotropic matter in EiBI. This result agrees and generalises the findings in the literature. For instance, Bianchi I solutions within the EiBI theory were studied in \cite{Harko:2014nya} and it was found that the deformation matrix was indeed isotropic for an isotropic fluid despite having a Bianchi I ansatz for $q_{\mu\nu}$ and $g_{\mu\nu}$. The spherically symmetric configurations of EiBI theory coupled to an anisotropic fluid have also been studied in \cite{Menchon:2017qed} with an isotropic deformation matrix. Again, when going to the isotropic case, the obtained solutions for the deformation matrix also become isotropic (in fact, they are proportional to the identity matrix, which is a consequence of having considered a cosmological constant-like fluid).

The result that no anisotropic solutions exist within EiBI gravity can be generalised in a straightforward manner to the functional extensions of the EiBI theory considered in \cite{Odintsov:2014yaa}, where the action is given by an arbitrary function $f$ of the scalar $\det(\Id+\hat{P})$. In that case, the above condition generalises to
\beq
f'\det(\Id+\hat{P})\left(\frac{P_i}{1+P_i}-\frac{P_j}{1+P_j}\right)=0,\quad i\neq j,
\eeq
which again, given that $f'\det(\Id+\hat{P})$ must be nonvanishing and $P_i>0$ to have a well behaved deformation matrix, implies that the only possible solution is the isotropic solution with $P_i=P_j$. Therefore, no anisotropic solutions exist in presence of isotropic matter in EiBI inspired theories either.
%%%%%%%%%%%%%%%%%%%%%%%%%%%%%
\subsection{$F(X_1,X_n)$ theories}\label{sec:FX1Xn}
%%%%%%%%%%%%%%%%%%%%%%%%%%%%%
General results can also be obtained for theories that have a Lagrangian defined in terms of $X_1$ and only one of the higher order scalars $X_n$ with $n=2,3$ or $4$. The presence of $X_1$ is imposed in order to guarantee the existence of one branch of solutions continuously connected with the EH Lagrangian at low curvatures. For these particular cases, the two independent conditions \eqref{eq:anisotropiccondition3} are
\begin{align}
\begin{split}
&a_1(P_1-P_2)+a_n(P^n_1-P^n_2)=0,\\
&a_1(P_1-P_3)+a_n(P^n_1-P^n_3)=0.
\label{eq:anisotropicconditionX1Xn}
\end{split}
\end{align}
For the axisymmetric case, we can choose $P_2=P_1$ so that the first equation is trivially satisfied, and we have a relation $P_3(P_1)$. For a completely anisotropic solution without axisymmetry, the equations (\ref{eq:anisotropicconditionX1Xn}) imply a relation $P_3(P_1,P_2)$ of the form
\beq
\frac{a_n}{a_1}=\frac{P_1^n-P_2^n}{P_1-P_2}=\frac{P_1^n-P_3^n}{P_1-P_3}.
\eeq
For $n=2$, this relation can be reduced to $P_1+P_2=P_1+P_3$ which in turn implies $P_2=P_3$ and, consequently, only axisymmetric solutions are allowed. For $n=3$ we instead obtain two branches of solutions, the axisymmetric one, and a second branch with $P_1+P_2+P_3=0$ so the completely anisotropic solutions for $n=3$ must have $\hat{P}$ with traceless spatial part. Finally, for $n=4$ we again have the axisymmetric branch and possibly another completely anisotropic branch defined by the relation
\beq
(P_1^2+P_2^2)(P_1+P_2)=(P_1^2+P_3^2)(P_1+P_3).
\eeq
As can be seen by writing the explicit solutions for $P_3$
\beq
P_3=-\frac{P_1+P_2\pm\sqrt{-((P_ 1+ P_2)^2+2P_1^2+2P_2^2)}}{2}\quad\text{or}\quad P_2=P_3,
\eeq
this equation has no real solutions other than $P_2=P_3$ which is also an axisymmetric solution. Thus, for $n=4$ there can be no completely anisotropic branches. 

Solving the space of potentially anisotropic solutions for the general case is very cumbersome so in the next section we will focus on the quadratic theory which encodes the lowest-order corrections to the Einstein-Hilbert term. This will be relevant for the general theories with solutions that are perturbatively close to those of the Einstein-Hilbert action at low energies, so that it will be possible to extract information for the general theories from our analysis of the quadratic one.
%%%%%%%%%%%%%%%%%%%%%%%%%%%%%%
\subsection{General quadratic theory}\label{sec:GenQuad}
%%%%%%%%%%%%%%%%%%%%%%%%%%%%%%
The Lagrangian for a general quadratic metric-affine RBG theory with projective symmetry, if parametrised in terms of the invariants $X_n$, reads 
\beq
 F=X_{1}+\alpha X_{1}^{2}+\beta X_{2}.
 \label{Fquadratic}
\eeq
 Note that the parameter that would have gone with $X_{1}$ is fixed to $1$ in order to recover the Einstein Hilbert action at low curvatures. Although this theory may seem to have 2 independent dimensionless parameters $\alpha$ and $\beta$, one of them can be absorbed into the mass scale $ {\mg^{2}}$ and there is only one free parameter (besides the new scale ${\mg}$). Thus (\ref{Fquadratic}) is the most general quadratic Lagrangian within the RBG family that reduces to the metric-affine Einstein Hilbert action in the low curvature limit and captures the perturbative effects of any nonlinear theory in that regime. Of course, there could be nonperturbative effects that are not properly captured by \eqref{Fquadratic}, although this would typically imply strong departures from GR in the low energy regime which could be observationally accessible. In order to obtain the dependence of the curvatures $P_{i}$  in terms of the energy content we make use of ($\ref{eq:P0}$) and ($\ref{eq:Pi}$), which particularised for the general quadratic action \eqref{Fquadratic} read
\begin{equation}\label{sysquadratic}
    \begin{split}
    &P_{0}-\Tr(\hat{P}_s)+\alpha\left[3P_0^2+2P_0\Tr(\hat{P}_s)-\Tr^2(\hat{P}_s)\right]+\beta\left[3P_0^2-\Tr(\hat{P}_s^2)\right]+2\rho=0,\\
    &2P_i-\Tr(\hat{P})+\alpha\left[4P_i\Tr(\hat{P})-\Tr^2(\hat{P})\right]+\beta\left[4P_i^2-Tr(\hat{P}^2)\right]-2p=0,
    \end{split}
\end{equation}
where $i=1,2,3$ and $\hat{P}_s$ is the spatial $3\times3$ sub-matrix of $\hat{P}$.  From here on, we will drop the bar in $\bar \rho$ and $\bar p$ to ease the notation, but all the $\rho$'s appearing in the text should be understood as normalised by $1/({\mg^2}\mpl^2)$. According to what has been discussed in section \ref{sec:FX1Xn},  a quadratic theory can only have isotropic or axisymmetric solutions, but not completely anisotropic solutions are allowed. Thus, in order to look for solutions to the above system of equations (\ref{sysquadratic}) we might first impose isotropy or axisymmetry. In the former case, with $P_1=P_2=P_3$ the above equations \eqref{sysquadratic} reduce to
\begin{align} \label{sysquadraticiso}
\begin{split}
&P_0+3P_2+3p-\rho=0,\\
&3\alpha\Big[2P_0 P_2-3P_2^2+P_0^2\Big]+3\beta(P_0^2-P_2^2)+(P_0-3P_2)+2\rho=0,
\end{split}
\end{align}
and in the axisymmetric case, we find
\begin{align} \label{sysquadraticaxi}
&P_0+2P_1+P_2+3p-\rho=0,\nonumber \\
&(P_1-P_2)\Big[1+2\alpha(P_0+2P_1+P_2)+2\beta(P_1+P_2)\Big]=0,\\
&\alpha\Big[2P_0 (2P_1+ P_2)-(2P_1+P_2)^2+3P_0^2\Big]+\beta(3P_0^2-2P_1^2-P_2)+(P_0-2P_1-P_2^2)+2\rho=0,\nonumber
\end{align}
where we have chosen $P_1=P_3\neq P_2$ (note that the physical solutions will not distinguish between this choice and $P_1=P_2\neq P_3$ or $P_1=P_3\neq P_1$). Due to the nonlinearities of the systems,both the isotropic and axisymmetric cases have two branches of solutions. Assuming a barotropic fluid with $p=\omega\rho$, the first isotropic branch (that we will call iso-I) is given by
\begin{equation}\label{isobranch1}
\begin{split}
&P_{0}(\rho)=\resizebox{0.8\hsize}{!}{$\frac{ (3 \omega -1) (6 \alpha +\beta ) \rho -3 \left(1-\sqrt {1-[4\alpha  (3 \omega -1)+ 2\beta (1+5\omega ) ] \rho +(1-3 \omega )^2 (2 \alpha +\beta )^2\rho ^2  }\right)}{8 \beta }$}\\
&P_{1}(\rho)=\resizebox{0.8\hsize}{!}{$\frac{1+(1-3 \omega) (2 \alpha +3 \beta )\rho-\sqrt {1-[4\alpha  (3 \omega -1)+ 2\beta (1+5\omega ) ] \rho +(1-3 \omega )^2 (2 \alpha +\beta )^2\rho ^2  }}{8 \beta }$}
\end{split}
\end{equation}
and the second isotropic branch (iso-II) is given by the functions
\begin{align}\label{isobranch2}
\begin{split}
&P_{0}(\rho)=\resizebox{0.8\hsize}{!}{$\frac{(3 \omega -1) (6 \alpha +\beta ) \rho -3 \left(1+\sqrt {1-[4\alpha  (3 \omega -1)+ 2\beta (1+5\omega ) ] \rho +(1-3 \omega )^2 (2 \alpha +\beta )^2\rho ^2  } \right)}{8 \beta}$}\\
&P_{1}(\rho)=\resizebox{0.8\hsize}{!}{$\frac{1+  (1-3 \omega) (2 \alpha +3 \beta ) \rho +\sqrt {1-[4\alpha  (3 \omega -1)+ 2\beta (1+5\omega ) ] \rho +(1-3 \omega )^2 (2 \alpha +\beta )^2\rho ^2  } }{8 \beta }$}
\end{split}
\end{align}
For the axisymmetric case, both branches have the same solution for $P_0(\rho)$, which is given by
\begin{align}
&P_{0}(\rho)=\frac{-2 \rho ^2 (1-3 \omega )^2 (\alpha +\beta ) (2 \alpha +\beta )+2 \rho  (\alpha  (6 \omega -2)+\beta  (5 \omega -1))-1}{4 \beta  \rho  (3 \omega -1) (2 \alpha +\beta )-4 \beta }
\end{align}
and the two branches differ in their solutions for $P_1(\rho)$ and $P_2(\rho)$. The first axisymmtric branch (axi-I) is described by
\begin{align}\label{aniso1}
\begin{split}
&P_{1}(\rho)=\frac{-2 \rho ^2 (1-3 \omega )^2 (\alpha +\beta ) (2 \alpha +\beta )+\rho  (-4 (\alpha +\beta )+12 \alpha  \omega +8 \beta  \omega )-1}{4 \beta  \rho  (3 \omega -1) (2 \alpha +\beta )-4 \beta}\\
&P_2(\rho)=\frac{2 \rho ^2 (1-3 \omega )^2 (2 \alpha +\beta ) (3 \alpha +\beta )+2 \rho  (\alpha  (6-18 \omega )+\beta  (3-7 \omega ))+3}{4 \beta  \rho  (3 \omega -1) (2 \alpha +\beta )-4 \beta}
\end{split}
\end{align}
and the second axisymmetric branch (axi-II) is described by the functions
\begin{align}\label{aniso2}
\begin{split}
&P_{1}(\rho)=\resizebox{0.8\hsize}{!}{$\frac{6 \rho ^2 (1-3 \omega )^2 (\alpha +\beta ) (2 \alpha +\beta )^2-4 \rho  (2 \alpha +\beta ) (\alpha  (9 \omega -3)+\beta  (12 \omega -5))+6 \alpha +11 \beta}{12 \beta  \rho  (3 \omega -1) (2 \alpha +\beta )^2-12 \beta  (2 \alpha +\beta )}$}\\
&P_2(\rho)=\resizebox{0.8\hsize}{!}{$\frac{2 \rho  \left(6 \alpha ^2 (3 \omega -1)+\alpha  \beta  (33 \omega -17)+\beta ^2 (15 \omega -7)\right)-6 \rho ^2 (1-3 \omega )^2 (\alpha +\beta )^2 (2 \alpha +\beta )-3 \alpha -5 \beta}{12 \beta  \rho  (3 \omega -1) (\alpha +\beta ) (2 \alpha +\beta )-12 \beta  (\alpha +\beta )}$}.
\end{split}
\end{align}
As far as the deformation matrix $\hat{\Omega}$ is concerned, it can be written in terms of the $P$'s by means of \eqref{eq:OmegaMatrixMatrixFormAniso}. For the general quadratic Lagrangian given by (\ref{Fquadratic}) we find
\begin{equation}\label{Omegaquadratic}
    \Omega_{\mu}=\frac{\left[\prod_{\nu=0}^{3}\left(1+2\beta P_{\nu}+2\alpha \mathrm{Tr}(\hat P)\right)\right]^{1/2}}{1+2\beta P_{\mu}+2\alpha \mathrm{Tr}(\hat P)}.
\end{equation} 
Let us analyse the behaviour of these solutions for radiation and matter fluids. The first thing to notice here is that while the eigenvalues of $\hat P$, and therefore of $\hat \Omega$, depend on both parameters $\alpha$ and $\beta$ for a matter fluid ($\omega=0$), they do not depend on $\alpha$ for a radiation fluid ($\omega=1/3$) except for the axi-II branch, thus $\beta$ is the only relevant parameter that controls the behaviour of isotropic radiation fluids. Then, while for a radiation fluid $\beta\mapsto-\beta$ is equivalent to $\rho\mapsto-\rho$, for a matter fluid we find an equivalence between $(\alpha,\beta)\mapsto(-\alpha,-\beta)$ and $\rho\mapsto-\rho$. Thus, qualitatively, we have one kind of behaviour for radiation fluids, and two different behaviours for matter fluids, depending on the sign of $\alpha\beta$.

\subsubsection{Isotropic solutions in the quadratic theory}
\begin{figure}[h]
    \begin{minipage}{\textwidth}
        \centering
        \includegraphics[width=\textwidth]{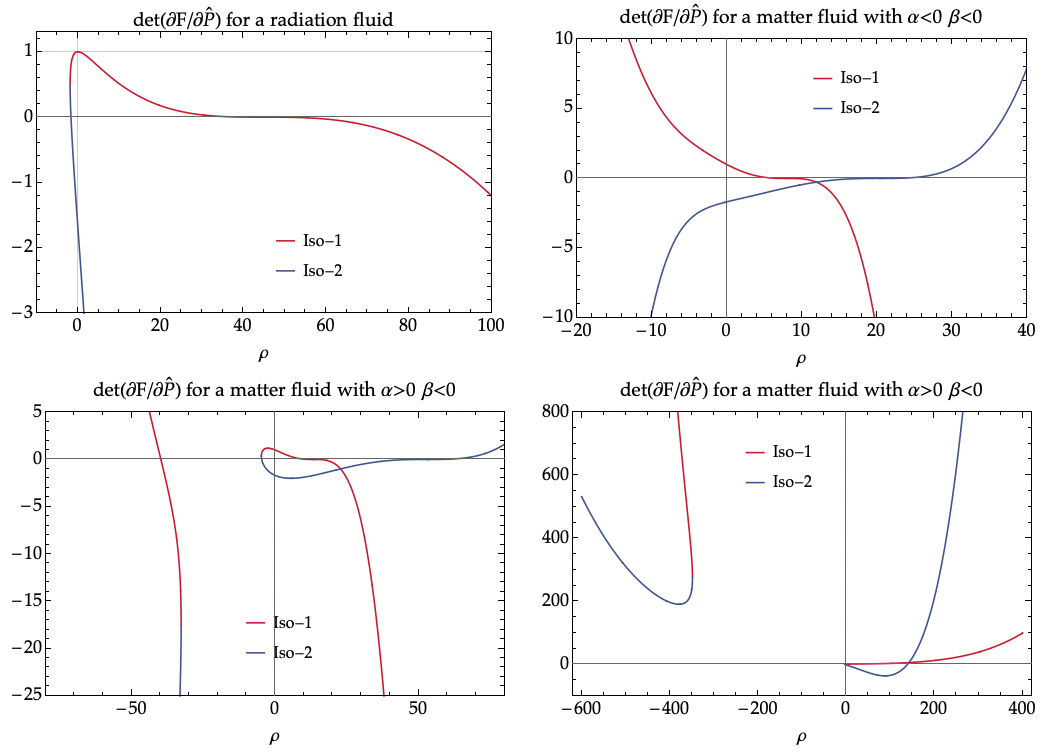} % second figure itself
    \end{minipage}\hfill
     \caption{The determinant of $\partial F/\partial\hat P$ is plotted for both isotropic branches and  $\beta=-0.1$. The plot above in the right is plotted for $\alpha=-0.01$, and the two below for  $\alpha=0.01$ and $\alpha=0.0345$ (left and right respectively).  It can be seen how $\det(F_{\hat P})=1$ in vacuum for iso-I in all the cases, but that is never the case for iso-II. } 
    \label{figdetFPiso}
\end{figure}

Isotropic solutions (figure \ref{figdetFPiso}) have already been studied in \cite{Barragan:2010qb}, where asymptotically Minkowski solutions and bouncing solutions were found. Let me review the behaviour of the deformation matrix for these solutions as obtained from our analysis. Given that $\Omegah$ is proportional to the square root of $\det(\cF_{\hat P})$ (here $\cF_{\hat P}=\partial\cF/\partial\hat P$), we must begin by studying the sign of this determinant for the different solutions (we will assume $\beta<0$). The qualitatively distinct cases for $\det(\cF_{\hat P})$ in isotropic solutions are plotted in figure \ref{figdetFPiso}. 

For a radiation fluid we have that $\det(F_{\hat P})$ is positive in the interval $\rho\in(\frac{3}{16\beta},-\frac{9}{2\beta})$ and negative for $\rho>-\frac{9}{2\beta}$ in the iso-I branch, and it is positive in the interval $\rho=(\frac{3}{16\beta},\frac{1}{6\beta})$ and negative for $\rho>\frac{1}{6\beta}$ in the iso-II. At $\rho=\frac{3}{16\beta}$ both branches give the same value for $\det(F_{\hat P})$, and it becomes complex (in both branches) for $\rho<\frac{3}{16\beta}$. Thus the two branches come from one single solution in the complex plane. 

The analysis becomes more involved in the general case for the matter dominated case. We find two different qualitative behaviours that depend on the relative sign between $\alpha$ and $\beta$. We see that each both branches have a zero in $\det(F_{\hat P})$ at positive values of $\rho$ for the case $\alpha\beta>0$. The zeros are given by 
\beq
\rho^I_0=\frac{-\sqrt{3 \alpha  \beta +\beta ^2}-6 \alpha -2 \beta }{12 \alpha ^2+7 \alpha  \beta +\beta ^2} \qquad\text{and}\qquad \rho^{II}_0=\frac{\sqrt{3 \alpha  \beta +\beta ^2}-6 \alpha -2 \beta }{12 \alpha ^2+7 \alpha  \beta +\beta ^2}
\eeq
 in the iso-I and iso-II branches respectively. In this cases both branches have $\det(F_{\hat P})\in\mathbb{R}$ for all values of $\rho$, being monotonically decreasing in the iso-I branch, and  monotonically increasing in the iso-II branch. Only the iso-I branch satisfies $\det(F_{\hat P})=1$ in vacuum, thus recovering GR. The case with opposite signs of $\alpha$ and $\beta$ is more involved due to the fact that there are more possible values of $\rho$ at which $\det(F_{\hat P})$ has zeroes or poles, as well as intervals in which it becomes complex. These depend, in general, on the particular values of the parameters $\alpha$ and $\beta$. In figure \ref{figdetFPiso} we plotted two of the possible cases. Note that in these cases, the richer structures of zeros and plots of $\det(F_{\hat P})$ gives rise to disconnected (in a continuity sense) subbranches within the two isotropic branches. Each of the sub-branches of one of the branches always connects smoothly with one of the sub-branches of the other branch, thus implying again that each of the branches comes from a unique solution in the complex plane.

\begin{figure}[h]
		\centering
		\begin{minipage}{\textwidth}
        \centering
        \includegraphics[width=\textwidth]{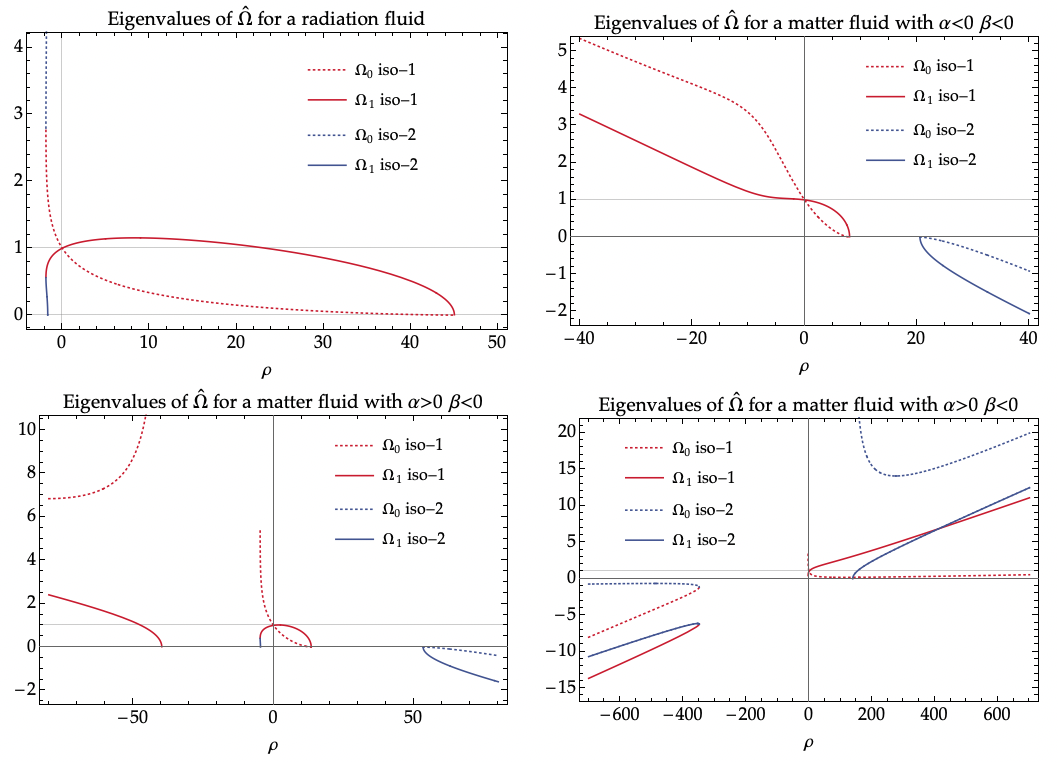} % first figure itself
    \end{minipage}\hfill
       \caption{The eigenvalues of the deformation matrix are plotted for both isotropic branches and  $\beta=-0.1$. The plot above in the right is plotted for $\alpha=-0.01$, and the two below for  $\alpha=0.01$ and $\alpha=0.0345$ (left and right respectively).  It can be seen how the deformation matrix reduces to the identity in vacuum for iso-I in all the cases, but that is never the case for iso-II. } 
    \label{fig:Omegaiso}
\end{figure}

A feature worth to note is that, for isotropic branches, the value of $\det(F_{\hat P})$ in vacuum is independent of $\alpha$ and $\beta$, and it evaluates to $1$ for iso-I and to $-27/16$ for iso-II. Given that the deformation matrix is proportional to $\sqrt{\det(F_{\hat P})}$, this implies that the iso-II branch does not have a well defined Einstein frame in vacuum. Regarding the properties of the deformation matrix, a remarkable feature for the isotropic solutions is that the value of the deformation matrix in vacuum does not depend on the values of the parameters $\alpha$ and $\beta$, and it is the identity for iso-I, whereas for iso-II we find $\Omegah_{\rho\to0}=i\sqrt{3}/2\; {\textrm{diag}}(-3,1,1,1)$. This implies that while for iso-I the nonlinearities fade out smoothly in the infrared, this is not the case for iso-II. The consequence is that the iso-II branch does not have a well defined Einstein frame in vacuum, since there are no real solutions for the deformation matrix in this case. These properties can be verified in figure \ref{fig:Omegaiso}, where we plot the eigenvalues of the deformation matrix for the different cases. From the plots we can also see how, except for the radiation solutions, matter solutions with $\alpha\beta>0$ and one of the subcases of matter solutions with $\alpha\beta<0$ (corresponding to $3\alpha+\beta<0$), the deformation matrix becomes singular at some maximum density, thus jeopardising the construction of the Einstein frame at higher densities. Physically, this is associated to an actual upper bound for the energy density allowed in these branches of the theory, a property with the potential to regularise both black hole and cosmological solutions and, consequently, the avoidance of singularities by generating a wormhole throat or a bounce when the energy densities reach this critical value \cite{Koivisto:2005yc,Barragan:2009sq,Koivisto:2010jj,Barragan:2010qb,Scargill:2012kg,Olmo:2012nx,Olmo:2015dba,Olmo:2015bya,Olmo:2016fuc,Olmo:2016tra,Olmo:2017fbc,Menchon:2017qed}. It is important to stress, however, that these solutions can also present other pathologies (instabilities, violations of energy conditions, superluminalities, etc.). For the $3\alpha+\beta > 0$ subcase of the $\alpha\beta>0$ solutions\footnote{The sign of $3\alpha+\beta$ is related to the structures of zeroes of $\det(F_{\hat P})$. } the deformation matrix does not become critical at any positive value of $\rho$.

\subsubsection{Axisymmetric solutions in the quadratic theory}

\begin{figure}[h]
		\begin{minipage}{\textwidth}
        \centering
        \includegraphics[scale=0.42]{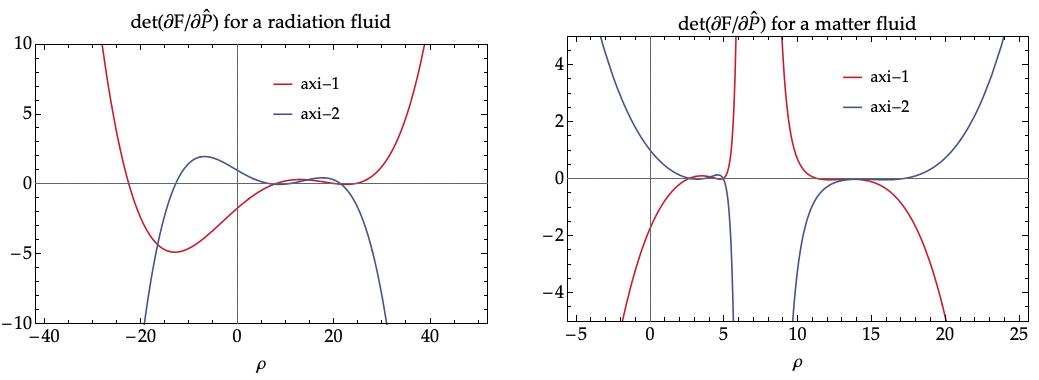} % first figure itself
    \end{minipage}\hfill
          \caption{Plots of $\det(F_{\hat P})$ for axisymmetric solutions, both with values $\beta=-0.1$ and $\alpha$ is chosen so that $\det(F_{\hat P})_{axi-II}=1$ in vacuum for that value of $\beta$ $(\alpha\approx-0.0213)$. The left plot is for a radiation fluid while the right one is for a matter fluid. } 
    \label{fig:detaniso}
\end{figure}

Let us now turn to the analysis of the axisymmetric solutions, focusing on whether there is any viable mechanism of isotropisation at low densities for any of the axisymmetric branches of the general quadratic theory. Axisymmetric branches are characterised by $P_{i}=P_k\neq P_j$. We will assume $P_1=P_3\neq P_2$ without loss of generality through this section. As for the isotropic case, there are two branches of anisotropic solutions, namely axi-I  and axi-II, described by \eqref{aniso1} and \eqref{aniso2} respectively. When coupled to a radiation fluid, the axi-I branch does not depend on the values of $\alpha$. Concerning the determinant of $F_{\hat P}$ in vacuum, it is independent of the model parameters for axi-I, and takes the same value as in iso-II (namely $-27/16$), suggesting that iso-II might be an isotropic limit of axi-I. However for axi-II, it does depend on the values of $\alpha$ and $\beta$ as 
\begin{equation}
\lim_{\rho\to0}\det(F_{\hat P})_{axi-II}=\frac{(3 \alpha +5 \beta )^2 (6 \alpha +11 \beta ) \left(6 \alpha ^2+13 \alpha  \beta +9 \beta ^2\right)}{1296 (\alpha +\beta )^3 (2 \alpha +\beta )^2}
\end{equation}
Thus, the parameters could in principle be tuned so that $\det(F_{\hat P})_{axi-II}=1$ in vacuum. Generally, $\det(F_{\hat P})$ has several roots, the number depending on the relations between $\alpha$ and $\beta$ except for the radiation case in axi-I, where it vanishes when $\det(F_{\hat P})\propto(4 \beta  \rho -9) (4 \beta  \rho +3) (4 \beta  \rho +9)^2$. In figure  \ref{fig:detaniso} we show plots of $\det(F_{\hat P})$ for both branches in the matter and radiation dominated cases.

 As for the properties of the deformation matrix in vacuum, it is complex for axi-I, taking the value $\Omegah_{axi-I}\overset{\rho=0}{=}i\sqrt{3}/2\, {\textrm{diag}}(1,1,-3,1)$,  which is different from that of iso-II, hence implying that one branch cannot be the isotropisation of the other, as neither can be axi-II due to the dependence on $\alpha$ and $\beta$ of $\Omegah$ in vacuum. This suggests that axisymmetric and isotropic branches are in general nonperturbatively different even at low densities in the general quadratic theory, and no isotropisation mechanism takes place in general. However, although it is not possible to find particular combinations of $\alpha$ and $\beta$ such that the deformation matrix becomes the identity in vacuum, we can indeed find particular combinations such that it isotropises in vacuum. Nonetheless, for axi-II, some of its eigenvalues are always negative in vacuum, thus jeopardising the hyperbolic nature of the corresponding field equations.  
 \begin{figure}[h]
		\centering
   \begin{minipage}{\textwidth}
        \centering
        \includegraphics[width=\textwidth]{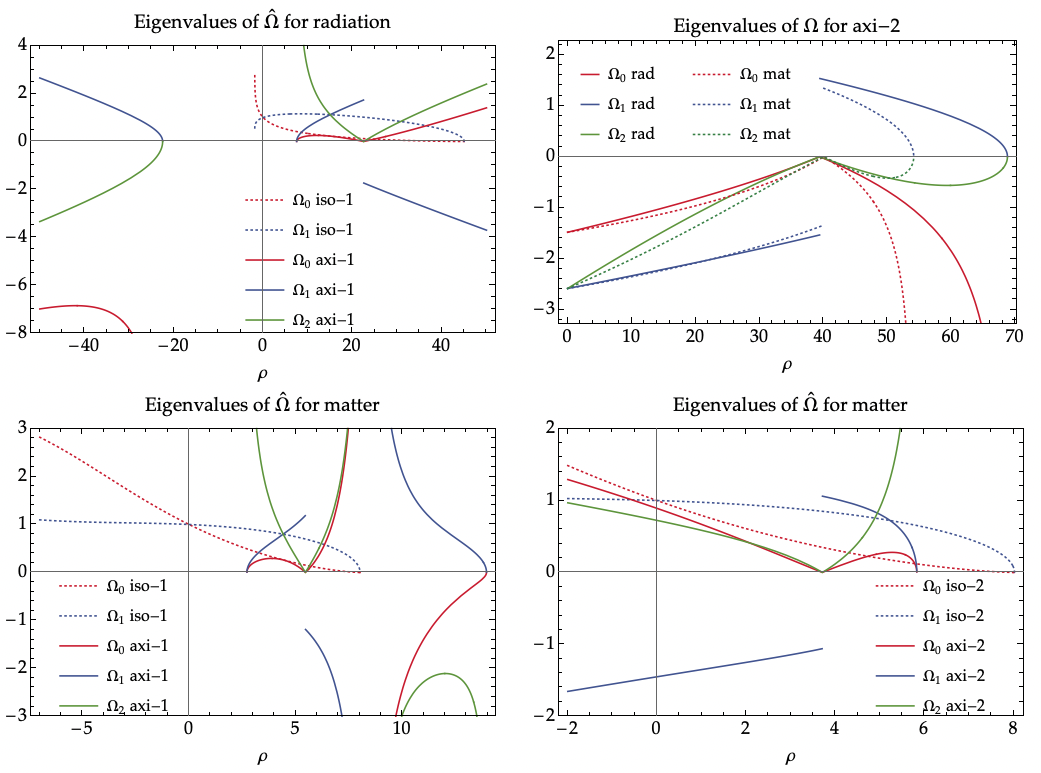}
   \end{minipage}
        \caption{Plots of the eigenvalues of $\Omegah$. The top-left graphic is plotted for the values of $\beta=-0.1$ and the value of $\alpha$ such that $\det(F_{\hat P})_{axi-II}=1$ in vacuum ($\alpha\approx-0.0213$), the top-right is plotted for $\alpha=-0.01$ and the value of $\beta$ such that $\Omegah$ isotropizes in vacuum ($\beta\approx 0.1297$), and both on the bottom are plotted for $\alpha=-0.01$ and $\beta=-0.1$ respectively. The axi-I-branch always isotropizes to iso-I at some nonzero density for both fluids and in a nonsmooth way for the spatial eigenvalues, but the axi-II-branch isotropizes but not to the iso-I (neither 2) except for a particular value of the parameters. In this case, the spatial eigenvalue does not isotropize at the same value of $\rho$ as the temporal one.}\label{fig:Omegaaxi} 
\end{figure}
Apart from not having a well-defined vacuum, the deformation matrix for axisymmetric solutions is, in general, rather involved, as can be seen with the examples plotted in figure \ref{fig:Omegaaxi}. There is always a point for which the axi-I isotropizes and then become anisotropic again as the density grows. At this isotropization point, the eigenvalues of $\Omegah$ of axi-I coincide with those of iso-I both for matter and radiation. Nevertheless, the derivatives of the eigenvalues are never the same for isotropic and axisymmetric solutions at that point. The hope that an anisotropic solution could then isotropize in a smooth (and thus predictable) way is in vain. For axi-II, although it isotropizes, it does not meet the iso-I (which recall that it is the only isotropic branch giving the correct low-density limit).

The above analysis shows that, for the general quadratic theory, even though some branches of solutions correspond to anisotropic deformations, they are generically pathological at low densities where the branches do not exist. Obviously, these branches are disconnected from the solution that continuously connects with GR at low densities. Despite deriving this result only for quadratic theories, they have far reaching consequences, strongly suggesting that branches of any theory that are perturbatively close to GR at low densities do not admit smooth anisotropic deformations. Thus, the anisotropic branches of more general (nonlinear) theories with a smooth behaviour at low densities, if they exist, must be nonperturbative, \ie they must strongly rely on their nonlinear nature. 

\section{Anisotropic deformation matrix in physical scenarios}\label{sec:PhysicalScenarios}

Having understood which are the necessary conditions for a given RBG theory to have solutions with anisotropic deformation matrix, we can now analyse the consequences in scenarios with physical interest, such as cosmological or black hole spacetimes.

\subsection{Cosmological scenarios}
\begin{figure}[h]
		\centering
   \begin{minipage}{\textwidth}
        \centering
        \includegraphics[scale=0.3]{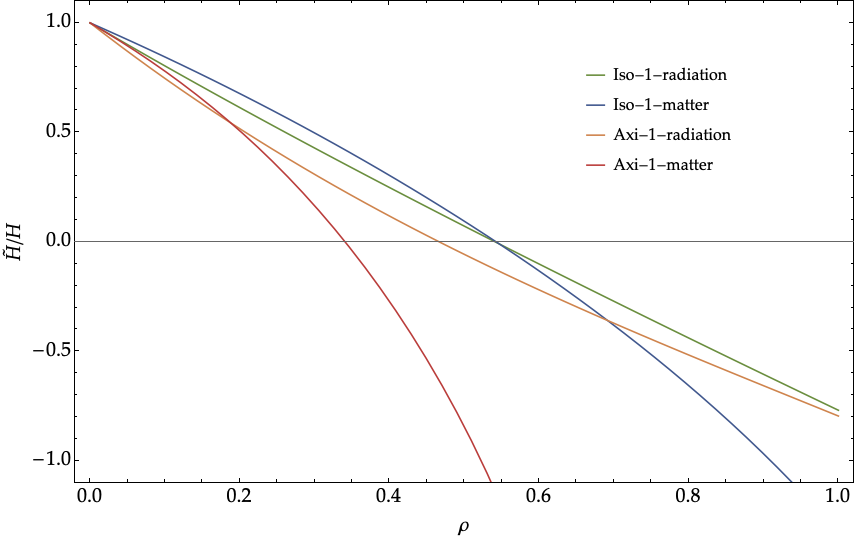}   \end{minipage}
        \caption{Here we plot the ratio between the Hubble factor of the RBG frame ({\it i.e.} that associated to $g_{\mu\nu}$) and the averaged Hubble factor of the Einstein frame ({\it i.e.} that associated to $q_{\mu\nu}$) for the general quadratic theory given by \eqref{Fquadratic}. $\rho$ is normalised by $1/M^2\mpl^2$ and we have chosen $\alpha=-0.2$ and $\beta=-0.1$. We can see how there is a density above which a $g_{\mu\nu}$ expanding phase corresponds to a $q_{\mu\nu}$ contracting phase and viceversa for both isotropic and axisymmetric branches.}\label{fig:quotientHs}
\end{figure}

The results obtained in the previous section apply to general spacetimes filled with a perfect fluid. We will now focus on a cosmological context where the fluid is also homogeneous, \ie which have a symmetry under spatial translations. Our interest here is to study a scenario where the spacetime metric is isotropic but the $q_{\mu\nu}$ metric is not, so that matter fields do indeed see an isotropic universe but gravitational waves propagate in a nonanisotropic background.\footnote{Recall that minimally coupled matter fields propagate in the background of the RBG frame metric in RBG theories, whereas gravitational waves do so according to the background of $q_{\mu\nu}$ \cite{Jimenez:2015caa}.} The spacetime metric will thus have an FLRW form
\beq
\dif s_g^2=-N^2(t)\dif t^2+a^2(t)\dif\vec{x}^2
\eeq
where we have assumed vanishing curvature of the spatial sections. Since we are exploring solutions where the deformation matrix is not isotropic, the metric $q_{\mu\nu}$ will be of the Bianchi I form
\beq
\dif s_q^2=-  N_q^2(t)\dif t^2+\sum_{i=1}^3a_i^2(t)(\dif x^i)^2.
\label{eq:metricqBianchi}
\eeq
We can define the isotropic scale factor $\tilde a=\left[a_1a_2a_3\right]^{1/3}$ and encode the anisotropic expansion in $\gamma_{ij}(t)=e^{2\beta_{i}(t)}\delta_{ij}$, with $\beta_i=\log{(a_i/\tilde a)}$ (no summation over $i$ in the definition of $\gamma_{ij}$ is understood). The functions $\beta_i$ describing the anisotropic expansion are subject to the constraint
\beq\label{betaconstraint}
    \sum_{i=1}^{3}\beta_{i}=0.
\eeq
We can now use the relations between $a_i$ and $a$ to define the function $\cA\equiv\tilde a/a=\big(\Omega_1\Omega_2\Omega_3\big)^{1/6}$
that relates the isotropic scale factor of the $q$-metric and the scale factor of $g_{\mu\nu}$. Using this definition, we can write $\beta_i$ and $\tilde a_i$ in the form
\beq
\beta_i=\frac12\log\frac{\Omega_i}{\cA^2}\qquad\text{and}\qquad\tilde{a}_{i}=\frac{\sqrt{\Omega_{i}}}{\mathcal{A}}\hspace{1mm}\tilde{a}.
\eeq
Furthermore, In Bianchi I, one can define 3 Hubble rates and an averaged one as  $\tilde{H}_{i}=\dot\tilde a_i/a_i$ and $\tilde{H}=\dot\tilde a/\tilde a$ respectively, which by using the continuity equation can be written as
\begin{equation}\label{tildeH}
    \tilde{H}=H\left[1-3(\rho+p)\left(\partial_{\rho}\log \mathcal{A}+c_{s}^{2}\partial_{p}\log \mathcal{A}\right)\right].
\end{equation}
This shows how the sign of $H$ and $\tilde H$ can be the opposite, so that when the RBG metric $g_{\mu\nu}$ is in an expanding phase, the Einstein frame metric $q_{\mu\nu}$ can be in a stationary or contracting phase (see figure \ref{fig:quotientHs}). By performing some calculations, we find the following equation for $H^{2}$
\begin{equation}\label{Hnoniso}
    \frac{3H^{2}}{N^{2} M^2}=\frac{\frac12\left(\sum_{i}\frac{\Omega_{0}}{\Omega_{i}}P_{i}-P_{0}\right)}{\left[1-3(\rho+p)\left(\partial_{\rho}\log \mathcal{A}+c_{s}^{2}\partial_{p}\log \mathcal{A}\right)\right]^{2}
    -\frac{1}{6}\sum_{i=1}^{3}\left[(\partial_{\rho}\beta_{i}+c_{s}^{2}\partial_{p}\beta_{i})(-3(\rho+p))\right]^{2}},
\end{equation}
where the right hand side can be written as a function of $\rho$ and $p$ by solving the field equations (\ref{eq:P0}) and (\ref{eq:Pi}). We see that the nonlinearities that permit the existence of the anisotropic solutions also complicate the structure of the corresponding Friedman equation. \\

%%%%%%%%%%%%%%%%%%%%%%%%
\subsection{Static spherically symmetric geometries}
%%%%%%%%%%%%%%%%%%%%%%%%
Another relevant physical scenario is that of spherically symmetric solutions. We can then study what kind of metric $q_{\mu\nu}$ we can get from an arbitrary static spherically symmetric spacetime metric $g_{\mu\nu}$. A general static and spherically symmetric metric can be written as (see {\it e.g.} \cite{Padmanabhan:2010zzb}), 
\begin{equation}\label{sphericallysymmetric}
ds_g^2=-C(r)dt^2+B^{-1}(r)dr^2+r^2\left(d\theta^2+\sin^2(\theta)d\phi^2\right),
\end{equation}
where $r$ measures the area of the $2-$spheres. Since $\hat\Omega$ can be written in vacuum as an analytic function of $\hat q$ or $\hat g$ and the matter fields, we can assume an arbitrary but diagonalised $\hat\Omega=diag(\Omega_t,\Omega_r,\Omega_\theta,\Omega_\phi)$. Using \eqref{eq:RelationBothMetricsAniso} and \eqref{sphericallysymmetric} we can then write 
\begin{equation}\label{sphericallysymmetric}
ds_q^2=-\Omega_t \bar{C}dt^2+\Omega_r\bar B^{-1}dr^2+\tilde{r}^2\left(d\theta^2+\frac{\Omega_\phi}{\Omega_\theta}\sin^2(\theta)d\phi^2\right),
\end{equation}
where 
\beq
\bar{C}= C\lr{\frac{\tilde r}{\Omega_\theta^{1/2}}},\quad \bar B = B\lr{\frac{\tilde r}{\Omega_\theta^{1/2}}}\quad\text{ and } \quad \tilde{r}^2=\Omega_{\theta}r^2.
\eeq
This is, in general, not spherically symmetric, unless $\Omega_\theta=\Omega_\phi$ and all the eigenvalues $\Omega_\mu$ depend only on $r$, which means that $r$ can be written in terms of $\tilde r$ only. In that case we can write $r(\tilde r)$ and, without assuming $\Omega_\theta=\Omega_\phi$, we can write
\begin{equation}\label{sphericallysymmetric}
ds_q^2=-\tilde{C}(\tilde r)dt^2+\tilde B^{-1}(\tilde r)d\tilde r^2+\tilde{r}^2\left(d\theta^2+\frac{\Omega_\phi\lrsq{r(\tilde r)}}{\Omega_\theta\lrsq{r(\tilde r)}}\sin^2(\theta)d\phi^2\right),
\end{equation}
where 
\beq
\resizebox{0.9\hsize}{!}{$\tilde{C}(\tilde r)=\Omega_t\lrsq{r(\tilde r)}C\lrsq{\frac{\tilde r}{\Omega_\theta^{1/2}\lrsq{r(\tilde r)}}},\quad \tilde B^{-1}(\tilde r) =\Big(1-\frac{d \ln \lr{\Omega_\theta\lrsq{r(\tilde r)}}}{d\ln\tilde r}\Big) \frac{\Omega_r\lrsq{r(\tilde r)}}{\Omega_\theta\lrsq{r(\tilde r)}} B^{-1}\lrsq{\frac{\tilde r}{\Omega_\theta^{1/2}\lrsq{r(\tilde r)}}}.$}
\eeq
 In this case, the coordinate $\tilde r$ also measures the area of the 2-spheres as given by the Einstein frame metric, and $q_{\mu\nu}$ will have a spherically symmetric form provided that $\Omega_\theta=\Omega_\phi$. If this condition is not met, the angular coordinates $\phi$ is periodic in $(\Omega_\theta/\Omega_\phi)^{-1/2}2\pi$, and thus $q_{\mu\nu}$ will describe a conical singularity due to a deficit in angle proportional to $1-(\Omega_\theta/\Omega_\phi)^{-1/2}$, thus spoiling the symmetry. On the other hand, if the condition is met and we have spherical symmetry in the Einstein frame, we can ask ourselves wether the presence (or absence) of horizons is modified in both frames. Usually, a divergence of the ${rr}$ of the metric signals the presence of event horizons. In this case, we see that a divergence in $g_{rr}$ at $r_h$ is also translated as a divergence of the $q_{rr}$ component due to analyticity of the deformation matrix, but now at $\tilde r_h=r_h\Omega^{1/2}(r_h)$ (note that the prefactor of $\tilde B$ does not vanish). Note as well that the $tt$ and $rr$ components of the Einstein frame metric will not generally be inverse of each other in the case that those of the RBG frame metric are (unless $\Omega_{\theta}$ is constant and $\Omega_r=\Omega_\theta\Omega_t^{-1}$). Thus, because Birkoff's theorem applies to the Einstein frame, in vacuum, no deformation matrix that preserves spherical symmetry can occur except if it satisfies $\Omega_r=\Omega_\theta\Omega_t^{-1}$ and $\Omega_\theta=\Omega_\phi$ are constants.
%%%%%%%%%%%%%%%%%%%%%%%%
\section{Anisotropy in the Einstein frame}\label{sec:AnisoEinstein}
%%%%%%%%%%%%%%%%%%%%%%%%
After exploring the possibility of having an anisotropic deformation for an isotropic matter source, it is illuminating to look at the problem from the Einstein frame perspective directly, where as explained in section \ref{sec:EinsteinFrame} the field equations for RBG theories can be recast into
\beq
G^\mu{}_\nu(q)={\mpl^{-2}}\tilde{T}^\mu{}_\nu.
\eeq
From this perspective, it is hard to evade the question of how to square the obtained anisotropic deformations with the (cosmological) no-hair theorems of GR \cite{Wald:1983ky}. This becomes even more pressing in view of the form of the source of the Einstein equations for $q_{\mu\nu}$, given in \eqref{eq:RelationStressEnergyTensorsFramesRBG}), which is isotropic provided both $T_{\mu\nu}$ and $g_{\mu\nu}$ are. Then, how do we reconcile the general result that the shear decays with the persistent anisotropic solutions obtained in the precedent sections? The resolution to this dichotomy again comes from the nonlinearity of the Einstein equations that allows to have anisotropic solutions even if the source is isotropic. The no-hair theorems for cosmological solutions, for instance, states that the anisotropic shear typically decays during the expansion. In our case, we have obtained that it is possible to have an anisotropic deformation, which is equivalent to having a Bianchi I metric for $q_{\mu\nu}$ even if $g_{\mu\nu}$ is of the FLRW type. That the anisotropy can be maintained can be understood from the fact that an expanding solution for the matter fields requires that the metric $g_{\mu\nu}$ describes a growing scale factor, but the evolution for the metric $q_{\mu\nu}$, besides being anisotropic, does not need to correspond to an expanding phase, as can be seen in figure \ref{fig:quotientHs}. For instance, if this anisotropic evolution describes a contracting phase, the shear corresponding to $q_{\mu\nu}$ can actually grow substantially while the metric $g_{\mu\nu}$ describes an isotropic expanding phase. On the other hand, even if the evolution also corresponds to an expanding phase, the effective expansion of the metric $q^{\mu\nu}$ can be slower than the one experienced by matter fields, namely that of $g_{\mu\nu}$, so that it can persist after many e-folds of the matter fields expansion.

An interesting example to consider in some detail is that of a cosmological constant or, more generally, matter sectors that are able to support maximally symmetric backgrounds in the RBG frame. A quick glance at \eqref{eq:RelationStressEnergyTensorsFramesRBG} reveals that a cosmological constant in the RBG frame also gives a cosmological constant in the Einstein frame. If we assume $T_{\mu\nu}=\Lambda g_{\mu\nu}$, then we find that
\beq\label{eq:Einstenlikeeqs}
\tilde{T}^\mu{}_\nu=-\frac{\mathcal{L}_G+\Lambda}{\sqrt{\det\Omegah}}\delta^\mu{}_\nu\equiv\tilde{\Lambda}\delta^\mu{}_\nu.
\eeq
By virtue of the Bianchi identities associated to diffeomorphisms, we find that $\tilde{\Lambda}$ must also be a constant so that the solution for $q_{\mu\nu}$ will also correspond to a maximally symmetric metric. It can happen however that a positive $\Lambda$ can lead to a negative or vanishing $\tilde{\Lambda}$. However, drawing any physical conclusion from this is of limited interest since in the absence of propagating matter fields, the only physically relevant object is the metric $q_{\mu\nu}$ that describes the characteristics of the propagation of gravitational waves. In this respect, it should be noticed that what one would call vacuum configuration in the RBG frame is different from the vacuum configuration in the Einstein frame. For instance, if we have a vacuum configuration with $T_{\mu\nu}=0$, in the Einstein frame this configuration could give rise to a cosmological constant. Likewise, if we define the vacuum in the RBG frame as the configuration with trivial matter fields, we can have a cosmological constant, but the value of the cosmological constant in both frames can be different.

The physical effect that could be measured comes when we compare the propagation of gravitational waves and some matter fields. As explained in section \ref{sec:StructureRBG}, in the minimally coupled case that we are considering, the matter fields follow the geodesics of $g_{\mu\nu}$ while gravitational waves see the metric $q_{\mu\nu}$ (see also \cite{Jimenez:2015caa}). To illustrate this, let us assume that $g_{\mu\nu}=\eta_{\mu\nu}$ and $\hat{\Omega}$ is anisotropic so we have $q_{\mu\nu}={\textrm{diag}}(N,a,b,c)$ and, for simplicity, we will assume that they are constant (i.e. we are considering vacuum configurations). If we now compare the trajectories of photons and gravitons, they respectively follow the null geodesics of the metrics:
\begin{eqnarray}
&&\dif s^2_g=-\dif t^2+\dif \vec{x}^2,\\
&&\dif s^2_q=-N\dif t^2+a\dif x^2+b\dif y^2 +c\dif z^2.
\end{eqnarray}
If we emit a graviton and a photon at $t=t_0$ from the origin along the $z-$direction, we will have
\beq
z_{\textrm{photon}}=t-t_0,\quad z_{\textrm{graviton}}=\frac{N}{c}(t-t_0)
\eeq
so their trajectories differ as $\Delta z= \left(1-\frac{N}{c}\right)(t-t_0)$. This would of course be tightly constrained by the observations of the neutron star merger GW170817 \cite{TheLIGOScientific:2017qsa}. An important point to realise is that this effect of the anisotropic $\hat{\Omega}$ cannot be absorbed into a coordinate redefinition, since that would affect the propagation of the matter fields and the relative separation would remain. In the standard case, the fact that all fields follow the same metric is what allows to absorb the anisotropic solutions of vacuum Einstein equations  that we have considered into a redefinition of the coordinates so that it does not have any physical effect. Furthermore, notice that this effect does not depend on the deformation matrix being anisotropic, but it will arise whenever $\hat{\Omega}\neq\Id$. The fact of having an anisotropic deformation matrix will further introduce polarisation and direction dependent effects. \\

Let us end our discussion on the Einstein frame by explaining another subtle point that usually arises when going to this frame. This subtlety is related to the need of solving the nonlinear equation for the deformation matrix that has been the core of this chapter. The Einstein frame formulation of the RBG theories can be achieved directly working at the level of the equations, in which case one ends up with Eq. \eqref{eq:Einstenlikeeqs}. In those equations, the right hand side depends on the metric $g_{\mu\nu}$ so, in order to properly have the differential equations determining $q_{\mu\nu}$, one needs to solve the equation for the deformation matrix $\Omega^\mu{}_\nu$. It is then usually assumed that the solution can be written as a covariant expression of the stress-energy tensor. As explained in section \ref{sec:EinsteinFrame}, by virtue of the Cayley-Hamilton theorem, one is then entitled to make the ansatz
\beq
\hat{\Omega}=\sum_{n=0}^3 c_n\hat{T}^n
\label{eq:OmegaTn}
\eeq
with $c_n$ some scalar functions of the invariants of $T^\mu{}_\nu$. However, though this is a very reasonable and natural guess for the branch that is perturbatively close to GR in vacuum, it does not (always) cover the full space of solutions. This should be clear from our results above and, owed to the nonlinear nature of the matrix equation satisfied by $\hat{\Omega}$, more general solutions are possible where the explicit covariant relation exhibited in \eqref{eq:OmegaTn} is spontaneously broken. For example, in vacuum, one can have solutions where $\hat{\Omega}$ is not proportional to the identity so that Lorentz invariance is spontaneously broken. The same can happen for nonvacuum situations. In the construction of the Einstein frame at the level of the action directly, the same situation occurs when one has to integrate out the metric $g_{\mu\nu}$. Again, this has been done in section \eqref{sec:EinsteinFrame} by solving its algebraic equation, which is nonlinear and allows for branches of solutions that do not explicitly preserve covariance. After plugging these solutions in the action, the matter sector will then contain the effects of those nontrivial branches. {In this respect, it is interesting to notice that the equivalence to GR must be understood in a broader sense, since the branches with broken symmetries will give rise to matter sectors where these symmetries are also broken. Interestingly, the loss of symmetries in the matter sector could alter the number of propagating degrees of freedom. 

As a conclusion, we see that the usual isotropic ansatz employed for the deformation matrix in physical applications within RBGs with an isotropic matter sector, besides being a natural choice, it may be necessary to avoid the pathologies that we have discussed in the evolution. We should notice however that the suitability of the isotropic deformation was not guaranteed a priori. As an example we can mention the cosmological isotropic bouncing solutions that can be unstable due to the growth of the shear in the contracting phase and something along these lines (barring the obvious differences) might have happened for the solutions with isotropic deformation in RBGs. Our analysis then provides a strong support for the physical motivation of the isotropic ansatz for the deformation matrix in projective invariant RBG theories.

%%%%%%%%%%%%%%%%%%%%%%%%%
%%%%%%%%%%%%%%%%%%%%%%%%%
%%%%%%%%%%%%%%%%%%%%%%%%%

			%NEWCHAPTER%

%%%%%%%%%%%%%%%%%%%%%%%%%
%%%%%%%%%%%%%%%%%%%%%%%%%
%%%%%%%%%%%%%%%%%%%%%%%%%

\chapter{Absorption by black bole remnants in metric-affine gravity}\label{sec:Absorption}

\initial{I}n the previous chapter, the general structure of RBG theories with and without projective symmetry were presented. Both cases were seen to admit an Einstein frame where the gravitational sector is described by metric-affine GR and Nonsymmetric Gravity Theory \cite{Moffat:1994hv} respectively. Moreover, in the case with projective symmetry, we saw that the solution space has nontrivial branches of solutions in which the symmetries of the RBG frame may not be the same a those of the Einstein frame even when the matter fields of the RBG frame also satisfy them. This chapter will be devoted to the study of a general class of exotic compact objects with interesting properties that arise as spherically symmetric solutions of RBG theories with projective symmetry\footnote{Through this chapter, I will only be referring to RBG theories with projective symmetry. However, I will drop the explicit statement {\it with projective symmetry} and write simply RBG theories in order to facilitate the information flow.} coupled to a free Maxwell field in the branch that connects with GR at low energies. We will mainly be concerned with their absorption properties when scalar waves are scattered off them, though we will also perform a preliminary geodesic analysis that will correspond to the eikonal approximation of the scalar absorption profile.

 In the last years there has been increasing interest in the study of compact objects which may figure as astrophysical alternatives to classical black holes (BHs) or exhibit unconventional features, such as hair or signs of new high-energy physics \cite{Johnson-McDaniel:2018uvs,Cunha:2017wao, Herdeiro:2017phl,Cardoso:2016oxy,Herdeiro:2014goa,Liebling:2012fv}. This interest has grown in parallel with the development of gravitational wave detectors, which have provided convincing evidence that collisions between massive astrophysical-size compact objects occur frequently \cite{Cardoso:2016oxy,TheLIGOScientific:2017qsa,GBM:2017lvd,Abbott:2017dke,Abbott:2017gyy,Barack:2018yly,} and, together with the first images of supermassive black holes \cite{Akiyama:2019cqa,Akiyama:2019eap}, can be used to unveil properties of the strong field regime of the gravitational interaction. However, the current capabilities of such observatories are yet insufficient to confirm or rule out the existence of the BH event horizon itself and we will have to wait for future developments in order to have a chance to settle this issue, as well as other related questions. Therefore, the possibility to test subtle details of the strong gravity regime is still beyond our current techniques, and we must do our best to scrutinise the spectrum of phenomenological possibilities from a theoretical perspective. 
 
 Among the various open questions posed by BH investigations, understanding whether spacetime singularities \cite{Earman:1995fv,Geroch:1968ut,Berger:2002st,Wald:1997wa} are real, or an artefact of our mathematical models, is one of the most challenging problems both from technical and philosophical perspectives. Though the BH event horizon is taken by some authors as a possibility to minimise this issue, adopting an {\it out of sight, out of mind} attitude, a lot of effort has been devoted to the construction of nonsingular alternatives for BH interiors. In this sense, the physical nature of singularities 
has been attacked from different perspectives in the literature, including non-linear corrections on the matter fields \cite{BardeenBH,AyonBeato:1998ub,Bronnikov:2000vy,Ansoldi:2008jw,DiazAlonso:2009ak,DiazAlonso:2010eh,DiazAlonso:2012mb} for a general analysis of this issue), as well as non-perturbative effects \cite{Barcelo:2017lnx}, fully dynamical models of BH formation and evaporation \cite{Fabbri:2005mw,Hayward:2005gi,Zhang:2014bea,Liu:2014kra,Malafarina:2017csn,Arrechea:2020jtv},  quantum-gravitational pressure counter-effects preventing the formation of the singularity \cite{Rovelli:2014cta,Spallucci:2017aod,Abedi:2015yga,Ashtekar:2018lag,Ashtekar:2018cay}, or via the replacement of the event horizon by a compact surface mimicking the Schwarzschild radius as seen from far away observers \cite{Cardoso:2016oxy,Bueno:2017hyj}. 

We are interested in exploring some properties of a family of nonsingular BH solutions which arise generically in RBG theories coupled to regular matter fields. These solutions were first found by exploring semiclassical gravity effects on Reissner-Nordstr\"om BHs of quadratic RBG theories \cite{Olmo:2012nx,Olmo:2011np}, and were later seen to be solutions of Eddington-inspired Born-Infeld gravity \cite{Olmo:2012nx,Olmo:2011npOlmo:2013gqa} (see section \ref{sec:EiBIEMMapping}). The most remarkable property of these new solutions is that they represent geodesically complete spacetimes with wormhole structure \cite{Olmo:2012nx,Olmo:2011np,Olmo:2013gqaOlmo:2016fuc,Olmo:2015bya,Olmo:2015dba}, where a spherical throat replaces the central singularity  found in GR when coupled to a Maxwell field due to the higher-order curvature terms of the RBG action. Among the various families of solutions of this electrovacuum theory, there is a subset which is completely regular, in the sense that curvature invariants are bounded everywhere, even at the wormhole throat \cite{Olmo:2012nx,Olmo:2011np,Olmo:2013gqa,Olmo:2016fuc,Olmo:2015bya,Olmo:2015dbaOlmo:2013gqa,Olmo:2012nx,Olmo:2011np}. These solutions smoothly interpolate between Schwarzschild-like BHs (when their mass is sufficiently high) and naked solitons (when their mass approaches the Planck scale), always having a wormhole of finite area at their center. For this reason, because they smoothly connect massive BH solutions with Minkowski spacetime, they can be regarded as natural candidates for BH remnants \cite{Lobo:2013ufa,Lobo:2014fma}. Thus, this unconventional family of massive topological entities offers an interesting environment to study qualitative new features of BH remnants. With this idea in mind, a first step to understand their properties can be taken by studying their interaction with scalar waves. Given that their BH phase is essentially identical to that corresponding to Schwarzschild BHs \cite{Olmo:2012nx,Olmo:2011np}, here we focus on the horizonless configurations (naked solitonic phase), which can be seen as two copies of Minkowski spacetime connected by a spherical wormhole, where the energy density concentrates. Due to the fact that these objects are horizonless alternatives to standard BHs and that they usually present photospheres, they fit well into the classification of extreme/exotic compact objects (ECOs) found in \cite{Cardoso:2017cqb}. ECOs can be further characterised into subclassses, namely UCOs (ultra-compact objects) and ClePhOs (clean photosphere objects) \cite{Cardoso:2017cqb}. UCOs are compact objects with a photosphere and ClePhOs are UCOs with an effective radius very close to the Schwarzschild radius. It was recently found in \cite{Macedo:2018yoi} that, due to an ``effective cavity" between ClePhOs' effective surface and its photosphere, ClePhOs present an absorption spectrum characterised by Breit-Wigner like resonances which could allow for experimental searches. In this chapter, we will be mainly devoted to replicate the analysis that was done in \cite{Macedo:2018yoi} in order to study the absorption properties of other types of ECOs existent in alternative theories, in search of characteristic signatures that could distinguish them from regular BHs or other ECOs \cite{Macedo:2015ywb}. As we will see, the absorption spectrum of the family of regular solutions studied here exhibits a pattern associated to a rich structure of quasibound states in the remnant phase similar to that found for ClePhOs in \cite{Macedo:2018yoi}, thus allowing to tell them apart from regular BHs of the same mass. Throughout this chapter, we will use the metric signature $(-,+,+,+)$ and natural units, such that $G=\hbar=c=1$.

%%%%%%%%%%%%%%%%%%%%%%%%%%%%%%%%%%%%%%%%%%%%%%
\section{Spherically symmetric electrovacuum solutions}\label{sec:framework}
%%%%%%%%%%%%%%%%%%%%%%%%%%%%%%%%%%%%%%%%%%%%%%

The BH solutions we are going to study arise naturally in RBG theories by coupling them to a spherically symmetric and static Maxwell electric field. They are characterised by a line element of the form
\beq\label{metric}
ds^2=-A(x) dt^2+\frac{1}{A(x) \mathcal{Z}_+^2(x)}dx^2+r^2(x)\left(d\theta^2+\sin^2\theta d\varphi^2\right),
\eeq
where
\begin{align}\label{eq:DefinitionsAbsorption}
\begin{split}
&A(x) \equiv \frac{1}{\mathcal{Z}_+(x)}\left[1-\frac{r_S}{r_c}\frac{(1+\delta_1 \, H(x))}{ z(x) \, \mathcal{Z}_-^{1/2}(x)}\right],\qquad z(x) \equiv \frac{r(x)}{r_c}, \qquad  \mathcal{Z}_\pm(x) \equiv 1\pm\frac{1}{z^4(x)} \\
&r^2(x)=\frac{1}{2}\left(x^2+\sqrt{x^4+4r_c^2}\right),\qquad r_c \equiv \sqrt{l_{\textrm{G}} r_q},\qquad \delta_1 \equiv \frac{1}{2r_S}\left[\frac{r_q^3}{l_{\textrm{G}}}\right]^{1/2},\qquad r_q^2 \equiv 2 q^2 .
\end{split}
\end{align}
Here the $x$ coordinate, defined through (\ref{eq:DefinitionsAbsorption}), takes values in the whole real axis $(-\infty, +\infty)$. The parameter $r_S$  defines the Schwarzschild mass $r_S=2M$. The length $l_{\textrm{G}}$ is  related to the mass scale ${\mg}$ by $l_{\textrm{G}}=\lr{\sqrt{2}{\mg}}^{-1}$, and controls the nonlinear deformations of the matter sector in the Einstein frame of the corresponding RBG (see section \ref{sec:EinsteinFrame}). The function  $H(x)$ is given by
\beq
H(x)=-\frac{1}{\delta_c}+\frac{1}{2}\sqrt{z^4(x)-1}[f_{3/4}(x)+f_{7/4}(x)],
\eeq
where
\beq
f_\lambda(x)={_2F}_1[1/2,\lambda,3/2,1-z^4(x)]
\eeq
are hypergeometric functions, and $\delta_c\approx 0.572069$ is an integration constant needed to find the correct behavior at spacelike infinity. The different parameters appearing in the line element \eqref{metric} can be rewritten as functions of the dimensionless parameters $N_q \equiv q/e$ (with $e$ being the proton charge) and the charge-to-mass ratio $\delta_1$, defined in \eqref{eq:DefinitionsAbsorption}. Let us write these relations explicitly 
\begin{equation}
q=eN_q \ , \ r_q=2l_{\textrm{P}} N_q/N_c \ , \ r_S=\frac{r_c^3}{2\delta_1 l_{\textrm{G}}^2} \ ,
\end{equation}
where $l_{\textrm{P}}$ is the Planck length and $N_c\equiv \sqrt{2/\alpha_{em}}\approx 16.55$ is a critical number of charges, which represents the transition from BH ($N_q>N_c$) to naked wormhole ($N_q<N_c$). In the definition of $N_{c}$, $\alpha_{em}$ is the fine structure constant. These definitions show how the line element \eqref{metric} is totally specified by the two dimensionless parameters $(\delta_1, N_q)$ plus the scales $l_{\textrm{G}}$  and $l_{\textrm{P}}$. This family of metrics leads to three qualitatively different types of spacetime, depending on the relative values of $\delta_1$ and $N_q$. with respect to the critical values. These three types are:
\begin{itemize}
\item[1]{\textbf{Schwarzschild like solutions}}: characterised by $\delta_1 < \delta_c$, they possess an event horizon 
(on each side of the wormhole) for all values of $N_q$.
\item[2]{\textbf{Reissner-Nordstr\"om like solutions}}: With $\delta_1 > \delta_c$, they may exhibit; on each side of the wormhole;  two, one (degenerate), or no horizons, like in the usual Reissner-Nordstr\"om (RN) solution of GR. 
\item[3]{\textbf{Regular solutions}}: With $\delta_1 = \delta_c$, if $N_q > N_c$, one finds one horizon on each side
of the wormhole (similar to the Schwarzschild case). If $N_q = N_c$, the two symmetric horizons 
meet at the wormhole throat, $r = r_c$ (or $ x= 0$). For $N_q < N_c$  the horizons disappear yielding
a wormhole that connects two asymptotically Minkowskian universes. We will refer to these solutions as BH remnants, as they are continuously connected with BH configurations. The existence of
such remnants, which may arise at the end of BH evaporation or due to large density fluctuations in the early universe \cite{Hawking:1971ei}, might be of special relevance for the understanding
of the information loss problem \cite{Chen:2014jwq} and may also have potential observational consequences \cite{Lobo:2013prg}.  It is important to note that when the charge-to-mass ratio $\delta_1$ is set to the value $\delta_c$, the mass spectrum of the solutions is completely determined by the charge parameter $N_q$, through the relation \cite{Olmo:2013gqa}
\begin{equation}
M=m_P \left(\frac{N_q}{N_c}\right)^{3/2}\left(\frac{l_{\textrm{P}}}{l_{\textrm{G}}}\right)^{1/2} n_{BI} \ ,
\end{equation}
where $n_{BI}= \pi^{3/2}/(3\Gamma[3/4]^2)\approx 1.23605$. Up to a $\sqrt{2}$ numerical factor, this mass/energy expression is identical to the one found for point charges in the Born-Infeld electromagnetic theory. 
\end{itemize}
From here on we will refer to these as Type I, II and III solutions. All the above cases rapidly tend to the standard GR solutions just a few $r_c$ units away from the wormhole throat (located at $r=r_c$).
%%%%%%%%%%%%%%%%%%%%%%%%%%%%%%%
\section{Light rays and scalar waves}
\label{sec:Abs}
%%%%%%%%%%%%%%%%%%%%%%%%%%%%%%%

We now shift our attention to the properties of geodesics and propagation of scalar waves in the spacetime defined by the line element (\ref{metric}). Since it represents a  static and spherically symmetric geometry, we have two Killing fields, $\xi_1=\partial_t$ and $\xi_2=\partial_\varphi$, which satisfy $\nabla \big(g(u,\xi_i)\big)[u]=0$ along metric geodesics with tangent vector $u^\mu$. Due to spherical symmetry, we may restrict our attention to geodesics at the equatorial plane ($\theta=\pi/2$), without loss of generality.  
The two Killing fields give the following conserved quantities along equatorial geodesics
\beq\label{Killing}
\begin{split}
&E=-A\dot{t} , \\
&L=r^2(x)\dot{\varphi},
\end{split}
\eeq
where a dot over a quantity means its derivative with respect to the affine parameter of the corresponding geodesic.

%%%%%%%%%%%%%%%%%%%%%%%%%%%%%%%
\subsection{Capture of null geodesics}
%%%%%%%%%%%%%%%%%%%%%%%%%%%%%%%

For metric geodesics we also have conservation of the norm of their tangent vector, $g(u,u)=-k$, where $k=0$ for null geodesics and $k=1$ for (affinely-parametrized) timelike geodesics. Using \eqref{metric} and \eqref{Killing} we can therefore write

\beq\label{eqforzeta}
\frac{1}{2}\tilde{m}(x)\dot{x}^2+V_{eff}(x)=E^2,
\eeq
where we have defined
\beq
\label{pot}
V_{\textrm{eff}}(x)\equiv A(x)\left(\frac{L^2}{r^2(x)}+k\right) ,\qquad\text{and}\qquad \tilde{m}(x)\equiv 4/\mathcal{Z}^2_+(x) .
\eeq
The above equation \eqref{eqforzeta} is similar to that of a Newtonian particle of variable mass $m(x)$ and energy $E^2$ in a central effective potential $V_{eff}$. 
As we can see in figure  \ref{fig:claspotentials}, the effective potential associated to null geodesics presents a well at the wormhole throat, with one maxima on each side, defining two unstable photospheres. The minimum of the potential, at the wormhole throat, is related to a stable photosphere. We can see that the depth of the potential well increases as the normalised number of charges $n_s$ increases. Intuitively, concerning scalar waves,  this is telling us that the wormholes will be more absorptive the more charged they are, because their area grows linearly with the charge. Moreover, from the presence of a potential well, we can anticipate the existence of quasibound modes around the throat in the wave regime. Indeed, the absorption spectrum of scalar waves, 
computed in Sec. \ref{sec:TM}, shows the existence of these modes.

\begin{figure}[h]
\includegraphics[width=0.5\textwidth]{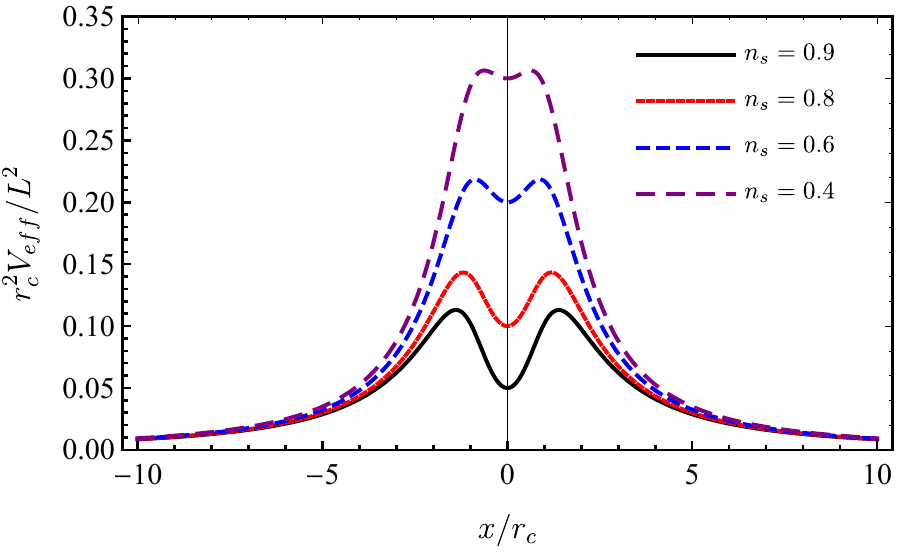}
\includegraphics[width=0.5\textwidth]{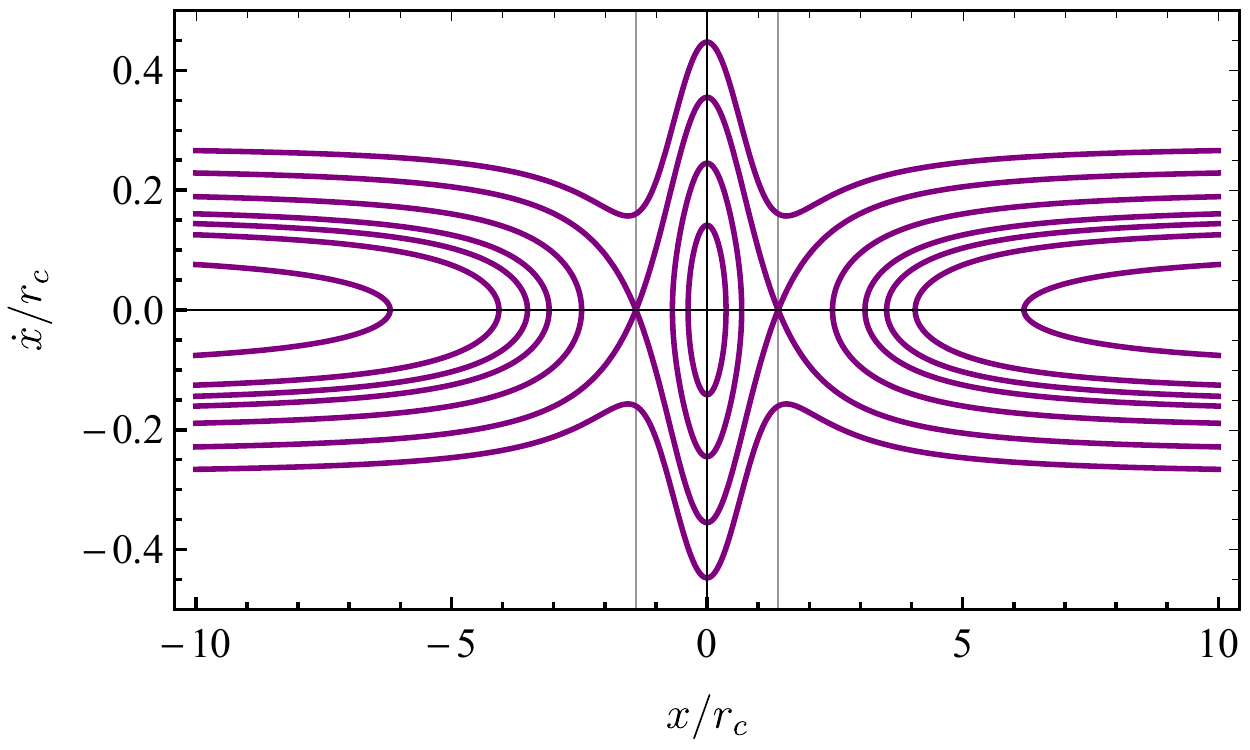}
\caption{The left image is the effective potential $V_{\textrm{eff}}(x)$, given by \eqref{pot} and normalised by $r_c^2/L^2$, for null geodesics ($k=0$) in BH remnant spacetimes. Here $n_s = N_q / N_c$. Note that for values of $n_s\approx 1$ we have a more pronounced potential well at $x=0$, and the well disappears as $n_s\rightarrow 0$ (or $N_q\rightarrow 0$). The right image is the corresponding phase portrait for $n_s=0.9$. Vertical lines show the maxima of the effective potential. We see two unstable equilibrium points at the maxima of $V_{\textrm{eff}}$ and a stable equilibrium point at the central minimum $x=0$.}
\label{fig:claspotentials}
\end{figure}

In order to calculate the absorption cross section of light rays by naked wormholes, we need to find the position of the circular orbits, \ie the photospheres. Though, strictly speaking, the position dependence of $\tilde{m}(x)$  breaks the equivalence with the particle of mass $m$ in a central potential, by definition, the photospheres are the trajectories satisfying $\ddot{x}=0$ with the initial condition $\dot{x}_0=0$. Let us see how, as in the constant mass case, the photospheres also correspond to the maxima of the effective potential. In order to prove this, we can write \eqref{eqforzeta} as
\beq\label{LinearODEzeta}
\dot{x}=\sqrt{\frac{2\lrsq{E-V_{\textrm{eff}}(x)}}{\tilde{m}(x)}} \; ,
\eeq where the right hand side can be written as $f(x)$. The above equation can be mapped into the autonomous system 
\begin{align}
&\dot{\zeta}=f(x,\theta)=\theta\label{phasesys}\\
&\dot{\theta}=g(x,\theta)=\sqrt{\frac{\tilde{m}}{2}}\theta\lr{\frac{\tilde{m}'\sqrt{E-V_{\textrm{eff}}}}{2\tilde{m}}-\frac{V_{\textrm{eff}}'}{2(E-V_{\textrm{eff}})}}\nonumber\end{align}
with the constraint $\theta=\sqrt{2\lrsq{E-V_{\textrm{eff}}(x)}/\tilde{m}(x)}$, where $\dot{E}=0$ has been used. As usual, the equilibrium points $(x,\theta_0)$ of this system are given by the two conditions $f(x_0,\theta_0)=0$ and $g(x_0,\theta_0)=0$. The first condition is $\theta_0=0$, which together with the constraint equation, and given that $\tilde{m}(x)$ is bounded, implies $E=V_{\textrm{eff}}(x_0)$. If the first condition holds, with the use of the constraint equation the second condition can be written as 
\beq
\frac{V'_{eff}(x_0)}{2\sqrt{E-V_{\textrm{eff}}(x_0)}}=0,
\eeq
which implies that $V'_{eff}(\zeta_0)$ has to vanish quicker than $\sqrt{E-V_{\textrm{eff}}}$ when approaching $x_0$. The conditions for equilibrium points together with the constraint equation yield $\theta_0=\dot x_0=0\;$ and  $V'_{eff}(x_0)=0$. Stable points are always associated to constant equilibrium solutions, which in this case describe circular orbits $(x=x_0, \theta=\dot{x}=0)$ related with extrema of the effective potential. In figure \ref{fig:claspotentials} we plot the phase portrait of the system, which is qualitatively equivalent to the constant mass case in a central potential, and allows to quickly grasp the stability properties of the orbits for different initial conditions. There we can see how light signals with $E>V_{\textrm{eff}}(\zeta_0)$ will go from one asymptotic region to another. Light signals with $E<V_{\textrm{eff}}(\zeta_0)$ emitted in the region  $|x|>|x_0|$ will bounce back to infinity in their corresponding asymptotic region. More interestingly, light signals with $E<V_{\textrm{eff}}(x_0)$ emitted in the region  $|x|<|x_0|$ would stay in that region bouncing back and forth. As stated above, the effects of this region will later be shown to generate quasi-normal modes for scalar waves. Null geodesics impinging from infinity, which reach and stay at the maximum of the potential are called \textit{critical},
and they are characterised  by $V_{\textrm{eff}}(x_{\textrm{max}})=E^2$. This relation fixes their impact parameter, $b \equiv L/E$, to be 
\begin{equation}\label{Bcritclas}
b_c=\sqrt{\frac{L^2}{V_{\textrm{eff}, max}}}=\frac{r_{\textrm{max}}}
{\sqrt{A_{\textrm{max}}}},
\end{equation} 
where the subindex $\textrm{max}$ denotes evaluation of the corresponding function at $x_{\textrm{max}}$. The critical impact parameter is related to the frequency of the unstable circular null geodesic by
\beq
\Omega_l=b_c^{-1}.
\label{eq:frequency_null}
\eeq
Null geodesics with $b>b_c$ are scattered by the BH remnant and stay 
in Region I (defined in Subsec. \ref{scalar_absorption}), 
whereas those with $b<b_c$ overcome $V_{\textrm{eff,max}}$ and cross the wormhole throat 
to Region II. 
The classical absorption cross section for BH remnants is then given by

\begin{equation}\label{ClasAbsCrossBHremnants}
\sigma_{c} = \pi b_c^2 = \frac{\pi r_{\textrm{max}}^2}{A_{\textrm{max}}}.
\end{equation}

Despite that in our model it is not possible to solve $V'_{\textrm{eff}, max}=0$ analytically, it is always possible to find  $ x_{max}$ through a numerical approach.  In figure  \ref{fig:ClassAbsCros} we present a plot of the total absorption cross section for null rays absorbed by a naked wormhole as a function of $n_s$. We can see that the absorption cross section increases monotonically with the (normalised) number of charges. Therefore, for an observer at infinity, $n_s$ can be regarded as an effective dissipative coefficient. As it will be seen later, this analogy can be extended to the analysis of scalar wave absorption by the wormhole.

\begin{figure}[h]
\centering
\hspace{-1.5cm}\includegraphics[scale=0.7]{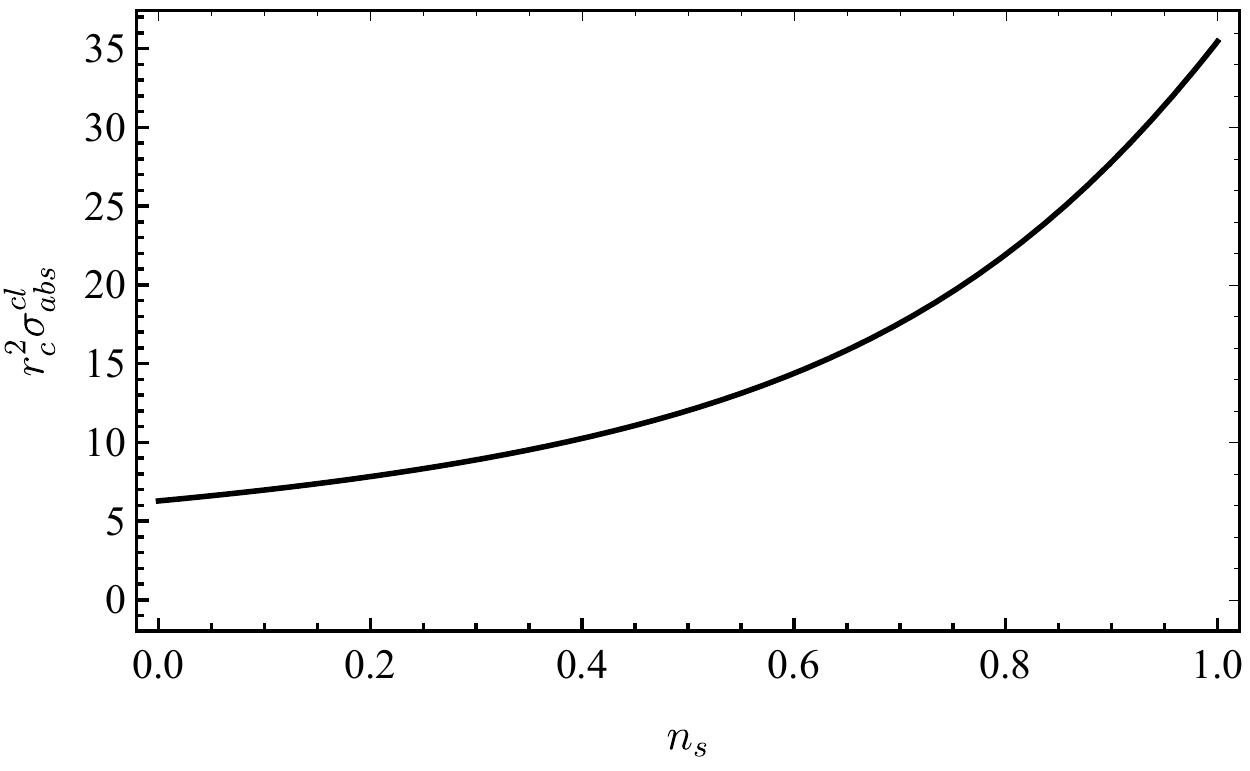}
\caption{Null geodesics absorption cross section of BH remnants for different values of $n_s$. Recall that BH remnant solutions are characterised by $\delta_1=\delta_c$ and  $n_s=N_q/N_c\in (0,1)$.}
\label{fig:ClassAbsCros}
\end{figure}

%%%%%%%%%%%%%%%%%%%%%%%%%%%%%%%%%%%%%%%%%%%%%%
\subsection{Absorption of massless scalar waves}
\label{scalar_absorption}
%%%%%%%%%%%%%%%%%%%%%%%%%%%%%%%%%%%%%%%%%%%%%%

As it is well known, the absorption of null geodesics is associated to the high-frequency limit (geometric optics approximation) of scattering planar massless waves \cite{Castineiras:2005ww,Macedo:2014uga}. The geodesic analysis, however, is not sensitive to the full range of phenomena that waves can experience, providing incomplete information about the absorption and scattering spectra, as well as the modal structure of the spacetime. These characteristics are also strongly dependent on the spin of the waves considered \cite{Crispino:2010fd,Oliveira:2011zz,Nakamura:1976nc,Benone:2014qaa,Leite:2017zyb,Leite:2018mon}. As a first approach to this problem, we consider massless scalar waves, which provide interesting insights on the features of the spacetime beyond the geodesic approximation \cite{Olmo:2015bya,Benone:2015bst,Leite:2017hkm} . \\

Let us consider a minimally coupled massless scalar field $\Phi$ described by \eqref{scalaraction}. The corresponding field equations are (see section \ref{sec:MinCoupScalar})
\beq
\Box_g \Phi=\frac{1}{\sqrt{-g}}\partial_\mu\left[\sqrt{-g}\partial^{\mu}\Phi\right]=0.
\label{eq:scalarwave}
\eeq
Its Einstein-frame action will generally feature self-interaction terms which will be of ${\cal O}(\partial\Phi^3)$ or higher. Since we are interested in linear perturbations, we will neglect these corrections and solving \eqref{eq:scalarwave} in the background described by the line element (\ref{metric}) with appropriate boundary conditions will suffice. Given that the background is spherically symmetric, we use separation of variables to decompose the field as
\beq
\Phi=\frac{\phi(t,x)}{r(x)}Y_{\ell m}(\theta,\phi),
\label{eq:scalardecom}
\eeq
where $Y_{\ell m}(\theta,\phi)$ are the scalar spherical harmonics. Plugging \eqref{eq:scalardecom} into \eqref{eq:scalarwave} we obtain the $1+1$-dimensional wave equation
\beq
\left(\frac{\partial^2}{\partial r_\star^2}-\frac{\partial^2}{\partial t^2}-V_\varphi(r_\star)\right)\phi(x, t)=0,
\label{eq:timeeq}
\eeq
where the effective potential $V_\varphi$ is given by
\beq
V_\varphi(r_\star)=\frac{A\ell(\ell+1)}{r^2}+\frac{d^2r}{d{r^2_\star}} \ ,
\label{Vvarphi}
\eeq
and we have defined a tortoise-like coordinate $r_\star$ by
\beq
{dr_{\star}}\equiv \frac{dx}{A\,\mathcal{Z}_+} \ .
\eeq
and satisfies
\begin{equation}
\frac{d^2r}{d{r^2_\star}}= A\mathcal{Z}_+\frac{d}{dx}\lr{A\mathcal{Z}_+\frac{dr}{dx}}\, .
\end{equation}
Equation \eqref{eq:timeeq} 
can be reduced to an ordinary differential equation by resorting to separation of variables, decomposing the function $\varphi(t,x)$ as $\phi(t,x)=\varphi(x) e^{-i\omega t}$, which leads to
\beq
\left[\frac{d^2}{d r_\star^2}+\omega^2-V_\varphi(r_\star)\right]\varphi(x)=0.
\label{eq:freqeq}
\eeq

In figure  \ref{fig:potentialsphi} we show the effective scalar potential $V_\varphi$ for different choices of $n_s$. For BH remnants, we have that a potential well may appear at $r=r_c$, showing different features from the BH case. The potential is consistent with the one from the geodesic analysis, given by (\ref{pot}).
\begin{figure*}
\includegraphics[width=0.5\textwidth]{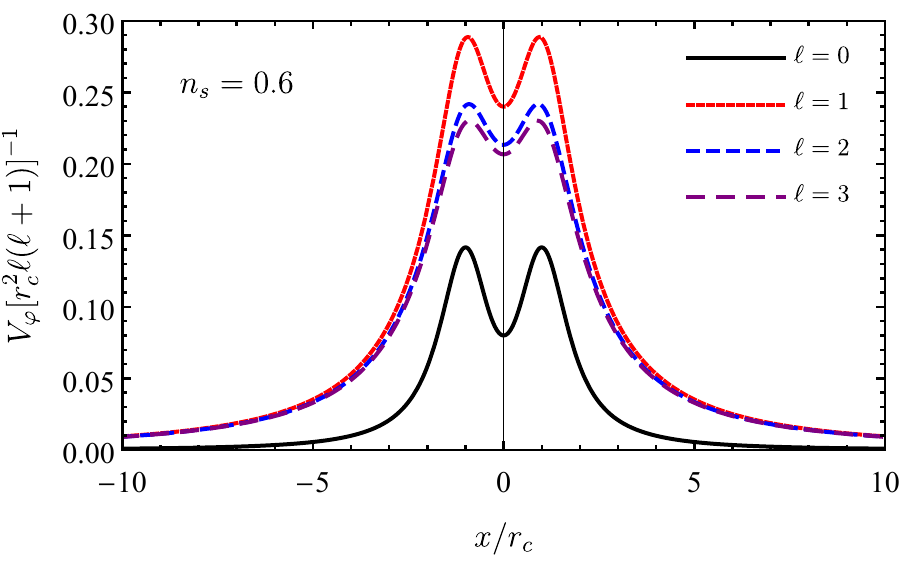}
\includegraphics[width=0.5\textwidth]{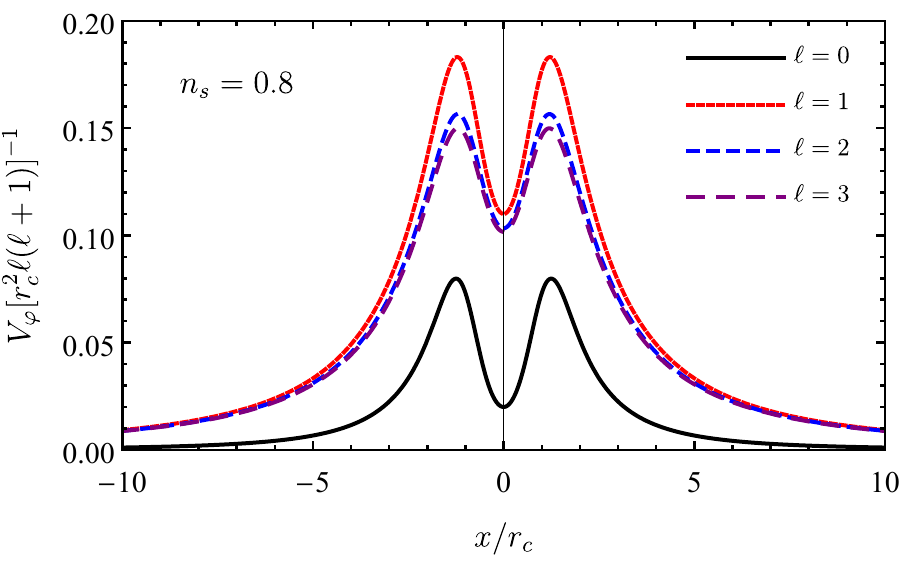}
\includegraphics[width=0.5\textwidth]{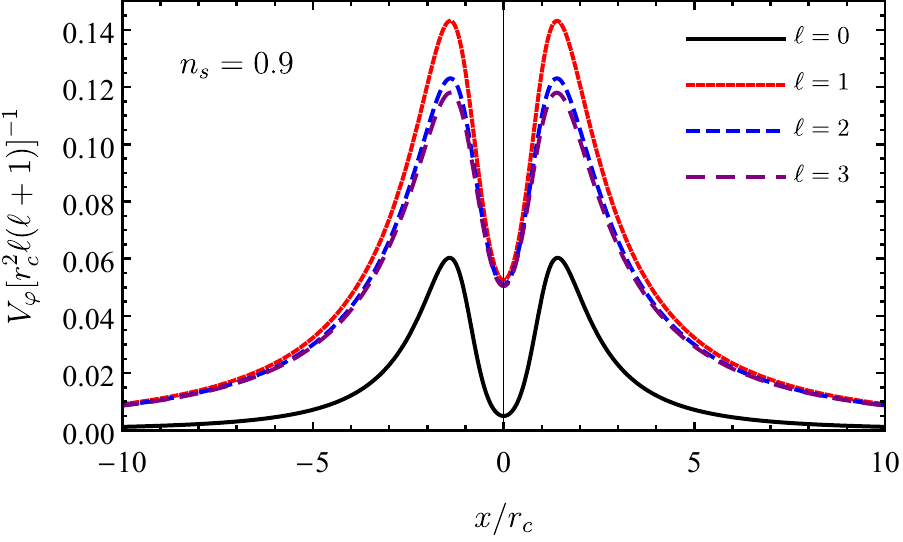}
\includegraphics[width=0.5\textwidth]{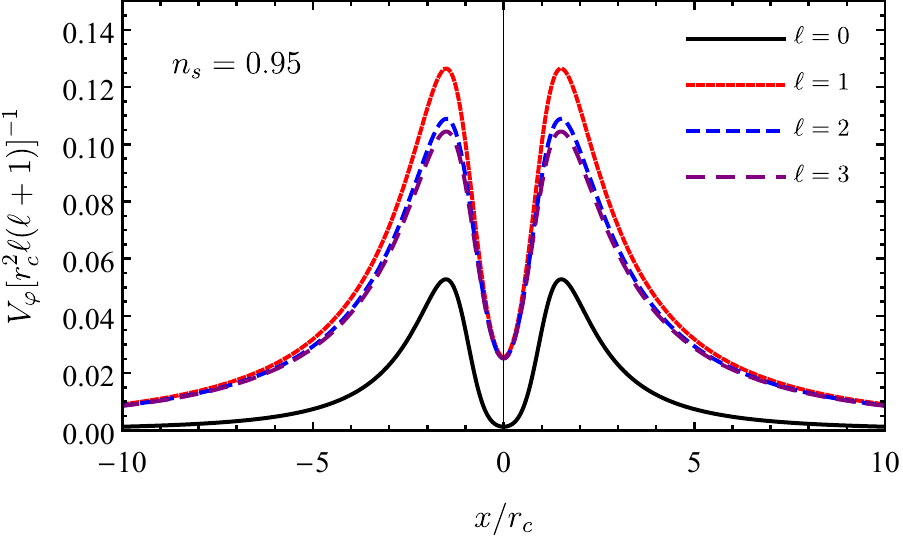}
\caption{Effective scalar potential $V_\varphi$ [given by \eqref{Vvarphi}] for BH remnants. 
The plots with $\ell>1$ are normalised by $\ell(\ell+1)$, for better visualization. 
We note the presence of a well centered at $x=0$, which gets deeper as the limit $n=1$ is approached, and is shallower for higher multipoles. We also note the similarity of $V_\varphi$ with the potential obtained in the null geodesic analysis, $V_{\textrm{eff}}$ [given by \eqref{pot}], plotted in figure  \ref{fig:claspotentials}.}%
\label{fig:potentialsphi}%
\end{figure*}
Proper boundary conditions should be supplemented to \eqref{eq:freqeq}. The Penrose diagram of remnant configurations and the corresponding illustration of the scattering problem is depicted in figure  \ref{fig:PenDiagNWH}. The right and left hand sides of the diagram are identified as Regions I and II, respectively. We are interested in planar waves incoming from past (null) infinity on the bottom right part of the diagram, ${\cal J}_{\textrm{I}}^-$, being reflected to ${\cal J}_{\textrm{I}}^+$ and transmitted to ${\cal J}_{\textrm{II}}^+$. 
In Region I, asymptotically ($r \to +\infty$), we have
\beq
\varphi_{R_I}(x)\approx{\mathcal{A}}_{\ell m} e^{-i \omega r_\star}+{\mathcal{R}}_{\ell m} e^{i \omega r_\star},
\label{eq:phiari}
\eeq
where ${\cal A}_{\ell m}$ is the amplitude of the incoming wave and ${\cal R}_{\ell m}$ the amplitude of the reflected one. To write (\ref{eq:phiari}) we have used the fact that the potential vanishes asymptotically. 
The wave coming from ${\cal J}_{\textrm{I}}^-$ is scattered by the compact object, leading to a phase difference between ${\cal A}_{\ell m}$ and ${\cal R}_{\ell m}$. The compact object can also partially absorb the wave, resulting in a difference in the moduli of the amplitude of the incoming and reflected waves. In the case of BHs, the absorption is associated to a purely ingoing wave into the horizon. For wormholes, which is the case of the BH remnant treated here, we identify the absorption with the part of the wave that is transmitted through the throat to the other side (Region II). 
\begin{figure}
\centering
\includegraphics[scale=0.5]{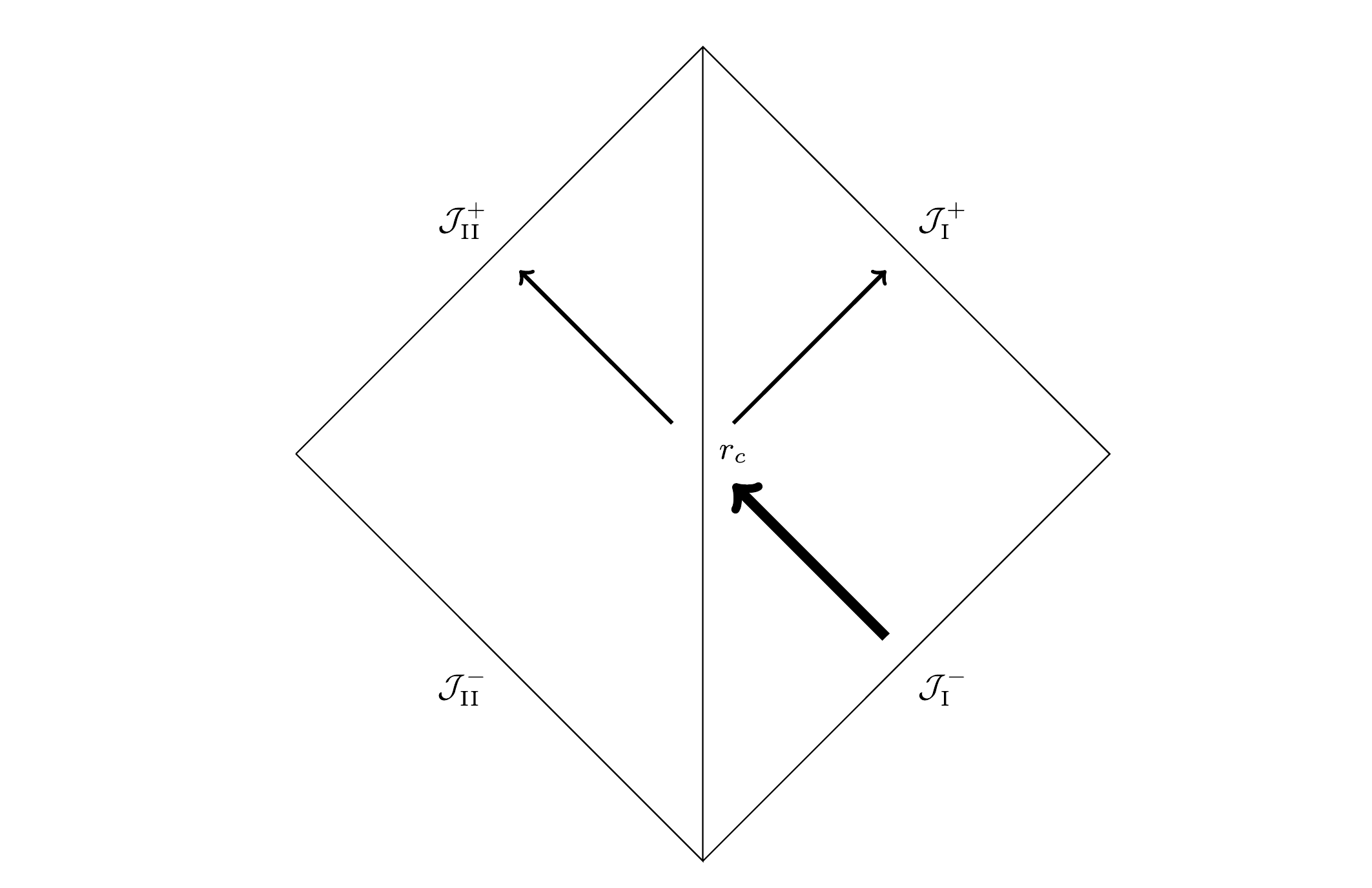}
\caption{Penrose diagram for a BH remnant configuration ($\delta_1=\delta_c$ and $N_q<N_c$). The wormhole is represented by the vertical timelike trajectory denoted by $r_c$. The triangular sectors on each side represent two asymptotically Minkowskian universes. The arrows represent the scattering of the massless scalar waves.}
\label{fig:PenDiagNWH}
\end{figure}
To describe the scattering phenomenology, we have to compute the phase-shift  $\delta_{\omega \ell}$, which is related to the reflection coefficient by
\beq
e^{2i\delta_{\omega \ell}}=(-1)^{\ell+1}\frac{{\cal R}_{\omega \ell}}{{\cal A}_{\omega \ell}}.
\eeq
In general, the phase-shift is complex whenever $|{\cal R}_{\omega \ell}|\neq|{\cal A}_{\omega \ell}|$, \ie when there is  dissipation in the system. The absorption cross section is given by \cite{Futterman:1988ni}
\beq
\sigma=\sum_{\ell=0}^{\infty}\sigma_{\ell},
\eeq
where $\sigma_l$ is the partial absorption cross section for each multipole, given by
\begin{align}
\sigma_l=\frac{\pi}{\omega^2}(2\ell+1) \Gamma_{\omega \ell}\label{sigmagamma},
\end{align}
with
\beq
\Gamma_{\omega \ell}=1-\left|\frac{{\cal R}_{\omega \ell}}{{\cal A}_{\omega \ell}}\right|^2
\eeq
being the transmission coefficients. To compute the reflection coefficient, we must impose that the boundary conditions in the asymptotic limit of Regions I and II are satisfied, resulting in equations for the amplitude of the wave in those limits. This can be done by analytical approximations of the wave function or by numerically integrating it from the asymptotic limit of Region II to the asymptotic limit of Region I, and comparing the result with the asymptotic form given by \eqref{eq:phiari}.
%%%%%%%%%%%%%%%%%%%%%%%%%%%%%%%%%%%%%%%%%%%%%%
\subsubsection{Trapped modes}
\label{sec:TM}
%%%%%%%%%%%%%%%%%%%%%%%%%%%%%%%%%%%%%%%%%%%%%%
Due to the shape of the potential, quasibound states can exist, associated to the potential well located at $r=r_c$. These quasibound states are similar to the trapped modes arising in ultracompact stars \cite{Kokkotas:1999bd}, and in the eikonal limit they are related to the stable null-geodesics existing at $r=r_c$ \cite{Cardoso:2014sna}. The modes are complex, having small imaginary part due to the tunneling to the asymptotic regions of spacetime. They are determined by the boundary conditions
\beq
\varphi=\left\{
\begin{array}{ll}
	e^{-i\omega r_\star},& x\to-\infty,\\
	e^{i\omega r_\star},& x\to\infty,
\end{array}\right.
\eeq
which generates an eigenvalue problem for the frequency $\omega$. 
The existence of trapped modes in the BH remnant case is a crucial difference from the BH spacetime, where the imaginary part is associated to the timescale of the unstable null geodesic \cite{Cardoso:2008bp}. 
The quasibound modes generate a signature in the absorption spectrum, leading to narrow spectral lines in it. In fact, this signature has been also found in weakly dissipative ultracompact stars, where the trapped modes give rise to structures similar to the Breit-Wigner resonances in nuclei scattering  \cite{Macedo:2018yoi}. The position and the structure of the spectral lines depend on the nature of the compact object and, therefore, they may be used to tell them apart.

In the eikonal limit, the real part of the trapped modes $\omega_r$ can be found through the Born-Sommerfeld quantization rule \cite{Gurvitz:1988zz}
\beq
\int_{r_{\star a}}^{r_{\star b}}dr_\star\sqrt{\omega_r^2-V_{\varphi}(r_\star)}= \pi(n+1/2),
\label{eq:quantization}
\eeq
where ${\omega_r}^2<V_{\varphi}$, $n$ is a positive integer, and  $r_{\star a,b}$ are the inner turning points, defined through ${\omega_r}^2-V_{\varphi} =0$. As previously mentioned, the imaginary part of these modes is usually very small, what leads to the presence of resonant narrow peaks in the transmission coefficient. From \eqref{eq:quantization} we can find the position of the resonant peaks, and we also find the relation  \cite{Cardoso:2014sna}
\beq
\omega_r\sim a\ell+b,
\label{eq:bound_cond}
\eeq
where $b$ is a constant that depends on the overtone number (see figure  \ref{omega-r}), and
\beq
a=\lim_{x\to 0}\frac{A(x)^{1/2}}{r(x)}
\eeq
is the angular frequency of the stable null geodesic. The above result tells us that the frequency of the trapped modes is evenly spaced with the overtone number. Such characteristic generates interesting patterns in the transmission coefficient. 

In addition to the trapped modes, an approximation based on the Breit-Wigner expression for nuclei scattering can be used to describe the absorption cross section. Essentially, when $\omega\approx \omega_r$, we have \cite{Macedo:2018yoi}
\beq
\left.\Gamma_{\omega \ell}\right|_{\omega\approx\omega_r}\propto\frac{1}{(\omega-\omega_r)^2+\omega_i^2},
\eeq
where $\omega_i$ is the imaginary part of the mode. We can see that the transmission factor peaks at $\omega=\omega_r$ with a height that depends on the imaginary part of the mode. Conversely, the above expression can also be used to extract the frequencies of the trapped modes from the computation of the transmission factor.

\begin{figure}
\centering
\hspace{-1cm}\includegraphics[scale=0.8]{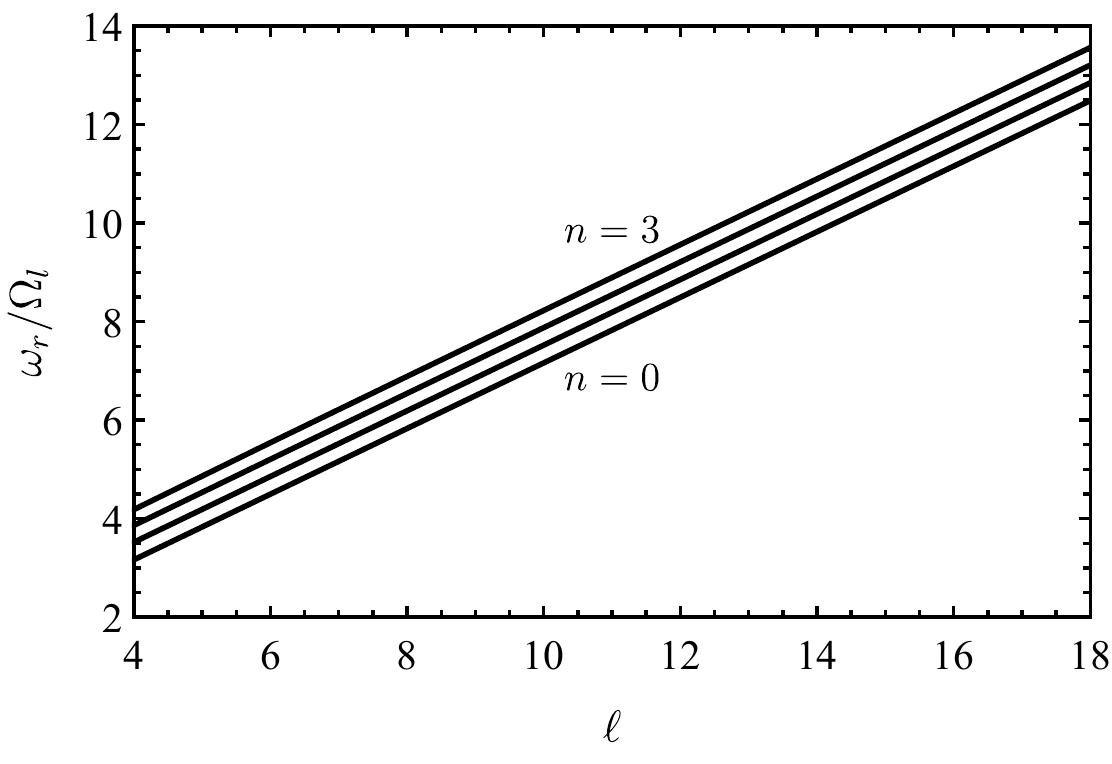}%
\caption{Real part of the fundamental ($n=0$) and first three overtones ($n=1,2,3$) frequencies of the trapped modes, obtained through (\ref{eq:quantization}), as a function of $\ell$, for the case $n_s=0.9$.}%
\label{omega-r}%
\end{figure}

%%%%%%%%%%%%%%%%%%%%%%%%%%%%%%%%%%%%%%%%%%%%%%
\section{Absorption cross section and phenomenological implications }
\label{sec:NR}
%%%%%%%%%%%%%%%%%%%%%%%%%%%%%%%%%%%%%%%%%%%%%%

We can now analyse the numerical results for the absorption cross section of planar massless scalar waves by BH remnants. The absorption properties are intrinsically related to the geodesic quantities, as noted before. Therefore we choose to normalize the absorption cross section by its corresponding classical limit. Such normalization brings our results closer to observational quantities, and it also makes easier to compare them with those obtained for BHs within GR.

\begin{figure*}[h]
\centering 
\hspace{-0.5cm}\includegraphics[scale=0.8]{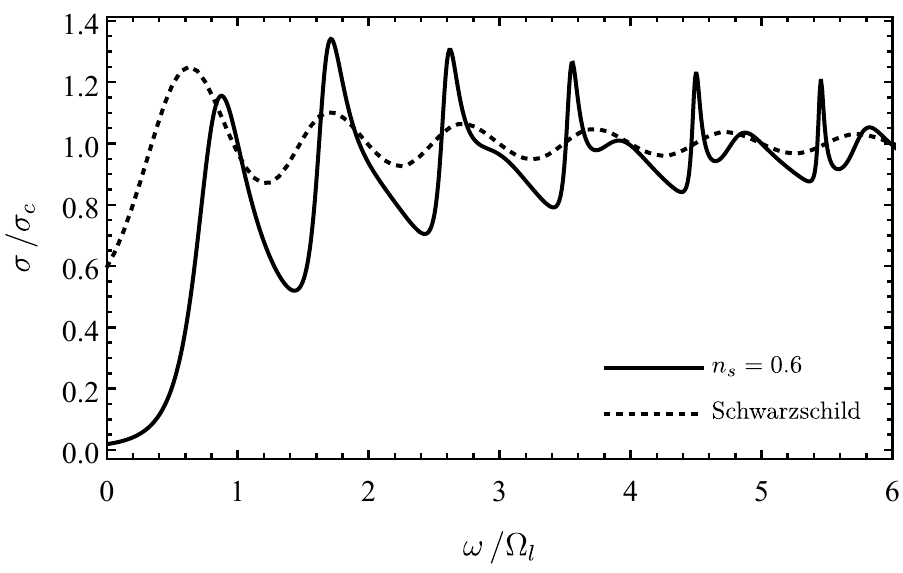}\includegraphics[scale=0.8]{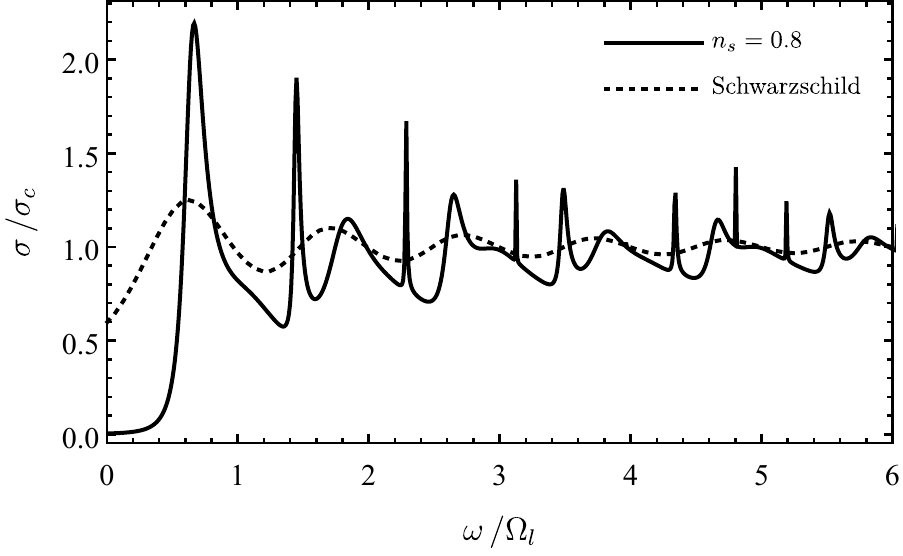}\\
\hspace{-0.5cm}\includegraphics[scale=0.8]{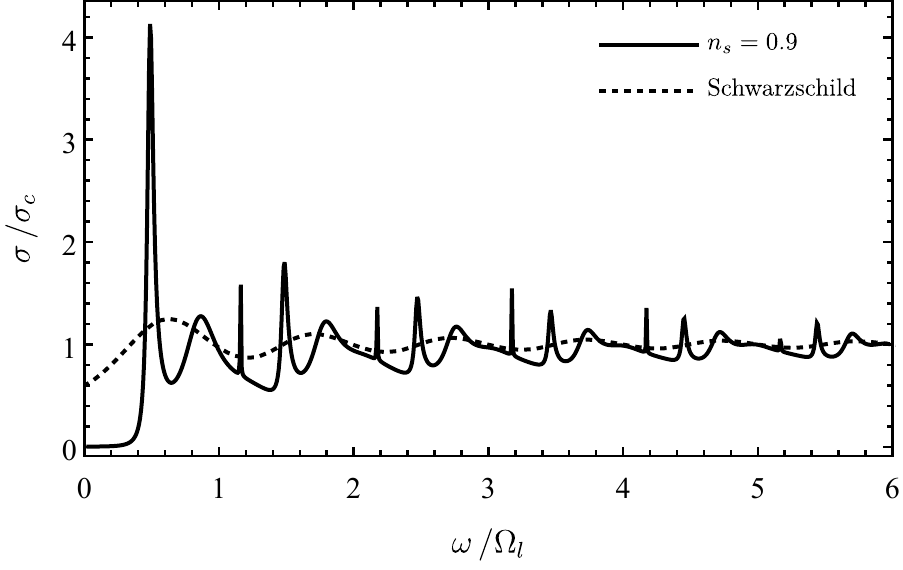}\includegraphics[scale=0.8]{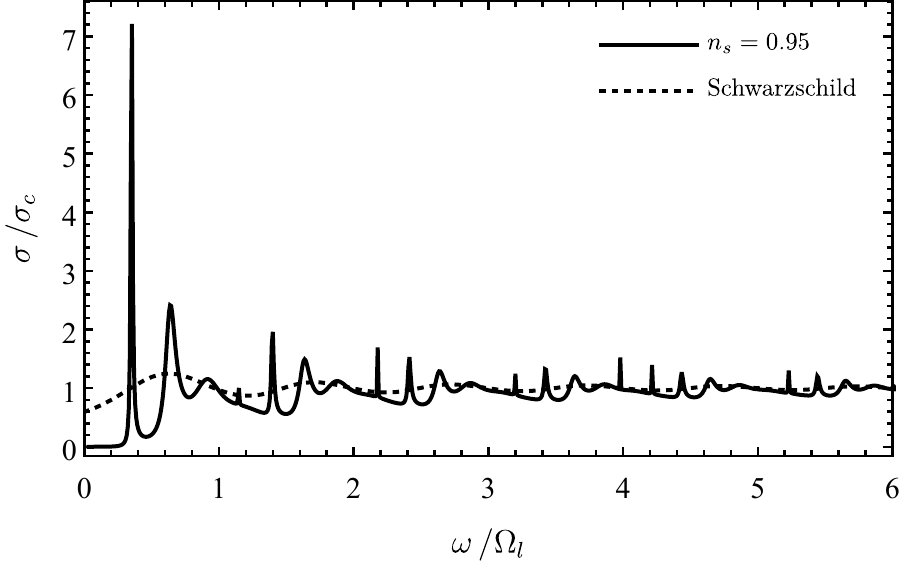}
\caption{Scalar absorption cross section of BH remnants for different values of $n_s$. The absorption cross section is normalised by the classical cross section and the frequency by the light-ring frequency. Narrow peaks arise when trapped modes exist in the potential well. The dotted lines correspond to the Schwarzschild BH case.}%
\label{fig:ScalarAbsCross}%
\end{figure*}

In figure  \ref{fig:ScalarAbsCross} we plot the absorption cross section for massless planar scalar fields as a function of the frequency, which we normalised by the light-ring frequency value $\Omega_l$ given by \eqref{eq:frequency_null}. The absorption cross section is normalised by its classical counterpart, so that the plots in figure  \ref{fig:ScalarAbsCross} tend to unity in the high-frequency regime. We note that the absorption in the low-frequency regime is different from the Schwarzschild BH result, showing a Breit-Wigner type resonant behavior for some given frequencies, indicating the presence of trapped modes in the potential well around the wormhole throat. This result is analog to recent findings regarding the absorption spectrum of ClePhOs, as reported in  \cite{Macedo:2018yoi}. We note that the (normalised) number of charges $n_s$ is analog to the absorption parameter $\mathcal{K}$ of  \cite{Macedo:2018yoi}. figure  \ref{fig:TransfCoeff} is a plot of the transmission coefficient of BH remnants as a function of the frequency. 
From the left panel of figure  \ref{fig:TransfCoeff}, it can be seen that for a fixed multipole $\ell$, the number of peaks increases as $n_s$ approaches the unity. Moreover, the number of peaks for a fixed value of $n_s$ increases as we increase the multipole number $\ell$, as it can be seen in the right panel of figure  \ref{fig:TransfCoeff}. These different peaks enter at different frequency regimes, as it can be seen in the absorption plots of figure  \ref{fig:ScalarAbsCross}. For a peak to be pronounced in the absorption spectrum it has to have a frequency high enough to penetrate the potential barrier, \ie $\omega^2\sim V_{\textrm{eff},max}$. We note from (\ref{sigmagamma}) that the absorption cross section contains a multiplicative factor of $\omega^{-2}$.

 \begin{figure*}[h]
 \centering
\hspace{-0.7cm}\includegraphics[scale=0.97]{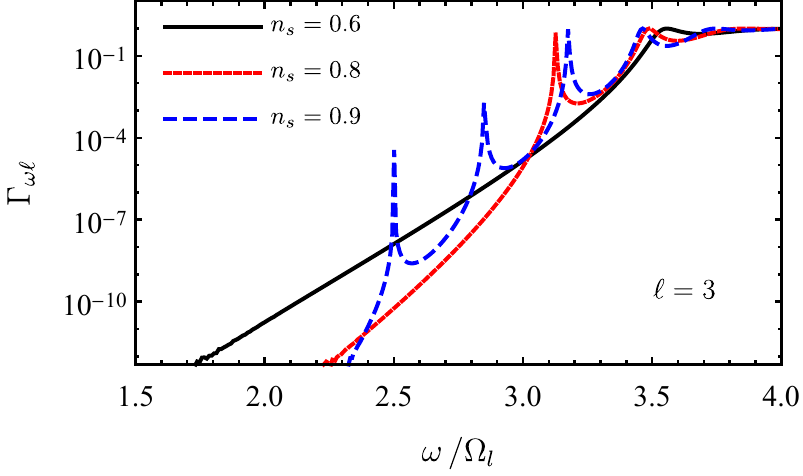}\includegraphics[scale=0.7]{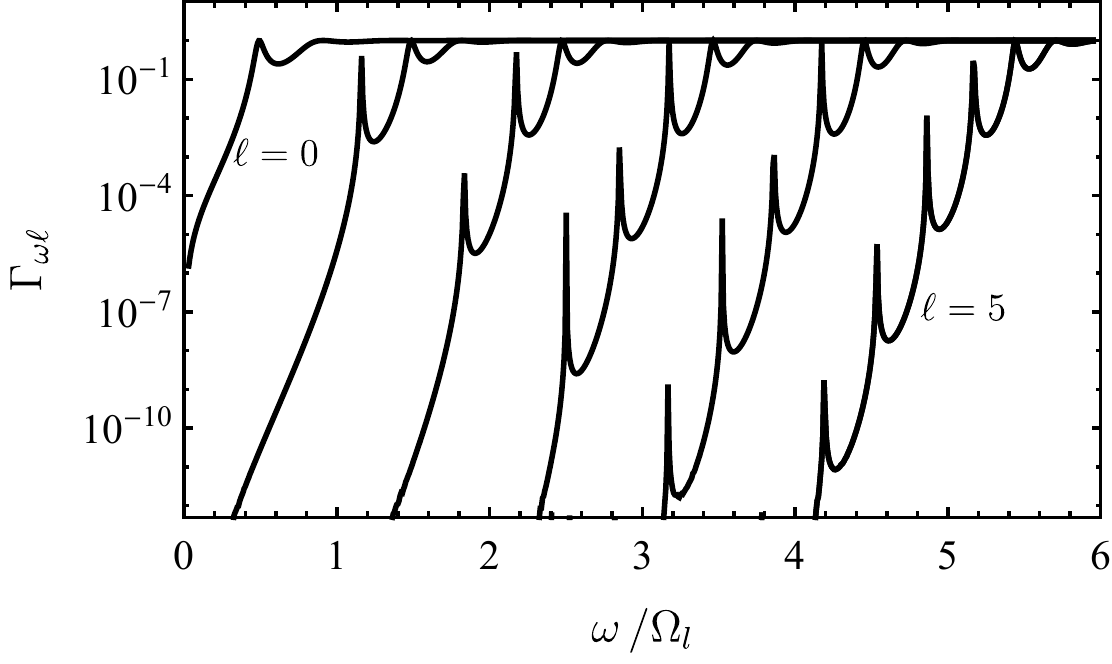}%
\caption{Representative cases for the transmission coefficient in BH remnants. \textit{Left panel:} As noted in the behavior of the effective scalar potential $V_\varphi$, trapped modes are more likely to arise for $n_s\approx 1$, and this feature impacts the transmission coefficients, generating resonant peaks. \textit{Right panel:} In addition to the $n_s$ dependence, more resonant peaks appear for higher values of $\ell$, as illustrated here for the $n_s=0.9$ case.}%
\label{fig:TransfCoeff}%
\end{figure*}

The results presented in the previous section indicate that the family of objects considered in this work has similar absorptive properties as ClePhOs. The existence of a Breit-Wigner like structure is due to the fact that the remnants, like ClePhOs, have an inner structure with two characteristic surfaces on which the waves can resonate. Nonetheless, though the peaks in the absorption spectrum of dissipative star-like ClePhOs are similar to the case of the BH remnants explored here, the overall absorption is different, what may provide an observational discriminator for the existence of event horizons in different kinds of compact objects. We notice that at higher frequencies, the absorption by BH remnants tends to the Schwarzschild BH result, being equivalent to the capture cross section of null geodesics. This is not the case for weakly dissipative star-like ClePhOs, for which the high-frequency absorption cross section tends to $\sigma_c(1-|{\cal K}|^2)$, with $| {\cal K}| \approx 1$. Therefore, there is clearly a distinctive signature of star-like ClePhOs that allows to discriminate them from BH remnants. \footnote{Recall that for $n_s=1$ our solutions develop a horizon, so that we are restricted to $0<n_s<1$.}

Another feature of these BH remnants that could be analysed in order to find observable discriminators from standard BHs and/or other ECOS is  their emission spectrum.  In this regard, note that the emission spectrum of ECOs will also have characteristic lines  described by $\Gamma_{\omega l}$ and, therefore, their emission spectrum will probably also be similar to that of ClePhOs, what further hinders their distinguishability within the ECO family. We also note that, since these features depend on the geometric properties of the objects, electromagnetic and gravitational perturbations may present similar characteristics. The work presented through this chapter represents a first step in understanding the phenomenological implications tied to these BH remnants regarding its interaction with external perturbations, and we have focused on the scalar case for simplicity. While this is important to get a grasp on more complex structures, it should be clear that an analysis of the full gravitational wave perturbations is needed to further quantify the physical phenomena explored here. In this sense, we expect that a Breit-Wigner like spectrum will still be present for gravitational waves and will converge to the scalar field one in the high-frequency limit, where the geodesic approximation is valid. However, at lower frequencies it is difficult to anticipate quantitative results. Though such an analysis is not yet available for the RBG family of theories, some general conclusions can be extracted from the basic properties of these theories. In particular, given that Ricci-Based gravity theories recover Einstein's equations in vacuum, the propagation of gravitational perturbations only involves two polarizations that travel at the speed of light, which is an important viability test for modified theories of gravity, especially after the simultaneous observation of gravitational and electromagnetic radiation from a neutron star merger \cite{Lombriser:2015sxa,Lombriser:2016yzn,Baker:2017hug,Sakstein:2017xjx,Creminelli:2017sry,Ezquiaga:2017ekz,Ezquiaga:2018btd}. Additionally, since the modified dynamics of RBGs manifests itself via nonlinearities induced by on the matter sector, the coupling between gravitational and matter modes must be important, potentially leading to new observational features. Thus, in the general case, a dedicated analysis of the perturbation equations for gravitational waves and their phenomenological implications should be carried out for each gravity theory that generates the line element (\ref{metric}). In this respect, since the remnants considered here have an electric charge, gravitational perturbations couple with electromagnetic ones generating new modes and more spectral lines in the absorption spectrum, which could be analised by extending the methods previously developed in the literature  \cite{Macedo:2014uga,Crispino:2010fd, Zerilli:1974ai,Crispino:2008zz,Crispino:2014eea,Crispino:2015gua}. Therefore, ideally, one can potentially observe the gravitational sector through  its imprint on the electromagnetic one, and the new couplings generated by the modified dynamics could help to discriminate between GR and other theories. In this sense, we notice that the effects of the nonlinearities of RBG theories have only been studied explicitly in microscopic systems \cite{Latorre:2017uve,Delhom:2019wir,BeltranJimenez:2021iqs}, and astrophysical scenarios shall reveal new physical implications of these nonlinear couplings. 

The work that has led to the elaboration of this chapter aimed to be the first step in the characterisation of the interactions between wormhole ECOs and matter fields, revealing that they present absorptive spectral features very similar to those of star-like ClePhOs \cite{Macedo:2018yoi}. The implications of such result are two-folded: (i) They allow to distinguish ECOs from standard GR BHs at the observational level, and could also be used in order to discriminate between the different modified gravity approaches that are studied today and do not predict the existence of ECOs; (ii)  They can be used to distinguish wormhole and star-like ClePhOs, since their absorption spectra have distinctive features, like the high-frequency limit. Following the classification of the solutions provided in Sec. \ref{sec:framework}, our study has focused on spherically symmetric compact objects with $\delta_1=\delta_c$ and $0<n_s<1$, for which there is no event horizon (which are called BH remnants). The mass spectrum of this set of solutions is bounded above by (approximately) the Planck mass \cite{Olmo:2013gqa}, limiting their astrophysical motivation. \footnote{More massive alternative BH remnants are possible if nonlinear effects in the matter sector are taken into account \cite{Olmo:2013mla}.}  Nonetheless, our analysis paves the road to the study of the spectral properties of  other types of solutions with higher astrophysical relevance.  In particular, the solutions studied here are geodesically complete and possess bounded curvature scalars everywhere. But there exists another branch of solutions, with different charge-to-mass ratio, $\delta_1>\delta_c$, for which curvature scalars diverge at the wormhole throat, despite being geodesically complete as well. In this part of the spectrum we find what could be seen as naked divergences, \footnote{The geodesic completeness of such naked objects would make cosmic censorship hypotheses unnecessary.} as opposed to naked singularities (which are geodesically incomplete). 
Since the interaction of matter fields with regions of extreme (even divergent) curvature is well defined in scenarios with wormholes (see  \cite{Olmo:2015dba} for a concrete example involving the model considered here and \cite{Giveon:2004yg,Olmo:2017fbc} for different models in GR), it is important to evaluate in detail the observable impact that such curvature divergences might have on the absorption and emission spectra of such objects.
%%%%%%%%%%%%%%%%%%%%%%%%%%%%%%%%%%%%%%%%%%%%%%%%%%%%%%%
%%%%%%%%%%%%%%%%%%%%%%%%%%%%%%%%%%%%%%%%%%%%%%%%%%%%%%%
\chapter{Ghosts in metric-affine theories of gravity}\label{sec:UnstableDOF}

\initial{I}n field theories, physical observables are related to functions over spacetime which we call fields and which obey some set of partial differential equations which are generally linear or quasilinear and well-posed. These equations will admit a set of exact (\ie nonperturbative) solutions called backgrounds or vacua. We are here interested in field theories which admit perturbative wave-like solutions propagating on top of such vacua. Particularly on the stability of such vacua under initially small perturbations.

Although mathematically acceptable, solutions that grow unboundedly are pathological from the physical point of view, as they typically predict divergences on some observables which have catastrophic consequences that are not observed. Of course, if the rate of growth of perturbations can be made small enough, then such predictions can be made compatible with current observations. As well, if the solutions that grow unboundedly are calculated using some approximations, then the best we can conclude is that the approximation is not physically valid. In particular, if one finds that perturbations on top of a particular background grow unboundedly,\footnote{Typically, perturbations that grow unboundedly are also called unstable degrees of freedom.} this does not  generally allow to conclude that the theory is physically meaningless, but rather that such background is unstable.

From a classical point of view, though they are legitimate solutions, unstable backgrounds are seen as pathological because the set of initial conditions that leads to this solution has zero phase space volume, \ie perfectly fine tuned conditions are required to reach such a physical situation. From the quantum mechanical point of view, the situation is even more dramatic as, even if having perfect fine tuning in the initial conditions, quantum fluctuations would always destabilise the background. Therefore, unstable backgrounds are only acceptable if the timescale of the instabilities is big enough to be compatible with observations.

The presence of unstable degrees of freedom in gravitational theories has played a prominent role in determining physically viable theories beyond GR, and particularly in the search for a UV complete theory of gravity. To begin with, the presence of ghostly instabilities already shows up when building a general kinetic term for a (massless or massive) spin-2 field. To see this, let us accept that spin-2 fields are naturally described by a rank-2 symmetric tensor field\footnote{Note that, in general, these fields have 10 independent components, and therefore can describe also additional degrees of freedom.A massless spin-2 field carries 2 degrees of freedom while a massive spin-2 field carries 5.} $h_{\mu\nu}$. The most general local and Lorentz invariant kinetic term for this type of field (around a Minkowskian background) is of the form
\begin{equation}
\cl_2^K=\frac{1}{2}\partial^{\alpha} h^{\mu \nu}\left(b_{1} \partial_{\alpha} h_{\mu \nu}+2 b_{2} \partial_{\mu} h_{\nu \alpha}+b_{3}\eta_{\mu \nu} \partial_{\alpha} h +2 b_{4}  \eta_{\alpha\mu}\partial_{\nu} h\right),
\end{equation}
and it is well known that, unless the coefficients satisfy $b_1=b_4=-b_2=-b_3$, there will be Ostrogradski ghosts (see section \ref{sec:Ostrogradski} below). This can be seen, for instance, by decomposing $h_{\mu\nu}$ as
\beq
h_{\mu\nu}=\hT_{\mu\nu}+2\partial_{(\mu}\xi_{\nu)}
\eeq
where $\partial^\nu\hT_{\mu\nu}=0$. Then $\hT_{\mu\nu}$ has 6 independent components and $\xi_\mu$ the other 4. Now, with this decomposition, it is apparent that the above terms will give rise to  Ostrogradskian instabilities unless the coefficients $b_i$ are tuned so as to avoid second-order derivatives for $\xi_\mu$ in the Lagrangian. Therefore, the requirement of absence of ghostly degrees of freedom in a theory for a Lorentz invariant symmetric rank-2 tensor field uniquely fixes the form of the kinetic term, which is the linearised version of the Einstein-Hilbert term\footnote{Let us point out that the Maxwellian kinetic term is also the unique Lorentz invariant and local kinetic term for a vector field that guarantees the stability of the field around Minkowski.}. Note that this form of the kinetic term is oblivious to $\xi_\mu$,  which implies that it is invariant under gauge transformations $h_{\mu\nu}\mapsto h_{\mu\nu}+\partial_{(\mu}\xi_{\nu)}$. This has the consequence that $h_{\mu\nu}$ only propagates 2 degrees of freedom corresponding to a massless spin-2 field \cite{deRham:2014zqa,Hinterbichler:2011tt}.

If we now try to add a mass term for the tensor field, to maintain Lorentz invariance it must be proportional to $h_{\mu\nu}h^{\mu\nu}-a h^2$ where $h=\eta^{\mu\nu}h_{\mu\nu}$. Because this mass term breaks the gauge symmetry of the kinetic term, it will generally provide a kinetic term for the $\xi$ modes which, unless $a=1$, can be seen to propagate a ghostly scalar degree of freedom already at the linear level (see \eg \cite{deRham:2014zqa,Hinterbichler:2011tt}). Hence, we see that absence of ghosts at the linear level also fixes a possible mass term for the graviton to be the Fierz-Pauli mass term, which (for $b_1=-1$) leads to the Fierz-Pauli Lagrangian 
\beq
\cl_2=-\frac{1}{2}\partial^{\alpha} h^{\mu \nu}\partial_{\alpha} h_{\mu \nu}+\frac{1}{2}\partial^{\alpha} h^{\mu \nu}\partial_{\mu} h_{\nu \alpha}- \partial_{\mu} h^{\mu\nu}\partial_{\nu} h+\frac{1}{2}\partial^{\alpha} h^{\mu \nu}\eta_{\mu \nu} \partial_{\alpha} h -\frac{1}{2} m^{2}\left(h_{\mu \nu} h^{\mu \nu}-h^{2}\right)
\eeq
as the unique Lagrangian for a massive spin-2 field that does not propagate ghosts at the linear level and around Minkowski spacetime. By decomposing the vector field $\xi_\mu$, which plays the role of a St\"ueckelberg field (see section \ref{sec:Stueckelberg} below), as $\xi_\mu\mapsto A_\mu+\partial_\mu\pi$, we see that the Fierz-Pauli mass term provides dynamics to both $A_\mu$ and $\pi$ which propagate a helicity-1 and a scalar degrees of freedom respectively. Thus, the Fierz-Pauli theory for a massive spin-2 field propagates 5 degrees of freedom corresponding to the helicity-2, 1 and 0 modes.

Nevertheless, the linear level is not the end of the story.  The nonlinear terms introduced by the Fierz-Pauli mass terms can be seen to provide higher-order derivatives for $\pi$ that, though not relevant around trivial backgrounds, can become relevant around some nonlinear background configuration. This implies that there will always be backgrounds around which the higher derivative terms excite a sixth degree of freedom that will be an Ostrogradski ghost known as the Boulware-Deser ghost \cite{Boulware:1973my,deRham:2014zqa,Hinterbichler:2011tt}. We then see that one of the more immediate modifications to GR from the field theory point of view, \ie giving a mass to the graviton, already gives rise to unstable degrees of freedom easily. Remarkably, recent findings show that there are nontrivial ways to evade such instabilities leading to ghost-free massive gravity \cite{deRham:2010ik,deRham:2010kj} (see also \cite{deRham:2014zqa} and references within).

From the geometrical perspective, there is a central result known as Lovelock theorem that essentially goes in a similar direction. In 4 spacetime dimensions there is only one geometric object that can be built off the metric and its first and second-order derivatives which is divergence free and symmetric \cite{Lovelock:1972vz}. By the Bianchi identity under diffeomorphisms, this implies that in 4 dimensions there is only one diffeomorphism invariant Lagrangian (up to boundary terms) that is built solely from the metric and its (first\footnote{Note that though the Einstein-Hilbert term contains second derivatives of the metric, these can be seen to be a boundary term. Indeed, the original form of the action for GR by Einstein did not contain the derivative terms \cite{EinsteinOriginalAction}.}) derivatives that gives rise to second-order field equations for the metric. In turn, this leads to the finding of the $k$-th order Lovelock terms which generalise the Einstein-Hilbert term and become a boundary term in $2k$ spacetimes dimensions. The Einstein-Hilbert term is the first-order Lovelock term and is a boundary term in 2 spacetime dimensions, and in 4 spacetime dimensions, the only nontrivial Lovelock terms are the Einstein-Hilbert term and the Gauss-Bonnet term
\beq
R_{\alpha\beta\mu\nu}R^{\alpha\beta\mu\nu}-4R_{\mu\nu}R^{\mu\nu}+R^2,
\eeq
which is a boundary term in $D=4$. Hence the only diffeomorphism invariant action that gives second-order field equations in 4 spacetimes dimension is (dynamically equivalent to) the Einstein-Hilbert action. This leads to the conclusion that no modifications of GR which contain only a metric field and that keep diffeomorphism invariance can be formulated in 4 spacetime dimensions. Any diffeomorphism invariant modification would thus include extra fields (\ie degrees of freedom). Particularly, if new diffeomorphism invariant terms (different from the Gauss-Bonnet term) are added to the Einstein-Hilbert action, the resulting field equations for the metric will be of higher-order, leading to the propagation of Ostrogradski ghosts. 

These results contrast with the motivation that stems both from quantum field theory in curved spacetimes and from considering quantum corrections for the gravitational field. On the one hand, the renormalizability of matter fields in curved spaces was seen to require the presence of quadratic curvature terms in the effective action \cite{Utiyama:1962sn,Parker:2009uva}. On the other hand, the nonrenormalizability of GR \cite{tHooft:1974toh,Donoghue:1994dn} motivated the exploration of theories with higher-order curvature invariants in the quest for a UV complete theory of gravity. A key result in this direction is the classic work by Stelle \cite{Stelle:1976gc} in which a theory with quadratic curvature corrections to the Einstein-Hilbert term was proven to be renormalizable. Though this was the first positive result in finding a UV complete theory of the gravitational interaction, it was seen that the theory suffers from a fatal drawback for it to make physical sense as a fundamental theory: it was shown to either be nonunitary or contain a massive spin-2 ghost in its spectrum. This result was questioned \cite{Tomboulis:1983sw,Antoniadis:1986tu} due to the identification of the ghost from the bare propagator, which was not correct in the presence of unstable particles, such as the ghost was. The criticisms argued that if the correct (dressed) propagator was used, the ghost poles were gauge dependent and thus not physical. However, later work contradicted such claim showing the gauge independence of the ghost poles \cite{Johnston:1987ue}. On another line, there are recent results suggesting that there might be a way of quantising higher-order derivative theories in a way which avoids both ghost degrees of freedom and loss of unitarity. This is achieved by requiring only an antilinear Hamiltonian instead of a hermitian one, which has implications on what is the Hilbert space of physical states for the quantised theory \cite{Bender:2007wu,Mannheim:2018ljq,Mannheim:2020ryw,Mannheim:2021xrg}. If these results are correct, extending them to Stelle's renormalizable model would be a major achievement in the field.

Due to the results mentioned above, the idea that the appearance of ghosts in renormalizable theories of gravity was due to the higher-order derivative terms in the field equations introduced by the higher-order (metric) curvature terms permeated the community. Then, researchers in the field of metric-affine (also Palatini or 1st order) gravity theories realised that these higher derivatives of the metric do not appear when higher-order curvature invariants are considered in the metric-affine approach, where the connection is {\it a priori} independent of the metric and the Riemann has only first-order derivatives of the connection. Apparently, the idea that metric-affine theories would be free of ghosts due to this property spread through the (more geometrically oriented) community. This idea was reinforced by the discovery that several theories that contain ghosts when formulated in the metric formalism, such as \eg Born-Infeld or Stelle's quadratic gravity, are ghost-free when formulated in the metric-affine approach. However, although higher-order derivatives are sufficient to have ghosts as shown by Ostrogradski, their presence is not a necessary condition for a theory to contain unstable degrees of freedom. 

In this chapter we will present the results of a joint work with Jose Beltr\'an Jim\'enez in which we disprove the widespread belief that metric-affine theories of gravity do not contain ghosts \cite{BeltranJimenez:2019acz,Jimenez:2020dpn}. We show that a class of metric-affine theories whose action is an arbitrary function of the metric and the Ricci tensor generally contains ghosts unless projective symmetry is imposed. These findings allowed us to argue why generic metric-affine theories will contain ghost degrees of freedom in their spectrum, and that care should be taken in their formulation if one wants to avoid their presence. Our findings point in the same direction as other recent works \cite{Percacci:2019hxn,Aoki:2019rvi}.

%%%%%%%%%%%%%%%%%%%%%%%%%%%%%%%%%%%%%%%%%%%%%%%%%%%%%%%
\section{Instabilities and their physical implications}\label{sec:TypesOfInstabilities}
%%%%%%%%%%%%%%%%%%%%%%%%%%%%%%%%%%%%%%%%%%%%%%%%%%%%%%%
We are interested in classifying different type of instabilities that can arise in field theories admitting wave-like perturbations around their vacua according to their physical implications. For our purpose, it will suffice to consider scalar perturbations around a nontrivial vacuum which varies with a characteristic time $T$ and length scale $L$. That vacuum could be an exact solution for the same scalar field, a gravitational background, or any exact solution for the fields in the theory. We will only worry about perturbations of the scalar field, described by the scalar degree of freedom $\phi$. The results presented here can be found with more detail in \eg \cite{Rubakov:2014jja,Cline:2003gs,Carroll:2003st,Joyce:2014kja}.

On such a background, the leading order perturbations of the real scalar field are described by a Lagrangian of the form
\begin{equation}
\cl_\phi=\frac{1}{2}\big(a \dot\phi^2- b (\partial_i\phi)^2-\mu \phi^2\big),
\end{equation}
where $a$, $b$, $\mu$ are coefficients that vary on the characteristic scales of the background and we use the mostly minus signature. Neglecting the variations of the background, the energy density and field equations are
\begin{equation}
T_{00}=\frac{1}{2}\big(a \dot\phi^2+b(\vec{\na}\phi)^2+\mu \phi^2\big)\qquad\quad\text{and}\qquad\quad\ddot \phi-\frac{b}{a}\na^2\phi+\frac{\mu}{a}\phi=\mathcal{O}(T^{-1},L^{-1})
\end{equation}
and
\begin{equation}\label{eq:SolutionKleinGordonInstabilities}
\phi=A e^{i\lr{\omega t-\sqrt{\frac{b}{a}}\vec{k}\cdot\vec{x}}}+B e^{-i\lr{\omega t-\sqrt{\frac{b}{a}}\vec{k}\cdot\vec{x}}}\quad\qquad\text{where}\quad\qquad \omega^2=a^{-1}\lr{b\vec{k}^2+\mu}
\end{equation}
are a basis of solutions and corresponding dispersion relation up to $\mathcal{O}(T^{-1},L^{-1})$ corrections. Note that the perturbations have an effective mass $m_{\textrm{eff}}= \sqrt{|\mu|/a}$. Depending on the sign of $a$, $b$ and $\mu$ the perturbations remain small or become unstable in different ways. For illustrative purposes, it will suffice to consider a spatially homogeneous background slowly varying in time, though the arguments generalise in a straightforward manner.  Let us depict all the possibilities.

\noindent\textbf{Stable case:} $a>0$, $b>0$ and $\mu\geq0$

In this case $\omega$ is always real and the energy density is always positive. The perturbations remain bounded propagate at speed $\sqrt{b/a}$ in natural units. If $b\leq a$ then the perturbations travel at subliminal speeds. If $b>a$ then they are superluminal which, although it is not a problem regarding their stability, it signals that the theory is not the low energy description of a Lorentz invariant UV complete quantum theory \cite{Adams:2006sv}. If $b=0$ the perturbations travel at the speed of light, but one has to be careful for even the smallest modification in the background (\eg the backreaction of the perturbations) could make them be superluminal.

\noindent\textbf{Tachyonic instability:} $a>0$, $b>0$ and $\mu<0$

Although high-momentum perturbations are fine, $\omega$ becomes imaginary for sufficiently low momentum, namely for $|\vec{k}_{\textrm{low}}|<|\mu|/b$. This has the consequence that, at late times, low-momentum modes become dominated by an exponential growth as $\phi\sim e^{|\omega_{\textrm{low}}|t}$ where $|\omega_{\textrm{low}}|\leq m_{\textrm{eff}}$.  The characteristic time of the instability, namely the time when the exponential growth becomes dominant will roughly be $t_c\sim  m_{\textrm{eff}}^{-1}$. We then have to distinguish two cases: 1) when $t_c\ll T$ the exponential growth of the perturbations destabilises the background driving the system nonperturbatively far from it. That vacuum is therefore unstable. 2) when $t_c\gg T$ we have that the background evolution is much quicker than the time required for the instability to develop and therefore the background is stable for times $t\ll t_c$, when perturbation theory gives valid predictions provided that there is a regime where $m_{\textrm{eff}}\ll T^{-1} \ll \omega$ (or $\omega^{-1} \ll T \ll t_c $), \ie when the high energy modes that are stable are also insensitive to the background evolution. Given that this instability was derived from perturbation theory over an almost constant background, the results cannot be trusted and one needs to perform a nonperturbative analysis of the full system to know its stability properties at late times.

\noindent\textbf{Gradient or Laplacian instability:} $a>0$ and $b<0\quad$ or $\quad a<0$ and $b>0$

This kind of instability is always problematic since for high-momentum modes $\omega$ becomes imaginary and the perturbations are dominated by an exponential growth $\phi\sim e^{k t}$ with an arbitrarily fast growth rate and the background is therefore unstable. Note that for the cases with $a\mu>0$ the modes with low-enough momentum $|\vec{k}|\leq \sqrt{|\mu/b|}$ do not develop instabilities. Hence one might think that in this case, an effective theory for the low momentum with a suitable cutoff $\Lambda$ modes could be physically sound over such background. However note that in this case, low-momentum modes have a characteristic time far above the cutoff of the EFT $t_{\textrm{low}}>>t_\Lambda\sim\Lambda^{-1}$ and therefore will be sensitive to the instability, while the modes with $t_k<<t_\Lambda$ are far above the cutoff of the EFT ($k\gg\Lambda$). Hence a theory with a gradient instability is physically meaningless, \ie it makes no reliable predictions.

\noindent\textbf{Ghostly instability:} $a<0$ and $b<0$

In this case, high momentum modes are stable and at the classical level, we should only worry about a tachyonic instability for the low-momentum modes in the case that $\mu>0$. Nevertheless, this case is highly problematic if the perturbations are quantised, for they violate either conservation of probability or carry negative energy (see \eg \cite{Cline:2003gs}). Accepting that we do not want to deal with nonunitary theories, let us elaborate on what would happen in the case of the quanta carrying negative energy. In the best case scenario, these quanta couple only to gravity.\footnote{Recall that, for consistency reasons, a massless spin-2 field must couple universally as explained in chapter \ref{sec:GravityAsGeometry}} Given that they carry negative energy, the process $\ket{0}\rightarrow \phi\phi + \gamma\gamma$ mediated by a graviton is kinematically allowed. Moreover, given that the final momentum of the particles can be arbitrarily high while keeping energy conservation due to the negative energy of the ghosts, the phase space integral diverges and the creation rate is arbitrarily large unless the theory is treated as an EFT with a Lorentz-violating cutoff $\Lambda$. In such case, the decay rate of the vacuum due to this process will be of order $\Lambda^8{\mpl}^{-4}$. Arguing in this manner, an upper limit for the cutoff of an EFT with ghosts was set from the observations of the gamma ray spectrum coming from the universe \cite{Cline:2003gs}. Thus, we see that although ghosts do not necessarily predict classical instabilities, their existence in a quantum theory precludes its viability unless it is an EFT and the cutoff is sufficiently low. The possibility that nonperturbative physics might stabilise the vacuum through a ghost condensate (a vacuum expectation value for the ghost field) has also been discussed \cite{ArkaniHamed:2003uy}.

These arguments can be generalised for systems with more degrees or freedom as follows. Consider the Lagrangian describing perturbations for $N$ degrees of freedom encoded in the spacetime functions $\phi^1,...,\phi^N$
\begin{equation}\label{eq:LagSeveralDOF}
\cl_\phi=\frac{1}{2}\big(a_{IJ} \dot\phi^I\dot\phi^J- b_{IJ} (\partial_i\phi^I)(\partial_i\phi^J)-\mu_{IJ} \phi^I \phi^J\big),
\end{equation}
by diagonalising $a_{IJ}$, $b_{IJ}$ and $\mu_{IJ}$ we can know whether any of the degrees of freedom is unstable and how. Negative eigenvalues of $b_{IJ}$ and /or $a_{IJ}$ imply the existence of ghost and/or gradient instabilities, and negative eigenvalues of $\mu_{IJ}$ signal the presence of tachyonic instabilities. If we want to identify which are the pathological degrees of freedom, then we must perform field redefinitions such that they lead to canonical and diagonal kinetic matrix. If it is possible to do so, then it will be possible to identify the pathological degrees of freedom and the nature of their pathologies. 

As a remark, let us point out that, here, each $\phi^I$ should be a {\it truly propagating} degree of freedom. Though this is strictly redundant, sometimes the word degree of freedom is misused for fields whose dynamics may be constrained by the field equations. We are assuming that in \eqref{eq:LagSeveralDOF} the $\phi^I$ are all dynamical, and the constraints have already been integrated out. For instance, in a Proca theory described by $A_{\mu}$, there are only three propagating degrees of freedom, given that the $0$ component of the 1-form field is constrained. Thus $I$ runs only from 1 to 3, and one cannot assume $\phi^{I_\mu}=A_{\mu-1}$ for all values of $I$. This usually complicates a full stability analysis in theories that contain fields which are not scalars, where the correspondence between the field components and the propagated degrees of freedom is not straightforward.

%%%%%%%%%%%%%%%%%%%%%%%%%%%%%%%%%%%%%%%%%%%%%%%%%%%%%%%

\subsection{Ostrogradski ghosts}\label{sec:Ostrogradski}

There is a powerful theorem by Ostrogradski \cite{Ostrogradski:1850fid} that poses a major restriction to the allowed Lagrangians that can describe a fundamental theory. The key consequence of this theorem is that any nondegenerate\footnote{By nondegenerate, we mean that the higher-order time derivatives cannot be integrated by parts and written in terms of first-order time derivatives alone plus a boundary term.} Lagrangian with time derivatives of higher order than one describes a system with ghost instabilities. This restricts any Lagrangian that is a candidate to describe a fundamental theory to have at most first-order time derivatives, or to be dynamically equivalent to a Lagrangian of such class. This restriction becomes extremely powerful if combined with Lorentz symmetry: only first-order derivatives can enter the Lagrangian of a fundamental theory, as any temporal derivative comes in hand spatial derivatives if the Lagrangian is to be Lorentz invariant (or generally covariant). Thus, this theorem plays a key role in illuminating the path towards a UV complete theory of gravity, for we know that any (purely metric) theory with higher-order curvature invariants which is not of the Lovelock or $f(R)$ forms will have ghostly degrees of freedom in its spectrum which (in principle, though see \cite{Hawking:2001yt,Bender:2007wu,Mannheim:2018ljq,Mannheim:2020ryw,Mannheim:2021xrg}) cannot be tolerated in a fundamental theory. As well, the theorem also forces us to be careful on how we couple matter fields nonminimally to curvature invariants, since generic couplings between curvature and matter fields will `excite' the piece of the curvature tensor with second order derivatives for the metric. Indeed, there has been much research on finding out the allowed couplings for different types of matter fields, see \eg \cite{Horndeski:1974wa,Kobayashi:2019hrl,Trodden:2011xh,Gleyzes:2014dya,Heisenberg:2018vsk,Babichev:2020tct}

Given its major importance, let us give a brief review of the theorem and its consequences. We will prove the theorem only for systems with a finite number of degrees of freedom, closely following the nice review \cite{Woodard:2015zca}, but see \eg \cite{deUrries:1998obu} for an analysis in field theories. To prove the theorem, let us consider a system of one degree of freedom $q(t)$ described by a Lagrangian $L(q,q^{(1)},...,q^{(N)},t)$ which is a function of $q$ and its first $N$ time-derivatives (and possibly of time). The physical trajectories are the solutions to
\beq\label{eq:EulerLagrangeOstrogradski}
\sum_{i=0}^N \lr{-\frac{d}{dt}}^i\frac{\partial L}{\partial q^{(i)}}=0
\eeq
which using the chain rule can be rewritten as
\beq
(-1)^N q^{(2N)}\frac{\partial^2L}{(\partial q^{(N)})^2}+F(q,q^{(1)},...,q^{(2N-1)})=0\;,
\eeq
leading to a $2N$-th order differential equation for $q(t)$ provided that
\beq\label{eq:NonDegeneracyN}
\frac{\partial^2L}{(\partial q^{(N)})^2}\neq0.
\eeq
If this condition does not hold, the proof still holds if there is any $n>1$ such that the above condition with the replacement $N\rightarrow n$ does hold. In that case, the system is described by a $2n$-th order differential equation, and there exists a dynamically equivalent Lagrangian which can be reached from the original one by integrating by parts and which is a function of $q$ and its first $n$ time-derivatives. The proof of the theorem that follows can be applied to that Lagrangian without loss of generality. A Lagrangian satisfying the condition \eqref{eq:NonDegeneracyN} is usually called non degenerate. However, we will use the term non-$N$-degenerate\footnote{As a remark, let us point out that the non$n$-degeneracy condition for systems with more than one particle is that the determinant of $\frac{\partial^2L}{\partial q^{(n)}_i\partial q^{(n)}_j}\neq0$.}
 for a Lagrangian satisfying  \eqref{eq:NonDegeneracyN}, so that we can use the term nondegenerate for Lagrangians that are non-$n$-degenerate for some $n>1$. Let us also point out that the non-$n$-degeneracy condition is coordinate independent. This is easily seen for a coordinate change $x(q)$, which leads to
\beq
\frac{\partial^2 \tilde L}{(\partial x^{(n)})^2}=\frac{\partial L}{(\partial q^{n})^2}\lr{\frac{d q}{dx}}^n
\eeq
and we have $d q/dx\neq0$ for any well defined coordinate change. 

Since a $2N$-th order ODE requires $2N$ pieces of initial data to determine a solution, the phase space of the system will be $2N$-dimensional and therefore we need to find $2N$ canonical coordinates $Q_i$ and $P_i$. A suitable choice of canonical coordinates is
\beq
Q_i=q^{(i-1)}\qquad\text{and}\qquad P_i=\sum_{j=i}^N  \lr{-\frac{d}{dt}}^{j-i}\frac{\partial L}{\partial q^{(j)}}=0.
\eeq
Note that $P_N=\partial L/\partial q^{(N)}$ which is a function of $(Q_1,...,Q_N,q^{(N)})$. Hence, by the inverse function theorem, non$N$-degeneracy implies that there exists a function $A(Q_1,...,Q_N,P_N)$ such that 
\beq
P_N=\left.\frac{\partial L}{\partial q^{(N)}}\right|_{\tiny{\begin{matrix}q^{(N)}=A\\q^{(i-1)}=Q_i \end{matrix}}}
\eeq
is an identity. This choice of canonical coordinates leads to the Hamiltonian
\beq\label{eq:OstrogradskiHamiltonian}
H(Q_i,P_i,t)=\left.\sum_{i=1}^{N}P_i q^{(i)}-L\right|_{\tiny{\begin{matrix}q^{(N)}=A \\ q^{(i-1)}=Q_i \end{matrix}}}
=\sum_{i=1}^{N-1}P_i Q_{i+1}+P_N A-L(Q_1,...,Q_N,A,t)
\eeq
which generates time evolution in the sense that for any observable $f(Q_i,P_i)$ we have $\dot f=\{f,H\}$. Particularly, the dynamics of the system given by \eqref{eq:EulerLagrangeOstrogradski} is recovered by 
\beq
\dot P_1=-\frac{\partial H}{\partial Q_1}=-P_N\frac{\partial A}{\partial Q_1}+\frac{\partial L}{\partial Q_1}+\left.\frac{\partial L}{\partial q^{(N)}}\right|_{\tiny{\begin{matrix}q^{(N)}=A \\ q^{(i-1)}=Q_i \end{matrix}}}\frac{\partial A}{\partial Q_1}=\frac{\partial L}{\partial Q_1},
\eeq
and the rest of the equations give the definitions of $Q_{i}$ and $P_{i<N}$ so that the two systems of equations are equivalent.

Now, notice that all the momenta enter linearly in the Hamiltonian \eqref{eq:OstrogradskiHamiltonian} except possibly $P_N$. This result is coordinate independent, as it only relies in the nondegeneracy condition, and implies that the Hamiltonian is unbounded from below. In particular, there will be at least $N-1$ degrees of freedom whose energy is not positive definite. Hence, even if the Lagrangian does not depend explicitly on time, so that the Hamiltonian is conserved, this has the consequence that the degrees of freedom carrying positive energy can be infinitely excited by exciting negative energy degrees of freedom while keeping the total energy conserved. Although this might not be a problem for a single particle system, whenever we have an interacting quantum theory which suffers from the Ostrogradskian instability (\ie a ghost), the vacuum will decay with infinite decay rate as explained above. Therefore, a theory suffering from Ostrogradski instabilities can only make physical sense as a low energy theory valid up to a cutoff scale.

Let us note that these results have also been extended to systems with odd-order equations of motion \cite{VillasenorTesis,Weldon:2003by} and have been seen to survive canonical quantisation\footnote{Although the gosts can be avoided at the cost of loosing unitarity.} \cite{Weldon:2003by,Motohashi:2020psc}, while path integral quantisation may yield a theory that recovers unitarity at low energies \cite{Hawking:2001yt}. As well, there are recent results that suggest that an alternative quantisation method that requires an antilinear Hamiltonian instead of a hermitian one could render healthy higher-derivative quantum theories \cite{Bender:2007wu,Mannheim:2018ljq,Mannheim:2020ryw,Mannheim:2021xrg}. As a final remark, let us mention a recent generalisation of this result achieved by looking into the relation between the existence of ghosts and the constraints of a system\footnote{Recall that any higher-order nondegenerate Lagrangian can always be cast equivalent to a first-order one with auxiliary fields and constraints.} \cite{Aoki:2020gfv}.

 %%%%%%%%%%%%%%%%%%%%%%%%%%%%%%%%%%%%%%%%%%%%%%%%%%%%%%%

%%%%%%%%%%%%%%%%%%%%%%%%%%%%%%%%%%%%%%%%%%%%%%%%%%%%%%%

\section{Ghosts in curvature-based metric-affine theories}\label{sec:GhostsInCurvatureBased}
%%%%%%%%%%%%%%%%%%%%%%%%%%%%%%%%%%%%%%%%%%%%%%%%%%%%%%%

In metric-affine theories the connection is an independent field and therefore higher-order curvature invariants do not introduce higher derivatives of the metric. This fuelled the hope that metric-affine higher-order curvature theories (and general metric-affine theories) could be ghost-free. Here we prove that, although there are subclasses of metric-affine theories which are ghost-free, this hope is not fulfilled when general metric-affine theories are considered. To that end we consider a particular class among all the metric-affine theories of gravity for which we know how to solve the connection (at least formally), which allows to unveil the presence of ghosts in the full nonlinear theory. Our results also show that these ghosts, which are present in the spectrum of generic metric-affine theories, cannot be cured in general by considering nonminimal matter couplings.

We will show the presence of these ghosts in two different ways, which will also clarify the instability problems of Non Symmetric Gravity theories \cite{Moffat:1978tr}. One of these ways will consist on resorting to the Stueckelberg trick, which is a nice construction that allows to take a massless limit of the action of a would-be gauge invariant massive field (the mass breaks the gauge symmetry) without loosing degrees of freedom in the process. To simplify the understanding of our results, we will begin by introducing this nice technique and showing a simplified example of how the ghosts emerge in the theories that we considered.

\subsubsection{St\"ueckelberg's trick: a warmup proxy with scalar and spin-1 fields }\label{sec:Stueckelberg}
One of the ways in which the presence of ghosts in RBGs without projective symmetry will be shown is by resorting to the decoupling limit of the St\"ueckelberg modes of a 2-form field. As we have seen above, a massless 2-form propagates a scalar degree of freedom while a massive 2-form propagates three degrees of freedom. We will see below that in the action of RBGs without projective symmetry, there appears a massive 2-form field and a vectorial projective mode which does not have a proper kinetic term, and which couples gravitationally (their coupling is $\propto{\mpl}$) to the massive 2-form. The St\"ueckelberg mechanism allows to separate two of the modes related to a massive 2-form from the third one, which can be associated to the scalar mode propagated by a massless 2-form.\footnote{See the analog result for a massive vector field in \eg \cite{Hinterbichler:2011tt}.} Then, the decoupling limit, where the mass of the 2-form vanishes, allows to decouple the scalar degree of freedom propagated by the massless 2-form from the other two. In this limit, it will become apparent that the coupling between the projective mode and the massive 2-form hides the presence of two ghostly degrees of freedom propagated by a massless vector field, which can naturally be associated to the projective mode.

 For illustrative purpose, let us consider a simpler system of a scalar and massive spin-1 field which serves as an analogy for the behaviour of (part of) the ghostly sector of an RBG without projective symmetry. This system will be described by the Lagrangian
\beq\label{eq:ExampleLagrangianStueckelberg}
\cL=-\frac14 F_{\mu\nu}F^{\mu\nu}+\lambda A^\mu\partial_\mu\varphi+\frac12m^2A^2,
\eeq
where the spin-1 and the scalar field play analog roles to the massive 2-form and the projective mode that appear in RBGs without projective symmetry respectively. The scalar imposes the constraint
\beq
\partial_\mu A^\mu=0,
\eeq
while the spin-1 field equations are
\beq
\partial_\mu F^{\mu\nu}+\lambda\partial^\nu\varphi + m^2A^\nu=0.
\eeq
Let us start by counting degrees of freedom so that we can explicitly see that the St\"ueckelberg trick preserves the number of propagating degrees of freedom after taking the decoupling limit. The expected number of propagating degrees of freedom is 4: the 3 polarisations propagated by the massive vector field and the scalar one. Although it may seem that the constraint imposed by the scalar field could alter this counting, the counting is indeed correct. To see that, let us see the number of initial Cauchy data that we need to provide as initial conditions for field equations. In principle we should give the values of all the dynamical fields and their first (time) derivatives. However, we have constraints, and some of them can actually be expressed in terms of the others. Concretely, the constraint provided by the scalar field equation can be written as
\beq
\dot{A}_0+\partial_i A^i=0
\eeq
which tells us that the time derivative of $A_0$ on the Cauchy surface is determined by the initial values of $A^i$. On the other hand, the temporal component of the vector field equations gives
\beq
-\partial_i\dot{A}^i+\lambda\dot{\varphi} + m^2A^0=0
\eeq
that allows to express the initial value of $A_0$ in terms of the values of $\dot\varphi$ and $\dot{A}^i$ on the Cauchy surface. We have exhausted all the constraints and we obtain that we only need to give the initial values of $A^i$, $\dot{A}^i$, $\varphi$ and $\dot\varphi$, what corresponds to 8 phase space conditions, \ie there are 4 dynamical degrees of freedom, in agreement with our expectations.

An alternative way of counting the number of propagating modes which at the same time sheds some light on their stability properties is realised by resorting to the St\"ueckelberg trick and taking the decoupling limit. As explained in \eg \cite{Hinterbichler:2011tt}, for a Proca field described by the  Lagrangian
\beq\label{eq:ProcaActionStueckelber}
\cL_{\textrm{P}roca}=-\frac14 F_{\mu\nu}F^{\mu\nu}+\frac12m^2A^2,
\eeq
where $F=\dif A$, the St\"ueckelberg trick consists on restoring the $U(1)$ gauge-invariance of the vector field by introducing a new scalar degree of freedom called St\"ueckelberg field through the replacement $A\rightarrow A+\frac{1}{m}\dif\chi$, which leads to the St\"ueckelbergised Proca action
\beq
\cL=-\frac14F_{\mu\nu}F^{\mu\nu}+\frac12\partial^\mu\chi\partial_\mu\chi+\frac12m^2A^2+mA^\mu\partial_\mu\chi.
\eeq
Contrary to what happens in the massless limit of the Proca action, which leads to the loss of the longitudinal degree of freedom propagated by the massive vector field, the massless (or decoupling) limit of the St\"ueckelbergised Proca action, namely
\beq
\cL=-\frac14F_{\mu\nu}F^{\mu\nu}+\frac12\partial^\mu\chi\partial_\mu\chi\;, 
\eeq
still describes the 3 degrees of freedom propagated by the Proca field: 2 encoded in the massless vector field and one encoded in the St\"ueckelberg field. Furthermore, this massless limit decouples two of the degrees propagated by the vector from the scalar one, which is the reason why it is called the \textbf{decoupling limit}. Thus the decoupling limit of the St\"ueckelbergised Proca action describes a free spin-1 gauge field and a free scalar field which can be put into correspondence with the transverse and longitudinal polarisations propagated by the massive vector field. Hence, the St\"ueckelberg trick allows to somehow isolate the transverse modes of a massive vector field from its longitudinal mode.

The St\"ueckelbergisation of the above Lagrangian \eqref{eq:ExampleLagrangianStueckelberg} leads to
\beq
\cL=-\frac14F_{\mu\nu}F^{\mu\nu}+\partial^\mu\chi\partial_\mu\bar{\varphi}+\frac12\partial^\mu\chi\partial_\mu\chi+\frac12m^2A^2+mA^\mu(\partial_\mu\bar{\varphi}+\partial_\mu\chi),
\eeq
where we have performed the field redefinition $\varphi\mapsto\bar{\varphi}=\frac{\lambda}{m}\varphi$. If we now take the decoupling limit $m\rightarrow0$, we find
\beq\label{eq:ExampleLagrangianStueckelbergDecoupling}
\cL=-\frac14F_{\mu\nu}F^{\mu\nu}+\partial^\mu\chi\partial_\mu\bar{\varphi}+\frac12\partial^\mu\chi\partial_\mu\chi.
\eeq
Note that, although the St\"ueckelberg field decouples from the gauge spin-1 field as usual, it still couples to the original scalar present in \eqref{eq:ExampleLagrangianStueckelberg}. In this limit, it becomes much more apparent that the theory progates 4 degrees of freedom corresponding to the 2 transverse modes, the longitudinal polarisation and the original scalar field. As well, this limit allows to clearly see the pathological behaviour of the scalar field due to the absence of a proper kinetic term $(\partial\bar{\varphi})^2$. As explained in \ref{sec:TypesOfInstabilities}, the presence of a ghost can be seen computing the eigenvalues of the associated matrices $a^{IJ}$ and $b^{IJ}$, which in this case are both the same due to Lorentz invariance of the background (it is Minkowski space). Let us define $K^{IJ}=a^{IJ}=b^{IJ}$, where $I$ and $J$ run through $(\chi,\varphi)$, and call it the kinetic matrix of the scalar sector. This matrix is\footnote{That $\hat K$ in \eqref{eq:KineticMatrixExample} is the kinetic matrix stems from the fact that $(\partial^\mu\chi,\partial^\mu\bar{\varphi})\hat K(\partial_\mu\chi,\partial_\mu\bar{\varphi})^\top$ gives the scalar kinetic terms in \eqref{eq:ExampleLagrangianStueckelbergDecoupling}.} 
\begin{equation}\label{eq:KineticMatrixExample}
\hat{K}=\begin{pmatrix}
1/2 & 1/2 \\
1/2 & 0 
\end{pmatrix},
\end{equation} 
 and its eigenvalues are $(1\pm\sqrt{5})/4$. Given that there is a negative eigenvalue in both $a^{IJ}$ and $b^{IJ}$ (namely in the kinetic matrix), there is a ghostly degree of freedom around Minkowski. To correctly identify the ghost, note that by means of the field redefinition $\chi\mapsto \chi-\bar{\varphi}$, the above Lagrangian \eqref{eq:ExampleLagrangianStueckelbergDecoupling} reads
 \beq
\cL=-\frac14F^2-\frac12\partial^\mu\bar{\varphi}\partial_\mu\bar{\varphi}+\frac12\partial^\mu\chi\partial_\mu\chi,
\eeq
where now the kinetic matrix has canonical eigenvalues for scalar fields although one of them (namely the one corresponding to $\bar\varphi$)  with the wrong sign. We then see the unavoidable presence of a scalar ghost associated to $\bar\varphi$ and therefore $\varphi$ in the original Lagrangian \eqref{eq:ExampleLagrangianStueckelberg}.

%%%%%%%%%%%%%%%%%%%%%%%%%%%%%%%%%%%%%%%%%%%%%%%%%%%%%%%%%%%%
%%%%%%%%%%%%%%%%%%%%%%%%%%%%%%%%%%%%%%%%%%%%%%%%%%%%%%%%%%%%

\subsection{Ghosts in RBG theories without projective symmetry}\label{sec:GenRBG}

Projective symmetry has played a crucial role in the development of the general framework of RBG theories, with prominent examples like \eg Eddington-inspired Born-Infeld gravity, whose implications seem to have been partially unnoticed by the community until recently. Historically,\footnote{Let us use this word though the story of these developments is as recent as (approximately) these past two decades.} the different RBG models that were considered featured only the symmetric part of the Ricci tensor in the action\footnote{With a few particular examples of very simple dependences on the antisymmetric part, see \eg \cite{Buchdahl:1979ut,Vitagliano:2010pq,Olmo:2013lta}.} due to the fact that this restriction allowed to solve the connection easily as the Levi-Civita connection of an auxiliary metric. This apparently {\it ad hoc} restriction can be seen to actually be a consequence of requiring projective symmetry to a general RBG action, as explained in section \ref{sec:RBGTheory}, though this fact appears not to have been relevant to the eyes of the researchers until recently. Indeed, a recent work by Afonso and collaborators \cite{Afonso:2017bxr} put the spotlight on this symmetry, which allowed them to conclude that torsion does not play any physical role if both the gravitational and matter sectors respect this symmetry, as it is described by a spurious projective mode. Following this line, Jose Beltr\'an-Jm\'enez (one of the authors of  \cite{Afonso:2017bxr}) encouraged me to study what would happen with these theories if the projective symmetry was dropped, thus allowing the presence of the antisymmetric part of the Ricci tensor in the action. Our expectations were that the explicit breaking of projective symmetry limited its consequences to give dynamics to the spurious projective mode, promoting it to a (likely) massive vector field which propagates 3 new degrees of freedom and we would end up with an Einstein-Proca-like system. Nevertheless, it soon became apparent that the explicit breaking of projective symmetry in RBG theories has more subtle and deeper consequences than expected, as it ended up unleashing 5 new ghostly degrees of freedom. In what follows, we present a detailed account of the arguments leading to these conclusions.

In section \ref{sec:EinsteinFrame} we showed how RBG theories (with or without projective symmetry) admit an Einstein frame so that if projective symmetry is enforced, their gravitational sector is equivalent to that of GR in the sense that they both propagate a massless spin-2 degree of freedom\footnote{Recall from chapter \ref{sec:GravityAsGeometry} that a massless spin-2 field is universal, and there is a unique consistent nonlinear theory for massles spin-2 fields.}. This becomes clear from the fact that, in the projectively invariant case,  the Einstein frame metric $q^{\mu\nu}$, defined by \eqref{eq:DefinitionMetricq}, is symmetric; its dynamics is described by the Einstein equations coupled to the stress-energy tensor of a given matter source; and the connection is the Levi-Civita connection of this metric $q^{\mu\nu}$. The explicit breaking of projective symmetry in the RBG Lagrangian allows the full Ricci tensor to appear in the action, thus jeopardising the symmetric nature of the corresponding $q^{\mu\nu}$. This crucially changes the situation  and the Einstein frame representation of the theory is no longer GR, as it resembles the Nonsymmetric Gravity Theory (NGT) introduced by Moffat \cite{Moffat:1978tr}, which has already been explored in different versions. Although the nonsymmetric frame of generalised RBGs does not exactly reproduce Moffat's nonsymmetric gravity, it does so in certain limits. A crucial difference is the coupling to matter fields, although even this can be made equivalent by {\it ad hoc} choices of the matter couplings in Moffat's theory. This points to the fact that the pathologies that plague Moffat's theory \cite{Damour:1991ru,Damour:1992bt} (see also \cite{Moffat:1993gi,Moffat:1994hv,Clayton:1995yi,Clayton:1996dz,Moffat:1996hf,Prokopec:2005fb,Poplawski:2006ev,Janssen:2006nn,Janssen:2006jx,BeltranJimenez:2012sz,Golovnev:2014aca}) will also be a feature of RBGs without projective symmetry. We will provide a detailed analysis of the pathologies that plague RBGs without projective symmetry and, as a by-product, this will contribute to an alternative understanding of the origin of the pathologies on NGT.

 Let us start by considering vacuum solutions, so that no matter fields are present\footnote{We allow the appearance of a cosmological constant like term $\bar\cU$ that accounts for a possible nontrivial dependence of $\cU$ on the background $q^{\mu\nu}$ solution.} and the analysis of the gravitational sector becomes cleaner. As shown in section \ref{sec:EinsteinFrame}, the action for vacuum RBG theories without projective symmetry in their Einstein frame is written as
\beq\label{eq:RBGNoProjectiveEinsteinFrame}
\cS=\frac12{{\mpl^2}}\int \dd^Dx \Big[\sqrt{-q}q^{\mu\nu}R_{\mu\nu}+\bar\cU\Big],
\eeq
where $R_{\mu\nu}$ is the Ricci tensor of a general affine connection (see section \ref{sec:GeneralAffineConnection}), $q^{\mu\nu}$ is a metric with an antisymmetric par, which will be encoded in the 2-form $B^{\mu\nu}\equiv q^{[\mu\nu]}$, and $\cU$ is some potential for the nonsymmetric object $q_{\mu\nu}$. Of course, in the case of a symmetric $q_{\mu\nu}$, this term can only contribute a cosmological constant by virtue of covariance, but for the nonsymmetric case, it can have a nontrivial structure  with relevant consequences. In fact, such a term was invoked in \cite{Damour:1992bt} in an attempt to resolve the pathologies of Moffat's theory. However, different methods seem to conclude that the instabilities that plague this theory around arbitrary backgrounds cannot be healed in this way. We will start by analysing the problem in the decoupling limit of the St\"ueckelbergised action for the 2-form as, in our opinion, is the simplest and most transparent procedure to show presence of pathologies. We will then proceed to show the same results through the exploration of the field equations of the theory, which will allow us to include matter and show that it does not help in curing the pathologies of the theory.
 
\subsubsection{A detour: Identifying the degrees of freedom propagated by a 2-form field}\label{sec:Appendix2-form}

Before starting our analysis of the pathologies of RBGs without projective symmetry, due to the appearance of the 2-form $B^{\mu\nu}=q^{[\mu\nu]}$, it will prove useful to identify the degrees of freedom carried by a massless and massive 2-form field.  To that end, let us consider the action for a massless 2-form in $D=4$, namely
\beq
\cS=-\frac{1}{12}\int\dd^4x\sqrt{-g} H_{\alpha\beta\gamma}H^{\alpha\beta\gamma}
\eeq
with $H_{\al\be\ga}\equiv\partial_{[\al}B_{\be\ga]}$. The number and type of degrees of freedom contained in the 2-form can be  identified, for instance, by dualising the above action. To that end, let us rewrite it in its first-order form
\beq
\cS=-\frac16\int\dd^4x\sqrt{-g}\left(\Pi^{\alpha\beta\gamma}\partial_{[\alpha} B_{\beta\gamma]}-\frac12\Pi_{\alpha\beta\gamma}\Pi^{\alpha\beta\gamma}\right),
\eeq
where $\Pi^{\alpha\beta\gamma}$ are the conjugate momenta of the 2-form $B_{\mu\nu}$. Upon variation with respect to the conjugate momenta we obtain
\beq\label{mom2form}
\Pi_{\alpha\beta\gamma}=\partial_{[\alpha} B_{\beta\gamma]},
\eeq
while the 2-form field equations give
\beq
\partial_\alpha \Pi^{\alpha\beta\gamma}=0,
\eeq
which are of course the Hamilton equations of a Kalb-Rammond field. Since the conjugate momentum is divergence-free as imposed by the 2-form field equation, namely $\star\;\dif\star \Pi=0$ (see section \ref{sec:MetricStructure}), in 4 dimensions this constraint is realised through $\Pi=\star\;\dif\phi$ where $\star$ stands for the Hodge dual and $\phi$ a scalar field. By definition of the Hodge dual \eqref{DefHodgeDual}, we then have $\Pi^{\alpha\beta\gamma}=H^{\alpha\beta\gamma}\propto \sqrt{-g}\epsilon^{\alpha\beta\gamma\mu}\partial_{\mu}\phi$, and plugging the solution of the constraint imposed by the 2-form into the above action we end up with
\beq
\cS=\int\dif^4x\sqrt{-g}\partial_\mu\phi\partial^\mu\phi,
\eeq
showing that a Kalb-Rammond field indeed propagates a scalar degree of freedom. This procedure can formally be done at the level of the path integral $Z=\int [D\Pi] [DB] e^{i\cS}$ by integrating out the 2-form field. Since the action is a quadratic form in the momentum, the integration can straightforwardly be performed  giving a functional delta that imposes the constraint.

Let us now turn to the case of a massive 2-form field. Again, we will resort to its first-order formulation
\beq
\cS=-\frac16\int\dd^4x\sqrt{-g}\left(\Pi^{\alpha\beta\gamma}\partial_{[\alpha} B_{\beta\gamma]}-\frac12\Pi^2+{\frac{1}{2}}m^2B^2\right),
\eeq
where we see that the 2-form now becomes an auxiliary field instead of a Lagrange multiplier imposing the divergence-free constraint on the conjugate momentum as in the massless case. The field equation for the momentum again gives its relation with the derivatives of the 2-form as \eqref{mom2form}. The 2-form equations however now give
\beq
\partial_\alpha \Pi^{\alpha\beta\gamma}- m^2 B^{\be\ga}=0.
\eeq
Using this equation to solve for the 2-form   in terms of the conjugate momentum and plugging it into the action (notice that this is an algebraic equation for $B_{\mu\nu}$), we can rewrite the action as
\beq
\cS=\frac{1}{12}\int\dd^4x\sqrt{-g}\left({\frac{1}{m^2}}\partial_\alpha\Pi^{\alpha\beta\gamma}\partial^{\lambda}\Pi_{\lambda\beta\gamma}+\Pi^2\right).
\eeq
We can now dualise this action\footnote{Note that the Hodge dual of a p-form has the same number of independent components than the p-form, as they live in vector spaces of the same dimension $\binom{n}{p}=\binom{n}{n-p}$, and therefore it does not alter the counting of degrees of freedom.} by means of $H^{\alpha\beta\gamma}\propto \epsilon^{\alpha\beta\gamma\mu}A_\mu$ and after canonically normalising $A_\mu$ by $A_\mu\mapsto (m/2) A_\mu$, the above action can equivalently be expressed as
\beq
\cS=\int\dd^4x\sqrt{-g}\left(-\frac14 F_{\mu\nu}F^{\mu\nu}-\frac12 M^2 A^2\right),
\eeq
where $M^2=3m^2$. We have then that our original massive 2-form field action is dual to the Proca action and, consequently, it propagates three degrees of freedom. Note that the scalar degree of freedom propagated by the massless 2-form can be identified with the longitudinal polarisation of the Proca field, and the two extra degrees of freedom carried by the massive 2-form correspond to the transverse polarisations. Indeed, we will see below how when considering the decoupling limit of the St\"uecklebergised  action for a massive 2-form, the St\"ueckleberg fields that restore the gauge symmetry of the 2-form come in the form of a gauge spin-1 field which naturally propagates the two transverse polarisations.\\ 

For completeness, let us also discuss what happens for more general interacting 2-form fields. Again, starting from its first-order formulation, we can write 
\beq
\cS=\int\dd^4x\sqrt{-g}\Big[\Pi^{\alpha\beta\gamma}\partial_{[\alpha} B_{\beta\gamma]}-\mathcal{H}(B_{\mu\nu},\Pi^{\alpha\beta\gamma})\Big],
\eeq
where $\mathcal{H}$ is the Hamilton function that defines the interacting theory. Following the same procedure, we can obtain the 2-form field equations as
\beq
\partial_\alpha \Pi^{\alpha\mu\nu}=-\frac{\partial\mathcal{H}}{\partial B_{\mu\nu}}.
\label{Eq:eqdH}
\eeq
This equation now can be algebraically solved (at least formally) for the 2-form field to obtain $B^{\mu\nu}=B^{\mu\nu}(\Pi,\partial\cdot\Pi)$. We can then integrate out the 2-form field by plugging this solution into the action so we obtain $\cS=\cS[\Pi,\partial\cdot\Pi]$. If we dualise this theory to a vector field as above, we finally get that our original action can be rewritten as $\cS=\cS[F_{\mu\nu},A_\alpha]$, \ie as an interacting massive vector field. If the Hamiltonian function does not explicitly depend on the 2-form field, \ie if we have a gauge 2-form field, then Eq. (\ref{Eq:eqdH}) is instead solved by $H=\star\;\dif\phi$ as in the massless case above, so the action can instead be expressed as $\cS=\cS[(\dif\phi)^2]$ that describes an interacting shift-symmetric scalar field.  

This dualisation procedure can also be applied to cases when the 2-form field is coupled to some matter fields and even for nonAbelian 2-form fields with some internal group structure. After having presented a detailed account of the number and type of degrees of freedom carried by a massless and massive 2-form, and having as well presented the St\"ueckleberg trick, we have now all the necessary tools to approach the study of the number and stability of degrees of freedom in RBGs without projective symmetry.

%%%%%%%%%%%%%%%%%%%%%%%%%%%%%%%%%%%%%%%%%%%%%%%%%%%%%%%%%%%%
%%%%%%%%%%%%%%%%%%%%%%%%%%%%%%%%%%%%%%%%%%%%%%%%%%%%%%%%%%%%

\subsubsection{Ghosts in the decoupling limit of the 2-form}\label{sec:InstDecLim}

To study the decoupling limit of the St\"ueckelbergised 2-form, let us consider the antisymmetric sector perturbatively up to quadratic order so that
\beq\label{eq:AntisymmetricMetricQuadraticFieldRedefinitions}
q_{\mu\nu}=\sq_{\mu\nu}+\frac{\sqrt{2}}{{\mpl}} \left(B_{\mu\nu}+\alpha B_{\mu\alpha} B^\alpha{}_\nu+\beta B^2\sq_{\mu\nu}\right),
\eeq
with $\sq_{\mu\nu}$ an arbitrary symmetric metric, $B_{\mu\nu}$ a 2-form field corresponding to the antisymmetric part of $q_{\mu\nu}$, and where the parameters $\alpha$ and $\beta$ account for the possibility of field redefinitions at quadratic order (see e.g. \cite{Damour:1992bt}). The numerical factor and the Planck mass have been introduced for convenience. When expanding the action of RBGs without projective symmetry in the Einstein frame \eqref{eq:RBGNoProjectiveEinsteinFrame} around such a background at second-order in $B_{\mu\nu}$ we find\footnote{Here we will stick to the $D=4$ case for simplicity. In arbitrary dimensions, the analysis can be carried in a similar fashion, although taking into account that the degrees of freedom carried by each field might change with the dimension.}:
\begin{align}
\cS^{(2)}=\int\dd^4x\sqrt{-\sq}\Big[&\frac12{{\mpl^2}}R^{\sq}-\frac{1}{12} H_{\mu\nu\rho} H^{\mu\nu\rho}-\frac14 m^2 B^2-\frac{\sqrt{2}{\mpl}}{3}B^{\mu\nu}\partial_{[\mu}\Gamma_{\nu]}\nonumber\\
&+\frac{1-2\alpha+4\beta}{4} R^{\sq} B^2+\alpha R^{\sq}_{\mu\nu}B^{\mu\alpha}B^\nu{}_\alpha-R^{\sq}_{\mu\nu\alpha\beta}B^{\mu\alpha}B^{\nu\beta}\Big]
\label{eq:Actquad}
\end{align}
where $H_{\mu\nu\rho}=3\partial_{[\mu} B_{\nu\rho]}$ the field strength of the 2-form field, $m^2$ is the mass generated from $\bar\cU$, and $\Gamma_\mu$ is the projective mode of the connection. In order to make apparent the presence and nature of the instabilities, let us first consider a flat background so the couplings to curvature in (\ref{eq:Actquad}) disappear and it becomes clear that the pathologies arise already around Minkowskian backgrounds.

As well as for the vector field, the St\"ueckelbergisation of the 2-form is realised through the introduction of the St\"uckelberg fields $b_\mu$ which restore its gauge symmetry via the replacement $B_{\mu\nu}\rightarrow \hat{B}_{\mu\nu}+\frac2m \partial_{[\mu} b_{\nu]}$, which in a Minkowskian background $\bar{q}=\eta$ leads to
\begin{align}
\begin{split}
\cS_{\textrm{Stuck., flat}}^{(2)}=\int\dd^4x\Big[&-\frac{1}{12} H_{\mu\nu\rho} H^{\mu\nu\rho}-\frac14 m^2 B^2-\partial_{[\mu}b_{\nu]} \partial^{[\mu}b^{\nu]}-m \hat{B}_{\mu\nu}\partial^{[\mu}b^{\nu]}\\
&\hspace{2.5cm}-\frac{\sqrt{2}{\mpl}}{3}B^{\mu\nu}\partial_{[\mu}\Gamma_{\nu]}-\frac{2\sqrt{2}{\mpl}}{3m}\partial^{[\mu}b^{\nu]}\partial_{[\mu}\Gamma_{\nu]}\Big]
\label{eq:Actquad}
\end{split}
\end{align}
In order to properly take the decoupling limit, we have to redefine $\Gamma_\mu\mapsto \frac{3m}{\sqrt{2}{\mpl}}\Gamma_\mu$ and keep it finite (\ie take ${\mpl}\to\infty$).  Taking then the decoupling limit $m\rightarrow 0$, we see how the scalar mode of the gauge invariant 2-form sector described by $\hat{B}_{\mu\nu}$ decouples from both the St\"ueckelberg field (which corresponds to the extra modes carried by the 2-form when it is massive) and the projective mode. The relevant sector of the action in the decoupling and on a Minkowskian background is then
\begin{align}\label{ActionDec}
\cS^{(2)}_{\textrm{dec,flat}}=\int\dd^4x\left(-\frac{1}{12} \hat{H}_{\mu\nu\rho} \hat{H}^{\mu\nu\rho}-\partial_{[\mu}b_{\nu]} \partial^{[\mu}b^{\nu]}-2 \partial_{[\mu}b_{\nu]} \partial^{[\mu}\Gamma^{\nu]} \right)
\end{align}
We see that the decoupling limit shows the presence of five degrees of freedom: one associated to the massless 2-form $\hat{B}_{\mu\nu}$ and two associated to each of the helicity-1 modes described by the St\"ueckelberg field $b_\mu$ and the projective mode $\Gamma^\mu$ respectively. This is of course the expected counting for \eqref{eq:Actquad} corresponding to a massive 2-form and a gauge spin-1 field. In this decoupling limit it is then apparent that the theory is plagued by ghost-like instabilities owed to the mixing  $\partial_{[\mu}b_{\nu]} \partial^{[\mu}\Gamma^{\nu]}$ that comes in without the diagonal $ \partial_{[\mu}\Gamma_{\nu]} \partial^{[\mu}\Gamma^{\nu]}$ element. Indeed, the kinetic matrix 
\begin{equation}
\hat{K}=\begin{pmatrix}
-1 & -1 \\
-1 & 0 
\end{pmatrix},
\end{equation} 
has eigenvalues $(-1\pm\sqrt{5})/2$, one of which is negative, which in a Lorentz invariant background ($a^{IJ}=b^{IJ}=\hat{K}^{IJ}$) implies the presence of ghostly degrees of freedom as explained in section \ref{sec:TypesOfInstabilities}. More explicitly, if we diagonalise the kinetic matrix by means of the field redefinition\footnote{Note that the coefficients are chosen so that the redefined fields are canonically normalised.} $b_\mu=A_\mu+\xi_\mu$, $\Gamma_\mu=-2 \xi_ \mu$  the action \eqref{ActionDec} reads
\beq
\cS^{(2)}_{\textrm{dec,flat}}=\int\dd^4x\left[-\frac{1}{12} \hat{H}_{\mu\nu\rho} \hat{H}^{\mu\nu\rho}-\partial_{[\mu} A_{\nu]} \partial^{[\mu} A^{\nu]}+\partial_{[\mu} \xi_{\nu]} \partial^{[\mu} \xi^{\nu]}\right],
\eeq
 showing that $\xi_\mu$ is a ghost around a Minkowskian background as it has a Maxwellian kinetic term with the wrong sign. 
 
Let us now turn on the symmetric sector by dropping the Minkowskian condition on $q^{(\mu\nu)}$ and allow for an arbitrary curved $\bar{q}$-background. It should then be clear that while the ghosts that we uncovered around a Minkowskian background will generally persist, the nonminimal couplings to the curvature in (\ref{eq:Actquad}) will present additional pathologies. Within our approach we can readily see and interpret the nature of these pathologies as Ostrogradskian instabilities (see section \ref{sec:Ostrogradski}) associated to having higher-order derivatives in the Lagrangian for the metric $\bar{q}$, which leads to higher-order equations of motion for the Stueckelberg fields that will propagate Ostrogradski ghosts. The presence of Ostrogradski instabilities  within NGT has  not been properly identified, and it represents yet another problem for NGT besides the pathological asymptotic fall off behaviour discussed in \cite{Damour:1992bt}.

To show the presence of the Ostrogradskian instabilities through the St\"ueckelberg trick, we now must take care that the decoupling limit now needs to take into account that the curvature scales as $R\sim{\mpl^{-2}}$, and the appropriate limit to be taken is $m\rightarrow0$ and ${\mpl}\rightarrow\infty$ with $\Lambda\equiv m {\mpl}$ fixed. In this limit, the St\"uckelberg fields $b_\mu$ will feature nonminimal couplings with the schematic form $\sim \frac{1}{\Lambda^2}R^{\sq}\dif b\,\dif b$. These couplings feature second-order time derivatives of the metric which generically give rise to higher-order equations of motion and Ostrogradskian instabilities as explained in section \ref{sec:Ostrogradski}. An exceptional case is provided by the Horndeski vector-tensor interaction found in \cite{Horndeski:1976gi}. Having the two free parameters $\alpha$ and $\beta$ in \eqref{eq:AntisymmetricMetricQuadraticFieldRedefinitions} that allow for field redefinitions at quadratic order, one would be tempted to say that the pathology is not physical since the Horndeski interaction could be reached by an appropriate local field redefinition. Nevertheless, it is worth noticing that even this Horndeski interaction presents pathologies around relevant backgrounds \cite{Jimenez:2013qsa}. Furthermore, we need to remember that this is the quadratic action and it is expected that going to higher perturbative orders, new higher-order nonminimal couplings will be generated. Since there are no healthy such terms beyond the Horndeski interaction in four dimensions, these will need to be trivial modulo field redefinition to avoid re-introducing the pathologies. 

At this point, the pathological character of these theories should be unequivocal taken at face value. One could argue that interpreted as effective field theories, there could be a certain regime of validity at low energies. However, the very presence of the ghosts already around a Minkowskian background shown above makes this hope difficult to realise. In this respect, this ghost could be stabilised easily by introducing a healthy kinetic term for the projective mode, namely $\partial_{[\mu}\Gamma_{\nu]}\partial^{[\mu}\Gamma^{\nu]}$. Although such a term cannot be generated from RBGs, within the EFT approach, it must appear unless projective symmetry is assumed to be a symmetry of the EFT, along with a bunch of other terms accompanying it. An EFT approach to the restricted class of Poincar\'e gauge theories has been pursued in \cite{Aoki:2019snr}, and a recent work shows the presence of ghostly degrees of freedom for generic quadratic metric-affine EFTs \cite{Percacci:2019hxn} around Minkowskian backgrounds. The nonminimal couplings however, being (irrelevant) higher dimension operators, should typically be perturbative and, consequently, the associated ghosts would only come at a scale beyond the cutoff. As well, it might be possible to tune some coefficients to push the ghosts to higher scales so that the corresponding irrelevant operators could have nonperturbative effects on the low-energy phenomenology.  

Let us mention that a potential caveat of our analysis (up to now) is that we have neglected the matter sector, but this should not worry us too much since including matter fields will hardly render the theories stable. Rather, one could expect a more pathological behaviour. We will address this point later to show it explicitly.

%%%%%%%%%%%%%%%%%%%%%%%%%%%%%%%%%%%%%%%%%%%%%%%%%%%%%%%%%%%%
%%%%%%%%%%%%%%%%%%%%%%%%%%%%%%%%%%%%%%%%%%%%%%%%%%%%%%%%%%%%

\subsubsection{Another view on the problem with additional degrees of freedom}\label{sec:ProblemAdditionalDOFs}

In the previous section we have shown how vacuum RBG without a projective symmetry (or vacuum NGT for that matter) are plagued by ghostly instabilities arising from two sectors, namely: the dynamical projective mode whose mixing with the 2-form leads to the necessary presence of a spin-1 ghost and the nonminimal couplings of the massive 2-form field that gives rise to Ostrogradski instabilities. This has been neatly shown in the decoupling limit of the St\"ueckelbergised action of the 2-form field. Here we will show the appearance of these pathologies in an alternative manner, namely by tracing the appearance of the new degrees of freedom in the action and taking into account the nature of their field equations. Let us consider our family of theories described by the action
\beq\label{GeneralAction2}
\cS[g_{\mu\nu},\Ga]=\frac12\int \dd^Dx \sqrt{-g}\,F\big(g^{\mu\nu},R_{\mu\nu}\big),
\eeq
where we again consider vacuum RBGs without projective symmetry. Let us now separate a metric contribution to the connection from the rest, \ie let us perform the following field redefinition 
\beq
\Gamma^\alpha{}_{\mu\beta}=\Gamh^\alpha{}_{\mu\beta}+\Upsilon^\alpha{}_{\mu\beta}
\label{Eq:GammasplittingChO}
\eeq
where $\Gamh^\alpha{}_{\mu\beta}$ are the Christoffel symbols \eqref{eq:Christoffel} of the (symmetric) metric $h^{\mu\nu}$, which is defined through the splitting of the nonsymmetric metric as
\beq\label{metricsplitting}
\sqrt{-q}q^{\mu\nu}=\sqrt{-h}h^{\mu\nu}+\sqrt{-h}B^{\mu\nu}
\eeq
with $\sqrt{-h}h^{\mu\nu}=\sqrt{-q}q^{(\mu\nu)}$ and $\sqrt{-h}B^{\mu\nu}=\sqrt{-q}q^{[\mu\nu]}$. Using the transformation properties of the Ricci tensor under an arbitrary change in the connection \eqref{eq:TransfRicciCoRicciHomothetic} for the field redefinition (\ref{Eq:GammasplittingChO}), we can write the generalised RBG action in its Einstein frame \eqref{eq:RBGNoProjectiveEinsteinFrame} as
\beq\label{eq:RBGNoProjectiveEinsteinFrameDecomposed}
\begin{split}
\cS=\frac12\int\dd^Dx\sqrt{-h}\Big[&R^h-\Upsilon^{\lambda\alpha\mu}\Upsilon_{\alpha\mu\lambda}+\Upsilon^\alpha{}_{\alpha\lambda}\Upsilon^\lambda{}_{\kappa}{}^\kappa-\Upsilon^\alpha{}_{\alpha\lambda}\Upsilon^\lambda{}_{\mu\nu}B^{\mu\nu}\\&
-\Upsilon^\alpha{}_{\nu\lambda}\Upsilon^\lambda{}_{\alpha\mu}B^{\mu\nu}-B^{\mu\nu}\na^h_\alpha\Upsilon^\alpha{}_{\mu\nu}-B^{\mu\nu}\na^h_\nu\Upsilon^\alpha{}_{\alpha\mu}+\cU(B)\Big].
\end{split}
\eeq
where $\na^h$ is the covariant derivative with respect to the Levi-Civita connection of $h^{\mu\nu}$, we have used the fact that the connection $\Gamh^\alpha{}_{\mu\beta}$ is torsion-free, and we have dropped a boundary term. Notice that we have used (and will use in the subsequent manipulations) $h_{\mu\nu}$ as the metric so we will raise and lower indices\footnote{Recall that any metric defines a canonical isomorphism between the tangent space and its dual that in practice corresponds to raising and lowering indices as defined in \eqref{eq:MetricIsomorphism}.} with $h^{\mu\nu}$ and its inverse $h_{\mu\nu}$. The field equations for the free part of the connection $\Upsilon^\alpha{}_{\mu\nu}$ obtained by variation of the above action are
\beq\label{Eq:Omegaeq}
\begin{split}
\na^h_{\alpha}B^{\mu\nu}+\delta_{\alpha}{}^{\mu} \na^h_{\beta}B^{\nu\beta}&-\Upsilon^{\mu\nu}{}_{\alpha}-\Upsilon^{\nu}{}_{\alpha}{}^{\mu}+h^{\mu\nu} \Upsilon^{\beta}{}_{\beta\alpha}+
\delta_{\alpha}{}^{\mu} \Upsilon^{\nu\beta}{}_{\beta}\\
&+B^{\mu}{}_{\beta} \Upsilon^{\nu}{}_{\alpha}{}^{\beta}-B^{\nu}{}_{\beta} \Upsilon^{\mu\beta}{}_{\alpha}-B^{\mu\nu} \Upsilon^{\beta}{}_{\beta\alpha}- \delta_{\alpha}{}^{\mu} B_{\beta\lambda}\Upsilon^{\nu\beta\lambda}=0.
\end{split}
\eeq
Taking the trace with respect to $\alpha$ and $\nu$ of the above equation we obtain
\beq\label{DivConstB}
\na^h_\mu B^{\mu\nu}=0
\eeq
which constrains the 2-form field $B^{\mu\nu}$ to be divergence-free. Enforcing this constraint into the connection field equations \eqref{Eq:Omegaeq} we arrive at
\beq\label{Eq:Omegaeqdivfree}
\begin{split}
\na^h_{\alpha}B^{\mu\nu}&-\Upsilon^{\mu\nu}{}_{\alpha}-\Upsilon^{\nu}{}_{\alpha}{}^{\mu}+h^{\mu\nu} \Upsilon^{\beta}{}_{\beta\alpha}+
\delta_{\alpha}{}^{\mu} \Upsilon^{\nu\beta}{}_{\beta}\\
&+B^{\mu}{}_{\beta} \Upsilon^{\nu}{}_{\alpha}{}^{\beta}-B^{\nu}{}_{\beta} \Upsilon^{\mu\beta}{}_{\alpha}-B^{\mu\nu} \Upsilon^{\beta}{}_{\beta\alpha}- \delta_{\alpha}{}^{\mu} B_{\beta\lambda}\Upsilon^{\nu\beta\lambda}=0.
\end{split}
\eeq
The divergence-free constraint on the 2-form in \eqref{DivConstB} can also be written as $\star\dif\star B=0$ where $\ast$ is the Hodge dual associated to $h_{\mu\nu}$, see sections \ref{sec:MetricStructure} and \ref{sec:MetricStructure}. The solution to this equation is given by $B=\star\hspace{-3pt}\dif A$ where $A$ is a $(D-3)$-form if the spacetime dimension is $D$. Particularizing to $D=4$ we have that $B^{\mu\nu}$ can be expressed as the dual of the field strength of some 1-form $A_\mu$, so that the constraint \eqref{DivConstB} derived from the connection field equations implies
\begin{equation}
B^{\mu\nu}=-\frac{1}{2\sqrt{-h}}\epsilon^{\mu\nu\alpha\beta}\partial_{[\alpha}A_{\beta]}.
\end{equation}
Notice that this is an exact constraint so that it becomes clear that the 2-form can propagate at most the same number of degrees of freedom as a vector field, according to what we saw in section \ref{sec:Appendix2-form}. 

It is also easy to see that a projective mode $\Upsilon^\alpha{}_{\mu\nu}=\xi_\mu\delta^\alpha{}_\nu$ is a solution when $B_{\mu\nu}=0$. This was indeed expected since for vanishing $B_{\mu\nu}$ we recover the usual projective-invariant theory whose connection is the Levi-Civita connection of $h_{\mu\nu}$ up to a projective mode. As a matter of fact, in the case of RBG where projective symmetry is explicitly broken, this projective mode is the only dynamical component of the connection and the remaining components of $\Upsilon$ can be expressed in terms of $B_{\mu\nu}$ by solving (\ref{Eq:Omegaeq}). We will later show this in detail and we will find a perturbative solution up to lowest order in $B^{\mu\nu}$.

Since the equations are linear in $\Upsilon^\al{}_{\mu\nu}$, the projective mode can be regarded as a homogeneous solution for $\Upsilon^\alpha{}_{\beta\gamma}$ in the general case, \ie it belongs to the kernel of \eqref{Eq:Omegaeqdivfree}. In order to isolate this projective mode (homogeneous solution) from the remaining nondynamical part of the connection (nonhomogeneous solution), it proves useful to introduce the shifted connection
\beq\label{Eq:Omegah}
\Upsilonh^\alpha{}_{\mu\nu}=\Upsilon^\alpha{}_{\mu\nu}+\frac{1}{D-1}\Upsilon_\mu\delta^\alpha{}_\nu
\eeq
with $\Upsilon_\mu=2\Upsilon^\alpha{}_{[\alpha\mu]}$. This shifted connection satisfies $\Upsilonh^\alpha{}_{[\alpha\mu]}=0$ and it is invariant under a projective transformation of $\Upsilon^\al{}_{\mu\nu}$. In terms of these variables the action \eqref{eq:RBGNoProjectiveEinsteinFrameDecomposed} can be written as
\beq
\begin{split}
\cS=\frac12\int\dd^Dx\sqrt{-h}\Big[R^h-&\frac{2}{D-1}B^{\mu\nu}\partial_{[\mu}\Upsilon_{\nu]}-B^{\mu\nu}\na^h_\alpha\Upsilonh^\alpha{}_{\mu\nu}-B^{\mu\nu}\na^h_\nu\Upsilonh^\alpha{}_{\alpha\mu}+\Upsilonh^\alpha{}_{\alpha\lambda}\Upsilonh^\lambda{}_{\kappa}{}^\kappa\\-&\Upsilonh^{\alpha\mu\lambda}\Upsilonh_{\lambda\alpha\mu}-\Upsilonh^\alpha{}_{\alpha\lambda}\Upsilonh^\lambda{}_{\mu\nu}B^{\mu\nu}-\Upsilonh^\alpha{}_{\nu\lambda}\Upsilonh^\lambda{}_{\alpha\mu}B^{\mu\nu}+\cU(B)\Big].\label{RBGactionExpanded}
\end{split}
\eeq
We then see that the projective mode $\Upsilon_\mu$ is in fact the responsible for the divergence-free constraint on the 2-form field. From this form of the action we can already understand the root of the pathologies. Firstly, the absence of a pure kinetic term for the projective mode will render this sector unstable on arbitrary $B_{\mu\nu}$ backgrounds. To show this, let us consider a background where the 2-form develops a nontrivial profile. On such a background, and leaving out kinetic terms and/or nonminimal couplings that will not affect our argument here, the relevant sector is described by
\beq
\cS\supset\int\dd^Dx\sqrt{-h}\Big(B^{\mu\nu}\partial_{[\mu}\Upsilon_{\nu]}-m^2M^{\alpha\beta\mu\nu}B_{\alpha\beta}B_{\mu\nu}\Big),
\eeq
where $m^2$ is some mass parameter and $M^{\alpha\beta\mu\nu}$ the mass tensor that depends on the background configuration with the obvious symmetries of being antisymmetric in the first and second pair of indices and symmetric under the exchange $(\alpha\beta)\leftrightarrow (\mu\nu)$. If the background 2-form field is trivial, the mass tensor reduces to $M^{\alpha\beta\mu\nu}=h^{\alpha[\mu}h^{\nu]\beta}$, so in that case we have
\beq
\cS\supset\int\dd^Dx\sqrt{-h}\Big(B^{\mu\nu}\partial_{[\mu}\Upsilon_{\nu]}-m^2B_{\mu\nu}B^{\mu\nu}\Big).
\eeq
We can diagonalise this sector by performing the field redefinition $B^{\mu\nu}=\hat{B}^{\mu\nu}+\frac{1}{2m^2}\partial^{[\mu}\Upsilon^{\nu]}$, which turns the above action into
\beq
\cS\supset\int\dd^Dx\sqrt{-h}\left(\frac{1}{4m^2}\partial_{[\mu}\Upsilon_{\nu]}\partial^{[\mu}\Upsilon^{\nu]}-m^2\hat{B}_{\mu\nu}\hat{B}^{\mu\nu}\right),
\eeq
which after the field redefinition $\Upsilon_\mu\mapsto 2m\Upsilon_\mu$ becomes
\beq\label{eq:GhostProjectiveTrivialBackground}
\cS\supset\int\dd^Dx\sqrt{-h}\left(\partial_{[\mu}\Upsilon_{\nu]}\partial^{[\mu}\Upsilon^{\nu]}-m^2\hat{B}_{\mu\nu}\hat{B}^{\mu\nu}\right),
\eeq
Once this sector of the gravitational action has been diagonalised, it becomes apparent that the projective mode acquires the usual gauge-invariant Maxwellian kinetic term for a vector field, but with the wrong sign. Hence, we clearly see that the presence of a ghost around a trivial $B^{\mu\nu}$ background is unavoidable.\footnote{One could argue that by changing the sign of \eqref{eq:GhostProjectiveTrivialBackground} the ghost can be transformed into a tachyonic instability. However, one should keep in mind that \eqref{eq:GhostProjectiveTrivialBackground} is a piece of the action and what would look like a tachyon in \eqref{eq:GhostProjectiveTrivialBackground} would actually be a ghost when the full action is considered, as the signs of the eigenvalues of the kinetic matrix tell us.} However, there is the possibility that within a nontrivial $B^{\mu\nu}$ background the 2-form field behaves as a ghost condensate \cite{ArkaniHamed:2003uy}. In order to see if this is possible, notice that in a general $B^{\mu\nu}$ background, the diagonalisation requires a field redefinition of the form 
\beq
B^{\mu\nu}=\hat{B}^{\mu\nu}+\frac{1}{2m^2}\Lambda^{\mu\nu\alpha\beta}\partial_{[\alpha}\Upsilon_{\beta]}
\eeq
with $\Lambda^{\mu\nu\alpha\beta}$ satisfying generally
\beq
M^{\alpha\beta\lambda\kappa}\Lambda_{\lambda\kappa}{}^{\mu\nu}=h^{\alpha[\mu}h^{\nu]\beta}.
\label{Eq:LambdaM}
\eeq
In this case, the relevant sector of the gravitational action can be written as
\beq
\cS\supset\int\dd^Dx\sqrt{-h}\left(\frac{1}{4m^2}\Lambda^{\alpha\beta\mu\nu}\partial_{[\alpha}\Upsilon_{\beta]}\partial_{[\mu}\Upsilon_{\nu]}-m^2M^{\alpha\beta\mu\nu}B_{\alpha\beta}B_{\mu\nu}\right).
\eeq
or
\beq
\cS\supset\int\dd^Dx\sqrt{-h}\left(\Lambda^{\alpha\beta\mu\nu}\partial_{[\alpha}\Upsilon_{\beta]}\partial_{[\mu}\Upsilon_{\nu]}-m^2M^{\alpha\beta\mu\nu}B_{\alpha\beta}B_{\mu\nu}\right).
\eeq
after the field redefinition $\Upsilon_\mu\mapsto 2m\Upsilon_\mu$. As explained in section \ref{sec:TypesOfInstabilities}, in order to see whether the ghost persists in the general case, we have to look at the eigenvalues of $\Lambda^{\alpha\beta\mu\nu}$, which now plays the role of the kinetic matrix. The ghostly nature of the projective mode is avoided if $\Lambda^{\alpha\beta\mu\nu}$ is a super-metric with negative eigenvalues, \ie if it has the same {\it signature} as $-h^{\alpha[\mu}h^{\nu]\beta}$ being $h^{\mu\nu}$ a Lorentzian metric (as then it would have a negative eigenvalue corresponding to the kinetic term of $\Upsilon_\mu$). On the other hand, stability of the 2-form sector requires a mass matrix with positive eigenvaluess, \ie with the {\it signature} of $+h^{\alpha[\mu}h^{\nu]\beta}$. These two conditions are however inconsistent with each other by virtue of the relation (\ref{Eq:LambdaM}) and therefore no ghost condensation can stabilise the theory. However, we find that the presence of a ghost in the projective sector of generalised RBGs is unavoidable and occurs in an arbitrary background. This is the ghost found in \ref{sec:InstDecLim} in the decoupling limit of the corresponding St\"ueckelbergised action.

It is interesting to notice that the redefinition of the 2-form field that diagonalises the quadratic action for the trivial background configuration corresponds to a gauge-like transformation for the 2-form, hence, its field strength will be oblivious to such redefinition. In particular, this means that kinetic terms with the correct gauge invariant form $H^2$ will not be affected by the diagonalisation and, therefore, cannot change our conclusion about the presence of a ghost. The same reasoning applies to nontrivial backgrounds that vary weakly as compared to $m^2$. If this is not the case, one might envision that sufficiently strongly varying backgrounds could give rise to a stabilisation \`a la ghost condensate. Even without taking into account couplings to gravity, it should be apparent that there will always be UV modes with a sufficiently high frequency for which the background is effectively constant and, therefore, our discussion above will also apply, thus showing the pathological character of these modes. A natural way around this problem is to assume that those modes are beyond the regime of validity of the theory and, consequently, it does not pose an actual problem. In that case however, the full EFT approach should be taken from the very beginning. Moreover, there will also be nonminimal couplings to the curvature, which after diagonalisation will introduce yet additional pathologies arising from that sector so our hopes stand on shaky grounds anyways. To understand this, we must look at the connection equations \eqref{Eq:Omegaeq}, from where it is apparent that the solution for $\Upsilonh$ will have the schematic form
\beq
\Upsilonh\sim \frac{\na^h B}{1+B}.
\eeq
Plugging this solution back into the RBG action, written as \eqref{RBGactionExpanded}, and integrating out the nondynamical piece of the connection $\Upsilonh$, additional terms like $(\na^h B)^2$ and $B(\na^h)^2B$ will arise. The latter can be integrated by parts to be put in the form of the former. Doing this however can result in nongauge invariant derivative terms and/or nonminimal couplings arising from commuting covariant derivatives. Both of such terms are potentially dangerous and the source of Ostrogradskian instabilities. Finally, let us comment on the remarkable fact that the quadratic derivative terms generated in the action can be brought into the standard gauge-invariant kinetic term of a two form. However, this is an accident of the leading order solution and it is broken at higher-orders. Let us see this explicitly by finding a solution for the connection.

%%%%%%%%%%%%%%%%%%%%%%%%%%%%%%%%%%%%%%%%%%%%%%%%%%%%%%%%%%%%%%%%%%%%%%%%%%%%%%%%%%%%%%%%%%%%%%%%%%%%%%%%%%%%%%%%%%%%%%%%%%%%%%%%%%%%%%%%%%%%%%%%%%%%%%%%%%%%%%%%%%%%%%%%%%%%%%%%%%%%

\subsubsection{Formal and perturbative solutions for the connection}\label{sec:SolConn}

Let us proceed to find solutions for the connection, so that we can explicitly see that they have the schematic form shown above. We will first illustrate the form of the solutions for the connection by considering vacuum configurations, so that the action is given by
\beq
\cS=\frac12\int \dd^Dx \Big[\sqrt{-q}q^{\mu\nu}R_{\mu\nu}+\cU(q)\Big].
\eeq
The connection equations for this action are the same as we obtained in \eqref{eq:GeneralRBGConnectuonEquations} or (\ref{eq:Gammah}), \ie the connection deprived of its projective mode satisfies
\beq\label{conneq2}
\partial_\la(\sqrt{-q}q^{\mu\nu})+\hat{\Gamma}^{\mu}{}_{\la\al}\sqrt{-q}q^{\al\nu}+\hat{\Gamma}^{\nu}{}_{\al\la}\sqrt{-q}q^{\mu\al}-\hat{\Gamma}^{\al}{}_{\la\al}\sqrt{-q}q^{\mu\nu}=0.
\eeq
This equation does allow, at least formally, to algebraically solve for the connection in terms of $q^{\mu\nu}$. With this in mind, let us again decompose the connection as in \eqref{Eq:GammasplittingChO}, so that we extract the Levi-Civita connection of the symmetric component of $h^{\mu\nu}$. The projectively transformed connection is therefore given by 
\beq
\hat\Gamma^\alpha{}_{\mu\beta}=\Gamh^\alpha{}_{\mu\beta}+\Upsilonh^\alpha{}_{\mu\beta}
\label{Eq:GammasplittingChO1}
\eeq
where $\Upsilonh$ is defined as in \eqref{Eq:Omegah}. We can now introduce the above splitting \eqref{Eq:GammasplittingChO1} into the connection equations equations \eqref{conneq2}. By performing the usual trick of adding and subtracting the resulting equation with suitably permuted indices, we can write a formal solution for the connection as
\begin{equation}\label{formalsolconn}
    \Upsilonh^{\alpha}{}_{\mu\nu}=\left[\frac{1}{2} h^{\kappa \lambda}\left(\na_{\beta}^{h} B_{\ga \lambda}+\na_{\ga}^{h} B_{\lambda \beta}-\na_{\lambda}^{h} B_{\beta \ga}\right)\right]\left(A^{-1}\right)_{\kappa}{}^{\alpha}{}_{\mu\nu}{}^{\beta \gamma},
\end{equation}
where by definition ${A}^\kappa{}_{\al'}{}^{\mu'\nu'}{}_{\be\ga} ({A}^{-1})_\kappa{}^\al{}_{\mu\nu}{}^{\be\ga}\equiv \delta^{\al'}{}_{\al} \delta^{\mu'}{}_{\mu} \delta^{\nu'}{}_{\nu}$. Here ${A}^\kappa{}_{\al'}{}^{\mu'\nu'}{}_{\be\ga}$ is linear in $B_{\mu\nu}$ and is given by
\begin{align}\label{Operatorsolcon}
\begin{split}
 &{A}^\kappa{}_{\al}{}^{\mu\nu}{}_{\be\ga} \equiv \delta^\kappa{}_\al \delta^\mu{}_\be \delta^\nu{}_\ga +\frac{1}{2} \delta^\mu{}_{\alpha}\left(h^{\nu\kappa} h_{\beta \gamma}-2 \delta^{\nu}{}_{(\beta} \delta^\kappa{}_{\gamma)}\right)+b^\kappa{}_{\al}{}^{\mu\nu}{}_{\be\ga}{}^{\rho\sigma}B_{\rho\sigma}\\
 &b^\kappa{}_{\al}{}^{\mu\nu}{}_{\be\ga}{}^{\rho\sigma}=\frac{1}{2}\left[h_{\alpha\gamma} h^{\mu\sigma} \delta^\nu{}_{\beta} h^{\rho\kappa}+\delta^{\beta}{}_\rho h_{\alpha\gamma} h^{\mu\kappa} h^{\nu \sigma}+\delta^{\rho}{}_\gamma \delta^\mu{}_{\alpha} \delta^\nu{}_{\beta} h_{\kappa \sigma}-h^{\rho\kappa} \delta^{\mu}{}_\gamma h^{\nu\sigma} h_{\alpha \beta}\right.\\
 &\hspace{0.4cm}\left.-\delta^{\rho}{}_ \beta h^{\sigma\kappa} \delta^\mu{}_{\alpha}\delta^{\nu}{}_\gamma-\delta^{\rho}{}_ \beta \delta^{\sigma}{}_\gamma\delta^{\mu}{}_\alpha h^{\nu\kappa}-\delta^{\rho}{}_\gamma h_{\alpha \beta} h^{\mu{\sigma}} h^{\nu{\alpha}}-\delta^{\rho}_{\gamma} \delta^{\kappa}{}_{\alpha} \delta^{\mu}{}_\beta h^{\nu\sigma}+\delta^{\rho}{}_\beta \delta^{\kappa}{}_ \alpha h^{\mu \sigma} \delta^{\nu}{}_\gamma\right].
\end{split}
\end{align}
In order to explicitly show the appearance of problematic couplings, it will suffice to give a perturbative solution to leading-order in $B$. To that end, let us consider a trivial 2-form background and expand around it, leaving the symmetric sector $h^{\mu\nu}$ completely general. The only task then is either to compute the $\mathcal{O}(B^0)$ term of $({A}^{-1})_\kappa{}^\al{}_{\mu\nu}{}^{\be\ga}$ or to directly solve the equations \eqref{Eq:Omegaeq} for $\Upsilonh$ expanded as a power series. Let us proceed with the second method by expanding $\Upsilonh$ as a power series of the 2-form $B$ in the form
\beq\label{eq:seriesom}
\Upsilonh^{\al}{}_{\mu\nu}=\sum_{n=0}^{\infty}\Upsilonh_{(n)}{}^\al{}_{\mu\nu},
\eeq
where the sub-index $n$ implies that the quantity $\Upsilonh_{(n)}{}^\al{}_{\mu\nu}$ is of order $\cO(B^n)$. We can now use \eqref{Eq:GammasplittingChO1} to split the connection symbols that appear in \eqref{conneq2}. Plugging the above expansion of $\Upsilonh^{\al}{}_{\mu\nu}$ into the resulting equation, we obtain
\beq
\begin{split}
\na^h_\la B^{\mu\nu}&-B^{\mu\nu}\Upsilonh_{(0)}{}^{\al}{}_{\la\al}-B^\nu{}_\al\Upsilonh_{(0)}{}^{\mu\al}{}_{\la}+B^\mu{}_\al\Upsilonh_{(0)}{}^{\nu}{}_\la{}^\al\\&+h^{\mu\nu}\Upsilonh_{(1)}{}^\al{}_{\la\al}-\Upsilonh_{(1)}{}^{\mu\nu}{}_\lambda-\Upsilonh_{(1)}{}^{\nu}{}_\la{}^\mu-\Upsilonh_{(0)}{}^{\mu\nu}{}_\la-\Upsilonh_{(0)}{}^{\nu}{}_\lambda{}^\mu+h^{\mu\nu}\Upsilonh_{(0)}{}^{\al}{}_{\la\al}=\cO(B^2).\label{conneqpert}
\end{split}
\eeq
Notice that this equation is consistent with substituting the perturbative series \eqref{eq:seriesom} in \eqref{Eq:Omegaeqdivfree}, as it should be.\footnote{To see this explicitly, one should take into account the equation resulting from contracting $\alpha$ and $\mu$ in  \eqref{Eq:Omegaeqdivfree} together with the identity $\Upsilonh^\al{}_{[\al\be]}=0$, wich leads to $\Upsilonh^{\nu\beta}{}_{\beta}- B_{\beta\lambda}\Upsilonh^{\nu\beta\lambda}=0$.} The zeroth order term gives the equation
\beq
\Upsilonh_{(0)}{}^{\mu\nu}{}_\la+\Upsilonh_{(0)}{}^{\nu}{}_\lambda{}^\mu-h^{\mu\nu}\Upsilonh_{(0)}{}^{\al}{}_{\la\al}=0,
\eeq
which after contracting with $h_{\mu\nu}$ gives $\Upsilonh_{(0)}{}^\al{}_{\la\al}=0$ for $D\neq2$. This leaves us with the equation $\Upsilonh_{(0)}{}^{\mu\nu}{}_\la+\Upsilonh_{(0)}{}^{\nu}{}_\lambda{}^\mu=0$. Doing the usual permutation trick we arrive to the unique solution (the equations are linear)
\beq
\Upsilonh_{(0)}{}^\al{}_{\mu\nu}=0,
\eeq
which ensures that the Levi-Civita connection of $h^{\mu\nu}$ is, up to a projective mode, the solution for the affine connection for a symmetric metric. Indeed, this was expected given that for vanishing $B^{\mu\nu}$ we are just solving for GR, which has the Levi-Civita connection of the metric as the only solution up to a projective mode (see \eg \cite{Bernal:2016lhq,Janssen:2019htx}). Plugging the 0-th order result into \eqref{conneqpert}, we arrive to the equation for the $\cO(B)$ term, which reads
\beq
\na^h_\la B^{\mu\nu}-\Upsilonh_{(1)}{}^{\mu\nu}{}_\la-\Upsilonh_{(1)}{}^\nu{}_\la{}^\mu-h^{\mu\nu}\Upsilonh_{(1)}{}^\al{}_{\la\al}=0.
\eeq
Contracting with $h_{\mu\nu}$ we arrive at the condition  $\Upsilonh_{(1)}{}^\al{}_{\la\al}=0$ for $D\neq2$, which leads to the equation 
\beq
\na^h_\la B^{\mu\nu}-\Upsilonh_{(1)}{}^{\mu\nu}{}_\la-\Upsilonh_{(1)}{}^{\nu}{}_\lambda{}^\mu=0.
\eeq
Again, this equation can be uniquely solved by performing the permutation trick, which yields
\beq\label{pertsolcon}
\Upsilonh_{(1)}{}^\al{}_{\mu\nu}=\frac{1}{2}h^{\al\la}\left(\na^h_\mu B_{\nu\la}+\na^h_\nu B_{\la\mu}-\na^h_\la B_{\mu\nu}\right)
\eeq
in agreement with previous results in NGT obtained in \cite{Damour:1991ru}. As well, this also agrees with the formal solution \eqref{formalsolconn} given above as, though the form of $a^\kappa{}_{\al}{}^{\mu\nu}{}_{\be\ga}$ in \eqref{Operatorsolcon} suggests that the formal solution has more contributions to first-order in $B$ than \eqref{pertsolcon}, it can be seen that ${a}^\kappa{}_{\al}{}^{\mu\nu}{}_{\be\ga}\hat\Upsilon_{(1)}{}^{\al}{}_{\mu\nu}=\Upsilon_{(1)}{}^{\kappa}{}_{\be\ga}+\mathcal{O}(B^2)$, which implies that the formal solution \eqref{formalsolconn} and the first-order perturbative one \eqref{pertsolcon} are consistent

 As stated in the end of the previous section, and analogously to the results on NGT in \cite{Damour:1991ru}, the dependence of $\Upsilonh$ on the derivatives of $B^{\mu\nu}$ will introduce additional pathologies in the 2-form field. As a matter of fact, upon substitution of this solution into \eqref{RBGactionExpanded} and integration by parts, we arrive at the desired action, similar to \eqref{eq:Actquad}, which features a gauge-invariant kinetic term for the 2-form together with the nonminimal couplings advertised above. Again, the gauge invariance of the derivative operators for the 2-form is accidental of this order, but it is broken at cubic and higher-orders. It is possible, although tedious, to obtain the solution for $\Upsilon$ at arbitrary order by following this perturbative scheme. Obtaining a full solution in closed form appears to be a more challenging task.

\subsection{Matter couplings cannot exorcise the ghosts}\label{sec:Mattercoupling}

In the precedent sections we have only considered matter fields which do not couple to the connection. However, our conclusions on the presence of pathological dof's  do not change substantially by coupling the connection to the matter sector. Couplings to matter fields in a metric-affine framework is an interesting issue by itself, specially when it involves spinor fields (see e.g. \cite{Hehl:1994ue,Delhom:2020hkb,BeltranJimenez:2020sih}). It is not the scope of this section to carefully go through the different coupling prescriptions to matter nor their consistency, as this was already discussed in chapter \ref{sec:MinimalCoupling}. Instead, our aim is to show how our results above are not substantially affected in the presence of matter fields both with minimal and nonminimal couplings in general. We will mostly discuss minimal-couplings, although we will also elaborate on the extension of these results when nonminimal couplings are also included. 

\subsubsection{The nonsymmetric gravity frame for nonminimally coupled fields}\label{sec:NGTframemat}

Curvature couplings to the matter sector include derivatives of the affine connection in the matter Lagrangian. This further complicates the connection equation \eqref{GeneralisedConneq} by adding extra terms on the right hand side. However, there is a class of couplings for which, while adding technical complications, the qualitative results remain the same with just some minor adjustments with respect to the minimally coupled fields. 

We will start by considering bosonic fields whose nonminimal couplings are through the Ricci tensor. To illustrate this point, we can consider a scalar field $\varphi$ as a proxy for the matter sector. If we restrict to only first derivatives of the scalar, we can use for instance $R^{\mu\nu}\partial_\mu\varphi\partial_\nu\varphi$ or $R(\partial\varphi)^2$ in our action. In the usual metric formalism, these two terms are only allowed if they enter through the specific combination $(R^{\mu\nu}-\frac12 Rg^{\mu\nu})\partial_\mu\varphi\partial_\nu\varphi$ and accompanied by the appropriate second derivative interactions of the scalar field in order to avoid Ostrogradski instabilities (see \eg \cite{Heisenberg:2018vsk}). In the metric-affine formalism however, this is not necessary and the dependence on said terms is completely arbitrary. Let us note that these interactions will not break the projective symmetry since they only depend on the symmetric part of the Ricci tensor. Interestingly, it has been suggested in \cite{Aoki:2019rvi} that the projective symmetry could also play a crucial role to guarantee the absence of ghosts for theories containing up to second-order covariant derivatives of a scalar field. The authors of \cite{Aoki:2019rvi} also found that in the stable theories the connection is devoid of any propagating mode as a consistency condition as we argued above.

Our reasoning can be straightforwardly extended to other fields such as vector fields $A_\mu$ where interactions like $R_{\mu\nu}A^\mu A^\nu$ or $R_{\mu\nu} F^{\mu\alpha}F^\nu{}_\alpha$ also respect the projective symmetry and are permitted. The crucial point of all these interactions is that an Einstein frame still exists where it is apparent that the connection remains an auxiliary field \cite{Afonso:2017bxr}. In the absence of the projective symmetry, we will encounter the same pathologies as exposed for the pure gravitational sector and the inclusion of a contrived matter sector cannot remedy it.

In section \ref{sec:RBGTheory} we showed how to go to the Einstein frame of RBG theories for minimally coupled matter fields. Let us see here how to proceed in the presence of nonminimally coupled matter fields. In this case the action reads
\beq\label{GeneralActionHyp}
\cS[g_{\mu\nu},\Ga,\Psi]=\frac12\int \dd^D x \sqrt{-g}\,F\big(g^{\mu\nu},R_{\mu\nu}\big)+\cS_\textrm{m}[g,\Psi,\Gamma].
\eeq
Parallel to \ref{sec:ProblemAdditionalDOFs}, we now go to the Einstein frame of the above theory, and after splitting the corresponding auxiliary metric as in \eqref{metricsplitting} and the connection as in \eqref{Eq:GammasplittingChO}, and also isolating the projective mode from $\Upsilon^\al{}_{\mu\nu}$ as in \eqref{Eq:Omegah}, we get 

\beq
\begin{split}
\cS=\frac12\int&\dd^Dx\sqrt{-h}\Big[R^h-\frac{2}{D-1}B^{\mu\nu}\partial_{[\mu}\Upsilon_{\nu]}+\Upsilonh^\alpha{}_{\alpha\lambda}\Upsilonh^\lambda{}_{\kappa}{}^\kappa-\Upsilonh^{\alpha\mu\lambda}\Upsilonh_{\lambda\alpha\mu}-\Upsilonh^\alpha{}_{\alpha\lambda}\Upsilonh^\lambda{}_{\mu\nu}B^{\mu\nu}\\
&-\Upsilonh^\alpha{}_{\nu\lambda}\Upsilonh^\lambda{}_{\alpha\mu}B^{\mu\nu}-B^{\mu\nu}\na^h_\alpha\Upsilonh^\alpha{}_{\mu\nu}
-B^{\mu\nu}\na^h_\nu\Upsilonh^\alpha{}_{\alpha\mu}+\cU(B)\Big]+\tilde\cS_\textrm{m}[h,B,\Psi,\hat\Upsilon,\Upsilon].\label{RBGactionExpandedHyp}
\end{split}
\eeq
where now $\tilde\cS_\textrm{m}$ is the matter action in the Einstein frame, and the variables inside square brackets means that the matter action can depend on those fields and their derivatives in general. Concretely $\Upsilon$ stands for the dependence of the matter action on the projective mode, so it will be absent for projectively invariant matter. It is apparent that the gravitational sector features the same pathological terms as in vacuum. Obviously, a trivial matter background will not modify those terms. A nontrivial matter background contributing to the background of the symmetric part of the metric could in principle help in healing the ghost associated to the projective mode by providing a healthy kinetic term for it. However, the nonminimal couplings to the curvature for the 2-form that are generated after integrating $\hat{\Upsilon}$ out can hardly be cured. In any case, this would require very specific choices of the matter sector. To make this statement more explicit, let us consider a particular class of matter sector coupled to the connection.

\subsubsection{Ultralocal matter couplings}\label{sec:SolConmat}

For mater actions which do not include curvature couplings (\ie no derivatives of the connection), we already know that the projective mode will be problematic due to the absence of a proper kinetic term for it. In order to understand if the inclusion of a general coupling between matter and connection can solve the instability problems we can now compare the above action \eqref{RBGactionExpandedHyp} to \eqref{RBGactionExpanded}. First notice that the divergence-free constraint of the 2-form \eqref{DivConstB} that came from the field equations of the projective mode gets modified if nonprojectively invariant matter actions are taken into account, and the trace of the hypermomentum acts now as a source for $B$
\beq
\na^h_\mu B^{\mu\nu}=\frac{D-1}{4}\Delta_\al{}^{[\mu\al]}
\eeq

where $\Delta_{\lambda}^{\mu \nu}$ is the hypermomentum defined as

\begin{equation}\label{DefHypermomentum}
\left.\Delta_{\lambda}{}^{\mu \nu} \equiv 2 \frac{\delta \mathcal{S}_{m}}{\delta \Gamma^{\lambda}{}_{\mu \nu}}\right|_{g_{\mu \nu}}=2 \left.\frac{\delta \mathcal{S}_{m}}{\delta \Upsilon^{\lambda}{}_{\mu \nu}}\right|_{g_{\mu \nu}}
\end{equation}
and which vanishes for matter fields that do not couple to the connection. Looking at the form of this action, we can see that the projective mode will in general feature the same problems as in the previous case when the matter and connection did not couple. The Ostrogradski instabilities that arise from the couplings between the 2-form $B^{\mu\nu}$ and the curvature of $h_{\mu\nu}$ will still be there no matter what matter action we choose. Therefore, we see that allowing for an arbitrary coupling between matter and connection is not helpful in solving any of the instabilities listed above. To explicitly see what kind of couplings arise, we have to solve the connection equation now with a hypermomentum. Since generally an analytic solution is not possible, even if it is not very illuminating we will attempt to find a perturbative solution which will already give us a clear picture of the problem. 

The connection equations when a coupling between matter and connection is present are, in general, given by
\begin{equation}\label{GeneralisedConneq}
\begin{split}
{\na_{\lambda}\left[\sqrt{-q} q ^{\nu\mu}\right]-\del^\mu{}_\la \na_{\rho}\left[\sqrt{-q} q ^{ \nu\rho}\right]}{=\Delta_{\lambda}{}^{\mu \nu}+\sqrt{-q}\left[\mathcal{T}^{\mu}{}_{\lambda \alpha} q ^{\nu\al}+\mathcal{T}^{\alpha}{}_{\alpha \lambda} q ^{\nu\mu}-\delta_{\lambda}^{\mu} \mathcal{T}^{\alpha}{}_{\alpha \beta} q ^{\nu\be}\right]}.
\end{split}
\end{equation}
  In order to remain as close as possible to the previous analysis in section \ref{sec:SolConn}, it is necessary to use the shifted connection \eqref{Eq:Omegah} and find the relation between the hypermomentum of the original connection $\Delta_\al{}^{\mu\nu}$ and the shifted hypermomentum  $\Deltah_\al{}^{\mu\nu}$, which reads
\beq
\Delta_\al{}^{\mu\nu}=\Deltah_\al{}^{\mu\nu}+\frac{2}{D-1}\delta_\al{}^{[\mu}\Deltah_\be{}^{\nu]\be},
\eeq
where the shifted hypermomentum is defined in an analogous manner as \eqref{DefHypermomentum}. This implies that the hypermomentum of projectively invariant matter fields satisfies $\Deltah_\be{}^{\mu\be}=0$. We can now recast \eqref{GeneralisedConneq} in the form of \eqref{conneq2} by doing the same manipulations, thus finding
\beq
\partial_\la(\sqrt{-q}q^{\mu\nu})+\hat{\Gamma}^{\mu}{}_{\la\al}\sqrt{-q}q^{\al\nu}+\hat{\Gamma}^{\nu}{}_{\al\la}\sqrt{-q}q^{\mu\al}-\hat{\Gamma}^{\al}{}_{\la\al}\sqrt{-q}q^{\mu\nu}=\Deltah_\al{}^{\mu\nu}+\frac{2}{D-1}\delta_\al{}^{[\mu}\Deltah_\be{}^{\nu]\be}.\label{conneqhyperm}
\eeq
 As in the vanishing hypermomentum case, we can obtain a formal solution for the full connection in the case of arbitrary hypermomentum as
\begin{equation}
\begin{split}
\Upsilonh^{\alpha}{}_{\mu\nu}=\frac{1}{2} h^{\kappa \lambda}\Bigg[&\left(\na_{\beta}^{h} B_{\ga \lambda}+\na_{\ga}^{h} B_{\lambda \beta}-\na_{\lambda}^{h} B_{\beta \ga}\right)\\
&+\frac{1}{\sqrt{-h}} h^{\kappa \lambda}\left(\Deltah_{\beta\ga \lambda}+\Deltah_{\ga\lambda \beta}+\Deltah_{\lambda\beta \ga}+\frac{2}{D-1}h_{\lambda[\gamma}\Deltah^\al{}_{\be]\al}\right)\Bigg]\left(A^{-1}\right)_{\kappa}{}^{\alpha}{}_{\mu\nu}{}^{\beta \gamma},
\end{split}
\end{equation}
where $\left(A^{-1}\right)_{\kappa}{}^{\alpha}{}_{\mu\nu}{}^{\beta \gamma}$ is the same operator as in the vanishing hypermomentum case, which is specified in \eqref{Operatorsolcon}. Notice that the above formula points to the fact that the addition of hypermomentum does not solve any of the instabilities due to the dependence of $\Upsilonh$ on the derivatives of $B_{\mu\nu}$. To see that this is the case, let us find a perturbative solution to the connection in an analogous way to that of \ref{sec:SolConn}. First we need to write $\Deltah_ \al{}^{\mu\nu}=\Deltah^{(0)}_ \al{}^{\mu\nu}+\Deltah^{(1)}_ \al{}^{\mu\nu}+...$ as a power series in $B^{\mu\nu}$, where the superscript $(n)$ indicates that such term is of order $\mathcal{O}(B^n)$. Then, after splitting the shifted connection as in \eqref{Eq:GammasplittingChO1} and then writing $\Upsilonh^\al{}_{\mu\nu}$ as a power series in $B^{\mu\nu}$ as in \eqref{eq:seriesom}, we can write \eqref{GeneralisedConneq} in an analogous way to \eqref{conneqpert} as
\beq
\begin{split}
\na^h_\la B^{\mu\nu}&-B^{\mu\nu}\Upsilonh_{(0)}{}^{\al}{}_{\la\al}-B^\nu{}_\al\Upsilonh_{(0)}{}^{\mu\al}{}_{\la}+B^\mu{}_\al\Upsilonh_{(0)}{}^{\nu}{}_\la{}^\al-\Upsilonh_{(0)}{}^{\mu\nu}{}_\la-\Upsilonh_{(0)}{}^{\nu}{}_\lambda{}^\mu+h^{\mu\nu}\Upsilonh_{(0)}{}^{\al}{}_{\la\al}\\
&+h^{\mu\nu}\Upsilonh_{(1)}{}^\al{}_{\la\al}-\Upsilonh_{(1)}{}^{\mu\nu}{}_\lambda-\Upsilonh_{(1)}{}^{\nu}{}_\la{}^\mu\\
&-\Deltah^{(0)}_\al{}^{\mu\nu}+\frac{2}{D-1}\delta_\al{}^{[\mu}\Deltah^{(0)}_\be{}^{\nu]\be}-\Deltah^{(1)}_\al{}^{\mu\nu}+\frac{2}{D-1}\delta_\al{}^{[\mu}\Deltah^{(1)}_\be{}^{\nu]\be}=\cO(B^2).
\end{split}
\eeq
 Notice that in general, $\Deltah^{(n)}_\al{}^{\mu\nu}$ might have a complicated dependence on the affine connection, and thus on $\Upsilonh^\al{}_{\mu\nu}$, which may complicate further the solution of the above equation for $\Upsilonh^\al{}_{\mu\nu}$ order by order in $B^{\mu\nu}$. Thus, in general, one could make a further expansion of each $\Deltah^{(n)}_\al{}^{\mu\nu}=\Deltah^{(0,n)}_\al{}^{\mu\nu}+\Deltah^{(1,n)}_\al{}^{\mu\nu}+...$ where the superscript $(m,n)$ denotes a term of order $\mathcal{O}(\Upsilonh^m)$ and $\mathcal{O}(B^n)$. Since the completely general case is rather cumbersome, and is not particularly illuminating, let us focus on the case where the hypermomentum does not depend on the affine connection, \ie where the connection couples to matter only linearly. This would be the case, for instance, of minimally coupled spin-1/2 fields, which have a hypermomentum of the form 
 \beq
 \Delta^{(\Psi)}_\al{}^{\mu\nu}=-i\sqrt{-h}h_{\al\rho}\epsilon^{\rho\sigma\mu\nu}\lrsq{\bar{\Psi}\gamma_\sigma\gamma_5\Psi}.
 \eeq
Under such assumption, we can expand $\Upsilonh^\alpha{}_{\mu\nu}$ only in terms of $B^{\mu\nu}$.  Assuming thus no dependence of the hypermomentum on the connection\footnote{The equation of zeroth order would still be formally valid for $\hat\Delta_{\al}{}^{\mu\nu}$ that depends on the connection, altough in that case it will be harder to isolate $\Upsilon^\al{}_{\mu\nu}$.} (\ie the matter couples to the connection only linearly), we can proceed exactly as in section \ref{sec:SolConn} to obtain the following zeroth-order solution
\beq
\begin{split}
\Upsilonh_{(0)}{}^\al{}_{\mu\nu}=&\frac{1}{\sqrt{-h}}\Big[\Deltah^{(0)}{}_{\mu\nu}{}^\al+\Deltah^{(0)}_\nu{}^\al{}_\mu-\Deltah^{(0)}{}^\al{}_{\mu\nu}+\frac{2}{D-1}\delta^\la{}_{[\nu}\Deltah^{(0)}{}^\al{}_{\mu]\la}\\
&\hspace{6cm}+\frac{1}{2(D-2)}\lr{h_{\mu\nu}\Deltah^{(0)}{}^{\al\la}{}_\la-2\delta^\al{}_{(\mu}\Deltah^{(0)}{}_{\nu)\la}{}^\la}\Big],
\end{split}
\eeq
where $\Deltah^{(0)}{}_\al{}^{(\al\be)}=0$ and $\Upsilonh_{(0)}{}^{\al\be}{}_{\be}=0$ must be satisfied as can be shown from the connection field equations and the identity $\Upsilonh_{(n)}{}^\al{}_{[\al\be]}=0$. The first-order solution is obtained equally, leading to
\beq
\begin{split}
\Upsilonh_{(1)}{}^\al{}_{\mu\nu}&=\frac{1}{2}h^{\al\la}\Big(\na^h_\la B_{\nu\mu}+\na^h_\mu B_{\nu\la}+\na^h_\nu B_{\la\mu}\Big)-\frac{1}{\sqrt{-h}}\Bigg[\frac12\lr{\Deltah^{(1)}{}^\al{}_{\mu\nu}+\Deltah^{(1)}{}_{\mu\nu}{}^\al+\Deltah^{(1)}{}_\nu{}^\al{}_{\mu}}\\
&-\frac{1}{D-2}\delta^\al{}_{(\mu}\Deltah^{(0)}{}_{\nu)}{}^{\ga\sigma}B_{\ga\sigma}+\frac{2}{(D-1)(D-2)}\delta^\al{}_{(\mu}B_{\nu)\ga}\Deltah^{(0)}{}_\sigma{}^{\sigma\ga}+\frac{1}{D-2}\delta^\al{}_{(\mu}\Deltah^{(1)}{}_{\nu)\sigma}{}^{\sigma}\\
&+\frac{1}{D-2}\delta^\al{}_{[\mu}B_{\nu]\ga}\Deltah^{(0)}{}^{\ga\sigma}{}_\sigma+\frac{2}{D-1}\delta^\al{}_{[\mu}\Deltah^{(1)}{}^{\sigma}{}_{\nu]\sigma}+\frac12\lr{ B_{\mu\sigma}\Deltah^{(0)}{}^{\sigma}{}_{\nu}{}^{\al}-B_{\nu\sigma}\Deltah^{(0)}{}^{\sigma\al}{}_{\nu}}\\
&+\frac{1}{2(D-2)}h_{\mu\nu}B_{\ga\sigma}\Deltah^{(0)}{}^{\al\ga\sigma}-\frac{1}{D-2}h_{\mu\nu}B^\al{}_\sigma\Deltah^{(0)}{}_{\ga}{}^{\ga\sigma}-\frac{1}{2(D-2)}h_{\mu\nu}\Deltah^{(1)}{}^{\al\sigma}{}_{\sigma}\Bigg],
\end{split}
\eeq
As we can see, besides obtaining the problematic $\Upsilonh\sim\na^h B+\mathcal{O}(B^2)$ terms that we obtained in the vanishing hypermomentum case, we here obtain also a bunch of terms that couple nonminimally the matter fields with themselves and with the 2-form $B^{\mu\nu}$ through their hypermomentum. It is apparent that the addition of these new terms cannot heal the problematic behaviour of the $\na^h B$ terms by themselves, thus clarifying why the addition of nonminimal couplings to matter fields would not solve the instability problem. Indeed, if the theory is regarded as an EFT, the extra couplings between the unstable 2-form and the matter fields potentially reduce the time-scale in which the 2-form instability manifests physically through its decay to lighter particles.

%%%%%%%%%%%%%%%%%%%%%%%%%%%%%%%
%%%%%%%%%%%%%%%%%%%%%%%%%%%%%%%%%%%%%%%%%%%%%%%

\subsection{Constrained geometries can exorcise (some of) the ghosts}\label{sec:GeoConst}

In the precedent sections we have seen that abandoning the projective symmetry in the higher-order curvature sector of a metric-affine theory results in the appearance of ghost-like pathologies precisely related to the projective mode and an extra 2-form field that comes from the excitations of the antisymmetric part that the Einstein frame metric develops if the projective symmetry is dropped. As is well known, in some cases, in addition to imposing symmetries, there are other mechanisms that can {\it freeze} degrees of freedom\footnote{In this sense, from the Hamiltonian point of view, gauge symmetries are understood as first-class constraints that reduce the apparent number of degrees of freedom of the system \cite{Henneaux:1992ig}.} by imposing suitable constraints, which can then cure the ghosts if the constraints are the right ones.

We will now discuss the different frameworks where RBG theories with explicitly broken projective symmetry can be rendered stable, not by imposing additional symmetries, but by enforcing suitable constraints on the connection, \ie by restricting to some specific geometries. In this respect, besides the projectively invariant RBG class, it is known that there are families of theories which contain higher-order curvature corrections and are stable (ghost-free) for some particular classes of geometries. We will review some of these known examples where the connection is deprived of specific components of the nonmetricity and/or torsion. We will finally show a general result that imposing a vanishing torsion stabilises RBG without projective symmetry theories transforming it into a theory with an extra interacting massive vector field. We will see as well how imposing a vanishing nonmetricity is not able to heal the theories.

\subsubsection{Torsion-free theories exorcise the ghosts}\label{sec:TorsionFree}

 We will start by showing how imposing a vanishing torsion avoids the presence of ghosts. The implementation of this constraint can be performed either by only allowing for variations of the symmetric part of the connection (\ie assuming a symmetric connection from the beginning) or by introducing a set of Lagrange multipliers that enforce the constraint $T^\alpha{}_{\mu\nu}=0$ as dictated by their field equations. The resulting connection equations after taking into account this constraint reads
\beq\label{ConnectionFieldEqsTorsionFree}
\na_\la\lrsq{\sqrt{-q}q^{(\mu\nu)}}-\na_\rho\lrsq{\sqrt{-q}q^{\rho(\mu}}\del^{\nu)}_\la=0.
\eeq
Notice that the only difference with respect to the equations for the unconstrained connection is precisely the trivialisation of their antisymmetric part in $\mu$ and $\nu$. Let us decompose $q^{\mu\nu}$ again into its symmetric and antisymmetric parts as in \eqref{metricsplitting}. Due to the vanishing of the torsion tensor, the general decomposition of the connection \eqref{eq:ContortionDistortion} lacks the contortion tensor. Thus, the connection can here be split in a Levi-Civita connection of $h^{\mu\nu}$ and a disformation part that depends on the nonmetricity $N_{\lambda\mu\nu}\equiv \na^h_\lambda h_{\mu\nu}$ as\footnote{This splitting allows us to write a general affine connection in terms of the torsion, an arbitrary invertible symmetric 2-tensor, its first derivatives and its covariant derivative (\ie its nonmetricity with respect to $\Ga$).}
\beq\label{ConnectionDeccomposition}
\Ga_{\mu\nu}^\al=\bar{\Ga}_{\mu\nu}^\al(h)+L^\al{}_{\mu\nu}(N)
\eeq
without loss of generality, where the disformation tensor is now built with the nonmetricity of $h^{\mu\nu}$. The above splitting allows to obtain the following relations that will be of use below
\begin{align}
\na_\lambda\big(\sqrt{-h}h^{\lambda\nu}\big)&=\sqrt{-h}\bL^\nu,\label{L1}\\
\na_\lambda\big(\sqrt{-h}B^{\lambda\nu}\big)&=\sqrt{-h}\na^h_\lambda B^{\lambda\nu},\label{L2}
\end{align}
where $\bL^\nu\equiv L^\nu{}_{\alpha\beta} h^{\alpha\beta}$ is one of the two independent traces of the disformation tensor. The trace of the connection equation (\ref{ConnectionFieldEqsTorsionFree}) together with \eqref{L1} yields 
\beq
\na^h_\lambda B^{\lambda \nu}=\frac{1-D}{1+D}\bL^\nu,
\label{eq:constraint1}
\eeq
which implies the dynamical constraint\footnote{As explained in section \ref{sec:MinimalCoupling}, the above equation can also be written as ${}\star\dif\star B\propto L$ where $\ast$ is the Hodge dual with respect to $h_{\mu\nu}$,  and since ${}\star{}\star \alpha\propto\alpha$ for any p-form $\alpha$, we have that ${}\star\dif\star L\propto{}\star\dif\star{}\star\dif\star B\propto{}\star\dif\dif\star B=0$ because $\dif^2=0$ on any p-form.}
\beq
\na^h_\nu \bL^\nu=0.
\label{eq:constraint2}
\eeq
On the other hand, contracting the connection equation (\ref{ConnectionFieldEqsTorsionFree}), with $h_{\mu\nu}$ defined as the inverse of $h^{\mu\nu}$, leads to 
\beq\label{eq:RelationTracesDisformationNoTorsion}
L_\mu=\frac{2}{(2-D)(1+D)}\bL_\mu,
\eeq
where $L_{\mu}\equiv L^\alpha{}_{\mu\alpha}$ and indices are raised and lowered with $h_{\mu\nu}$. Thus, we see that there is only one independent trace of the disformation tensor. Using the above relations in the connection equation \eqref{ConnectionFieldEqsTorsionFree}, we are led to
\beq
2h^{\alpha(\mu} L^{\nu)}{}_{\lambda\alpha}=l_{\textrm{G}} h^{\mu\nu}+(2-D) L_\alpha h^{\alpha(\mu}\delta^{\nu)}{}_\lambda.
\label{eq:connectionTfree2}
\eeq
Given that the nonmetricity tensor of the auxiliary metric
is given by $N_\lambda{}^{\mu\nu}\equiv -\na_\lambda h^{\mu\nu}=-2h^{\alpha(\mu}L^{\nu)}{}_{\lambda\alpha}$, which implies the identity $L_\mu=-\frac12h_{\alpha\beta}N_\mu{}^{\alpha\beta}\equiv-\frac12 \tilde{N}_\mu$, the above equation can be used to re-write the connection equation (\ref{eq:connectionTfree2}) as a constraint for the nonmetricity tensor
\beq\label{relationtracesL}
N_\lambda{}^{\mu\nu}=\frac12\Big[ \tilde{N}_\lambda h^{\mu\nu}+(2-D)\tilde{N}_\alpha h^{\alpha(\mu}\delta^{\nu)}_\lambda\Big],
\eeq
which becomes completely specified by its Weyl component (although it is not Weyl-like). Thus we see that the connection field equations can be fully solved explicitly, and the connection is given by a disformation piece given by the nonmetricity tensor \eqref{relationtracesL} added to the Levi-Civita of $h^{\mu\nu}$. Given that this disformaton piece is completely determined by $\tilde{N}_\mu$ (the Weyl trace of the nonmetricity of $h^{\mu\nu}$), which is divergenceless by the constraint equations \eqref{eq:constraint2} and  \eqref{eq:RelationTracesDisformationNoTorsion}, the field equations of the connection describe only the propagation of one additional vector component, instead of a vector field plus a 2-form as in the most general case. Moreover, from  the divergence-free constraint (\ref{eq:constraint2}) obtained above, this new vectorial component must be a Proca field, thus propagating only three extra degrees of freedom. The corresponding metric equations of the system will allow to solve algebraically for $h^{\mu\nu}$ as a function of the matter fields and (possibly) the new vector field $\tilde{N}_\mu$, which ensures the absence of the pathologies that were found in the most general case. 

To illustrate this, let us re-consider a particular example that has already been treated in the literature. Assume a metric-affine gravitational Lagrangian of the form 
\begin{equation}
 \mathcal{L}=R+c_1 R_{[\mu\nu]}R^{[\mu\nu]}.   
\end{equation} 
As explained above (see section \ref{sec:GeneralAffineConnection}), this theory explicitly breaks  projective symmetry  due to the presence of the antisymmetric part of the Ricci in the action. Therefore pathologies should arise in the general case unless further constraints are imposed. However, as shown in past works \cite{Buchdahl:1979ut,Vitagliano:2010pq},  the torsion-free version of this model reduces to an Einstein-Proca system , where the Proca field arises from the connection sector. For more general examples with violation of projective symmetry but where the torsion-free constraint is imposed, the Proca field will in general develop nontrivial interactions, as was already discussed in  \cite{Olmo:2013lta} for the Ricci-Based sub-family $F(g^{\mu\nu}, R^{\mu\nu}R_{\mu\nu})$ with the torsion-free constraint.

To enlighten the mechanism that renders the torsion-free version of RBG theories without projective symmetry ghost-free, let us resort to the Einstein frame of the theory making explicit the torsion-free constraint. The action of the theory can be written as
\begin{align}\label{NoTorsionEinstein}
\cS=&\frac12\int\dd^Dx\sqrt{-g}\Big[F(\Sigma,A)+\frac{\partial F}{\partial \Sigma_{\mu\nu}}\big(R_{(\mu\nu)}-\Sigma_{\mu\nu}\big)+\frac{\partial F}{\partial A_{\mu\nu}}\big(R_{[\mu\nu]}-A_{\mu\nu}\big)+\frac{1}{\sqrt{-g}}\lambda_\alpha{}^{\mu\nu}T^\alpha{}_{\mu\nu}\Big],
\end{align}
where $\lambda_\alpha{}^{\mu\nu}$ is a Lagrange multiplier that enforces the torsion-free constraint $T^\alpha{}_{\mu\nu}=0$ and $A_{\mu\nu}$ and $\Sigma_{\mu\nu}$ are auxiliary fields that are antisymmetric and symmetric respectively. In an analogue manner to what we did for general RBG theories in section \ref{sec:RBGTheory}, we can perform field redefinitions which allow us to algebraically solve for the space-time metric $g^{\mu\nu}$ in terms of $h^{\mu\nu}$, $B^{\mu\nu}$ and the matter fields; thus integrating $g^{\mu\nu}$ out. We can then write the Einstein frame action for torsion-free RBG theories without projective symmetry as
\begin{align}
\cS=&\frac12\int\dd^Dx\Big[\sqrt{-h}h^{\mu\nu}R_{(\mu\nu)}+\sqrt{-h}B^{\mu\nu}R_{[\mu\nu]}+\cU(h,B,T)+\lambda_\alpha{}^{\mu\nu}T^\alpha{}_{\mu\nu}\Big].
\end{align}
This action gives the same connection equations that we solved above \eqref{ConnectionFieldEqsTorsionFree}, so we can take the above solution (basically the splitting (\ref{ConnectionDeccomposition}) and equation (\ref{relationtracesL}) together)  and plug it back into the above action. As it can be seen, the solution for the connection satisfies the relations 
\begin{align}
R_{[\mu\nu]}=&-\frac{1}{2}\partial_{[\mu}\tilde{N}_{\nu]},\nonumber \\
R_{(\mu\nu)}=&R^h_{\mu\nu}+\frac{(D-2)(D-1)}{16}\tilde{N}_\mu \tilde{N}_\nu-\frac{(D-1)}{4}h_{\mu\nu}\na^h_\alpha \tilde{N}^\alpha
\end{align}
which, after dropping the surface term $\na^h_\mu \tilde{N}^\mu$, allow us to re-express the action \eqref{NoTorsionEinstein} in terms of the metric $h_{\mu\nu}$, the 2-form $B_{\mu\nu}$ and the vector field $\tilde{N}^\mu$ as
\begin{align}
\begin{split}
\cS=\frac12\int\dd^Dx\Big[\sqrt{-h}\Big(&R^h+\frac{(D-2)(D-1)}{16}\tilde{N}^2-\frac{1}{2}B^{\mu\nu}\partial_{[\mu}\tilde{N}_{\nu]}\Big)+\cU(h,B,T)\Big],
\end{split}
\label{eq:finalaction}
\end{align}
Notice that this form of the action reproduces the constraint on the 2-form (\ref{eq:constraint1}) as the field equations of the vector field $\tilde{N}^\mu$ (which correspond to the connection equations in the original frame of the theory), which read
\beq
\na^h_\mu B^{\mu\nu}=-\frac{(D-2)(D-1)}{4}\tilde{N}^\nu,
\eeq
and imply the constraint $\na^h_\alpha \tilde{N}^\alpha=0$. At the same time the 2-form field equations yield a nonlinear relation among the 2-form, the field-strength of the vector field $\tilde{N}^\mu$, and the matter fields given by 
\beq\label{ProcaEqPotential}
\partial_{[\mu}\tilde{N}_{\nu]}=\frac{2}{\sqrt{-h}}\frac{\partial{\cU}}{\partial B^{\mu\nu}}.
\eeq
 This stems from the fact that our final action (\ref{eq:finalaction}) is nothing but the first-order form of a self-interacting massive vector field coupled to the matter sector. 
 
 Going back to the particular case $F=R+c_1 R_{[\mu\nu]}R^{[\mu\nu]}$, we can reproduce the above results, which agree with \cite{Buchdahl:1979ut,Vitagliano:2010pq,Olmo:2013lta}. For this particular example, the metric $h^{\mu\nu}$ is exactly $g^{\mu\nu}$, the 2-form is given by $B^{\mu\nu}=2c_1R^{[\mu\nu]}$, and therefore the effective potential reads 
 \beq
 \cU=-(\sqrt{-h}/4c_1) B^{\mu\nu}B_{\mu\nu},
 \eeq
  which leads to an equation \eqref{ProcaEqPotential} of the form
 \beq
 \text{d}\tilde{N}=\frac{2}{c_1}B
 \eeq
showing that (\ref{eq:finalaction}) is indeed the first-order description of a free Proca field $\tilde{N}_\mu$ with field-strength proportional to $B_{\mu\nu}$. To summarize, we have shown in this section that imposing a torsion-free geometry cures the instabilities for RBG theories without projective symmetry. It does so by turning the theory equivalent to a healthy Einstein-Proca theory instead of a Non Symmetric Gravity theory with a ghostly projective mode. 

\subsubsection{Weyl geometries}
Let us now briefly comment on another paradigmatic extension of the Riemannian framework introduced by Weyl shortly after the GR inception which has been analised widely in the literature (see e.g. the nice survey in \cite{Scholz:2011za}). This geometry is characterised by local scale (gauge) invariance and a torsion-free connection so the only nontrivial part of the affine connection is the so-called Weyl trace of the nonmetricity $A_\alpha=-\frac{2}{D}g^{\mu\nu}Q_{\alpha\mu\nu}$. This allows to replace the metric compatibility condition $\na^g_\alpha g_{\mu\nu}=0$ by $\na_\alpha g_{\mu\nu}\equiv(\na^g_\alpha-A_\alpha)g_{\mu\nu}=0$, which is the $C_{(1,3)}$-covariant derivative in the associated bundle to $\tb$ (see chapter \ref{sec:DifferentialGeometry}), where $C_{(1,3)}$ is the group of conformal transformations, which contains the Poincar\'e group as a subgroup as well as the group of scale transformations
\beq
g_{\alpha\beta}\rightarrow e^{2\alpha(x)} g_{\alpha\beta}.
\eeq
$A_\mu$ is the corresponding connection 1-form that transforms as $A_\mu\rightarrow A_\mu-\partial_\mu \alpha$ as required by \eqref{eq:TransformationConnection}. $A_\mu$ is usually called dilaton field, as it is the gauge field associated to dilatations or scale transformations.

Theories whose actions are constructed in terms of quadratic curvature invariants for a Weyl connection trivially admit ghost-free formulations and, consequently, imposing the connection to be of the Weyl form evidently avoids the ghostly pathologies of the general RBG theories. This constraint can be implemented either by imposing the connection to be Weyl-like from the beginning or by adding suitable Lagrange multipliers. Now we should impose a vanishing torsion and also vanishing of all the nonmetricity irreducible components except for the Weyl trace. Since for the torsion-free case there are no ghostly degrees of freedom, it is clear that for Weyl geometries, since they are a sub-class of the torsion-free ones, which also feature additional constraints (nonmetricity is forced to be vectorial), there will be no ghosts either. General quadratic theories in Weyl geometries have been studied in e.g. \cite{Jimenez:2014rna} where it was shown that some interesting nontrivial interactions for the Weyl vector can be generated.

\subsubsection{Geometries with a general vector distortion of the connection}\label{sec:Vecdir}

The affine connection in Weyl geometries is characterised by a vector field that controls the departure from the Levi-Civita connection. A natural generalisation is to include not only this vector part, but a general vector piece of the connection in both the torsion and the nonmetricity sectors. Such a general connection was considered in \cite{Aringazin_1991} in the absence of torsion and was extended to include the torsion trace in \cite{Jimenez:2015fva,Jimenez:2016opp}. The connection in these geometries can be parameterised as
\beq
\Gamma^\alpha{}_{\mu\nu}=\bar{\Gamma}^\alpha{}_{\mu\nu}-b_1A^\alpha g_{\beta\gamma}+b_2\delta^\alpha_{(\beta} A_{\gamma)}+b_3\delta^\alpha_{[\beta} A_{\gamma]}+b_4\epsilon^\alpha{}_{\mu\nu\rho} S^\rho.
\eeq
This is the minimal field content to describe the desired geometrical setup. It is necessary to have at least two different vector fields with opposite transformation properties under parity in order to account for the axial part of the torsion. The remaining vector pieces, \ie the two nonmetricity traces and the torsion trace, have been identified (up to some proportionality constant) so that this sector is fully described by one single vector field. It would be interesting to study the geometries where the different vector pieces are not identified and if the presence of internal symmetries in that sector plays any role (see \cite{Iosifidis:2019fsh} related to this point). The present framework however allows to substantially simplify the analysis. Within the framework of curvature-based metric-affine  theories, the general quadratic action can be written as
 \begin{equation}\label{SquadraticVD}
 \begin{split}
\cS_{\textrm{VD}}  = M_{2} \int &\dd^D x \sqrt{-g}\Big[R^2 + R_{\alpha\beta\gamma\delta}\Big( d_1 R^{\alpha\beta\gamma\delta} + d_2 R^{\gamma\delta\alpha\beta} -  d_3 R^{\alpha\beta\delta\gamma}\Big) -  4\Big( c_1 R_{\mu\nu}R^{\mu\nu} +  c_2 R_{\mu\nu}R^{\nu\mu} \Big)\\
& -4  P_{\mu\nu}\left( c_3 P^{\mu\nu} + c_4 P^{\nu\mu} - c_5 R^{\mu\nu} - c_6 R^{\nu\mu}\right) -4  H_{\mu\nu}(c_7 H^{\mu\nu} + c_8 R^{\mu\nu}+ c_9 P^{\mu\nu})\Big)\,  
\Big]\,.
 \end{split}
\end{equation}
where $d_i$ and $b_i$ are some dimensionless constants and $M_2$ some scale and $P_{\mu\nu}$ and $H_{\mu\nu}$ are the co-Ricci and homothetic tensors (see \eqref{eq:RicciHomotheticCoRicci} for their definition). This action will generically lead to instabilities, once again along the lines of what one would expect as discussed in detail above. In order to guarantee a ghost-free pure graviton sector, it is convenient to impose that the theory reduces to a Gauss-Bonnet theory in the Riemannian limit, \ie when $A_\mu\rightarrow 0$ and $S^\rho\rightarrow 0$. 

A remarkable result is that, for this general class of theories, it is sufficient to restrict the geometrical framework rather than the parameters in the action in order to obtain a ghost-free vector-tensor theory \cite{Jimenez:2015fva,Jimenez:2016opp}. These ghost-free geometries are characterised by $2b_1-b_2-b_3=0$ and the resulting action reduces to
\begin{align}
\begin{split}
\cS_{\textrm{VD}}=\mu\int\dd^Dx\sqrt{-g}&\Big[(R^g){}^2-4R^g_{\mu\nu}(R^g){}^{\mu\nu}+R^g_{\mu\nu\rho\sigma}(R^g){}^{\mu\nu\rho\sigma}\\
&-\frac\alpha 4 F_{\mu\nu} F^{\mu\nu}+\xi A^2\na_\mu A^\mu -\lambda A^4-\beta \lr{(R^g){}^{\mu\nu}-\frac{1}{2}R^g g^{\mu\nu}} A_\mu A_\nu\Big]
\end{split}
\end{align}
where $\alpha$, $\xi$, $\lambda$ and $\beta$ are some constants that are given in terms of the parameters in \eqref{SquadraticVD} and  $F_{\mu\nu}=2\partial_{[\mu} A_{\nu]}$. The noteworthy property of this action is that the vector field features derivative nongauge invariant interactions and a nonminimal coupling to the curvature. However, this nonminimal coupling precisely belong to the class of ghost-free interactions \cite{Heisenberg:2018vsk}. 

Thus, the general result regarding the ghostly pathologies has been resolved in the vector distorted geometries by two conditions, namely: $i)$ imposing the recovery of the safe Gauss-Bonnet quadratic gravity in the absence of distortion and $ii)$ restricting the class of geometries. The singular property of the selected ghost-free geometries is that they generalise the Weyl connection by including a nontrivial trace of the torsion sector while maintaining the Weyl invariance of the metric (in)-compatibility condition. This can be easily understood by noticing that the nonmetricity for this restricted class of geometries is $Q_{\mu\alpha\beta}=(b_3-b_2) A_\mu g_{\alpha\beta}$ which is of the Weyl type. However the torsion is nonvanishing and given by $T^\alpha{}_{\mu\nu}=2b_3\delta^\alpha_{[\mu}A_{\nu]}$. We refer to \cite{Jimenez:2015fva,Jimenez:2016opp} for a detailed discussion on the interesting geometrical properties of these geometries.

%%%%%%%%%%%%%%%%%%%%%%%%%%%%%%%%%%%%%%%%%%%
%%%%%%%%%%%%%%%%%%%%%%%%%%%%%%%%%%%%%%%%%%

\subsubsection{Riemann-Cartan geometries: vanishing nonmetricity does not suffice}\label{sec:NMfree}
Let us now consider the extension of the Riemannian framework to the so-called Riemann-Cartan geometry, where the connection is allowed to have a torsion component while keeping a trivial nonmetricity. This can be achieved by introducing a suitable Lagrange multiplier in the general action for RBG theories without projective symmetry as
\beq\label{GeneralActionwithNMmult}
\cS[g_{\mu\nu},\Ga,\lambda]=\frac12\int \dd^Dx \sqrt{-g}\,\lrsq{F\big(g^{\mu\nu},R_{\mu\nu}\big)+\lambda^{\alpha}{}_{\mu\nu}\na_\al g^{\mu\nu}}+\cS_\textrm{m}[g_{\mu\nu},\Psi].
\eeq
It is not hard to see that, while the torsion-free constraint heals the instabilities of generalised RBGs, this is not the case for a constraint imposing the vanishing of the  nonmetricity tensor. Given that the full analysis is rather cumbersome in this case, we will simply highlight the main differences between the vanishing nonmetricity and vanishing torsion constraints, emphasising which are the conditions that improve the pathological behaviour of generalised RBGs in their torsion-free versions that do not occur when the nonmetricity free constraint is imposed. 

First of all notice that varying the above action with respect to $\lambda^\al{}_{\mu\nu}$ one gets the constraint $\na_\al g^{\mu\nu}=-Q_\al{}^{\mu\nu}=0$. Now an infinitesimal variation of the above action with respect to the connection yields
\beq
\begin{split}\label{VarActionNMfree}
\delta_\Ga\cS=&\frac12\int \dd^Dx \sqrt{-g}\frac{\partial F}{\partial R_{\mu\nu}}\delta_\Ga R_{\mu\nu}=-\frac12\int \dd^Dx \sqrt{-q}q^{\mu\nu}\lr{\na_\al\delta\Ga^\al{}_{\nu\mu}-\na_\nu\delta\Ga^\al{}_{\al\mu}-T^\lambda{}_{\nu\al}\delta\Ga^\al{}_{\la\mu}}
\end{split}
\eeq
where the conditions $Q^\al{}_{\mu\nu}=0$ and $\delta_\Ga Q_{\al}{}^{\mu\nu}=0\rightarrow \delta_\Ga L^{\al}{}_{\mu\nu}=0$ are imposed by the Lagrange multiplier field equation after integrating it out. The root of the difference between the two cases is the third term in the variation of the Ricci tensor \eqref{eq:TransfRicciCoRicciHomothetic}. In the above variation of the action, that term vanishes in the torsion-free case (after integrating out the vanishing torsion field), while this does not occur in the nonmetricity case. As a consequence, the connection field equations for the vanishing nonmetricity case are
\begin{eqnarray}
\na_\la \lrsq{\sqrt{-q}q^{\nu\mu}}-\del^\mu{}_\la\na_\rho\lrsq{ \sqrt{-q}q^{\nu\rho}}
&=&\sqrt{-q}\lrsq{T^\mu{}_{\la\al} q^{\nu\al}+T^\al{}_{\al\la} q^{\nu\mu}-\del^\mu{}_\la T^\al{}_{\al\be} q^{\nu\be}},
\end{eqnarray}
thus having the same tensorial structure as the ones in the general case\footnote{Notice that here we could drop the $\sqrt{-q}$ from the connection field equations by defining $q^{\mu\nu}\equiv\partial F/R_{\mu\nu}$. However since it does not introduce any advantage, we will not do it to facilitate the comparison with the torsion-free case.} \eqref{eq:GeneralRBGConnectuonEquations}, which does not happen in the torsion-free case \eqref{ConnectionFieldEqsTorsionFree}. This difference will have consequences in the number of degrees of freedom propagated in the different cases, as well as in their stability properties. To make this clearer, let us first decompose the nonmetricity free connection as $\Gamma^\al{}_{\mu\nu}=\Gamh^\al{}_{\mu\nu}+L^\al{}_{\mu\nu}+K^\al{}_{\mu\nu}$. Notice that although the covariant derivative $\na_\al g_{\mu\nu}$ vanishes, this is not true for $\na_\al h_{\mu\nu}$, and thus the distortion tensor corresponding to $h_{\mu\nu} $ in the connection decomposition is nonvanishing. We thus see that the nonmetricity free condition does not have as nice an implementation as the torsion-free condition does, and the structure of the equations is identical to the general case, having also the divergence-free constraint of the 2-form
\begin{equation}\label{ConstraintBNM}
\na^h_\la B^{\la\mu}=0.
\end{equation}
 In the torsion-free case, we found instead that the divergence of the 2-form was proportional to one of the traces of the distortion tensor $\bL_\mu$. Thus, in both the torsion-free and the nonmetricity-free cases the divergence of the 2-form can be eliminated from the field equations. Another important point is that the absence of $K^\al{}_{\mu\nu}$ in the torsion-free case and the index symmetries of $B^{\mu\nu}$ and $L^\al{}_{\mu\nu}$ yield the relations \eqref{L1} and \eqref{L2}. While \eqref{L1} is still occurring in this case, the analogue relation to \eqref{L2} is now
\begin{align}
\na_\lambda (\sqrt{-h}B^{\lambda\nu})=\sqrt{-h}\na^h_\lambda B^{\lambda\nu}+\sqrt{-h}\lr{t_\al B^{\al\nu}+\frac{1}{2}T^{\nu}{}_{\al\be}B^{\al\be}},\label{L2NM}
\end{align}
where $t_\al\equiv T^\be{}_{\be\al}$ and the first term on the right hand side vanishes due to \eqref{ConstraintBNM}. Thus, while in the torsion-free case these relations together with the divergence of the two-form \eqref{eq:constraint1} allow to write $\na_\al(\sqrt{-h}h^{\al\mu})$ and $\na_\al(\sqrt{-h}B^{\al\mu})$ in terms of the vector field $\bL_\mu$, this is not the case in the nonmetricity free scenario. 

The differences between the torsion-free and nonmetricity free cases that have been outlined rely only on the decomposition of the connection that we have been able to perform in each case. This in turn allows us to understand the differences in the tensorial structure of the connection field equations in both cases, which plays a crucial role in determining the constraints to the degrees of freedom of the full theory due to each of the geometrical setups. In turn, these constraints could be responsible (in the torsion-free case) of stabilising the theory, but it will not necessarily be the case in general. 

To proceed with the argument, let us split $q^{\mu\nu}$ into its symmetric and antisymmetric parts. On the one hand, due to the symmetrization of $\mu$ and $\nu$ in the connection field equations in the torsion-free case, only the contraction $\na_\al B^{\al\mu}$ enters the connection field equations. As explained above, this can be substituted by $\bL_\mu$ in the torsion-free case and, together with the relations \eqref{L1} and \eqref{L2}, it allows to find a relation between both traces of the distortion tensor. Then, since $\na_\al  B^{\mu\nu}$ does not appear in the equations, and $\na_\al  h^{\mu\nu}$ can be written only in terms of $L^\al{}_{\mu\nu}$, the connection equation \eqref{ConnectionFieldEqsTorsionFree} allows to find a solution for the full connection as the Levi-Civita conection of the auxiliary metric plus a distortion part characterised only by the vector field $\bL_\mu$. On the other hand, as well as in the general case, the symmetrization of $\mu$ and $\nu$ does not occur in the connection field equations in the vanishing nonmetricity case. Hence, not only its trace, but also the full covariant derivative of $B^{\mu\nu}$ enters the connection field equations. As a consequence, the constraint on the 2-form that rendered it as an auxiliary field in the torsion-free case no longer applies, and $B_{\mu\nu}$ still has dynamics, as in the general case. This difference makes it impossible to solve the connection only in terms of a new vector field related to the projective mode.

 Indeed, it can be seen that the torsion tensor in this case has the schematic form $\na B/(1+B)$ as happened to $\Upsilonh$ in section \ref{sec:GenRBG}, which will generally lead to the presence of the Ostrogradskian instabilities propagated by the 2-form. The Einstein frame version of this theory will be formally identical to the one of the general case, since the distortion of $h^{\mu\nu}$ is not vanishing here. Hence, the number of propagating degrees of freedom is the same as in the general case, and we are forced to conclude that the constraint of vanishing nonmetricity does not heal the instabilities of the full theory: the 5 new degrees of freedom corresponding to the projective mode and the 2-form will in general also propagate the instabilities found in the general case.

As a final remark, let us note that the Poincar\'e gauge theories of gravity \cite{Kibble:1961ba} are formulated in a Riemann-Cartan geometry. It is known that the general quadratic theories of this class present pathologies and only very specific choices of parameters give rise to healthy theories (see e.g. \cite{Hayashi:1979wj,Sezgin:1979zf,Yo:1999ex,Blagojevic:2018dpz,Vasilev:2017twr,Jimenez:2019tkx,Jimenez:2019qjc}). However, it is possible to have phenomenologically viable theories by interpreting them as effective field theories as done in \cite{Aoki:2019snr,Aoki:2020zqm}.

%%%%%%%%%%%%%%%%%%%%%%%%%%%%%%%%%%%%%%%%%%
%%%%%%%%%%%%%%%%%%%%%%%%%%%%%%%%%%%%%%%%%%

\subsection{Ghosts in Hybrid theories}\label{sec:Hybrid}
So far we have considered RBG in the pure metric-affine formalism, so that only the curvature of the full connection enters the action. As explained in chapter \ref{sec:DifferentialGeometry}, every spacetime endowed with a metric tensor has a canonical connection given by the Christoffel symbols of the metric. Thus, in any spacetime with a general affine connection, there is a coexistent affine structure provided by the Levi-Civita connection. The hybrid formalism \cite{Harko:2011nh, Capozziello:2015lza} steps outside the purely metric-affine framework and embraces these two coexisting affine structures so that the action contains the curvatures of the two connections. 

As we will see, rather than improving the situation of the pure metric-affine formalism, delving into the hybrid framework generically introduces even more pathologies. This may not be too surprising since the hybrid formalism is prone to the independent pathologies of the metric and metric-affine formalisms separately from the outset and hence it is natural to expect the same pathologies at the very least. The existence of pathologies in the hybrid formalism was analysed in \cite{Koivisto:2013kwa} by looking at the propagator on flat spacetime and identifying the presence of ghosts for a class of hybrid theories whose action is an arbitrary function of the two Ricci scalars $R^g$ and $R$ and the hybrid Ricci squared term $R^g_{\mu\nu}R^{\mu\nu}$.  We will generalise these results to a more general class of hybrid theories which is the natural hybrid extension of purely metric-affine RBG theories without projective symmetry and without assuming any background.

To that end, let us consider the following hybrid action
\beq
\cS_{\textrm{hybrid}}=\int\dd^Dx\sqrt{-g}f(g^{\mu\nu},R_{\mu\nu},R^g_{\mu\nu}).
\eeq
We will then proceed analogously to the pure metric-affine formalism to write the action as
\beq
\cS_{\textrm{hybrid}}=\int\dd^Dx\Big[\sqrt{-q} q^{\mu\nu}R_{\mu\nu}+\cU(R^g_{\mu\nu},q,g)\Big]
\label{Eq:hybrid2}
\eeq
where we have defined
\beq
\cU\equiv\sqrt{-g}\left[f-\frac{\partial f}{\partial\Sigma_{\mu\nu}} \Sigma_{\mu\nu}\right],\quad\text{and}\quad \sqrt{-q}q^{\mu\nu}\equiv \sqrt{-g}\frac{\partial f}{\partial\Sigma_{\mu\nu}},
\label{eq:defUq}
\eeq
and here $f$ is understood as a function of $g^{\mu\nu}$, $R^g_{\mu\nu}$ and the auxiliary field $\Sigma_{\mu\nu}$, which plays the same role as the one introduced in metric-affine RBG theories in section \ref{sec:RBGTheory}, \ie it is constrained to be $R_{\mu\nu}$ on-shell. 

The general hybrid action written in the form \eqref{Eq:hybrid2} is sufficient to understand the multiple sources of instabilities. Since we have linearised in the Ricci of the connection $R_{\mu\nu}$, that sector alone already reproduces the pathologies associated to the projective mode and the additional 2-form field that we have extensively discussed in precedent sections. Furthermore, even if we impose a projective symmetry in an attempt to avoid those pathologies, we can then straightforwardly integrate out the connection and obtain the equivalent action
\beq
\cS_{\textrm{hybrid}}=\int\dd^Dx\Big[\sqrt{-q} q^{\mu\nu}R^q_{(\mu\nu)}+\cU(R_{\mu\nu},q,g)\Big],
\label{Eq:hybrid3}
\eeq
so we have an Einstein-Hilbert term to describe the dynamics of the (now symmetric) field $q_{\mu\nu}$ as in the healthy RBG theories with projective symmetry. That pure metric-affine sector is then fine. However, the hybrid couplings introduce yet two additional sources of pathologies. 

On the one hand, if we have an arbitrary dependence on the metric Ricci tensor $R^g_{\mu\nu}$, the theory will be prone to the usual Ostrogradski instabilities in the metric sector. Furthermore, even if we avoid those problems by (for instance) utilising only the Ricci scalar of the metric, that is known to represent a safe higher-order curvature of the metric formalism, the potential $\cU$ will introduce arbitrary interactions between $q_{\mu\nu}$ and $g_{\mu\nu}$ so we will have an interacting bi-metric theory that will again introduce ghostly modes unless much care is taken in the construction of the interactions (see \eg \cite{Hassan:2011zd}). We can understand this a bit better by considering a simplified theory where the metric and metric-affine sectors are split as
\beq
\cS_{\textrm{hybrid}}=\int\dd^Dx\sqrt{-g}\left[\frac12 R^g+\mathcal{F}(g^{\mu\nu},R_{(\mu\nu)})\right]
\eeq
where we have separated the pure metric sector described by the Einstein-Hilbert action and the metric-affine sector on which we have imposed a projective symmetry. Each of these sectors by itself would seem perfectly fine. However, they can talk to each other through the $\sqrt{-g}$ factor in the volume element and this will be the source of the problems. In view of our results above and neglecting matter fields for simplicity, we can expect to have two Einstein-Hilbert terms once we integrate the connection out. This is in fact the case, but we also generate a potential so the action reads
\beq
\cS_{\textrm{hybrid}}=\int\dd^Dx\left[\frac{\sqrt{-q}}{2}q^{\mu\nu}R_{\mu\nu}+\frac{\sqrt{-g}}{2} R^g+\mathcal{U}(q,g)\right],
\label{Eq:hybridbi-metric3}
\eeq
where the dependence on the general potential term in \eqref{Eq:hybrid2} can be separated as the $R(g)$ term in the above action. The resulting action is then a bi-metric theory where the two metrics interact through the potential $\cU$ and it will suffer from a Boulware-Deser ghost \cite{Boulware:1973my}. Since this potential is determined by the function $f$, only functions that generate the known ghost-free potentials \cite{deRham:2010kj,Hassan:2011zd} have a chance to be stable.

 It is then clear that resorting to a hybrid action not only cannot cure the found instabilities in RBG theories, but makes things even worse by introducing yet new sources of ghosts. A way around this general no-go result for stable hybrid theories results in theories where the bi-metric construction fails. This happens for theories where only the Ricci scalars are allowed, \ie theories described by the action
\beq
\cS_{\textrm{hybrid}}=\int\dd^Dx\sqrt{-g}f(R,R^g).
\eeq
We can proceed analogously by performing the corresponding Legendre transformations to linearise in $R^g$ and $R$, but now we only need to introduce two auxiliary scalar fields instead of the tensor $\Sigma_{\mu\nu}$ so we can rewrite the action as
\beq
\cS_{\textrm{hybrid}}=\int\dd^Dx\Big[\sqrt{-g}\,\chi\, g^{\mu\nu}R_{\mu\nu}+\sqrt{-g}\varphi g^{\mu\nu} R^g_{\mu\nu}+\cU(\varphi,\chi)\Big].
\eeq
From this action we see that now the connection is nothing but the Levi-Civita connection of a metric that is conformally related to $g_{\mu\nu}$. In other words, the definition of $q^{\mu\nu}$ in \eqref{eq:defUq} yields $q_{\mu\nu}=\tilde{\chi} g_{\mu\nu}$, with $\tilde{\chi}=\chi^{\frac{2}{D-2}}$, so we only introduce an extra scalar instead of the full symmetric $q_{\mu\nu}$. The action then takes the form
\beq
\cS_{\textrm{hybrid}}=\int\dd^Dx\sqrt{-g}\Big[(\varphi+\chi)R^g+2(1-D)\chi\Big(\Box\log\tilde{\chi}+\frac{D^2-4D-4}{2}(\partial\log\tilde{\chi})^2\Big)+\cU(\varphi,\chi)\Big].
\eeq
It is then apparent that these theories propagate two additional scalars and avoid the Boulware-Deser ghosts of the general case. Nevertheless, it was found in \cite{Koivisto:2013kwa} that even these theories seem to present tachyonic or ghostly instabilities around a flat Minkowski background so that it is unavoidable to have some kind of instabilities.

%%%%%%%%%%%%%%%%%%%%%%%%%%%%%%%%%%%%%%%%%%%%%%%%%%%%%%%
%%%%%%%%%%%%%%%%%%%%%%%%%%%%%%%%%%%%%%%%%%%%%%%%%%%%%%%

\section{On ghostly instabilities in general metric-affine theories}\label{sec:OnMoreGeneral}

So far we have focused on theories constructed in terms of the Ricci tensor alone as a simplified proxy to prove the pathological character of general metric-affine theories described by higher-order curvature actions. Our results should suffice to clearly identify the origin for the potential pathologies in more general metric-affine theories where not only the Ricci tensor appears in the action, but also arbitrary nonlinear terms constructed with the Riemann curvature tensor.

 In general, if we have an action with an arbitrary dependence on the Riemann tensor formulated in a metric-affine geometry, we can always introduce the splitting of the connection into its Levi-Civita part, the torsion and the nonmetricity. That way, it is possible to re-formulate the theory in a purely Riemannian setup with additional nonminimally coupled matter fields.\footnote{We will expand on this argument in section \ref{sec:PertNM}.} These fields, \ie the torsion and the nonmetricity, can be decomposed into their irreducible representations under some appropriate group, GL(4,$\mathbb{R}$) or ISO(1,3) (the Poincar\'e group) being natural choices (see e.g. \cite{Hehl:1994ue}), and they will feature nonminimal couplings to the curvature and, quite generically, these will involve either derivatives of the fields or couplings of spin higher than zero. In both cases, as it is well-known, such interactions are prone to be pathological, as they typically excite Ostrogradskian instabilities. In the precedent sections we have explicitly shown how these expected pathologies come about for a particular class of metric-affine theories, but it is clear that the same problems will persist for more general actions. Particularly, only when the extra fields drop from the spectrum due to the imposition of projective symmetry could we have stable theories\footnote{Except in exceptional cases with constrained geometries, but these cases do not fit into the purely metric-affine framework, as they involve {\it a priori} assumptions on the affine structure of spacetime. Furthermore, the resulting theories always propagate a massless spin-2 field that can be identified as GR, and the extra fields, like the Proca field in the torsion-free case, can be regarded as matter feilds.} and, in that case, the gravitational sector simply reduces to GR.
 
It is important to emphasise that we are providing a general argument against some commonly quoted statements\footnote{From a field theoretic perspective it is evident that having second-order field equations is not a sufficient reason to guarantee the absence of Ostrogradski instabilities, a straightforward argument being that it is always possible to reduce the order of the equations by introducing auxiliary fields. However, in the community with a stronger geometrical approach to gravity this seems to be less clear.} that the metric affine theories avoid instabilities because the field equations remain of second-order. This does not mean however that {\it all} metric-affine theories with higher-order curvature terms featuring additional propagating degrees of freedom (other than the graviton) will be pathological, but rather that one should be careful on how these theories are constructed and not give for granted that the very fact of using a metric-affine formulation prevents the appearance of ghosts from operators involving arbitrary powers of the Riemann tensor. 

Of course, nonpathological theories exist and they can be constructed in a variety of manners (some of which we have discussed in section \ref{sec:GeoConst}, usually introducing additional symmetries, constraints or geometrical identities. However, it should be clear from our discussion that one should in general be careful when constructing theories in a metric-affine framework. Indeed, there are recent works that point in the same direction as our discussion. Particularly, the general problematic character of metric-affine gravity theories can be seen from the analysis of the perturbative degrees of freedom of the most general quadratic metric-affine theory around Minkowski performed in \cite{Percacci:2019hxn}. There, it was shown that already at that level, wise choices of parameters must be taken to avoid instabilities. Indeed, when all metric-affine covariant terms (not only curvature-based) are considered in the action, even imposing projective symmetry was not enough to generally stabilise the theory. 

It is important however to stress that our analysis above goes beyond the linear regime around Minkowski and, in fact, some of the diagnosed instabilities cannot be seen from such a perturbative analysis. Thus, though the perturbative analysis gives necessary conditions for stability, these are not sufficient to ensure the nonlinear stability of the theories. For an example of how a perturbative analysis may not be sufficient to ensure the stabiility of the full nonlinear theory see \eg \cite{Yo:1999ex,Jimenez:2019qjc,Jimenez:2019tkx} within the context of Poincar\'e gauge theories of gravity.

Let us finally briefly comment on how our results can be relevant from a purely effective field theorist approach to the metric-affine theories. This approach has been thoroughly pursued in \cite{Aoki:2019snr} within the class of Riemann-Cartan geometries including up to dimension 4 operators. We have seen that higher-order powers of the Riemann tensor generically introduce ghost-like instabilities in the metric-affine formalism very much like in the metric approach and essentially for the same reasons. It is possible however to adopt an EFT approach where these would just be irrelevant operators with perturbative effects below the cutoff of the theory. In this view, the ghosts are not really part of the perturbative spectrum of the theory because their masses lie beyond the domain of validity of the EFT so they are harmless. If the gravitational cutoff  is assumed to be the Planck mass and the Wilsonian coefficients are $\mathcal{O}(1)$ according to naturalness arguments, then the resulting EFT will be similar to the usual EFT approach to GR but containing additional modes that (at least form a field theoretic perspective) can be regarded as matter fields. On the other hand, if we assume that the Planck scale only represents the cutoff for the purely metric sector and the metric-affine sector comes in with another cutoff scale $\Lambda<{\mpl}$, then one would expect a breakdown of the effective theory at that scale. As commented also in section \ref{sec:MetricAffineEFT}, this implies that classical solutions where the curvature becomes larger than $\Lambda$ cannot be generically trusted in the sense that they are mathematical solutions to a set of field equations that make physical sense only when the energies (curvatures) are kept below $\Lambda$.

%%%%%%%%%%%%%%%%%%%%%%%%%%%%%%%%%%%%%%%%%%%%%%%%%%%%%%%
%%%%%%%%%%%%%%%%%%%%%%%%%%%%%%%%%%%%%%%%%%%%%%%%%%%%%%%
%%%%%%%%%%%%%%%%%%%%%%%%%%%%%%%%%%%%%%%%%%%%%%%%%%%%%%%
\chapter{Metric-affine gravity through the EFT lens}\label{sec:MetricAffineEFT}
%%%%%%%%%%%%%%%%%%%%%%%%%%%%%%%%%%%%%%%%%%%%%%%%%%%%%%%
%%%%%%%%%%%%%%%%%%%%%%%%%%%%%%%%%%%%%%%%%%%%%%%%%%%%%%%
\initial{W}e have discussed in the previous chapters several features that are particular to RBG theories, signalling in due case how these features generalise to more general metric-affine theories. In this chapter we will digress on how RBG theories can be understood as effective theories, even though they do not fit well into the principles of the Effective Field Theory (EFT) framework. We will also discuss some generic features expected on an EFT of a metric-affine gravitational sector and discuss some of the perturbative features of generic metric-affine theories. To begin with the discussion, let us review the crucial features of RBG theories that are relevant to it (see chapter \ref{sec:RBGTheory} for details). The action of RBG theories is of the form 
\beq\label{RBGaction}
\cS=\frac{1}{2}{\mq^{-4}}\int \text{d}^4x \sqrt{-g}\cl(g^{\mu\nu},\mg^{-2}R_{(\mu\nu)}) +\cS_\textrm{m}\ ,
\eeq
where $\cl$ is any analytic function of the metric and the symmetrized Ricci tensor, and $\cS_\textrm{m}$ represents the matter action where the matter fields $\Psi_m$ couple either algebraically to the connection or through the symmetrised Ricci tensor. The reason why only the symmetric part of the Ricci tensor is considered is because, as seen in chapter \ref{sec:UnstableDOF}, including the antisymmetric part unleashes ghostly degrees of freedom. This restriction can be enforced through a symmetry principle, namely by imposing symmetry under projective transformations
\beq
\Ga^\al{}_{\mu\nu}\mapsto \Ga^\al{}_{\mu\nu}+\xi_\mu\delta^\al{}_\nu.
\eeq
By computing the connection field equations it is possible to verify that the connection enters as an auxiliary field in all RBGs. This is even more apparent in the Einstein frame of the theory \eqref{eq:EHframe1}, where the gravitational sector is described by metric-affine GR for a metric $q^{\mu\nu}$ defined in terms of $\partial(\cl+\cl_\textrm{m})/\partial R_{(\mu\nu)}$ and on-shell related to the metric $g^{\mu\nu}$ and the matter fields. This on-shell relation is indeed what allows to build the Einstein frame by a field redefinition of the metric $g^{\mu\nu}$ in terms of $q^{\mu\nu}$ and the matter fields, which now yields a matter sector coupled to $q^{\mu\nu}$ with additional interactions\footnote{If there is any term that is nonlinear in the symmetrised Ricci tensor present in the gravitational sector of the action, these new interactions couple every one of the degrees of freedom of the matter sector. If the only symmetrised Ricci terms, besides from the Einstein-Hilbert term, appear in the matter action coupling nonminimally to some of the matter fields, the role of this terms is to couple these matter fields to all of the matter fields in the theory.} among the matter fields.  In the case that the matter fields couple to the connection only through the symmetrised Ricci tensor, the equations for the connection in this frame are algebraic, so that the connection is an auxiliary field that obeys a constraint equation. Its solution is uniquely given by the Levi-Civita connection \footnote{If there are matter fields that couple to the connection algebraically (\ie without derivatives), then these couplings will not change the auxiliary character of the connection, but will introduce modifications in the connection that make it depart from the Levi-Civita connection of $q^{\mu\nu}$, just as what happens in metric-affine GR when one considers Dirac fields, where the connection has a term quadratic in the fermions.} of $q^{\mu\nu}$ up to a projective mode, which is unphysical in the case with projective symmetry. Thus, we see that RBG theories coupled to matter are just constrained theories which, when (some of) the constraints are solved, turn out to be GR coupled to a nonlinearly modified matter sector with the same matter degrees of freedom. 

These nonlinear modifications of the matter sector can be encoded into the deformation matrix ${\Omega}^\mu{}_\nu$,  defined by \eqref{eq:DeformationMatrixDefinition}, and which relates the metrics $g^{\mu\nu}$ and $q^{\mu\nu}$ by \eqref{eq:DeformationMatrixRelation}, and on-shell for the metric field equation, can be written perturbatively in terms of the matter fields as
\beq\label{eq:ExpansionOmegaEFT}
\Omega^\mu{}_\nu=\sum_{n=0}^{3}C_n\lr{\frac{T}{{\mq}^{4}}}\frac{(T^\mu{}_\nu)^n}{{\mq}^{4n}}
\eeq
where the $C_n$ are model dependent scalar functions and $T$ is the trace of the stress-energy tensor $T^\mu{}_\nu$. Strictly speaking, the above on-shell relation \eqref{eq:ExpansionOmegaEFT} holds for the (likely more physical) branch of solutions that reduce to GR in the low energy limit. This can be important because the deformation matrix satisfies a nonlinear equation in terms of the stress-energy tensor and, besides the solution \eqref{eq:ExpansionOmegaEFT} that could be obtained by imposing Lorentz covariance in both frames, there could be other branches with spontaneous symmetry breaking (see chapter \ref{sec:SolutionsDeformation} for details). However, these nonstandard branches are seemingly pathological and, since we want to recover GR at low energies, \eqref{eq:ExpansionOmegaEFT} is the relevant series expansion for the solution.  As a remark, we note that one could be concerned with the fact that, in the non-perturbative regime of the theory, the field redefinition that leads to the Einstein frame might be singular. However, that does not necessarily mean that the solutions written in the Einstein frame variables do not have physical sense, but it can rather be interpreted as the RBG field variables not being physically sound field variables in those regimes of the theory. In any case, these concerns escape the realm of effective theories, which is where the discussion of this chapter takes place. Indeed, from the EFT perspective, the field redefinition is given by the above perturbative series \eqref{eq:ExpansionOmegaEFT}, which provides in fact the relevant regime. An important feature of the above expansion of the deformation matrix that will be relevant for the later discussion on the embedding on the EFT framework of these theories is that, given that the stress-energy tensor usually satisfies the symmetry of the matter action, then the symmetries of the original matter sector are generally preserved when going to the Einstein frame at least at the perturbative level, which is the relevant one within the EFT framework.

Of course, one can build more general metric-affine theories involving arbitrary functions of the Riemann, nonmetricity, and torsion tensors as well as their derivatives in the action. As outlined in chapter \ref{sec:ObservableTraces}, such theories generically propagate a massless spin-2 mode encoded in some metric $q^{\mu\nu}$ plus other degrees of freedom that are not present in GR. We first focus on pure RBG corrections to GR because they provide a particular kind of effects which will also be present in these more general theories, given that a piece of their action will be of the RBG type, and may be constrained well before those new degrees of freedom may become observable. Hence, we will postpone the discussion of the most general case
after discussing in detail on the RBG-type corrections, their interpretation as effective theories, and their possible embedding in the EFT framework.

\section{Ricci-Based corrections and EFTs}\label{sec:RBGandEFT}

When the matter sector is interpreted within the realm of the effective field theory  framework (EFT), the fact that RBG theories can be written as GR coupled to a nonlinearly modified matter sector leads to a curious consequence: there is no observational difference between an RBG and GR if both are coupled to a matter sector described within the EFT framework. In the EFT framework (see {\it e.g.} \cite{Pich:1998xt}) one intends to give a description of the phenomena occurring below the cutoff scale ${\textrm{M}}$ where UV physics is accounted for by higher dimensional operators suppressed by powers of this mass scale.\footnote{Here UV/IR (or high/low energy) refers to energies greater/lower than the EFT scale ${\textrm{M}}$. By construction, the EFT will only be physically meaningful at energy scales below ${\textrm{M}}$, as unitarity is typically broken above it.} The power of this framework stems from the fact that this can be done in great generality without knowing at all the details of the particular UV theory. This is done as follows: 1) identifying the set of degrees of freedom (or fields) that describe the spectrum of the theory in the IR regime. 2) Choose a preferred set of symmetries that is assumed to be satisfied by the UV theory, which will be referred to as the symmetries of the EFT. 3) Construct the most general Lagrangian that can be built with all the low energy fields that is consistent with the chosen symmetries. With this simple algorithm, one parametrises all the possible observable effects that can be seen by doing experiments involving only low energy degrees of freedom in the case that the UV theory satisfied those symmetries, no matter what its spectrum or dynamics would be.

Let us outline how one proceeds in order to build such a Lagrangian. Given a set of asymptotic states described by the fields $\Psi_i$, and a set of symmetries of the EFT, the corresponding EFT Lagrangian is defined by $\mathcal{L}_{\textrm{eff}}=\mathcal{L}_0+\mathcal{L}_{\textrm{eff}}^{\textrm{d>4}}$ where $\mathcal{L}_0$ contains all the relevant and marginal operators ($d\leq 4$) and $\mathcal{L}_{\textrm{eff}}^{\textrm{d>4}}$ contains all the possible irrelevant operators ($d>4$) that can be built with the fields $\Psi_i$ and their derivatives which respect the EFT symmetries. As is well known, the set of mass dimension-$d$ operators of a given quantum field theory forms a vector space $\mathcal{A}_d$. Furthermore, two operators are said to be equivalent, in the sense that they contribute equally to physical observables, if they differ (up to a total derivative) by an on-shell constraint\footnote{The field equations of the EFT can be calculated order by order.  To a given order, an on-shell constraint is an operator proportional to the lower order field equations, which by definition vanish on-shell (see {\it e.g.} \cite{Weinberg:1995mt,Weinberg:1996kr}).}. Hence, $\mathcal{A}_d$ can be split into the equivalence classes defined by this relation, and it suffices to consider one operator of each class to have a physically complete basis of $\mathcal{A}_d$ (see {\it e.g.} \cite{Einhorn:2013kja}). Thus, without loss of generality we can write	
\begin{equation}\label{EFTLag}
\mathcal{L}_{\textrm{eff}}^{\textrm{d>4}}=\sum_{n=5}^\infty\sum_{i_n}\frac{\alpha_{(n,i_n)}}{{\textrm{M}}^{n}}\mathcal{O}_{(n,i_n)},
\end{equation}
where the set $\{\mathcal{O}_{(n,i_n)}\}_{i_n}$ is a basis of $\mathcal{A}_n$ and $i_n$ runs from 1 to its dimension. The dimensionless constant $\alpha_{(p,q)}$ is called Wilson coefficient of the operator $\mathcal{O}_{(p,q)}$. Assuming naturalness, the Wilson coefficients will be of $\mathcal{O}(1)$ and the EFT defined by \eqref{EFTLag} will be generically valid at energies below ${\textrm{M}}$, as unitarity will typically be violated above this scale for natural Wilson coefficients. As a remark, it is interesting to notice that the violation of tree-level unitarity does not imply the necessity of including new physics at that precise scale as shown for instance with the self-healing mechanism discussed in \cite{Aydemir:2012nz}. 
 
Let us now couple a matter sector described by an  EFT with Lagrangian $\mathcal{L}_{\textrm{eff}}$ to a gravitational sector given by a particular RBG theory. This might appear as an odd construction because the EFT philosophy should also be employed in the construction of the gravity sector. However, there is nothing a priori inconsistent with considering an RBG gravity sector and our interest here is precisely to discuss how these gravity theories can fit into the EFT framework. Since a perturbative field redefinition allows to write the theory as GR coupled to a matter sector with the same degrees of freedom and symmetries as the original one below the scale ${\mq}$, the mapped EFT retains its structure, {\it i.e.}, the basis of operators of the original EFT is still a basis of the mapped EFT. In other words, the `new' operators that appear after the field redefinition that allows to go to the Einstein frame were already present in the original matter Lagrangian. Before proceeding further, it may be in order to digress a bit here again on the preservation of the symmetries when going to the Einstein frame. In the EFT matter sector there will be gauge symmetries that have to do with massless particles and are fundamental for the correct number of degrees of freedom. We do not expect the Einstein frame formulation of the theory to change this because the stress-energy tensor of the gauge fields will also be gauge invariant. Regarding global symmetries, these typically arise as accidental symmetries of the low energy theory but can be broken by higher dimension operators. Thus, when going to the Einstein frame in the RBGs no new operators will be generated.
  
In view of the above discussion, we note the following: if the original matter Lagrangian was already an EFT with cutoff scale ${\textrm{M}}$, so that its Lagrangian was a linear combination defined by the arbitrary Wilson coefficients $\alpha_{(n,i_n)}$ of the operators $\{\mathcal{O}_{(n,i_n)}\}_{i_n}$, which provide a physical basis of the space of operators of the theory, it should be clear that the effect of the nonlinearities introduced in the matter sector after the mapping can be reabsorbed into a redefinition of the Wilson coefficients\footnote{Note that by the structure of \eqref{eq:ExpansionOmegaEFT}, the new operators that will enter the Lagrangian after the mapping are of mass dimension $4n$, and therefore only the Wilson coefficients corresponding to $4n$-dimensional operators will be nontrivially changed.} $\alpha_{(n,i_n)}\mapsto\tilde{\alpha}_{(n,i_n)}$ which schematically looks like 
\begin{equation}\label{redefWilson}
\tilde{\alpha}_{(4n,i_n)}=\alpha_{(4n,i_n)}+\beta_{(n-4,i_n)}\lr{\frac{\textrm{M}}{\mq}}^{n-4}.
\end{equation}
 where the $\beta_{(n,i_n)}$ are related with the $C_n$ coefficients and the $(T^\mu{}_\nu)^n$ tensorial structures appearing in \eqref{eq:ExpansionOmegaEFT}. Since by construction the Wilson coefficients are arbitrary in an EFT\footnote{Of course, they can be constrained by experiments, thus ruling out regions of parameter space. But regarding the theoretical construction of the EFT, these are arbitrary coefficients}, this redefinition is not relevant, and the matter sector coupled to the Einstein frame metric $q^{\mu\nu}$ describes exactly the same EFT as the one coupled to the RBG frame metric $g^{\mu\nu}$. There is a subtlety that can arise here: If the scale ${\mq}$ is much below the matter EFT scale ${\textrm{M}}$, then the redefinitions in \eqref{redefWilson} spoil the naturalness of the Wilson coefficients, which would generally restrict the range of validity of the EFT to energies below ${\mq}$.
 
 As a consequence of the above discussions we arrive at the following conclusion: though the predictions of a given RBG coupled to a given matter sector will differ in general from those of GR coupled to the same matter sector, if the matter sector coupled to the RBG is built within the EFT framework, its predictions will be indistinguishable from those of GR coupled to the same EFT. This has been illustrated explicitly in section \ref{sec:Mapping} by the coupling of EiBI gravity to Maxwell, which is equivalent to the coupling of GR to Born-Infeld electromagnetism rather than of GR to Maxwell. However, if instead of Maxwell we hd considered an EFT for the electromagnetic field, then the matter sector would be the same in both frames. 
 
If the EFT approach is extended to the gravitational interactions, it is straightforward to see that the EFT for a general metric-affine sector cannot be reduced to an RBG by imposing any symmetries involving the metric and/or affine connection, even at the lowest order. To see this explicitly, let us review the symmetries satisfied by RBG theories and explore the consequences to their imposition to an EFT of a metric-affine sector at lowest order. Starting with symmetries related to the metric, RBG theories need not satisfy none appart from invariance under diffeomorphisms, hence these symmetries would not (in principle) be of any help in reducing the general metric-affine EFT to an RBG theory. Concerning the affine sector, there is only one known symmetry that symmetric RBG operators enjoy, namely a projective symmetry. The lowest order (quadratic) EFT Lagrangian for a general metric-affine sector with diffeomorphism symmetry is given by\footnote{These 12 operators form a basis (in the sense described above) of the dimension 2 diffeomorphism invariant operators that can be built with a metric and a connection. } \cite{Jimenez:2019ghw}
\begin{equation}\label{eq:QuadraticMAG}
\begin{split}
\cl^{(2)}_{MAG}=& R+ a_{1} T_{\alpha \mu \nu} T^{\alpha \mu \nu}+a_{2} T_{\alpha \mu \nu} T^{\nu \mu \alpha}+a_{3} T_{\mu} T^{\mu}+b_{1} Q_{\alpha \mu \nu} T^{\nu \alpha \mu}+b_{2} Q_{\mu} T^{\mu}+b_{3} \bar{Q}_{\mu} T^{\mu} \\
&+c_{1} Q_{\alpha \mu \nu} Q^{\mu \alpha \nu}+c_{2} Q_{\alpha \mu \nu} Q^{\mu \nu \alpha}+c_{3} Q_{\mu} Q^{\mu}+c_{4} \bar{Q}_{\mu} \bar{Q}^{\mu}+c_{5} Q_{\mu} \bar{Q}^{\mu}
\end{split}
\end{equation}
where $\tilde Q_\mu$, $Q_\mu$ and $T_\mu$ are suitale traces of the nonmetricity and torsion tensors. A starting point to see whether this can be reduced to its RBG piece (the Einstein-Hilbert action) by enforcing symmetries of the affine sector would be to try projective symmetry.  Invariance of the above action under an infinitesimal or finite projective transformation in an arbitrary spacetime dimension D leads to the following rank 3 system of linear equations for the 11 coefficients 
\begin{equation}
\begin{split}
&b_1+(D-1)b_2-4c_1-4D c_3-2c_5=0\\
&b1-(D-1)b_3+4 c_2+4c_4+2D c_5=0\\
&4a_1+2a_2+2(D-1)a_3-2 b_1 -2 D b_2 -2 b_3=0
\end{split}
\end{equation}
which clearly does not force the general theory to be of the RBG kind. Given that RBGs are not (in general) invariant under other transformations in the affine sector, it appears hopeless that symmetries can restrict a general metric-affine quadratic Lagrangian to be of the symmetric RBG type, even if one has to sacrifice some operators of the symmetric RBG class as well. We could go on by trying to make sense only of a subclass of the RBG theories by finding a larger set of symmetries which happen to constrain the general metric-affine quadratic action to lie within this subclass. For instance, we could consider the three kinds of vectorial transformations of the connection (see {\it e.g.}\cite{Jimenez:2015fva}), but this would only fix five of the coefficients, while actually forbidding any operator within the symmetric RBG class. Indeed, neither the Einstein-Hilbert Lagrangian nor the general teleparallel equivalent of GR are invariant under vectorial symmetries other than the projective one, which suggests that such a theory would not propagate a desired massless spin-2 mode at low energies even if constraints to the geometry are imposed. Given that symmetries are unable to make sense of RBG theories within the EFT framework, let us now elaborate on whether a restriction on the new degrees of freedom that the affine sector can propagate could select RBG theories among the class of general metric-affine theories. As we know, apart from the matter degrees of freedom, RBGs only propagate a massless spin-2 mode. Hence, in order to restrict a general metric-affine Lagrangian to be of the RBG type by restricting the allowed degrees of freedom, we would end up building the effective field theory of a massless spin-2 field, which is a purely metric theory that describes GR at low energies \cite{Donoghue:1994dn,Donoghue:1995cz}, and where the affine sector needs not play any role at all. Indeed, the symmetric RBG operators are irrelevant in vacuum and redundant if one considers an EFT of the matter sector as well. Therefore, it appears that RBGs cannot be consistently embedded into the EFT framework this way. Actually, this line of thinking can be reversed, leading to stronger consequences: if there existed a set of symmetries that would reduce a general metric-affine EFT to be of the RBG type, this would require that the resulting effective theory only propagates a massless spin-2 mode in vacuum, which in turn implies that such theory would be a particular case of an effective theory of a massless spin-2 field coupled to an effective matter sector, though it would not be the most general possible effective theory compatible with the symmetries, as required by the principles of the EFT framework. As a conclusion, we state that RBG theories cannot be properly embedded within the EFT framework of a metric-affine sector. Nonetheless, this does not preclude RBG theories from being regarded as physically sound effective theories for perturbative phenomena at energy scales $E<\mq$, but then its effects are identical to those of (some of) the irrelevant operators of the matter sector. These results suggest the possibility of finding other metric-affine operators which are redundant when a full metric-affine EFT is considered.

From a nonperturbative perspective, there are RBGs able to remove cosmological and black hole singularities
 at the classical level, restoring geodesic completeness via the emergence of wormholes or cosmological bounces. However, quantum corrections in the matter sector of those theories are likely to strongly backreact onto those backgrounds, potentially rendering them as unphysical \cite{BeltranJimenez:2017uwv}. In this regard, it is important to recall that there are effective theories with irrelevant operators that satisfy nonrenormalization theorems. This property is important at a phenomenological level because it allows to have nonperturbative classical effects from nonrenormalisable operators while maintaining the quantum corrections under control. The paradigmatic example is of course General Relativity where the Planck mass in front of the Einstein-Hilbert term does not get renormalised by graviton loops.\footnote{{There is a simple argument that explains why this is the case. If we consider the so-called $\Gamma\Gamma$ Einstein form for the GR action (i.e. the Einstein-Hilbert Lagrangian deprived of the total derivative term), then diffeomorphisms are only realised up to a boundary term. Since Feynman diagrams realise the symmetries in an exact form, all loops can only generate quantum corrections to the higher order terms.}} It receives quantum corrections from matter loops, but these are typically $\mathcal{O}(m^2/\mpl^2)\lesssim 10^{-30}$ for the standard model particles, so that GR is actually an excellent quantum EFT (see e.g. \cite{Donoghue:1995cz,Dobado:1997jx,Burgess:2003jk})\footnote{My colleague an mentor Jose Beltr\'an-Jim\'enez can hardly resist referring to Weinberg's words on the topic \cite{Weinberg:2009bg} whenever the topic is up to discusion.}. Of course, there is the problem of the cosmological constant, but this is a naturalness problem rather than a breakdown of the EFT. Similarly, theories like nonlinear electrodynamics, $K$-essence, or Galileon theories exhibit analogous properties for their quantum corrections. Thus, if we take RBG theories and the matter sector is composed by {\it e.g.} a massless scalar field solely, the Einstein frame version of the theory after integrating out the connection will give rise to a $K$-essence model with a Lagrangian of the form $\mathcal{L}=\Lambda^4_\phi K(\partial_\mu\phi\partial^\mu\phi/M^4)$, with $\Lambda_\phi$ and $M$ some mass scales. The nonperturbative regime of the original RBG theory can then be mapped into a phenomenological effect arising from nonrenormalisable operators in $K$ where $M$ corresponds to the scale of nonlinearities and $\Lambda$ is parameterically given by $\Lambda^2\sim\sqrt{\mpl M}$ (thus playing the role of $\mq$). The structure of these scalar theories, in particular their shift symmetry, guarantees that the quantum corrections will come in with derivatives of the Lagrangian $\sim \partial K$ so it is plausible to have a regime where $\partial_\mu\phi\partial^\mu\phi/M^4\gtrsim 1$ while quantum corrections are kept under control and the irrelevant operators in the classical action are technically natural (see e.g. \cite{deRham:2014wfa,Brax:2016jjt} for explicit derivations of these statements).Thus, the nonperturbative classical solutions have a chance of surviving and we can trust their predictions. Similarly, the quantum corrections in nonlinear electrodynamics can be kept small even in regions where the electromagnetic fields are above the scale of nonlinearities, namely $\vert F_{\mu\nu}\vert\gtrsim M^2$, being Born-Infeld electromagnetism a paradigmatic example.
 
 Regarding the posibility of the Einstein frame effective operators of the matter sector being nonrenormalisable operators, so that they can lead to observable nonperturbative effects, an important concern with RBG theories is that the irrelevant operators generated after integrating out the nondynamical connection are somewhat universal, in the sense that they will involve all fields in the matter sector. Therefore, when coupled to the Standard Model, this universal nature of the RBG theories will typically lead to an EFT in the Einstein frame without any underlying symmetry or structure guaranteeing any nonrenormalisation result or naturalness of the resulting interactions like in the case of pure K-essence or nonlinear electrodynamics. Quite the opposite, the very presence of the Higgs field and its potential already points towards the impossibility of having nonperturbative classical solutions based on irrelevant operators without going beyond the regime of validity of the would-be EFT. Leaving aside these issues with quantum corrections, one should not forget that part of the interest on these theories stems from their nonperturbative properties as classical field theories, as they accommodate a plethora of exact solutions with interesting features which can expand our dictionary of viable spacetimes, such as singularity free cosmological and spherically symmetric backgrounds \cite{Barragan:2009sq,Banados:2010ix,Olmo:2013gqa,Maso-Ferrando:2021ngp,Benisty:2021laq,Guerrero:2020uhn}, as well as wormholes and other compact objects which behave in interesting and unexpected ways \cite{Lobo:2013adx,Lobo:2013vga,Lobo:2014fma,Lobo:2014zla,Lobo:2020vqh,Afonso:2019fzv,Afonso:2020giy,Orazi:2020mhb}.

\section{On a general metric-affine sector}\label{sec:PertNM}
 
We have just argued why RBG theories do not fit the EFT framework because one cannot reduce a general metric-affine Lagrangian to an RBG one by means of enforcing symmetries. This suggests to take a closer look at the EFT of a general metric-affine sector to understand what properties would it have in the perturbative regime, and whether there is any subclass of theories that do fit well within the EFT framework. To that end, it would be helpful to understand the general dynamics of these theories, or at least relate some of its aspects to those occurring in particular subsets among the general metric-affine class that we understand better such as RBG theories.

\subsection{Aspects of general metric-affine field equations}
Let us then consider a general diffeomorphism invariant metric-affine, theory where the fundamental fields are a metric $g^{\mu\nu}$ and an affine connection $\Gamma^\alpha{}_{\mu\nu}$. Diffeomorphism invariance forces the action to depend on scalars built with the metric, the Riemann and torsion tensors of the connection, and covariant derivatives of these objects.\footnote{Note that the nonmetricity appears as covariant derivatives of the metric} Let us split the dependence on the connection in terms of two variables, namely the symmetric part of the Ricci tensor of the connection $R_{(\mu\nu)}$, and the rest, which we will denote by the dimensionless quantity\footnote{Each of the terms inside $\tgar$ will be suppressed by powers of a heavy mass scale $\mg$ to render it dimensionless.} $\tilde\Gamma_{\textrm{Ric}}$ and can depend on the metric as well. Though this splitting, strongly inspired in RBG theories, might seem quite arbitrary and meaningless, as we will see, the $R_{(\mu\nu)}$ terms contribute to the connection equation in a way that can be absorbed into a new metric $q^{\mu\nu}$ related to the metric $g^{\mu\nu}$ by a field redefinition. Due to this fact, some properties of the theory become more apparent when described with this choice of field variables. In this regard, we will also see that this redefinition allows to write a piece of the nonmetricity tensor $\nabla g$ in a particular form that is related to a generic effect of the symmetric RBG corrections. We will thus consider an action of the form
\beq\label{eq:GeneralmetricaffineAction}
\cS=\frac{\mq^4}{2}\int d^D x\sqrt{-g} \cl \left[ g^{\mu\nu},{\mg^{-2}}R_{(\mu\nu)}, \tgar, \Psi_i \right],
\end{equation}
Where $\Psi_i$ denotes an arbitrary collection of matter fields, usually separated into a matter sector described by $\cl_\textrm{m}$, which couple to the metric at least through their kinetic terms, and in an arbitrary way to the connection. ${\mg}$ is a heavy mass scale that controls the deviations from the metric-affine Einstein-Hilbert (EH) term, and $\mq$ is the geometrical mean of this scale and the Planck scale, namely ${\mq}=\sqrt{{\mpl\mg}}$. There are several ways to ensure that we recover GR at low energies for the metric $g^{\mu\nu}$. For instance, one may impose geometric restrictions so that the curvature vanishes and the Lagrangian reduces to the general teleparallel equivalent to GR or associated \cite{Aldrovandi:2013wha,Nester:1998mp,BeltranJimenez:2017tkd,Jimenez:2019ghw}. However, the idea of the metric-affine formalism is to avoid any {\it a priori} assumption on the affine structure such as vanishing curvature, torsion, nonmetricity or combinations. Hence, the only way to ensure that GR is recovered at low energies for the metric $g^{\mu\nu}$ is to impose that the lowest order term in the Lagrangian be the EH term, thus having
\beq\label{eq:GravLagDeviations}
\cl=\frac{1}{{\mg^2}}g^{\mu\nu} R_{(\mu\nu)}+F\left[ g^{\mu\nu},\cR_{(\mu\nu)}, \tgar, \Psi_i \right],
\eeq
where $\cR_{(\mu\nu)}={\mg^{-2}}R_{(\mu\nu)}$ is a normalised dimensionless symetrised Ricci tensor and $F$ is a dimensionless scalar function that encodes higher-order corrections to the Einstein-Hilbert (EH) term. The vanishing of this action for arbitrary infinitesimal variations of the connection gives the field equations for the connection, which can be written as
\beq\label{eq:GeneralConnectionEq}
 \resizebox{.9\hsize}{!}{$\nabla_\lambda \lrsq{\sqrt{-g}\lr{g^{\mu\nu}+\frac{\partial F}{\partial \cR_{(\mu\nu)}}}}-\del^\mu{}_\la \nabla_{\rho}\left[\sqrt{-g} \lr{g^{\nu\rho}+\frac{\partial F}{\partial \cR_{(\nu\rho)}}}\right]={\mg^{2}}\tilde F_{\lambda}{}^{\mu\nu}(g^{\mu\nu},\cR_{(\mu\nu)},\tgar,\Psi_i)$}
\eeq
where $\tilde F_{\lambda}{}^{\mu\nu}$ (2,1)-tensor with mass dimension $-1$, so that the leading order terms are of order ${\mg^{-2}}$ and correspond to the linear torsion terms of \eqref{eq:GeneralRBGConnectuonEquations}. In general $\tilde F_{\lambda}{}^{\mu\nu}$ will be a function of the metric, $\tgar$, the collection of matter fields, though the functional dependence of this (2,1)-tensor on the matter fields does not arise in the case that they couple to the connection only through $R_{(\mu\nu)}$ or do not couple at all. Note that the hypermomentum term appearing in \eqref{eq:GeneralRBGConnectuonEquations} is included in this tensor, but it is negligible in a perturbative expansion with respect to the other corrections, as it is of order ${\mpl^{-2}}$. Defining now 
\beq\label{eq:DefGeneralq}
\sqrt{-g}\lr{g^{\mu\nu}+\frac{\partial F}{\partial \cR_{(\mu\nu)}}}=\sqrt{-q}q^{\mu\nu}
\eeq
we have that $q^{\mu\nu}(g^{\alpha\beta},\cR_{(\alpha\beta)},\tgar,\Psi_i)$. Assuming invertibility of this relation, and inverting it with respect to the metric, allows to write $g^{\alpha\beta}(q^{\mu\nu},\cR_{(\mu\nu)},\bgar,\Psi_i)$, where a bar over a previously tilded quantity indicates that its dependence on $g^{\mu\nu}$ has been removed in favour of $q^{\mu\nu}$ through the above algebraic relation. We can then write the connection field equations \eqref{eq:GeneralConnectionEq} as
\beq\label{eq:GeneralConnEqDiffInv}
\nabla_\lambda \lrsq{\sqrt{-q}q^{\mu\nu}}-\del^\mu{}_\la \nabla_{\rho}\left[\sqrt{-q} q ^{ \nu\rho}\right]={\mg^{2}}\bar F_{\lambda}{}^{\mu\nu}(q^{\rho\sigma},\cR_{(\rho\sigma)},\bgar,\Psi_i).
\eeq
On the other hand, the metric field equations read\footnote{Note that here the dependence of the matter fields is inside $\cl$ so that if decomposed onto a gravitational plus a matter sector $\cl_\textrm{m}$, the first term of \eqref{eq:GeneralMetricEqs} contains the usual stress-energy tensor of the matter sector that appears usually in the right hand side of the metric field equations.}
\beq\label{eq:GeneralMetricEqs}
\frac{\partial \cl}{\partial g^{\mu\nu}}-\frac{1}{2}\cl \,g_{\mu\nu}=0,
\eeq
which can schematically be written as the vanishing of some function $\tilde f_{\mu\nu}(g^{\rho\sigma},R_{(\rho\sigma)}, \tgar, \Psi_i )$ which, using the algebraic relation between $g^{\mu\nu}$ and $q^{\mu\nu}$ provided by \eqref{eq:DefGeneralq} leads to writing the metric field equations as 
\beq\label{eq:AlgebraicRic}
\bar f_{\mu\nu}(q^{\rho\sigma},\cR_{(\rho\sigma)}, \bgar, \Psi_i )=0.
\eeq
Again, these equations in turn provide an algebraic relation $\cR_{(\rho\sigma)}(q^{\alpha\beta},\bgar,\Psi_i)$, which allows to write the connection field equations \eqref{eq:GeneralConnEqDiffInv} as
\beq\label{eq:GeneralConnEq2}
\nabla_\lambda \lrsq{\sqrt{-q}q^{\mu\nu}}-\del^\mu{}_\la \nabla_{\rho}\left[\sqrt{-q} q^{\nu\rho}\right]={\mg^2}\hat F_{\lambda}{}^{\mu\nu}\lr{q^{\rho\sigma},\bgar,\Psi_i},
\eeq
where a hat over a previously tilded quantity indicates that its dependence on $\cR_{(\mu\nu)}$ has been removed in favour of $q^{\alpha\beta}$, $\bgar$ and the matter fields through the algebraic relation \eqref{eq:GeneralMetricEqs}. Now, we know that the homogeneous version of $\eqref{eq:GeneralConnEq2}$ is exactly the connection field equation for metric-affine GR, which is an algebraic equation for the connection. Hence, for the homogeneous case, the connection is an auxiliary field that can be written in terms of $q^{\mu\nu}$ and its first derivatives as the Levi-Civita connection of $q^{\mu\nu}$ as a solution up to a projective mode. However, the inhomogeneous part, due to the dependence in $\bgar$, will generally depend on derivatives of the connection, turning it into a dynamical field and unleashing potentially dangerous extra degrees of freedom to the theory, as shown in chapter \ref{sec:UnstableDOF}. In any case, we know that in the low energy/curvature limit of the theory, the function $F$ controlling deviations from the Einstein-Hilbert Lagrangian in \eqref{eq:GravLagDeviations} must vanish (or contribute at most with a cosmological constant), so that in that limit, $\hat F_\mu{}^{\alpha\beta}$ is perturbatively close to a projective mode that in such case is spurious due to the projective symmetry of the EH term. Thus, in the low energy limit, we will have that the connection is given by the Levi-Civita connection of the metric $q^{\mu\nu}$. In virtue of the decomposition of a general connection \eqref{eq:ConnectionDecomposition}, this suggests to write the full solution of the connection equation \eqref{eq:GeneralConnEq2} as
\beq\label{eq:GeneralSolConn}
\Gamma^\alpha{}_{\mu\nu}={}^q\Gamma^\alpha{}_{\mu\nu}+\Upsilon^\alpha{}_{\mu\nu}\lr{q^{\rho\sigma},\bgar,\Psi_i},
\eeq
where $\Upsilon$ must be a solution to the inhomogeneous piece of the connection equation, namely
\beq\label{eq:NonHomogeneousConnEq}
\Upsilon^\alpha{}_{\mu\sigma}q^{\sigma\beta}+\Upsilon^\beta{}_{\mu\sigma}q^{\alpha\sigma}-\Upsilon^\sigma{}_{\mu\sigma}q^{\alpha\beta}=\frac{{\mg^2}}{\sqrt{-q}}\hat F_{\mu}{}^{\alpha\beta}\lr{q^{\rho\sigma},\bgar,\Psi_i},
\eeq
and where $\bgar$ will generally contain up to $n$-th order derivatives of $q^{\mu\nu}$ and up to $(n-1)$-th order derivatives of $\Upsilon^\alpha{}_{\mu\nu}$.

\subsection{Perturbative solutions and a general metric-affine EFT}

From the structure of general metric-affine theories, we see that if there are terms in the action that depend of the symmetrised Ricci besides the EH term, then there is an effective metric which solves the homogeneous equation for the connection as in RBG theories. Though the connection will exhibit nontrivial dynamics in the general case as opposed to being an auxiliary field, we know that at the perturbative level it will be the Levi-Civita connection of $q^{\mu\nu}$ up to corrections suppressed in inverse powers of $\mg$. As well, $q^{\mu\nu}$ and $g^{\mu\nu}$ will also differ by perturbative corrections. To see this, note that the leading order term in $F$ that depends on the symmetrised Ricci tensor will be at least of order ${\mg^{-4}}$, as it is associated to the quadratic Ricci term in F. Therefore  $\partial F/\partial\cR$ is at most of order ${\mg^{-2}}$. We can use this to invert the definition of $q^{\mu\nu}$ given in \eqref{eq:DefGeneralq} perturbatively in inverse powers of $\mg$ to obtain
\beq\label{eq:PertrbativeMetric}
g^{\alpha\beta}=q^{\alpha\beta}-\frac{\partial F}{\partial \cR_{(\alpha\beta)}}+\mathcal{O}\lr{{\mg}^{-3}}.
\eeq
We can now write explicitly the ${\mg^{-2}}$ dependence of the correction by defining $\Theta^{\mu\nu}$ as the leading order term of the expansion of $\partial F/\partial R_{(\mu\nu)}$ in inverse powers of ${\mg}$ when this derivative is written in terms of $q^{\mu\nu}$, $\partial q^{\mu\nu}$, $\Upsilon^\alpha{}_{\mu\nu}$ and the matter fields $\Psi_i$ through the algebraic relations \eqref{eq:DefGeneralq} and \eqref{eq:AlgebraicRic} and after substituting the solution for the connection \eqref{eq:GeneralSolConn} taking into account only the relevant terms up to order ${\mg^{-2}}$. By doing so, we find the perturbative on-shell relation
\beq\label{eq:RelationMetricPer}
g^{\alpha\beta}=q^{\alpha\beta}-{\mg^{-2}}\Theta^{\mu\nu}(q^{\mu\nu},\partial q^{\mu\nu},\Upsilon^\alpha{}_{\mu\nu},\Psi_i)+\mathcal{O}\lr{{\mq}^{-3}},
\eeq
where $\Theta$ is of 0-th order in ${\mg}$ and of mass dimension 2. In virtue of \eqref{eq:GeneralConnEq2}, \eqref{eq:GeneralSolConn}, the perturbative expression that we obtain for the nonmetricity tensor $Q_\alpha{}^{\mu\nu}=-\nabla_\alpha g^{\mu\nu}$ from the above relation is
\beq\label{eq:PertNMGeneralTheories}
Q_{\alpha}{}^{\mu\nu}={\mg^{-2}}\nabla{}_\alpha\Theta^{\mu\nu}-\Upsilon^{\sigma}{}_{\alpha\beta}q^{\beta\nu}-\Upsilon^{\sigma}{}_{\alpha\beta}q^{\beta\nu}+\mathcal{O}({\mg}^{-3}).
\eeq

Note that in the limit ${\mg}\to\infty$, the above theory becomes GR, $g^{\mu\nu}=q^{\mu\nu}$, and  $\Upsilon^\alpha{}_{\mu\nu}$ becomes an unphysical projective mode, so that the residual nonmetricity can be gauged away with an appropriate choice of projective gauge. Thus, we see that the scale $\mg$ controls both the piece of the nonmetricity tensor with physical relevance and the difference between the metrics $g^{\mu\nu}$ and $q^{\mu\nu}$. Hence, in any metric-affine theory yielding the EH Lagrangian at low energies and where the symmetrised Ricci tensor appears in the action in higher order corrections, will have a connection which, perturbatively, can be written as the Levi-Civita connection of a metric $q^{\mu\nu}$ plus corrections suppressed by a high-energy scale $\mg$ that controls the deviations from the EH Lagrangian. Perturbatively, this metric $q^{\mu\nu}$ will differ from $g^{\mu\nu}$ due to the $\mg$ suppressed corrections that contain explicitly the symmetrised part of the Ricci tensor through the $\Theta^{\mu\nu}$. This goes hand in hand with a nonmetricity tensor that is also controlled by $\mq$ and proportional to the covariant derivative of this $\Theta^{\mu\nu}$ tensor.  Indeed, this nonmetricity tensor necessarily features two pieces: 1) a piece containing derivatives of the connection that will generally excite new degrees of freedom associated to it, and vanishes unless the torsion tensor, the nonmetricity tensor, other irreducible pieces of the Riemann tensor apart from $R_{(\mu\nu)}$,  appear explicitly in the Lagrangian \eqref{eq:GravLagDeviations}. 2) A piece that is perturbatively related to the $\Theta^{\mu\nu}$ tensor, which is directly related to the $\mq$ suppressed corrections that contain explicitly the symmetrised part of the Ricci tensor, perturbatively relates $q^{\mu\nu}$ and $g^{\mu\nu}$, and vanishes in the absence of these corrections. 

Note that the matter fields in \eqref{eq:GravLagDeviations} couple to the metric $g^{\mu\nu}$ at least through their kinetic terms and the volume element. Thus, by means of \eqref{eq:RelationMetricPer}, we can perform a perturbative field redefinition of the metric so that the matter fields couple to $q^{\mu\nu}$ and the action \eqref{eq:GravLagDeviations} will feature new interactions among the matter fields that were not present in the original matter Lagrangian. Moreover, these interactions are intimately related to the higher-order symmetrised Ricci corrections to the EH piece of the action, and thus to the presence of a nonmetricity tensor of the form \eqref{eq:PertNMGeneralTheories}. This redefinition may seem arbitrary, but it is motivated by what happens in RBG theories. There, $\bgar$ is not present in the action, and the result is that the theory described in terms of the field variables $q^{\mu\nu}$ is GR coupled to a modified matter sector, which is modified precisely due to the $\Theta$ tensor which, in that case, depends only on the metric $q^{\mu\nu}$ and the matter stress-energy tensor multiplied by the inverse squared of the Planck mass, and introduces new effective interactions through the stress energy tensor suppressed by powers of ${\mq}$ as we will see below more explicitly. In the RBG frame of the theory, the nonmetricity tensor is also given by the covariant derivative of this $\Theta$ tensor and is as well controlled by the heavy mass scale $\mq$, having as well a close relation to the appearance of effective interactions in the Einstein frame of the theory.  

\subsection{Einstein-like frame and ghosts in general metric-affine theories}

The above discussion can also be done in a cleaner and more systematic way by building a generalised Einstein-like frame for generic metric-affine theories, which will also clarify why the apparently arbitrary field redefinition of the metric $g^{\mu\nu}$ introduced in \eqref{eq:DefGeneralq} is useful. In an analog manner as for RBG theories, we proceed by linearising the above action \eqref{eq:GeneralmetricaffineAction} with respect to the symmetrised Ricci tensor. Thus, mimicking the procedure, let us linearise the general Lagrangian for a metric-affine theory \eqref{eq:GeneralmetricaffineAction} with respect to the symmetrised Ricci tensor leads to
\beq\label{eq:ModifiedActiongeneral}
\cS=\frac{{\mq}^{4}}{2}\int d^D x \sqrt{-g}\left[\cL\left(g^{\mu\nu},\Sigma_{\mu\nu},\tgar, \Psi_i\right)+\frac{\partial \cL}{\partial \Sigma_{\mu\nu}}\Big({\mg^{-2}}R_{(\mu\nu)}-\Sigma_{\mu\nu}\Big)\right]
\eeq
where $\Sigma_{\mu\nu}$ is an auxiliary field\footnote{We will not write explicitly the dependence of $\cL$ but in this section it should be assumed that $\cL$ means $\cL(g^{\mu\nu},\Sigma_{\mu\nu})$} whose field equation is the constraint ${\mg}^{-2}R_{(\mu\nu)}$ provided that the hessian of the Lagrangian with respect to $\Sigma_{\mu\nu}$ does not vanish. When the constraint for the auxiliary field is implemented, it can be integrated out and the resulting action is exactly \eqref{eq:GeneralmetricaffineAction}, so that the theories are equivalent. We can now introduce a new field variable $q^{\mu\nu}$ through the following field redefinition
\beq\label{eq:DefMetricqSigma}
\sqrt{-q}q^{\mu\nu}=\sqrt{-g}\frac{\partial \cL}{\partial \Sigma_{\mu\nu}},
\eeq
which, by solving algebraically with respect to $\Sigma$ allows to express the auxiliary field as a function of the   $q^{\mu\nu}$, $g^{\mu\nu}$, $\tgar$ and the matter fields $\Psi_i$ through the solutions $\Sigma(q,g,\tgar,\Psi_i)$. Note that once the constraint stemming from the field equations of the auxiliary field is implemented, the above field redefinition looks exactly like the definition for $q^{\mu\nu}$ given in \eqref{eq:DefGeneralq}. After this field redefinition, we can then express a general metric-affine action as
\beq\label{eq:NonSymmetricActionTwoMetrics}
\cS=\frac{\mpl^2}{2}\int d^D x\lrsq{\sqrt{-q}q^{\mu\nu}R_{\mu\nu}+{\mg^{2}}\,\cU\lr{q,g,\tgar,\Psi_i}},
\eeq
where recall that ${\mq}=\mpl{\mg}$, and we have introduced the dimensionless generalised potential term
\beq
\cU\lr{q,g,\tgar,\Psi_i}=\sqrt{-g}\lrsq{\cL-\frac{\partial \cL}{\partial\Sigma_{\mu\nu}}\Sigma_{\mu\nu}}_{\Sigma=\Sigma(q,g,\tgar,\Psi_i)}.
\eeq
The action \eqref{eq:NonSymmetricActionTwoMetrics} already features the standard Einstein-Hilbert term in the first order formalism, but for the object $q^{\mu\nu}$ instead of the metric $g^{\mu\nu}$. Indeed, note that $g^{\mu\nu}$ appears algebraically in the potential $\cU$ and, therefore, it is an auxiliary field with this choice of field variables whose field equations are an algebraic constraint 
\beq
\frac{\partial \cU}{\partial g^{\mu\nu}}=0
\eeq
that can be solved for $g^{\mu\nu}$ to obtain it in terms of $q^{\mu\nu}$, $\bgar$ and the matter fields $\Psi_i$; where the bar over $\bgar$ replacing the tilde means that the possible dependence on the metric in $\tgar$ has been replaced by the solution. Note that, usually, the matter fields appear in  the matter Lagrangian so that we could split $\partial \cU/\partial g$ into a piece containing only $g$, $q$, and $\tgar$; and another one containing the matter fields and $g$ which would be the corresponding stress-energy tensor \eqref{eq:StressEnergyTensor} multiplied by $\mpl^{-2}$. Hence, implementing the constraint equation for $g^{\mu\nu}$ back into the action we end up with
\beq
\cS=\frac{\mpl^2}{2}\int d^D x\lrsq{\sqrt{-q}q^{\mu\nu}R_{\mu\nu}+{\mg^2}\,\hat\cU\lr{q,\bgar,\Psi_i}},
\eeq
where the hat over $\cU$ accounts for the substitution of $g$ via the constraint. We see that the lowest order term is just the EH action for $q^{\mu\nu}$ and $\hat \cU$ contains the deviations to GR and the matter sector. The metric field equations are now
\beq\label{eq:MetricFieldEqEFgen}
\cG_{\mu\nu}(q,\Gamma)=-{\mg^2}\,\frac{\delta\hat\cU}{\delta q^{\mu\nu}}
\eeq
where $\cG^\mu{}_\nu(q,\Gamma)$ is the object on the left hand side of \eqref{eq:GeneralRBGMetricFieldEquationsEinsteinTensor} with an index lowered with the inverse of $q^{\mu\nu}$; and the term on the right hand side depends on $q^{\mu\nu}$, $\bgar$, and contains the stress-energy tensor associated to the matter sector coupled to the metric $q^{\mu\nu}$. On the other hand, the connection field equations will now look
\beq\label{eq:ConnEqGenMetricAffine}
 \resizebox{.9\hsize}{!}{${\nabla_{\lambda}\left[\sqrt{-q} q ^{\mu\nu}\right]-\del^\mu{}_\la \nabla_{\rho}\left[\sqrt{-q} q ^{ \nu\rho}\right]}=\sqrt{-q}\left[\mathcal{T}^{\mu}{}_{\lambda \alpha} q ^{\nu\al}+\mathcal{T}^{\alpha}{}_{\alpha \lambda} q ^{\nu\mu}-\delta_{\lambda}^{\mu} \mathcal{T}^{\alpha}{}_{\alpha \beta} q ^{\nu\be}\right]+\frac{\delta\hat\cU}{\delta\Gamma^\lambda{}_{\mu\nu}}.$}
\eeq
The derivative of $\hat\cU$ with respect to the connection corresponds to a piece of tensor $\hat F_{\lambda}{}^{\mu\nu}$ on the right hand side of \eqref{eq:NonHomogeneousConnEq} and, when the matter fields appear in a the matter Lagrangian so that we could split the $\delta\hat\cU/\delta\Gamma$ term into a piece containing only $q$ and $\tgar$; and another one consisting on the hypermomentum, defined by \eqref{eq:DefHypermomentum}. To keep with the analogy, following \eqref{Eq:Gammaheq}, we can define a connection $\check\Gamma$ shifted by a projective mode as
\beq
\check{\Gamma}^\alpha{}_{\mu\nu}=\Gamma^\alpha{}_{\mu\nu}+\frac{2}{D-1}\check\Gamma^\lambda{}_{[\lambda\mu]}\delta^\alpha{}_\nu.
\eeq
Using the trace of \eqref{eq:ConnEqGenMetricAffine} and, in terms of the redefined connection, which allows to remove the torsion terms, we can write the connection equation in a more compact way, namely
\beq
\check\nabla_{\lambda}\left[\sqrt{-q} q ^{\mu\nu}\right]=\frac{\delta\check\cU}{\delta\check\Gamma^\lambda{}_{\mu\nu}}+\frac{1}{D-1}\lr{\delta_\al{}^{\mu}\frac{\delta\check\cU}{\delta\check\Gamma^\be{}_{\nu\be}}-\delta_\al{}^{\nu}\frac{\delta\check\cU}{\delta\check\Gamma^\be{}_{\mu\be}}},
\eeq
where $\check\cU$ is $\hat\cU$ with the dependence on the connection rewritten in terms of the redefined connection $\check\Gamma$. The above equation is the analogous to \eqref{eq:GeneralRBGConnectuonEquationsSplitted} with a symmetric $q^{\mu\nu}$, and is exactly equal in the limit where the gravitational Lagrangian contains only $R_{\mu\nu}$ as, in that case, the derivatives of $\check\cU$ are just the hypermomentum. Hence, we see that in a general metric-affine theory reducing to the EH action at low energies, due to the structure of the connection field equations, it is this metric $q^{\mu\nu}$ the one which should be compared to the GR metric, as happens also in RBG theories. This is the reason why the apparently arbitrary redefinition introduced in \eqref{eq:DefGeneralq} turns out to be useful to build a generalised Einstein-frame. Indeed, this choice has the advantage of making apparent that the effects of the higher order corrections related to symmetrised RBG terms in the action of general metric-affine theories do no introduce new degrees of freedom. Instead, the above form for a general metric-affine theory shows that symmetrised RBG operators just introduce new interactions through the generalised potential term $\check\cU$ among virtually all the degrees of freedom of the theory. At a perturbative level, these effects are encoded in the corrections associated to the $\Theta$ tensor, which introduce effective interactions among all the degrees of freedom of the theory that couple to the metric $g^{\mu\nu}$, coupling in this way all the original matter fields among themselves, as well as, potentially, to the new degrees of freedom of the metric-affine sector. Hence, when building an EFT of a metric-affine sector, one expects that operators pertaining to the RBG sub-class will be redundant.\footnote{Namely they would only contribute to a redefinition of the respective Wilson coefficients, see section \ref{sec:RBGandEFT}.} If all the operators allowed by the symmetries are considered for all the propagating degrees of freedom of the theory below the cutoff scale, then those terms (which we included in $\bgar$) will also couple  these new degrees of freedom to the curvature of the metric $q^{\mu\nu}$, as can be seen explicitly by splitting the connection $\check\Gamma$ as
\begin{equation}
\Gamma^\alpha{}_{\mu\nu}={}^q\Gamma^\alpha{}_{\mu\nu}+L^\alpha{}_{\mu\nu}+K^\alpha{}_{\mu\nu}
\end{equation}
by means of \eqref{eq:ConnectionDecomposition}, and then rewriting the field equations for the metric $q^{\mu\nu}$, namely \eqref{eq:MetricFieldEqEFgen}, as
\beq
G^\mu{}_\nu(q)=-{\mg^2}\,\frac{\delta U}{\delta q^{\mu\nu}},
\eeq
where $U$ now stands for $\check\cU$ where the dependence on the connection has been written in terms of $\partial q$, $\nabla q$ and the torsion tensor.  As seen in chapter \ref{sec:UnstableDOF} for the particular case where $\bgar$ depends only on the antisymmetrised Ricci tensor, around arbitrary $q$ backgrounds, these couplings are generally prone to excite Ostrogradski ghosts and possibly other instabilities unless the coefficients are fine tuned to avoid it. In the EFT regime, these instabilities could be pushed above the cutoff scale of the theory, so that they remain valid effective theories below this cutoff, though it is not clear that this can be done while keeping the new degrees of freedom within the spectrum of the low energy theory over generic backgrounds. The take-home message is that, if one is trying to build a viable metric-affine theory free of these pathologies, then care must be taken in avoiding these pathological couplings, either by including only the symmetrised Ricci tensor into the action, or by fine tuning the coefficients to guarantee the stability of the theory, though this will be generally a difficult task \cite{Jimenez:2020dpn,Percacci:2019hxn}.

This question is more precisely formulated in the EFT language, where it reduces to finding all the metric-affine operators that are redundant\footnote{If the reader is not familiarised with the concept of redundant operator, see the discussion in section \ref{sec:RBGandEFT}.} in a metric-affine EFT once all the allowed operators of the matter sector have been allowed.

%%%%%%%%%%%%%%%%%%%%%%%%%
%%%%%%%%%%%%%%%%%%%%%%%%%
%%%%%%%%%%%%%%%%%%%%%%%%%

			%NEWCHAPTER%

%%%%%%%%%%%%%%%%%%%%%%%%%
%%%%%%%%%%%%%%%%%%%%%%%%%
%%%%%%%%%%%%%%%%%%%%%%%%%

\chapter{Observable traces of nonmetricity}\label{sec:ObservableTraces}

\initial{F}irst works in gravitation which took into account post-Riemannian geometries dealt with the possibility of including the torsion tensor in the description of gravitation, and it was seen that fermions naturally generate torsion when coupled to GR \cite{Kibble:1961ba}. Afterwards, it was  discovered that one can obtain the same results by constructing the gauge theory of the Poincar\'e group\footnote{Known as Einstein-Cartan-Sciamma-Kibble or ECKS theory.} \cite{Hehl:1976kj,Hehl:1994ue}. Some authors tried then to understand what could be the observable consequences of torsion and used them to place experimental constraints to the possible existence of a nonvanishing torsion tensor in different contexts \cite{Shapiro:2001rz,Mao:2006bb,March:2011sa,Lehnert:2013jsa,Lucchesi:2015mga,Boos:2016cey}. Despite the effort put in understanding the observable effects of torsion, the existence (or lack) of observables related to nonmetricity is not yet well understood. The first works that included nonmetricity, only took into account a special case of it, namely Weyl-like nonmetricity given by $Q_{\al\mu\nu}=2A_\al g_{\mu\nu}$ \cite{Weyl:1918ib}. Later on, nonmetricity was studied as a gauge potential arising in the gauge theory of the group of general affine transformations \cite{Hehl:1976kj,Hehl:1994ue}, generalising the gauging of the Poincar\'e group. More recently, modifications of the GR Lagrangian including (or based on) nonmetricity, such as $f(Q)$ theories, RBG theories, or general metric-affine theories \cite{Aldrovandi:2013wha,Nester:1998mp,BeltranJimenez:2017tkd,Jimenez:2019ghw,Zhao:2021zab,Hohmann:2021fpr,Afonso:2017bxr,Afonso:2018bpv,Delhom:2019zrb,Iosifidis:2020dck,Hehl:1994ue,Jimenez-Cano:2020lea,Jimenez-Cano:2020chm,1867432} have been widely considered. Though the observable effects that torsion may have have been studied in relative depth, the search for physical effects associated to the presence of nonmetricity, if any, has been historically overlooked. Indeed, though it was pointed out that a second clock effect would arise in theories with nonmetricity already as a criticism to Weyl's ideas, in order for this to be true, one would need to be able to build clocks which measure a proper time that is not associated to metric geodesics and is sensitive to these nonmetric effects, which might not be possible with the field content of the universe, hence invalidating the criticisms unless such clocks are found to exist (see chapter \ref{sec:GeneralisedTime} for details).

 In this chapter we will focus on a particular kind of effects that arise in RBG theories that, from the geometrical perspective, can be traced back to the particular form of the nonmetricity tensor in these theories. Though these effects will be {\it universal} across the matter sector in a precise sense explained below, they are in general not universal enough so as to respect the Equivalence Principle, in the sense that freely falling particles might not all follow the same paths. However, the induced violations will be suppressed by a high energy scale controlling deviations from the GR Lagrangian. From the field theoretic perspective, the Equivalence Principle is understood as a consistency condition to couple to a massless spin-2 field, which must couple universally (see chapter \ref{sec:GravityAsGeometry}). From this viewpoint, the Einstein frame of the theory suggests to see these effects as effective interactions in the matter sector coupled to GR. Hence, the violations of the Equivalence Principle can be understood as deviations from the geodesic trajectories due to these effective interactions, which act as a fifth-force.
 
  Aside from the ontological interpretation of these effects either in terms of field interactions or as purely geometrical effects, our aim here will be to unveil the general conditions that a metric-affine theory has to satisfy in order to predict these corrections, explaining how they arise naturally in RBG theories due to the existence of a metric whose canonical connection is the solution to the RBG connection equations (namely the Einstein frame metric, see section \ref{sec:StructureRBG}). We will then explicitly compute this effects when the matter sector consists on a minimally coupled spin-0, spin-1/2 and spin-1 field and arbitrary combinations of them. Then we will constrain the presence of this effects by considering the corrections to the Standard Model (SM) to the scattering processes ${\textrm{e}}^{+}{\textrm{e}}^{-} \to {\textrm{e}}^{+}{\textrm{e}}^{-}$, $\ga\ga\to\ga\ga$ and $e^{-}\ga \to e^{-}\ga$, which will allow us to set experimental constraints to the RBG scale $\mq$ once a particular RBG model is chosen. We will also argue why these effects can also be expected in generic theories of metric-affine gravity, based on the fact that they can always be formulated as metric theories coupled to an involved matter sector containing terms with the nonmetricity and torsion tensors. From the geometrical perspective, this bunch of other terms contain all the post-Riemannian features of the spacetime geometry, which would be a metric geometry in their absence. Thus, if one finds physical effects that are associated to these types of terms in the action, one could say that one has found observable traces of nonmetricity, torsion, or both. On the other hand, from the field theoretic perspective, this bunch of terms involving these two tensorial objects that the geometrists would call nonmetricity and torsion can be rightfully interpreted as two additional matter fields that couple in a particular way to the degrees of freedom described by the metric sector of the theory, be it a massless spin-2 field, or any other spectrum possibly associated to the corresponding metric sector. From this perspective, the geometrical viewpoint only makes sense if there is universality, and only in the case that these two tensorial objects lead to effects in the matter sector that are universal in some sense could one accept to associate this tensorial objects to the spacetime geometry in a meaningful way. On the other hand, from the geometrical perspective where gravitation is tied to the spacetime geometric structure, there is no trouble in accepting that each matter field can couple differently to the spacetime geometry, though this would lead to violations of universality and the WEP. When I started my PhD, one of my main goals was to try to shed light into the question of whether spacetime nonmetricity has some kind of universal physical effects that would be observable in any theory where it is nonvanishing. I have arrived to the conclusion that, in general, the effects of the nonmetricity tensor depend on its couplings to the rest of the degrees of freedom of the theory, so that there are no model-independent (or universal) effects that arise from the explicit coupling of nonmetricity to matter fields, as one could have expected from the geometrical viewpoint. However, nonmetricity is a curious object and, as we will see in this chapter, there are some quite generic effects due to the presence of $R_{(\mu\nu)}$ terms in the action beyond the EH term that, from the geometrical viewpoint, could be related to a piece of the nonmetricity which takes a somewhat universal form in presence of these corrections but vanishes in their absence. Remarkably, the appearance of these effects does not seem to depend on the coupling of nonmetricity to mater so that, in a sense, it provides a partial answer to the original question. However, as noted above, these effects have a more natural interpretation in terms of effective interactions among the degrees of freedom of the theory which are suppressed by a UV scale controlling deviations from GR and will generally jeopardise perturbative unitarity of the theories.

\section{Effective interactions below nonmetricity scale}\label{sec:PertNM}

Inspired by the known results in RBG theories, in the previous chapter we elaborated on how, even in general metric-affine theories, the role of the $R_{(\mu\nu)}$ corrections reduces to introduce new couplings among the matter degrees of freedom. In order to see explicitly how these couplings appear due to these higher-order symmetrised Ricci corrections, let us turn to the simplest possible example, namely theories built only with this kind of corrections out of all the possible diffeomorphism covariant objects that can be built from the connection. Namely, we will analyse the particular case of Ricci-Based gravity with projective symmetry, which have an action of the form.
\beq
\cS=\frac{\mq^4}{2}\int d^4 x\sqrt{-g} \cl \left[ g^{\mu\nu},{\mg^{-2}}R_{(\mu\nu)}\right]+\cS_\textrm{m}[g^{\mu\nu}, T_{[\alpha\mu\nu]}, \Psi_i],
\end{equation}
where the matter fields $\Psi_i$ will be assumed to be minimally coupled spin-0, 1/2 and 1 fields for simplicity, though this is not necessary for the conclusions of this section to hold. Hence, the connection can only enter the matter action through the totally antisymmetric piece of the torsion tensor.\footnote{The analysis generalises in a straightforward manner if the matter fields couple nonminimally to the symmetrised Ricci tensor. In that case, one has to add the term $\sqrt{-g}\partial \cl_\textrm{m}/\partial R_{(\mu\nu)}$ to the auxiliary metric. An example of this will be shown in chapter \ref{sec:SLSB}.} In this case, we know from chapter \ref{sec:StructureRBG} that the connection will be the Levi-Civita connection of $q^{\mu\nu}$ up to an unphysical projective mode and a hypermomentum term that is algebraic in the fermionic fields. On the other hand, after integrating out the connection, and performing the field redefinition which allows to write $g^{\mu\nu}$ in terms of $q^{\mu\nu}$ and the matter fields, the metric field equations become Einstein's equations for the metric $q^{\mu\nu}$ coupled to a modified matter sector \eqref{eq:EinsteinEquationsq}. The relation between both metrics is encoded in the deformation matrix, which on-shell is given in terms of one of the metrics and the matter fields by \eqref{eq:OmegaExpansionTmunu}. The perturbative expansion of this expression for the deformation matrix leads to
\beq
g^{\mu\nu} = q^{\mu\nu} - {\mq^{-4}}\big( \al T q^{\mu\nu}+\beta T^{\mu\nu}\big)+\cO({\mq^{-8}}), \label{metricstressenergy}
\eeq
and
\beq\label{eq:NMperturbative}
Q_{\alpha}{}^{\mu\nu}={\mq^{-4}}\big[ \al (\dif T)_\alpha q^{\mu\nu}+\beta \nabla_\alpha T^{\mu\nu}\big]+\cO({\mq^{-8}}),
\eeq
where $\alpha$ and $\beta$ are dimensionless coefficients that depend on the particular RBG model under consideration, $T^{\mu\nu}$ the matter stress energy tensor \eqref{eq:StressEnergyTensor} and $T$ its trace. We see that, in this case, the lowest order deviations between both metrics as well as the leading contributions to the nonmetricity tensor are controlled by the heavy mass scale ${\mq}$. Comparing this to \eqref{eq:RelationMetricPer}, we obtain 
\beq
\Theta_{\textrm{RBG}}^{\mu\nu}={\mpl^{-2}}\big( \al T q^{\mu\nu}+\beta T^{\mu\nu}\big).
\eeq
It is interesting to note that the $\alpha$ term in (\ref{eq:NMperturbative}) yields a contribution to the Weyl trace of the nonmetricity tensor that can be eliminated from the nonmetricity tensor with the appropriate projective transformation. Note, however, that it cannot be removed from the metric, and it leads to specific observable effects. The possibility of being able to gauge away this term from $Q_{\mu\alpha\beta}$ just tells us that its effects cannot be associated to the nonmetricity tensor in a theory with projective symmetry in a gauge-independent way. For the particular subclass of RBG theories consisting on $f(R)$ theories, only the $\alpha$ term is present in the above relations, which makes sense taking into account that the presence of nonmetricity in the RBG frame (also known as Jordan frame) of metric-affine $f(R)$ theories is a matter of projective gauge choice.\footnote{Projective symmetry is a symmetry of $f(R)$ theories (unless broken by the matter sector).} On the other hand, a projective transformation cannot eliminate the $\beta$ term from $Q_{\mu\alpha\beta}$ so that, from the RBG frame perspective, its observable effects can be linked to a nonmetricity tensor of the form \eqref{eq:NMperturbative}.

As explained in chapter \ref{sec:RBGTheory}, and is apparent because it satisfied the Einstein Equations coupled to a matter stress-energy tensor, perturbations to the metric $q^{\mu\nu}$ describe a massless spin-2 excitation. In other words, the metric $q^{\mu\nu}$ is the responsible of controlling the long-range gravitational force according to GR. Therefore, in virtue of the Equivalence Principle, the $q^{\mu\nu}$ metric can be made locally Minkowskian by a suitable choice of coordinates. On the contrary, the deviations of $g^{\mu\nu}$ from $q^{\mu\nu}$, which are sensitive to the local distributions of energy-momentum, cannot be eliminated in this way, which opens the door for apparent violations of the Equivalence Principle(s) in these theories due to fifth-force like effects as viewed from the RBG frame. From the Einstein frame perspective, these violations amount to the new interactions on the matter sector that deviate the associated particles from following the geodesics of the metric. Thus, from this perspective, no violation of the Equivalence Principle(s) occurs, as the deviations are due to the fact that the matter fields are not free. As a remark, let us comment a bit more precisely on what we mean by {\it apparent} violations of the Equivalence Principle(s). The word apparent here means that this would be a violation of the WEP only if one insists on interpreting these effects as having geometrical origin and is willing to assume that they are part of the gravitational interaction because of that reason. From a field theoretic perspective, the massless spin-2 still couples universally and the effects are just seen as effective interactions among the matter sector. In this view gravity is still mediated only by a massless spin-2 and therefore there are no violations of the WEP or SEP.

In light of the above discussion it is clear that the role played by the metric $g^{\mu\nu}$ in RBG theories is nontrivial. In fact, it is associated to two different kinds of phenomena, namely: 1) the propagation of a masless spin-2 excitation, which is responsible for a long-range force consisting on Newtonian and post-Newtonian corrections, as well as nonperturbative gravitational phenomena. These effects are generated by the total amount of energy-momentum within a spacetime region, namely, through integration over the matter sources. 2) new effects associated to the local distribution of energy-momentum which are related to the existence of a nonmetricity tensor of the form \eqref{eq:NMperturbative} in the RBG frame. Thus we see that, in RBG theories (as well as in general metric-affine theories containing the symmetrised Ricci tensor in the action as discussed in the begining of the section), the long-range interaction associated to a spin-2 field is only a part of the physical content of the metric $g^{\mu\nu}$, which in general features also new terms that are sensitive to the local distribution of energy-momentum (and other matter-related quantities in theories more general than RBG theories). Moreover, from the structure of the field equations, the existence of these corrections is tied to the on-shell form of the nonmetricity tensor $\nabla g$, and we shall call them $Q$-induced interactions.

Knowing that the observable effects of the $\Theta$ term can be encoded into a series of new $Q$-induced effective interactions that couple all the degrees of freedom of the theory (besides possibly the spin-2 field associated to $q^{\mu\nu}$), our aim is now to compute the relevance of these corrections in high-energy observables such as particle collisions in order to constrain the presence of these corrections. In that context, note that even if our Earth-based Laboratory is not an inertial frame, the gravitational field of the Earth is extremely weak, so that the Einstein frame metric can be approximated by $q^{\mu\nu}\approx\eta^{\mu\nu}$ up to Newtonian and post-Newtonian corrections. We will focus here on the particular case of RBG theories, though the generic case should not be expected to yield milder constraints (in fact, it will possibly be the opposite). 

\section{$Q$-induced interactions for spin-0, 1/2 and 1}\label{sec:NMLag}

In order to derive the explicit $Q$-induced corrections to a matter sector consisting of minimally coupled spin-0, 1/2 and 1 fields on the Earth surface, we can start from  \eqref{metricstressenergy} and, neglecting Newtonian and post-Newtonian corrections, approximate $q^{\mu\nu}\approx\eta^{\mu\nu}$, which leads to
 \begin{equation}\label{MinkowskiMetricPerturbed}
 \begin{split}
&g^{\mu\nu} = \eta^{\mu\nu} -{\mq^{-4}}\lr{ \al T \eta^{\mu\nu}+\beta T^{\mu\nu}}+\cO({\mq^{-8}}),\\
&\sqrt{-g}=1+\frac{4\al+\be}{2{\mq^4}}T+\cO({\mq^{-8}}).
\end{split}
\end{equation}
Given that the connection is the Levi-Civita connection of $q_{\mu\nu}$, in suitable coordinates, the connection symbols will vanish up to Newtonian and post-Newtonian corrections which we are neglecting, unless spin-1/2 are considered, in which case Planck-scale suppressed torsion terms that can be algebraically written in terms of the fermion fields will also appear. These terms will only generate interactions among the fermions of the theory, as minimally coupled bosons do not couple to the connection (see chapter \ref{sec:MinimalCoupling}). In order to compute the $Q$-induced interaction for these cases, we need to substitute the metric in the matter Lagrangians by the expressions in \eqref{MinkowskiMetricPerturbed}. It will also prove useful to write down the form of the Minkowskian stress-energy tensor for these fields.
\subsubsection{Scalar field}
The Lagrangian\footnote{Through this section, we include the volume element in the Lagrangian density, so it is a $D$-form instead of a scalar function.} for a (complex) minimally coupled scalar field in an arbitrary post-Riemannian spacetime, with an arbitrary potential, and which can in principle interact with gauge bosons through a corresponding exterior covariant differential\footnote{Namely, the scalar field is a 0-form section of some vector $G$-bundle with a corresponding $G$-connection.} denoted by $\cD$, reads
\begin{equation}\label{scalaraction}
\cL^{(0)}=\sqrt{-g}\lrsq{g^{\al\be}\cD_\al\phi^* \cD_\be\phi+V_0}. 
\end{equation}
where $\cL^{(s)}$ and $V_s$ correspond to the Lagrangian and an arbitrary potential (so other fields can appear in $V_s$) for a field with spin $s$ respectively, with $s=0$ in the scalar case. By means of \eqref{MinkowskiMetricPerturbed}, we can expand the above Lagrangian around its Minkowskian version as\begin{equation}\label{scalarmink}
\cL^{(0)}=\eta^{\al\be}\cD_{\al}\phi^*\cD_\be\phi+V^{(0)}_0+{\mq^{-4}}\cL^{(0)}_Q \, , 
\end{equation}
where, given that $V_s$ can depend on the metric in the most general case, we have defined 
\beq
V_s=\sum_{n=0}^\infty {\mq^{-4n}}\,V_s^{(n)},
\eeq 
so that $V_s^{(n)}$ do not depend on ${\mq}$. The first two terms are the Minkowskian version of the original Lagrangian \eqref{scalaraction}, and  $\cL^{(0)}_{ Q}$ contains the $Q$-induced effective interactions between the scalar field and the stress-energy tensor, and to leading order in inverse powers of $\mq$ takes the form
\begin{align}\label{scalarpert}
\begin{split}
\cL^{(0)}_Q=\lrsq{\frac{2\al+\be}{2}T_{\textrm{M}}\eta^{\mu\nu}-\be T_{\textrm{M}}{}^{\mu\nu}}\cD_\mu\phi^*\cD_{\nu}\phi+\lrsq{\frac{4\al+\be}{2}V_0^{(0)} T_{\textrm{M}}+ V^{(1)}_0}+\cO({\mq^{-4}}),
\end{split}
\end{align}
where the subscript M stands for Minkowskian. These interaction terms mix the scalar field with all the other matter fields in the theory through the stress-energy tensor, including self-interactions. Given that the interacting terms are a product of the matter stress-energy tensor and some piece of the scalar Lagrangian, they will respect all the symmetries of the original matter sector. Then any minimally coupled spin-0 field in an RBG theory can be identified with a scalar field with the same quantum numbers that interacts with all the fields in the matter Lagrangian but evolves according to GR. The Minkowskian stress-energy tensor associated to \eqref{scalaraction} is given by
\beq
T_{\textrm{M}}^{(0)}{}_{\mu\nu}=\eta_{\mu\nu}\lrsq{\cD^\al\phi^* \cD_\al\phi+V_0}-2\lrsq{\cD_{(\mu}\phi^*\cD_{\nu)}\phi+\frac{\partial V_0{}_{\textrm{M}}}{\partial \eta^{\mu\nu}}}. \label{scalartmunu}
\eeq

\subsubsection{Dirac field}
The Lagrangian for a minimally coupled Dirac field in an arbitrary post-Riemannian spacetime, with an arbitrary potential, and which can in principle interact with gauge bosons through a corresponding exterior $G$-covariant differential\footnote{Namely, the spinor field is a 0-form section of the product bundle $\cS\cm\times\cB$ where $\cB$ is a vector $G$-bundle with a corresponding $G$-connection.} reads
\begin{equation}\label{BDlag}
\cL^{(1/2)}=\sqrt{-g}\lrsq{\frac{1}{2}\fr_a{}^\mu\lr{\bpsi\ga^a({\na}_\mu\psi)-({\na}_\mu\bpsi)\ga^a\psi}+V_{1/2}},
\end{equation}
where ${\na}$ accounts for the covariant derivative of the spinor bundle and the $G$-bundle. From $g_{\mu\nu}=\df^a{}_\mu\df^b{}_\nu\eta_{ab}$, by means of \eqref{MinkowskiMetricPerturbed} the solder forms are given by
\begin{equation}\label{vierbeinexp}
\fr_a{}^\mu={\delta_a}^\mu-\frac{1}{2{\mq^4}}\lr{\al T{\delta_a}^\mu+\be {T_a}^\mu}+\cO({\mq^{-8}}),
\end{equation}
which allows to write the Einstein frame spinor Lagrangian perturbatively as 
\begin{equation}\label{BDlagpert}
\cL^{(1/2)}=\frac{1}{2}\lrsq{\bpsi\ga^\mu(\cD_{\mu}\psi)-(\cD_{\mu}\bpsi)\ga^\mu\psi}+V^{(0)}_{1/2}+{\mq^{-4}}\cl_{\textrm{G}}^{(1/2)},
\end{equation}
where $\cD$ is the $G$-covariant differential which accounts for the gauge interactions. Here we have already neglected the Planck-suppresed torsion contributions which source a 4-fermion effective interaction, as they are irrelevant compared to the $Q$-induced interactions unless ${\mq}$ approaches the Planck mass. Again, the first two terms of the above Lagrangian are the Minkowskian version of \eqref{BDlag}, and  $\cL^{(1/2)}_{\textrm{Q}}$ contains the $Q$-induced effective interactions between the Dirac field and the stress-energy tensor, and to leading order in inverse powers of $\mq$ it takes the form
\begin{equation}\label{spinorpert}
\resizebox{.9\hsize}{!}{$\cL^{(1/2)}_{ Q}=\lrsq{\frac{3\al+\be}{4}T_{\textrm{M}}\eta^{\mu\nu}-\frac{\be}{4} {T_{\textrm{M}}}^{\mu\nu}}\lrsq{\bpsi\ga_\mu(D_{\nu}\psi)-(D_{\nu}\bpsi)\ga_\mu\psi}+\lrsq{\frac{4\al+\be}{2}T_{\textrm{M}}V^{(0)}_{1/2}+V_{1/2}^{(1)}}+\cO({\mq^{-4}}).$}
\end{equation}
These interaction terms mix the Dirac field with all the other matter fields in the theory through the stress-energy tensor, including self-interactions, respecting the original symmetries of the action as in the previous cases. Then any minimally coupled spin-1/2 field in an RBG theory can be identified with a spin-1/2 field with the same quantum numbers that interacts with all the fields in the matter sector but evolves according to GR. The Minkowskian stress-energy tensor associated to \eqref{BDlag} is given by
 \beq
 \resizebox{.9\hsize}{!}{$T_{\textrm{M}}^{(1/2)}{}_{\mu\nu}=\eta_{\mu\nu}\lrsq{\frac{1}{2}\lr{\bpsi \ga^\al (D_{\al}\psi)-(D_{\al}\bpsi)\ga^\al\psi}+V_{1/2}}-\lrsq{\bpsi\ga_{(\mu}(D_{\nu)}\psi)-(D_{(\nu}\bpsi)\ga_{\mu)}\psi+2\frac{\partial V_{1/2}{}_{\textrm{M}}}{\partial \eta^{\mu\nu}}}.$} \label{spinortmunu}
 \eeq
\subsubsection{1-form field}
The Lagrangian for a minimally coupled spin-1 field field in an arbitrary post-Riemannian spacetime, described by a $G$-connection 1-form coupled to an arbitrary potential that might generally break $G$-covariance (gauge symmetry) is given by
\begin{equation}\label{VectorLag}
\cL^{(1)}=\sqrt{-g}\lrsq{\frac{1}{4}g^{\mu\nu}g^{\al\be}{F}^\dagger_{\mu\al}{F}_{\nu\be}+V_1},
\end{equation}
where ${F}=\cD A=\dif A+A\wedge A$. By means of \eqref{MinkowskiMetricPerturbed} we can write the corresponding Einstein frame Lagrangian perturbatively as
\begin{align}\label{Vectormink}
\cL^{(1)}=\frac{1}{4}\eta^{\mu\nu}\eta^{\al\be}F^\dagger_{\mu\al}F_{\nu\be}+V^{(0)}_1+{\mq^{-4}}\cL^{(1)}_{ Q}.
\end{align}
As in the two previous cases, the two first terms are the Minkowskian version of \eqref{VectorLag}, and $\cL^{(1)}_{ Q}$ contains the $Q$-induced effective interactions between the spin-1 field and the stress-energy tensor, which to leading order in inverse powers of $\mq$ takes the form
\begin{align}\label{vectorpert}
\begin{split}
\cL^{(1)}_{ Q}=\lrsq{\frac{2\al+\be}{4}T_{\textrm{M}}\eta^{\al\be}-\frac{\be}{2} T_{\textrm{M}}{}^{\al\be}}F^\dagger{}^{\mu}{}_\al F_{\mu\be}+\lrsq{\frac{4\al+\be}{2}T_{\textrm{M}} V_1^{(0)}+V^{(1)}_1}+\cO({\mq^{-4}}).
\end{split}
\end{align}
These interaction terms mix the spin-1 field with all the other matter fields in the theory through the stress-energy tensor, including self-interactions, respecting the original symmetries of the action as in the scalar case. Then any minimally coupled spin-1 field in an RBG theory can be identified with a spin-1 field with the same quantum numbers that interacts with all the fields in the matter Lagrangian but evolves according to GR. The Minkowskian stress-energy tensor associated to \eqref{VectorLag} is given by
\beq
T_{\textrm{M}}^{(1)}{}_{\mu\nu}=\eta_{\mu\nu}\lrsq{\frac{1}{4}F^\dagger_{\al\be}F^{\al\be}+V_1}-\lrsq{F^\dagger_{(\mu|\al|}{F_{\nu)}}^\al+2\frac{\partial V_{1{\textrm{M}}}}{\partial \eta^{\mu\nu}}}. \label{vectortmunu}
\eeq

As a final remark, note that this is just a perturbative version of the explicit building of the Einstein frame of the theory, namely the mapping procedure exemplified in section \ref{sec:Mapping} with a 1-form field. The $Q$-induced effective interactions computed here represent the leading order corrections of the Einstein frame matter Lagrangian with respect to the original one. These corrections couple any matter field to the stress-energy tensor, namely to all the matter fields including itself, in a way which respects the symmetries of the original matter Lagrangian. In general, the implications of these $Q$-induced interactions are the following: below the scale $\mq$, they describe a series of perturbative interactions that, from the geometrical point of view, can be directly linked to the nonmetricity tensor. In this view, the scale $\mq$ characterises the scale at which nonmetricity becomes nonperturbative. From the field theoretic perspective, the Einstein frame matter Lagrangian is an effective theory which breaks down at the scale $\mq$ (or suitable combinations of $\mg$ and $\mpl$ in more general theories) so that  predictions of the theory cannot be trusted above the cutoff scale. These corrections enter through the metric $g^{\mu\nu}$ and are sensitive to the local distribution of energy-momentum. Notably these departures are different in nature from those corresponding to the post-Newtonian behaviour of RBG models which, as usual, are associated to integrated energy-momentum within the relevant region, instead of feeling the local distribution of energy-momentum, and are characterised by the Planck scale $\mpl$.

\section{Collider constraints to $Q$-induced interactions}\label{sec:ExpCons}

Following the path of \cite{Latorre:2017uve,Delhom:2019wir}, where the existence of $Q$-induced interactions was first noticed within RBG theories, we will now try to derive their corrections to particle scattering processes measured at high energy colliders such as ${\textrm{e}}^{+}{\textrm{e}}^{-} \to {\textrm{e}}^{+}{\textrm{e}}^{-}$ and $e^{-}\gamma \to {\textrm{e}}^{-}\gamma$. These computations will allow us to constrain the parameters regulating $Q$-induced interactions within RBG theories (though the order of magnitude is expected to be similar for more general ones) by looking at data from experiments in high energy colliders.\footnote{Forthcoming work will also test these interactions using ultra-high energy neutrino detections at IceCube \cite{JoanAdri}. Also in \cite{CANTATA:2021ktz} a quick review of microscopic effects in metric-affine theories will include the  results in this chapter.}

\subsubsection{Leading contributions to ${\textrm{e}}^+{\textrm{e}}^-\to{\textrm{e}}^+{\textrm{e}}^-$, $\ga\ga\to\ga\ga$ and ${\textrm{e}}^-\ga\to{\textrm{e}}^-\ga$}

To find the corrections to the Standard Model (SM) operators contributing to these processes, we need to compute the self-interaction terms due to the $Q$-induced corrections of an electron and a photon field. Starting with the electron, this implies plugging the Minkowskian stress-energy tensor for the electron field \eqref{spinorpert} into its own $Q$-induced Lagrangian \eqref{spinorpert}. Given that this process occurs at tree level, we can use on-shell identities. Moreover, since this  process has ben measured at LEP for energies of order 100 GeV, the electron masses can be neglected. Taking this into account, the leading order $Q$-induced contribution to Bhabha scattering is given by the operator\cite{Latorre:2017uve}
\beq\label{Bhabhaop}
\mathcal{O}^{{\textrm{e}}^{+}{\textrm{e}}^{-}}_{Q}=-\frac{\beta}{{\mq^4}}\left[\bar{\psi}_{e}\left(\gamma_{\nu} \overleftrightarrow{\partial}^{\mu}+\gamma^{\mu}\overleftrightarrow{\partial}_{\nu}\right) \psi_{e}\right]\left[\bar{\psi}_{e} \gamma^{\nu} \overleftrightarrow{\partial}_{\mu} \psi_{e}\right].
\eeq
Doing the same for a Maxwell photon, we find \cite{Delhom:2019wir}
\begin{equation}\label{4VLag}
\cO^{2\gamma}_Q = -\frac{\be}{8{\mq^4}}\lrsq{F_{\mu\nu}F^{\mu\nu}F_{\al\be}F^{\al\be}-4F_{\mu\nu}F^{\nu\al}{F^\mu}_{\sigma}{F^\sigma}_{\al}},
\end{equation}
which is a particular case of the well known $C$, $P$, Lorentz and gauge invariant effective Lagrangian describing photon-photon collisions below the mass scale of some charged fermion. Notice that while the Euler-Heisenberg Lagrangian \cite{Pich:1998xt,Euler:1936oxn,Heisenberg:1935qt} obtained by integrating out a massive lepton in the QED path integral gives a relation $b/a=-14/5$ , the above Lagrangian satisfies $b/a=-4$.

Regarding the contribution to Compton scattering, to obtain the full contribution we must compute the correction to the electron Lagrangian due to the photon stress-energy tensor, namely inserting \eqref{vectortmunu} into \eqref{spinorpert} and the correction to the photon Lagrangian due to the electron stress energy tensor, namely plugging \eqref{spinortmunu} into \eqref{vectorpert}. using again on-shell identities and throwing away electron masses we find \cite{Delhom:2019wir}
\begin{equation}\label{vector-fermion}
\begin{split}
\cO_Q^{\gamma{\textrm{e}}^{-}}=-\frac{9\be}{4{\mq^4}}F^{\mu\al}{F^\nu}_\al\lrsq{\bpsi_e\ga_\mu(\cD_{\nu}\psi)-(\cD_\nu\psi_e)\ga_\mu\psi}.
\end{split}
\end{equation}

Even though almost every process is sensitive to contributions appearing in Ricci-Based gravity theories,  obtaining constraints for the scale $\Lambda_Q$ is not a straightforward procedure in general. Corrections induced in the vertices and in the partition distributions functions of gluons and quarks make it very difficult to study processes in which particles are produced via $p\bar{p}$ production. This makes high-energy data from LHC not convenient for this study and requires to consider experimental bounds at lower energies. Thus we will use for that purpose current data on light-by-light and Compton scattering.

\subsubsection{Experimental constraints to the nonmetricity scale}

Let us consider first Bhabha and Compton scattering as a probe for the $Q$-induced interactions, as both processes have been observed at clean high energy colliders such as LEP since decades ago. The highest energy probes of Bhabha scattering come from LEP \cite{Abbiendi:2003dh,Schael:2013ita}, where the experimental cross section at a center of mass energy of $\sqrt{s}=207$ GeV, and for $\theta_{\textrm{acol}}<10^\circ$ and $|\cos\theta_{{\textrm{e}}^\pm}|<0.96$, was measured to be $\sigma^{\textrm{e}xp}_{{\textrm{e}}^+{\textrm{e}}^-\to {\textrm{e}}^+{\textrm{e}}^-}=256.9\pm1.4\pm1.3$ pb. The lowest order $Q$-induced contribution to this cross section due to \eqref{Bhabhaop} comes from the mixing with the SM operators. At that center of mass energy, the correction is roughly $\delta_Q\sigma\simeq 0.35 \beta{\mq^{-4}}$, and compatibility with the measurements leads to the order of magnitude constraint \cite{Latorre:2017uve}
\beq
{\mq}\gtrsim 0.6 \beta^{1/ 4} \mathrm{TeV}.
\eeq
In turn, the most recent data for the cross-section of Compton scattering comes from the L3 collaboration \cite{Achard:2005cs}, where the process was measured at different energies as  in Tab. \ref{tab:expcompt}. The leading order $Q$-induced correction to the SM differential cross section for Compton scattering in RBG theories can be obtained from \eqref{vector-fermion}, and is given by
\begin{equation}
 \resizebox{.9\hsize}{!}{$\frac{d \sigma^{Q}_{e^{-} \gamma \to e^{-} \gamma}}{d \Omega} = \frac{1}{256\pi^2 s} \left[\frac{9}{2}\left(\frac{\beta}{\Lambda_Q^4}\right)\left(3\cos^2\theta+2\cos\theta+11\right)Q_e^2 s^2 + 4 Q_e^4 \frac{\cos^2\theta + 2\cos\theta + 5}{\cos\theta +1}  \right].$}
\label{eq:comptondiff}
\end{equation}

where $Q_e$ is the electron charge. Note that in the measurements leading to table \ref{tab:expcompt}, only the region of the phase space in which $|\cos\theta|<0.8$ is considered. This will be taken into account when placing the bounds on $\mq$. As in table \ref{tab:expcompt} we have the experimental value measured at 12 different energies, it is convenient to combine all these measurements performing a $\chi^2$ test with 10 degrees of freedom. 
As figure  \ref{fig:chi_Compton} shows, by studying the probability of the resulting $\chi^2$ function we can constrain the values of $\beta {\mq^{-4}}$ up to a certain probability.  The green  
and blue bands contain the $1\sigma$ and $2 \sigma$ probability respectively.\\

\begin{table}[h]
\setlength{\tabcolsep}{20pt}
\centering
%\begin{center}
\begin{tabular}{c c c}
\hline
$\sqrt{s}$ (GeV) & $\sigma_{e^-\gamma \rightarrow e^-\gamma}^{exp}$  (GeV$^{-2}$) & $\sigma_{e^-\gamma \rightarrow e^-\gamma}^{QED}$ (GeV$^{-2}$)\\
\hline 
21 & 771.2$\pm$21.6 & 764.8  \\ 
29.8 & 370.6$\pm$11.3 & 381.1  \\
39.7 & 213.2$\pm$5.4 & 214.7  \\
49.7 & 128.7$\pm$3.9 & 136.7 \\
59.8 & 95.0$\pm$3.5 & 94.6  \\
69.8 & 70.6$\pm$2.9 & 69.4  \\
79.8 & 55.2$\pm$2.6 & 53.1  \\
92.2 & 38.8$\pm$2.2 & 39.8  \\
107.2 & 27.3$\pm$2.2 & 29.4  \\
122.3 & 20.0$\pm$2.1 & 22.6  \\
137.3 & 17.3$\pm$2.1 & 17.9  \\
159.3 & 9.1$\pm$2.0 & 13.3  \\
\hline
\end{tabular}
\caption{Experimental values and SM prediction of the cross section for Compton scattering taking from \cite{Achard:2005cs}.} 
\label{tab:expcompt} 
%\end{center}
\end{table}

\begin{figure}
    \centering
    \includegraphics[scale=1.]{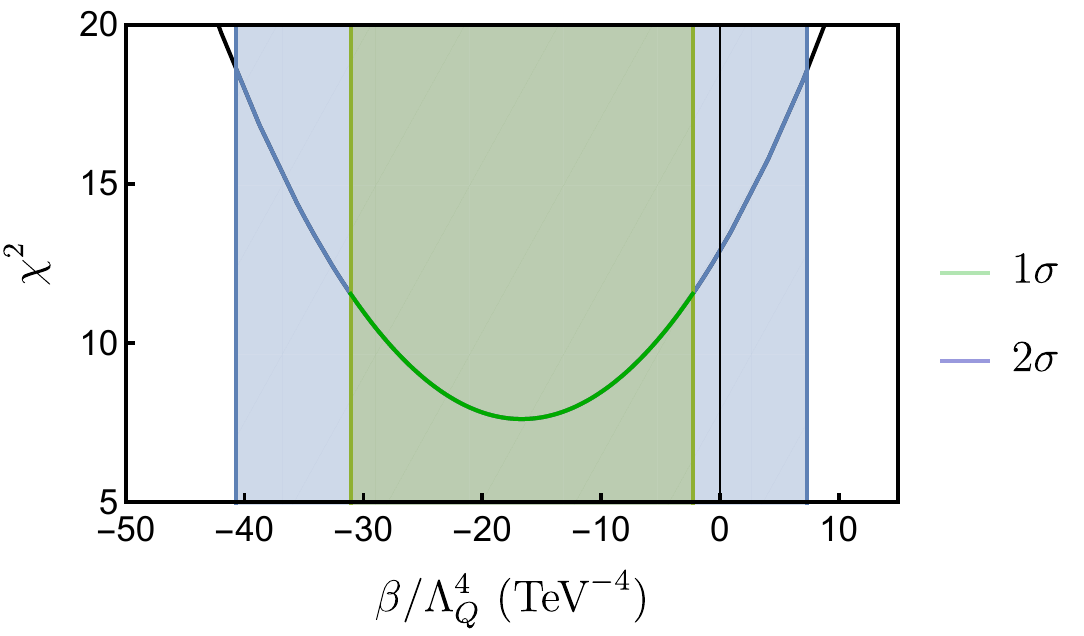}
    \caption{ Value of the $\chi^2$ for different values of $\beta {\mq^{-4}}$. The green and blue bands indicate the allowed values of $\beta {\mq^{-4}}$ at $1\sigma$  and $2\sigma$ probability respectively.}
    \label{fig:chi_Compton}
\end{figure}
In figure  \ref{fig:chi_Compton} the full $1\sigma$ probability is in the region $\beta < 0$. While the precise implications of the sign of this parameter might depend on the model and particular physical scenario under consideration, in some relevant models this is well understood in cosmological as well as black hole scenarios. At $2\sigma$  we get the constraints
\begin{align}
 & |\beta|^{-1/4} {\mq}   > 385 \text{ GeV}  \qquad \text{for the plus sign in the Lagrangian \eqref{BIgrav}},\label{2scomptonneg}\\
 & |\beta|^{-1/4} {\mq}  > 606 \text{ GeV}  \qquad \text{for the minus sign in the Lagrangian \eqref{BIgrav}}.\label{2scomptonneg}
\end{align}

These are bounds for a general RBG theory, though the order of magnitude serves for constraining $Q$-induced interactions in more general theories as well. Once a specific RBG model is chosen, the value of $\beta$ is set and the bound is translated to the heavy mass scale $\mq$ (or $\mg$ for that matter). Note that in this case the SM,  corresponding to $\beta {\mq^{-4}} = 0$ is already in the $2\sigma$ probability region. That means that at $1\sigma$ the values of  $\beta {\mq^{-4}}$ giving a lower value of the $\chi^2$ (higher probability) will be negative compensating the SM contribution. As mentioned before, at $2\sigma$ the SM is already in agreement with the data so positive values of $\beta$ give bounds in this region.

As a particular example, let us consider the widely discussed RBG model named Eddington-inspired Born-Infeld (see section \ref{sec:EiBIEMMapping}), $\beta=\pm 1$ for the  $\mp$ choice of sign in front of the $R_{(\mu\nu)}$ term of the Lagrangian \eqref{BIgrav}. In EiBI, while $\beta=1$ leads to a bouncing cosmology, $\beta=-1$ describes a cosmology in which an asymptotically Minkowski past region connects with the present contracting branch \cite{Banados:2010ix,BeltranJimenez:2017uwv}.  Interestingly, both solutions avoid the Big Bang singularity\footnote{Nonetheless, a potential Big Rip singularity could arise if phantom dark energy is considered within EiBI \cite{Bouhmadi-Lopez:2013lha,Bouhmadi-Lopez:2014jfa}, though quantum effects could remove the singularity \cite{Albarran:2018mpg}.}, though as found in \cite{BeltranJimenez:2017uwv}, the propagation of gravitational waves (GWs) generally presents instabilities in these cosmological models. On the one hand, Beltran \textit{et. al.} show that for a massless scalar field with $\beta>0$ GWs develop instabilities at the bounce due to the fact that the propagation speed diverges and the friction term vanishes, pointing a strong coupling problem. On the other hand, for the asymptotically Minkowski solution where $\beta<0$, they show that the pathologies are due to the vanishing of the propagation speed, which could in principle be avoided by including higher derivative terms. Regarding spherically symetric solutions, while $\beta=-1$ are generally singular , the $\beta=1$ branch remarkably admits nonsingular wormhole space-times when coupled to Maxwell electrodynamics \cite{BeltranJimenez:2017doy}. The above bounds for a general RBG model can be easily translated to the EiBI theory for the two signs, so that
 \begin{align}
 &{\mq^{\mathrm{EiBI}} }> 385 \text{ GeV}  \qquad \text{for the plus sign in the Lagrangian \eqref{BIgrav}}
\label{2scomptonneg}\\
 &{\mq^{\mathrm{EiBI}} } > 606  \text{ GeV}  \qquad \text{for the minus sign in the Lagrangian \eqref{BIgrav}}.\label{eq:2scomptonpos}
\end{align}
In some works, it is common to use the parametrisation $\kappa=2 c^{7} \hbar^{3} \mathrm{M}_{\mathrm{Q}}^{-4}$, which would be constrained by \begin{equation}
    |\kappa|<3.5\times10^{-14}\text{m}^5 \text{kg}^{-1}\text{s}^{-2}.
\end{equation}
This bound, which is of the same order of magnitude as the one obtained from Bhabha scattering, improves in 6 orders of magnitude the bound for the scale $\mq$ (and 12 orders of magnitude for the $\kappa$ parameter) as compared to other constraints obtained from astrophysical or nuclear physics \cite{Avelino:2012ge,Avelino:2012qe,Avelino:2019esh} phenomena.

The two previous processes are present in the SM at tree level, and therefore the $Q$-induced interactions are expected to produce a small correction to the observed values. However, light-by-light scattering occurs at loop level in the SM, hence being strongly suppressed \cite{Fichet:2014uka,Witten:1977ju,Terazawa:1973tb}. 
Therefore, it could be used in principle to obtain stringent bounds even from experiments searching photon self-interactions at lower energies. This has been done with X-ray pulses \cite{Inada:2014srv} obtaining an upper bound for the cross section which can be used to constrain $\mq$. The differential and total cross sections for $\ga\ga \to \ga\ga$ that one obtains from the leading order $Q$-induced corrections in RBG theories \eqref{4VLag} at tree level are given by
\begin{eqnarray}
\frac{d\sigma^Q_{\gamma\ga\rightarrow\ga\ga}}{d\Omega} &=& \lr{\frac{\be}{8 {\mq^4}}}^2\frac{1}{256\pi^2}s^3\lrsq{512+32\lr{\lr{1-\cos{\theta}}^4+ \lr{1+\cos{\theta}}^4}} \label{photonphotondiffcrossec} , \\
\sigma^Q_{\gamma\ga\rightarrow\ga\ga} &=&\lr{\frac{\be}{8 {\mq^4}}}^2\frac{56}{5\pi}s^3, \label{photonphotoncrossec}
\end{eqnarray}
By demanding \eqref{photonphotoncrossec} to be in agreement with the current experimental limit of $\gamma\ga \to \ga\ga$ at 6.5 keV, $\sigma^{\textrm{e}xp}_{\gamma\ga \to \ga\ga} < 1.9 \times 10^{-27}$ m$^2 $ \cite{Yamaji:2016xws}, we can set a lower bound
\beq\label{Lightbylightbound}
 |\be|^{-1/4} \mq > 23.3 \; \text{keV}.
\eeq
Where the value of $\beta$ is fixed in each particular RBG theory, allowing to constrain directly the energy scale $\mq$. Due to the difference in energies at which Bhabha or Compton and photon-photon scattering are currently tested, and the unobservability of photon self interactions in the keV range with current experimental precision, the bound obtained in \eqref{Lightbylightbound} is considerably weaker than the one obtained from Bhabha or Compton scattering in \cite{Latorre:2017uve,Delhom:2019wir}. However, future experiments searching for light-by-light scattering in the keV range could help in tightening current constraints to electromagnetic self interactions provided that a substantial increase in the experimental resolution is achieved, and therefore to $Q$-induced interactions in RBG models and beyond. If the experimental precision is not improved, higher-energy experiments will allow us to obtain stringent bounds to RBG. In figure \ref{fig:photonsprospects} we can see how the limit would change if the precision is improved or the energy scale changes. For instance, keeping the same upper limit $\sigma_{\text{bound}}$ while increasing the energy scale of the experiment in an order of magnitude, bounds will improve roughly in one order of magnitude. Recently, light-by-light scattering has been measured by ATLAS at $\sqrt{s}\sim\mathcal{O}$(TeV) in LHC ultraperipheric collisions involving pairs of Pb ions \cite{Aaboud:2017bwk}. After an involved analysis that deals with the complexity behind the dirtiness of the measurements at LHC, these data allow to set a lower bound to the mass scale of Born-Infeld electrodynamics through its lowest order corrections to Maxwell electrodynamics \cite{Ellis:2017edi}. Though we cannot use this analysis for generic $Q$-induced interactions, we can take advantage of the results in section \ref{sec:EiBIEMMapping}, where we were able to go beyond the perturbative terms in \eqref{metricstressenergy} and obtain the full form of the Einstein frame matter Lagrangian of EiBI gravity coupled to Maxwell electrodynamics, which turns out to be BI electrodynamics if we identify the heavy scale $\mq$ with the $\beta$ parameter of the BI action\footnote{Do not confuse it with the $\beta$ parameter controlling $Q$-induced interactions} as written in \cite{Ellis:2017edi} as $(2\beta^2)^{1/4}=\mq$. Therefore, from their bounds $\beta\gtrsim 10^4$ GeV$^2$, we obtain the bound 

\begin{align}\label{constraint}
{\mq^{EiBI}}\gtrsim 120 \text{ GeV}
\end{align}

\begin{figure}
\centering
\includegraphics[scale=0.8]{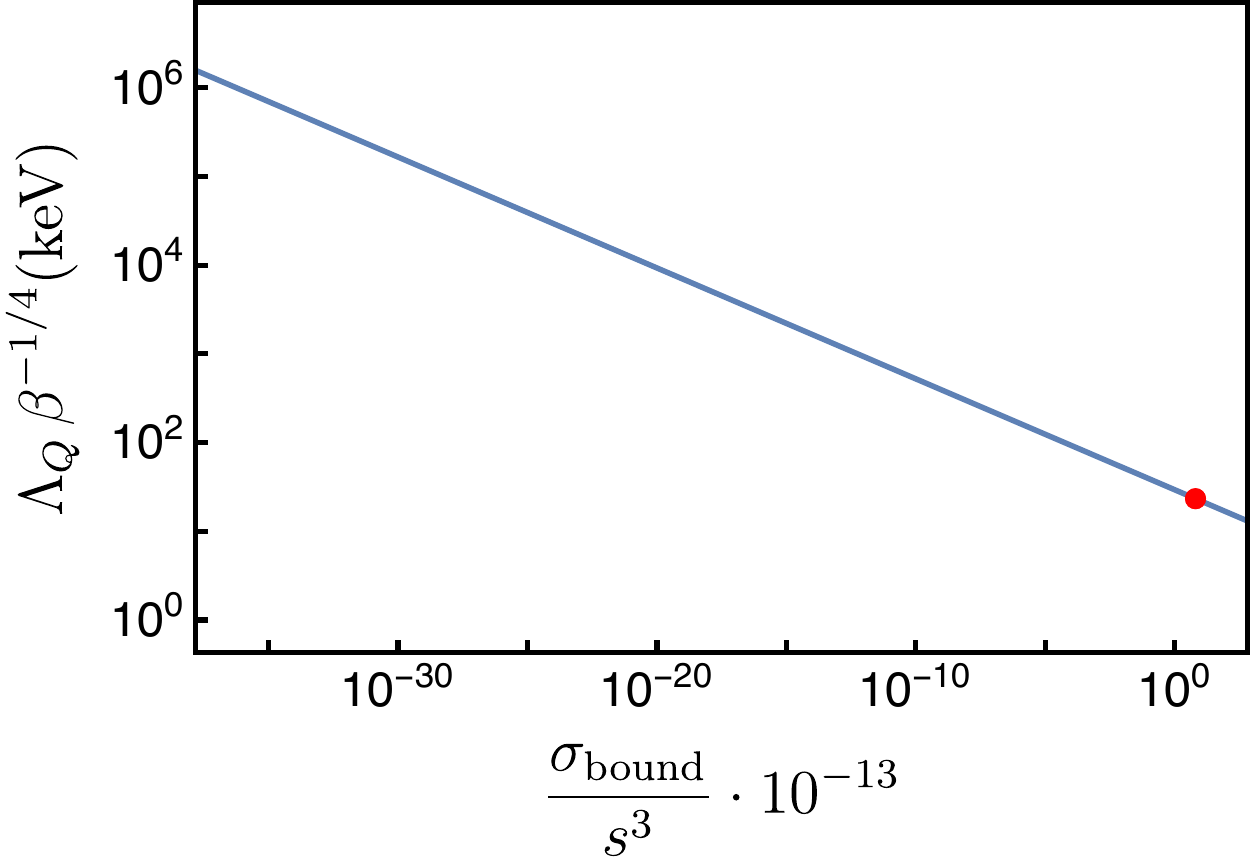}
\caption{Expected bounds on $\Lambda_Q$ for different values of the ratio $\frac{\sigma_{\text{bound}}}{s^3}$ in logarithmic scale. Our bound is denoted by a red point.}
\label{fig:photonsprospects}
\end{figure}

As a take-home message from this chapter, one should bear in mind that metric-affine theories with higher-order corrections or nonminimal couplings to matter featuring the symmetrised Ricci tensor $R_{(\mu\nu)}$ lead to a particular kind of corrections in the matter sector which we dubbed as $Q$-induced interactions as their presence is related to the form of a piece of the nonmetricity tensor $\nabla g$ of this theories. For the case without higher-order curvature corrections but with linear nonminimal couplings of the matter through $R_{(\mu\nu)}$, the corresponding $Q$-induced interactions couple each of the nonminimally coupled fields with all the matter fields in the theory (including itself). If there are any terms that are nonlinear in $R_{(\mu\nu)}$, either in the matter or in the gravitational sectors, then the corresponding $Q$-induced interactions couple all the matter fields to the stress-energy tensor, so that there appear couplings between all the degrees of freedom in the theory perturbatively in inverse powers of the heavy mass scale $\mg$ or $\mq$ that control the nonlinear terms. Within RBG theories, the interactions are safe, and the theory is a valid effective theory at energies below the mass scale $\mq$. For other more general theories, the higher order corrections generally excite new degrees of freedom. If there are nonlinear $R_{(\mu\nu)}$ terms in the action, these new degrees of freedom will couple, not only to the other fields, but generically to higher-order derivatives of the metric $q^{\mu\nu}$, potentially introducing Ostrogradskian instabilities in the theory which will destroy the validity of the EFT unless these degrees of freedom are sufficiently above the cutoff scale. Thus, it appears that, if dealing with generic metric-affine theories of gravity, one should either use only this covariant object from all the possible covariant geometric objects to build the action of the theory or be extremely careful in the way that the $R_{(\mu\nu)}$ terms couple the new degrees of freedom, as this could yield pathological couplings. in line with the results in chapter \ref{sec:UnstableDOF}

%=========================================================
\clearemptydoublepage
%
%
% File: chap01.tex
% Author: Victor F. Brena-Medina
% Description: Introduction chapter where the biology goes.
%
\let\textcircled=\pgftextcircled

\part{Funhouse}

\markboth{Part III}{}

{\noindent\Large\textbf{Part III - Outline}}
\vspace{0.5cm}

This is the last part of the thesis. The name is just a pun which reminds us that we do science because we have fun with it, just as The Stooges did with their music. However, this should not be interpreted as being less rigorous material. Here we develop some interesting ideas that sprouted in more relaxed environments, such as interesting conversations with friends (and collaborators, or a linear combination of both), reflections about some peculiar issues, or speculative ideas that, in one or the another way, have led to some curious and interesting results that could or could not lead to deeper insights on the properties of gravitation. Thus, in contrast with the two previous parts, the reader should expect to find a miscellanea of works which need not be in close connection to each other, but where interesting results have been achieved.

\chapter{Spontaneous Lorentz symmetry breaking in the metric-affine framework}\label{sec:SLSB}

The consistent inclusion of Lorentz symmetry breaking in a curved space is currently an open problem, though it is motivated by several approaches to find the UV completion of GR \cite{Colladay:1996iz,Gambini:1998it,Amelino-Camelia:2001dbf,Alfaro:2004aa,Bojowald:2004bb,Calcagni:2016zqv}. The usual way of including explicit Lorentz symmetry breaking by introducing a constant privileged direction cannot be generalised in a straightforward manner in presence of a nontrivial gravitational background while respecting diffeomorphism covariance. This is due to the fact that the partial derivative operator is tied to a given coordinate system, and it is not covariant when acting on generic tensor fields (see chapter {\ref{sec:DifferentialGeometry}}). Covariant differential operators that could be used to implement the constant condition could be $\dif$ or $\nabla$. If the privileged direction is defined by a vector $b$, imposing the vanishing of $\dif b$ would lead to $b=\dif \phi$ for a given scalar $\phi$, which is not a Lorentz violating constraint. By imposing $\nabla b=0$ we end up with a constraint equation for the connection, or the metric if we do it in the metric formalism, which is not generally acceptable from the physical viewpoint.

Another possibility is thus to resort to a spontaneous breaking of Lorentz symmetry via a vacuum expectation value of some field which belongs to some representation of the Lorentz group which is not the trivial one. The easiest way to go is by considering a vector field with a potential that leads to the existence of stable nontrivial vacua so that the vacuum expectation value (VEV) of the field breaks Lorentz symmetry spontaneously. This was done in presence of gravity by Kostelecky in what is known as the Bumblebee model \cite{Kostelecky:2003fs}, where a coupling of the vector field to the Ricci tensor is considered. Several aspects of this model have been studied over the years, all within the Riemannian framework \cite{Bertolami:2005bh,Seifert:2009gi,Maluf:2014dpa,Santos:2014nxm,Casana:2017jkc,Oliveira:2018oha,Ovgun:2018ran,Ovgun:2018xys,Ding:2019mal,Kanzi:2019gtu,Chen:2020qyp,Li:2020wvn,Maluf:2020kgf,Maluf:2021lwh,Oliveira:2021abg,Assuncao:2019azw}. As well, experimental tests of Lorentz violating extensions of the Standard Model and GR have led to stringent constraints on Lorentz breaking parameters \cite{Bailey:2006fd,Kostelecky:2008ts,Kostelecky:2015dpa,Kostelecky:2016kfm}. As a first step to study spontaneous symmetry breaking of Lorentz symmetry within the metric-affine framework, in this chapter we will consider a metric-affine formulation of the Bumblebee model \cite{Delhom:2019wcm,Delhom:2020gfv}. 

As we will see, this fits perfectly into the framework of RBG theories already developed in chapter \ref{sec:RBGTheory}. Indeed, due to the fact that the Ricci tensor appears only linearly in the Lagrangian, this will be a particularly easy example of Ricci-Based theory with projective symmetry and nonminimal couplings between matter and geometry. As we will see, the resulting theory admits an exact formal solution for the independent connection which leads to the emergence of a nonmetricity tensor generated by the nonminimal Bumblebee coupling to the geometry, which in the Einstein frame will couple to the rest of the matter fields present in the theory due to $Q$-induced interactions. We will first analyse briefly the structure of the theory, building the corresponding Einstein frame.  In the weak gravitational field limit, and perturbatively in the nonminimal coupling constant, we will study the stability of the Bumblebee vacua. Then, assuming a generic vacuum for the Bumblebee field, which should be particularised to the stable one for practical applications, we study the resulting field equations for scalar and Dirac fields, focusing on the Lorentz violating parameters and their modified dispersion relations.

\section{The Metric-Affine Bumblebee model}

We will consider a metric-affine Bumblebee model where the only nonminimal coupling to the geometry occurs through the Bumblebee field. This model is described by a Lagrangian of the form 
\begin{eqnarray} 
\nonumber\mathcal{L}=R+\frac{2\xi}{{\mq^2}} B^{\mu} B^{\nu} R_{\mu\nu}+ \frac{2}{{\mpl^2}}\lrsq{-\frac{1}{4}B^{\mu\nu}B_{\mu\nu}-V(B^{\mu}B_{\mu}\pm b^2) +\mathcal{L}^{\textrm{MC}}_{\textrm{m}}\big(g_{\mu\nu},\Psi_i)}.
\label{Bumblebee}
\end{eqnarray}
Here $B_{\mu\nu}=(\mbox{d}B)_{\mu\nu}$, and  we have written the Bumblebee piece of the action separated from the minimally coupled piece of the matter sector $\mathcal{L}^{\textrm{MC}}_{\textrm{m}}$, and we have neglected the fermion couplings to torsion so that $\mathcal{L}^{\textrm{MC}}_{\textrm{m}}$ does not depend on the connection and the effects of the Bumblebee nonminimal coupling appear more transparent\footnote{Indeed, adding the fermionic minimal coupling to torsion only adds a linear coupling to the connection in the spinorial sector of $\mathcal{L}^{\textrm{MC}}_{\textrm{m}}$. This would lead to a Planck-scale suppresed four-fermion contact interaction which we will neglect for the same reasons as in chapter \ref{sec:ObservableTraces}.\label{foot1Bumb}}.  The potential of the Bumblebee field $B_\mu$ guarantees that there are nontrivial Bumblebee vacua which spontaneously break Lorentz symmetry by introducing a privileged direction. 

The above action is within the class described by \eqref{eq:GeneralRBGAction} with a nonminimal coupling of the matter sector to the geometry through the Bumblebee field controlled by the dimensionless coefficient $\xi$. Following the reasoning in chapter \ref{sec:RBGTheory}, we find that the connection field equations of \eqref{Bumblebee} are an algebraic constraint, so that the connection is an auxiliary field that can be algebraically solved as the canonical connection of the Einstein frame metric
\begin{align}
 \sqrt{-q}q^{\mu\nu}=g^{\mu\nu}+\frac{2\xi}{{\mq^2}} B^\mu B^\nu.\label{eq:AuxMetricBumb}
\end{align}
This relation allows us to fin  $\beta_g^2=g^{\mu\nu}B_{\mu}B_{\nu}$, in terms of $\beta_q^2=q^{\mu\nu}B_{\mu}B_{\nu}$ as
\beq\label{eq:SolutionSquaredBumb}
\beta^2_q=\beta^2_g\sqrt{1+\frac{2\xi}{{\mq^2}}\beta_g}
\eeq
 which can be solved algebraically to find $\beta_g^2(\beta_q^2)$, leading to
 the expression of $g_{\mu\nu}$ in terms of $q_{\mu\nu}$ and the Bumblebee field as
\begin{equation}\label{eq:gmn_0}
\begin{split}
g_{\mu\nu}=\frac{1}{\sqrt{1+\frac{2\xi}{{\mq^2}} \bar\beta^2}}q_{\mu\nu}+\frac{2\xi}{{\mq^2}}\frac{1}{1+\frac{2\xi}{{\mq^2}} \bar\beta^2}B_{\mu}B_{\nu} \ ,\end{split}
\end{equation}
where we have called $\bar\beta^2$ to the solution $\beta_g^2(\beta_q^2)$ of \eqref{eq:SolutionSquaredBumb} and we can find $g^{\mu\nu}$ by inverting the above equation as a matrix equation. Proceeding as in section \ref{sec:EinsteinFrame}, we arrive to the Einstein frame form of the Bumblebee action \eqref{Bumblebee}, which reads
\begin{eqnarray} 
\nonumber\mathcal{L_{\textrm{EF}}}=R+\frac{2}{{\mpl^2}}\bar{\mathcal{L}}^\xi_{\textrm{m}}\big(q_{\mu\nu},B_\mu,\Psi_i\big)
\label{EinsteinBumblebee}
\end{eqnarray}
where any barred tensor indicates that its indices are raised and lowered with $q^{\mu\nu}$, so that the $g_{\mu\nu}$ metrics have been substituted by $q_{\mu\nu}$ through \eqref{eq:gmn_0} and its inverse. We have now included the bumblebee terms within $\bar{\mathcal{L}}^\xi_{\textrm{m}}$, which is denoted by the superscript $\xi$. In the above form, it is apparent that, due to the fact that it belongs to the (nonminimally coupled) RBG subclass of metric-affine theories, the Bumblebee model can be interpreted as GR coupled to a modified matter sector in which all the matter fields couple to the Bumblebee through $Q$-induced interactions with coupling strength $\xi/{\mq^2}$, including new self-interactions that modify the Bumblebee potential. 

According to the above form of the Bumblebee action, the metric $q_{\mu\nu}$ satisfies the Einstein equations coupled to a highly non-linear matter sector. Therefore, $q_{\mu\nu}$ will depart from the Minkowski metric only in regions where the Newtonian and post-Newtonian effects are expected to be relevant, \ie regions with a strong gravitational field. As a result, as it follows from (\ref{eq:gmn_0}), the metric $g_{\mu\nu}$ will not only describe the two propagating degrees of freedom corresponding to a massless spin-2 field described by $q_{\mu\nu}$, but it will also encode information on the local value of the Bumblebee field via a conformal factor and a disformal term proportional to $B_{\mu}B_{\nu}$ both of which source nonmetricity as can be seen by noting that the connection field equations are solved by the constraint  $\nabla_\alpha q^{\mu\nu}=0$. Thus, the nonmetricity tensor is nontrivial, controled by $\xi/{\mq^2}$, and entirely due to the covariant derivatives of the Bumblebee field. Since this field is expected to have a nontrivial VEV that spontaneously breaks Lorentz invariance, this is an example of a gravitationally generated nonmetricity tensor that can develop a VEV. In contrast, in RBGs with minimally coupled matter, the nonmetricity is associated to derivatives of the stress-energy tensor of the matter fields, which vanish in vacuum. Within the assumption that there is a constant nonmetricity background around the Earth, experimental constraints to all the possible effective couplings between fermions and photons and nonmetricity were derived from Lorentz violation searches within Earth laboratories in \cite{Foster:2016uui}. Since minimally coupled matter fields do not couple explicitly to nonmetricity, these constraints do not apply to our model. However, this model provides an explicit example of a gravitational model with a nontrivial nonmetricity VEV that the author knows of. Furthermore, note that constraints on Lorentz-violating couplings such as those in the Standard Model Extension \cite{Bailey:2006fd} could translate into constraints on the Bumblebee nonminimal coupling $\xi$. Below, we will facilitate the translation of these bounds by deriving the effective Klein-Gordon and Dirac equations in presence of a nontrivial Bumblebee VEV. 
%%%%%%%%%%%%%%%%%%%%%%
%%%%%%%%%%%%%%%%%%%%%%
%%%%%%%%%%%%%%%%%%%%%%
\section{Weak gravitational field limit}
%%%%%%%%%%%%%%%%%%%%%%
%%%%%%%%%%%%%%%%%%%%%%
%%%%%%%%%%%%%%%%%%%%%%
We can now explore the weak gravitational field limit of the model, so that the effects of the Bumblebee couplings to matter appear transparently. This limit would be applicable in \eg non-gravitational experiments on Earth's surface, where all Newtonian and post-Newtonian corrections to the Minkowski metric can be safely neglected. Given that $q_{\mu\nu}$ satisfies Einstein's equations, this amounts to the approximation $q_{\mu\nu}\approx\eta_{\mu\nu}$, in the same spirit as in chapter \ref{sec:ObservableTraces}. Considering also $\xi$ as a small coupling, we can write the following perturbative relations between both metrics
\begin{equation}
\begin{split}
g^{\mu\nu}=\eta^{\mu\nu}-\frac{2\xi}{{\mq^2}}\lr{B^{\mu}B^{\nu}-\frac{1}{2}B^{\alpha}B_{\alpha}\eta^{\mu\nu}}+\mathcal{O}\lr{\frac{\xi^2}{{\mq^4}}}
\label{efmet}
\end{split}
\end{equation}
where we now raise and lower indices with the Minkowski metric consistently with our approximations. This expression will allow us to write the Einstein frame effective Lagrangians and field equations for the Bumblebee field, as well as for scalar and Dirac matter fields. 
%%%%%%%%%%%%%%%%%%%%%%
\subsection{Effective dynamics for matter fields}
%%%%%%%%%%%%%%%%%%%%%%
Proceeding in similar lines as in section \ref{sec:NMLag}, we can find the effective Lagrangians describing scalar and Dirac fields perturbatively in $\xi/{\mq^2}$, which are given by
\begin{align}
&\resizebox{.9\hsize}{!}{$\mathcal{L}_{(0)}=-\frac{1}{2}\Phi(\Box_\eta+m^2)\Phi+\frac{\xi}{{\mq^2}}\Phi\Big[B^{\mu}B^{\nu}\partial_{\mu}\partial_{\nu}+ \big(B^\mu (\partial_\nu B^\nu)+B^\nu (\partial_\nu B^\mu)\big)\partial_\mu+\frac{m^2}{2} B^2\Big]\Phi+\mathcal{O}\lr{\frac{\xi^2}{{\mq^4}}}$},\label{31}\\
&\resizebox{.9\hsize}{!}{$\mathcal{L}_{(1/2)}=\bar\Psi(i \gamma^\mu \partial_\mu-m) \Psi-\frac{\xi}{{\mq^2}} \bar\Psi\Bigg[\frac{i}{2} B^{2} \gamma^\mu \partial_{\mu}+i B^\mu B^{\nu} \gamma_{\mu} \partial_{\nu}+\frac{i}{2}\Big(B_{\alpha}\left(\partial_{\mu} B^{\alpha}\right)+B^{\nu}\left(\partial_{\nu} B_{\mu}\right)+(\partial_\alpha B^\alpha) B_{\mu}\Big) \gamma^{\mu}-m B^{2}\Bigg] \Psi+\mathcal{O}\lr{\frac{\xi^2}{{\mq^4}}}$}.\label{SpinorLagPert}
\end{align}
where $Q$-induced interactions couple the matter field to the Bumblebee and we have neglected the spinors coupling to torsion as explained above in footnote \ref{foot1Bumb}.

As the Bumblebee field develops a nontrivial VEV, the $Q$-induced interactions with the Bumblebee in the above effective Lagrangians can carry coefficients for Lorentz violation. However, for these coefficients to be observable, they have to be nonvanishing for a stable nontrivial Bumblebee vacuum. Thus, before analysing these coefficients, we must study the vacuum structure of the bumblebee field in search for stable nontrivial VEVs. To that end, we have to specify first a particular form for the Bumblebee potential which provides spontaneous breaking of Lorentz symmetry. We will choose the usual Mexican hat potential, given by
\begin{equation}
V(B^\mu B_\mu\mp b^2)=\frac{\lambda}{4}\big(B^\mu B_\mu\mp b^2\big)^2,
\end{equation}
where $\lambda$ is a positive weak coupling. Here $b^2>0$ and the $\mp$ sign accounts for the possibility of having a spacelike or timelike Bumblebee VEV respectively.  With this choice of potential, the Einstein frame Lagrangian for the Bumblebee reads
\begin{equation}
\resizebox{.9\hsize}{!}{${\cal L}_{BEF}=-\frac{1}{4}B_{\mu\nu}B^{\mu\nu}+\frac{M^2}{2}B^2-\frac{\Lambda}{4}(B^2)^2+\frac{\xi}{{\mq^2}}\Big[B^{\mu\nu}B^\alpha{}_{\nu}B_\mu B_\alpha-\frac{1}{4}B_{\mu\nu}B^{\mu\nu}B^2-\frac{3}{4}\Lambda(B^2)^3\Big]+\mathcal{O}\lr{\frac{\xi^2}{{\mq^4}}}$}\label{BumbLagPert}
\end{equation} 
respectively, where the Bumblebee effective mass is given by
$M^2= \lambda b^2\lr{\pm 1+\frac{\xi}{2{\mq^2}}b^2}$ and $\Lambda\equiv \lambda\lr{1\pm \frac{4\xi}{\mq^2}b^2}$. Given that we are only interested in the qualitative details of the vacuum structure, and we are considering $\xi$ as a perturbative coupling, it will suffice to unveil the vacuum structure in the $\xi\rightarrow0$ limit. In other words, the perturbative modifications introduced by $\xi$ will not change the number and nature of the vacua, so that we can study the vacuum structure of the $\xi\rightarrow0$ case.

\subsection{Stability of the Bumblebee vacua}

The explicitly covariant form of the Bumblebee Lagrangian \eqref{BumbLagPert} is a constrained theory, as $B_0$ is non-dynamical due to the gauge invariant kinetic term (see section \ref{sec:GhostsInCurvatureBased}). Indeed, the corresponding field equations for $B_0$ in the $\xi\to0$ limit are
\begin{equation}
\lambda B_0\left(B_0^2-{\textbf{B}}^2-\frac{\mu}{\lambda}\right)-\partial_i \dot B^i=0,
\end{equation}
where ${\textbf{B}}^2=\delta^{ij}B_iB_j$ and $\mu=\pm M^2$. This is an algebraic equation for $B_0$ and, already for the $\xi\to0$ case, there will be different branches of the theory corresponding to one of the three solutions to the above constraint equation. Since we are doing the analysis perturbatively in $\xi$, we will just be interested in the vacuum structure of the $\xi\to0$ theory, and the stability properties of each of the vacua will not change perturbatively in $\xi$. The above constraint equation has three solutions that can be parametrised in the general form
\begin{equation}\label{B0solutions}
B_0^{(k)}=a_k\frac{\bar{B}}{3 A}+a_k^\ast A
\end{equation}
where
\begin{equation}
\bar{B}\equiv {\textbf{B}}^2+\frac{M^2}{2},\quad A\equiv\left(\frac{\partial_i \dot B^i}{2 \lambda }+\sqrt{\left(\frac{\partial_i \dot B^i}{2 \lambda }\right)^2-\left(\frac{\bar{B}}{3}\right)^3}\right)^{1/3}
\end{equation}
and the three solutions are given by
\begin{equation}
a_1=1,\qquad a_2=-\frac{1+{\textrm{i}}\sqrt{3}}{2} \quad\text{and}\quad a_3=a_2^\ast,
\end{equation}
so that $B_0^{(3)}=B_0^{(2)}{}^\ast$. The only branch admitting a Lorentz invariant vacuum, namely $B_0=0$ and $B_i=0$ is the $B_0^{(1)}$ branch, which admits it for both signs. To see that, take $B_i=0$ (then $\partial_i \dot B^i=0$ too). The values for three branches are
\begin{equation}\label{LIvacuum}
B_0^{(1)}|_{B_i\to 0}=0,\qquad B_0^{(2)}|_{B_i\to 0}\propto B_0^{(3)}|_{B_i\to 0}\propto \sqrt{\frac{|\mu|}{\lambda}},
\end{equation}
the proportionality factors are given by $\pm 1$ for the $+|M^2|$ case and $\pm \mathrm{i}$ for the $-|M^2|$ case. Note however that the fact that the $B^{(0)}$ solution for the constraint is compatible with the trivial configuration for $B_\mu$ does not necessarily imply that this configuration is a (Lorentz invariant) vacuum of the theory. This must be analysed by checking the vacuum structure corresponding to the Lagrangian \eqref{BumbLagPert}, which will allow us to see whether the different branches have Lorentz violating (meta)stable vacua where we can compute the quantum corrections associated to the different fields of the theory. To that end, one has two options: 1) compute the quadratic actions for perturbations to $B_\mu=\beta_\mu+\tilde{B}_\mu$ around any background, then integrate out the non-dynamical $\tilde{B}_0$ by solving the corresponding (linear) constraint equation, and study the eigenvalues of the kinetic matrix for the perturbations $\tilde{B}_i$ on top of each of the vacua, or 2) resort to the Hamiltonian formalism. Let us proceed with 2). To that end, after solving the constraint equation for $B_0$, the Hamiltonian to analyse reads
\begin{align}
H_k=&\frac{1}{2}\vec{E}^2+\frac{1}{4}B_{ij}B^{ij}-\left(a_k \frac{\bar B^2}{3A}+a_k^\ast A\right)\partial_i E^i+\frac{\lambda}{4}\Bigg[a_k \left(\frac{\bar B^2}{3A}\right)^4\nonumber\\
&+a_k^\ast A^4-2 a_k^\ast\left(\frac{\bar B^2}{3}\right)^3\frac{1}{A^2}-a_k\frac{2\bar B^2}{3A} A^2+\frac{(\bar B^2)^2}{3}\Bigg]
\end{align}
where we have dropped a constant term and $\vec{E}^2=\delta^{ij}E_iE_j$, where we have defined 
\begin{equation}
E^i\equiv\left.\frac{\partial\mathcal{L}_{BEF}}{\partial \dot B_i}\right|_{\xi\to 0}
\end{equation}
as the conjugate momenta to $B_i$. The constraint equation is now solved in terms of $B_i$ and the conjugate momenta as in \eqref{B0solutions} with $A$ now given by
\begin{equation}
A\equiv\left(\frac{\partial_i E^i}{2 \lambda }+\sqrt{\left(\frac{\partial_i E^i}{2 \lambda }\right)^2-\left(\frac{\bar{B}}{3}\right)^3}\right)^{1/3}.
\end{equation}

\begin{figure*}[h]
\centering
\includegraphics[width=\textwidth]{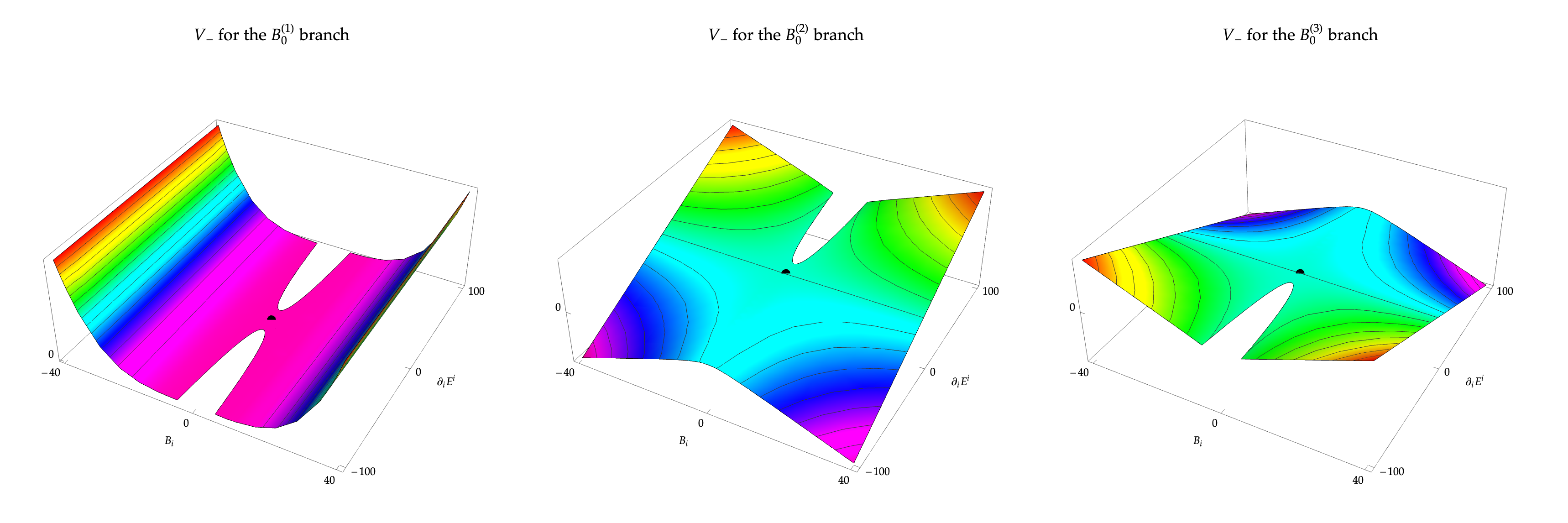}
\includegraphics[width=\textwidth]{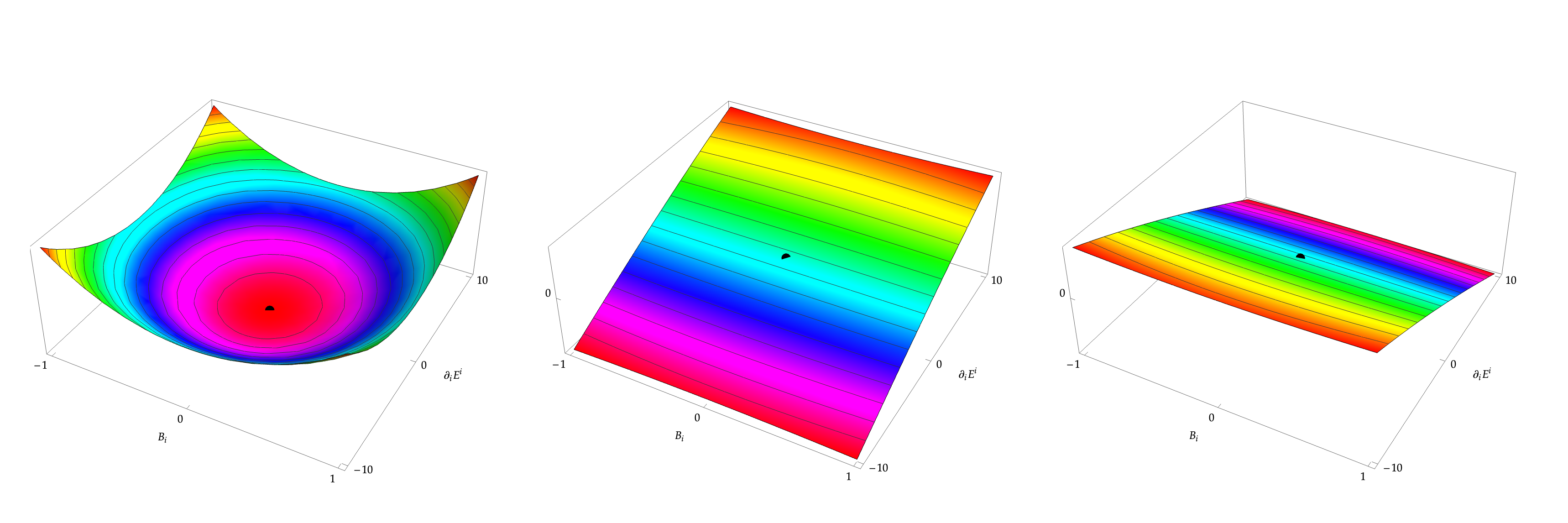}
\caption{\label{FigPotpos} In the upper row we plot the phase space potential for the Bumblebee field as a function $V(\textbf{B},\partial_i E^i)$ for the three $B_0$ branches for the $-|M^2|$ case. The black dot highlights the point $(0,0,V(0,0))$. We can see how only the $B_0^{(1)}$ branch has a critical point for the potential, namely a minimum, which corresponds to the trivial vacuum. This minimum is global within the $B_0^{(0)}$ branch, but it is local if the full solution space is considered, given that the potential is not bounded for below for the other branches $B_0^{(i\neq1)}$. Thus, it corresponds to a classically stable Lorentz invariant vacuum that can decay by quantum tunneling. In the lower row, the same potentials are plotted close to the trivial configuration $\textbf{B}=0$ and $\partial_i E^i=0$. We can appreciate clearly how the $B_0^{(1)}$ branch has a minimum at this point, while the two other branches do not have any critical point there.}
\end{figure*}

Now we must consider the two possible signs for $M^2$, which lead to different vacuum structures. Let us start first with the $-|M^2|$ case.  Here, we see that $\bar B$ is strictly positive for all values of $B_i$, and we can write the Hamiltonian perturbatively for each of the three branches as
\begin{equation}
\begin{split}
H&^{-}_k=\frac{1}{2}\vec{E}^2+\frac{1}{4}B_{ij}B^{ij}+\frac{\lambda}{4}(1-n_{k}^2)\left({\textbf{B}}^2+\frac{|M^2|}{\lambda}\right)^2\\
&+n_k\sqrt{{\textbf{B}}^2+\frac{|M^2|}{\lambda}}\partial_i E^i+\frac{(\partial_i E^i)^2}{2(1-2n_{k}^2)\lambda\left({\textbf{B}}^2+\frac{|M^2|}{\lambda}\right)}
\end{split}
\end{equation}
up to $\mathcal{O}\left((\partial_i E^i)^3\right)$ corrections, where
\begin{equation}
n_1=0,\qquad n_2=1 \quad\text{and}\quad n_3=-1
\end{equation}
characterise each branch. Now, note that the $B_0^{(1)}$ branch does not have a linear term in $\partial_i E^i$, and the quadratic term is positive definite, as are all the remaining ones. The configuration $B_i=0$ and $E^i=0$ is a classically stable vacuum with a value for the Hamiltonian of $M^4/4\lambda$ which, in this branch, is Lorentz-invariant since, $B^{(1)}_0$ is also vanishing for the trivial configuration due to its constraint equation. For the other two branches, this field configuration leads to a vanishing Hamiltonian, so that the $B_i=E^i=0$ vacuum of the $B_0^{(1)}$ branch is a local minimum and, though classically stable, it can decay through tunnelling processes turning it into a metastable vacuum. In figure \ref{FigPotpos} we plot the relevant piece of the full Hamiltonian for each of the branches. As expected, the $B_0^{(1)}$ branch has a classically stable vacuum at $B_i=E^i=0$. However, we see that the $B_0^{(2)}$ and $B_0^{(3)}$ branches do not have any extremal points due to the fact that its derivative in the $\partial_i E^i$ direction never vanishes.\\

\begin{figure*}[h]
\centering
\includegraphics[width=\textwidth]{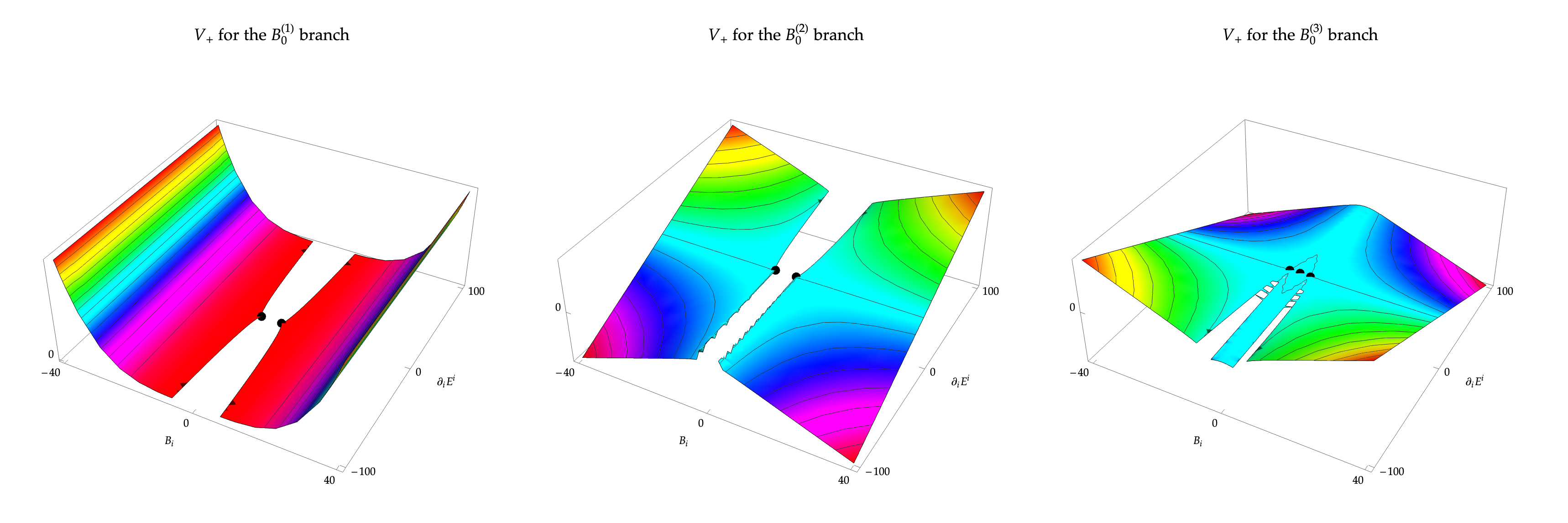}
 \includegraphics[width=\textwidth]{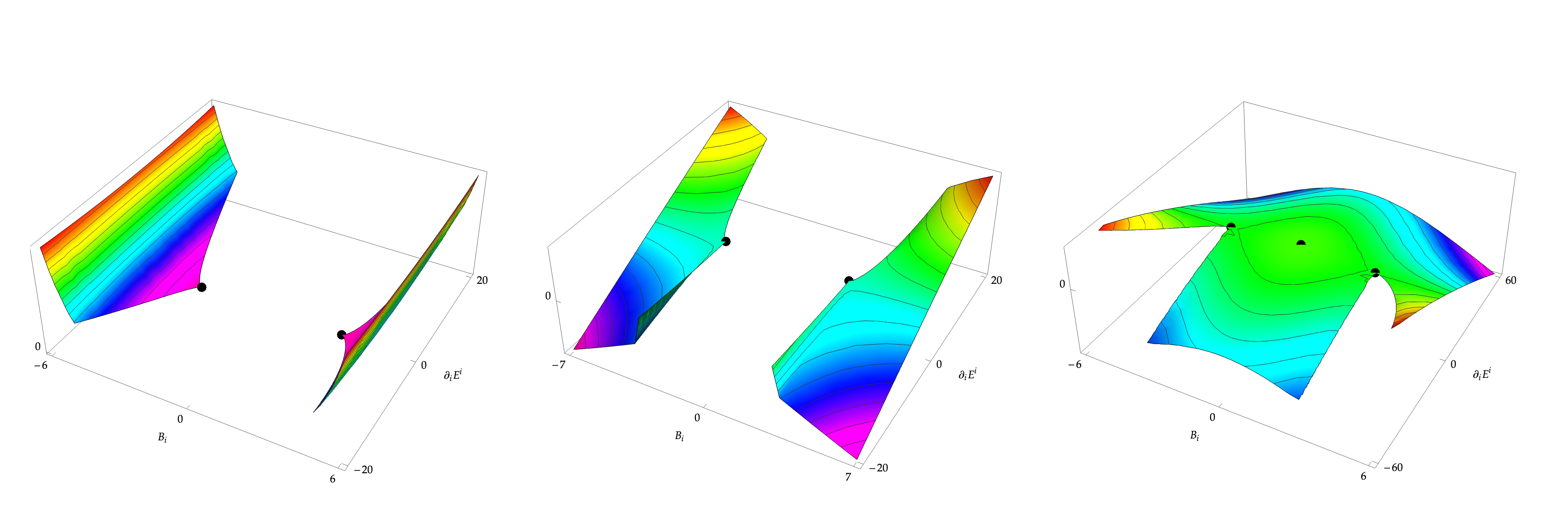}
\caption{\label{FigPotneg} In the upper row we plot the phase-space potential for the Bumblebee field as a function $V({\textbf{B}},\partial_i E^i)$ for the three $B_0$ branches in the case $\mu=-M^2<0$. The black dot highlights the point $(0,0,V(0,0))$ in the $B_0^{(1)}$ branch plot and the points $(\pm\sqrt{M^2/\lambda},0,0)$. We can see how in the $B_0^{(1)}$ branch the trivial Lorentz invariant vacuum is unstable, as expected for a tachyonic mass sign. As opposed to the $\mu>0$ case, now there is a pair of (degenerate) nontrivial vacua in the $B_0^{(3)}$ branch which spontaneously breaks Lorentz symmetry spatially, as the Bumblebee condenses to $(0,B_i)$ with ${\textbf{B}}=\sqrt{M^2/\lambda}$. This vacuum is a global minimum within the $B_0^{(3)}$ branch, though the potential is not bounded from below for the other branches $B_0^{(i\neq1)}$, turning it into a pair of metastable Lorentz violating vacua. The potential for the $B_0^{(2)}$ branch does not have any critical points. In the lower row, the same potentials are plotted close to the trivial configuration ${\textbf{B}}=0$ and $\partial_i E^i=0$ to allow for better appreciation of these features.}
\end{figure*}

For the $+|M^2|$ case, the expansion of the Hamiltonian around the trivial configuration reads
\begin{equation}
\begin{split}
H&^{+}_k=\frac{1}{2}\vec{E}^2+\frac{1}{4}B_{ij}B^{ij}+\frac{\lambda}{4}(1-n_{k}^2)\left({\textbf{B}}^2-\frac{|M^2|}{\lambda}\right)^2\\
&- n_k\,{\textrm{i} }\sqrt{-{\textbf{B}}^2+\frac{|M^2|}{\lambda}}\partial_i E^i+\frac{(\partial_i E^i)^2}{2(1-2n_{k}^2)\lambda\left({\textbf{B}}^2-\frac{|M^2|}{\lambda}\right)}
\end{split}
\end{equation}

which has an imaginary coefficient for the linear term in the $B^{(2)}_0$ and $B^{(3)}_0$ branches if $\textbf{B}^2<|M^2|/\lambda$. Thus we have to analyse the full Hamiltonian for (at least) these branches. Indeed, though the above expansion of the Hamiltonian is real around the trivial configuration for the $B^{(1)}_0$ branch, this is an artefact of the expansion, and higher order terms become complex around the trivial configuration except in the $\partial_iE^i=0$ direction, as can be seen in figures \ref{FigPotpos} and \ref{DerPotZero}. We thus have to proceed with the analysis of the full Hamiltonian for the $+|M^2|$ case. In figure \ref{FigPotpos} we plot the relevant piece of the Hamiltonian to analyse the vacuum structure of each of the branches. We see that, unlike expected from the perturbative expansion, the $B^{(3)}_0$ branch leads to a real Hamiltonian around the trivial configuration $B_i=E^i=0$, where it has a temporally broken unstable vacuum with $B_0=\sqrt{|M^2|/\lambda}$. The $B^{(2)}_0$ is complex around $B_i=E^i=0$, and its only vacua is a saddle point at $\textbf{B}^2=|M|/\sqrt{\lambda}$ and $\partial_i E^i=0$, which is also present in the $B^{(3)}_0$ branch. The behavior of the $B^{(1)}_0$ branch is surprising due to the fact that it becomes complex for
\begin{equation}
\left(\frac{\partial_i E^i}{2 \lambda}\right)^2-\left(\textbf{B}^2-\frac{|M^2|}{\lambda}\right)^3>0,
\end{equation}
which is solved by
\begin{equation}
|\textbf{B}|>\sqrt{3\left(\frac{\partial_i E^i}{2 \lambda}\right)^{2/3}+\frac{M^2}{\lambda }},
\end{equation}
except in the $\partial_i E^i=0$ direction, where it is constant and equal to zero from $\textbf{B}^2=|M^2|/\lambda$ up to $\textbf{B}^2=0$. This can be verified by plotting the directional derivative of the (relevant part of the) Hamiltonian in the $\textbf{B}$ direction for fixed values of $\partial_i E^i$, as done in figure \ref{DerPotZero}. In figures \ref{FigPotpos} and \ref{DerPotZero}, we see that for $\partial_i E^i=0$, there is a continuum and degenerated set of minima defined by $\textbf{B}^2\in(|M^2|/\lambda,0)$ which break Lorentz symmetry except for the trivial one. We can also see that for other branches the Hamiltonian is not bounded from below so, although these degenerated vacua are classically stable, thecan also decay through quantum tuneling, being thus metastable Lorentz-breaking vacua.

To sum up, we see that for the $-|M^2|$ case, the only classically stable vacuum is the trivial one. On the other hand, for the $+|M^2|$ case, there is a degenerate set of local minima which corresponds to metastable Lorentz breaking vacua characterised by $\partial_i E^i=0$ and $\textbf{B}^2\in(|M^2|/\lambda,0)$. As a remark, let us comment that, though everything seems to indicate that there is a continuum of Lorentz-breaking metastable vacua connected to the trivial one in the $+|M^2|$ case, in agreement with previous literature on the Bumblebee model (see e.g. \cite{Bluhm:2008yt}), the regions where the Hamiltonian becomes complex deserve a more detailed analysis to better understand their properties.\\

\begin{figure*}[h]
\centering
\includegraphics[width=0.45\textwidth]{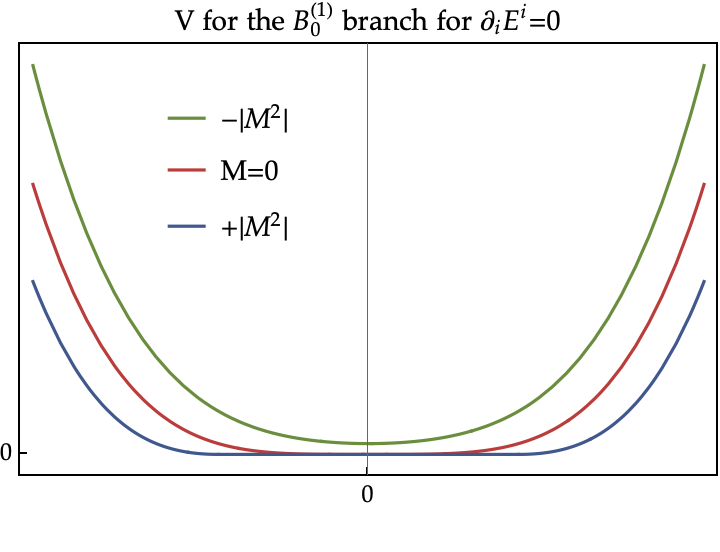}
\includegraphics[width=0.45\textwidth]{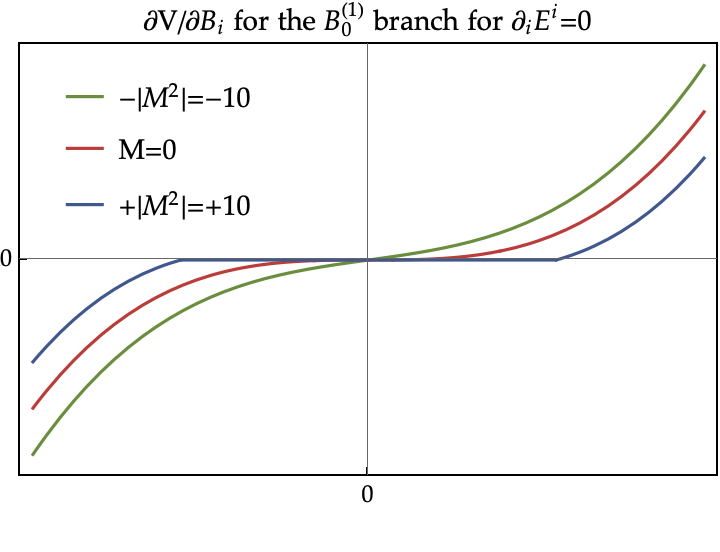}
\caption{\label{DerPotZero} Here we plot the phase space potential for a fixed value $\partial_i E^i=0$ for the $B^{(1)}_0$ branch and for the two choices in $\pm|M^2|$ as well as for a vanishing mass. There, it can be seen how, despite developing a complex part for any $\partial_i E^i>0$ if $\textbf{B}^2$ is small enough, the Hamiltonian is real for any value of $\textbf{B}^2$ if $\partial_i E^i$ is fixed to zero. We can also see how the directional derivative in the $\textbf{B}$ direction vanishes from $\textbf{B}^2=|M^2|/\lambda$ up to $\textbf{B}^2=0$, showing a continuous set of (classically stable) Lorentz breaking vacua characterised by from $\textbf{B}^2\in(M^2|/\lambda,0)$ and $\partial_i E^i=0$.}
\end{figure*}

\section{Lorentz-violating coefficients}

We now know that there is a classically stable nontrivial timelike vacuum for the bumblebee field in the $\mu<0$ case. We will first consider an approximately\footnote{In presence of a nontrivial gravitational field, this condition is not well defined, but we are neglecting those effects.} constant timelike VEV  $\langle B_\mu\rangle=b_\mu=(0,b_i)$ with $\delta^{ij}b_i b_j=b^2=M^2/\lambda$ in a weak gravitational field. Later we will generalise for arbitrary timelike VEVs. Generally, observables which couple to $b_{\mu}$ will be sensitive to the spontaneous breaking of Lorentz symmetry by the Bumblebee field. Since the present model displays non-minimal couplings between the Bumblebee field and the matter sources through the non-Riemannian part of the connection, there will arise several Lorentz violating (LV) coefficients in the Einstein frame effective matter sector. 
 
\subsubsection{Scalar field}
The Lagrangian for a scalar field propagating on top of the nontrivial Bumblebee background  takes the form
\begin{eqnarray}
\nonumber\mathcal{L}_{sc}&=&-\frac{1}{2}\Phi(\Box+m^2)\Phi+\frac{\xi}{{\mq^2}}\Phi\Big[(s^{\mu\nu}\partial_{\mu}\partial_{\nu})+\frac{1}{2}m^2 b^2\Big]\Phi+\mathcal{O}\lr{\frac{\xi^2}{{\mq^4}}},\label{ScalarLagBackground}
\end{eqnarray}
where $s^{\mu\nu}= b^{\mu}b^{\nu}$. The $\mathcal{O}(\xi)$ terms will typically induce LV coefficients through the VEV of the Bumblebee field. The $s^{\mu\nu}$ term constitutes a modification of the standard kinetic term which can be encoded in an effective metric for the scalar field of the form $g_{\textrm{eff}}^{\mu\nu}=\eta^{\mu\nu}-(2\xi/{\mq^2}) s^{\mu\nu}$ for a generic background. However, in the classically stable background, $s^{00}=s^{0i}=0$, which modifies only the coupling with the spacial derivatives. Hence, a ``wrong'' signature of the LV coefficient $s^{\mu\nu}$ could trigger Laplacian instabilities around strong enough Bumblebee backgrounds (see chapter \ref{sec:UnstableDOF}). Note, however, that in such case, the perturbative expansion would break down given that $(\xi/{\mq^2}) b^2$ would be $\mathcal{O}(1)$, and a full non-perturbative analysis would be required. The correction to the mass term in (\ref{31}) can also be encoded in an effective mass of the form $m_{\textrm{eff}}^2=m^2(1-(\xi/{\mq^2}) b^2)$ which could also trigger tachyonic-like instabilities for a space-like Bumblebee VEV (again non-perturbative effects could play a non-negligible role). 

In order to explore potential instabilities in more detail, let us analyse the dispersion relation of the classically stable vacuum, which reads
\begin{equation}\label{disprelscalarspacelike}
E^2=\vec{p}^2+m^2+\frac{2\xi}{\mq^2} \left(\frac{1}{2} m^2 b^2+(\vec{b}\cdot\vec{p})^2\right) +\mathcal{O}\lr{\frac{2\xi^2}{\mq^2}}.
\end{equation}
This dispersion relation is healthy for positive values of $\xi$. For negative values of $\xi$ a tachyonic-like instability as well as a Laplacian instability  (in directions which are non-orthogonal to $\vec{b}$) could potentially arise. In case that these instabilities appear, we should check their persistence in a full nonperturbative analysis of the theory.

\subsubsection{Dirac field}

Let us now turn our attention to the spin $1/2$ fields. To explore the physics of our interest in a more convenient way, we will work with the decomposition of LV coefficients that is more commonly used in the literature \cite{Colladay:1996iz,Colladay:1998fq}. To that end, let us write the weak-field spinor action on top of a nontrivial Bumblebee background as \eqref{SpinorLagPert} as 
 \begin{equation}
\mathcal{L}_{sp}=\bar{\Psi}\left(i\Gamma^{\mu}\partial_{\mu}- \hat M\right)\Psi,
\label{SpinorLagShort}
\end{equation}
where ${\Gamma}^{\mu}$ and ${M}$ are elements of the 16-dimensional Clifford algebra defined by the Dirac gamma matrices. We can thus expand them in the usual basis of this algebra as
\begin{equation}
\begin{split}
{\Gamma}^{\mu}&={e}^{\mu}I+(\delta^\mu{}_\alpha+{c}^{\mu}{}_{\alpha})\gamma^{\alpha}+{d}^{\mu}{}_{\alpha}\gamma_{5}\gamma^{\alpha}+i{f}^{\mu}\gamma_{5}+\frac{1}{2}{g}^{\mu}{}_{\lambda\alpha}\sigma^{\lambda\alpha},\\
{\hat M}&={m}_{\textrm{eff}}I+{a}_{\mu}\gamma^{\mu}+{k}_{\mu}\gamma^{\mu}\gamma_{5}+\frac{1}{2}l_{\mu\nu}\sigma^{\mu\nu},
\label{LVSpinorDecomp}
\end{split}
\end{equation} 
where ${c}^{\mu}{}_{\alpha}$, ${d}^{\mu}{}_{\alpha}$, ${e}^{\mu}$, ${f}^{\mu}$, ${g}^{\mu}{}_{\lambda\alpha}$, ${a}_{\mu}$, ${k}_{\mu}$ and $l_{\mu\nu}$ are LV coefficients. Comparing \eqref{SpinorLagPert} to \eqref{SpinorLagShort} and \eqref{LVSpinorDecomp}, we find the non-zero LV coefficients
\begin{eqnarray}
{c}^{\mu}{}_{\alpha}&=&\frac{\xi}{{2\mq^2}}\lr{\frac{M^2}{\lambda}\delta^{\mu}{}_\alpha+2b^\mu b_\alpha}+\mathcal{O}\lr{\frac{2\xi^2}{\mq^2}},\\
{m}_{\textrm{eff}}&=&m\left(1+\frac{\xi}{{\mq^2}} \frac{M^2}{\lambda}\right)+\mathcal{O}\lr{\frac{2\xi^2}{\mq^2}}.
\label{coeff}
\end{eqnarray}
We see that within the metric-affine Bumblebee model, the LV coefficients that appear provide a modification of the fermionic mass through $m_{\textrm{eff}}$ {and} a modification of the standard kinetic term through ${c}^{\mu}_{}{\alpha}$. In general, these will introduce modifications in the dispersion relation of spin $1/2$ fields. To find them let us write the Dirac equation derived from \eqref{SpinorLagShort} as
 \begin{equation}\label{SpinorEqShort}
\left(i\Gamma^{\mu}\partial_{\mu}- \hat M\right)\Psi=0\ , 
\end{equation}
 and multiplying on the left by $\left(i\Gamma^{\mu}\partial_{\mu}+M\right)$ we arrive at
\begin{align}
\begin{split}
\lrsq{\Gamma^\mu\Gamma^\nu\partial_{\mu}\partial_{\nu}-i[\hat M,\Gamma^\mu]\partial_{\mu}+\hat M^2} \Psi=0 \ .
\end{split}
\end{align} 
By using now the relations
\begin{align}
&\{\Gamma^\mu,\Gamma^\nu\}=2\eta^{\mu\nu}+\frac{2\xi}{{\mq^2}}\lr{\frac{M^2}{\lambda}\eta^{\mu\nu}-2b^\mu b^\nu}+\mathcal{O}\lr{\frac{2\xi^2}{\mq^2}},\\
&[ \hat M,\Gamma^\mu]=\mathcal{O}\lr{\frac{2\xi^2}{\mq^2}},\\
& \hat M^2=m^2\left(1+\frac{2\xi}{{\mq^2}} \frac{M^2}{\lambda}\right)+\mathcal{O}\lr{\frac{2\xi^2}{\mq^2}},
\end{align}
we find the following dispersion relation
\begin{align}
%\begin{split}
\nonumber&\lrsq{E^{2}\lr{1+\frac{2\xi}{{\mq^2}}\frac{M^2}{\lambda}}-\left(1+\frac{2\xi}{{\mq^2}} \frac{M^2}{\lambda}\right)\left(\vec{p}^{2}+m^{2}\right)-\frac{2\xi}{{\mq^2}}(\vec{p} \cdot \vec{b})^{2}} \Psi=\mathcal{O}\lr{\frac{2\xi^2}{\mq^2}} \ .
%\end{split}
\end{align}
For a (constant) spacelike VEV $b_{\mu}=[0,\vec{b}]$ we obtain
\begin{eqnarray}
E^{2}=\vec{p}^2+m^2+\frac{2\xi}{{\mq^2}} (\vec{p}\cdot\vec{b})^2+\mathcal{O}\lr{\frac{2\xi^2}{\mq^2}} \ .
\end{eqnarray}
For any sign of $\xi$, the $\cO(\xi)$ term could trigger Laplacian instabilities at the perturbative level, though a full analysis should be carried out in order to check that the instabilities are not an artefact of the perturbative expansion. See \cite{Kostelecky:2000mm} for a discussion on the typical energy scales at which these instabilities become relevant. Available constraints on LV parameters, as well as on the characteristic scale of instabilities if these are developed, could be used to constrain the allowed values of the nonminimal coupling sourcing LV in the matter sector.

\subsubsection{Arbitrarily varying $b_\mu$}
\label{app1}
If we drop the assumption of constant $b_\mu$, we find the effective Lagrangian for scalar fields
\begin{eqnarray}
\nonumber\mathcal{L}_{sc}=-\frac{1}{2}\Phi(\Box+m^2)\Phi+\frac{\xi}{{\mq^2}}\Phi\Big[(s^{\mu\nu}\partial_{\mu}\partial_{\nu})+ t^\mu \partial_\mu-\frac{M^2}{2\lambda}m^2 \Big]\Phi+\mathcal{O}\lr{\frac{2\xi^2}{\mq^2}},
\end{eqnarray}
with $t^\mu=b^\mu(\partial_\nu b^\nu)+b^\nu(\partial_\nu b^\mu)$. Note  the presence of the additional coefficient $t_\mu$ in relation to the standard case (it vanishes for constant Bumblebee VEVs). This coefficient introduces an imaginary term in the scalar dispersion relation. The modified scalar dispersion relations now looks
\begin{align}
\nonumber&E^2 -\vec{p}^2-\lrsq{1-\frac{\xi}{{\mq^2}}\frac{M^2}{\lambda}}m^2-\frac{2\xi}{{\mq^2}} (\vec{b}\cdot\vec{p})^2+\frac{2\xi}{{\mq^2}}(-\vec{b}+i\vec{t})\cdot\vec{p} =\mathcal{O}\lr{\frac{\xi^2}{\mq^2}}.\label{disprelscalar}
\end{align}
Note the existence of imaginary terms in the above dispersion relation, which only vanish if  one considers a frame where $t^{\mu}p_{\mu}=0$. This undesired property leads to complex eigenvalues of the Hamiltonian operator,  which  turns out to be non-Hermitian, and potential Laplacian instabilities.   

For spinor fields, the dispersion relation is modified in relation to the constant $b_{\mu}$ as 
\begin{align}
\big(\Gamma^\mu\Gamma^\nu\partial_{\mu}\partial_{\nu}+\Gamma^{\mu}(\partial_{\mu}\Gamma^{\nu})\partial_{\nu}+i\Gamma^{\mu}\partial_{\mu}\hat M-i[\hat M,\Gamma^\mu]\partial_{\mu}+M^2\big) \Psi=0.
\end{align} 
Now the relevant relations are
\begin{align}
&\{\Gamma^\mu,\Gamma^\nu\}=2\eta^{\mu\nu}+\frac{2\xi}{{\mq^2}}\lr{\frac{M^2}{\lambda}\eta^{\mu\nu}+2b^\mu b^\nu}+\mathcal{O}\lr{\frac{2\xi^2}{\mq^2}},\\
&[ M,\Gamma^\mu]=\frac{i}{2} a_\alpha \sigma^{\alpha\mu}+\mathcal{O}\lr{\frac{2\xi^2}{\mq^2}},\\
& M^2=m^2\left(1-\frac{2\xi}{{\mq^2}} b^2\right)+2m  a_\mu \gamma^\mu+\mathcal{O}.\lr{\frac{2\xi^2}{\mq^2}},
\end{align}
We find the following dispersion relation
\begin{eqnarray}
&E^{2}\lr{1+\frac{2\xi}{{\mq^2}} \frac{M^2}{\lambda}}+ \frac{2\xi}{{\mq^2}}E \lr{\frac{i}{2} a_i \sigma^{i 0}-i\gamma^{\mu}\gamma^{\beta}\partial_{\mu}c_{0\beta}}-\left(1-\frac{2\xi}{{\mq2}} \frac{M^2}{\lambda}\right)\left(\vec{p}^{2}+m^{2}\right)\\
&+\frac{2\xi}{{\mq^2}}\lr{-(\vec{p} \cdot \vec{b})^{2}+i\gamma^{\mu}\gamma^{\alpha}\partial_{\mu}a_{\alpha}+\frac{i}{2} a_\alpha \sigma^{\alpha i} p_i-2m a_\mu \gamma^\mu}=\nonumber \mathcal{O}\lr{\frac{2\xi^2}{\mq^2}}.
\end{eqnarray}
Although we can see in the form of the terms that Laplacian instabilities could arise, the analysis of the instabilities in this case appears more intricate than in the constant one. However, we conclude that a non-constant $b_{\mu}$ produces an effective Minkowskian theory which will in general present instability issues.

%%%%%%%%%%%%%%%%%%%%%%%%%
%%%%%%%%%%%%%%%%%%%%%%%%%
%%%%%%%%%%%%%%%%%%%%%%%%%

			%NEWCHAPTER%

%%%%%%%%%%%%%%%%%%%%%%%%%
%%%%%%%%%%%%%%%%%%%%%%%%%
%%%%%%%%%%%%%%%%%%%%%%%%%

\chapter{A scale invariant notion of time in presence of arbitrary nonmetricity}\label{sec:GeneralisedTime}

\section{Introduction}

From a geometrical perspective, the torsion tensor measures the failure to close for infinitesimal loops built by parallel transport,  while the nonmetricity tensor measures how parallel transport modifies lengths and angles. Concretely, if we decompose the nonmetricity into its irreducible components, its Weyl component controls the change in length of a parallely transported vector. spacetimes where the nonmetricity is fully specified by its Weyl component are named Weyl spacetimes, honoring the first work where this kind of nonmetricity was taken into account by Weyl \cite{Weyl:1918ib}. This irreducible component of the nonmetricity tensor transforms as a gauge 1-form under scale transformations of the metric, {\it i.e.,} it is the gauge field associated to scale transformations (usually called dilaton field). This fact fostered the interest in Weyl geometries, since they provide a natural way of introducing scale transformations without changing the affine structure (which cannot be done in Riemannian geometries). However, though a Weyl-like nonmetricity is necessary for defining scale transformations that do not change the affine structure, the usual restriction on the nonmetricity to be of this form in conformal invariant theories is unnecessary, and this can be achieved with general nonmetricity \cite{Delhom:2019yeo}. In this case only the vectorial irreducible components of nonmetricity transform as a gauge 1-form, while the tensorial irreducible components transform trivially by a conformal factor\footnote{Concretely under a conformal transformation  $g\mapsto\tilde g=e^\phi g$ in 4 spacetime dimensions, the different irreducible components listed in \eqref{decompNM} transform as: $\tilde Q_{1\mu}=Q_{1\mu}+4 (d\phi)_\mu$ , $\tilde Q_{2\mu}=Q_{1\mu}+ (d\phi)_\mu$, $\tilde S_{\mu\al\be}=e^\phi S_{\mu\al\be}$ and $M_{\mu\al\be}=e^\phi M_{\mu\al\be}$}.

 Motivated by the discussion of Perlick  \cite{PerlickTime} about how to construct a suitable Weyl-invariant notion of proper time that reduces to the standard definition of proper time in the Riemannian limit, and following  \cite{Lobo:2018zrz} in its analysis of the physical role played by the Weyl 1-form, in this chapter we will be concerned about finding a suitable definition of proper time that respects scale invariance in the presence of arbitrary nonmetricity (or generalised Weyl invariance in the sense of \cite{Delhom:2019yeo}), and also with the physical consequences of having nontrivial nonmetricity if there were physical clocks which were sensitive to this notion of time. To that end we will generalise the parametrisation for generalised proper time (GPT) found in \cite{Avalos:2016unj} to the case of arbitrary nonmetricity and find the existence of a conformally invariant second clock effect related to Weyl component of the nonmetricity tensor. We will then discuss GPT in light of the  M\"arkze-Wheeler construction (see \eg \cite{MarzkeWheeler,Desloge1989-DESATD,Pauri:2000cr}), which allows to operationally define the notion of clock in GR, elaborating on what kind of matter fields shall one need to build a generalised clock by means of this construction. 

%%%%%%%%%%%%%%%%%%%%%%%%%%%%%%%%%%%%%%%%%%%%%%%%%%%%%%%%%%%%%%%%%%%%%%%%%%%%%%%%%%%%%%%%%%%%%%%%%%%%%%%%%%%%%%%%%%%%%%%%%%%%%%%%%%%%%%%%%%%%%%%%%%%%%%%%%%%%%%%%%%%%%%%%%%%%%%%%%%%%%%%%%%%%%%%%%%%%%%%%%%%%%%%%

%%%%%%%%%%%%%%%%%%%%%%%%%%%%%%%%%%%%%%%%%%%%%%%%%%%%%%%%%%%%%%%%%%%%%%%%%%%%%%%%%%%%%%%%%%%%%%%%%%%%%%%%%%%%%%%%%%%%%%%%%%%%%%%%%%%%%%%%%%%%%%%%%%%%%%%%%%%%%%%%%%%%%%%%%%%%%%%%%%%%%%%%%%%%%%%%%%%%%%%%%%%%%%%%

\section{Generalised proper time}\label{sec:GenTime}

The fact that the usual Riemannian proper time, which is defined as the arclength of timelike curves, is not invariant under scale (or Weyl) transformations might be uncomfortable for those who expect UV physics to be scale invariant. To solve this issue, Perlick found a canonical\footnote{Canonical here meaning that uses only the ingredients available in a Weyl structure, or Weyl spacetime.} Weyl-invariant notion of proper time in a Weyl spacetime which reduces to the standard Riemannian proper time in the appropriate limit \cite{PerlickTime}. We will call this Perlick time. Recently, it was shown \cite{Avalos:2016unj} that GPT coincides with the operational time given by Ehlers, Pirani and Schild in \cite{EPS}, where they deduced from an operational point of view, and under certain assumptions regarding freely falling trajectories, that the spacetime manifold could be described by a Weyl spacetime. In \cite{Avalos:2016unj} it was shown that the Perlick time interval between two events $\ga(t_0)$ and $\ga(t)$ belonging to a timelike curve $\ga: t\in I\subset\bbr \mapsto \ga(t)\in \cm$ is given by
\begin{align}\label{PerlickTimeEq} 
\Delta\tau(t)= \frac{\left.\frac{d\tau}{dt}\right|_{(t=t_0)}}{\sqrt{g\lr{\dot\ga(t_0),\dot\ga(t_0)}}}\int_{t_0}^t ds \sqrt{g\lr{\dot\ga(u),\dot\ga(u)}}e^{-\frac{1}{2}\int_{u_0}^u  du \;\om\lr{\dot\ga(u)}} \;,
\end{align}
where $\dot\ga=d\ga (t)/dt$, and in a Weyl space the nonmetricity tensor is given by $\na g=\om\otimes g$. Our aim is to generalise this notion giving a definition of proper time in scale-invariant spacetimes with a general form of the nonmetricity tensor $Q$, {\it i.e.}, in generalised Weyl spacetimes in the sense of \cite{Delhom:2019yeo}. To do so, let us give the definition of Perlick time presented in \cite{PerlickTime}.

A $\tau$-parametrised timelike curve $\ga: \tau\in I \mapsto \ga(\tau)\in M$ is a \textbf{generalised clock} if
\beq\label{DefGenTime}
g\lr{\dot\ga,\nabla_{\dot\ga}\dot\ga}=0 \quad \forall\tau\in I.
\eeq
The parameter $\tau$ parametrising a generalised clock is the \textbf{generalised proper time} (GPT) measured by the clock. It can be shown that every timelike path\footnote{Note that the causal character of a path does not depend on the parametrisation that one uses to describe it as a curve.} $\ga$ admits a parametrisation with GPT \cite{Avalos:2016unj}. In their argument the authors start with a timelike curve $\ga(t)$ and show that a reparametrisation $\mu : t \in I \mapsto \mu(t)=\tau\in I'$ satisfying

\beq\label{ConditionRepClock}
\frac{d^2\mu}{dt^2}-\frac{g\lr{\dot\ga(t),\nabla_{\dot\ga}\dot\ga}}{g\lr{\dot\ga(t),\dot\ga(t)}}\frac{d\mu}{dt}=0
\eeq
has the property that $\tilde{\ga}(\tau)=\ga\circ\mu^{-1}(\tau)$ is a generalised clock. As \eqref{ConditionRepClock} always has a unique solution for timelike paths, every observer can be a generalised clock. The proof outlined in \cite{Avalos:2016unj} is independent of the relation between the metric and affine structure of the spacetime, which allows us to use their result and follow the steps of \cite{Avalos:2016unj} to find a general solution for \eqref{ConditionRepClock} in a spacetime with arbitrary $Q$. In order to derive this solution, let us start from the identity
\beq\label{CompatibiltyCondition}
\lr{\na_X g}(Y,Z)= X\lr{g\lr{Y,Z}}-g\lr{\na_X Y,Z}-g\lr{Y,\na_XZ},
\eeq
where $X,\, Y,\, Z$ are three arbitrary vector fields. Notice that by definition of the nonmetricity tensor, $Q\lr{X,Y,Z}\equiv\lr{\na_X g}(Y,Z)$. Using \eqref{CompatibiltyCondition} with $X=Y=Z=\dot\ga(t)$ and dividing by the tangent vetor squared we find
\beq
\frac{g\lr{\dot\ga,\na_{\dot\ga}\dot\ga}}{g\lr{\dot\ga,\dot\ga}}=\frac{1}{2}\lr{\dot\ga\left[ \ln\lr{g\lr{\dot\ga,\dot\ga}}\right]-\frac{Q\lr{\dot\ga,\dot\ga,\dot\ga}}{g\lr{\dot\ga,\dot\ga}}}\:,
\eeq
which is analogous to Eq.(9) of \cite{Avalos:2016unj} after the substitution\footnote{Notice that this substitution is a particularization of a general nonmetricity tensor to a Weyl-type nonmetricity tensor.} $Q(\dot\ga,\dot\ga,\dot\ga)\mapsto \om\lr{\dot\ga}g\lr{\dot\ga,\dot\ga}$ and taking into account that their $d/dt$ is a derivative in the direction of the curve, so that on a scalar function it is the action of $\dot\gamma$ on that scalar function. Combining this equation with \eqref{ConditionRepClock} yields the following condition for a reparametrisation to lead to a GPT parametrisation in presence of arbitrary nonmetricity
\beq\label{cond_add}
\frac{d\mu}{dt}=\frac{d\mu(t_0)}{dt}\left[\frac{g(\dot\gamma(t),\dot\gamma(t))}{g(\dot\gamma(t_0),\dot\gamma(t_0))}\right]^{1/2}e^{-\frac{1}{2}\int_{t_0}^t ds \frac{Q(\dot\gamma(s),\dot\gamma(s),\dot\gamma(s))}{g(\dot\gamma(s),\dot\gamma(s))}}.
\eeq
Integrating this equation for $\mu=\tau$ leads to an operational expression for computing the GPT elapsed between two events $A=\ga(t_0)$ and $B=\ga(t)$ for the observer $\ga(t)$ in spacetimes with general nonmetricity. This expression reads
\beq\label{GeneralProperTimeEq}
\Delta\tau(t)=\frac{\left.\frac{d\tau}{dt}\right|_{t=t_0}}du{\sqrt{g\lr{\dot\ga(t_0),\dot\ga(t_0)}}}\int_{t_0}^t ds\sqrt{g\lr{\dot\ga(u),\dot\ga(u)}}e^{-\frac{1}{2}\int_{u_0}^u\frac{Q\lr{\dot\ga(s),\dot\ga(s),\dot\ga(s)}}{g\lr{\dot\ga(s),\dot\ga(s)}}},
\eeq
and reduces to the one found for Perlick time \eqref{PerlickTimeEq} if the nonmetricity tensor is specified to be of the Weyl kind. 

A desirable feature for a notion of time is additivity, namely, that the time passed in going from event $A$ to event $B$ through a path $\ga_{AB}$ added to the time passed in going from event $B$ to event $C$ through a path $\ga_{BC}$ is the same as the time passed in going from event $A$ to event $C$ through the path $\ga_{AB}+\ga_{BC}$ where the sum here must be understood as concatenation of paths. The proof of the additivity of GPT in Weyl spacetimes given in \cite{Avalos:2016unj} also works in presence of general nonmetricity, and therefore the additivity is guaranteed. Moreover, due to the fact that \eqref{GeneralProperTimeEq} reduces to \eqref{PerlickTimeEq} for Weyl nonmetricity and, as proven in \cite{Avalos:2016unj}, \eqref{PerlickTimeEq} has the correct Weyl-Integrable-spacetime (WIST) and Riemannian limits, GPT will also have the correct WIST and Riemannian limits, thus being a sensible generalisation of Riemannian proper time in presence of arbitrary nonmetricity.
\par
In post-Riemannian manifolds, the scale invariance of the affine structure implies that the nonmetricity tensor must transform in a particular way under scale transformations. In fact, one can verify that the simultaneous transformations
\begin{equation}\label{conf-trans}
\tilde{\boldsymbol{g}}=e^{\phi}\boldsymbol{g} \ \ \ \ \text{and} \ \ \ \ \tilde{Q}=e^{\phi}(Q+\boldsymbol{d}\phi\otimes \boldsymbol{g})
\end{equation}
leave invariant the affine connection (as scale transformations should), where $\phi$ is any arbitrary smooth scalar function \cite{Delhom:2019yeo}. Since conformal transformations do not modify the orthogonality conditions, from its definition (\ref{DefGenTime}), GPT is scale invariant independently of the affine structure. This can also be verified by using \eqref{conf-trans} on the operational expression \eqref{GeneralProperTimeEq} that allows to explicitly compute the GPT within a given timelike path. Hence GPT is a sensible conformal invariant notion of time not only in Weyl spacetimes, but also in spacetimes with arbitrary nonmetricity.

%%%%%%%%%%%%%%%%%%%%%%%%%%%%%%%%%%%%%%%%%%%%%%%%%%%%%%%%%%%%%%%%%%%%%%%%%%%%%%%%%%%%%%%%%%%%%%%%%%%%%%%%%%%%%%%%%%%%%%%%%%%%%%%%%%%%%%%%%%%%%%%%%%%%%%%%%%%%%%%%%%%%%%%%%%%%%%%%%%%%%%%%%%%%%%%%%%%%%%%%%%%%%%%%

\section{Relation between GPT and Ehlers-Pirani-Schild proper time}\label{sec:EPS}

In the framework introduced by Ehlers, Pirani and Schild (EPS) in \cite{EPS}, one of the key assumptions that lead to the conclusion that the universe should be a Weyl spacetime\footnote{Note that from the metric-affine point of view, a Riemannian spacetime is a particular instance of a Weyl space with $\om=0$. However, the canonical affine structure of a Riemannian spacetime is not conformal invariant.} was the compatibility between the projective structure defined by the trajectories of freely falling particles and the conformal structure defined by the trajectories of light rays. They also define a notion of proper time within this framework which is Weyl invariant and coincides with GPT in Weyl spaces \cite{Avalos:2016unj}. In other words, under the restriction to the nonmetricity tensor of being Weyl-like, the GPT boils down to EPS proper time. In the following, we will be concerned with finding the most general kind of nonmetricity such that equivalence between EPS and GPT holds. To that end, we will proceed by following the proof given in \cite{Avalos:2016unj} for the equivalence of EPS and Perlick clocks but leaving nonmetricity completely arbitrary.

By definition \cite{EPS}, a timelike curve $\ga(\tau)$ is an \textbf{EPS clock}, {\it i.e.}, it is parametrised by \textbf{EPS time}, if there exists a vector field $V_\ga$ parallel along $\ga(\tau)$ which satisfies 
\beq\label{EPSclockDef}
g(\dot\ga(\tau),\dot\ga(\tau))=g(V_\ga(\tau),V_\ga(\tau))
\eeq
along the curve. Differentiating this condition and using that $V_\ga(\tau)$ is parallelly transported along $\ga(\tau)$, namely  $\na_{\dot\ga}V_\ga(\tau)=0$, the following relation follows from \eqref{CompatibiltyCondition}
\begin{align}\label{ConditionEPSisPerlick}
2g\lr{\na_{\dot\ga}V_\ga,\dot\ga}=Q(\dot\ga,V_\ga,V_\ga)-Q(\dot\ga,\dot\ga,\dot\ga)
\end{align}
where al the quantities are evaluated at a point $\ga(\tau)$. This condition is valid for any EPS clock in presence of arbitrary nonmetricity. Hence, by definition of generalised clock \eqref{DefGenTime}, for any EPS clock to be also a generalised clock, the condition 
\beq
Q(\dot\ga,\dot\ga,\dot\ga)=Q(\dot\ga,V_\ga,V_\ga),
\eeq
must be satisfied along any timelike curve $\ga(\tau)$, where $V_\ga(\tau)$ is the vector field satisfying the EPS clock condition \eqref{EPSclockDef} along $\ga(\tau)$. 

Let us now try to answer the opposite question, namely, under which conditions any generalised clock is an EPS clock. By definition, a timelike curve $\ga(\tau)$ is a generalised clock if $\dot\ga(\tau)$ and $\na_{\dot\ga}\dot\ga$ are orthogonal along the curve. Define (locally) a parallel vector field $V_\ga$ along $\ga(\tau)$ as the unique solution to the the initial value problem
\beq\label{InitialValueProblem}
\na_{\dot\ga}V_\ga=0\qquad\text{with initial condition}\qquad V_\ga(\tau_0)=\dot\ga(\tau_0).
\eeq
Using parallelism of $V_\ga$ along $\ga(\tau)$ and the orthogonality between $\dot\ga(\tau)$ and $\na_{\dot\ga}\ga$, subtracting \eqref{CompatibiltyCondition} applied to $X=\dot\gamma$ and $Y,Z=V_\ga,\dot\ga$, one finds that the initial value problem
\begin{align}
\frac{d}{d\tau}\Big(g\lr{V_\ga,V_\ga}-g\lr{\dot\ga,\dot\ga}\Big)=Q(\dot\ga,V_\ga,V_\ga)-Q(\dot\ga,\dot\ga,\dot\ga)\quad\text{with }\quad V_\ga(\tau_0)=\dot\ga(\tau_0)
\end{align}
must be solved along $\ga(\tau)$. The initial condition guarantees the vanishing of both sizes at $\tau=\tau_0$. Thus, given the form of the above initial value problem, and the uniqueness of its solution, the initial condition will remain true along $\ga(\tau)$, and therefore any EPS clock $\ga(\tau)$ will also be a generalised clock, if and only if
\beq\label{ConditionPerlickisEPS}
Q(\dot\ga,V_\ga,V_\ga)=Q(\dot\ga,\dot\ga,\dot\ga)
\eeq
is satisfied along any timelike curve $\ga(\tau)$. Therefore, the condition \eqref{ConditionPerlickisEPS} is a necessary and sufficient condition for any generalised clock to be an EPS clock. 

The natural next step is to find out what is the most general kind of nonmetricity tensor that satisfies the above condition. To that end, let us proceed as follows. For every timelike curve $\ga(\tau)$, define a symmetric  (0,2) tensor by $q^{(\ga)}=\na_{\dot\ga} g$. With this definition, and given an atlas covering $\gamma$, the condition \eqref{ConditionPerlickisEPS} can be written as 
\begin{equation}\label{CondQEPS=Perlick}
q^{(\ga)}{}_{\al\be}\lr{V_{\ga}{}^\al V_{\ga}{}^\be-\dot\ga^\al\dot\ga^\be}=0,
\end{equation}
which has to be satisfied for all timelike paths for some parametrisation. The above condition gives only one algebraic constraint given by \eqref{ConditionPerlickisEPS} for 10 independent components of $q^{(\ga)}{}_{\al\be}$. Thus the system is indeterminate, which makes sense, given that we know that any Weyl spacetime satisfies this condition. However, let us try to solve the system step by step to confirm this. Since it is homogeneous, there is the trivial solution $q^{(\ga)}=0$ for all $\ga(\tau)$ which implies that the nonmetricity must vanish, \ie a Riemannian spacetime is a particular case of spacetimes where EPS and generalised clock coincide. There is another solution that can be found by looking at the definition of an EPS clock \eqref{EPSclockDef}. From that definition, it is  apparent that for any spacetime that satisfies ${q}^{(\ga)}=\phi_\ga g$ where $\phi_\ga$ is an arbitrary scalar function that depends on the particular path. If that is the case, at each point $p\in\cm$, we can define a 1-form $\omega\in\ctb$ such that $\omega_{p}[\dot\gamma_p]=\phi_\ga(p)$ for each timelike path through $p$, so that for any pair of vectors fields $A,B$ and timelike path parametrised by $\ga(\tau)$ we have
\beq
q^{(\gamma)}(A,B)=\omega[\dot\gamma]g(A,B), 
\eeq
which implies that $Q=\omega\otimes g$, \ie all Weyl spacetimes are a solution to the above system as expected.
In order to see whether there are any other solutions to the system, we can look at the above conditions when written in terms of the irreducible components of the nonmetricity tensor. In 4 spacetime dimensions, the nonmetricity tensor can be decomposed in its Lorentz-irreducible pieces as
\begin{align}\label{NMdecomposed}
\resizebox{0.9\hsize}{!}{$Q_{\mu\alpha\beta}=\frac{1}{18}(5Q_{1\mu} g_{\alpha\beta}-Q_{1\alpha} g_{\beta\mu}-Q_{1\beta} g_{\mu\alpha}-2Q_{2\mu} g_{\alpha\beta}+4Q_{2\alpha} g_{\beta\mu}+4Q_{2\beta} g_{\mu\alpha})+S_{\mu\alpha\beta}+M_{\mu\alpha\beta},$}
\end{align}
where the different objects are
\begin{align}
\begin{split}\label{decompNM}
&Q_{1\mu}\equiv  g^{\alpha\beta}Q_{\mu\alpha\beta} \ \ \  , \ \ \ Q_{2\mu}\equiv  g^{\alpha\beta}Q_{\alpha\mu\beta},\\
&\resizebox{0.9\hsize}{!}{$S_{\mu\alpha\beta}\equiv \frac{1}{3}(Q_{\mu\alpha\beta}+Q_{\alpha\beta\mu}+Q_{\beta\mu\alpha})-\frac{1}{18}(Q_{1\mu} g_{\alpha\beta}+Q_{1\alpha} g_{\beta\mu}+Q_{1\beta} g_{\mu\alpha})-\frac{1}{9}(Q_{2\mu} g_{\alpha\beta}+Q_{2\alpha} g_{\beta\mu}+Q_{2\beta} g_{\mu\alpha})$},\\
&\resizebox{0.9\hsize}{!}{$M_{\mu\alpha\beta}\equiv \frac{1}{3}(2Q_{\mu\alpha\beta}-Q_{\alpha\beta\mu}-Q_{\beta\mu\alpha})-\frac{1}{9}(2Q_{1\mu} g_{\alpha\beta}-Q_{1\alpha} g_{\beta\mu}-Q_{1\beta} g_{\mu\alpha})+\frac{1}{9}(2Q_{2\mu} g_{\alpha\beta}-Q_{2\alpha} g_{\beta\mu}-Q_{2\beta} g_{\mu\alpha})$}.
\end{split}
\end{align}

After some algebra, we can see that \eqref{CondQEPS=Perlick} leads to a relation between vectorial and tensorial components of nonmetricity that reads
\begin{equation}\label{condEPSPerlick2}
\dot\ga^\mu\lr{S_{\mu\al\be}+M_{\mu\al\be}+\frac{1}{9}g_{\mu(\al}\lr{Q_{1\be)}-4Q_{2\be)}}}=0.
\end{equation}
Given the tracelessness of both $M$ and $S$ in any pair of indices, taking the trace of this equation in the two free indices, we find that 
\beq
\dot\ga^\mu\lr{Q_{1\mu}-4Q_{2\mu}}=0
\eeq
has to be satisfied if \eqref{condEPSPerlick2} is satisfied. This last equation is satisfied for every timelike path only if $Q_1=4Q_2$. Plugging this into the general equation \eqref{condEPSPerlick2}, we find that $M=-S$ should hold for it to be true for every timelike path. Thus $Q_1=4Q_2$ and $M=-S$ must be satisfied by the most general kind of nonmetricity satisfying that any EPS clock is a generalised clock. Plugging these conditions into \eqref{NMdecomposed} we see that a nonmetricity tensor satisfying these conditions is
\begin{equation}
Q=Q_2\otimes g=\frac{1}{4}Q_1\otimes g,
\end{equation}
namely, the most general kind of spacetime in which any EPS clock is a generalised clock and vice versa. In fact, a conclusion of the EPS paper is that from its construction based on the compatibility of the projective and conformal structures defined by freely falling massive and massless particles, one is led univocally to a Weyl geometry \cite{EPS}, though some subtleties have been  addressed in \cite{Trautman2012,Matveev:2013fpl,Scholz:2019tif,Matveev:2020wif}. That is the reason why the EPS time is irrevocably connected to the geometrical definition of time that uses the metric and the connection introduced by Perlick, and if one wants to develop a notion of proper time in a spacetime where free particles follow autoparallels of an affine connection with nontrivial nonmetricity, GPT can be a good candidate from the theoretical point of view, since it suitably generalises Riemannian and Perlick times\footnote{Though recall that, in general, one cannot find a variational principle for matter fields leading to them following the autoparallel curves of an arbitrary connection, it can be found for some particular connections.}. Without having much hopes that any real observer could be sensitive to this notion of time, but driven by pure curiousity, let us explore an important physical effect that such an observer would feel from the use of this geometrical time. Namely, what is known as the second clock effect. 

%%%%%%%%%%%%%%%%%%%%%%%%%%%%%%%%%%%%%%%%%%%%%%%%%%%%%%%%%%%%%%%%%%%%%%%%%%%%%%%%%%%%%%%%%%%%%%%%%%%%%%%%%%%%%%%%%%%%%%%%%%%%%%%%%%%%%%%%%%%%%%%%%%%%%%%%%%%%%%%%%%%%%%%%%%%%%%%%%%%%%%%%%%%%%%%%%%%%%%%%%%%%%%%%

\section{GPT and the second clock effect}\label{sec:SecondClock}

\begin{figure}
\centering
\includegraphics[scale=0.2]{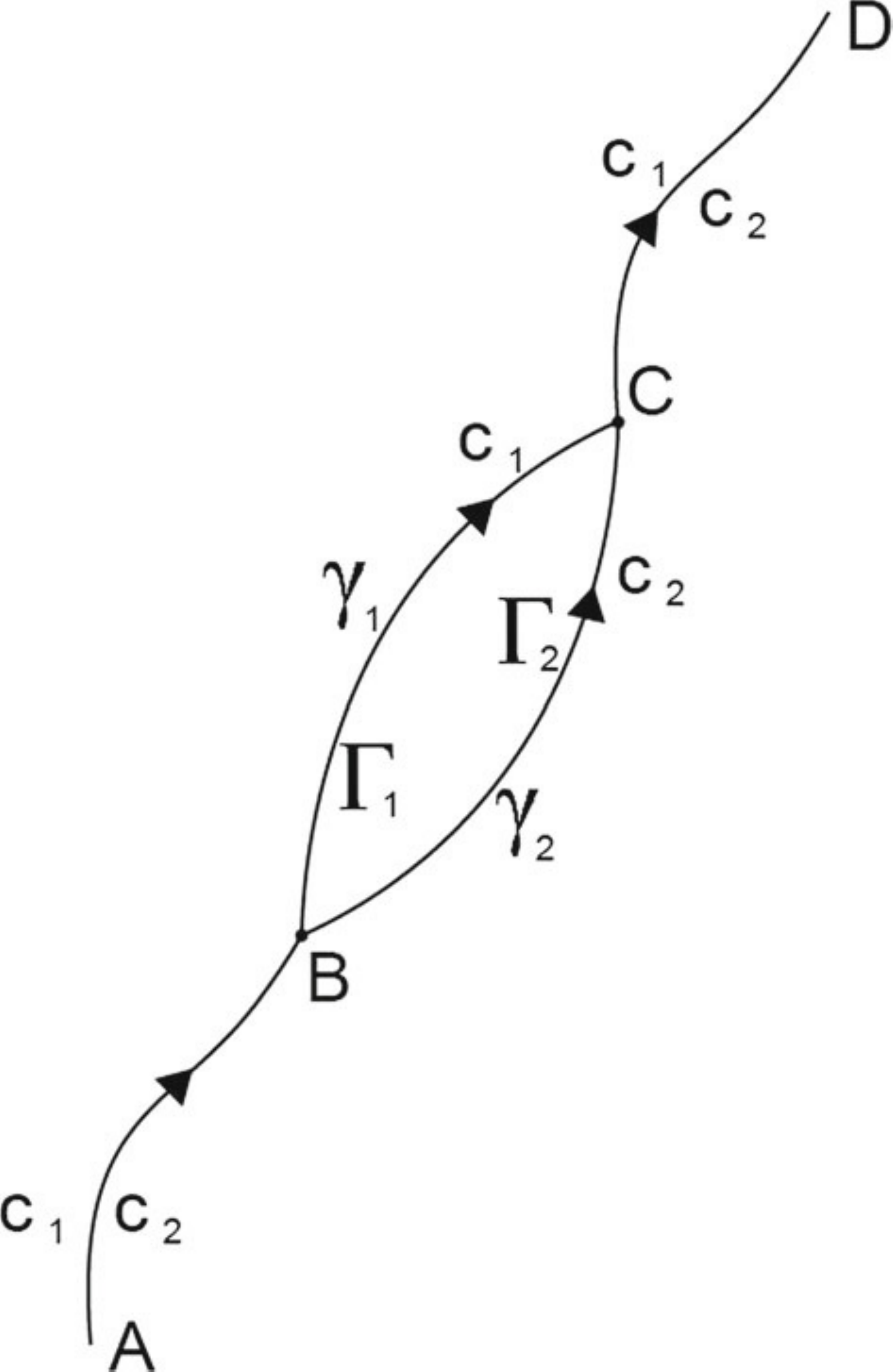}
\caption{Synchronized clocks $c_{1}$ and $c_{2}$ follow world lines
$\gamma_{1}$ and $\gamma_{2}$, which are coincident from point $A$ to $B$,
where they are separated to follow the parts $\Gamma_{1}$ and $\Gamma_{2}$ of
these lines until point $C$, where they are once again joined and continue
together until point $D$.}%
\label{fig:fig1Clock}%
\end{figure}

As a postscript to the original paper by Weyl \cite{Weyl:1918ib}, in which he introduced his geometrisation of electromagnetism and gravitation by means of a Weyl-like nonmetricity tensor, Einstein criticised the proposal by stating that the theory would suffer from an unpleasant effect due to the nonintegrability of lengths, namely, the clock rate of the clocks in the theory would depend on their past histories, which would have imprints in \eg the spectral lines of atoms which have never been observed. This effect was later coined as the \textbf{second clock effect} \cite{brown2000}. In order to illustrate it, consider two clocks $c_{1}$ and $c_{2}$ synchronised at event $A$ as in figure \ref{fig:fig1Clock}, which are transported following the same curve until event $B$, then separated and transported along two different curves, $\Gamma_{1}$ and $\Gamma_{2}$, until event $C$, where they are rejoined and transported to event $D$ following the same curve again. The proper time $\tau_i$ measured by the clock number $i$ after going from event $C$ to event $D$ is given by 
\begin{align}\label{time1}
\tau_i =\frac{\dot\tau_i(u_{iC})}{\sqrt{g\lr{\dot\gamma_i(u_{iC}),\dot\gamma_i(u_{iC})}}}\int_{u_{iC}}^{u_{iD}}du \sqrt{g\lr{\dot\gamma_i(u),\dot\gamma_i(u)}} \;e^{-\frac{1}{2}\int_{u_{iC}}^u ds\frac{Q\lr{\dot\gamma_i(s),\dot\gamma_i(s),\dot\gamma_i(s)}}{g\lr{\dot\gamma_i(s),\dot\gamma_i(s)}}},
\end{align}
The times from $A$ to $C$ can be computed using Eq.(\ref{cond_add}) as
\begin{align}
&\tau_i(u_{iC})=\dot\tau_i(u_A)\sqrt{\frac{g\lr{\dot\gamma_i(u_{iC}),\dot\gamma_i(u_{iC})}}{g\lr{\dot\gamma_i(u_A),\dot\gamma_i(u_A)}}}e^{-\frac{1}{2}\int_{u_A}^{u_{iC}}ds\frac{Q\lr{\dot\gamma_i(s),\dot\gamma_i(s),\dot\gamma_i(s))}}{g\lr{\dot\gamma_i(s),\dot\gamma_i(s)}}}
\end{align}
where $u_i$ is the parameter of the generalised clock $\gamma_i$. As in \cite{Avalos:2016unj}, after a reparametrisation from $u_2$ to $u_1$, and using synchronisation at event $A$, namely $\dot\ga_1(u_A)=\dot\ga_2(u_A)$, $\dot\ga_1(u_{1C})=\dot\ga_2(u_{2C})=\dot\ga(u_C)$, from the above equations we find
\begin{align}
&{\tau}_2(u_D)=\tau(u_D)\frac{\dot\tau_2(u_A)}{\dot\tau(u_A)}e^{\frac{1}{2}\lrsq{\int_{u_A}^{u_{1C}}\frac{Q(\dot\gamma_1(s),\dot\gamma_1(s),\dot\gamma_1(s))}{g(\dot\gamma_1(s),\dot\gamma_1(s))}ds-\int_{u_A}^{{u}_{2C}} \frac{Q(\dot\gamma_2(s),\dot\gamma_2(s),\dot\gamma_2(s))}{g(\dot\gamma_2(s),\dot\gamma_2(s))}ds}}.
\end{align}
Since both clocks have the same scale at the event $A$, {\it i.e.}, ${\dot\tau}_1(u_A)=\dot\tau_2(u_A)$, we conclude that a clock that measures GPT will measure a second clock effect in presence of a general nonmetricity tensor given by
\beq\label{SecondClockEffect}
\bar{\tau}=\tau \exp\left[\frac{1}{2}\int_{\Gamma_1-\Gamma_2}\frac{Q\lr{\dot\gamma(s),\dot\gamma(s),\dot\gamma(s)}}{g\lr{\dot\gamma(s),\dot\gamma(s)}}ds\right].
\eeq
This expression reduces to the result found in \cite{Avalos:2016unj} in Weyl spacetimes. and it is invariant under the action of conformal transformations \eqref{conf-trans}, $\int_{\Gamma_1-\Gamma_2} d\phi=0$ ( $\Gamma_1-\Gamma_2$ is a closed path). Therefore our construction describes a conformally invariant second clock effect. The observability of this event depends on whether clocks sensitive to this notion of time can be built with the available matter content in the universe. In \cite{Delhom:2020vpe}, assuming that a muon proper time is the GPT\footnote{Yes, Iarley and me were still young and bold enough to assume that, and we were surprisingly allowed by our supervisors!} allowed us to set bounds on the irreducible components of the nonmetricity tensor background at the Muon Storage Ring of $|Q|\lesssim 10^{-14}cm$.

\section{Is it possible to build a clock that measures GPT?}

As a final discussion, let us quickly elaborate on the possibility of measuring this invariant notion of proper time. Namely, let us digress on the question of whether one can actually construct a generalised clock with the content of our universe, or whether it is just a funny theoretical concept. To that end, recall that a physical clock as a clock that measures time as experienced by physical observers, namely massive bodies. By means of a generalised M\"arkze-Wheeler construction, one could in principle build clocks by using light rays and autoparallels of the connection, instead of Riemannian ones, as done by M\"arkze and Wheeler in \cite{MarzkeWheeler,Desloge1989-DESATD}, and then study whether this construction bears any relation with GPT. Notice that if it is to be related with GPT, scale invariance should play a central role on this generalised M\"arkze-Wheeler construction. Nonetheless, even if the construction of generalised clocks is possible in this way, that does not guarantee that it is possible to find such clocks in nature. Particularly, one would need to find massive particles that follow autoparallel curves of an affine connection with nontrivial nonmetricity, thus implying a violation of the Einstein Equivalence Principle, which is strongly constrained experimentally \cite{Will:2014kxa}. Furthermore, as we discussed in  section \ref{sec:geodesics}, it is generally not possible to find an action principle for matter fields leading to autoparallel curves of an arbitrary connection as the ones followed by freely falling bodies\footnote{Though this is possible for some particular kinds of connections.}. 
 General relativity is constructed in such a way  that free test particles follow Riemannian geodesics due to the conservation of the energy-momentum tensor, which in turn is required by diffeomorphism invariance \cite{Geroch:1975uq}. In \ref{sec:geodesics}, we argued that the characteristic curves of the solutions to the field equations of minimally coupled fields do not follow autoparallel paths of a connection with nonmetricity, but rather Riemannian geodesics or autoparallels of a connection with some particular form of torsion in the case of minimally coupled fermions \cite{HEHL1971225,Rumpf:1979vh,Audretsch:1981xn,Nomura:1991yx,Cembranos:2018ipn}. Hence, they are not well suited to build generalised clocks within the M\"arkze-Wheeler construction. Particularly, unless muons are seen to couple nontrivially to the nonmetricity, were this a feature of our spacetime, they would not be measuring GPT. However, some proposals for different couplings have arisen recently. For instance, the case of integrable Weyl spacetimes was addressed in \cite{Romero:2012hs}, in which a coupling that obeys the gauge invariance of the geometry and makes free particles follow Weyl geodesics in this theory was proposed. This issue was also addressed in \cite{deCesare:2016mml} for non-integrable Weyl geometry, were the authors concluded that free particles should follow Riemannian geodesics. In the context of $f(Q)$-gravity, it has been recently proposed  \cite{Harko:2018gxr} a coupling with matter where an extra force arises in the equation describing propagation of freely falling particles when written as a Riemannian geodesic equation. More recently, the case of a non-minimal coupling between matter and geometry in manifolds endowed with a nonmetric connection has gained some attention \cite{Gomes:2018sbf,SanomiyaProceedings,Sanomiya:2020svg}. Thus, being optimistic, the issue of whether GPT can be regarded as physical might depend on the particular model, since the possibility of constructing a generalised clock within every model depends on its particular geometry-matter coupling. However, a case-by-case analysis is required to ascertain whether these clocks exist in each particular theory.

%%%%%%%%%%%%%%%%%%%%%%%%%%%%%%%%%%%%%%%%%%%%%%%%%%%%%%%
%%%%%%%%%%%%%%%%%%%%%%%%%%%%%%%%%%%%%%%%%%%%%%%%%%%%%%%
%%%%%%%%%%%%%%%%%%%%%%%%%%%%%%%%%%%%%%%%%%%%%%%%%%%%%%%

\chapter{4D Einstein-Gauss-Bonnet theory is not well defined}\label{sec:4DEGB}

In a recent work \cite{Glavan:2019inb} it was claimed that there exists a theory of gravitation in four spacetime dimensions which fulfils all the assumptions of the Lovelock theorem \cite{Lovelock:1972vz} yet not its conclusions. The authors formulated the usual Einstein-Gauss-Bonnet (EGB) theory in an arbitrary dimension $\dimM$ with a coupling constant for the Gauss-Bonnet term rescaled by a $1/(\dimM-4)$ factor, as defined by the following action
\begin{equation}\label{action}
    \cS=\int \mathrm{d}^{\dimM}x \sqrt{|-g|}\left[ \frac{{\mpl^2}}{2}R+\frac{\alpha}{\dimM-4}\mathcal{G}-\CosmC\right].
\end{equation}
Here $\Lambda_{0}$ is a cosmological constant term, $R$ is the (metric) Ricci scalar, and $\mathcal{G}$ the (metric) Gauss-Bonnet (GB) term. As it is well known, the GB term is a topological invariant only in $\dimM=4$, but not in higher dimensions, thus generally yielding a nontrivial contribution to the field equations in arbitrary $\dimM>4$. In \cite{Glavan:2019inb} it is claimed that the contribution of the Gauss-Bonnet (GB) term to the equations of motion is always proportional to a $(\dimM-4)$ factor, which in principle compensates the divergence introduced in the coupling constant, thus allowing for a well defined $\dimM\to4$ \textit{limit} at the level of the field equations. It was argued that a nontrivial correction to General Relativity due to the GB term in \eqref{action} remains in $\dimM=4$. The proposal in \cite{Glavan:2019inb} is now known as 4-dimensional Einstein-Gauss-Bonnet theory (4DEGB).

The above action is one of the celebrated Lovelock actions in arbitrary $\dimM$. In \cite{Glavan:2019inb} it is claimed that all the assumptions of Lovelock theorem hold for this action after the $D\rightarrow4$ prescription is enforced, though it was also claimed that the resulting field equations do violate the conclusions of the Lovelock theorem. This was supposed to be accomplished by defining a 4-dimensional diffeomorphism-invariant theory satisfying the metricity condition and having second-order field equations which differ from those of General Relativity (GR). The authors of \cite{Glavan:2019inb} then proceeded to show the consequences of these modifications to GR in some scenarios with a high degree of symmetry. These results gained an astonishing amount of attention shortly after publication, with many works studying the properties of the solutions presented in \cite{Glavan:2019inb}, and other works criticising several aspects of the 4DEGB proposal \cite{Bonifacio:2020vbk,Hennigar:2020lsl,Mahapatra:2020rds,Ai:2020peo,Gurses:2020ofy} while, in some cases, relating them to other well defined theories \cite{Lu:2020iav,Kobayashi:2020wqy,Tian:2020nzb,Fernandes:2020nbq,Shu:2020cjw}. In this chapter, we will present several arguments that debunk many of the claims in \cite{Glavan:2019inb}, relating our results to some of the above works.

Let us sum up some of the main ideas of the chapter. It was claimed in \cite{Glavan:2019inb} that the contribution of the GB term  to the field equations (and not just its trace) is proportional to $(\dimM-4)$, and that this would imply the GB contribution to the field equations vanishes in four spacetime dimensions. The authors of \cite{Glavan:2019inb} then consider a coupling constant with a $1/(\dimM-4)$ factor that would \textit{regularise} the otherwise vanishing GB contribution, now yield a finite correction to the four dimensional field equations. We will show that, besides a term proportional to $(\dimM-4)$, the GB term contributes to the field equations with an additional part from which no power of $(\dimM-4)$ can be factorized, but which nonetheless vanishes identically in $\dimM=4$. 

Regarding tensor perturbations in D4EGB we will reproduce the results of \cite{Glavan:2019inb} for linear perturbations around a maximally symmetric background. This allows to find that, at the linear level and around maximally symmetric backgrounds, the proposal only propagates a massless graviton and that the corrections to GR provided by the \textit{regularised} GB term only enter through a global $\alpha$-dependent factor multiplying the linear perturbation equations in GR. Nonetheless we will see that the field equations describing second-order perturbations contain ill-defined terms even around a Minkowskian background. Indeed, we will argue that unless one is considering solutions with enough symmetry so as to force a specific combination of Weyl tensors to vanish in arbitrary dimensions, the term that is not proportional to $(\dimM-4)$ in the field equations renders the full D4EGB field equations ill-defined.

Finally, we will comment on the geometries presented in \cite{Glavan:2019inb} as the $\dimM\to4$ \textit{limit} of the spherically symmetric solutions for EGB theory in $\dimM\geq5$ found in \cite{Boulware:1985wk}. We will see that the claim made in \cite{Glavan:2019inb} that no particle can reach the central curvature singularity in a finite proper time within these geometries does not hold for freely-falling trajectories with vanishing orbital angular momentum. Furthermore, we will show that the \textit{regularised} D4EGB field equations are not well defined in spherically symmetric spacetimes unless the contribution which is not proportional to $(\dimM-4)$ is artificially stripped away from the field equations. Also, in the case that this term is removed, we will see that the spherically symmetric geometries presented in \cite{Glavan:2019inb} are not solutions of the remaining field equations in $\dimM=4$.

\section{The $\dimM\to4$ prescription}\label{sec:Dto4}

Let us first comment on whether the $\dimM\to4$ prescription taken in \cite{Glavan:2019inb} corresponds to a well-defined continuous limiting process in a topological sense. To that end, consider the $k$-th order Lovelock term in an arbitrary dimension $\dimM$
\begin{align}
S^{(k)} 
& = \int \dfR^{a_1 a_2}\wedge...\wedge\dfR^{a_{2k-1} a_{2k}}\wedge \star (\fr_{a_1}\wedge...\wedge \fr_{a_{2k}}) 
 \label{eq: Lovelock general}
\end{align}
where Greek indices refer to a coordinate basis and Latin indices to a frame in which the metric is Minkowskian (see chapter \ref{sec:DifferentialGeometry}). If we analyse the problem in differential form notation, when varying the action with respect to the dual frame $\df^a$ we find
\begin{equation}\label{var1}
\star\frac{\delta S^{(k)}}{\delta \df^a} = (\dimM-2k)(\dimM-2k-1)!\ J^{(k)}_{ac}\ \df^c\,,
\end{equation}
where $J^{(k)}_{ac}$ is a regular tensor built from combinations of the Riemann tensor that differ for each $k$.
 
The second factor in the above equation comes from the contraction of two Levi-Civita symbols. Therefore, it is of combinatorial nature, namely, it essentially has to do with the counting of the number of possible antisymmetric permutations of a bunch of indices. Notice that this counting process is not a continuous process in which the number of indices being counted (or equivalently the dimension) can take any value, but it ought to be an integer one. Indeed, for \eqref{var1} to be valid, $\dimM$ must be greater than $2k$ because a $(-1)!$ cannot arise from counting possible permutations. Since \eqref{var1} is not valid for $\dimM=2k$, it cannot be stated that the factor $(\dimM-2k)$ is the responsible for the vanishing of \eqref{var1} in $\dimM=2k$. The reason under its vanishing can actually be traced back to the properties of $2k$-forms in $2k$ dimensions. Indeed, by explicitly writing the Hodge star operator\footnote{Recall section \ref{sec:MetricStructure}} in \eqref{eq: Lovelock general}, in the critical dimension we obtain
\begin{equation}
\star_g (\fr_{a_1}\wedge...\wedge \fr_{a_{2k}})\big|_{\scriptsize \dimM=2k} = F \varepsilon_{a_1 ... a_{2k}}\,,
\end{equation}
where $\varepsilon_{a_1 ... a_{2k}}$ is the Levi-Civita symbol, $F$ is a non-zero constant that depends on $k$. As a consequence of this, and the well-known fact that the curvature factors in the action do not contribute (via spin connection) to the dynamics in Lovelock theories \cite{Mardones:1990qc}, the equations of motion for the soldering forms are identically satisfied. Observe that this is no longer true if $\dimM>2k$, since, in that case, the Hodge dual of $\cofr_{a_1}\wedge...\wedge \cofr_{a_{2k}}$ is not a $0$-form and gives a nontrivial contribution to the equation of motion of the soldering form.

It is clarifying as well to consider \eqref{var1} as a metric variation, {\it i.e.} avoiding the differential form notation and working directly with the metric components. We can rewrite the general $k$-th order Lovelock term \eqref{eq: Lovelock general} in an arbitrary dimension $\dimM\geq 2k$ as\beq
\cS^{(k)}=\tfrac{(2k)!}{2^k}\int R_{\nu_1\nu_2}{}^{\mu_1 \mu_2}...R_{\nu_{2k-1}\nu_{2k}}{}^{\mu_{2k-1}\mu_{2k}} \delta_{\mu_1}^{[\nu_1}...\delta_{\mu_{2k}}^{\nu_{2k}]} \sqrt{|g|}\mathrm{d}^\dimM x \,, 
\eeq
and its variation with respect to the metric is not proportional to $(\dimM-4)$, but rather of the form
\begin{equation}
\frac{1}{\sqrt{|g|}}\frac{\delta S^{(k)}}{\delta g^{\mu\nu}} = (\dimM-2k) A_{\mu\nu} + W_{\mu\nu}, \label{eq: AW decomposition}
\end{equation}
where no $(\dimM-2k)$ factor can be extracted from $W_{\mu\nu}$. For instance, the first-order Lovelock term (the Einstein-Hilbert action) leads to $A^\text{EH}_{\mu\nu}=0$ and $W^\text{EH}_{\mu\nu}=G_{\mu\nu}$, which vanishes by algebraic reasons in $\dimM=2$. Analogously, by decomposing the Riemann tensor into its irreducible pieces (see {\it e.g.} \cite{Ortin:2004ms}), the second-order Lovelock term, {\it i.e.} the Gauss-Bonnet term, leads to\footnote{The calculations have been checked with \href{http://www.xact.es/}{xAct }. There is a Mathematica notebook in the supplementary material of \cite{Arrechea:2020evj} where the calculations are explicitly made.}
\begin{align}
&\resizebox{.9\textwidth}{!}{$A^\text{GB}_{\mu\nu}  =  \frac{\dimM-3}{(\dimM-2)^2} \Big[\frac{2\dimM}{\dimM-1}R_{\mu\nu} R -\frac{4(\dimM-2)}{\dimM-3}R^{\rho\lambda}C_{\mu\rho\nu\lambda}-4R_{\mu}{}^\rho R_{\nu\rho}+2g_{\mu\nu}R_{\rho\lambda}R^{\rho\lambda}-\frac{\dimM+2}{2(\dimM-1)}g_{\mu\nu} R^2 \Big]\,,$} \label{eq: Amn GaussBonnet}  \\
&W^\text{GB}_{\mu\nu} =  2 \left[ C_{\mu}{}^{\rho\lambda\sigma}C_{\nu\rho\lambda\sigma}-\frac{1}{4} g_{\mu\nu} C_{\tau\rho\lambda\sigma}C^{\tau\rho\lambda\sigma}\right] \,,
\end{align}
where we have introduced the Weyl tensor $C_{\mu\nu\rho\lambda}$. Taking this into account, the field equations arrived at from \eqref{action} in arbitrary dimension are\footnote{Since the trace of $W^\text{GB}_{\mu\nu}$ is proportional to $(\dimM-4)$, the divergence disappears from the trace of the equation of motion. This does not occur for the traceless piece of the equation.}
\begin{equation}\label{fieldeqs}
G_{\mu\nu}+\frac{1}{{\mpl^2}}\CosmC g_{\mu\nu}+\frac{2\alpha}{{\mpl^2}}\left(A^\text{GB}_{\mu\nu}+\frac{W^\text{GB}_{\mu\nu}}{\dimM-4}\right)=0\,.
\end{equation}
The \textit{regularization} prescription given in \cite{Glavan:2019inb} consisted on evaluating $\dimM=4$ after calculating the equations of motion in arbitrary $\dimM$. If this is done, while the $A^\text{GB}_{\mu\nu}$ term indeed provides a finite nontrivial correction to the Einstein field equations if the coupling constant of the GB term is $\alpha/(\dimM-4)$, the $W^\text{GB}_{\mu\nu}$ term will be ill-defined in this case since, in general, $W^\text{GB}_{\mu\nu}$ does not go to zero asymptotically as $(\dimM-4)$. Indeed, the reason for $W^\text{GB}_{\mu\nu}$ to vanish in $\dimM=4$ is that the Riemann tensor loses independent components as one lowers the dimension and, in $\dimM=4$, this loss of components imply that $W^\text{GB}_{\mu\nu}$ necessarily vanishes by algebraic reasons, analogously to what happens to the Einstein tensor in $\dimM=2$. In other words, the reason for these expressions to be zero in certain dimensions is that they are \textit{algebraic} identities fulfilled by the curvatures of all possible metrics in the critical dimension, as opposed to \textit{analytic} identities at which one could arrive by a continuous limiting process given a suitable topology. A somewhat simpler example of the fact that the vanishing of the GB variation is due to algebraic reasons is provided by Galileon or interacting massive vector field theories. There, it can be seen that due to the Cayley-Hamilton theorem, the interaction Lagrangian of a given order $k$ identically vanishes for dimensions higher than the critical dimension associated to $k$ \cite{BeltranJimenez:2016rff}. 

The authors of \cite{Glavan:2019inb} appeal to an analogy between their method and the method of dimensional regularisation commonly employed in quantum field theory. The dimensional regularization method allows to extract the divergent and finite contributions from integrals that are divergent in $\dimM=4$ but non-divergent for higher $\dimM$. It consists on considering the analytic continuation of such integrals to the complex plane as a function of the complexified dimension $\dimM$, and then taking the limit $\dimM\to 4$ in a manner that allows to separate the divergent and finite contributions of the integrals. A key aspect that ensures the well-definiteness of dimensional regularization as an analytic continuation is that the regularised integrals are scalar functions which have no algebraic structure sensitive to the number of dimensions of the space they are defined in.\footnote{Typically the tensorial structures within the integrals are extracted from them by employing Lorentz-covariance arguments, and therefore the integrals to regularise are scalar functions.} Note however that this is not the case for the Gauss-Bonnet term, which has a nontrivial tensorial structure that is not well defined for non-integer dimensions. Thus, although the process of dimensional regularization can be rigorously defined by using the smooth $\dimM\to4$ limit of the appropriate analytic continuation of the scalar integrals, this fails to be a continuous limiting process when the quantities involved have a nontrivial algebraic structure, such as tensors or $p$-forms do. It would also be interesting to attempt to find a precise mathematical meaning to the {\it limiting} procedure in the presence of tensor fields which satisfy certain algebraic identities only in a particular number of dimensions. This could be done, for instance, by introducing a formal limit (see {\it e.g.} \cite{Mazur:2001aa}) and studying its properties, though it looks like a highly nontrivial task.

\section{Perturbations around maximally symmetric backgrounds}\label{sec:maxsim}

Despite the above considerations note that, even though the {\it regularisation} method proposed in \cite{Glavan:2019inb} will not work in general, it suffices for finding solutions that satisfy enough symmetries so as to render the $W^\text{GB}_{\mu\nu}$ identically {zero} in arbitrary dimension. Thus, by symmetry-reducing the action before enforcing $\dimM=4$, we can get rid of the problematic $W^\text{GB}_{\mu\nu}$ term and arrive to well-defined equations of motion. This is the case, for instance, of all conformally flat geometries, which have an identically vanishing Weyl tensor in $D\geq4$, thus {satisfying} the desired property that $W^\text{GB}_{\mu\nu}=0$ in $\dimM\geq4$ which makes the $D\to4$ {\it limit} of the symmetry-reduced D4EGB field equations \eqref{fieldeqs} well defined. Maximally symmetric geometries, or FLRW spacetimes, are conformally flat, and therefore the $D\rightarrow4$ prescription yields well defined field equations in such cases. Let us analyse the maximally symmetric solutions of \eqref{fieldeqs} studied in \cite{Glavan:2019inb}. In these geometries, the Riemann tensor is given by
\begin{equation}\label{eq: max symm}
R_{\mu\nu}{}^{\rho\sigma}=\frac{\KMaxSym}{{\mpl^2}(\dimM-1)}\left(\delta^{\rho}_{\mu} \delta^{\sigma}_{\nu}-\delta^{\sigma}_{\mu} \delta^{\rho}_{\nu}\right),
\end{equation}
and $W^\text{GB}_{\mu\nu}$ vanishes in arbitrary dimensions as explained above. In this case, the variation of the GB term is indeed proportional to $(\dimM-4)$ and, therefore, after this symmetry-reduction of the action \eqref{action}, the field equations \eqref{fieldeqs} read
\begin{equation}\label{fieldeqssym}
   G_{\mu\nu}+\frac{1}{{\mpl^2}}\CosmC g_{\mu\nu}+\frac{2\alpha}{{\mpl^2}}A^\text{GB}_{\mu\nu}=0 \,,
\end{equation}\label{regularisedEOM}
where $A^\text{GB}_{\mu\nu}$ provides a regular,  $\alpha-$dependent correction to GR. Although these properties will be shared by all conformally flat solutions, we should bear in mind that arbitrary perturbations around these backgrounds will be sensitive to the ill-defined contributions that come from the $W^\text{GB}_{\mu\nu}$ dependence of the full D4EGB field equations \eqref{fieldeqs}. Remarkably, the ill-defined corrections which enter the equations of motion through the $\alpha W^\text{GB}_{\mu\nu}/(\dimM-4)$ term do not contribute to linear order in perturbation theory around a maximally symmetric background which, presumably, is the reason why these ill-defined contributions were not noticed in \cite{Glavan:2019inb}, where only linear perturbations were considered. Nonetheless, as we will see, the divergent terms related to $W^\text{GB}_{\mu\nu}$ will enter the perturbations at second-order. To show this, let us consider a general perturbation around a maximally symmetric background by splitting the full metric as
\begin{equation}
    g_{\mu\nu}=\bar{g}_{\mu\nu}+\epsilon h_{\mu\nu}
\end{equation}
where $\bar{g}_{\mu\nu}$ is a maximally symmetric solution of \eqref{fieldeqs}. Therefore, the left hand side of \eqref{fieldeqs} can be written as a perturbative series in $\epsilon$ of the form
\begin{equation}
     E^{(0)}{}_{\mu\nu}+\epsilon E^{(1)}{}_{\mu\nu}+\epsilon^2 E^{(2)}{}_{\mu\nu}+\ldots\,,
\end{equation}
where $E^{(0)}{}_{\mu\nu}=0$ are the background field equations, $E^{(1)}{}_{\mu\nu}=0$ are the equations for linear perturbations, and so on. Using the zeroth-order equation, the linear perturbations in $\dimM$ dimensions and around a maximally symmetric background are described by\footnote{Although \eqref{linearpert} and the equations for linear perturbations in \cite{Glavan:2019inb} differ by the ordering of the covariant derivatives of the $\nabla_{\rho}\nabla_{\nu}h^{\mu\rho}$ term and the sign in the mass term, our equation \eqref{fieldeqs} coincides with those in {\it e.g.} \cite{Ortin:2004ms} for linearized perturbations around a maximally symmetric background. In any case the difference is not physically relevant, as can be seen by choosing a particular gauge.}
\begin{align}
\resizebox{.87\textwidth}{!}{$\nabla^{\rho}\nabla^{\mu}h_{\nu\rho}+\nabla_{\rho}\nabla_{\nu}h^{\mu\rho}-\nabla^{\rho}\nabla_{\rho} h_{\mu\nu}-\nabla^{\mu}\nabla_{\nu}h  +\delta^{\mu}{}_{\nu}(\nabla^{\sigma}\nabla_{\sigma}h -\nabla_{\rho}\nabla_{\sigma}h^{\rho\sigma})-\frac{\KMaxSym}{{\mpl^2}} (\delta^\mu_\nu h-2h^\mu{}_\nu)=0\,,$}\label{linearpert}
\end{align}
where $h\equiv h^\sigma{}_\sigma$, the indices are risen and lowered with $\bar g^{\mu\nu}$, and the covariant derivatives are those associated to the Levi-Civita connection of the background metric $\bar{g}_{\mu\nu}$. By inspection, we can see that this equation is regular in $\dimM=4$. Furthermore, as noted in \cite{Glavan:2019inb}, the equation governing linear perturbations \eqref{linearpert} are those ocuring in GR when linearised around an arbitrary background. Let us now go to quadratic order in the perturbations. For our purpose it will be sufficient to consider perturbations around a Minkowskian background. By using the zeroth- and first-order perturbation equations, and enforcing a vanishing background curvature $\KMaxSym=0$, we can write the second-order perturbation equations $E^{(2)}{}_{\mu\nu}=0$ as (see the supplementary material of \cite{Arrechea:2020evj})
\begin{align}\label{quadraticpert}
\resizebox{.85\textwidth}{!}{$[\text{GR terms of }\mathcal{O}(h^{2})]_{\mu\nu} +\frac{\alpha}{{\mpl}^{2}(\dimM-4)} \times\Big\{
 -2\nabla_{\gamma}\nabla_{\alpha}h_{\nu\beta}\nabla^{\gamma}\nabla^{\beta}h_{\mu}{}^{\alpha}
 +2\nabla_{\gamma}\nabla_{\beta}h_{\nu\alpha}\nabla^{\gamma}\nabla^{\beta}h_{\mu}{}^{\alpha}$}\nonumber \\
\resizebox{.85\textwidth}{!}{$+2\nabla^{\gamma}\nabla^{\beta}h_{\nu}{}^{\alpha}\nabla_{\mu}\nabla_{\alpha}h_{\beta\gamma} 
 +2\nabla^{\gamma}\nabla^{\beta}h_{\mu}{}^{\alpha}\nabla_{\nu}\nabla_{\alpha}h_{\beta\gamma}-2\nabla^{\gamma}\nabla^{\beta}h_{\mu}{}^{\alpha}\nabla_{\nu}\nabla_{\beta}h_{\alpha\gamma} 
 -2\nabla^{\gamma}\nabla^{\beta}h_{\nu}{}^{\alpha}\nabla_{\mu}\nabla_{\beta}h_{\alpha\gamma}$}\nonumber \\
\resizebox{.85\textwidth}{!}{$-2\nabla_{\mu}\nabla^{\gamma}h^{\alpha\beta}\nabla_{\nu}\nabla_{\beta}h_{\alpha\gamma}
 +2\nabla_{\mu}\nabla^{\gamma}h^{\alpha\beta}\nabla_{\nu}\nabla_{\gamma}h_{\alpha\beta}+g_{\mu\nu}\big(
 2\nabla_{\delta}\nabla_{\beta}h_{\alpha\gamma}\nabla^{\delta}\nabla^{\gamma}h^{\alpha\beta}
 -\nabla_{\delta}\nabla_{\gamma}h_{\alpha\beta}\nabla^{\delta}\nabla^{\gamma}h^{\alpha\beta}$}\nonumber \\
  \qquad\qquad -\nabla_{\beta}\nabla_{\alpha}h_{\gamma\delta}\nabla^{\delta}\nabla^{\gamma}h^{\alpha\beta}\big) \Big\}=0 \,.
\end{align}

In view of the above expression,\footnote{Note that we have kept the covariant derivatives in \eqref{quadraticpert} since our result is not restricted to a particular coordinate choice.} it is apparent that even around a flat background, the $W^{\text{GB}}_{\mu\nu}$ piece of \eqref{fieldeqs} contributes to the second-order perturbation equations with a term that is ill-defined in $\dimM=4$. These findings provide a clear example which serves to show that the D4EGB field equations \eqref{fieldeqs} are generally ill-defined. As a remark, let us point out that our results are somewhat in the line to those found in \cite{Bonifacio:2020vbk}, where it was seen that the amplitudes of GB in the $\dimM\to4$ \textit{limit} correspond to those of a scalar-tensor theory where the scalar is infinitely strongly coupled. Hence, they concluded that this new pathological degree of freedom would only show up beyond linear order in perturbations.

Going beyond a Minkowskian background, perturbations around arbitrary maximally symmetric backgrounds \eqref{eq: max symm} pick up additional terms which diverge as $\Lambda/(\dimM-4)$ (see the supplementary material of \cite{Arrechea:2020evj}). Concretely, up to second-order in $h_{\mu\nu}$, there are $\mathcal{O}(\Lambda)$ terms of the form $h(\nabla^2 h)$ and $\mathcal{O}(\Lambda^2)$ terms of the form $h^2$. Consequently, $\Lambda$-proportional terms provide additional divergences which make de Sitter and anti-de Sitter backgrounds also ill-defined beyond linear order in perturbation theory.

\section{An action for the regularised equations?}\label{sec:action}

We have seen that, unless the $W^\text{GB}_{\mu\nu}$ term is stripped away from the field equations \eqref{fieldeqs} after taking the variation of the D4EGB action \eqref{action}, they will be, in general, ill-defined. Let us now comment on the possibility of finding a diffeomorphism-invariant action whose field equations in $\dimM\geq4$ yield the {\it stripped} equations, namely field equations of the form \eqref{fieldeqssym}.\footnote{Even though the $\dimM\to4$ process, if understood as a limit, will have the same conceptual problems described in section \ref{sec:Dto4}, in this case they might be swept under the rug since the $1/(\dimM-4)$ dependence actually disappears from the field equations.} To find such an action starting from the EGB one, we should be able to subtract a scalar from the EGB action so that the contribution $W^\text{GB}_{\mu\nu}$ disappears after taking the variation with respect to the metric, yet the diffeomorphism symmetry of the EGB action is not lost. In trying to find such a term, we are immediately led to an inconsistency due to the form of the Bianchi identities related to diffeomorphism invariance. To see this, note that the Bianchi identities associated to diffeomorphism-invariant actions imply that its variation with respect to the metric must be identically divergenceless  \cite{Ortin:2004ms}. Hence, given that the Gauss-Bonnet term, {\it i.e.} \eqref{eq: Lovelock general} with $k=2$, is diffeomorphism invariant, it must satisfy this identity. Hence, by using the A-W decomposition \eqref{eq: AW decomposition} and substituting $A^\text{GB}_{\mu\nu}$ with \eqref{eq: Amn GaussBonnet}, it follows that the off-shell relation
\begin{align}\label{DivW}
      \nabla^\mu W^\text{GB}_{\mu\nu} = - \frac{4(\dimM-4)}{\dimM-2}C_{\nu\rho\lambda\mu}\nabla^\mu R^{\rho\lambda}\,.
\end{align}
must be satisfied. Observe that the right-hand side of this equation is not identically zero in an arbitrary dimension, as can be seen by considering the following counterexample in five dimensions,
\begin{equation}
    {\textrm{d}}s^2={\textrm{d}}t^2 - {\textrm{e}}^{2t}{\textrm{d}}x^2- {\textrm{e}}^{4t}({\textrm{d}}y^2+{\textrm{d}}z^2+{\textrm{d}}w^2)\,,
\end{equation}
for which the equation \eqref{DivW} reads
\begin{equation}
    \nabla^\mu W^\text{GB}_{\mu\nu} = -4 \delta^t_\nu\neq0\,.
\end{equation}

Together with the fact that the variation with respect to the metric of any diffeomorphism-invariant action is identically divergence-free, the above result implies that the term $W^{\text{GB}}_{\mu\nu}$ does not come from an action that is a scalar under diffeomorphisms and built only with a metric. Consequently, there does not exist any term that can be added to the action \eqref{action} to cancel the $W^{\text{GB}}_{\mu\nu}$ contribution in the D4EGB field equations \eqref{fieldeqs} without spoiling its diffeomorphism-invariance. 
Other authors have proposed alternative ways to regularise the action \eqref{action}, generally leading to a scalar-tensor theory  of  the  Horndeski  family \cite{Lu:2020iav,Bonifacio:2020vbk,Fernandes:2020nbq,Hennigar:2020lsl},  thus leaving the Lovelock theorem intact.

We thus conclude that no diffeomorphism-invariant action can give the desired {\it stripped} field equations \eqref{fieldeqssym} in $\dimM\geq4$. Nevertheless, nothing prevents the existence of a non-diffeomorphism-invariant action having \eqref{fieldeqssym} as its field equations. Should it be possible to find such action, however, the absence of diffeomorphism invariance would potentially unleash the well known pathologies that occur in massive gravity (see {\it e.g.} \cite{Hinterbichler:2011tt,deRham:2014zqa}), thus propagating a Boulware-Deser ghost \cite{Boulware:1973my}. 

\section{Geodesic analysis of the spherically symmetric geometries}\label{sec:geodesic}

In addition to maximally symmetric and FLRW spacetimes, spherically symmetric geometries claimed to be solutions of D4EGB were also considered in \cite{Glavan:2019inb}, where it was stated that they are described by the 4-dimensional metric 
\begin{equation}\label{sphericallysymmetricmetric}
    {\textrm{d}}s^2 = A_{\pm}(r){\textrm{d}}t^2 -A_{\pm}^{-1}(r){\textrm{d}}r^2-r^2{\textrm{d}}\Omega^2_2\,,
\end{equation}
where $A_{\pm}(r)$ has the form
\begin{equation}\label{eq:metric}
A_{\pm}(r)=1+\frac{r^2}{32\pi\alpha G}\left[1\pm\sqrt{1+\frac{128\pi\alpha G^2 M}{r^3}}\right].
\end{equation}
First of all, let us point out that $\dimM-$dimensional spherically symmetric geometries described by metrics of the form \cite{Ortin:2004ms}
\begin{equation}\label{sphericalmetricD}
{\textrm{d}}s^2 = A(r){\textrm{d}}t^2 -A^{-1}(r){\textrm{d}}r^2-r^2{\textrm{d}}\Omega^2_{\dimM-2}\,,
\end{equation}
do not in general satisfy that $W^{\text{GB}}_{\mu\nu}=0$ in arbitrary $\dimM\geq4$. To see this, it suffices to restrict us to the 5-dimensional case, where there is a nontrivial condition for $W^{\text{GB}}_{\mu\nu}$ to vanish for metrics of the above form that reads
\begin{equation}
    r^2\frac{{\textrm{d}}^2A}{{\textrm{d}}r^2}-2r\frac{{\textrm{d}}A}{{\textrm{d}}r}+2A-2=0\,.
\end{equation}
This only happens for the particular case \mbox{$A=1+C_1 r+C_2 r^2$} where $C_i$ are integration constants. This suggests that \eqref{eq:metric} cannot be regarded as a solution of the D4EGB field equations, given that \eqref{fieldeqs} is not well-defined for $\dimM$-dimensional spherically symmetric metrics \eqref{sphericalmetricD} in the $\dimM\to4$ \textit{limit}. Indeed, as the authors of \cite{Glavan:2019inb} explain, the 4-dimensional spherically symmetric geometries  \eqref{eq:metric} were obtained by first re-scaling $\alpha$ with a $1/(\dimM-4)$ factor in the sherically symmetric solutions obtained in \cite{Boulware:1985wk} for EGB in $\dimM\geq5$, and then taking the $\dimM\to4$ \textit{limit} of these metrics, instead of solving the $\dimM\to4$ \textit{limit} of \eqref{fieldeqs}.

Nevertheless, it could be that the spherically symmetric geometries of \cite{Glavan:2019inb} are solutions of \eqref{fieldeqssym}, namely the {\it stripped}  field equation, after being stripped away of the pathological $W^{\text{GB}}_{\mu\nu}$ term. In the supplementary material of \cite{Arrechea:2020evj}, it can be seen that \eqref{fieldeqssym} has four different branches of solutions for $\alpha>0$. Two of them are exactly the Schwarzschild and Schwarzschild-(anti-)de Sitter
\begin{align}
    &A_1=1-\frac{2GM}{r}\,,\nonumber\\
    &A_2=1+\frac{r^{2}}{16\pi G \alpha}-\frac{2GM}{r}\,.
\end{align}
and the other two cannot be solved analytically, though their asymptotic behavior near the origin can be seen to be $A\cong r^{-3-2\sqrt{3}}+\mathcal{O}(r^0)$. Thus these solutions can neither be the ones found in \cite{Glavan:2019inb}, although they approach the Schwarzschild and Schwarzschild-(anti-)de Sitter solutions at spatial infinity.

Let us now turn to the behavior of the spherically symmetric geometries presented in \cite{Glavan:2019inb}. As noted in \cite{Glavan:2019inb}, the $\alpha<0$ branch of the above solution is not well defined {for values of the radial coordinate below $r<(-128\pi\alpha G^2 M)^{-1/3}$}, so their analysis focused on the $\alpha>0$ branches, showing that the above metric describes solutions which behave asymptotically as Schwarzschild or Schwarzschild-de Sitter solutions by choosing the negative and positive signs respectively. Concerning the former branch of solutions, it was shown in \cite{Glavan:2019inb} that {its} causal structure {(namely, the presence or absence of event horizons)} depends on the ratio between the {\it mass} {parameter} $M$ and a new mass scale $M_*=\sqrt{16\pi\alpha/G}$ that characterizes the D4EGB corrections to GR. From \eqref{sphericallysymmetricmetric} and \eqref{eq:metric} it can be shown that the $g_{tt}$ component of the metric {vanishes at the spherical surfaces}
\begin{equation}\label{eq:horizons}
    r_\pm=GM\left[1\pm\sqrt{1-\left(\frac{M_*}{M}\right)^2}\right].
\end{equation}
In view of this expression it becomes clear that solutions have no horizons for the $M<M_*$ case, outer and inner horizons if $M>M_*$ and one degenerate horizon if $M=M_*$. Interestingly, the mass scale $M_{*}$ plays a role similar to that of the electric (and magnetic) charges in the Reissner-Nordstr{\"o}m spacetime, with the exception that, in this case, the origin of such contributions comes exclusively from the gravitational field. The effect of the Gauss-Bonnet terms would then be that of making gravity repulsive at short distances, the magnitude of this repulsion being dictated by the strength of the GB coupling $\alpha$.

{Regarding the presence of singularities in the solutions, we see that despite the metric components \eqref{eq:metric} being finite at the origin} \begin{equation}
A(r)=1-\sqrt{\frac{2M}{GM_*^2}}r^{1/2}+\mathcal{O}(r^{3/2})\,,  
\end{equation}
curvature invariants diverge as $R\propto r^{-3/2}$, $R_{\mu\nu}R^{\mu\nu}\sim R_{\mu\nu\alpha\beta}R^{\mu\nu\alpha\beta}\propto r^{-3}$. {In \cite{Glavan:2019inb} it is argued that an observer could never reach this curvature singularity given the repulsive effect of gravity at short distances.} {This would imply} that the spacetime described by \eqref{sphericallysymmetricmetric} is complete in the sense that no (classical) physical observer ever reaches the curvature singularity at $r=0$ in a finite proper time. Nonetheless, there was no explicit proof in \cite{Glavan:2019inb} showing that this was indeed the case. We thus proceed to give a precise answer to the following question: does any (classical) physical observer reach the curvature singularity of \eqref{sphericallysymmetricmetric} in a finite proper time? To answer that question, it suffices to study the sub-class of radial and freely-falling (classical) observers, described by radial time-like geodesics. We will also consider radial null geodesics for illustrative purposes. 

\begin{figure}
    \centering
    \includegraphics[scale=0.7]{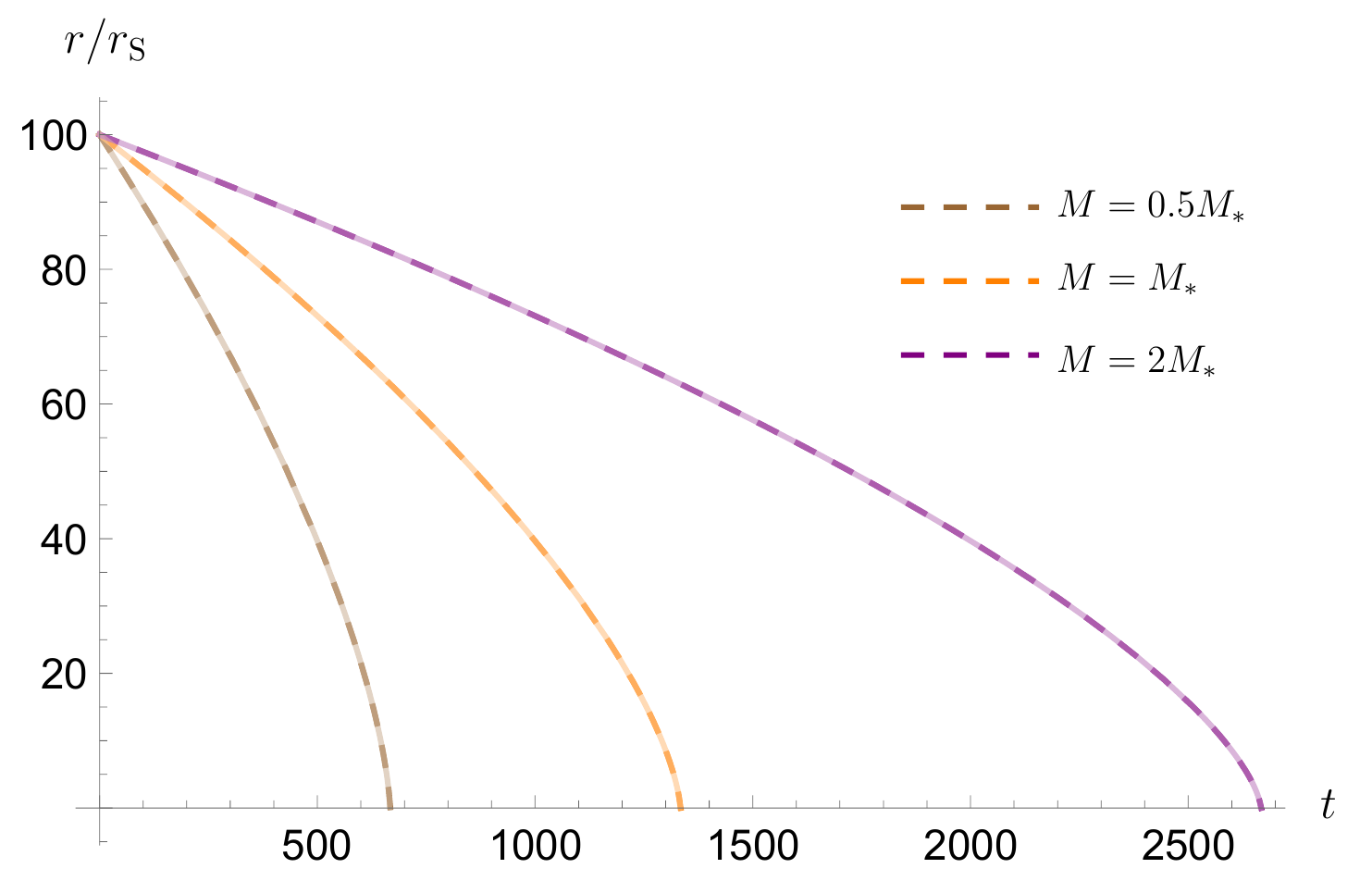}
    \caption{Plot of the radial ingoing trajectories (in units of $r_{\textrm{S}}=2M$) of a free-falling massive particle in the spacetime \eqref{sphericallysymmetricmetric} (dashed lines), and of the Schwarzschild solution (pastel colors) for different values of the $M$ parameter. At large distances trajectories are indistinguishable. We have chosen $M_{*}=G=1$, $r(0)=100~r_{\textrm{S}}$ and $E=1$ for visualisation purposes.}
    \label{IngoingGeodesics}
\end{figure}

 Consider the geodesic equation in the equatorial plane\footnote{Since spacetime is spherically symmetric, geodesics will lie in a plane, which can be chosen as the equatorial one in suitable coordinates. See {\it e.g.} \cite{Wald:1984rg} for details on the derivation of the geodesic equation and \cite{Olmo:2015bya} for the completeness analysis. } $\theta=\pi/2$ for the metric \eqref{sphericallysymmetricmetric},
\begin{equation}\label{geoeq}
\left(\frac{{\textrm{d}}r}{{\textrm{d}}\tau}\right)^2=E^2-V_{\text{eff}}(r), 
\end{equation}
with
\begin{equation}
V_{\text{eff}}(r)=A(r)\left(\frac{L^2}{r^2}-\kappa\right),
\end{equation}
and where $\tau$ is the proper time of the observer that moves along the solution $r(\tau)$. Here, $\kappa$ takes the values $\{-1,1,0\}$ for space-like, time-like and null geodesics respectively. $E$ and $L$ are constants of motion associated with time-translation and rotational symmetries respectively. It will suffice for our purpose to analyse radial geodesics, characterised by $L=0$. Firstly note that since radial photon trajectories are {insensitive} to the value of $A(r)$ in a spacetime described by any metric of the form \eqref{sphericallysymmetricmetric}, the trajectories stay the same as in GR. The solution to \eqref{geoeq} for time-like geodesics is plotted in Fig. \ref{IngoingGeodesics} for the cases with different causal structures. There, it can be seen that infalling massive particles starting in a region well beyond the Schwarzschild radius (where the space-time is effectively the same as in GR) reach the curvature singularity at $r=0$ in a finite proper time (no matter what its initial velocity is). Notice that, as can be seen in Fig. \ref{IngoingGeodesicsnear0}, the deviations from the GR trajectories are not relevant until the particle is at $r\cong r_{S}$. An asymptotic analysis of the geodesic equation reveals that, while in GR the curvature singularity at $r=0$ is reached with infinite velocity ${\textrm{d}}r/{\textrm{d}}\tau|_{\text{GR}}\propto r^{-1/2}+\mathcal{O}(r^{0})$, the geodesics described by \eqref{sphericallysymmetricmetric} reach it with finite velocity
\begin{equation}\label{eq:velocity}
  ({\textrm{d}}r/{\textrm{d}}\tau)^2|_{\text{D4EGB}}=E^{2}-1+\sqrt{2M/M_*^{2}}r^{1/2}+\mathcal{O}(r^{3/2})\,.
\end{equation}
It is interesting to note that in the case that the infalling particle starts at rest, no matter what its initial position is, it will reach the singularity with zero velocity (characterised by $E^2=1$): attractive and repulsive effects compensate each other along the trajectory of the particle.
\begin{figure}
    \centering
    \includegraphics[scale=0.7]{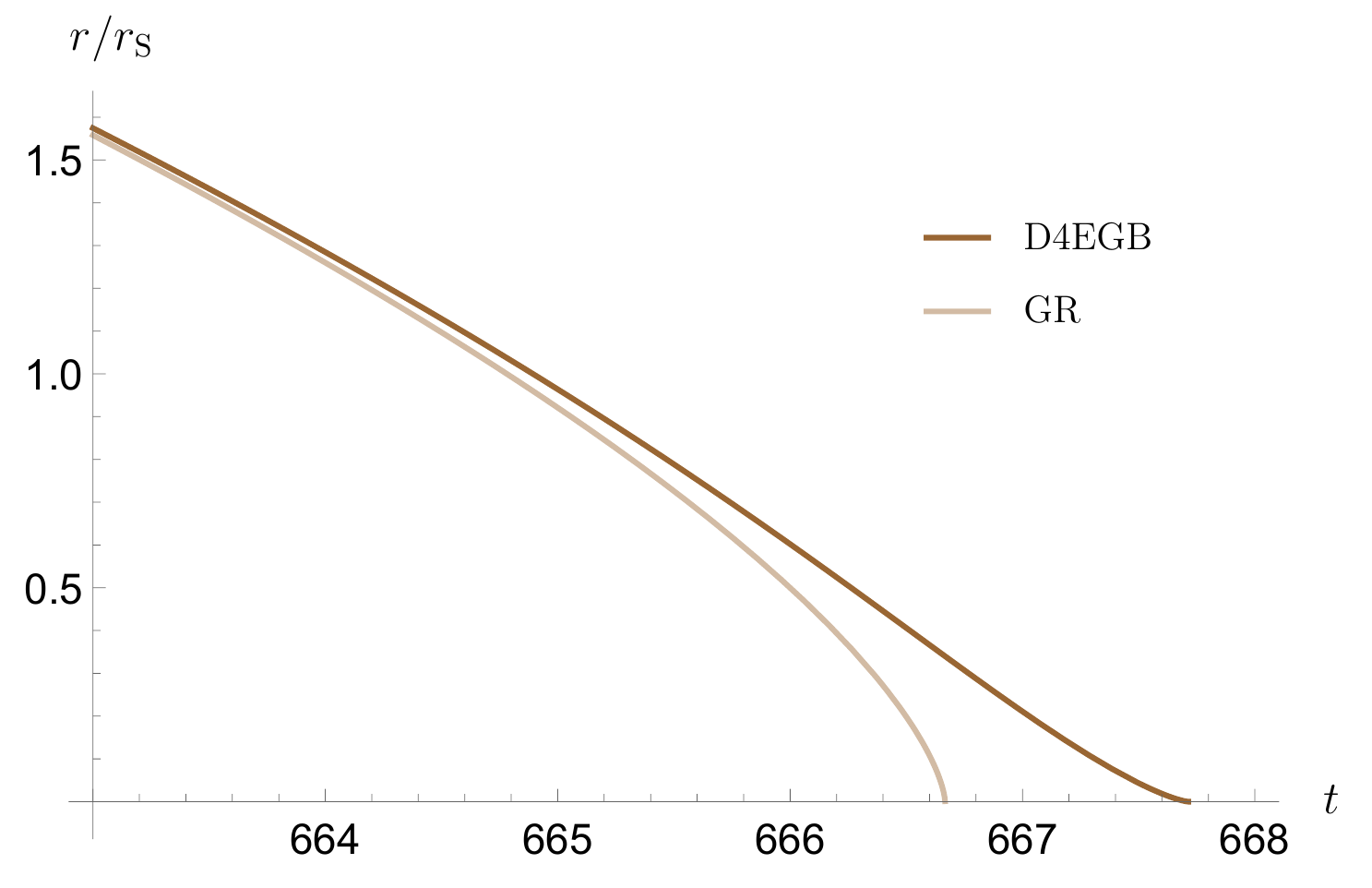}
    \caption{Figure showing how the trajectories of a massive particle in the spacetime described by \eqref{sphericallysymmetricmetric} (in units of \mbox{$r_{\textrm{S}}=2M$}) deviate from those of the Schwarzschild solution from GR near the central curvature singularity. Here, the $M=0.5M_*$ case is shown, although all timelike geodesics show the same behavior near the singularity. Units are $M_{*}=G=1$.}
    \label{IngoingGeodesicsnear0}
\end{figure}
The above proves that the statement made in \cite{Glavan:2019inb} that particles cannot reach the central singularity in spacetimes described by \eqref{sphericallysymmetricmetric} do not stand a rigorous analysis, as the singularity is reached in finite affine parameter.  Therefore, the hope that these solutions avoid the singularity problem is cast into serious doubt. Furthermore, the authors of \cite{Glavan:2019inb} also claim that under a realistic stellar collapse, matter would stop before reaching the singularity. This of course must be verified by a self-consistent analysis of the dynamical collapsing geometry, as was done in \cite{Malafarina:2020pvl}, revealing that the singularity indeed forms and gets covered by a horizon. Furthermore, the authors of \cite{Malafarina:2020pvl} also found that if the collapse is modelled {\it \`a la} Oppenheimer-Snyder, where dust is initially at rest, matter reaches the singularity with zero velocity, in agreement with our results.

We also note that, even if geodesic observers did never arrive at the singularity, the usual problems regarding curvature singularities would still remain: quantum corrections would be expected to become non-perturbative near the singularity and the background could not be treated classically anymore. However, the solutions would be {\it classically} singularity free in this case.

\section*{Final remarks} 

Here we have looked upon the idea of providing corrections to four dimensional General Relativity by means of the Gauss-Bonnet term devised in \cite{Tomozawa:2011gp} and recently revisited in \cite{Glavan:2019inb}. We have shown that this idea cannot be implemented for the Gauss-Bonnet ($k$-th order Lovelock) term in four ($2k$) spacetime dimensions by means of the procedure considered in \cite{Glavan:2019inb} without encountering inconsistencies. When considering solutions with a high degree of symmetry, such as maximally symmetric or general conformally flat solutions, this issue gets concealed at the level of the equations of motion due to the fact that the problematic terms $W^\text{GB}_{\mu\nu}$ in \eqref{fieldeqs} vanish for arbitrary $\dimM$ in these scenarios. Indeed we have shown that, when considering perturbations around a Minkowskian (or any maximally symmetric) background beyond linear order, such inconsistencies are immediately unveiled. This is also aligned with the conclusions at which the authors of \cite{Bonifacio:2020vbk} arrived by analysing the tree-level graviton scattering amplitudes in a Lagrangian-independent way by taking the four dimensional limit of the corresponding scattering amplitudes in EGB in higher dimensions. 

Regarding the spherically symmetric geometries presented in \cite{Glavan:2019inb}, we showed that they do not attain the required degree of symmetry as to make the problematic $W^\text{GB}_{\mu\nu}$ term vanishing in arbitrary dimension and, thus, bypass the pathologies encountered in the D4EGB field equations \eqref{fieldeqs}. By artificially removing the divergent $W^\text{GB}_{\mu\nu}$ term from \eqref{fieldeqs}, we encountered four spherically symmetric solutions, none of which coincides with those presented in \cite{Glavan:2019inb}. We also showed that the corresponding {\it stripped} equations cannot be derived from a diffeomorphism covariant action built only with the metric. Moreover, a geodesic analysis of the geometries from \cite{Glavan:2019inb} contradicts the observation about the singularity being unreachable by any observer in finite proper time.

To conclude, let us point out that the idea of extracting nontrivial corrections to the dynamics of a theory from topological terms by considering a divergent coupling constant is indeed very appealing, since its range of applicability extends far beyond gravitational contexts. For instance, it might serve to introduce parity-violating effects in Yang-Mills theories through the corresponding $F\tilde F$ terms that are topological in four dimensions. Indeed, a similar idea has been seen to lead to well-defined theories in the context of Weyl geometry \cite{BeltranJimenez:2014iie,BeltranJimenez:2015pnp,BeltranJimenez:2016wxw}. It could thus be interesting to explore various possibilities in this direction.

%%%%%%%%%%%%%%%%%%%%%%%%%%%%%%%%%%%%%%%%%%%%%%%%%%%%%%%
%%%%%%%%%%%%%%%%%%%%%%%%%%%%%%%%%%%%%%%%%%%%%%%%%%%%%%%
%%%%%%%%%%%%%%%%%%%%%%%%%%%%%%%%%%%%%%%%%%%%%%%%%%%%%%%

%=========================================================
\clearemptydoublepage
%
%
% File: chap01.tex
% Author: Victor F. Brena-Medina
% Description: Introduction chapter where the biology goes.
%
\let\textcircled=\pgftextcircled

\chapter{Outlook}

\initial{I}n this thesis we have progressed on the understanding on several theoretical and phenomenological aspects of metric-affine theories of gravity. This class of theories has its origin in the development by Cartan of a theory of connections \cite{Cartan1,Cartan2,Cartan3,Cartan4} right after the birth of GR, and in an attempt by Weyl of unifying gravity and electromagnetism through nonmetricity, which gave birth to the idea of gauge symmetry \cite{Weyl:1918ib}. After several decades, this gave rise to the birth of gauge theories of the Lorentz and Poincar\'e group as theories for the gravitational field \cite{Utiyama:1956sy,Kibble:1961ba,Sciama:1964wt,Sciama_1962}, which was later generalised to develop the gauge theory of the full affine group \cite{Hehl:1994ue}. Parallel to the development of gauge gravity, which naturally yields metric-affine geometries, other metric-affine theories, generally metric-affine formulations of higher-order curvature gravities, were being formulated. This bloomed with the discovery of the accelerated expansion of the universe, where metric-affine (or Palatini) theories of gravity were being considered to explain in a natural way the value of the cosmological constant. To do this, $1/R$ metric-affine corrections to GR were considered \cite{Vollick:2003aw}, but this model was quickly ruled out by accelerator experiments as an explanation for $\Lambda$ \cite{Flanagan:2003rb}. Then, other more general IR metric-affine corrections to GR, generally of the Ricci-Based type, were attempted \cite{Allemandi:2004wn,Allemandi:2004ca}, again ruled out as an explanation for $\Lambda$ due to their disastrous effects on the stability of the Hydrogen atom \cite{Olmo:2008ye}.

Soon after their failure in providing a natural explanation for the Cosmological Constant it was seen that these theories were also good in avoiding cosmological singularities at the classical level. First,  a covariant metric-affine $f(R)$ action mimicking the background evolution of LQC \cite{Olmo:2008nf}, followed by a systematic study of the conditions under which $f(R)$ theories present a bounce replacing the Big Bang singularity \cite{Barragan:2009sq}, which was seen to be unstable if anisotropies were included \cite{Koivisto:2010jj}. A similar analysis for other theories has followed this past decade \cite{Barragan:2010qb,Banados:2010ix,Bombacigno:2018tyw,Benisty:2021laq}. At the same time, it was found that singularity avoidance also occurs in spherically symmetric backgrounds within these theories \cite{Olmo:2016fuc,Olmo:2015bya,Olmo:2015dba,Olmo:2013gqa,Olmo:2012nx,Olmo:2011np,Afonso:2019fzv,Afonso:2020giy,Orazi:2020mhb,Rubiera-Garcia:2020gcl,Maso-Ferrando:2021ngp,Guerrero:2020uhn,Lobo:2013adx,Lobo:2013vga,Lobo:2014fma,Lobo:2014zla,Lobo:2020vqh,Olmo:2019flu,Olmo:2019qsj} which fostered their exploration. 

When I started my master thesis on corrections to the energy levels of the Hydrogen atom in spacetimes with nonmetricity, driven by Gonzalo's supportive advise, I became engaged with the question of understanding whether there are model-independent effects due to the presence of nontrivial nonmetricity in the spacetime structure. Though a posible answer came early in my first publication \cite{Latorre:2017uve}, I did not fully understood its implications until recently, when I built the generalised Einstein-like frame for generic metric-affine theories in the elaboration of chapter \ref{sec:MetricAffineEFT}. Then, I realised that though the effects in \cite{Latorre:2017uve} (and then \cite{Delhom:2019wir,JoanAdri}) were derived strictly within the RBG subclass of theories, they generalise to any metric-affine theories having $R_{(\mu\nu)}$ in the Lagrangian beyond the EH term. Furthermore, the presence of these terms appears to be linked to a particular form of the nonmetricity tensor when written in the right field variables. Thus, what first appeared to be an effect that arose only in a very limited subset of metric-affine theories, it has now been seen to arise in arbitrary ones, provided that they have these terms in their action. Though my appeal to the geometric view of metric-affine theories has declined since I started the PhD in favour of a more field theoretic one, from the former perspective these effects constitute a first answer to the question that stimulated my interest in these theories. And, what is more important, they allowed to set the stringent constraints up to date to RBG theories and, possibly, generic metric affine theories (with $R_{(\mu\nu)}$ operators beyond the EH term) by using data from high energy colliders. These results open a new avenue that explores whether these effective interactions arising in metric-affine theories are only due to $R_{(\mu\nu)}$ terms beyond the EH action, or other contributions to the action also have similar kinds of effects. This question is more precisely formulated in the EFT language, where it reduces to finding all the metric-affine operators that are redundant\footnote{If the reader is not familiarised with the concept of redundant operator, see the discussion in section \ref{sec:RBGandEFT}.} in a metric-affine EFT once all the allowed operators of the matter sector have been allowed. 

While scrutinising the properties of RBG theories and beyond, and with the wise guiding of Jose Beltr\'an-Jim\'enez, we found two nice results concerning these theories. One concerns the possibility of the deformation matrix, and therefore the Einstein frame metric, not having the same symmetries as the original $g^{\mu\nu}$ metric and the matter sector, which is often overlooked in the literature. This is allowed due to the nonlinear nature of the algebraic equations that determine the deformation matrix as a function of one of the metrics and the matter fields. Though not all RBG theories allow nontrivial solutions for the deformation matrix\footnote{Note that the trivial solution where it has the same symmetries as the metric and stress-energy tensor is always allowed.} given a set of symmetries of one of the metrics and the stress-energy tensor,\footnote{For instance, we showed that EiBI does not admit any nontrivial solution in presence of an isotropic and homogeneous metric and stress-energy tensor.} we showed that generic RBG theories admit these solutions for an isotropic and homogeneous metric and stress-energy tensor (see chapter \ref{sec:SolutionsDeformation}). Moreover, we saw that these branches of solutions were in general not perturbatively close to GR in vacuum, showing generally a pathological behaviour, and providing a strong argument in favor of the usual choice of solutions for the deformation matrix. In this line, some questions that arise are what are the properties of these other branches of solutions when linking backgrounds with other symmetries, or whether there are any branches which spontaneously break Lorentz symmetry after the mapping procedure is carried out.

The other result obtained with Jose Beltr\'an-Jim\'enez concerns the presence of ghosts in metric-affine theory. For some parts of the metric-affine community, there was a widespread belief hoping that the Ostrogradski ghosts present in higher-order metric theories could be exorcised by resorting to the metric-affine framework, due to the fact that metric-affine higher-order curvature corrections do not contain higher-order derivatives of the metric. Though there are known subclasses of metric-affine theories which do not suffer from these pathologies such as RBGs,\footnote{Recall that we use the acronym RBGs for the cases with projective symmetry unless the contrary is specified.} building on an explicit proof that RBG theories without projective  symmetry were plagued by ghostly instabilities, we found strong arguments suggesting the presence of ghostly instabilities in generic metric-affine theories, as explained in chapter \ref{sec:UnstableDOF}. We also explored the possibility of freezing the pathological degrees of freedom arising in RBG theories with projective symmetry by considering geometric constraints, with some positive conclusions. These results are a central part of the work developed during this thesis, and they suggest to explore the full metric-affine landscape in search for ghost-free islands in this vast sea of theories.\footnote{Let me remark the fine title of a talk by J. Beltr\'an-Jim\'enez named {\it The hazardous landscape of affinisea}.} This exploration could unveil relevant symmetries of the metric-affine sector which are necessary in order to avoid ghosts, such as the projective one. At the same time, this would restrict the allowed operators on a metric-affine EFT, thus making the task of finding all the redundant ones suggested by the results in chapter \ref{sec:MetricAffineEFT} less arduous.

The advent of gravitational wave astronomy headed by the LIGO-Virgo collaboration, as well as the birth of black hole imaging initiated by Event Horizon Telescope have fostered the quest for the understanding of the properties of Exotic Compact Objects (ECOs), within GR and beyond \cite{Liebling:2012fv,Cardoso:2016oxy,Cardoso:2017cqb,Cardoso:2014sna,Cardoso:2019rvt}. Thus, it is about time to provide a catalog of ECOs that can arise in metric-affine theories and explore their phenomenological properties. In this thesis we have taken a timid first step in this direction, exploring the phenomenology amid wormhole solutions arising in RBG theories. We found that, similar to what occurs with other compact objects \cite{Macedo:2018yoi}, the absorption spectrum of scalar waves is sensitive to the nontrivial structure provided by the wormhole throat, which generates a resonant absorption spectrum (see chapter \ref{sec:Absorption} and \cite{Delhom:2019btt}). These results should be expanded to include other types of perturbations and to the study of emission and reflection spectra as well. Furthermore, the rapidly growing catalog of exact solutions within RBG theories \cite{Afonso:2019fzv,Afonso:2020giy,Orazi:2020mhb,Olmo:2020fnk,Shao:2020weq,Guerrero:2020azx,Maso-Ferrando:2021ngp}, demands a prompt analysis of the phenomenological properties of these ECOs, in search for distinctive signatures that allow to discriminate them from compact objects arising in GR and beyond. In this direction, the mapping of RBG theories coupled to NEDs developed in section \ref{sec:Mapping} could relate possible nonlinearities that leave an imprint on electromagnetic signals produced in strong gravitational backgrounds (\eg neutron stars) to gravitational theories beyond GR.

Lastly, the positive results obtained in mimicking the background evolution of different models of isotropic and homogeneous Loop quantum cosmologies by a single family of metric-affine theories $f(R)$, granting robustness to previous results \cite{Olmo:2008nf}, open the possibility of exploring this relation beyond the background level, as well as in other more general scenarios such as anisotropic Loop quantum cosmologies \cite{Ashtekar:2009um,Ashtekar:2009vc,Wilson-Ewing:2010lkm,Wilson-Ewing:2015xaq} or singularity-free Loop quantised spherically symmetric spacetimes \cite{Olmedo:2017lvt,Ashtekar:2018lag,Ashtekar:2018cay}. To this end, we should be open to consider other families of theories beyond $f(R)$, given that, in these theories, the shear typically grows unboundedly through the bounce \cite{Barragan:2010qb}. As a long term program, it would also be interesting in diving into the possible relation of spacetime microstructure and metric-affine effective geometries. In this sense, with the known results that the continuum limit of crystalline structures with defects lead to torsion and nonmetricity \cite{ArashDefects,KupfermanDefects,DONGDefects,KronerDefects,KleinertDefects} in mind, it could be interesting to devise alternative ways of defining the continuum limit of QG theories that predict some kind of spacetime granularity in search for an effective description of our universe at scales where QG effects can be {\it integrated out} but might still manifest through post-Riemannian corrections.

%=========================================================
\clearemptydoublepage
% Apparently the guidelines don't say anything about citations or
% bibliography styles so I guess we can use anything.
%
% File: abstract.tex
% Author: V?ctor Bre?a-Medina
% Description: Contains the text for thesis abstract
%
% UoB guidelines:
%
% Each copy must include an abstract or summary of the dissertation in not
% more than 300 words, on one side of A4, which should be single-spaced in a
% font size in the range 10 to 12. If the dissertation is in a language other
% than English, an abstract in that language and an abstract in English must
% be included.

\chapter*{Resum}
\addcontentsline{toc}{chapter}{Resum}
\markboth{Resum}{}

\begin{SingleSpace}
\initial{L}a Gravitaci\'o combina un dels conjunts de fen\`omens m\'es intu\"itius per a l'\'esser hum\`a, i probablement altres esp\`ecies \cite{CambridgeDeclaration,BIRCH2020789}, amb el fet d'\'esser l'\'unica interacci\'o coneguda per a la qual no tenim una teor\'ia completa en l'ultravioleta satisfact\`oria a dia d'avui. La Relativitat General (RG) \'es la primera teor\'ia relativista de la gravitaci\'o concebuda a l'hist\`oria, i ha passat exit\'osament totes les proves experimentals dissenyades fins a dia d'avui, predient pel cam\'i diferents fen\`omens com la recent detecci\'o d'ones gravitacionals \cite{Abbott:2016blz,TheLIGOScientific:2017qsa,LIGOScientific:2021qlt} o el valor correcte de l'angle de deflexi\'o de la llum a l'eclipse de 1919 \cite{Crispino:2019yew,Crispino:2020txj}. La RG suggereix una interpretaci\'o de la gravitaci\'o en termes geom\`etrics, on les interaccions gravitat\`ories s\'on interpretades com la din\`amica de l'espaitemps on la resta de camps de matèria \'es propaguen. Des d'aquesta \'optica, el camp gravitatori es codifica en les propietats de la m\`etrica d'una varietat (pseudo-)Riemanniana. Tradicionalment, aquest ve descrit per la curvatura d'aquesta m\`etrica, encara que hi han interpretacions alternatives (aparentment) equivalents en termes d'altres objectes geom\'trics com la nometricitat o la torsi\'o associats a tipus particulars de connexions afins, com ocorre als marcs conceptuals de les teories teleparal\textperiodcentered leles \cite{Aldrovandi:2013wha,Nester:1998mp,BeltranJimenez:2017tkd,Jimenez:2019ghw,BeltranJimenez:2019tjy}.

Des d'aquesta perspectiva, la gravetat \'es una teor\'ia de la din\`amica de l'espaitemps en si, una visi\'o que va motivar desenvolupaments fruct\'ifers com el comen\c{c}ament de la cosmologia com a disciplina cient\'ifica amb els treballs pioners de Slipher, Lemaitre, i Hubble \cite{Slipher,Lemaitre,Hubble:1929ig}. A m\'es, acomoda de forma natural la m\`etrica de Friedman Lemaitre Robertson i Walker (FLRW), que a dia d'avu\'i proporciona la millor descripci\'o que tenim de les observacions cosmol\`ogiques a trav\'es del model est\`andard cosmol\'ogic, encara que la pres\`encia de components no observats al tensor d'energia-moment de l'univers es necess\`aria per tal que aquest model estiga d'acord amb les dades. Aix\'i mateix, tamb\'e va predir l'exist\`encia d'objectes compactes dels que res pot escapar una vegada dins de certa regi\'o de l'espaitemps, es a dir, forats negres i els seus horitzons. Tant, l'estudi de cosmologia com el dels objectes compactes s\'on, a dia d'avui, disciplines establertes i actives dins de la investigaci\'o en f\'isica gravitat\`oria, i els dos senyalen un dels majors problemes de la RG com a teor\'ia fonamental de la interacci\'o gravitat\`oria, constituit la pres\`encia de singularitats tant a l'univers primitiu (Big Bang) com al centre dels forat negres. 

Des del punt de vista cl\`assic, aquestes singularitats senyalen una ruptura de l'espaitemps a la que els observadors f\'isics poden arribar en un temps propi finit. Tot i que aix\`o no \'es inconsistent a nivell cl\`assic, ens resulta summament desagradable acceptar la idea de que els observadors f\'isics poden desapar\`eixer de l'univers si cauen a una singularitat. No obstant, des del punt de vista qu\`antic, aquest rebuig no es solament una q\"uesti\'o est\`etica, sin\'o que \'es inconsistent amb l'evoluci\'o temporal unit\`aria que resideix a la base de les teories qu\`antiques degut a que es perdria informaci\'o al final del proc\'es d'evaporaci\'o d'un forat negre a trav\'es de l'emissi\'o de radiaci\'o de Hawking. Acceptar el car\`acter qu\`antic del camp gravitatori ofereix una soluci\'o a aquest problema. De fet, la RG, encara que no \'es renormalitzable, \'es una teor\'ia qu\`antica de camps efectiva per al camp gravitatori ben comportada fins energies de l'escala de Planck \cite{Burgess:2003jk,Donoghue:2012zc,Ruhdorfer:2019qmk}, on perd l'unitarietat. Per aquest motiu, les solucions cl\`assiques amb diverg\`encies de curvatura no tenen sentit f\'isic a dist\`ancies m\'es curtes que la longitud de Planck, escala a partir de la qual s'espera que els efectes qu\`antics de la gravitaci\'o siguen dominants i canvien la estructura no pertorbativa de la teor\'ia. Aix\'i, les solucions singulars de la RG es diferenciarien el suficient de les solucions exactes de la teor\'ia completa a escales per baix de la longitud de Planck, deixant aquestes singularitats fora del r\`egim de validessa de la teor\'ia cl\`assica. En aquest sentit, hi ha la creen\c{c}a generalitzada que la compleci\'o ultravioleta de la RG solucionar\`a el problema de les singularitats degut als efectes qu\`antics. De fet, a\c{c}\`o pareix oc\'orrer en algunes teories candidates a completar la RG a l'ultravioleta com Loop Quantum Gravity (LQG) \cite{Agullo:2016tjh,Li:2021mop,Gambini:2013hna,Ashtekar:2018lag,Ashtekar:2018cay,Gambini:2020qhx,Ashtekar:2020ckv,Kelly:2020lec,Kelly:2020uwj}.

 Encara que hi han raons per estudiar modificacions a la RG a baixes energies per tal d'explicar alguns dels efectes usualment atribu\"its a la mat\`eria o energia fosques, la motivaci\'o m\'es forta per estudiar desviacions de la RG es trobar una teor\'ia qu\`antica de la gravitaci\'o que puga tindre sentit f\'isic a energies arbitr\`ariament altes, ja que sabem que la RG deixa de tenir sentit a l'escala de Planck si s'accepta que la gravitaci\'o es un fenomen qu\`antic a escales petites. Una de les vies per obtenir informaci\'o sobre el comportament qu\`antic de la gravitaci\'o es estudiar els possibles efectes residuals detectables a m\'es baixes energies. En aquesta l\'inia hi ha varies possibilitats. D'una banda, l'acci\'o d'Einstein i Hilbert (EH) ha de modificar-se amb correccions semicl\`assiques per garantir renormalitzabilitat dels camps de mat\`eria a espaitemps corbats \cite{Parker:2009uva}. Correccions quadr\`atiques en la curvatura quadr\`atica donen lloc a una teor\'oa de la gravitaci\'o renormalitzable a costa de sacrificar la seva unitarietat \cite{Stelle:1976gc,Antoniadis:1986tu,Johnston:1987ue} degut a que aquestes correccions donen lloc a derivades d'ordre superior de la m\`etrica, desencadenant la propagaci\'o de inestabilitats d'Ostrogradski (veure cap\'itol 7). Una possible forma d'evitar aquestes inestabilitats es rec\'orrer a la formulaci\'o metric-aff\'i d'aquestes teories, on la connexi\'o i la m\`etrica son independents, i els termes no lineals en el tensor de Riemann ja no contenen segones derivatives de la m\`etrica. Aquest formalisme consisteix en extendre la RG permitint la aparici\'o de geometries espaitemporals m\'es generals que les (pseudo-)Riemannianes degut a la introducci\'o d'una connexi\'o af\'i independent, que dona lloc a una visi\'o de l'espaitemps com una varietat post-Riemanniana, es a dir, una varietat diferenciable amb una estructura af\'i i una m\`etrica independents. Aquesta independ\`encia queda codificada en dos 
 objectes geom\`etrics anomenats tensors de nometricitat i torsi\'o, que mesuren les desviacions de la {\it Riemannianitat} a l'espaitemps. Aquest formalisme t\'e orige als treballs de Cartan on es va formular una primera teor\'ia de connexions independentment de l'estructura m\`etrica \cite{Cartan1,Cartan2,Cartan3,Cartan4}, aix\'i com els de Weyl on s'intenta unificar la gravitaci\'o i l'electromagnetisme per mitj\`a de teories gauge \cite{Weyl:1918ib}. Dins d'aquest formalisme trobem, per una banda, teories gauge de la gravitaci\'o, on aquesta es descrita per camps gauge associats a una simetria local del grup de Poincar\'e que dona lloc a l'aparici\'o de torsi\'o, treball desenvolupat inicialment per Utiyama, Kibble i Sciamma \cite{Utiyama:1956sy,Kibble:1961ba,Sciama:1964wt}, o altres m\'es generals com la teor\'ia gauge del grup af\'i, on apareixen curvatura, la nometricitat i la torsi\'o relacionats amb els camps gauge, principalment desenvolupat per Hehl i col$\cdot$laboradors \cite{Hehl:1976kj,Hehl:1994ue}. Per altra banda, trobem altres teories que no entren dins d'aquesta formulaci\'o gauge dels grups de simetr\'ies locals a l'espaitemps. Aquestes segones tradicionalment s'han anomenat teories de Palatini o de primer ordre, i solen vindre descrites per Lagrangians no lineals en el tensor de Riemann de la connexi\'o espaitemporal. Aquesta formulaci\'o de teories no lineals en la curvatura va generar certa esperan\c{c}a en que la teor\'ia quadr\`atica i renormalitzable de Stelle f\'ora unit\`aria degut a la possible ausencia d'inestabilitats d'Ostrogradski en aquesta formulaci\'o, i a la vegada mantinguera la seua renormalitzabilitat. Un dels resultat centrals d'aquesta tesi es una prova de que a les teories amb termes no lineals en la curvatura formulades al formalisme metricaf\'i tamb\'e presenten inestabilitats d'Ostrogradski i altres graus de llibertat inestables de tipus fantasma de forma gen\`erica, aix\'i com teories metricaf\'ins m\'es generals.
 
En aquesta tesi tractem diferents aspectes del formalisme , aix\'i com algunes propietats te\`oriques i fenomenol\`ogiques d'aquestes teories. A la primera part, comencem amb una discusi\'o d'aspectes interpretatius (entre filos\`ofics i te\`orics), aix\'i com aspectes matem\`atics subtils del formalisme. Concretament, al cap\'itol \ref{sec:GravityAsGeometry} amb una discusi\'o sobre les diferents interpretacions possibles de la interacci\'o gravitatoria, es a dir, la interpretaci\'o geom\`etrica i la interpretaci\'o en termes de camp de for\c{c}a, com les altres interaccions conegudes. Intentarem posar el focus en les relacions entre aquestes, i les propietats observacionals dels fen\`omens gravitatoris que les permeten. Afegirem tamb\'e una xicoteta discussi\'o sobre les avantatges i els inconvenients de cadascuna de les interpretacions, i oferirem una idea de quina es l'enfocament de l'autor respecte als diferents resultat de la tesi en respecte a aquestes interpretacions. Despr\'es, als cap\'itosl \ref{sec:DifferentialGeometry} i \ref{sec:MinimalCoupling} oferirem una introducci\'o a q\"uestions formals relacionades amb el formalisme metricaf\'i, aix\'i com una discussi\'o de certes subtilesses matem\`atiques que hi apareixen. Concretament, al cap\'itol  \ref{sec:DifferentialGeometry} presentarem les ferramentes matem\`atiques necessaries partint del concepte de varietat diferenciable, i fent \'enfasi en la canonicitat de les diverses estructures. La idea es que el lector tinga clar al llegir quines estructures matem\`atiques venen donades de forma can\`onica respecte a altra estructura preexistent en el sentit que tenir eixa estructura preexistent garanteix l'existencia de l'altra de forma un\'ivoca, el que pot entendre's com a que s\' on part de la mateixa estructura en realiat. Creguem que aquestes q\"uestions son importants al formalisme metricaf\'i perqu\`e moltes de les estructures que s\'on can\`oniques al formalisme m\`etric, poden deixar de ser-ho al metricaf\'i, i mai est\`a de m\'es saber identificar-ho. En concret, la idea es poder arribar a la forma can\`onica de la connexi\'o al fibrat espinorial asociada a una connexi\'o af\'i al fibrat tangent de l'espaitemps (o a l'espaitemps mateix). Aquesta q\"uesti\'o es, amb certa freq\"u\`encia, poc clara en la literatura, docs hi han altres eleccions possibles per a la connexi\'o al fibrat espinorial a part de la can\`onica. Amb aquesta discussi\'o tenim la intenci\'o d'intentar aclarir quines d'aquestes eleccions s\'on arbitr\`aries i perqu\`e, aix\'i com donar una idea del grau d'arbitrarietat de cada elecci\'o. Una vegada assolit aquest objectiu, passem a discutir sobre la definici\'o d'acoblament m\'inim dels camps de mat\`eria a la geometria. Tot i que al formalisme m\`etric hi ha una recepta molt senzilla d'acoblament m\'inim que deixa aquesta q\"uesti\'o lliure d'ambig\"uitats, al intodu\"ir torsi\'o i nometricitat aquesta recepta introdueix certs acoblaments que poden entendre's com no-m\'inims en un sentit clar. Al cap\'itol \ref{sec:MinimalCoupling}, basat en \cite{Delhom:2020hkb}, discutirem els problemes d'aquesta recepta i mostrarem com en pres\`encia de torsi\'o i nometricitat, aquesta recepta \'es usualment formulada de na\"if, el que duu a la pres\'encia d'acoblaments no-m\'inims que arriben a trencar la invari\`ancia gauge del terme cin\'etic dels camps vectorials. Tanmateix, esclarirem una definici\'o precissa del que considerem acoblament m\'inim, i proposarem una recepta que es redueix a la usual en espaitemps (pseudo-)Riemannians per\`o evita els termes d'acoblament en pres\`encia de torsi\'o i nometricitat que apareixien amb la formulaci\'o na\"if de la recepta i que entraven en conflicte amb la definici\'o que donem d'acoblament m\'inim. Com a exemples, discutirem expl\'icitament les diferencies entre la recepta na\"if i la proposada a aquest treball per a camps escalars, de Dirac, i vectorials. Per acabar, argumentarem perqu\`e les geod\`esiques afins no son generalment traject\`ories de part\'icules lliures assumint que les toer\'ies fonamentals satisfan un principi de acci\'o extremal, al no poder dedu\"ir-se aquest tipus de traject\`ories de l'aproximaci\'o eikonal de camps de mat\`eria descrits pe una acci\'o. Aquesta punt \'es conf\'us a la literatura amb freq\"u\`encia.

Acabada la part primera, ja haurem mostrat les ferramentes i algunes subtilesses matem\`atiques del formalisme, i podem entrar en mat\`eria estudiant les propietats estructurals i din\`amiques de teories metricafins. Amb l'objectiu d'entendre aspectes generals de teories metricafins, ens centrarem primer en una sub-familia de teories anomenada Ricci Based gravity, ja que aquestes teories tenen propietats estructurals interessants que ens permetran despr\'es entendre propietats de teories metric afins gen\'eriques. Comen\c{c}arem el seu an\`alisi al cap\'itol \ref{sec:RBGTheory}, basat en \cite{Delhom:2019zrb,Jimenez:2020dpn}, on estudiarem la seua estructura general i les seues equacions de moviment, mostrant que existeix una elecci\'o de variables a l'espai de camps que dona lloc a una representaci\'o Einsteniana de les teories RBG. Vorem que la sub-familia sense simetria projectiva propaga nous graus de llibertat que donen lloc a inestabilitats de tipus fantasma, cas que ser\`a analitzat amb detall al cap\'itol \ref{sec:UnstableDOF}. Al contrari, la subfamilia amb simetria projectiva que sols cont\'e termes amb el tensor de Ricci simetritzat a l'acci\'o, a la que ens referirem a partir d'ara amb l'abreviatura RBG, tenen un Lagrangi\`a en la representaci\'o Einsteniana id\'entic al de Einstein i Hilbert. Vorem que a\c{c}\`o permet definir un procediment de mapeig de teories RBG acoblades a un sector de mat\`eria particular a la RG acoblada a un sector de mat\`eria que \'es una deformaci\'o no-lineal de l'original, amb noves interaccions, per\`o que no cont\'e nous graus de llibertat. Deduirem aquest proc\'es de mapeig expl\'icitament per a un camp electromagn\`etic, i mostrarem un exemple concret en el que la teor\'ia gravitat\`oria d'Eddington-inpired Born-Infeld (EiBI) acoblada a una electrodin\`amica de Maxwell ser\`a mapejada a la RG acoblada a una electrodin\`amica de Born-infeld, el que obri la possibilitat d'estudiar solucions exactes trobades en EiBI com a solucions exactes de la RG per a fonts de mat\`eria ex\`otiques.

Una vegada entesa l'estructura gen\`erica d'aquestes teories, seguim estudiant els aspectes no-trivials del seu espai de solucions al cap\'itol \ref{sec:SolutionsDeformation}, basat en \cite{BeltranJimenez:2020guo}. En aquest cap\'itol vorem que, degut a la naturalesa no-lineal de les equacions que relacionen la m\`etrica original amb la que descriu la teor\'ia a la representaci\'o Einsteniana, tot i que sempre hi ha una soluci\'o d'aquestes equacions en la que la representaci\'o Einsteniana es comporta exactament com la RG en el buit, poden haver-hi altres solucions possibles per a aquesta m\`etrica, donada la forma de la m\`etrica original (o viceversa). Estudiarem el cas particular amb m\`etrica i sector de mat\`eria homogenis i is\'otrops. Vorem com, per a aquest cas, a EiBI sols existeix la soluci\'o trivial, per\`o al cas general poden existir altres branques, mostrant-ho expl\'icitament per al cas quadrátic. Vorem tamb\'e que aquestes branques tenen un comportament en general patol\`ogic, el que ens duu a concloure que la elecci\'o de la soluci\'o trivial, que \'es usual en la literatura, est\`a justificada amb motius f\'isics, tot i que generalment a\c{c}\`o es un punt subtil que no estava ben analitzat. Seguim estudiant solucions particulars de la teor\'ia amb simetria esf\'erica. A aquestes teories apareixen objectes compactes que resolen el problema de la singularitat a nivell cl\`assic, com forats de cuc. Al cap\'itol \ref{sec:Absorption}, basat en \cite{Delhom:2019btt}, estudiarem les propietats de l'espectre d'absorci\'o de pertorbacions escalars per aquestos objectes, trobant resson\`ancies que es pareixen a les trobades a altres objectes compactes ex\`otics (ECOs) degut a la pres\'encia de la gola del forat de cuc. Aquestos resultats enceten l'estudi fenomenol\'ogic de les propietats d'aquestos objectes relacionades amb la interacci\'o amb pertorbacions, el que pot servir com a discriminant d'aquestes teories front a dades d'astronomia de m\'ultiples missatgers.

Al cap\'itol \eqref{sec:UnstableDOF}, basat en \cite{BeltranJimenez:2019acz,Jimenez:2020dpn}, tornem ara a estudiar propietats estructurals i fenomenol\'ogiques de teories m\`etric-afines. Comen\c{c}arem generalitzant les RBG al cas sense simetria projectiva, el que permet incloure la part antisim\`etrica del tensor de Ricci en l'acci\'o de la teor\'ia. a\c{c}\`o ens permetrá la q\"uesti\'o relativament antiga de si el formalisme metricaf\'i \'es lliure de inestabilitats de tipus fantasma degut a la abs\`encia de derivades d'ordre superior als Lagrangians amb correccions no-lineals. Els nostres resultats demostren que, al contrari, les inestabilitats de tipus fantasma ocorren gen\`ericament a teories metricafins. Comen\c{c}arem demostrant a\c{c}\`o expl\' icitament a la sub-familia de teories RBG on vorem que, al trencar la simetria projectiva expl\'icitament, el mode projectiu es torna un grau de llibertat que propaga inestabilitats de tipus fantasma, i vorem com tamb\'e apareix una 2-forma relacionada amb la part antisim\`etrica de la m\`etrica corresponent a la representaci\'o Einsteniana de la teor\'ia que excita inestabilitats de tipus Ostrogradski. Aix\'i vorem que, al trencar la simetria projectiva, apareixen cinc nous graus de llibertat inestables a aquestes teories. Vorem tamb\'e que a\c{c}\`o pot solventar-se afegint restriccions geom\`etriques a les teories.  Per \'ultim, mostrarem com aquestes inestabilitats tamb\'e apareixen al formalisme h\'ibrid, i argumentarem com serán un tret gen\'eric de teories metricafins a no ser que es construixquen expl\'icitament per evitar aquestos problemes.

Tot i que la visi\'o geom\`etrica es la predominant a la literatura metricaf\'i, deur\'iem recordar que aquestes teories poden tamb\'e entendre's des de la perspectiva de teor\'ia de camps, on la nometricitat i la torsi\'o son dos nous camps de mat\`eria que acoblen no-m\'inimament al camp d'esp\'i 2 sense massa. Aquest punt de vista \'es emprat  al llarg de la tesi, i permet un estudi sistem\`atic de les possibilitats del formalisme metricaf\'i per via del formalisme de teories efectives de camps (EFTs). Al cap\'itol \ref{sec:MetricAffineEFT}, basat en \cite{BeltranJimenez:2021iqs}, analitzarem la possibilitat d'incloure les teories de la sub-familia RBG, amb i sense simetria projectiva, al formalisme de les EFTs. Trobarem que, tot i que aquesta sub-familia admet una interpretaci\'o com a teories efectives per baix d'una escala t\'ipica d'altes energies, no casen b\'e amb el formalisme de les EFTs. Despr\'es analitzarem l'estructura d'una teor\'ia gen\`erica metricaf\'i argumentant que, al cas m\'es general, els operadors del Ricci sim\`etric son redundants en el sentit de les EFTs que sols introdueixen noves interaccions entre els graus de llibertat de la teor\'ia, sense introduir-hi nous. Per mostrar a\c{c}o, constru\"irem una repreentaci\'o generalitzada Einsteniana per a teories generals, el que ens permetr\`a tamb\'e mostrar m\'es expl\'icitament com apareixeran inestabilitats de tipus fantasma en teories metricafins gen\'eriques. Des d'aquesta representaci\'o Einsteniana derivem una forma general per al tensor nometricitat que pot expandir-se en termes de l'escala d'altes energies que controla desviacions respecte de la RG. En aquesta expansi\'o perturbativa, veiem que la nometricitat presenta un terme gen\'eric degut a la pres\`encia d'operadors amb la part sim\`etrica del tensor de Ricci a l'acci\'o a banda del terme d'Einstein i Hilbert. Aquestos termes donen lloc a interaccions efectives al sector de mat\`eria que estudiem amb detall al cap\'itol \ref{sec:ObservableTraces}, basat en \cite{Latorre:2017uve,Delhom:2019wir,JoanAdri}. En aquest cap\'itol estudiarem aquestes interaccions quan el sector de mat\`eria est\`a compossat pel Model Est\`andard, i les utilitzarem per obtenir cotes inferiors l'escala d'energia a la que modificacions a la RG de tipus metricaf\'i amb operadors que contenen la part sim\`etrica del tensor de Ricci m\'es enll\`a del terme d'Einstein-Hilberno entren en conflicte amb els experiments. Particularitzarem aquestes interaccions per al cas de teories RBG, obtenint les cotes m\'es restrictives a dia d'avui sobre aquesta classe de teories. Si be aquestos efectes s\'on clarament interpretats com a interaccions efectives des del punt de vista de teor\'ia de camps, des del punt de vista geom\`etric, aquestos poden relacionar-se amb una forma espec\'ifica del tensor nometricitat constituint, des d'aquest punt de vista, els primers efectes observables de la nometricitat que no depenen d'acoblaments espec\'ifics d'aquesta a la mat\`eria. Trobar aquest tipus d'efectes constitu\"ia una de les principals motivacions d'aquesta tesi doctoral al seu comen\c{c}ament. Des dels resultats obtinguts s'obrin noves vies per estudiar aquestes interaccions en detall en teories m\'es enll\`a dels models RBG aix\'i com trobar altre tipus de contribucions a un Lagrangi\`a metricaf\'i gen\'eric que tinguen efectes semblants.

Amb els resultats principals continguts a la segona part, la tercera part de la tesi cont\'e una miscel$\cdot$l\`ania de treballs que generalment est\`an relacionats amb el formalisme metricaf\'i per\`o que no tenen un fort vincle amb l'estudi de propietats gen\'eriques de l'estructura o la fenomenolog\'ia d'aquestes teories. Al cap\'itol \ref{sec:LQC}, basat en \cite{LSU}, estudiem el problema de trobar accions covariants efectives que descriguen la evoluci\'o de solucions cosmol\`ogiques que apareixen en models de quantitzaci\'o a la Loop. En aquest cap\'itol trobem una familia de teories $f(R)$ metricafins que s\'on capaces de dur a terme aquesta descripci\'o per a tres models de cosmologia qu\`antica a la Loop, anomenats com a LQC, mLQC-I i mLQC-II. Particularment, trobem una forma funcional del Lagrangi\`a amb tres par\`ametres tal que, amb una elecci\'o dels valors dels par\`ametres, pot descriure l'evoluci\'o de l'univers en aquestes cosmologies fins a l'escala de Planck, on es dona el rebot cosmol\'ogic. Aquestos resultats suggereixen noves vies d'exploraci\'o d'aquestes families en altres solucions quantitzades a la Loop on s'ha aconseguit millorar el caracter singular de la RG, com espaitemps amb simetria esf\'erica amb solucions de forat negre. Aix\'i mateix, aquestos resultats motiven la possible exploraci\'o de teories metricafins com a l\'imit al continu de la gravetat qu\`antica de lla\c{c}os, motivat per l'analog\'ia amb els cristalls amb defectes a la seva microestructura que en el l\'imit al continu s\'on descrits per varietats amb nometricitat i torsi\'o.

Al cap\'itol \ref{sec:SLSB} estudiem aspectes cl\`assics del fenomen de ruptura espont\`ania de la simetria Lorentz en espaitemps metricafins mitjan\c{c}ant una generalitzaci\'o del model de Bumblebee, originalment formulat al formalisme m\`etric, al formalisme metricaf\'i. Concretament, estudiarem aquesta teor\'ia com un cas de teor\'ia RBG amb simetria projectiva i acoblaments no-m\'inims entre la mat\`eria (el Bumblebee) i la geometria a trav\'es de la part sim\`etrica del tensor de Ricci. En aquesta teor\'ia, el bumblebee, un camp vectorial, adquireix un valor esperat en el buit notrivial degut a un potencial que trenca la simetria Lorentz espontániament. Estudiarem l'estructura de buit de la teor\'ia per a un potencial qu\`artic an\`aleg al del mecanisme de Higgs del Model Est\`andard perturbativament en l'acoblament no-m\'inim entre el bumblebee i al geometria, trobant un buit estable no trivial que trenca Lorentz espont\`aniament en el que el valor esperat del Bumblebee \'es de tipus espacial. Despr\'es estudiem les interaccions efectives de camps escalars i fermi\`onics amb aquesta soluci\'o que trenca espont\`aniament la simetria Lorentz, que indueixen termes efectius de violaci\'o de simetria Lorenz a les respectives equacions de moviment. Com a curiositat, veiem que el valor esperat no-trivial del Bumblebee genera, al cas general, un valor esperat al tensor nometricitat que proporciona una situaci\'o similar a la estudiada en \cite{Foster:2016uui}, on estudiaren efectes de violaci\'o de simetria Lorentz degut a un buit no-trivial del tensor no-metricitat. Aix\'i, aquesta teor\'ia representa el primer model conegut on la nometricitat pren un valor esperat al buit no-trivial, realitzant aquesta situaci\'o de forma din\`amica.

En el pen\'ultim cap\'itol de la tesi, estudiem el problema de trobar una definici\'o invariant d'escala per al temps propi en pres\`encia de nometricitat gen\`erica. La idea de trobar una noci\'o invariant d'escala ve motivada per la possibilitat de que l'univers presente eixa simetria a escales d'energia en l'ultravioleta profund. Els intervals de temps propi mesurat per un observador a la RG venen definits per la longitud espaitemporal de la seva l\'inia d'univers, que no \'es invariant d'escala. Perlick va trobar una definici\'o de temps pr\`opi que respectava la invarian\c{c}a d'escala en espaitemps de tipus Weyl, on la no metricitat pren una forma particularment simple, descrita sols per un vector. Al cap\'itol \ref{sec:GeneralisedTime}, basat en \cite{Delhom:2019yeo,Delhom:2020vpe} estudiem la generalitzaci\'o de la definici\'o donada per Perlick per al cas amb nometricitat gen\`erica, que pot fer-se de manera directa. Estudiem l'adequaci\'o de les propietats d'aquesta definici\'o a la definici\'o d'un temps propi, aix\'i com la relaci\'o amb el temps definit per Ehlers, Pirani, i Schild a \cite{EPS} demanant compatibilitat entre l'estructura conforma donada pels rajos de llum, i l'estructura af\'i donada per les part\'icules lliures massives, trobant les condicions que ha de complir la nometricitat per tal que aquestes dues definicions siguen equivalents.  Finalment mostrem la pres\`encia d'un segon efecte de rellotge per a aquesta definici\'o de temps, comentant sobre la improbable possibilitat de detectar-ho.

A l'\'ultim cap\'itol de la tercera part de la tesi, basat en els treballs \cite{Arrechea:2020gjw,Arrechea:2020evj}, presentem una cr\'itica a diferents aspectes del recent treball \cite{Glavan:2019inb}, on es presenta la teor\'ia Einstein-Gauss-Bonnet quadridimensional (D4EGB). En aquest cap\'itol argumentem que aquesta teor\'ia, tal i com est\`a presentada a \cite{Glavan:2019inb}, no est\`a ben definida, degut a que el {\it l\'imit} que els autors prenen per tal d'obtindre les equacions de moviment no es un proc\'es matem\`aticament ben definit per a estructures tensorials, a no ser que es consideren solucions particulars per a la m\`etrica amb simetria maximal. La conseq\"u\`encia es que les equacions de moviment presentades en \cite{Glavan:2019inb} no estan ben definides fora d'este tipus de m\`etriques. Ho mostrem expl\' icitament considerant pertorbacions de segon ordre al voltant d'un background maximalment simetric Minkowski\`a, i veiem que les equacions per a les pertorbacions no estan ben definides degut a una indeterminaci\'o de tipus $0/0$. Despres mostrem que, si regularitzem les equacions llevant els termes mal definits, aleshores no hi ha acci\'o invariant baix difeomorfismes que depenga sols d'una m\`etrica que done lloc a eixes equacions regularitzades. Finalment demostrem que les m\`etriques esf\'ericament  presentades a \cite{Glavan:2019inb} com a solucions de la teor\'ia no s\'on solucions ni de les equacions que no estan ben definides, ni de les regularitzades, i que, a m\'es, no s\'on geod\`esicament completes com s'afirma a \cite{Glavan:2019inb}. A\c{c}o darrer ho trobem al veure que trajectories en caiguda lliure de part\'icules massives radials arriben a la singularitat de curvatura central en un temps propi finit. Finalment, presentem les conclusions de la tesi amb futures l\'inies d'investigaci\'o obertes arr\`an dels resultats obtinguts.
  
\end{SingleSpace}
\clearpage
\clearemptydoublepage
\backmatter
\refstepcounter{chapter}
\bibliography{Bibliography.bib}
\bibliographystyle{JHEPmodplain}
\clearemptydoublepage
%
% Add index
%\printindex
%   
\end{document}